\documentclass[12pt,twoside]{article}
\usepackage{amsmath}
\usepackage{graphicx}
\usepackage{epsfig}
\usepackage[figuresright]{rotating}
\usepackage{cite}    
\usepackage{pstricks}
\usepackage{mciteplus}
\usepackage{authblk}
\usepackage{caption}
\usepackage{array}
\usepackage{longtable}

\usepackage[dvips]{hyperref}                              
\usepackage[all]{hypcap} 
\usepackage{relsize}                                      
\def\hfurl#1{http://hfag.phys.ntu.edu.tw/b2charm/#1} 
\newcommand{\mysection}[1]{\section{\boldmath #1}}
\newcommand{\mysubsection}[1]{\subsection[#1]{\boldmath #1}}
\newcommand{\mysubsubsection}[1]{\subsubsection[#1]{\boldmath #1}}
\newcommand{\mysubsubsubsection}[1]{\subsubsubsection{\boldmath #1}}

\newcommand{\lesssim}{\ensuremath{\raise-.5ex\hbox{$\buildrel<\over\sim$}\,}} 

\def\dof{{\rm dof}}

\newcommand\VCKM{{V}}
\newcommand\etacpf{{\eta_f}}
\newcommand\etacp{{\eta}}

\renewcommand\Im{{\rm Im}} 
\renewcommand\Re{{\rm Re}}

\newcommand\Abar{\kern 0.18em\overline{\kern -0.18em A}{}}
\newcommand\Af{A_f}
\newcommand\Abarf{\Abar_f}
\newcommand\Afbar{A_{\bar f}}
\newcommand\Abarfbar{\Abar_{\bar f}}
\newcommand\Acp{{\cal A}}
\newcommand\Adirnoncp{\ensuremath{\langle{\cal A}_{f\bar f}\rangle}\xspace}
\newcommand\mc{\multicolumn}

\newcommand {\cbf}{\ensuremath{{\cal B}}}

\newcommand {\vcb}{\ensuremath{|V_{cb}|}}
\newcommand {\vub}{\ensuremath{|V_{ub}|}}

\def\Bp      {\ensuremath{B^{+}}}
\def\Bm      {\ensuremath{B^{-}}}
\def\Bz      {\ensuremath{B^{0}}}
\def\Bs      {\ensuremath{B_{s}}}

\newcommand{\BzbDplnu}    {\ensuremath{\Bzb \to D^{+}\ell^{-}\nub}}
\newcommand{\BzbDstarlnu} {\ensuremath{\Bzb \to D^{*+}\ell^{-}\nub}}

\newcommand {\rhoz} {\ensuremath{\rho^0}\hbox{ }}

\usepackage{relsize}

\def\beq{\begin{equation}}
\def\eeq#1{\label{#1}\end{equation}}
\def\eeqn{\end{equation}}

\def\beqa{\begin{eqnarray}}
\def\eeqa#1{\label{#1}\end{eqnarray}}
\def\eeqan{\end{eqnarray}}

\let\bar=\overbar

\def\etal{{\it et al.}}
\def\ie{{\it i.e.}}
\def\eg{{\it e.g.}}
\def\etc{{\it etc.}}
\def\cf{{\it cf.}}

\def\D{{\cal D}}

\def\Dslash{\ensuremath{\not{\hbox{\kern-4pt $D$}}}\xspace}
\def\dslash{\not{\hbox{\kern-2pt $\del$}}}

\def\BR{\mbox{\rm BR}}
\def\ee{e^+e^-}

\def\alphas{\alpha_s}
\def\msb{{\bar{\ssstyle M \kern -1pt S}}}

\RequirePackage{xspace}

\usepackage{relsize}
\def\babar{\mbox{\slshape B\kern-0.1em{\smaller A}\kern-0.1em
    B\kern-0.1em{\smaller A\kern-0.2em R}}\xspace}
\def\belle{\mbox{\normalfont Belle}\xspace}

\def\dzero{\mbox{\normalfont D0}\xspace} 

\def\ee         {\ensuremath{e^-e^-}\xspace}

\def\mumu       {\ensuremath{\mu^+\mu^-}\xspace}

\def\nub        {\ensuremath{\overline{\nu}}\xspace}

\def\nub        {\ensuremath{\overline{\nu}}\xspace}

\def\nul        {\ensuremath{\nu_\ell}\xspace}

\def\Z      {\ensuremath{Z^0}\xspace}

\def\ubar  {\ensuremath{\overline u}\xspace}

\def\dbar  {\ensuremath{\overline d}\xspace}
\def\ddbar {\ensuremath{d\overline d}\xspace}

\def\sbar  {\ensuremath{\overline s}\xspace}

\def\b  {\ensuremath{b}\xspace}
\def\bbar  {\ensuremath{\overline b}\xspace}

\def\piz   {\ensuremath{\pi^0}\xspace}

\def\pip   {\ensuremath{\pi^+}\xspace}
\def\pim   {\ensuremath{\pi^-}\xspace}
\def\pipi  {\ensuremath{\pi^+\pi^-}\xspace}

\def\etapr {\ensuremath{\eta^{\prime}}\xspace}

\def\Kbar  {\kern 0.2em\overline{\kern -0.2em K}{}\xspace}

\def\Kpm   {\ensuremath{K^\pm}\xspace}
\def\Kmp   {\ensuremath{K^\mp}\xspace}
\def\Kp    {\ensuremath{K^+}\xspace}
\def\Km    {\ensuremath{K^-}\xspace}
\def\KS    {\ensuremath{K^0_{\scriptscriptstyle S}}\xspace} 
\def\KL    {\ensuremath{K^0_{\scriptscriptstyle L}}\xspace}

\def\Kstar   {\ensuremath{K^*}\xspace}

\def\Kstarpm   {\ensuremath{K^{*\pm}}\xspace}
\def\Kstarmp   {\ensuremath{K^{*\mp}}\xspace}
\def\Kz   {\ensuremath{K^0}\xspace}
\def\Kzb   {\ensuremath{\Kbar^0}\xspace}
\def\KzKzb {\ensuremath{K^0 \kern -0.16em \Kzb}\xspace}

\def\KorKstarpm {\ensuremath{K^{(*)\pm}}\xspace}

\def\Dz    {\ensuremath{D^0}\xspace}
\def\Dbar  {\kern 0.2em\overline{\kern -0.2em D}{}\xspace}

\def\Dzb   {\ensuremath{\Dbar^0}\xspace}
\def\DzDzb {\ensuremath{D^0 {\kern -0.16em \Dzb}}\xspace}

\def\Dstar   {\ensuremath{D^*}\xspace}

\def\Dstarp  {\ensuremath{D^{*+}}}
\def\Dstarm  {\ensuremath{D^{*-}}}
\def\DorDstar   {\ensuremath{D^{(*)}}\xspace}
\def\DorDstarz  {\ensuremath{D^{(*)0}}\xspace}
\def\DorDstarzb {\ensuremath{\Dbar^{(*)0}}\xspace}

\def\Ds    {\ensuremath{D^+_s}\xspace}

\def\Bz    {\ensuremath{B^0}\xspace}
\def\B     {\ensuremath{B}\xspace}
\def\Bbar  {\kern 0.18em\overline{\kern -0.18em B}{}\xspace}
\def\Bb    {\ensuremath{\Bbar}\xspace}
\def\Bzb   {\ensuremath{\Bbar^0}\xspace}
\def\Bu    {\ensuremath{B^+}\xspace}

\def\Bpm   {\ensuremath{B^\pm}\xspace}
\def\Bmp   {\ensuremath{B^\mp}\xspace}
\def\Bs    {\ensuremath{B_s}\xspace}
\def\Bsb   {\ensuremath{\Bbar_s}\xspace}
\def\BB    {\ensuremath{B\Bbar}\xspace} 
\def\BzBzb {\ensuremath{B^0 {\kern -0.16em \Bzb}}\xspace}

\def\jpsi  {\ensuremath{{J\mskip -3mu/\mskip -2mu\psi\mskip 2mu}}\xspace}

\mathchardef\Upsilon="7107
\def\Y#1S{\ensuremath{\Upsilon{(#1S)}}\xspace}

\mathchardef\Deltares="7101
\mathchardef\Xi="7104
\mathchardef\Lambda="7103
\mathchardef\Sigma="7106
\mathchardef\Omega="710A
\def\Deltabar   {\kern 0.25em\overline{\kern -0.25em \Deltares}{}\xspace}
\def\Lbar {\kern 0.2em\overline{\kern -0.2em\Lambda\kern 0.05em}\kern-0.05em{}\xspace}
\def\Sigbar{\kern 0.2em\overline{\kern -0.2em \Sigma}{}\xspace}
\def\Xibar{\kern 0.2em\overline{\kern -0.2em \Xi}{}\xspace}
\def\Obar{\kern 0.2em\overline{\kern -0.2em \Omega}{}\xspace}
\def\Nbar{\kern 0.2em\overline{\kern -0.2em N}{}\xspace}
\def\Xb{\kern 0.2em\overline{\kern -0.2em X}{}}

\def\BR{{\ensuremath{\cal B}}}

\newcommand{\tev}{\ensuremath{\mathrm{Te\kern -0.1em V}}\xspace}
\newcommand{\gev}{\ensuremath{\mathrm{Ge\kern -0.1em V}}\xspace}
\newcommand{\mev}{\ensuremath{\mathrm{Me\kern -0.1em V}}\xspace}
\newcommand{\kev}{\ensuremath{\mathrm{ke\kern -0.1em V}}\xspace}
\newcommand{\ev}{\ensuremath{\mathrm{e\kern -0.1em V}}\xspace}
\newcommand{\gevc}{\ensuremath{{\mathrm{Ge\kern -0.1em V\!/}c}}\xspace}
\newcommand{\mevc}{\ensuremath{{\mathrm{Me\kern -0.1em V\!/}c}}\xspace}
\newcommand{\gevcc}{\ensuremath{{\mathrm{Ge\kern -0.1em V\!/}c^2}}\xspace}
\newcommand{\mevcc}{\ensuremath{{\mathrm{Me\kern -0.1em V\!/}c^2}}\xspace}

\def\pb {\ensuremath{\rm \,pb}\xspace}

\def\fb   {\ensuremath{\mbox{\,fb}}\xspace}
\def\invfb   {\ensuremath{\mbox{\,fb}^{-1}}\xspace}
\def\mus  {\ensuremath{\rm \,\mus}\xspace}

\def\ps   {\ensuremath{\rm \,ps}\xspace}

\def\mus        {\ensuremath{\,\mu{\rm s}}\xspace}    
\def\ps         {\ensuremath{{\rm \,ps}}\xspace}  
\def\gsim{{~\raise.15em\hbox{$>$}\kern-.85em
          \lower.35em\hbox{$\sim$}~}\xspace}
\def\lsim{{~\raise.15em\hbox{$<$}\kern-.85em
          \lower.35em\hbox{$\sim$}~}\xspace}

\def\CP                 {\ensuremath{C\!P}\xspace}
\def\CPT                {\ensuremath{C\!PT}\xspace}
\def\ra                 {\ensuremath{\to}\xspace}

\def\pep2{PEP-II}

\def\rhobar {\ensuremath{\overline{\rho}}\xspace}
\def\etabar {\ensuremath{\overline{\eta}}\xspace}

\def\Vus  {\ensuremath{|V_{us}|}\xspace}

\def\Vub  {\ensuremath{|V_{ub}|}\xspace}

\def\stwob{\ensuremath{\sin\! 2 \beta   }\xspace}

\def\deltamd{\ensuremath{{\rm \Delta}m_d}\xspace}

\xspace
\newcommand{\fds}{\ensuremath{f_{D_s}}\xspace}

\def\jetset74   {\mbox{\tt Jetset \hspace{-0.5em}7.\hspace{-0.2em}4}}

\newcommand{\aerr}[4]   {\mbox{${{#1}^{+ #2}_{- #3}\pm #4}$}}
\newcommand{\berr}[4]   {\mbox{${{#1}\pm #2^{+ #3}_{- #4}}$}}
\newcommand{\cerr}[3]   {\mbox{${{#1}^{+ #2}_{- #3}}$}}
\newcommand{\aerrsy}[5] {\mbox{${{#1}^{+ #2 + #4}_{- #3 - #5}}$}}

\newcommand{\err}[3]   {\mbox{${{#1}\pm{#2}\pm{#3}}$}}

\newcommand{\nodata}{$$}
\newcommand{\vs}{\mbox{$vs.$}}
\def\etapr{{\eta^{\prime}}}

\def\sgline{\noalign{\vskip 0.10truecm\hrule\vskip 0.10truecm}}
\def\sglinespt{\noalign{\vskip 0.05truecm\hrule}}
\def\sglinespb{\noalign{\hrule\vskip 0.05truecm}}

\newcommand{\kz}    {\mbox{$K^0$}}

\newcommand{\RPP}{}

\renewcommand{\mysection}[1]{\section[#1]{#1}} 

\begin{document}

\setcounter{page}{1}
\thispagestyle{empty}
\renewcommand\Affilfont{\itshape\small} 

\title{
  Averages of $b$-hadron, $c$-hadron, and $\tau$-lepton properties 
  as of early 2012
\vskip0.20in
\large{\it Heavy Flavor Averaging Group (HFAG):}
\vspace*{-0.20in}}
\author[1]{Y.~Amhis}\affil[1]{LAL, Universit\'{e} Paris-Sud, France}
\author[2]{Sw.~Banerjee}\affil[2]{University of Victoria, Canada}
\author[3]{R.~Bernhard}\affil[3]{University of Freiburg, Germany}
\author[4]{S.~Blyth}\affil[4]{National United University, Taiwan}
\author[5]{A.~Bozek}\affil[5]{H. Niewodniczanski Institute of Nuclear Physics, Krakow, Poland}
\author[6]{C.~Bozzi}\affil[6]{INFN Ferrara, Italy} 
\author[7,8]{A.~Carbone}\affil[7]{INFN Bologna, Italy}\affil[8]{Universit\'{a} di Bologna, Italy}
\author[9]{A.~Oyanguren Campos}\affil[9]{IFIC, University of Valencia, Spain}
\author[10]{R.~Chistov}\affil[10]{Institute for Theoretical and Experimental Physics, Russia}
\author[6]{G.~Cibinetto} 
\author[11]{J.~Coleman} \affil[11]{University of Liverpool, UK}
\author[12]{J.~Dingfelder}\affil[12]{Bonn University, Germany}
\author[13]{W.~Dungel}\affil[13]{Austrian Academy of Sciences, Austria}
\author[14]{M.~Gersabeck}\affil[14]{European Organization for Nuclear Research (CERN), Switzerland}
\author[14,15]{T.~J.~Gershon}\affil[15]{University of Warwick, UK}
\author[16]{L.~Gibbons}\affil[16]{Cornell University, USA}
\author[17]{B.~Golob}\affil[17]{University of Ljubljana, Slovenia}
\author[18]{R.~Harr}\affil[18]{Wayne State University, USA}
\author[19]{K.~Hayasaka} \affil[19]{Nagoya University, Japan}
\author[20]{H.~Hayashii} \affil[20]{Nara Women's University, Japan}
\author[21]{O.~Leroy}\affil[21]{CPPM, Aix-Marseille Universit\'{e}, CNRS/IN2P3, Marseille, France}
\author[22]{D.~Lopes Pegna}\affil[22]{Princeton University, USA}
\author[23]{R.~Louvot}\affil[23]{Ecole Polytechnique F\'{e}d\'{e}rale de Lausanne (EPFL), Switzerland}

\author[24]{A.~Lusiani} \affil[24]{Scuola Normale Superiore and INFN, Pisa, Italy}
\author[25]{V.~L\"{u}th}\affil[25]{SLAC National Accelerator Laboratory, USA}
\author[26]{B.~Meadows} \affil[26]{University of Cincinnati, USA}
\author[27]{S.~Nishida}\affil[27]{KEK, Tsukuba, Japan}
\author[28]{M.~Patel}\affil[28]{Imperial College London, UK}
\author[29]{D.~Pedrini} \affil[29]{INFN Milano-Bicocca, Italy}
\author[30]{M.~Rama}\affil[30]{INFN Frascati, Italy}
\author[2]{M.~Roney}
\author[31]{M.~Rotondo}\affil[31]{INFN Padova, Italy}
\author[23]{O.~Schneider}
\author[13]{C.~Schwanda}
\author[25]{A.~J.~Schwartz}
\author[32]{B.~Shwartz}\affil[32]{Budker Institute of Nuclear Physics, Russia}
\author[33]{J.~G.~Smith}\affil[33]{University of Colorado, USA}
\author[34]{R.~Tesarek}\affil[34]{Fermilab, USA}
\author[14,34]{D.~Tonelli}
\author[26]{K.~Trabelsi}
\author[12]{P.~Urquijo}
\author[35]{R.~Van Kooten}\affil[35]{Indiana University, USA}

\date{May 9, 2013} 
\maketitle

\begin{abstract}
This article reports world averages of measurements of $b$-hadron, $c$-hadron, 
and $\tau$-lepton properties obtained by the Heavy Flavor Averaging Group (HFAG) 
using results available through the end of 2011. 
In some cases results available in the early part of 2012 are included.
For the averaging, common input parameters used in the various analyses 
are adjusted (rescaled) to common values, and known correlations are taken 
into account. 
The averages include branching fractions, lifetimes, 
neutral meson mixing parameters, \CP~violation parameters,
parameters of semileptonic decays and CKM matrix elements.
\end{abstract}

\newpage
\tableofcontents
\newpage


\mysection{Introduction}
\label{sec:intro}

Flavor dynamics is an important element in understanding the nature of
particle physics.  The accurate knowledge of properties of heavy flavor
hadrons, especially $b$ hadrons, plays an essential role for
determining the elements of the Cabibbo-Kobayashi-Maskawa (CKM)
weak-mixing matrix~\cite{Cabibbo:1963yz,Kobayashi:1973fv}. 
The operation of the \belle\ and \babar\ $e^+e^-$ $B$ factory 
experiments led to a large increase in the size of available 
$B$ meson, $D$ hadron and $\tau$ lepton samples, 
enabling dramatic improvement in the accuracies of related measurements.
The CDF and \dzero\ experiments at the Fermilab Tevatron 
have also provided important results in heavy flavor physics,
most notably in the $B^0_s$ sector.
The CERN Large Hadron Collider is now delivering high luminosity, 
enabling the collection of even higher statistics samples of $b$ 
and $c$ hadrons at the ATLAS, CMS, and (especially) LHCb experiments.
 
The Heavy Flavor Averaging Group (HFAG) was formed in 2002 to 
continue the activities of the LEP Heavy Flavor Steering 
group~\cite{Abbaneo:2000ej_mod,*Abbaneo:2001bv_mod_cont}. 
This group was responsible for calculating averages of 
measurements of $b$-flavor related quantities. HFAG has evolved 
since its inception and currently consists of seven subgroups:
\begin{itemize}
\item the ``$B$ Lifetime and Oscillations'' subgroup provides 
averages for $b$-hadron lifetimes, $b$-hadron fractions in 
$\Upsilon(4S)$ decay and $p\bar{p}$ collisions, and various 
parameters governing $B^0$-$\Bzb$ and $B_s^0$-$\Bsb^0$ mixing;

\item the ``Unitarity Triangle Parameters'' subgroup provides
averages for time-dependent $\CP$ asymmetry parameters and 
resulting determinations of the angles of the CKM unitarity triangle;

\item the ``Semileptonic $B$ Decays'' subgroup provides averages
for inclusive and exclusive $B$-decay branching fractions, and
subsequent determinations of the CKM matrix elements 
$|V_{cb}|$ and $|V_{ub}|$;

\item the ``$B$ to Charm Decays'' subgroup provides averages of 
branching fractions for $B$ decays to final states involving open 
charm or charmonium mesons;

\item the ``Rare Decays'' subgroup provides averages of branching 
fractions and $\CP$ asymmetries for charmless, radiative, 
leptonic, and baryonic $B$ meson decays;

\item the ``Charm Physics'' subgroup provides averages of branching 
fractions for $D$ meson hadronic and semileptonic decays, 
averages of $D^0$-$\Dzb$ mixing and $\CP$ and $T$ violation parameters, 
and an average value for the $D^{}_s$ decay constant~$f^{}_{D_s}$.
The subgroup also documents properties of charm baryons, and upper 
limits for rare and forbidden $D^0$, $D^+_{(s)}$, and $\Lambda_c^+$ 
decays.

\item the ``Tau Physics'' subgroup provides documentation and
averages for the $\tau$ lepton branching fractions
and the resulting determination of the CKM matrix element $|V_{us}|$,
and documents upper limits for $\tau$ lepton-flavor-violating decays.
\end{itemize}

The ``Lifetime and Oscillations'' and ``Semileptonic'' subgroups were formed from the merger of four LEP working groups.
The ``Unitary Triangle,'' ``$B$ to Charm Decays,'' and ``Rare Decays''
subgroups were formed to provide averages for new results obtained
from the $B$ factory experiments (and now also from the Fermilab 
Tevatron and CERN LHC experiments).
The ``Charm'' and ``Tau''  subgroups were formed more recently in 
response to the wealth of new data concerning $D$ and $\tau$ decays. 
Subgroup typically include representatives from \belle\ and \babar\ and, 
when relevant, CLEO, CDF, \dzero\ and LHCb. 

This article is an update of the last HFAG preprint,
which used results available at least through the end of 2009~\cite{Asner:2010qj}. 
Here we report world averages using results available at least through
the end of 2011. 
In some cases results available in the early part of 2012 have been
included.\footnote{
  The precise cut-off date for including results in the averages varies 
  between subgroups.}
In general, we use all publicly available results that have written documentation. 
These include preliminary results presented at conferences or workshops.
However, we do not use preliminary results that remain unpublished 
for an extended period of time, or for which no publication is planned. 
Close contacts have been established between representatives from
the experiments and members of subgroups that perform averaging 
to ensure that the data are prepared in a form suitable for 
combinations.  

In the case of obtaining a world average for which $\chi^2/\dof > 1$,
where $\dof$ is the number of degrees of freedom in the average
calculation, we do not scale the resulting error, as is presently 
done by the Particle Data Group~\cite{PDG_2010}. 
Rather, 
we examine the systematics of each measurement to better understand them. 
Unless we find possible systematic discrepancies between the measurements, 
we do not apply any additional correction to the calculated error. 
We provide the confidence level of the fit as an indicator for the 
consistency of the measurements included in the average. In case some
special treatment was necessary to calculate an average, or if an
approximation used in an average calculation might not be 
sufficiently accurate 
(\eg, assuming Gaussian errors when the likelihood function indicates 
non-Gaussian behavior), we include a warning message.

Chapter~\ref{sec:method} describes the methodology used for calculating
averages. In the averaging procedure, common input parameters used in 
the various analyses are adjusted (rescaled) to common values, and, 
where possible, known correlations are taken into account. 
Chapters~\ref{sec:life_mix}--\ref{sec:tau} present world 
average values from each of the subgroups listed above. 
A brief 
summary of the averages presented is given in Chapter~\ref{sec:summary}.   
A complete listing of the averages and plots,
including updates since this document was prepared,
are also available on the HFAG web site:
\vskip0.15in\hskip0.75in
\vbox{
  \href{http://www.slac.stanford.edu/xorg/hfag}{\tt http://www.slac.stanford.edu/xorg/hfag} 
}

\section{Methodology } 
\label{sec:method} 

The general averaging problem that HFAG faces is to combine 
information provided by different measurements of the same parameter
to obtain our best estimate of the parameter's value and
uncertainty. The methodology described here focuses on the problems of
combining measurements performed with different systematic assumptions
and with potentially-correlated systematic uncertainties. Our methodology
relies on the close involvement of the people performing the
measurements in the averaging process.

Consider two hypothetical measurements of a parameter $x$, which might
be summarized as
\begin{align*}
x &= x_1 \pm \delta x_1 \pm \Delta x_{1,1} \pm \Delta x_{2,1} \ldots \\
x &= x_2 \pm \delta x_2 \pm \Delta x_{1,2} \pm \Delta x_{2,2} \ldots
\; ,
\end{align*}
where the $\delta x_k$ are statistical uncertainties, and
the $\Delta x_{i,k}$ are contributions to the systematic
uncertainty. One popular approach is to combine statistical and
systematic uncertainties in quadrature
\begin{align*}
x &= x_1 \pm \left(\delta x_1 \oplus \Delta x_{1,1} \oplus \Delta
x_{2,1} \oplus \ldots\right) \\
x &= x_2 \pm \left(\delta x_2 \oplus \Delta x_{1,2} \oplus \Delta
x_{2,2} \oplus \ldots\right)
\end{align*}
and then perform a weighted average of $x_1$ and $x_2$, using their
combined uncertainties, as if they were independent. This approach
suffers from two potential problems that we attempt to address. First,
the values of the $x_k$ may have been obtained using different
systematic assumptions. For example, different values of the \Bz
lifetime may have been assumed in separate measurements of the
oscillation frequency $\deltamd$. The second potential problem is that
some contributions of the systematic uncertainty may be correlated
between experiments. For example, separate measurements of $\deltamd$
may both depend on an assumed Monte-Carlo branching fraction used to
model a common background.

The problems mentioned above are related since, ideally, any quantity $y_i$
that $x_k$ depends on has a corresponding contribution $\Delta x_{i,k}$ to the
systematic error which reflects the uncertainty $\Delta y_i$ on $y_i$
itself. We assume that this is the case and use the values of $y_i$ and
$\Delta y_i$ assumed by each measurement explicitly in our
averaging (we refer to these values as $y_{i,k}$ and $\Delta y_{i,k}$
below). Furthermore, since we do not lump all the systematics
together,
we require that each measurement used in an average have a consistent
definition of the various contributions to the systematic uncertainty.
Different analyses often use different decompositions of their systematic
uncertainties, so achieving consistent definitions for any potentially
correlated contributions requires close coordination between HFAG and
the experiments. In some cases, a group of
systematic uncertainties must be combined to obtain a coarser
description that is consistent between measurements. Systematic uncertainties
that are uncorrelated with any other sources of uncertainty appearing
in an average are lumped together with the statistical error, so that the only
systematic uncertainties treated explicitly are those that are
correlated with at least one other measurement via a consistently-defined
external parameter $y_i$. When asymmetric statistical or systematic
uncertainties are quoted, we symmetrize them since our combination
method implicitly assumes parabolic likelihoods for each measurement.

\begin{figure}
\begin{center}
\includegraphics[width=6.0in]{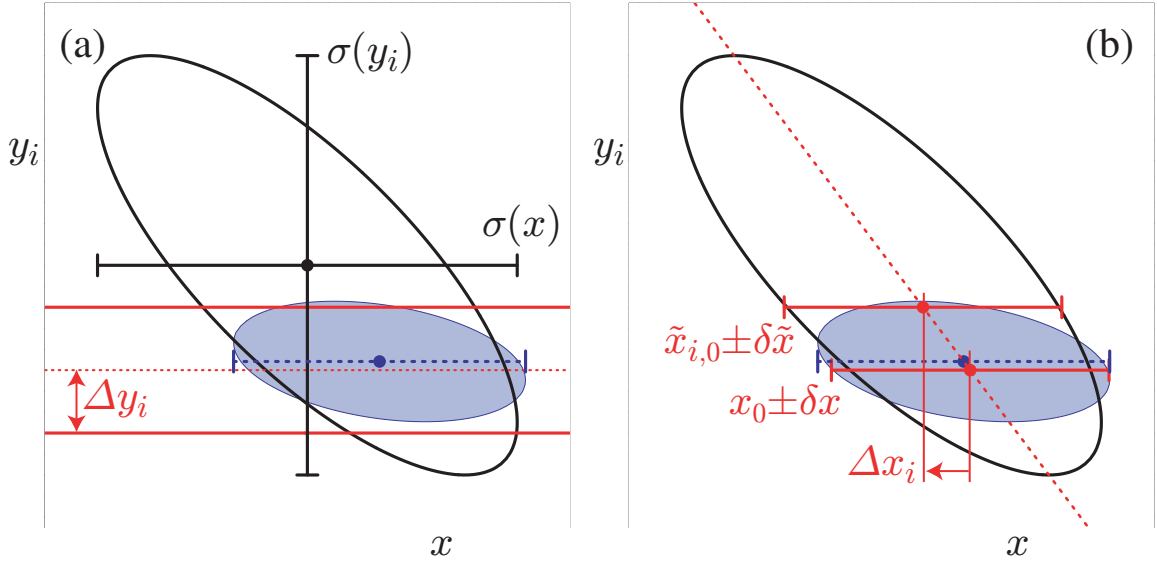}
\end{center}
\caption{The left-hand plot (a) compares the 68\% confidence-level
  contours of a
  hypothetical measurement's unconstrained (large ellipse) and
  constrained (filled ellipse) likelihoods, using the Gaussian
  constraint on $y_i$ represented by the horizontal band. The solid
  error bars represent the statistical uncertainties $\sigma(x)$ and
  $\sigma(y_i)$ of the unconstrained likelihood. The dashed
  error bar shows the statistical error on $x$ from a
  constrained simultaneous fit to $x$ and $y_i$. The right-hand plot
  (b) illustrates the method described in the text of performing fits
  to $x$ with $y_i$ fixed at different values. The dashed
  diagonal line between these fit results has the slope
  $\rho(x,y_i)\sigma(y_i)/\sigma(x)$ in the limit of a parabolic
  unconstrained likelihood. The result of the constrained simultaneous
  fit from (a) is shown as a dashed error bar on $x$.}
\label{fig:singlefit}
\end{figure}

The fact that a measurement of $x$ is sensitive to the value of $y_i$
indicates that, in principle, the data used to measure $x$ could
equally-well be used for a simultaneous measurement of $x$ and $y_i$, as
illustrated by the large contour in Fig.~\ref{fig:singlefit}(a) for a hypothetical
measurement. However, we often have an external constraint $\Delta
y_i$ on the value of $y_i$ (represented by the horizontal band in
Fig.~\ref{fig:singlefit}(a)) that is more precise than the constraint
$\sigma(y_i)$ from
our data alone. Ideally, in such cases we would perform a simultaneous
fit to $x$ and $y_i$, including the external constraint, obtaining the
filled $(x,y)$ contour and corresponding dashed one-dimensional estimate of
$x$ shown in Fig.~\ref{fig:singlefit}(a). Throughout, we assume that
the external constraint $\Delta y_i$ on $y_i$ is Gaussian.

In practice, the added technical complexity of a constrained fit with
extra free parameters is not justified by the small increase in
sensitivity, as long as the external constraints $\Delta y_i$ are
sufficiently precise when compared with the sensitivities $\sigma(y_i)$
to each $y_i$ of the data alone. Instead, the usual procedure adopted
by the experiments is to perform a baseline fit with all $y_i$ fixed
to nominal values $y_{i,0}$, obtaining $x = x_0 \pm \delta
x$. This baseline fit neglects the uncertainty due to $\Delta y_i$, but
this error can be mostly recovered by repeating the fit separately for
each external parameter $y_i$ with its value fixed at $y_i = y_{i,0} +
\Delta y_i$ to obtain $x = \tilde{x}_{i,0} \pm \delta\tilde{x}$, as
illustrated in Fig.~\ref{fig:singlefit}(b). The absolute shift,
$|\tilde{x}_{i,0} - x_0|$, in the central value of $x$ is what the
experiments usually quote as their systematic uncertainty $\Delta x_i$
on $x$ due to the unknown value of $y_i$. Our procedure requires that
we know not only the magnitude of this shift but also its sign. In the
limit that the unconstrained data is represented by a parabolic
likelihood, the signed shift is given by
\begin{equation}
\Delta x_i = \rho(x,y_i)\frac{\sigma(x)}{\sigma(y_i)}\,\Delta y_i \;,
\end{equation}
where $\sigma(x)$ and $\rho(x,y_i)$ are the statistical uncertainty on
$x$ and the correlation between $x$ and
$y_i$ in the unconstrained data.
While our procedure is not
equivalent to the constrained fit with extra parameters, it yields (in
the limit of a parabolic unconstrained likelihood) a central value
$x_0$ that agrees 
to ${\cal O}(\Delta y_i/\sigma(y_i))^2$ and an uncertainty $\delta x
\oplus \Delta x_i$ that agrees to ${\cal O}(\Delta y_i/\sigma(y_i))^4$.

In order to combine two or more measurements that share systematics
due to the same external parameters $y_i$, we would ideally perform a
constrained simultaneous fit of all data samples to obtain values of
$x$ and each $y_i$, being careful to only apply the constraint on each
$y_i$ once. This is not practical since we generally do not have
sufficient information to reconstruct the unconstrained likelihoods
corresponding to each measurement. Instead, we perform the two-step
approximate procedure described below.

\begin{figure}
\begin{center}
\includegraphics[width=6.0in]{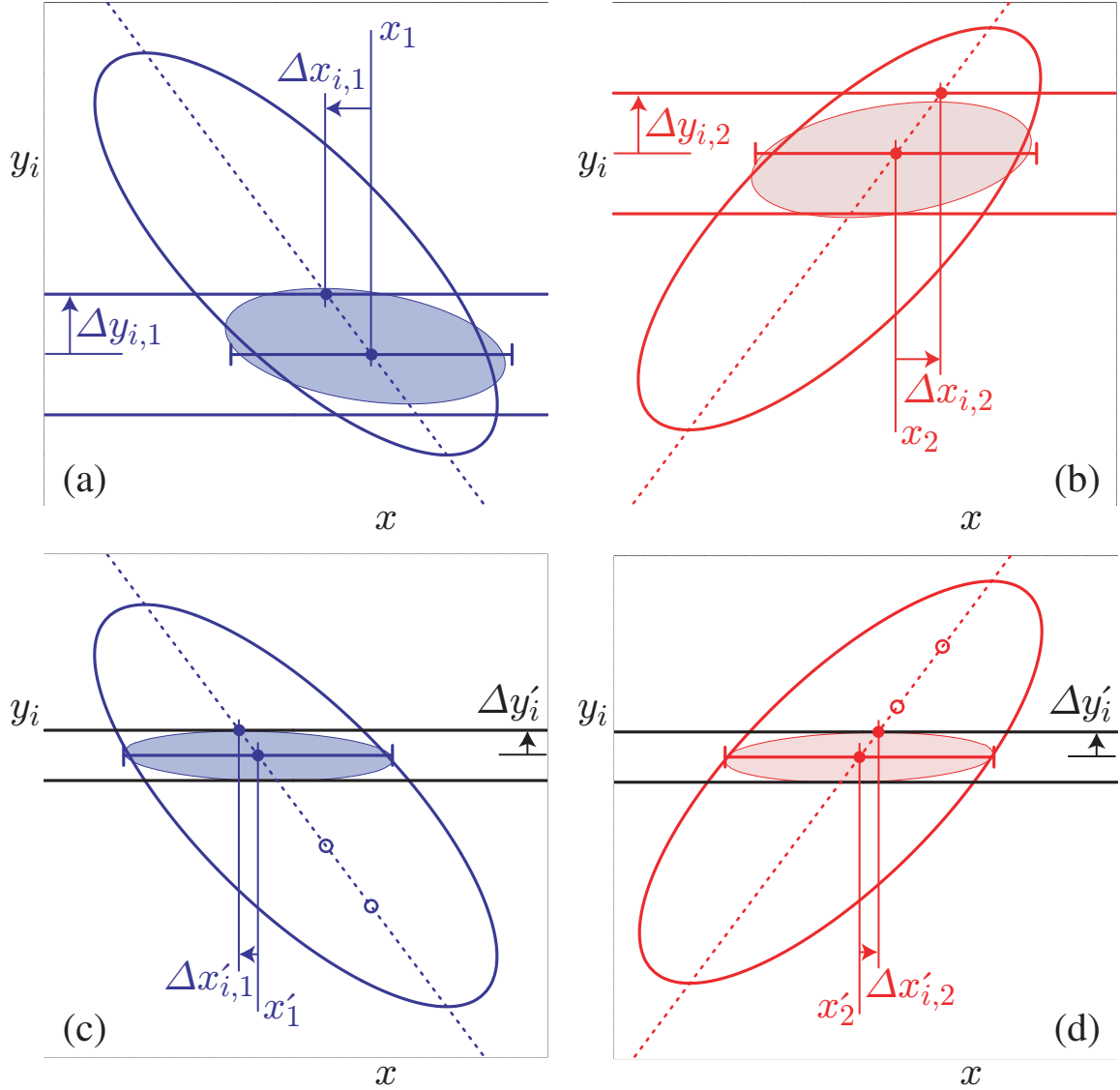}
\end{center}
\caption{The upper plots (a) and (b) show examples of two individual
  measurements to be combined. The large ellipses represent their
  unconstrained likelihoods, and the filled ellipses represent their
  constrained likelihoods. Horizontal bands indicate the different
  assumptions about the value and uncertainty of $y_i$ used by each
  measurement. The error bars show the results of the approximate
  method described in the text for obtaining $x$ by performing fits
  with $y_i$ fixed to different values. The lower plots (c) and (d)
  illustrate the adjustments to accommodate updated and consistent
  knowledge of $y_i$ as described in the text. Open circles mark the
  central values of the unadjusted fits to $x$ with $y$ fixed; these
  determine the dashed line used to obtain the adjusted values. }
\label{fig:multifit}
\end{figure}

Figs.~\ref{fig:multifit}(a,b) illustrate two
statistically-independent measurements, $x_1 \pm (\delta x_1 \oplus
\Delta x_{i,1})$ and $x_2\pm(\delta x_i\oplus \Delta x_{i,2})$, of the same
hypothetical quantity $x$ (for simplicity, we only show the
contribution of a single correlated systematic due to an external
parameter $y_i$). As our knowledge of the external parameters $y_i$
evolves, it is natural that the different measurements of $x$ will
assume different nominal values and ranges for each $y_i$. The first
step of our procedure is to adjust the values of each measurement to
reflect the current best knowledge of the values $y_i'$ and ranges
$\Delta y_i'$ of the external parameters $y_i$, as illustrated in
Figs.~\ref{fig:multifit}(c,b). We adjust the
central values $x_k$ and correlated systematic uncertainties $\Delta
x_{i,k}$ linearly for each measurement (indexed by $k$) and each
external parameter (indexed by $i$):
\begin{align}
x_k' &= x_k + \sum_i\,\frac{\Delta x_{i,k}}{\Delta y_{i,k}}
\left(y_i'-y_{i,k}\right)\\
\Delta x_{i,k}'&= \Delta x_{i,k}\cdot \frac{\Delta y_i'}{\Delta
  y_{i,k}} \; .
\end{align}
This procedure is exact in the limit that the unconstrained
likelihoods of each measurement is parabolic.

\begin{figure}
\begin{center}
\includegraphics[width=3.5in]{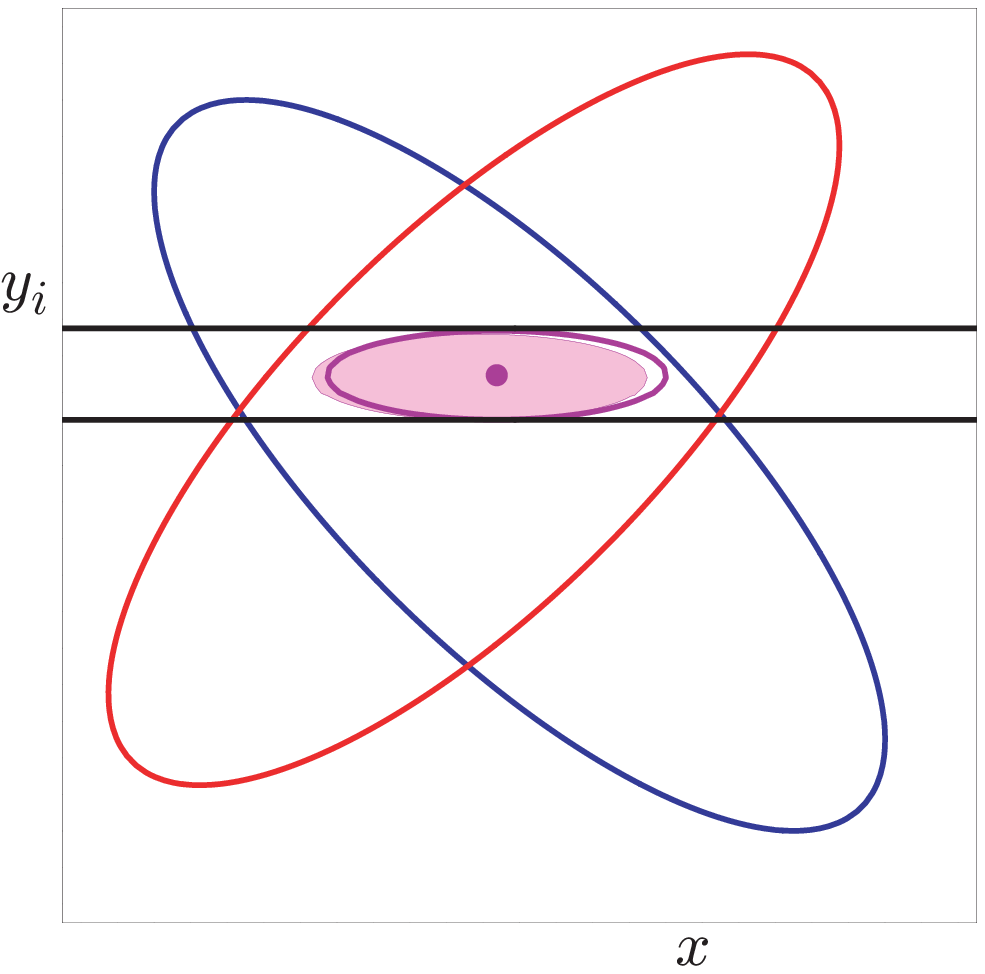}
\end{center}
\caption{An illustration of the combination of two hypothetical
  measurements of $x$ using the method described in the text. The
  ellipses represent the unconstrained likelihoods of each measurement,
  and the horizontal band represents the latest knowledge about $y_i$ 
  that is used to adjust the individual measurements. The filled small
  ellipse shows the result of the exact method using 
  ${\cal L}_{\text{comb}}$, and the hollow small ellipse and dot show 
  the result of the approximate method using $\chi^2_{\text{comb}}$.}
\label{fig:fit12}
\end{figure}

The second step of our procedure is to combine the adjusted
measurements, $x_k'\pm (\delta x_k\oplus \Delta x_{k,1}'\oplus \Delta
x_{k,2}'\oplus\ldots)$ using the chi-square 
\begin{equation}
\chi^2_{\text{comb}}(x,y_1,y_2,\ldots) \equiv \sum_k\,
\frac{1}{\delta x_k^2}\left[
x_k' - \left(x + \sum_i\,(y_i-y_i')\frac{\Delta x_{i,k}'}{\Delta y_i'}\right)
\right]^2 + \sum_i\,
\left(\frac{y_i - y_i'}{\Delta y_i'}\right)^2 \; ,
\end{equation}
and then minimize this $\chi^2$ to obtain the best values of $x$ and
$y_i$ and their uncertainties, as illustrated in
Fig.~\ref{fig:fit12}. Although this method determines new values for
the $y_i$, we do not report them since the $\Delta x_{i,k}$ reported
by each experiment are generally not intended for this purpose (for
example, they may represent a conservative upper limit rather than a
true reflection of a 68\% confidence level).

For comparison, the exact method we would
perform if we had the unconstrained likelihoods ${\cal L}_k(x,y_1,y_2,\ldots)$
available for each
measurement is to minimize the simultaneous constrained likelihood
\begin{equation}
{\cal L}_{\text{comb}}(x,y_1,y_2,\ldots) \equiv \prod_k\,{\cal
  L}_k(x,y_1,y_2,\ldots)\,\prod_{i}\,{\cal 
  L}_i(y_i) \; ,
\end{equation}
with an independent Gaussian external constraint on each $y_i$
\begin{equation}
{\cal L}_i(y_i) \equiv \exp\left[-\frac{1}{2}\,\left(\frac{y_i-y_i'}{\Delta
 y_i'}\right)^2\right] \; .
\end{equation}
The results of this exact method are illustrated by the filled ellipses
in Figs.~\ref{fig:fit12}(a,b) and agree with our method in the limit that
each ${\cal L}_k$ is parabolic and that each $\Delta
y_i' \ll \sigma(y_i)$. In the case of a non-parabolic unconstrained
likelihood, experiments would have to provide a description of ${\cal
  L}_k$ itself to allow an improved combination. In the case of
$\sigma(y_i)\simeq \Delta y_i'$, experiments are advised to perform a
simultaneous measurement of both $x$ and $y$ so that their data will
improve the world knowledge about $y$. 

 The algorithm described above is used as a default in the averages
reported in the following sections.  For some cases, somewhat simplified
or more complex algorithms are used and noted in the corresponding 
sections. Some examples for extensions of the standard method for extracting
averages are given here. These include the case where measurement errors
depend on the measured value, i.e. are relative errors, unknown
correlation coefficients and the breakdown of error sources.

For measurements with Gaussian errors, the usual estimator for the
average of a set of measurements is obtained by minimizing the following
$\chi^2$:
\begin{equation}
\chi^2(t) = \sum_i^N \frac{\left(y_i-t\right)^2}{\sigma^2_i} ,
\label{eq:chi2t}
\end{equation}
where $y_i$ is the measured value for input $i$ and $\sigma_i^2$ is the
variance of the distribution from which $y_i$ was drawn.  The value $\hat{t}$
of $t$ at minimum $\chi^2$ is our estimator for the average.  (This
discussion is given for independent measurements for the sake of
simplicity; the generalization to correlated measurements is
straightforward, and has been used when averaging results.) 
The true $\sigma_i$ are unknown but typically the error as assigned by the
experiment $\sigma_i^{\mathrm{raw}}$ is used as an estimator for it.
Caution is advised,
however, in the case where $\sigma_i^{\mathrm{raw}}$
depends on the value measured for $y_i$. Examples of this include
an uncertainty in any multiplicative factor (like
an acceptance) that enters the determination of $y_i$, i.e. the $\sqrt{N}$
dependence of Poisson statistics, where $y_i \propto N$
and $\sigma_i \propto \sqrt{N}$.
Failing to account for this type of
dependence when averaging leads to a biased average.
Biases in the average can be avoided (or at least reduced)
by minimizing the following
$\chi^2$:
\begin{equation}
\chi^2(t) = \sum_i^N \frac{\left(y_i-t\right)^2}{\sigma^2_i(\hat{t})} .
\label{eq:chi2that}
\end{equation}
In the above $\sigma_i(\hat{t})$ is the uncertainty
assigned to input $i$ that includes the assumed dependence of the
stated error on the value measured.  As an example, consider 
a pure acceptance error, for which
$\sigma_i(\hat{t}) = (\hat{t} / y_i)\times\sigma_i^{\mathrm{raw}}$ .
It is easily verified that solving Eq.~\ref{eq:chi2that} 
leads to the correct behavior, namely
$$ 
\hat{t} = \frac{\sum_i^N y_i^3/(\sigma_i^{\mathrm{raw}})^2}{\sum_i^N y_i^2/(\sigma_i^{\mathrm{raw}})^2},
$$
i.e. weighting by the inverse square of the 
fractional uncertainty, $\sigma_i^{\mathrm{raw}}/y_i$.
It is sometimes difficult to assess the dependence of $\sigma_i^{\mathrm{raw}}$ on
$\hat{t}$ from the errors quoted by experiments.  


Another issue that needs careful treatment is the question of correlation
among different measurements, e.g. due to using the same theory for
calculating acceptances.  A common practice is to set the correlation
coefficient to unity to indicate full correlation.  However, this is
not a ``conservative'' thing to do, and can in fact lead to a significantly
underestimated uncertainty on the average.  In the absence of
better information, the most conservative choice of correlation coefficient
between two measurements $i$ and $j$
is the one that maximizes the uncertainty on $\hat{t}$
due to that pair of measurements:
\begin{equation}
\sigma_{\hat{t}(i,j)}^2 = \frac{\sigma_i^2\,\sigma_j^2\,(1-\rho_{ij}^2)}
   {\sigma_i^2 + \sigma_j^2 - 2\,\rho_{ij}\,\sigma_i\,\sigma_j} ,
\label{eq:correlij}
\end{equation}
namely
\begin{equation}
\rho_{ij} = \mathrm{min}\left(\frac{\sigma_i}{\sigma_j},\frac{\sigma_j}{\sigma_i}\right) ,
\label{eq:correlrho}
\end{equation}
which corresponds to setting $\sigma_{\hat{t}(i,j)}^2=\mathrm{min}(\sigma_i^2,\sigma_j^2)$.
Setting $\rho_{ij}=1$ when $\sigma_i\ne\sigma_j$ can lead to a significant
underestimate of the uncertainty on $\hat{t}$, as can be seen
from Eq.~\ref{eq:correlij}.

Finally, we carefully consider the various sources of error
contributing to the overall uncertainty of an average.
The overall covariance matrix is constructed from a number of
individual sources, e.g.
$\mathbf{V} = \mathbf{V_{stat}+V_{sys}+V_{th}}$.
The variance on the average $\hat{t}$ can be written
\begin{eqnarray}
\sigma^2_{\hat{t}} 
 &=& 
\frac{ \sum_{i,j}\left(\mathbf{V^{-1}}\, 
\mathbf{[V_{stat}+V_{sys}+V_{th}]}\, \mathbf{V^{-1}}\right)_{ij}}
{\left(\sum_{i,j} V^{-1}_{ij}\right)^2}
= \sigma^2_{stat} + \sigma^2_{sys} + \sigma^2_{th} .
\end{eqnarray}
Written in this form, one can readily determine the 
contribution of each source of uncertainty to the overall uncertainty
on the average.  This breakdown of the uncertainties is used 
in the following sections.

Following the prescription described above, the central values and
errors are rescaled to a common set of input parameters in the averaging
procedures according to the dependency on any of these input parameters.
We try to use the most up-to-date values for these common inputs and 
the same values among the HFAG subgroups. For the parameters whose
averages are produced by HFAG, we use the values in the current 
update cycle.  For other external parameters, we use the most
recent PDG values available (usually Ref.~\cite{PDG_2010}). 
The parameters and values used are listed in each subgroup section.

\clearpage
%
%
%
%

%

%
%
%
%

%

\renewcommand{\floatpagefraction}{0.8}
\renewcommand{\topfraction}{0.9}

\newcommand{\comment}[1]{}

\newcommand{\auth}[1]{#1,}
\newcommand{\coll}[1]{#1 Collaboration,}
\newcommand{\authcoll}[2]{#1 \etal\ (#2 Collaboration),}
\newcommand{\authgrp}[2]{#1 \etal\ (#2),}
\newcommand{\titl}[1]{``#1'',} 
\newcommand{\J}[4]{{#1} {\bf #2}, #3 (#4)}
\newcommand{\subJ}[1]{submitted to #1}
\newcommand{\PRL}[3]{\J{Phys.\ Rev.\ Lett.}{#1}{#2}{#3}}
\newcommand{\subPRL}{\subJ{Phys.\ Rev.\ Lett.}}
\newcommand{\PRD}[3]{\J{Phys.\ Rev.\ D}{#1}{#2}{#3}}
\newcommand{\subPRD}{\subJ{Phys.\ Rev.\ D}}
\newcommand{\PREP}[3]{\J{Phys.\ Reports}{#1}{#2}{#3}}
\newcommand{\ZPC}[3]{\J{Z.\ Phys.\ C}{#1}{#2}{#3}}
\newcommand{\PLB}[3]{\J{Phys.\ Lett.\ B}{#1}{#2}{#3}}
\newcommand{\subPLB}{\subJ{Phys.\ Lett.\ B}}
\newcommand{\EPJC}[3]{\J{Eur.\ Phys.\ J.\ C}{#1}{#2}{#3}}
\newcommand{\NPB}[3]{\J{Nucl.\ Phys.\ B}{#1}{#2}{#3}}
\newcommand{\subNPB}{\subJ{Nucl.\ Phys.\ B}}
\newcommand{\NIMA}[3]{\J{Nucl.\ Instrum.\ Methods A}{#1}{#2}{#3}}
\newcommand{\subNIMA}{\subJ{Nucl.\ Instrum.\ Methods A}}
\newcommand{\JHEP}[3]{\J{J.\ of High Energy Physics }{#1}{#2}{#3}}
\newcommand{\JPG}[3]{\J{J.\ of Physics G}{#1}{#2}{#3}}
\newcommand{\ARNS}[3]{\J{Ann.\ Rev.\ Nucl.\ Sci.}{#1}{#2}{#3}}
\newcommand{\newref}{\\}

\newcommand{\particle}[1]{\ensuremath{#1}\xspace}
\renewcommand{\ee}{\particle{e^+e^-}}
\newcommand{\Ups}{\particle{\Upsilon(4S)}}
\newcommand{\Upsfive}{\particle{\Upsilon(5S)}}
\renewcommand{\b}{\particle{b}}
\renewcommand{\B}{\particle{B}}
\newcommand{\Bd}{\particle{B^0}}
\renewcommand{\Bs}{\particle{B^0_s}}
\renewcommand{\Bu}{\particle{B^+}}
\newcommand{\Bc}{\particle{B^+_c}}
\newcommand{\Bdbar}{\particle{\bar{B}^0}}
\newcommand{\Bsbar}{\particle{\bar{B}^0_s}}
\newcommand{\Lb}{\particle{\Lambda_b^0}}
\newcommand{\Xib}{\particle{\Xi_b}}
\newcommand{\Xibd}{\particle{\Xi_b^-}}
\newcommand{\Omegab}{\particle{\Omega_b^-}}
\newcommand{\Lc}{\particle{\Lambda_c^+}}

\newcommand{\fBs}{\ensuremath{f_{\particle{s}}}\xspace}
\newcommand{\fBd}{\ensuremath{f_{\particle{d}}}\xspace}
\newcommand{\fBu}{\ensuremath{f_{\particle{u}}}\xspace}
\newcommand{\fbb}{\ensuremath{f_{\rm baryon}}\xspace}
\newcommand{\fLb}{\ensuremath{f_{\Lambda_{b}}}\xspace}
\newcommand{\fXib}{\ensuremath{f_{\Xi_{b}}}\xspace}
\newcommand{\fOb}{\ensuremath{f_{\Omega_{b}}}\xspace}

\newcommand{\dmd}{\ensuremath{\Delta m_{\particle{d}}}\xspace}
\newcommand{\dms}{\ensuremath{\Delta m_{\particle{s}}}\xspace}
\newcommand{\xd}{\ensuremath{x_{\particle{d}}}\xspace}
\newcommand{\xs}{\ensuremath{x_{\particle{s}}}\xspace}
\newcommand{\yd}{\ensuremath{y_{\particle{d}}}\xspace}
\newcommand{\ys}{\ensuremath{y_{\particle{s}}}\xspace}
\newcommand{\chibar}{\ensuremath{\overline{\chi}}\xspace}
\newcommand{\chid}{\ensuremath{\chi_{\particle{d}}}\xspace}
\newcommand{\chis}{\ensuremath{\chi_{\particle{s}}}\xspace}
\newcommand{\Gd}{\ensuremath{\Gamma_{\particle{d}}}\xspace}
\newcommand{\DGd}{\ensuremath{\Delta\Gd}\xspace}
\newcommand{\DGGd}{\ensuremath{\DGd/\Gd}\xspace}
\newcommand{\Gs}{\ensuremath{\Gamma_{\particle{s}}}\xspace}
\newcommand{\DGs}{\ensuremath{\Delta\Gs}\xspace}
\newcommand{\DGGs}{\ensuremath{\Delta\Gs/\Gs}\xspace}
\newcommand{\ASLd}{\ensuremath{{\cal A}_{\rm SL}^\particle{d}}\xspace}
\newcommand{\ASLs}{\ensuremath{{\cal A}_{\rm SL}^\particle{s}}\xspace}
\newcommand{\ASLb}{\ensuremath{{\cal A}_{\rm SL}^\particle{b}}\xspace}

\newcommand{\DG}{\ensuremath{\Delta\Gamma}\xspace}
\newcommand{\phiccbars}{\ensuremath{\phi_s^{c\bar{c}s}}\xspace}

\renewcommand{\BR}[1]{\particle{{\cal B}(#1)}}
\newcommand{\CL}[1]{#1\%~\mbox{CL}}
\newcommand{\Qjet}{\ensuremath{Q_{\rm jet}}\xspace}

\newcommand{\labe}[1]{\label{equ:#1}}
\newcommand{\labs}[1]{\label{sec:#1}}
\newcommand{\labf}[1]{\label{fig:#1}}
\newcommand{\labt}[1]{\label{tab:#1}}
\newcommand{\refe}[1]{\ref{equ:#1}}
\newcommand{\refs}[1]{\ref{sec:#1}}
\newcommand{\reff}[1]{\ref{fig:#1}}
\newcommand{\reft}[1]{\ref{tab:#1}}
\newcommand{\Ref}[1]{Ref.~\cite{#1}}
\newcommand{\Refs}[1]{Refs.~\cite{#1}}
\newcommand{\Refss}[2]{Refs.~\cite{#1} and \cite{#2}}
\newcommand{\Refsss}[3]{Refs.~\cite{#1}, \cite{#2} and \cite{#3}}
\newcommand{\eq}[1]{(\refe{#1})}
\newcommand{\Eq}[1]{Eq.~(\refe{#1})}
\newcommand{\Eqs}[1]{Eqs.~(\refe{#1})}
\newcommand{\Eqss}[2]{Eqs.~(\refe{#1}) and (\refe{#2})}
\newcommand{\Eqssor}[2]{Eqs.~(\refe{#1}) or (\refe{#2})}
\newcommand{\Eqsss}[3]{Eqs.~(\refe{#1}), (\refe{#2}), and (\refe{#3})}
\newcommand{\Figure}[1]{Figure~\reff{#1}}
\newcommand{\Figuress}[2]{Figures~\reff{#1} and \reff{#2}}
\newcommand{\Fig}[1]{Fig.~\reff{#1}}
\newcommand{\Figs}[1]{Figs.~\reff{#1}}
\newcommand{\Figss}[2]{Figs.~\reff{#1} and \reff{#2}}
\newcommand{\Figsss}[3]{Figs.~\reff{#1}, \reff{#2}, and \reff{#3}}
\newcommand{\Section}[1]{Section~\refs{#1}}
\newcommand{\Sec}[1]{Sec.~\refs{#1}}
\newcommand{\Secs}[1]{Secs.~\refs{#1}}
\newcommand{\Secss}[2]{Secs.~\refs{#1} and \refs{#2}}
\newcommand{\Secsss}[3]{Secs.~\refs{#1}, \refs{#2}, and \refs{#3}}
\newcommand{\Table}[1]{Table~\reft{#1}}
\newcommand{\Tables}[1]{Tables~\reft{#1}}
\newcommand{\Tabless}[2]{Tables~\reft{#1} and \reft{#2}}
\newcommand{\Tablesss}[3]{Tables~\reft{#1}, \reft{#2}, and \reft{#3}}

\newcommand{\subsubsubsection}[1]{\vspace{2ex}\par\noindent {\bf\boldmath\em #1} \vspace{2ex}\par}


\newcommand{\definemath}[2]{\newcommand{#1}{\ensuremath{#2}\xspace}}

\definemath{\hfagCHIBARLEPval}{0.1259}
\definemath{\hfagCHIBARLEPerr}{\pm0.0042}
\definemath{\hfagTAUBDval}{1.519}
\definemath{\hfagTAUBDerr}{\pm0.007}
\definemath{\hfagTAUBUval}{1.642}
\definemath{\hfagTAUBUerr}{\pm0.008}
\definemath{\hfagRTAUBUval}{1.079}
\definemath{\hfagRTAUBUerr}{\pm0.007}
\definemath{\hfagTAUBSval}{1.466}
\definemath{\hfagTAUBSerr}{\pm0.031}
\definemath{\hfagRTAUBSval}{0.965}
\definemath{\hfagRTAUBSerr}{\pm0.021}
\definemath{\hfagTAULBval}{1.413}
\definemath{\hfagTAULBerr}{\pm0.030}
\definemath{\hfagTAUBBval}{1.378}
\definemath{\hfagTAUBBerr}{\pm0.027}
\definemath{\hfagRTAUBBval}{0.907}
\definemath{\hfagRTAUBBerr}{\pm0.018}
\definemath{\hfagTAUXBval}{1.49}
\definemath{\hfagTAUXBerp}{^{+0.19}}
\definemath{\hfagTAUXBern}{_{-0.18}}
\definemath{\hfagTAUBval}{1.566}
\definemath{\hfagTAUBerr}{\pm0.009}
\definemath{\hfagTAUXBDval}{1.56}
\definemath{\hfagTAUXBDerp}{^{+0.27}}
\definemath{\hfagTAUXBDern}{_{-0.25}}
\definemath{\hfagTAUOBval}{1.13}
\definemath{\hfagTAUOBerp}{^{+0.53}}
\definemath{\hfagTAUOBern}{_{-0.40}}
\definemath{\hfagTAUBCval}{0.458}
\definemath{\hfagTAUBCerr}{\pm0.030}
\definemath{\hfagTAUBSSLval}{1.463}
\definemath{\hfagTAUBSSLerr}{\pm0.032}
\definemath{\hfagTAUBSMEANCval}{1.509}
\definemath{\hfagTAUBSMEANCerr}{\pm0.012}
\definemath{\hfagTAUBSJFval}{1.430}
\definemath{\hfagTAUBSJFerr}{\pm0.050}
\definemath{\hfagRTAUBSSLval}{0.963}
\definemath{\hfagRTAUBSSLerr}{\pm0.022}
\definemath{\hfagRTAUBSMEANCval}{0.993}
\definemath{\hfagRTAUBSMEANCerr}{\pm0.009}
\definemath{\hfagRTAUBSMEANCsig}{0.7}
\definemath{\hfagONEMINUSRTAUBSMEANCpercent}{(0.7\pm0.9)\%}
\definemath{\hfagRTAULBval}{0.930}
\definemath{\hfagRTAULBerr}{\pm0.020}
\definemath{\hfagTAUBVTXval}{1.572}
\definemath{\hfagTAUBVTXerr}{\pm0.009}
\definemath{\hfagTAUBLEPval}{1.537}
\definemath{\hfagTAUBLEPerr}{\pm0.020}
\definemath{\hfagTAUBJPval}{1.516}
\definemath{\hfagTAUBJPerr}{\pm0.028}
\definemath{\hfagNSIGMATAULBCDFTWO}{3.4}
\definemath{\hfagSDGDGDval}{0.015}
\definemath{\hfagSDGDGDerr}{\pm0.018}
\definemath{\hfagTAUBSLONGval}{1.70}
\definemath{\hfagTAUBSLONGerr}{\pm0.12}
\definemath{\hfagTAUBSSHORTval}{1.463}
\definemath{\hfagTAUBSSHORTerr}{\pm0.042}
\definemath{\hfagBRDSDSval}{0.044}
\definemath{\hfagBRDSDSerr}{\pm0.014}
\definemath{\hfagDGSGSBRDSDSval}{+0.093}
\definemath{\hfagDGSGSBRDSDSerr}{\pm0.031}
\definemath{\hfagGSval}{0.6604}
\definemath{\hfagGSerr}{\pm0.0058}
\definemath{\hfagTAUBSMEANval}{1.514}
\definemath{\hfagTAUBSMEANerr}{\pm0.013}
\definemath{\hfagDGSGSval}{+0.159}
\definemath{\hfagDGSGSerr}{\pm0.023}
\definemath{\hfagRHODGSGSTAUBSMEAN}{0.}
\definemath{\hfagDGSval}{+0.105}
\definemath{\hfagDGSerr}{\pm0.015}
\definemath{\hfagRHODGSTAUBSMEAN}{0.}
\definemath{\hfagTAUBSLval}{1.403}
\definemath{\hfagTAUBSLerr}{\pm0.019}
\definemath{\hfagTAUBSHval}{1.645}
\definemath{\hfagTAUBSHerr}{\pm0.027}
\definemath{\hfagTAUBSMEANCOval}{1.520}
\definemath{\hfagTAUBSMEANCOerr}{\pm0.013}
\definemath{\hfagDGSGSCOval}{+0.152}
\definemath{\hfagDGSGSCOerr}{\pm0.021}
\definemath{\hfagRHODGSGSTAUBSMEANCO}{0.}
\definemath{\hfagDGSCOval}{+0.100}
\definemath{\hfagDGSCOerr}{\pm0.014}
\definemath{\hfagRHODGSTAUBSMEANCO}{0.}
\definemath{\hfagTAUBSLCOval}{1.412}
\definemath{\hfagTAUBSLCOerr}{\pm0.017}
\definemath{\hfagTAUBSHCOval}{1.644}
\definemath{\hfagTAUBSHCOerr}{\pm0.025}
\definemath{\hfagTAUBSMEANCONval}{1.509}
\definemath{\hfagTAUBSMEANCONerr}{\pm0.012}
\definemath{\hfagDGSGSCONval}{+0.144}
\definemath{\hfagDGSGSCONerr}{\pm0.021}
\definemath{\hfagRHODGSGSTAUBSMEANCON}{0.}
\definemath{\hfagDGSCONval}{+0.095}
\definemath{\hfagDGSCONerr}{\pm0.014}
\definemath{\hfagRHODGSTAUBSMEANCON}{0.}
\definemath{\hfagTAUBSLCONval}{1.408}
\definemath{\hfagTAUBSLCONerr}{\pm0.017}
\definemath{\hfagTAUBSHCONval}{1.626}
\definemath{\hfagTAUBSHCONerr}{\pm0.023}
\definemath{\hfagFCWval}{0.513}
\definemath{\hfagFCWerr}{\pm0.006}
\definemath{\hfagFNWval}{0.487}
\definemath{\hfagFNWerr}{\pm0.006}
\definemath{\hfagFFWval}{1.055}
\definemath{\hfagFFWerr}{\pm0.025}
\definemath{\hfagNSIGMAFFW}{2.2}
\definemath{\hfagFCNval}{0.513}
\definemath{\hfagFCNerr}{\pm0.013}
\definemath{\hfagFNNval}{0.487}
\definemath{\hfagFNNerr}{\pm0.013}
\definemath{\hfagFFNval}{1.053}
\definemath{\hfagFFNerr}{\pm0.054}
\definemath{\hfagFCval}{0.514}
\definemath{\hfagFCerr}{\pm0.007}
\definemath{\hfagFNval}{0.486}
\definemath{\hfagFNerr}{\pm0.007}
\definemath{\hfagFFval}{1.056}
\definemath{\hfagFFerr}{\pm0.028}
\definemath{\hfagNSIGMAFF}{2.0}
\definemath{\hfagFPRODval}{0.514}
\definemath{\hfagFPRODerr}{\pm0.019}
\definemath{\hfagFSUMval}{1.001}
\definemath{\hfagFSUMerr}{\pm0.030}
\definemath{\hfagFSFIVEOSval}{0.206}
\definemath{\hfagFSFIVEOSsta}{\pm0.010}
\definemath{\hfagFSFIVEOSsys}{\pm0.024}
\definemath{\hfagFSFIVEOSerr}{\pm0.027}
\definemath{\hfagFSFIVERLval}{0.215}
\definemath{\hfagFSFIVERLerr}{\pm0.032}
\definemath{\hfagFUDFIVEval}{0.759}
\definemath{\hfagFUDFIVEerp}{^{+0.027}}
\definemath{\hfagFUDFIVEern}{_{-0.040}}
\definemath{\hfagFSFIVEval}{0.199}
\definemath{\hfagFSFIVEerr}{\pm0.030}
\definemath{\hfagFSFUDFIVEval}{0.262}
\definemath{\hfagFSFUDFIVEerp}{^{+0.051}}
\definemath{\hfagFSFUDFIVEern}{_{-0.043}}
\definemath{\hfagFNBFIVEval}{0.042}
\definemath{\hfagFNBFIVEerp}{^{+0.046}}
\definemath{\hfagFNBFIVEern}{_{-0.006}}
\definemath{\hfagRBSTEVNOCONval}{0.140}
\definemath{\hfagRBSTEVNOCONerr}{\pm0.022}
\definemath{\hfagRLBTEVNOCONval}{0.290}
\definemath{\hfagRLBTEVNOCONerr}{\pm0.109}
\definemath{\hfagRBSLHCBNOCONval}{0.132}
\definemath{\hfagRBSLHCBNOCONerr}{\pm0.010}
\definemath{\hfagRLBLHCBNOCONval}{0.305}
\definemath{\hfagRLBLHCBNOCONerr}{\pm0.022}
\definemath{\hfagZFSFACTOR}{}
\definemath{\hfagZFBSNOMIXval}{0.087}
\definemath{\hfagZFBSNOMIXerr}{\pm0.014}
\definemath{\hfagZFBBNOMIXval}{0.099}
\definemath{\hfagZFBBNOMIXerr}{\pm0.016}
\definemath{\hfagZFBDNOMIXval}{0.407}
\definemath{\hfagZFBDNOMIXerr}{\pm0.009}
\definemath{\hfagWFSFACTOR}{}
\definemath{\hfagWFBSNOMIXval}{0.103}
\definemath{\hfagWFBSNOMIXerr}{\pm0.007}
\definemath{\hfagWFBBNOMIXval}{0.097}
\definemath{\hfagWFBBNOMIXerr}{\pm0.016}
\definemath{\hfagWFBDNOMIXval}{0.400}
\definemath{\hfagWFBDNOMIXerr}{\pm0.008}
\definemath{\hfagTFSFACTOR}{}
\definemath{\hfagTFBSNOMIXval}{0.094}
\definemath{\hfagTFBSNOMIXerr}{\pm0.016}
\definemath{\hfagTFBBNOMIXval}{0.262}
\definemath{\hfagTFBBNOMIXerr}{\pm0.073}
\definemath{\hfagTFBDNOMIXval}{0.322}
\definemath{\hfagTFBDNOMIXerr}{\pm0.032}
\definemath{\hfagLFSFACTOR}{}
\definemath{\hfagLFBSNOMIXval}{0.085}
\definemath{\hfagLFBSNOMIXerr}{\pm0.009}
\definemath{\hfagLFBBNOMIXval}{0.274}
\definemath{\hfagLFBBNOMIXerr}{\pm0.055}
\definemath{\hfagLFBDNOMIXval}{0.321}
\definemath{\hfagLFBDNOMIXerr}{\pm0.024}
\definemath{\hfagCHIBARTEVval}{0.127}
\definemath{\hfagCHIBARTEVerr}{\pm0.008}
\definemath{\hfagCHIBARval}{0.1260}
\definemath{\hfagCHIBARerr}{\pm0.0037}
\definemath{\hfagWFBSMIXval}{0.115}
\definemath{\hfagWFBSMIXerr}{\pm0.011}
\definemath{\hfagTFBSMIXval}{0.117}
\definemath{\hfagTFBSMIXerr}{\pm0.020}
\definemath{\hfagZFBSMIXval}{0.115}
\definemath{\hfagZFBSMIXerr}{\pm0.012}
\definemath{\hfagCHIDUval}{0.182}
\definemath{\hfagCHIDUerr}{\pm0.015}
\definemath{\hfagCHIDWUval}{0.1862}
\definemath{\hfagCHIDWUerr}{\pm0.0023}
\definemath{\hfagXDWval}{0.771}
\definemath{\hfagXDWerr}{\pm0.008}
\definemath{\hfagXDWUval}{0.770}
\definemath{\hfagXDWUerr}{\pm0.008}
\definemath{\hfagDMDWval}{0.507}
\definemath{\hfagDMDWsta}{\pm0.003}
\definemath{\hfagDMDWsys}{\pm0.003}
\definemath{\hfagDMDWerr}{\pm0.004}
\definemath{\hfagDMDWUval}{0.507}
\definemath{\hfagDMDWUerr}{\pm0.004}
\definemath{\hfagZFBSval}{0.103}
\definemath{\hfagZFBSerr}{\pm0.009}
\definemath{\hfagZFBBval}{0.090}
\definemath{\hfagZFBBerr}{\pm0.015}
\definemath{\hfagZFBDval}{0.403}
\definemath{\hfagZFBDerr}{\pm0.009}
\definemath{\hfagZRHOFBBFBS}{+0.036}
\definemath{\hfagZRHOFBDFBS}{-0.521}
\definemath{\hfagZRHOFBDFBB}{-0.871}
\definemath{\hfagWFBSval}{0.107}
\definemath{\hfagWFBSerr}{\pm0.005}
\definemath{\hfagWFBBval}{0.091}
\definemath{\hfagWFBBerr}{\pm0.015}
\definemath{\hfagWFBDval}{0.401}
\definemath{\hfagWFBDerr}{\pm0.007}
\definemath{\hfagWRHOFBBFBS}{-0.136}
\definemath{\hfagWRHOFBDFBS}{-0.224}
\definemath{\hfagWRHOFBDFBB}{-0.935}
\definemath{\hfagTFBSval}{0.103}
\definemath{\hfagTFBSerr}{\pm0.012}
\definemath{\hfagTFBBval}{0.236}
\definemath{\hfagTFBBerr}{\pm0.067}
\definemath{\hfagTFBDval}{0.330}
\definemath{\hfagTFBDerr}{\pm0.030}
\definemath{\hfagTRHOFBBFBS}{-0.530}
\definemath{\hfagTRHOFBDFBS}{+0.379}
\definemath{\hfagTRHOFBDFBB}{-0.986}
\definemath{\hfagLFBSval}{0.090}
\definemath{\hfagLFBSerr}{\pm0.008}
\definemath{\hfagLFBBval}{0.248}
\definemath{\hfagLFBBerr}{\pm0.051}
\definemath{\hfagLFBDval}{0.331}
\definemath{\hfagLFBDerr}{\pm0.023}
\definemath{\hfagLRHOFBBFBS}{-0.705}
\definemath{\hfagLRHOFBDFBS}{+0.614}
\definemath{\hfagLRHOFBDFBB}{-0.993}
\definemath{\hfagZFBSBDval}{0.256}
\definemath{\hfagZFBSBDerr}{\pm0.025}
\definemath{\hfagWFBSBDval}{0.266}
\definemath{\hfagWFBSBDerr}{\pm0.015}
\definemath{\hfagTFBSBDval}{0.311}
\definemath{\hfagTFBSBDerr}{\pm0.037}
\definemath{\hfagLFBSBDval}{0.273}
\definemath{\hfagLFBSBDerr}{\pm0.019}
\definemath{\hfagDMDHval}{0.496}
\definemath{\hfagDMDHsta}{\pm0.010}
\definemath{\hfagDMDHsys}{\pm0.009}
\definemath{\hfagDMDHerr}{\pm0.013}
\definemath{\hfagDMDBval}{0.508}
\definemath{\hfagDMDBsta}{\pm0.003}
\definemath{\hfagDMDBsys}{\pm0.003}
\definemath{\hfagDMDBerr}{\pm0.005}
\definemath{\hfagDMDTWODval}{0.509}
\definemath{\hfagDMDTWODsta}{\pm0.004}
\definemath{\hfagDMDTWODsys}{\pm0.004}
\definemath{\hfagDMDTWODerr}{\pm0.006}
\definemath{\hfagTAUBDTWODval}{1.527}
\definemath{\hfagTAUBDTWODsta}{\pm0.006}
\definemath{\hfagTAUBDTWODsys}{\pm0.008}
\definemath{\hfagTAUBDTWODerr}{\pm0.010}
\definemath{\hfagRHOstaDMDTAUBD}{-0.19}
\definemath{\hfagRHOsysDMDTAUBD}{-0.26}
\definemath{\hfagRHODMDTAUBD}{-0.23}
\definemath{\hfagQPDBval}{1.0002}
\definemath{\hfagQPDBerr}{\pm0.0028}
\definemath{\hfagQPDWval}{1.0004}
\definemath{\hfagQPDWerr}{\pm0.0019}
\definemath{\hfagQPDAval}{1.0005}
\definemath{\hfagQPDAerr}{\pm0.0019}
\definemath{\hfagASLDBval}{-0.0005}
\definemath{\hfagASLDBerr}{\pm0.0056}
\definemath{\hfagASLDWval}{-0.0009}
\definemath{\hfagASLDWerr}{\pm0.0038}
\definemath{\hfagASLDAval}{-0.0010}
\definemath{\hfagASLDAerr}{\pm0.0037}
\definemath{\hfagREBDBval}{-0.0001}
\definemath{\hfagREBDBerr}{\pm0.0014}
\definemath{\hfagREBDWval}{-0.0002}
\definemath{\hfagREBDWerr}{\pm0.0010}
\definemath{\hfagREBDAval}{-0.0002}
\definemath{\hfagREBDAerr}{\pm0.0009}
\definemath{\hfagASLSWval}{-0.0095}
\definemath{\hfagASLSWsta}{\pm0.0038}
\definemath{\hfagASLSWsys}{\pm0.0054}
\definemath{\hfagASLSWerr}{\pm0.0066}
\definemath{\hfagQPSWval}{1.0048}
\definemath{\hfagQPSWsta}{\pm0.0019}
\definemath{\hfagQPSWsys}{\pm0.0027}
\definemath{\hfagQPSWerr}{\pm0.0033}
\definemath{\hfagASLSval}{-0.0105}
\definemath{\hfagASLSerr}{\pm0.0064}
\definemath{\hfagQPSval}{1.0052}
\definemath{\hfagQPSerr}{\pm0.0032}
\definemath{\hfagASLDval}{-0.0033}
\definemath{\hfagASLDerr}{\pm0.0033}
\definemath{\hfagQPDval}{1.0017}
\definemath{\hfagQPDerr}{\pm0.0017}
\definemath{\hfagRHOASLSASLD}{-0.574}
\definemath{\hfagREBDval}{-0.0008}
\definemath{\hfagREBDerr}{\pm0.0008}
\definemath{\hfagDMSval}{17.719}
\definemath{\hfagDMSsta}{\pm0.036}
\definemath{\hfagDMSsys}{\pm0.023}
\definemath{\hfagDMSerr}{\pm0.043}
\definemath{\hfagXSval}{26.74}
\definemath{\hfagXSerr}{\pm0.22}
\definemath{\hfagCHISval}{0.499305}
\definemath{\hfagCHISerr}{\pm0.000011}
\definemath{\hfagRATIODMDDMSval}{0.02861}
\definemath{\hfagRATIODMDDMSerr}{\pm0.00026}
\definemath{\hfagVTDVTSval}{0.2110}
\definemath{\hfagVTDVTSexx}{\pm0.0009}
\definemath{\hfagVTDVTSthe}{\pm0.0055}
\definemath{\hfagVTDVTSerr}{\pm0.0055}
\definemath{\hfagXIval}{1.237}
\definemath{\hfagXIerr}{\pm0.032}
\definemath{\hfagBETASCOMBval}{+0.022}
\definemath{\hfagBETASCOMBerp}{^{+0.043}}
\definemath{\hfagBETASCOMBern}{_{-0.045}}
\definemath{\hfagPHISCOMBval}{-0.044}
\definemath{\hfagPHISCOMBerp}{^{+0.090}}
\definemath{\hfagPHISCOMBern}{_{-0.085}}
\definemath{\hfagDGSCOMBval}{+0.105}
\definemath{\hfagDGSCOMBerr}{\pm0.015}
\definemath{\hfagNSIGMASM}{0.8}
\definemath{\hfagBETASCOMBCONval}{+0.032}
\definemath{\hfagBETASCOMBCONerp}{^{+0.038}}
\definemath{\hfagBETASCOMBCONern}{_{-0.049}}
\definemath{\hfagPHISCOMBCONval}{-0.064}
\definemath{\hfagPHISCOMBCONerp}{^{+0.098}}
\definemath{\hfagPHISCOMBCONern}{_{-0.076}}
\definemath{\hfagDGSCOMBCONval}{+0.107}
\definemath{\hfagDGSCOMBCONerp}{^{+0.014}}
\definemath{\hfagDGSCOMBCONern}{_{-0.016}}
\definemath{\hfagNSIGMASMCON}{0.8}

\newcommand{\unit}[1]{~\ensuremath{\rm #1}\xspace}
\renewcommand{\ps}{\unit{ps}}
\newcommand{\invps}{\unit{ps^{-1}}}
\newcommand{\TeV}{\unit{TeV}}
\newcommand{\MeVcc}{\unit{MeV/\mbox{$c$}^2}}
\newcommand{\MeV}{\unit{MeV}}

\definemath{\hfagCHIBARLEP}{\hfagCHIBARLEPval\hfagCHIBARLEPerr}
\definemath{\hfagTAUBD}{\hfagTAUBDval\hfagTAUBDerr\ps}
\definemath{\hfagTAUBDnounit}{\hfagTAUBDval\hfagTAUBDerr}
\definemath{\hfagTAUBU}{\hfagTAUBUval\hfagTAUBUerr\ps}
\definemath{\hfagTAUBUnounit}{\hfagTAUBUval\hfagTAUBUerr}
\definemath{\hfagRTAUBU}{\hfagRTAUBUval\hfagRTAUBUerr}
\definemath{\hfagTAUBS}{\hfagTAUBSval\hfagTAUBSerr\ps}
\definemath{\hfagTAUBSnounit}{\hfagTAUBSval\hfagTAUBSerr}
\definemath{\hfagRTAUBS}{\hfagRTAUBSval\hfagRTAUBSerr}
\definemath{\hfagTAULB}{\hfagTAULBval\hfagTAULBerr\ps}
\definemath{\hfagTAULBnounit}{\hfagTAULBval\hfagTAULBerr}
\definemath{\hfagTAUBB}{\hfagTAUBBval\hfagTAUBBerr\ps}
\definemath{\hfagTAUBBnounit}{\hfagTAUBBval\hfagTAUBBerr}
\definemath{\hfagRTAUBB}{\hfagRTAUBBval\hfagRTAUBBerr}
\definemath{\hfagTAUXBerr}{\hfagTAUXBerp\hfagTAUXBern}
\definemath{\hfagTAUXB}{\hfagTAUXBval\hfagTAUXBerr\ps}
\definemath{\hfagTAUXBnounit}{\hfagTAUXBval\hfagTAUXBerr}
\definemath{\hfagTAUB}{\hfagTAUBval\hfagTAUBerr\ps}
\definemath{\hfagTAUBnounit}{\hfagTAUBval\hfagTAUBerr}
\definemath{\hfagTAUXBDerr}{\hfagTAUXBDerp\hfagTAUXBDern}
\definemath{\hfagTAUXBD}{\hfagTAUXBDval\hfagTAUXBDerr\ps}
\definemath{\hfagTAUXBDnounit}{\hfagTAUXBDval\hfagTAUXBDerr}
\definemath{\hfagTAUOBerr}{\hfagTAUOBerp\hfagTAUOBern}
\definemath{\hfagTAUOB}{\hfagTAUOBval\hfagTAUOBerr\ps}
\definemath{\hfagTAUOBnounit}{\hfagTAUOBval\hfagTAUOBerr}
\definemath{\hfagTAUBC}{\hfagTAUBCval\hfagTAUBCerr\ps}
\definemath{\hfagTAUBCnounit}{\hfagTAUBCval\hfagTAUBCerr}
\definemath{\hfagTAUBSSL}{\hfagTAUBSSLval\hfagTAUBSSLerr\ps}
\definemath{\hfagTAUBSSLnounit}{\hfagTAUBSSLval\hfagTAUBSSLerr}
\definemath{\hfagTAUBSMEANC}{\hfagTAUBSMEANCval\hfagTAUBSMEANCerr\ps}
\definemath{\hfagTAUBSMEANCnounit}{\hfagTAUBSMEANCval\hfagTAUBSMEANCerr}
\definemath{\hfagTAUBSJF}{\hfagTAUBSJFval\hfagTAUBSJFerr\ps}
\definemath{\hfagTAUBSJFnounit}{\hfagTAUBSJFval\hfagTAUBSJFerr}
\definemath{\hfagRTAUBSSL}{\hfagRTAUBSSLval\hfagRTAUBSSLerr}
\definemath{\hfagRTAUBSMEANC}{\hfagRTAUBSMEANCval\hfagRTAUBSMEANCerr}
\definemath{\hfagRTAULB}{\hfagRTAULBval\hfagRTAULBerr}
\definemath{\hfagTAUBVTX}{\hfagTAUBVTXval\hfagTAUBVTXerr\ps}
\definemath{\hfagTAUBVTXnounit}{\hfagTAUBVTXval\hfagTAUBVTXerr}
\definemath{\hfagTAUBLEP}{\hfagTAUBLEPval\hfagTAUBLEPerr\ps}
\definemath{\hfagTAUBLEPnounit}{\hfagTAUBLEPval\hfagTAUBLEPerr}
\definemath{\hfagTAUBJP}{\hfagTAUBJPval\hfagTAUBJPerr\ps}
\definemath{\hfagTAUBJPnounit}{\hfagTAUBJPval\hfagTAUBJPerr}
\definemath{\hfagSDGDGD}{\hfagSDGDGDval\hfagSDGDGDerr}
\definemath{\hfagTAUBSLONG}{\hfagTAUBSLONGval\hfagTAUBSLONGerr\ps}
\definemath{\hfagTAUBSLONGnounit}{\hfagTAUBSLONGval\hfagTAUBSLONGerr}
\definemath{\hfagTAUBSSHORT}{\hfagTAUBSSHORTval\hfagTAUBSSHORTerr\ps}
\definemath{\hfagTAUBSSHORTnounit}{\hfagTAUBSSHORTval\hfagTAUBSSHORTerr}
\definemath{\hfagBRDSDS}{\hfagBRDSDSval\hfagBRDSDSerr}
\definemath{\hfagDGSGSBRDSDS}{\hfagDGSGSBRDSDSval\hfagDGSGSBRDSDSerr}
\definemath{\hfagGS}{\hfagGSval\hfagGSerr\invps}
\definemath{\hfagGSnounit}{\hfagGSval\hfagGSerr}
\definemath{\hfagTAUBSMEAN}{\hfagTAUBSMEANval\hfagTAUBSMEANerr\ps}
\definemath{\hfagTAUBSMEANnounit}{\hfagTAUBSMEANval\hfagTAUBSMEANerr}
\definemath{\hfagDGSGS}{\hfagDGSGSval\hfagDGSGSerr}
\definemath{\hfagDGS}{\hfagDGSval\hfagDGSerr\invps}
\definemath{\hfagDGSnounit}{\hfagDGSval\hfagDGSerr}
\definemath{\hfagTAUBSL}{\hfagTAUBSLval\hfagTAUBSLerr\ps}
\definemath{\hfagTAUBSLnounit}{\hfagTAUBSLval\hfagTAUBSLerr}
\definemath{\hfagTAUBSH}{\hfagTAUBSHval\hfagTAUBSHerr\ps}
\definemath{\hfagTAUBSHnounit}{\hfagTAUBSHval\hfagTAUBSHerr}
\definemath{\hfagTAUBSMEANCO}{\hfagTAUBSMEANCOval\hfagTAUBSMEANCOerr\ps}
\definemath{\hfagTAUBSMEANCOnounit}{\hfagTAUBSMEANCOval\hfagTAUBSMEANCOerr}
\definemath{\hfagDGSGSCO}{\hfagDGSGSCOval\hfagDGSGSCOerr}
\definemath{\hfagDGSCO}{\hfagDGSCOval\hfagDGSCOerr\invps}
\definemath{\hfagDGSCOnounit}{\hfagDGSCOval\hfagDGSCOerr}
\definemath{\hfagTAUBSLCO}{\hfagTAUBSLCOval\hfagTAUBSLCOerr\ps}
\definemath{\hfagTAUBSLCOnounit}{\hfagTAUBSLCOval\hfagTAUBSLCOerr}
\definemath{\hfagTAUBSHCO}{\hfagTAUBSHCOval\hfagTAUBSHCOerr\ps}
\definemath{\hfagTAUBSHCOnounit}{\hfagTAUBSHCOval\hfagTAUBSHCOerr}
\definemath{\hfagTAUBSMEANCON}{\hfagTAUBSMEANCONval\hfagTAUBSMEANCONerr\ps}
\definemath{\hfagTAUBSMEANCONnounit}{\hfagTAUBSMEANCONval\hfagTAUBSMEANCONerr}
\definemath{\hfagDGSGSCON}{\hfagDGSGSCONval\hfagDGSGSCONerr}
\definemath{\hfagDGSCON}{\hfagDGSCONval\hfagDGSCONerr\invps}
\definemath{\hfagDGSCONnounit}{\hfagDGSCONval\hfagDGSCONerr}
\definemath{\hfagTAUBSLCON}{\hfagTAUBSLCONval\hfagTAUBSLCONerr\ps}
\definemath{\hfagTAUBSLCONnounit}{\hfagTAUBSLCONval\hfagTAUBSLCONerr}
\definemath{\hfagTAUBSHCON}{\hfagTAUBSHCONval\hfagTAUBSHCONerr\ps}
\definemath{\hfagTAUBSHCONnounit}{\hfagTAUBSHCONval\hfagTAUBSHCONerr}
\definemath{\hfagFCW}{\hfagFCWval\hfagFCWerr}
\definemath{\hfagFNW}{\hfagFNWval\hfagFNWerr}
\definemath{\hfagFFW}{\hfagFFWval\hfagFFWerr}
\definemath{\hfagFCN}{\hfagFCNval\hfagFCNerr}
\definemath{\hfagFNN}{\hfagFNNval\hfagFNNerr}
\definemath{\hfagFFN}{\hfagFFNval\hfagFFNerr}
\definemath{\hfagFC}{\hfagFCval\hfagFCerr}
\definemath{\hfagFN}{\hfagFNval\hfagFNerr}
\definemath{\hfagFF}{\hfagFFval\hfagFFerr}
\definemath{\hfagFPROD}{\hfagFPRODval\hfagFPRODerr}
\definemath{\hfagFSUM}{\hfagFSUMval\hfagFSUMerr}
\definemath{\hfagFSFIVEOS}{\hfagFSFIVEOSval\hfagFSFIVEOSerr}
\definemath{\hfagFSFIVEOSfull}{\hfagFSFIVEOSval\hfagFSFIVEOSsta\hfagFSFIVEOSsys}
\definemath{\hfagFSFIVERL}{\hfagFSFIVERLval\hfagFSFIVERLerr}
\definemath{\hfagFUDFIVEerr}{\hfagFUDFIVEerp\hfagFUDFIVEern}
\definemath{\hfagFUDFIVE}{\hfagFUDFIVEval\hfagFUDFIVEerr}
\definemath{\hfagFSFIVE}{\hfagFSFIVEval\hfagFSFIVEerr}
\definemath{\hfagFSFUDFIVEerr}{\hfagFSFUDFIVEerp\hfagFSFUDFIVEern}
\definemath{\hfagFSFUDFIVE}{\hfagFSFUDFIVEval\hfagFSFUDFIVEerr}
\definemath{\hfagFNBFIVEerr}{\hfagFNBFIVEerp\hfagFNBFIVEern}
\definemath{\hfagFNBFIVE}{\hfagFNBFIVEval\hfagFNBFIVEerr}
\definemath{\hfagRBSTEVNOCON}{\hfagRBSTEVNOCONval\hfagRBSTEVNOCONerr}
\definemath{\hfagRLBTEVNOCON}{\hfagRLBTEVNOCONval\hfagRLBTEVNOCONerr}
\definemath{\hfagRBSLHCBNOCON}{\hfagRBSLHCBNOCONval\hfagRBSLHCBNOCONerr}
\definemath{\hfagRLBLHCBNOCON}{\hfagRLBLHCBNOCONval\hfagRLBLHCBNOCONerr}
\definemath{\hfagZFBSNOMIX}{\hfagZFBSNOMIXval\hfagZFBSNOMIXerr}
\definemath{\hfagZFBBNOMIX}{\hfagZFBBNOMIXval\hfagZFBBNOMIXerr}
\definemath{\hfagZFBDNOMIX}{\hfagZFBDNOMIXval\hfagZFBDNOMIXerr}
\definemath{\hfagWFBSNOMIX}{\hfagWFBSNOMIXval\hfagWFBSNOMIXerr}
\definemath{\hfagWFBBNOMIX}{\hfagWFBBNOMIXval\hfagWFBBNOMIXerr}
\definemath{\hfagWFBDNOMIX}{\hfagWFBDNOMIXval\hfagWFBDNOMIXerr}
\definemath{\hfagTFBSNOMIX}{\hfagTFBSNOMIXval\hfagTFBSNOMIXerr}
\definemath{\hfagTFBBNOMIX}{\hfagTFBBNOMIXval\hfagTFBBNOMIXerr}
\definemath{\hfagTFBDNOMIX}{\hfagTFBDNOMIXval\hfagTFBDNOMIXerr}
\definemath{\hfagLFBSNOMIX}{\hfagLFBSNOMIXval\hfagLFBSNOMIXerr}
\definemath{\hfagLFBBNOMIX}{\hfagLFBBNOMIXval\hfagLFBBNOMIXerr}
\definemath{\hfagLFBDNOMIX}{\hfagLFBDNOMIXval\hfagLFBDNOMIXerr}
\definemath{\hfagCHIBARTEV}{\hfagCHIBARTEVval\hfagCHIBARTEVerr}
\definemath{\hfagCHIBAR}{\hfagCHIBARval\hfagCHIBARerr}
\definemath{\hfagWFBSMIX}{\hfagWFBSMIXval\hfagWFBSMIXerr}
\definemath{\hfagTFBSMIX}{\hfagTFBSMIXval\hfagTFBSMIXerr}
\definemath{\hfagZFBSMIX}{\hfagZFBSMIXval\hfagZFBSMIXerr}
\definemath{\hfagCHIDU}{\hfagCHIDUval\hfagCHIDUerr}
\definemath{\hfagCHIDWU}{\hfagCHIDWUval\hfagCHIDWUerr}
\definemath{\hfagXDW}{\hfagXDWval\hfagXDWerr}
\definemath{\hfagXDWU}{\hfagXDWUval\hfagXDWUerr}
\definemath{\hfagDMDW}{\hfagDMDWval\hfagDMDWerr\invps}
\definemath{\hfagDMDWnounit}{\hfagDMDWval\hfagDMDWerr}
\definemath{\hfagDMDWfull}{\hfagDMDWval\hfagDMDWsta\hfagDMDWsys\invps}
\definemath{\hfagDMDWnounitfull}{\hfagDMDWval\hfagDMDWsta\hfagDMDWsys}
\definemath{\hfagDMDWU}{\hfagDMDWUval\hfagDMDWUerr\invps}
\definemath{\hfagDMDWUnounit}{\hfagDMDWUval\hfagDMDWUerr}
\definemath{\hfagZFBS}{\hfagZFBSval\hfagZFBSerr}
\definemath{\hfagZFBB}{\hfagZFBBval\hfagZFBBerr}
\definemath{\hfagZFBD}{\hfagZFBDval\hfagZFBDerr}
\definemath{\hfagWFBS}{\hfagWFBSval\hfagWFBSerr}
\definemath{\hfagWFBB}{\hfagWFBBval\hfagWFBBerr}
\definemath{\hfagWFBD}{\hfagWFBDval\hfagWFBDerr}
\definemath{\hfagTFBS}{\hfagTFBSval\hfagTFBSerr}
\definemath{\hfagTFBB}{\hfagTFBBval\hfagTFBBerr}
\definemath{\hfagTFBD}{\hfagTFBDval\hfagTFBDerr}
\definemath{\hfagLFBS}{\hfagLFBSval\hfagLFBSerr}
\definemath{\hfagLFBB}{\hfagLFBBval\hfagLFBBerr}
\definemath{\hfagLFBD}{\hfagLFBDval\hfagLFBDerr}
\definemath{\hfagZFBSBD}{\hfagZFBSBDval\hfagZFBSBDerr}
\definemath{\hfagWFBSBD}{\hfagWFBSBDval\hfagWFBSBDerr}
\definemath{\hfagTFBSBD}{\hfagTFBSBDval\hfagTFBSBDerr}
\definemath{\hfagLFBSBD}{\hfagLFBSBDval\hfagLFBSBDerr}
\definemath{\hfagDMDH}{\hfagDMDHval\hfagDMDHerr\invps}
\definemath{\hfagDMDHnounit}{\hfagDMDHval\hfagDMDHerr}
\definemath{\hfagDMDHfull}{\hfagDMDHval\hfagDMDHsta\hfagDMDHsys\invps}
\definemath{\hfagDMDHnounitfull}{\hfagDMDHval\hfagDMDHsta\hfagDMDHsys}
\definemath{\hfagDMDB}{\hfagDMDBval\hfagDMDBerr\invps}
\definemath{\hfagDMDBnounit}{\hfagDMDBval\hfagDMDBerr}
\definemath{\hfagDMDBfull}{\hfagDMDBval\hfagDMDBsta\hfagDMDBsys\invps}
\definemath{\hfagDMDBnounitfull}{\hfagDMDBval\hfagDMDBsta\hfagDMDBsys}
\definemath{\hfagDMDTWOD}{\hfagDMDTWODval\hfagDMDTWODerr\invps}
\definemath{\hfagDMDTWODnounit}{\hfagDMDTWODval\hfagDMDTWODerr}
\definemath{\hfagDMDTWODfull}{\hfagDMDTWODval\hfagDMDTWODsta\hfagDMDTWODsys\invps}
\definemath{\hfagDMDTWODnounitfull}{\hfagDMDTWODval\hfagDMDTWODsta\hfagDMDTWODsys}
\definemath{\hfagTAUBDTWOD}{\hfagTAUBDTWODval\hfagTAUBDTWODerr\ps}
\definemath{\hfagTAUBDTWODnounit}{\hfagTAUBDTWODval\hfagTAUBDTWODerr}
\definemath{\hfagTAUBDTWODfull}{\hfagTAUBDTWODval\hfagTAUBDTWODsta\hfagTAUBDTWODsys\ps}
\definemath{\hfagTAUBDTWODnounitfull}{\hfagTAUBDTWODval\hfagTAUBDTWODsta\hfagTAUBDTWODsys}
\definemath{\hfagQPDB}{\hfagQPDBval\hfagQPDBerr}
\definemath{\hfagQPDW}{\hfagQPDWval\hfagQPDWerr}
\definemath{\hfagQPDA}{\hfagQPDAval\hfagQPDAerr}
\definemath{\hfagASLDB}{\hfagASLDBval\hfagASLDBerr}
\definemath{\hfagASLDW}{\hfagASLDWval\hfagASLDWerr}
\definemath{\hfagASLDA}{\hfagASLDAval\hfagASLDAerr}
\definemath{\hfagREBDB}{\hfagREBDBval\hfagREBDBerr}
\definemath{\hfagREBDW}{\hfagREBDWval\hfagREBDWerr}
\definemath{\hfagREBDA}{\hfagREBDAval\hfagREBDAerr}
\definemath{\hfagASLSW}{\hfagASLSWval\hfagASLSWerr}
\definemath{\hfagASLSWfull}{\hfagASLSWval\hfagASLSWsta\hfagASLSWsys}
\definemath{\hfagQPSW}{\hfagQPSWval\hfagQPSWerr}
\definemath{\hfagQPSWfull}{\hfagQPSWval\hfagQPSWsta\hfagQPSWsys}
\definemath{\hfagASLS}{\hfagASLSval\hfagASLSerr}
\definemath{\hfagQPS}{\hfagQPSval\hfagQPSerr}
\definemath{\hfagASLD}{\hfagASLDval\hfagASLDerr}
\definemath{\hfagQPD}{\hfagQPDval\hfagQPDerr}
\definemath{\hfagREBD}{\hfagREBDval\hfagREBDerr}
\definemath{\hfagDMS}{\hfagDMSval\hfagDMSerr\invps}
\definemath{\hfagDMSnounit}{\hfagDMSval\hfagDMSerr}
\definemath{\hfagDMSfull}{\hfagDMSval\hfagDMSsta\hfagDMSsys\invps}
\definemath{\hfagDMSnounitfull}{\hfagDMSval\hfagDMSsta\hfagDMSsys}
\definemath{\hfagXS}{\hfagXSval\hfagXSerr}
\definemath{\hfagCHIS}{\hfagCHISval\hfagCHISerr}
\definemath{\hfagRATIODMDDMS}{\hfagRATIODMDDMSval\hfagRATIODMDDMSerr}
\definemath{\hfagVTDVTS}{\hfagVTDVTSval\hfagVTDVTSerr}
\definemath{\hfagVTDVTSfull}{\hfagVTDVTSval\hfagVTDVTSexx\hfagVTDVTSthe}
\definemath{\hfagXI}{\hfagXIval\hfagXIerr}
\definemath{\hfagBETASCOMBerr}{\hfagBETASCOMBerp\hfagBETASCOMBern}
\definemath{\hfagBETASCOMB}{\hfagBETASCOMBval\hfagBETASCOMBerr}
\definemath{\hfagPHISCOMBerr}{\hfagPHISCOMBerp\hfagPHISCOMBern}
\definemath{\hfagPHISCOMB}{\hfagPHISCOMBval\hfagPHISCOMBerr}
\definemath{\hfagDGSCOMB}{\hfagDGSCOMBval\hfagDGSCOMBerr\invps}
\definemath{\hfagDGSCOMBnounit}{\hfagDGSCOMBval\hfagDGSCOMBerr}
\definemath{\hfagBETASCOMBCONerr}{\hfagBETASCOMBCONerp\hfagBETASCOMBCONern}
\definemath{\hfagBETASCOMBCON}{\hfagBETASCOMBCONval\hfagBETASCOMBCONerr}
\definemath{\hfagPHISCOMBCONerr}{\hfagPHISCOMBCONerp\hfagPHISCOMBCONern}
\definemath{\hfagPHISCOMBCON}{\hfagPHISCOMBCONval\hfagPHISCOMBCONerr}
\definemath{\hfagDGSCOMBCONerr}{\hfagDGSCOMBCONerp\hfagDGSCOMBCONern}
\definemath{\hfagDGSCOMBCON}{\hfagDGSCOMBCONval\hfagDGSCOMBCONerr\invps}
\definemath{\hfagDGSCOMBCONnounit}{\hfagDGSCOMBCONval\hfagDGSCOMBCONerr}


\mysection{\b-hadron production fractions, lifetimes and mixing parameters}
\labs{life_mix}


Quantities such as \b-hadron production fractions, \b-hadron lifetimes, 
and neutral \B-meson oscillation frequencies have been studied
in the nineties at LEP and SLC 
(\ee colliders at $\sqrt{s}=m_{\particle{Z}}$) 
as well as at the 
first version of the Tevatron
(\particle{p\bar{p}} collider at $\sqrt{s}=1.8\TeV$). 
Since then 
precise measurements of the \Bd and \Bu mesons
have also been performed at the 
asymmetric \B factories, KEKB and PEPII
(\ee colliders at $\sqrt{s}=m_{\Ups}$) while measurements related 
to the other \b-hadrons, in particular \Bs, \Bc and \Lb, 
have been performed at the upgraded Tevatron ($\sqrt{s}=1.96\TeV$)
and are continuing at the LHC ($pp$ collider at $\sqrt{s}=7\TeV$).
In most cases, these basic quantities, although interesting by themselves,
became necessary ingredients for the more complicated and 
refined analyses at the asymmetric \B factories, 
the Tevatron and the LHC,
in particular the time-dependent \CP asymmetry measurements.
It is therefore important that the best experimental
values of these quantities continue to be kept up-to-date and improved. 

In several cases, the averages presented in this chapter are 
needed and used as input for the results given in the subsequent chapters. 
Within this chapter, some averages need the knowledge of other 
averages in a circular way. This coupling, which appears through the 
\b-hadron fractions whenever inclusive or semi-exclusive measurements 
have to be considered, has reduced drastically in the past several years 
with increasingly precise exclusive measurements becoming available
and dominating practically all averages. 

In addition to \b-hadron fractions, lifetimes and 
mixing parameters, this chapter also deals with the 
\CP-violating phase $\phiccbars\simeq -2\beta_s$, which is the phase 
difference between the \Bs mixing amplitude and the 
$b\to c\bar{c}s$ decay amplitude. The angle $\beta$, 
which is the equivalent of $\beta_s$ for the \Bd 
system, is discussed in Chapter~\ref{sec:cp_uta}. 

\mysubsection{\b-hadron production fractions}
\labs{fractions}
 
We consider here the relative fractions of the different \b-hadron 
species found in an unbiased sample of weakly-decaying \b hadrons 
produced under some specific conditions. The knowledge of these fractions
is useful to characterize the signal composition in inclusive \b-hadron 
analyses, to predict the background composition in exclusive analyses, 
or to convert (relative) observe rates into (relative) branching fraction 
measurements. 
Many \B-physics analyses need these fractions as input. We distinguish 
here the following three conditions: \Ups decays, \Upsfive decays, and 
high-energy collisions (including \Z decays). 

\mysubsubsection{\b-hadron production fractions in \Ups decays}
\labs{fraction_Ups4S}

Only pairs of the two lightest (charged and neutral) \B mesons 
can be produced in \Ups decays, 
and it is enough to determine the following branching 
fractions:
\begin{eqnarray}
f^{+-} & = & \Gamma(\Ups \to \particle{B^+B^-})/
             \Gamma_{\rm tot}(\Ups)  \,, \\
f^{00} & = & \Gamma(\Ups \to \particle{B^0\bar{B}^0})/
             \Gamma_{\rm tot}(\Ups) \,.
\end{eqnarray}
In practice, most analyses measure their ratio
\begin{equation}
R^{+-/00} = f^{+-}/f^{00} = \Gamma(\Ups \to \particle{B^+B^-})/
             \Gamma(\Ups \to \particle{B^0\bar{B}^0}) \,,
\end{equation}
which is easier to access experimentally.
Since an inclusive (but separate) reconstruction of 
\Bu and \Bd is difficult, specific exclusive decay modes, 
${\Bu} \to x^+$ and ${\Bd} \to x^0$, are usually considered to perform 
a measurement of $R^{+-/00}$, whenever they can be related by 
isospin symmetry (for example \particle{\Bu \to \jpsi K^+} and 
\particle{\Bd \to \jpsi K^0}).
Under the assumption that $\Gamma(\Bu \to x^+) = \Gamma(\Bd \to x^0)$, 
\ie\ that isospin invariance holds in these \B decays,
the ratio of the number of reconstructed
$\Bu \to x^+$ and $\Bd \to x^0$ mesons is proportional to
\begin{equation}
\frac{f^{+-}\, \BR{\Bu\to x^+}}{f^{00}\, \BR{\Bd\to x^0}}
= \frac{f^{+-}\, \Gamma({\Bu}\to x^+)\, \tau(\Bu)}%
{f^{00}\, \Gamma({\Bd}\to x^0)\,\tau(\Bd)}
= \frac{f^{+-}}{f^{00}} \, \frac{\tau(\Bu)}{\tau(\Bd)}  \,, 
\end{equation} 
where $\tau(\Bu)$ and $\tau(\Bd)$ are the \Bu and \Bd 
lifetimes respectively.
Hence the primary quantity measured in these analyses 
is $R^{+-/00} \, \tau(\Bu)/\tau(\Bd)$, 
and the extraction of $R^{+-/00}$ with this method therefore 
requires the knowledge of the $\tau(\Bu)/\tau(\Bd)$ lifetime ratio. 

\begin{table}
\caption{Published measurements of the $\Bu/\Bd$ production ratio
in \Ups decays, together with their average (see text).
Systematic uncertainties due to the imperfect knowledge of 
$\tau(\Bu)/\tau(\Bd)$ are included. The latest \babar result~\cite{Aubert:2004rz}
supersedes the earlier \babar measurements \cite{Aubert:2001xs,Aubert:2004ur}.}
\labt{R_data}
\begin{center}
\begin{tabular}{lccll}
\hline
Experiment & Ref. & Decay modes & Published value of & Assumed value \\
and year & & or method & $R^{+-/00}=f^{+-}/f^{00}$ & of $\tau(\Bu)/\tau(\Bd)$ \\
\hline
CLEO,   2001 & \cite{Alexander:2000tb}  & \particle{\jpsi K^{(*)}} 
             & $1.04 \pm0.07 \pm0.04$ & $1.066 \pm0.024$ \\
\babar, 2002 & \cite{Aubert:2001xs} & \particle{(c\bar{c})K^{(*)}}
             & $1.10 \pm0.06 \pm0.05$ & $1.062 \pm0.029$\\ 
CLEO,   2002 & \cite{Athar:2002mr}  & \particle{D^*\ell\nu}
             & $1.058 \pm0.084 \pm0.136$ & $1.074 \pm0.028$\\
\belle, 2003 & \cite{Hastings:2002ff} & dilepton events 
             & $1.01 \pm0.03 \pm0.09$ & $1.083 \pm0.017$\\
\babar, 2004 & \cite{Aubert:2004ur} & \particle{\jpsi K}
             & $1.006 \pm0.036 \pm0.031$ & $1.083 \pm0.017$ \\
\babar, 2005 & \cite{Aubert:2004rz} & \particle{(c\bar{c})K^{(*)}}
             & $1.06 \pm0.02 \pm0.03$ & $1.086 \pm0.017$\\ 
\hline
Average      & & & \hfagFF~(tot) & \hfagRTAUBU \\
\hline
\end{tabular}
\end{center}
\end{table}

The published measurements of $R^{+-/00}$ are listed 
in \Table{R_data} together with the corresponding assumed values of 
$\tau(\Bu)/\tau(\Bd)$.
All measurements are based on the above-mentioned method, 
except the one from \belle, which is a by-product of the 
\Bd mixing frequency analysis using dilepton events
(but note that it also assumes isospin invariance, 
namely $\Gamma(\Bu \to \ell^+{\rm X}) = \Gamma(\Bd \to \ell^+{\rm X})$).
The latter is therefore treated in a slightly different 
manner in the following procedure used to combine 
these measurements:
\begin{itemize} 
\item each published value of $R^{+-/00}$ from CLEO and \babar
      is first converted back to the original measurement of 
      $R^{+-/00} \, \tau(\Bu)/\tau(\Bd)$, using the value of the 
      lifetime ratio assumed in the corresponding analysis;
\item a simple weighted average of these original
      measurements of $R^{+-/00} \, \tau(\Bu)/\tau(\Bd)$ from 
      CLEO and \babar (which do not depend on the assumed value 
      of the lifetime ratio) is then computed, assuming no 
      statistical or systematic correlations between them;


\item the weighted average of $R^{+-/00} \, \tau(\Bu)/\tau(\Bd)$ 
      is converted into a value of $R^{+-/00}$, using the latest 
      average of the lifetime ratios, $\tau(\Bu)/\tau(\Bd)=\hfagRTAUBU$ 
      (see \Sec{lifetime_ratio});
\item the \belle measurement of $R^{+-/00}$ is adjusted to the 
      current values of $\tau(\Bd)=\hfagTAUBD$ and 
      $\tau(\Bu)/\tau(\Bd)=\hfagRTAUBU$ (see \Sec{lifetime_ratio}),
      using the quoted systematic uncertainties due to these parameters;
\item the combined value of $R^{+-/00}$ from CLEO and \babar is averaged 
      with the adjusted value of $R^{+-/00}$ from \belle, assuming a 100\% 
      correlation of the systematic uncertainty due to the limited 
      knowledge on $\tau(\Bu)/\tau(\Bd)$; no other correlation is considered. 
\end{itemize} 
The resulting global average, 
\begin{equation}
R^{+-/00} = \frac{f^{+-}}{f^{00}} =  \hfagFF \,,
\labe{Rplusminus}
\end{equation}
is consistent with an equal production of charged and neutral \B mesons, 
although only at the $\hfagNSIGMAFF\,\sigma$ level.

On the other hand, the \babar collaboration has 
performed a direct measurement of the $f^{00}$ fraction 
using an original method, which does not rely on isospin symmetry nor requires 
the knowledge of $\tau(\Bu)/\tau(\Bd)$. Its analysis, 
based on a comparison between the number of events where a single 
$B^0 \to D^{*-} \ell^+ \nu$ decay could be reconstructed and the number 
of events where two such decays could be reconstructed, yields~\cite{Aubert:2005bq}
\begin{equation}
f^{00}= 0.487 \pm 0.010\,\mbox{(stat)} \pm 0.008\,\mbox{(syst)} \,.
\labe{fzerozero}
\end{equation}

The two results of \Eqss{Rplusminus}{fzerozero} are of very different natures 
and completely independent of each other. 
Their product is equal to $f^{+-} = \hfagFPROD$, 
while another combination of them gives $f^{+-} + f^{00}= \hfagFSUM$, 
compatible with unity.
Assuming\footnote{A few non-$\B\bar{B}$
decay modes of the $\Upsilon(4S)$ 
($\Upsilon(1S)\pi^+\pi^-$,
$\Upsilon(2S)\pi^+\pi^-$, $\Upsilon(1S)\eta$) 
have been observed with branching fractions
of the order of $10^{-4}$~\cite{Aubert:2006bm,*Sokolov:2006sd,*Aubert_mod:2008bv},
corresponding to a partial
width several times larger than that in the \ee channel.
However, this can still be
neglected and the assumption $f^{+-}+f^{00}=1$ remains valid
in the present context of the determination of $f^{+-}$ and $f^{00}$.}
 $f^{+-}+f^{00}= 1$, also consistent with 
CLEO's observation that the fraction of \Ups decays 
to \BB pairs is larger than 0.96 at \CL{95}~\cite{Barish:1995cx},
the results of \Eqss{Rplusminus}{fzerozero}
can be averaged (first converting \Eq{Rplusminus} 
into a value of $f^{00}=1/(R^{+-/00}+1)$) 
to yield the following more precise estimates:
\begin{equation}
f^{00} = \hfagFNW  \,,~~~ f^{+-} = 1 -f^{00} =  \hfagFCW \,,~~~
\frac{f^{+-}}{f^{00}} =  \hfagFFW \,.
\end{equation}
The latter ratio differs from one by $\hfagNSIGMAFFW\,\sigma$.

\mysubsubsection{\b-hadron production fractions in \Upsfive decays}
\labs{fraction_Ups5S}

\newcommand{\fsfive}{\ensuremath{f^{\Upsfive}_{s}}}
\newcommand{\fudfive}{\ensuremath{f^{\Upsfive}_{u,d}}}
\newcommand{\fnBfive}{\ensuremath{f^{\Upsfive}_{B\!\!\!\!/}}}

Hadronic events produced in $e^+e^-$ collisions at the \Upsfive energy
can be classified into three categories: 
light-quark ($u$, $d$, $s$, $c$) continuum events, $b\bar{b}$ continuum events,
and \Upsfive events. The latter two cannot be distinguished and will be called
$b\bar{b}$ events in the following. These $b\bar{b}$ events, which also include 
$b\bar{b}\gamma$ events because of possible initial-state radiation, 
can hadronize in different final states.
We define \fudfive\ as
the fraction of $b\bar{b}$ events with a pair of non-strange 
bottom mesons 
($B\bar{B}$, $B\bar{B}^*$, $B^*\bar{B}$, $B^*\bar{B}^*$,
$B\bar{B}\pi$, $B\bar{B}^*\pi$, $B^*\bar{B}\pi$,
$B^*\bar{B}^*\pi$, and $B\bar{B}\pi\pi$ final states, 
where
$B$ denotes a $B^0$ or $B^+$ meson and 
$\bar{B}$ denotes a $\bar{B}^0$ or $B^-$ meson), \fsfive\ as
the fraction of $b\bar{b}$ events with a pair of strange bottom mesons
($B_s^0\bar{B}_s^0$, $B_s^0\bar{B}_s^{0*}$, $B_s^{0*}\bar{B}_s^0$, and
$B_s^{0*}\bar{B}_s^{0*}$ final states), and 
\fnBfive\ as the fraction of $b\bar{b}$ events without 
bottom meson in the final state.
Note that the excited bottom-meson states decay via $B^* \to B \gamma$ and
$B_s^{0*} \to B_s^0 \gamma$.
These fractions satisfy
\begin{equation}
\fudfive + \fsfive + \fnBfive = 1 \,.
\labe{sum_frac_five}
\end{equation} 

\begin{table}
\caption{Published measurements of \fsfive.
All values have been obtained assuming $\fnBfive=0$. 
They are quoted 
as in the original publications, except for the most
recent measurement which is quoted as 
$1-\fudfive$, with \fudfive\ from \Ref{Drutskoy:2010an}.
The last line gives our average of \fsfive\ assuming $\fnBfive=0$.}
\labt{fsFiveS}
\begin{center}
\begin{tabular}{lll}
\hline
Experiment, year, dataset                 & Decay mode or method    & Value of \fsfive\  \\
\hline
CLEO, 2006, 0.42\invfb~\cite{Huang:2006em_mod}     & $\Upsfive\to D_{s}X$     & $0.168 \pm 0.026^{+0.067}_{-0.034}$  \\
             & $\Upsfive \to \phi X$    & $0.246 \pm 0.029^{+0.110}_{-0.053}$ \\
             & $\Upsfive \to B\bar{B}X$ & $0.411 \pm 0.100 \pm 0.092$ \\  
             & CLEO average of above 3  & $0.21^{+0.06}_{-0.03}$      \\  \hline
Belle, 2006, 1.86\invfb~\cite{Drutskoy:2006fg} & $\Upsfive \to D_s X$     & $0.179 \pm 0.014 \pm 0.041$ \\
             & $\Upsfive \to D^0 X$     & $0.181 \pm 0.036 \pm 0.075$ \\  
             & Belle average of above 2 & $0.180 \pm 0.013 \pm 0.032$ \\  \hline 
Belle, 2010, 23.6\invfb~\cite{Drutskoy:2010an} & $\Upsfive \to B\bar{B}X$ & $0.263 \pm 0.032 \pm 0.051$ 
\\ \hline
\multicolumn{2}{l}{Average of all above 
after adjustments to inputs of \Table{fsFiveS_external}} & 
\hfagFSFIVERL             \\  \hline 
\end{tabular}
\end{center}
\end{table}

\begin{table}
\caption{External inputs on which the \fsfive\ averages are based.}
\labt{fsFiveS_external}
\begin{center}
\begin{tabular}{lcl}
\hline
Branching fraction   & Value     & Explanation and reference \\
\hline
${\cal B}(B\to D_s X)\times {\cal B}(D_s \to \phi\pi)$ & 
$0.00374\pm 0.00014$ & derived from~\cite{PDG_2012} \\
${\cal B}(B^0_s \to D_s X)$ & 
$0.92\pm0.11$ & model-dependent estimate~\cite{Artuso:2005xw} \\
${\cal B}(D_s \to \phi\pi)$ & 
$0.045\pm0.004$ & \cite{PDG_2012} \\
${\cal B}(B\to D^0 X)\times {\cal B}(D^0 \to K\pi)$ & 
$0.0243\pm0.0011$ & derived from~\cite{PDG_2012} \\
${\cal B}(B^0_s \to D^0 X)$ & 
$0.08\pm0.07$ & model-dependent estimate~\cite{Drutskoy:2006fg,Artuso:2005xw} \\
${\cal B}(D^0 \to K\pi)$ & 
$0.0387\pm0.0005$ & \cite{PDG_2012} \\
${\cal B}(B \to \phi X)$ & 
$0.0343\pm0.0012$ & world average~\cite{PDG_2012,Huang:2006em_mod} \\
${\cal B}(B^0_s \to \phi X)$ &
$0.161\pm0.024$ & model-dependent estimate~\cite{Huang:2006em_mod} \\
\hline
\end{tabular}
\end{center}
\end{table}

The CLEO and Belle collaborations have published in 2006
measurements of several inclusive \Upsfive branching fractions, 
${\cal B}(\Upsfive\to D_s X)$, 
${\cal B}(\Upsfive\to \phi X)$ and 
${\cal B}(\Upsfive\to D^0 X)$, 
from which they extracted the
model-dependent estimates of \fsfive\
reported in \Table{fsFiveS}. This extraction was performed 
under the implicit assumption  
$\fnBfive=0$, using the relation 
\begin{equation}
\frac12{\cal B}(\Upsfive\to D_s X)=\fsfive\times{\cal B}(B_s^0\to D_s X) + 
\left(1-\fsfive-\fnBfive\right)\times{\cal B}(B\to D_s X) \,,
\labe{Ds_correct}
\end{equation}
and similar relations for
${\cal B}(\Upsfive\to D^0 X)$ and ${\cal B}(\Upsfive\to \phi X)$.
We list also in \Table{fsFiveS} 
the values of \fsfive\ derived from measurements of
$\fudfive={\cal B}(\Upsfive\to B\bar BX)$~\cite{Huang:2006em_mod,Drutskoy:2010an}, as well as our average value of  \fsfive,
all obtained under the assumption $\fnBfive=0$.

However, the assumption $\fnBfive=0$ is no longer valid since the 
observation of \Upsfive\ decays to $\Upsilon(1S)\pi^+\pi^-$,
$\Upsilon(2S)\pi^+\pi^-$,
$\Upsilon(3S)\pi^+\pi^-$ and
$\Upsilon(1S)K^+K^-$~\cite{Abe:2007tk},
and more recently to 
$h_b(1P)\pi^+\pi^-$ and 
$h_b(2P)\pi^+\pi^-$~\cite{Adachi:2011ji}.
The sum of these measured branching fractions, adding also the 
contributions of the 
$\Upsilon(1S)\pi^0\pi^0$, $\Upsilon(2S)\pi^0\pi^0$, $\Upsilon(3S)\pi^0\pi^0$,
$\Upsilon(1S)K^0\bar{K}^0$, $h_b(1P)\pi^0\pi^0$ and $h_b(2P)\pi^0\pi^0$ final states
assuming isospin conservation, amounts to
$$
{\cal B}(\Upsfive\to (\b\bar{\b})hh) = 0.042\pm0.006 \,,
~~~\mbox{for $(\b\bar{\b})=\Upsilon(1S,2S,3S),h_b(1P,2P)$ and $hh=\pi\pi,KK$}\,,
$$
which is to be considered as a lower bound for \fnBfive. 
Following the method described in \Ref{thesis_Louvot}, 
we perform a $\chi^2$ fit of the original 
measurements of the \Upsfive\ branching fractions of
\Refs{Huang:2006em_mod,Drutskoy:2006fg,Drutskoy:2010an},
using the inputs of \Table{fsFiveS_external},
the relations of \Eqss{sum_frac_five}{Ds_correct} and the
one-sided Gaussian constraint $\fnBfive \ge {\cal B}(\Upsfive \to (\b\bar{\b}) hh)$,
to simultaneously extract \fudfive, \fsfive\ and \fnBfive. Taking all known 
correlations into account, the best fit values are
\begin{eqnarray}
\fudfive &=& \hfagFUDFIVE \,, \labe{fudfive} \\
\fsfive  &=& \hfagFSFIVE  \,, \labe{fsfive}  \\
\fnBfive &=& \hfagFNBFIVE \,, \labe{fnBfive}
\end{eqnarray}
where the strongly asymmetric uncertainty on \fnBfive\ is due to the one-sided constraint
from the observed $(\b\bar{\b}) hh$ decays. These results, together with their correlation, 
imply
\begin{eqnarray}
\fsfive/\fudfive  &=& \hfagFSFUDFIVE  \,, \labe{fsfudfive} 
\end{eqnarray}
in fair agreement with the results of a \babar
analysis~\cite{Lees:2011ji} performed as a function 
of centre-of-mass energy\footnote{
This has not been included in the average, since 
no numerical value is given for $\fsfive/\fudfive$ in 
\Ref{Lees:2011ji}.}.

The production of $B^0_s$ mesons at the \Upsfive
is observed to be dominated by the $B_s^{0*}\bar{B}_s^{0*}$
channel, 
with $\sigma(e^+e^- \to B_s^{0*}\bar{B}_s^{0*})/%
\sigma(e^+e^- \to B_s^{0(*)}\bar{B}_s^{0(*)})
= (87.0\pm 1.7)\%$~\cite{Li:2011pg,Louvot:2008sc}.
The proportion of the various production channels 
for non-strange $B$ mesons have also been measured~\cite{Drutskoy:2010an}.

\mysubsubsection{\b-hadron production fractions at high energy}
\labs{fractions_high_energy}
\labs{chibar}

At high energy, all species of weakly-decaying \b hadrons 
may be produced, either directly or in strong and electromagnetic 
decays of excited \b hadrons. It is often assumed that the fractions 
of these different species are the same in unbiased samples of 
high-$p_{\rm T}$ \b jets originating from \particle{Z^0} decays, 
from \particle{p\bar{p}} collisions at the Tevatron, or from 
\particle{p p} collisions at the LHC.
This hypothesis is plausible under the condition that the square of
the momentum transfer to the produced \b quarks, $Q^2$, is large compared 
with the square of the hadronization energy scale, 
$Q^2 \gg \Lambda_{\rm QCD}^2$.
On the other hand, there is no strong argument to claim that the
fractions at different machines should be strictly equal, so 
this assumption should be checked experimentally. Although the 
available data is not sufficient at this time to perform a definitive
check, it is expected that more refined analyses of the Tevatron Run~II data 
and new analyses from LHC 
experiments may improve this situation and allow one to confirm or 
disprove this assumption with reasonable confidence. Meanwhile, the 
attitude adopted here is that these fractions are assumed to be equal 
at all high-energy colliders until demonstrated otherwise by 
experiment.
However, both CDF and LHCb report a $p_{\rm T}$ dependence for $\Lambda_b$
production relative to \Bu and \Bd; the number of $\Lambda_b$ baryons
observed at low $p_{\rm T}$ is enhanced with respect to that 
seen at LEP at higher $p_{\rm T}$.
Therefore we present 
three sets of complete averages: one set including only measurements 
performed at LEP, a second set including only measurements performed 
at the Tevatron, a third  set including measurements performed at LEP, 
Tevatron and LHCb.  The LHCb production fractions results, by themselves, 
are still incomplete, lacking measurements on the production of other weakly 
decaying heavy flavour baryons, $\Xi_b$ and $\Omega_b$, and a measurement of 
$\overline{\chi}$ giving an extra constraint between \fBd and \fBs.

Contrary to what happens in the charm sector where the fractions of 
\particle{D^+} and \particle{D^0} are different, the relative amount 
of \Bu and \Bd is not affected by the electromagnetic decays of 
excited ${\Bu}^*$ and ${\Bd}^*$ states and strong decays of excited 
${\Bu}^{**}$ and ${\Bd}^{**}$ states. Decays of the type 
\particle{{\Bs}^{**} \to B^{(*)}K} also contribute to the \Bu and \Bd rates, 
but with the same magnitude if mass effects can be neglected.  
We therefore assume equal production of \Bu and \Bd. We also  
neglect the production of weakly-decaying states
made of several heavy quarks (like \Bc and other heavy baryons) 
which is known to be very small. Hence, for the purpose of determining 
the \b-hadron fractions, we use the constraints
\begin{equation}
\fBu = \fBd ~~~~\mbox{and}~~~ \fBu + \fBd + \fBs + \fbb = 1 \,,
\labe{constraints}
\end{equation}
where \fBu, \fBd, \fBs and \fbb
are the unbiased fractions of \Bu, \Bd, \Bs and \b baryons, respectively.

The LEP experiments have measured
$\fBs \times \BR{\Bs\to\particle{D_s^-} \ell^+ \nu_\ell \mbox{$X$}}$~\cite{Abreu:1992rv,*Acton:1992zq,*Buskulic:1995bd}, 
$\BR{\b\to\Lb} \times \BR{\Lb\to\Lc\ell^-\bar{\nu}_\ell \mbox{$X$}}$~\cite{Abreu:1995me,Barate:1997if}
and $\BR{\b\to\Xib^-} \times \BR{\Xi_b^- \to \Xi^-\ell^-\overline\nu_\ell 
\mbox{$X$}}$~\cite{Buskulic:1996sm,Abdallah:2005cw}\footnote{The DELPHI result 
of \Ref{Abdallah:2005cw} is considered to supersede an older 
one~\cite{Abreu:1995kt}.} 
from partially reconstructed final states including a lepton, \fbb
from protons identified in \b events~\cite{Barate:1997ty}, and the 
production rate of charged \b hadrons~\cite{Abdallah:2003xp}. 
Ratios of \b-hadron fractions have been measured at CDF using 
lepton+charm final 
states~\cite{Affolder:1999iq,Aaltonen:2008zd,Aaltonen:2008eu}\footnote{CDF 
updated their measurement of \fLb/\fBd~\cite{Affolder:1999iq} to account 
for a measured $p_{\rm T}$ dependence between exclusively reconstructed 
$\Lambda_b$ and $B^0$~\cite{Aaltonen:2008eu}.} and double semileptonic decays 
with \particle{K^*\mu\mu} and \particle{\phi\mu\mu}
final states~\cite{Abe:1999ta}.  Measurements of the production of other heavy 
flavour baryons at the Tevatron are included in the determination of 
\fbb~\cite{Abazov_mod:2007ub,Abazov:2008qm,Aaltonen:2009ny}\footnote{\dzero reports 
$f_{\Omega_b^-}/f_{\Xi_b^-}$.  We use the CDF+\dzero average of 
$f_{\Xi_b^-}/f_{\Lambda_b}$ to obtain $f_{\Omega_b^-}/f_{\Lambda_b}$ and then 
combine with the CDF result.} using the constraint
\begin{eqnarray}
\fbb & = & f_{\Lambda_b} + f_{\Xi_b^0} + f_{\Xi_b^-} + f_{\Omega_b^-} 
     \nonumber \\
     & = & f_{\Lambda_b}\left(1 + 2\frac{f_{\Xi_b^-}}{f_{\Lambda_b}} 
           + \frac{f_{\Omega_b^-}}{f_{\Lambda_b}}\right),
\end{eqnarray}
where isospin invariance is assumed in the production of $\Xi_b^0$ and 
$\Xi_b^-$. Other \b-baryons are expected to decay strongly or 
electromagnetically to those baryons listed. For the production 
measurements, both CDF and \dzero\ reconstruct their \b-baryons exclusively 
to final states which include a $\jpsi$ and a hyperon 
($\Lambda_b\rightarrow \jpsi \Lambda$, 
$\Xi_b^- \rightarrow \jpsi \Xi^-$ and 
$\Omega_b^- \rightarrow \jpsi \Omega^-$).  
We assume that the partial decay width of a \b-baryon to a $\jpsi$ and the 
corresponding hyperon is equal to the partial width of any other \b-baryon to 
a $\jpsi$ and the corresponding hyperon.  LHCb has also measured
ratios of \b-hadron fractions, $\fBs/(\fBu+\fBd)$ and $\fLb/(\fBu+\fBd)$, in 
lepton+charm final states~\cite{Aaij:2011jp} and $\fBs/\fBd$ 
in fully reconstructed hadronic final states using theoretical
values for the branching fractions of two-body \Bs and \Bd 
decays~\cite{Aaij:2011hi}.

Both CDF and LHCb observe a $p_{\rm T}$ dependence in the relative fractions 
$\fLb/\fBd$~\cite{Aaltonen:2008eu,Aaij:2011jp}\footnote{CDF compares the 
$p_{\rm T}$ distribution of fully reconstructed 
$\Lambda_b\rightarrow\Lambda_c^+ \pi^-$ 
with $\overline{B^0}\rightarrow\D^+\pi^-$ which 
compares $\fLb/\fBd$ up to a scale factor. LHCb compares the $p_{\rm T}$ 
in the lepton+charm system between $\Lambda_b$ and \Bd and \Bu comparing 
$R_{\Lambda_b} = \fLb/(\fBu+\fBd) = \fLb/2\fBd$.}.
No $p_{\rm T}$ dependence is yet observed for $\fBs/(\fBu+\fBd)$.
CDF chose to correct an older result to account for the $p_{\rm T}$ dependence 
whereas LHCb chose to report a linear dependence of $\fLb/(\fBu + \fBd)$,
which yields unphysical results for $p_{\rm T} > 32~{\rm GeV}/c$.  In a second 
result, CDF binned their data in $p_{\rm T}$ of the electron+charm system.
\Figure{rlb_comb} shows the ratio $R_{\Lambda_b} = \fLb/(\fBu + \fBd)$
as a function of this $p_{\rm T}$, as measured by both CDF and LHCb.
Two fits are performed. The first fit using the LHCb
parameterization yields 
$R_{\Lambda_b} = (0.386\pm 0.21)\left[1 - (0.0270\pm 0.0056)\times p_{\rm T}\right]$.
A second fit using a simple exponential yields 
$R_{\Lambda_b} = \exp\left\{ (-0.928\pm 0.066) - (0.0344\pm 0.0086)\times p_{\rm T}\right\}$.
A common systematic uncertainty of 26\% on the scale of both results arises from the 
$\Lambda_c^+ \to p K^-\pi^+$ branching fraction.
The quality of the two fits are similar,
but the second parameterization gives a physical result for all $p_{\rm T}$.  A value of
$R_{\Lambda_b}$ is also calculated for LEP and placed at the approximate $p_{\rm T}$ for the lepton+charm
system, but this value does not participate in any fit.  Note that the $p_{\rm T}$ dependence
of $R_{\Lambda_b}$ combined with the constraint in \Eq{constraints} implies
a compensating $p_{\rm T}$ dependence in one or more of the production fractions, \fBu, \fBd,
or \fBs.

\begin{figure}
 \begin{center}
  \epsfig{figure=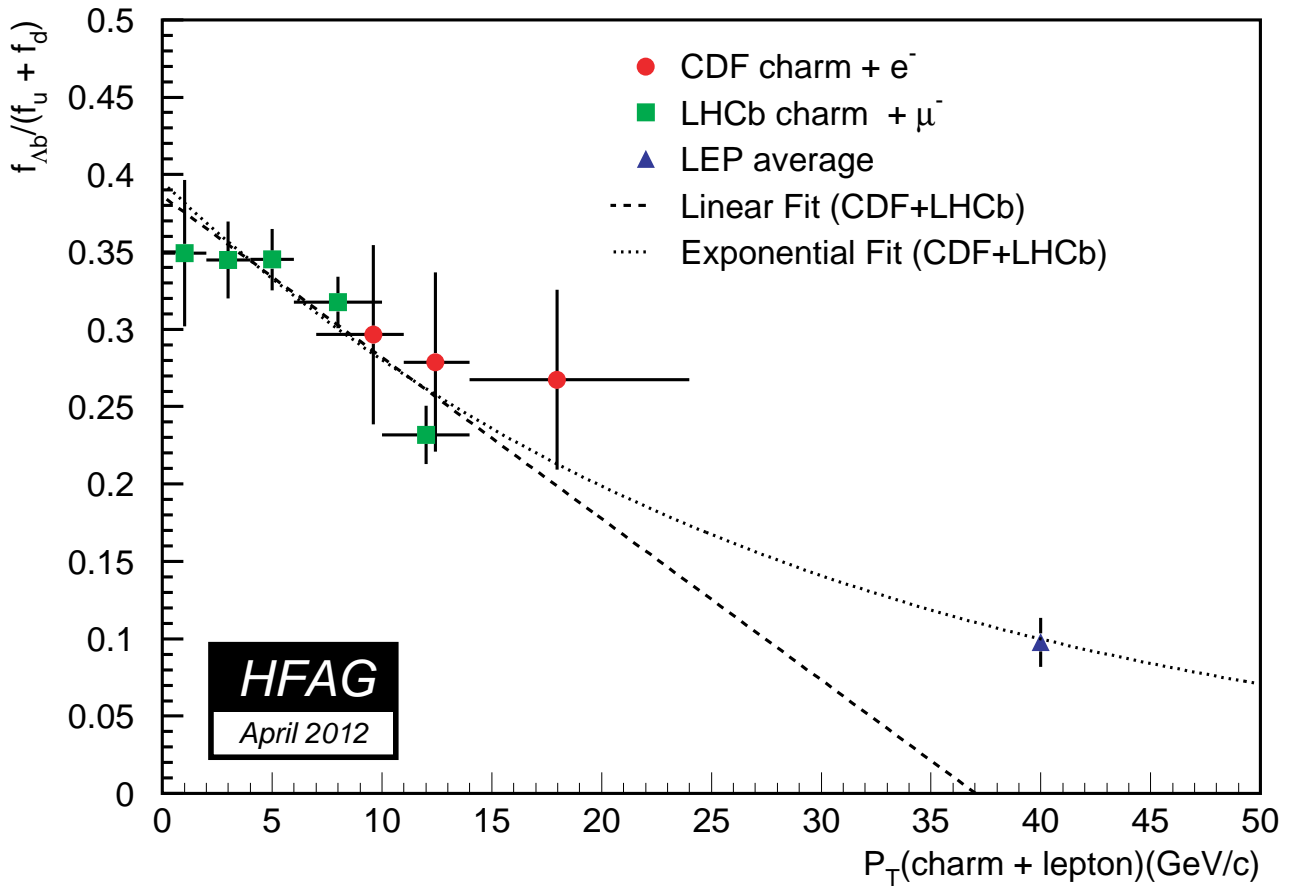,width=\textwidth}
  \caption{Ratio of production fractions $\fLb/(\fBu + \fBd)$ as 
   a function of $p_{\rm T}$ of the lepton+charm system for CDF~\cite{Aaltonen:2008eu}
   and LHCb~\cite{Aaij:2011jp} data. 
   A scale uncertainty due to the common systematic uncertainty 
   from the $\Lambda_c^+ \rightarrow pK^-\pi^+$ branching fraction
   is omitted. 
   The curves represent fits to the data:
   a linear fit using the LHCb parameterization (dashed), 
   and an exponential
   fit described in the text (dotted).
   The computed LEP ratio is included at an 
   approximate $p_{\rm T}$ in $Z$ decays, but does not participate in any fit.}
  \labf{rlb_comb}
 \end{center}
\end{figure}

In order to combine or compare LHCb results with other experiments,
the $p_{\rm T}$-dependent $\fLb/(\fBu + \fBd)$ is weighted by the $p_{\rm T}$ spectrum\footnote{In
practice the LHCb data are given in 14 bins in $p_{\rm T}$ and $\eta$ with a full covariance
matrix~\cite{Aaij:2011jp}. The weighted average is calculated as
$D^T C^{-1} M/\sigma$, where $\sigma = D^T C^{-1} D$, $M$ is a vector 
of measurements, $C^{-1}$ is the inverse covariance matrix and $D^T$ is the 
transpose of the design matrix (vector of 1's)}. \Table{LHCBcomp} compares 
the $p_{\rm T}$-weighted LHCb data with comparable averages from the CDF. 
The average CDF 
and LHCb data are in good agreement despite the \b hadrons being produced 
in different kinematic regimes.

\begin{table}
 \caption{Comparison of average production fraction ratios from CDF and LHCb.
 The kinematic regime of the lepton+charm system reconstructed in each
 experiment is also shown.}
 \labt{LHCBcomp}
 \begin{center}
  \begin{tabular}{lccc}
   \hline
   Quantity                         & CDF               & LHCb \\
   \hline
   $\fBs/(\fBu + \fBd)$             & \hfagRBSTEVNOCON  & \hfagRBSLHCBNOCON   \\
   $\fLb/(\fBu + \fBd)$             & \hfagRLBTEVNOCON  & \hfagRLBLHCBNOCON   \\
   Average lepton+charm $p_{\rm T}$ & $\sim 13$~GeV/$c$ & $\sim 7$~GeV/$c$ \\
   Pseudo-rapidity range             & $-1 < \eta < 1$   & $2 < \eta < 5$      \\
   \hline
  \end{tabular}
 \end{center}
\end{table}

All these published results have been combined following the procedure and 
assumptions described in \Ref{Abbaneo:2000ej_mod,*Abbaneo:2001bv_mod_cont},
to yield $\fBu=\fBd=\hfagWFBDNOMIX$, 
$\fBs=\hfagWFBSNOMIX$ and $\fbb=\hfagWFBBNOMIX$
under the constraints of \Eq{constraints}.  
Repeating the combinations, for LEP and the Tevatron, we obtain 
$\fBu=\fBd=\hfagZFBDNOMIX$,
$\fBs=\hfagZFBSNOMIX$ and $\fbb=\hfagZFBBNOMIX$ when using the LEP data only,
$\fBu=\fBd=\hfagTFBDNOMIX$, $\fBs=\hfagTFBSNOMIX$ 
$\fbb = \hfagTFBBNOMIX$ when using the Tevatron data only.  As noted previously,
the LHCb data are insufficient to determine a complete set of \b-hadron production
fractions. The world averages (LEP, Tevatron and LHCb) for the various fractions 
are presented here for comparison with previous averages.  Significant differences
exist between the LEP and Tevatron fractions, therefore use of the world averages
should be taken with some care.
For these combinations other external inputs are used, 
\eg\ the branching ratios of \B mesons to final states with a \particle{D}, 
\particle{D^*} or \particle{D^{**}} in semileptonic decays, which are needed 
to evaluate the fraction of semileptonic \Bs decays with a \particle{D_s^-} 
in the final state.

Time-integrated mixing analyses performed with lepton pairs 
from \particle{b\bar{b}} 
events produced at high-energy colliders measure the quantity 
\begin{equation}
\chibar = f'_{\particle{d}} \,\chid + f'_{\particle{s}} \,\chis \,,
\labe{chibar}
\end{equation}
where $f'_{\particle{d}}$ and $f'_{\particle{s}}$ are 
the fractions of \Bd and \Bs hadrons 
in a sample of semileptonic \b-hadron decays, and where \chid and \chis 
are the \Bd and \Bs time-integrated mixing probabilities.
Assuming that all \b hadrons have the same semileptonic decay width implies 
$f'_i = f_i R_i$, where $R_i = \tau_i/\tau_{\b}$ is the ratio of the lifetime 
$\tau_i$ of species $i$ to the average \b-hadron lifetime 
$\tau_{\b} = \sum_i f_i \tau_i$.
Hence measurements of the mixing probabilities
\chibar, \chid and \chis can be used to improve our 
knowledge of \fBu, \fBd, \fBs and \fbb.
In practice, the above relations yield another determination of 
\fBs obtained from \fbb and mixing information, 
\begin{equation}
\fBs = \frac{1}{R_{\particle{s}}}
\frac{(1+r)\overline{\chi}-(1-\fbb R_{\rm baryon}) \chid}{(1+r)\chis - \chid} \,,
\labe{fBs-mixing}
\end{equation}
where $r=R_{\particle{u}}/R_{\particle{d}} = \tau(\Bu)/\tau(\Bd)$.

The published measurements of \chibar performed by the LEP
experiments have been combined by the LEP Electroweak Working Group to yield 
$\chibar = \hfagCHIBARLEP$~\cite{LEPEWWG:2005ema_mod}.
This can be compared with the Tevatron average, $\chibar = \hfagCHIBARTEV$,
obtained from \dzero~\cite{Abazov:2006qw} and CDF~\cite{CDFnote10335:2011}
measurements with Run~II data.\footnote{
As explained in \Ref{CDFnote10335:2011}, a previous CDF analysis~\cite{Acosta:2003ie_mod}
performed with Run~I data overlooked a background component, so the corresponding result is not 
included in the average.}
The two averages agree, showing no evidence that the production fractions
of \Bd and \Bs mesons at the \particle{Z} peak or at the Tevatron are different.
We combine these two results in a simple weighted average,
assuming no correlations, and obtain 
$\chibar = \hfagCHIBAR$.

\begin{table}
\caption{Time-integrated mixing probability \chibar (defined in \Eq{chibar}), 
and fractions of the different \b-hadron species in an unbiased sample of 
weakly-decaying \b hadrons, obtained from both direct
and mixing measurements. The correlation coefficients between the fractions are 
also given.
The last column includes measurements performed at LEP, Tevatron and LHCb.}
\labt{fractions}
\begin{center}
\begin{tabular}{@{}llcccc@{}}
\hline
Quantity            &                      & $Z$ decays      & Tevatron       & LHCb~\cite{Aaij:2011jp} & all    \\
\hline
Mixing probability  & $\overline{\chi}$    & \hfagCHIBARLEP  & \hfagCHIBARTEV &         & \hfagCHIBAR \\
\Bu or \Bd fraction & $\fBu = \fBd$        & \hfagZFBD       & \hfagTFBD      &         & \hfagWFBD   \\
\Bs fraction        & $\fBs$               & \hfagZFBS       & \hfagTFBS      &         & \hfagWFBS   \\
\b-baryon fraction  & $\fbb$               & \hfagZFBB       & \hfagTFBB      &         & \hfagWFBB   \\
$\Bs/\Bd$ ratio     & $\fBs/\fBd$          & \hfagZFBSBD     & \hfagTFBSBD    & $0.267 ^{+0.021}_{-0.020}$ & \hfagWFBSBD \\
\multicolumn{2}{@{}l}{$\rho(\fBs,\fBu) = \rho(\fBs,\fBd)$} & \hfagZRHOFBDFBS & \hfagTRHOFBDFBS &         & \hfagWRHOFBDFBS \\
\multicolumn{2}{@{}l}{$\rho(\fbb,\fBu) = \rho(\fbb,\fBd)$} & \hfagZRHOFBDFBB & \hfagTRHOFBDFBB &         & \hfagWRHOFBDFBB \\
\multicolumn{2}{@{}l}{$\rho(\fbb,\fBs)$}                   & \hfagZRHOFBBFBS & \hfagTRHOFBBFBS &         & \hfagWRHOFBBFBS \\
\hline
\end{tabular}
\end{center}
\end{table}

Introducing the \chibar average in \Eq{fBs-mixing}, together with our world average 
$\chid = \hfagCHIDWU$ (see \Eq{chid} of \Sec{dmd}), the assumption $\chis= 1/2$ 
(justified by \Eq{chis} in \Sec{dms}), the 
best knowledge of the lifetimes (see \Sec{lifetimes}) and the estimate of \fbb given above, 
yields $\fBs = \hfagWFBSMIX$ 
(or $\fBs = \hfagZFBSMIX$ using only LEP data, 
or $\fBs = \hfagTFBSMIX$ using only Tevatron data),
an estimate dominated by the mixing information. 
Taking into account all known correlations (including the one introduced by \fbb), 
this result is then combined with the set of fractions obtained from direct measurements 
(given above), to yield the 
improved estimates of \Table{fractions}, 
still under the constraints of \Eq{constraints}.
As can be seen, our knowledge on the mixing parameters 
substantially reduces the uncertainty on \fBs.
It should be noted that the results 
are correlated, as indicated in \Table{fractions}.


%
%

\mysubsection{\b-hadron lifetimes}
\labs{lifetimes}

In the spectator model the decay of \b-flavoured hadrons $H_b$ is
governed entirely by the flavour changing \particle{b\to Wq} transition
($\particle{q}=\particle{c,u}$).  For this very reason, lifetimes of all
\b-flavoured hadrons are the same in the spectator approximation
regardless of the (spectator) quark content of the $H_b$.  In the early
1990's experiments became sophisticated enough to start seeing the
differences of the lifetimes among various $H_b$ species.  The first
theoretical calculations of the spectator quark effects on $H_b$
lifetime emerged only few years earlier.

Currently, most of such calculations are performed in the framework of
the Heavy Quark Expansion, HQE.  In the HQE, under certain assumptions
(most important of which is that of quark-hadron duality), the decay
rate of an $H_b$ to an inclusive final state $f$ is expressed as the sum
of a series of expectation values of operators of increasing dimension,
multiplied by the correspondingly higher powers of $\Lambda_{\rm
QCD}/m_b$:
\begin{equation}
\Gamma_{H_b\to f} = |CKM|^2\sum_n c_n^{(f)}
\Bigl(\frac{\Lambda_{\rm QCD}}{m_b}\Bigr)^n\langle H_b|O_n|H_b\rangle,
\labe{hqe}
\end{equation}
where $|CKM|^2$ is the relevant combination of the CKM matrix elements.
Coefficients $c_n^{(f)}$ of this expansion, known as Operator Product
Expansion~\cite{Shifman:1986mx,*Chay:1990da,*Bigi:1992su,*Bigi:1992su_erratum},
can be calculated perturbatively.  Hence, the HQE
predicts $\Gamma_{H_b\to f}$ in the form of an expansion in both
$\Lambda_{\rm QCD}/m_{\b}$ and $\alpha_s(m_{\b})$.  The precision of
current experiments makes it mandatory to go to the next-to-leading
order in QCD, {\em i.e.}\ to include correction of the order of
$\alpha_s(m_{\b})$ to the $c_n^{(f)}$'s.  All non-perturbative physics
is shifted into the expectation values $\langle H_b|O_n|H_b\rangle$ of
operators $O_n$.  These can be calculated using lattice QCD or QCD sum
rules, or can be related to other observables via the
HQE~\cite{Bigi:1995jr,*Bellini:1996ra}.  One may reasonably expect that powers of
$\Lambda_{\rm QCD}/m_{\b}$ provide enough suppression that only the
first few terms of the sum in \Eq{hqe} matter.

Theoretical predictions are usually made for the ratios of the lifetimes
(with $\tau(\Bd)$ chosen as the common denominator) rather than for the
individual lifetimes, for this allows several uncertainties to cancel.
The precision of the current HQE calculations (see
\Refs{Ciuchini:2001vx,*Beneke:2002rj,*Franco:2002fc,Tarantino:2003qw,*Gabbiani:2003pq,Gabbiani:2004tp} for the latest updates)
is in some instances already surpassed by the measurements,
\eg\ in the case of $\tau(\Bu)/\tau(\Bd)$.  Also, HQE calculations are
not assumption-free.  More accurate predictions are a matter of progress
in the evaluation of the non-perturbative hadronic matrix elements and
verifying the assumptions that the calculations are based upon.
However, the HQE, even in its present shape, draws a number of important
conclusions, which are in agreement with experimental observations:
\begin{itemize}
\item The heavier the mass of the heavy quark the smaller is the
  variation in the lifetimes among different hadrons containing this
  quark, which is to say that as $m_{\b}\to\infty$ we retrieve the
  spectator picture in which the lifetimes of all $H_b$'s are the same.
   This is well illustrated by the fact that lifetimes are rather
   similar in the \b sector, while they differ by large factors
   in the \particle{c} sector ($m_{\particle{c}}<m_{\b}$).
\item The non-perturbative corrections arise only at the order of
  $\Lambda_{\rm QCD}^2/m_{\b}^2$, which translates into 
  differences among $H_b$ lifetimes of only a few percent.
\item It is only the difference between meson and baryon lifetimes that
  appears at the $\Lambda_{\rm QCD}^2/m_{\b}^2$ level.  The splitting of the
  meson lifetimes occurs at the $\Lambda_{\rm QCD}^3/m_{\b}^3$ level, yet it is
  enhanced by a phase space factor $16\pi^2$ with respect to the leading
  free \b decay.
\end{itemize}

To ensure that certain sources of systematic uncertainty cancel, 
lifetime analyses are sometimes designed to measure a 
ratio of lifetimes.  However, because of the differences in decay
topologies, abundance (or lack thereof) of decays of a certain kind,
{\em etc.}, measurements of the individual lifetimes are more 
common.  In the following section we review the most common
types of the lifetime measurements.  This discussion is followed by the
presentation of the averaging of the various lifetime measurements, each
with a brief description of its particularities.



\mysubsubsection{Lifetime measurements, uncertainties and correlations}

In most cases lifetime of an $H_b$ is estimated from a flight distance
and a $\beta\gamma$ factor which is used to convert the geometrical
distance into the proper decay time.  Methods of accessing lifetime
information can roughly be divided in the following five categories:
\begin{enumerate}
\item {\bf\em Inclusive (flavour-blind) measurements}.  These
  measurements are aimed at extracting the lifetime from a mixture of
  \b-hadron decays, without distinguishing the decaying species.  Often
  the knowledge of the mixture composition is limited, which makes these
  measurements experiment-specific.  Also, these
  measurements have to rely on Monte Carlo for estimating the
  $\beta\gamma$ factor, because the decaying hadrons are not fully
  reconstructed.  On the bright side, these usually are the largest
  statistics \b-hadron lifetime measurements that are accessible to a
  given experiment, and can, therefore, serve as an important
  performance benchmark.
\item {\bf\em Measurements in semileptonic decays of a specific
  {\boldmath $H_b$\unboldmath}}.  \particle{W}from \particle{\b\to Wc}
  produces $\ell\nu_l$ pair (\particle{\ell=e,\mu}) in about 21\% of the
  cases.  Electron or muon from such decays is usually a well-detected
  signature, which provides for clean and efficient trigger.
  \particle{c} quark from \particle{b\to Wc} transition and the other
  quark(s) making up the decaying $H_b$ combine into a charm hadron,
  which is reconstructed in one or more exclusive decay channels.
  Knowing what this charmed hadron is allows one to separate, at least
  statistically, different $H_b$ species.  The advantage of these
  measurements is in statistics, which usually is superior to that of the
  exclusively reconstructed $H_b$ decays.  Some of the main
  disadvantages are related to the difficulty of estimating lepton+charm
  sample composition and Monte Carlo reliance for the $\beta\gamma$
  factor estimate.
\item {\bf\em Measurements in exclusively reconstructed hadronic decays}.
  These
  have the advantage of complete reconstruction of decaying $H_b$, which
  allows one to infer the decaying species as well as to perform precise
  measurement of the $\beta\gamma$ factor.  Both lead to generally
  smaller systematic uncertainties than in the above two categories.
  The downsides are smaller branching ratios, larger combinatoric
  backgrounds, especially in $H_b\rightarrow H_c\pi(\pi\pi)$ and
  multi-body $H_c$ decays, or in a hadron collider environment with
  non-trivial underlying event.  $H_b\to \jpsi H_s$ are relatively
  clean and easy to trigger on $\jpsi\to \ell^+\ell^-$, but their
  branching fraction is only about 1\%.
\item {\bf\em Measurements at asymmetric B factories}. 

In the $\Ups\rightarrow B \bar{B}$ decay, the \B mesons (\Bu or \Bd) are
essentially at rest in the \Ups frame.  This makes direct lifetime
measurements impossible in experiments at symmetric colliders producing 
\Ups at rest. 
At asymmetric \B factories the \Ups meson is boosted
resulting in \B and \particle{\bar{B}} moving nearly parallel to each 
other with the same boost. The lifetime is inferred from the distance $\Delta z$        
separating the \B and \particle{\bar{B}} decay vertices along the beam axis 
and from the \Ups boost known from the beam energies. This boost is equal to 
$\beta \gamma \approx 0.55$ (0.43) in the \babar (\belle) experiment,
resulting in an average \B decay length of approximately 250~(190)~$\mu$m. 
In order to determine the charge of the \B mesons in each event, one of the them is
fully reconstructed in a semileptonic or hadronic decay mode.
The other \B is typically not fully reconstructed, only the position
of its decay vertex is determined from the remaining tracks in the event.
These measurements benefit from large statistics, but suffer from poor proper time 
resolution, comparable to the \B lifetime itself. This resolution is dominated by the 
uncertainty on the decay vertices, which is typically 50~(100)~$\mu$m for a
fully (partially) reconstructed \B meson. 
With very large future statistics,
the resolution and purity could be improved (and hence the systematics reduced)
by fully reconstructing both \B mesons in the event. 
 
\item {\bf\em Direct measurement of lifetime ratios}.  This method has
  so far been only applied in the measurement of $\tau(\Bu)/\tau(\Bd)$.
  The ratio of the lifetimes is extracted from the dependence of the
  observed relative number of \Bu and \Bd candidates (both reconstructed
  in semileptonic decays) on the proper decay time.
\end{enumerate}

In some of the latest analyses, measurements of two (\eg\ $\tau(\Bu)$ and
$\tau(\Bu)/\tau(\Bd)$) or three (\eg\ $\tau(\Bu)$,
$\tau(\Bu)/\tau(\Bd)$, and \dmd) quantities are combined.  This
introduces correlations among measurements.  Another source of
correlations among the measurements are the systematic effects, which
could be common to an experiment or to an analysis technique across the
experiments.  When calculating the averages, such correlations are taken
into account per general procedure, described in
\Ref{LEPBOSC:1996}.

\mysubsubsection{Inclusive \b-hadron lifetimes}

The inclusive \b hadron lifetime is defined as $\tau_{\b} = \sum_i f_i
\tau_i$ where $\tau_i$ are the individual species lifetimes and $f_i$ are
the fractions of the various species present in an unbiased sample of
weakly-decaying \b hadrons produced at a high-energy
collider.\footnote{In principle such a quantity could be slightly
different in \particle{Z} decays and at the Tevatron, in case the
fractions of \b-hadron species are not exactly the same; see the
discussion in \Sec{fractions_high_energy}.}  This quantity is certainly
less fundamental than the lifetimes of the individual species, the
latter being much more useful in comparisons of the measurements with
the theoretical predictions.  Nonetheless, we perform the averaging of
the inclusive lifetime measurements for completeness as well as for the
reason that they might be of interest as ``technical numbers.''

\begin{table}[tp]
\caption{Measurements of average \b-hadron lifetimes.}
\labt{lifeincl}
\begin{center}
\begin{tabular}{lcccl} \hline
Experiment &Method           &Data set & $\tau_{\b}$ (ps)       &Ref.\\
\hline
ALEPH  &Dipole               &91     &$1.511\pm 0.022\pm 0.078$ &\cite{Buskulic:1993gj}\\
DELPHI &All track i.p.\ (2D) &91--92 &$1.542\pm 0.021\pm 0.045$ &\cite{Abreu:1994dr}$^a$\\
DELPHI &Sec.\ vtx            &91--93 &$1.582\pm 0.011\pm 0.027$ &\cite{Abreu:1996hv}$^a$\\
DELPHI &Sec.\ vtx            &94--95 &$1.570\pm 0.005\pm 0.008$ &\cite{Abdallah:2003sb}\\
L3     &Sec.\ vtx + i.p.     &91--94 &$1.556\pm 0.010\pm 0.017$ &\cite{Acciarri:1997tt}$^b$\\
OPAL   &Sec.\ vtx            &91--94 &$1.611\pm 0.010\pm 0.027$ &\cite{Ackerstaff:1996as}\\
SLD    &Sec.\ vtx            &93     &$1.564\pm 0.030\pm 0.036$ &\cite{Abe:1995rm}\\ 
\hline
\multicolumn{2}{l}{Average set 1 (\b vertex)} && \hfagTAUBVTXnounit &\\
\hline\hline
ALEPH  &Lepton i.p.\ (3D)    &91--93 &$1.533\pm 0.013\pm 0.022$ &\cite{Buskulic:1995rw}\\
L3     &Lepton i.p.\ (2D)    &91--94 &$1.544\pm 0.016\pm 0.021$ &\cite{Acciarri:1997tt}$^b$\\
OPAL   &Lepton i.p.\ (2D)    &90--91 &$1.523\pm 0.034\pm 0.038$ &\cite{Acton:1993xk}\\ 
\hline
\multicolumn{2}{l}{Average set 2 ($\b\to\ell$)} && \hfagTAUBLEPnounit &\\
\hline\hline
CDF1   &\particle{\jpsi} vtx&92--95 &$1.533\pm 0.015^{+0.035}_{-0.031}$ &\cite{Abe:1997bd} \\ 
ATLAS &\particle{\jpsi} vtx& 2010 & $1.489\pm 0.016 \pm 0.043$ & \cite{ATLAS-CONF-2011-145} \\
\hline
\multicolumn{2}{l}{Average set 3 (\particle{\b\to \jpsi})} && \hfagTAUBJPnounit & \\ 
\hline\hline
\multicolumn{2}{l}{Average of all above} && \hfagTAUBnounit & \\
\hline
\multicolumn{5}{l}{$^a$ \footnotesize The combined DELPHI result quoted in
\cite{Abreu:1996hv} is 1.575 $\pm$ 0.010 $\pm$ 0.026 ps.} \\[-0.5ex]
\multicolumn{5}{l}{$^b$ \footnotesize The combined L3 result quoted in \cite{Acciarri:1997tt} 
is 1.549 $\pm$ 0.009 $\pm$ 0.015 ps.}
\end{tabular}
\end{center}
\end{table}

In practice, an unbiased measurement of the inclusive lifetime is
difficult to achieve, because it would imply an efficiency which is
guaranteed to be the same across species.  So most of the measurements
are biased.  In an attempt to group analyses which are expected to
select the same mixture of \b hadrons, the available results (given in
\Table{lifeincl}) are divided into the following three sets:
\begin{enumerate}
\item measurements at LEP and SLD that accept any \b-hadron decay, based 
      on topological reconstruction (secondary vertex or track impact
      parameters);
\item measurements at LEP based on the identification
      of a lepton from a \b decay; and
\item measurements at the Tevatron based on inclusive 
      \particle{H_b\to \jpsi X} reconstruction, where the
      \particle{\jpsi} is fully reconstructed.
\end{enumerate}

The measurements of the first set are generally considered as estimates
of $\tau_{\b}$, although the efficiency to reconstruct a secondary
vertex most probably depends, in an analysis-specific way, on the number
of tracks coming from the vertex, thereby depending on the type of the
$H_b$.  Even though these efficiency variations can in principle be
accounted for using Monte Carlo simulations (which inevitably contain
assumptions on branching fractions), the $H_b$ mixture in that case can
remain somewhat ill-defined and could be slightly different among
analyses in this set.

On the contrary, the mixtures corresponding to the other two sets of
measurements are better defined in the limit where the reconstruction
and selection efficiency of a lepton or a \particle{\jpsi} from an
$H_b$ does not depend on the decaying hadron type.  These mixtures are
given by the production fractions and the inclusive branching fractions
for each $H_b$ species to give a lepton or a \particle{\jpsi}.  In
particular, under the assumption that all \b hadrons have the same
semileptonic decay width, the analyses of the second set should measure
$\tau(\b\to\ell) = (\sum_i f_i \tau_i^2) /(\sum_i f_i \tau_i)$ which is
necessarily larger than $\tau_{\b}$ if lifetime differences exist.
Given the present knowledge on $\tau_i$ and $f_i$,
$\tau(\b\to\ell)-\tau_{\b}$ is expected to be of the order of 0.01\ps.

Measurements by SLC and LEP experiments are subject to a number of
common systematic uncertainties, such as those due to (lack of knowledge
of) \b and \particle{c} fragmentation, \b and \particle{c} decay models,
\BR{B\to\ell}, \BR{B\to c\to\ell}, \BR{c\to\ell}, $\tau_{\particle{c}}$,
and $H_b$ decay multiplicity.  In the averaging, these systematic
uncertainties are assumed to be 100\% correlated.  The averages for the
sets defined above (also given in \Table{lifeincl}) are
\begin{eqnarray}
\tau(\b~\mbox{vertex}) &=& \hfagTAUBVTX \,,\\
\tau(\b\to\ell) &=& \hfagTAUBLEP  \,, \\
\tau(\b\to\particle{\jpsi}) &=& \hfagTAUBJP\,,
\end{eqnarray}
whereas an average of all measurements, ignoring mixture differences, 
yields \hfagTAUB.

\mysubsubsection{\Bd and \Bu lifetimes and their ratio}
\labs{taubd}
\labs{taubu}
\labs{lifetime_ratio}

\begin{table}[tp]
\caption{Measurements of the \Bd lifetime.}
\labt{lifebd}
\begin{center}
\begin{tabular}{lcccl} \hline
Experiment &Method                    &Data set &$\tau(\Bd)$ (ps)                  &Ref.\\
\hline
ALEPH  &\particle{D^{(*)} \ell}       &91--95 &$1.518\pm 0.053\pm 0.034$          &\cite{Barate:2000bs}\\
ALEPH  &Exclusive                     &91--94 &$1.25^{+0.15}_{-0.13}\pm 0.05$     &\cite{Buskulic:1996hy}\\
ALEPH  &Partial rec.\ $\pi^+\pi^-$    &91--94 &$1.49^{+0.17+0.08}_{-0.15-0.06}$   &\cite{Buskulic:1996hy}\\
DELPHI &\particle{D^{(*)} \ell}       &91--93 &$1.61^{+0.14}_{-0.13}\pm 0.08$     &\cite{Abreu:1995mc}\\
DELPHI &Charge sec.\ vtx              &91--93 &$1.63 \pm 0.14 \pm 0.13$           &\cite{Adam:1995mb}\\
DELPHI &Inclusive \particle{D^* \ell} &91--93 &$1.532\pm 0.041\pm 0.040$          &\cite{Abreu:1996gb}\\
DELPHI &Charge sec.\ vtx              &94--95 &$1.531 \pm 0.021\pm0.031$          &\cite{Abdallah:2003sb}\\
L3     &Charge sec.\ vtx              &94--95 &$1.52 \pm 0.06 \pm 0.04$           &\cite{Acciarri:1998uv}\\
OPAL   &\particle{D^{(*)} \ell}       &91--93 &$1.53 \pm 0.12 \pm 0.08$           &\cite{Akers:1995pa}\\
OPAL   &Charge sec.\ vtx              &93--95 &$1.523\pm 0.057\pm 0.053$          &\cite{Abbiendi:1998av}\\
OPAL   &Inclusive \particle{D^* \ell} &91--00 &$1.541\pm 0.028\pm 0.023$          &\cite{Abbiendi:2000ec}\\
SLD    &Charge sec.\ vtx $\ell$       &93--95 &$1.56^{+0.14}_{-0.13} \pm 0.10$    &\cite{Abe:1997ys}$^a$\\
SLD    &Charge sec.\ vtx              &93--95 &$1.66 \pm 0.08 \pm 0.08$           &\cite{Abe:1997ys}$^a$\\
CDF1   &\particle{D^{(*)} \ell}       &92--95 &$1.474\pm 0.039^{+0.052}_{-0.051}$ &\cite{Abe:1998wt}\\
CDF1  &Excl.\ \particle{\jpsi K^{*0}}&92--95 &$1.497\pm 0.073\pm 0.032$          &\cite{Acosta:2002nd}\\
CDF2   &Excl.\ \particle {\jpsi K_S}, \particle{\jpsi K^{*0}} &02--09 &$1.507\pm 0.010\pm0.008$           &\cite{Aaltonen:2010pj,*Abulencia:2006dr_mod_cont} \\
\dzero &Excl.\ \particle{\jpsi K^{*0}}&03--07 &$1.414\pm0.018\pm0.034$ &\cite{Abazov:2008jz,*Abazov:2005sa_mod_cont}\\ 
\dzero &Excl.\ \particle {\jpsi K_S} &02--11 &$1.508 \pm0.025 \pm0.043$  &\cite{Abazov:2012iy,*Abazov:2007sf_mod_cont,*Abazov:2004bn_mod_cont} \\
\babar &Exclusive                     &99--00 &$1.546\pm 0.032\pm 0.022$          &\cite{Aubert:2001uw}\\
\babar &Inclusive \particle{D^* \ell} &99--01 &$1.529\pm 0.012\pm 0.029$          &\cite{Aubert:2002gi,*Aubert:2002gi_erratum}\\
\babar &Exclusive \particle{D^* \ell} &99--02 &$1.523^{+0.024}_{-0.023}\pm 0.022$ &\cite{Aubert:2002sh}\\
\babar &Incl.\ \particle{D^*\pi}, \particle{D^*\rho} 
                                      &99--01 &$1.533\pm 0.034 \pm 0.038$         &\cite{Aubert:2002ms}\\
\babar &Inclusive \particle{D^* \ell}
&99--04 &$1.504\pm0.013^{+0.018}_{-0.013}$  &\cite{Aubert:2005kf} \\ 
\belle & Exclusive                     & 00--03 & $1.534\pm 0.008\pm0.010$        & \cite{Abe:2004mz}\\
ATLAS & Excl.\ \particle{\jpsi K^{*0}} & 2010 & $1.51 \pm0.04 \pm0.04$ & \cite{ATLAS-CONF-2011-092}$^p$ \\
LHCb  & Excl.\ \particle{\jpsi K^{*0}} & 2010 & $1.512 \pm0.032 \pm 0.042$ & \cite{LHCb-CONF-2011-001}$^p$ \\
LHCb  & Excl.\ \particle {\jpsi K_S}   & 2010 & $1.558 \pm0.056 \pm 0.022$ & \cite{LHCb-CONF-2011-001}$^p$ \\
\hline
Average&                               &        & \hfagTAUBDnounit & \\
\hline\hline           
\multicolumn{5}{l}{$^a$ \footnotesize The combined SLD result 
quoted in \cite{Abe:1997ys} is 1.64 $\pm$ 0.08 $\pm$ 0.08 ps.}\\[-0.5ex]
\multicolumn{5}{l}{$^p$ {\footnotesize Preliminary.}}
\end{tabular}
\end{center}
\end{table}

After a number of years of dominating these averages the LEP experiments
yielded the scene to the asymmetric \B~factories and
the Tevatron experiments. The \B~factories have been very successful in
utilizing their potential -- in only a few years of running, \babar and,
to a greater extent, \belle, have struck a balance between the
statistical and the systematic uncertainties, with both being close to
(or even better than) the impressive 1\%.  In the meanwhile, CDF and
\dzero have emerged as significant contributors to the field as the
Tevatron Run~II data flowed in, with CDF eventually providing the most precise results. 

\begin{table}[tbp]
\caption{Measurements of the \Bu lifetime.}
\labt{lifebu}
\begin{center}
\begin{tabular}{lcccl} \hline
Experiment &Method                 &Data set &$\tau(\Bu)$ (ps)                 &Ref.\\
\hline
ALEPH  &\particle{D^{(*)} \ell}    &91--95 &$1.648\pm 0.049\pm 0.035$          &\cite{Barate:2000bs}\\
ALEPH  &Exclusive                  &91--94 &$1.58^{+0.21+0.04}_{-0.18-0.03}$   &\cite{Buskulic:1996hy}\\
DELPHI &\particle{D^{(*)} \ell}    &91--93 &$1.61\pm 0.16\pm 0.12$             &\cite{Abreu:1995mc}$^a$\\
DELPHI &Charge sec.\ vtx           &91--93 &$1.72\pm 0.08\pm 0.06$             &\cite{Adam:1995mb}$^a$\\
DELPHI &Charge sec.\ vtx           &94--95 &$1.624\pm 0.014\pm 0.018$          &\cite{Abdallah:2003sb}\\
L3     &Charge sec.\ vtx           &94--95 &$1.66\pm  0.06\pm 0.03$            &\cite{Acciarri:1998uv}\\
OPAL   &\particle{D^{(*)} \ell}    &91--93 &$1.52 \pm 0.14\pm 0.09$            &\cite{Akers:1995pa}\\
OPAL   &Charge sec.\ vtx           &93--95 &$1.643\pm 0.037\pm 0.025$          &\cite{Abbiendi:1998av}\\
SLD    &Charge sec.\ vtx $\ell$    &93--95 &$1.61^{+0.13}_{-0.12}\pm 0.07$     &\cite{Abe:1997ys}$^b$\\
SLD    &Charge sec.\ vtx           &93--95 &$1.67\pm 0.07\pm 0.06$             &\cite{Abe:1997ys}$^b$\\
CDF1   &\particle{D^{(*)} \ell}    &92--95 &$1.637\pm 0.058^{+0.045}_{-0.043}$ &\cite{Abe:1998wt}\\
CDF1   &Excl.\ \particle{\jpsi K} &92--95 &$1.636\pm 0.058\pm 0.025$          &\cite{Acosta:2002nd}\\
CDF2   &Excl.\ \particle{\jpsi K} &02--09 &$1.639\pm 0.009\pm 0.009$          &\cite{Aaltonen:2010pj,*Abulencia:2006dr_mod_cont}\\ 
CDF2   &Excl.\ \particle{D^0 \pi}  &02--06 &$1.663\pm 0.023\pm0.015$           &\cite{Aaltonen:2010ta}\\
\babar &Exclusive                  &99--00 &$1.673\pm 0.032\pm 0.023$          &\cite{Aubert:2001uw}\\
\belle &Exclusive                  &00--03 &$1.635\pm 0.011\pm 0.011$          &\cite{Abe:2004mz}\\
LHCb  & Excl.\ \particle{\jpsi K} & 2010 & $1.689 \pm0.022 \pm 0.047$ & \cite{LHCb-CONF-2011-001}$^p$ \\
\hline
Average&                           &       &\hfagTAUBUnounit &\\
\hline\hline
\multicolumn{5}{l}{$^a$ \footnotesize The combined DELPHI result quoted 
in~\cite{Adam:1995mb} is $1.70 \pm 0.09$ ps.} \\[-0.5ex]
\multicolumn{5}{l}{$^b$ \footnotesize The combined SLD result 
quoted in~\cite{Abe:1997ys} is $1.66 \pm 0.06 \pm 0.05$ ps.}\\[-0.5ex]
\multicolumn{5}{l}{$^p$ {\footnotesize Preliminary.}}
\end{tabular}
\end{center}
\end{table}

At present time we are in an interesting position of having three sets
of measurements (from LEP/SLC, \B factories and the Tevatron) that
originate from different environments, obtained using substantially
different techniques and are precise enough for incisive comparison.


\begin{table}[tb]
\caption{Measurements of the ratio $\tau(\Bu)/\tau(\Bd)$.}
\labt{liferatioBuBd}
\begin{center}
\begin{tabular}{lcccl} 
\hline
Experiment &Method                 &Data set &Ratio $\tau(\Bu)/\tau(\Bd)$      &Ref.\\
\hline
ALEPH  &\particle{D^{(*)} \ell}    &91--95 &$1.085\pm 0.059\pm 0.018$          &\cite{Barate:2000bs}\\
ALEPH  &Exclusive                  &91--94 &$1.27^{+0.23+0.03}_{-0.19-0.02}$   &\cite{Buskulic:1996hy}\\
DELPHI &\particle{D^{(*)} \ell}    &91--93 &$1.00^{+0.17}_{-0.15}\pm 0.10$     &\cite{Abreu:1995mc}\\
DELPHI &Charge sec.\ vtx           &91--93 &$1.06^{+0.13}_{-0.11}\pm 0.10$     &\cite{Adam:1995mb}\\
DELPHI &Charge sec.\ vtx           &94--95 &$1.060\pm 0.021 \pm 0.024$         &\cite{Abdallah:2003sb}\\
L3     &Charge sec.\ vtx           &94--95 &$1.09\pm 0.07  \pm 0.03$           &\cite{Acciarri:1998uv}\\
OPAL   &\particle{D^{(*)} \ell}    &91--93 &$0.99\pm 0.14^{+0.05}_{-0.04}$     &\cite{Akers:1995pa}\\
OPAL   &Charge sec.\ vtx           &93--95 &$1.079\pm 0.064 \pm 0.041$         &\cite{Abbiendi:1998av}\\
SLD    &Charge sec.\ vtx $\ell$    &93--95 &$1.03^{+0.16}_{-0.14} \pm 0.09$    &\cite{Abe:1997ys}$^a$\\
SLD    &Charge sec.\ vtx           &93--95 &$1.01^{+0.09}_{-0.08} \pm0.05$     &\cite{Abe:1997ys}$^a$\\
CDF1   &\particle{D^{(*)} \ell}    &92--95 &$1.110\pm 0.056^{+0.033}_{-0.030}$ &\cite{Abe:1998wt}\\
CDF1   &Excl.\ \particle{\jpsi K} &92--95 &$1.093\pm 0.066 \pm 0.028$         &\cite{Acosta:2002nd}\\
CDF2   &Excl.\ \particle{\jpsi K^{(*)}} &02--09 &$1.088\pm 0.009 \pm 0.004$   &\cite{Aaltonen:2010pj,*Abulencia:2006dr_mod_cont}\\ 
\dzero &\particle{D^{*+} \mu} \particle{D^0 \mu} ratio
	                           &02--04 &$1.080\pm 0.016\pm 0.014$          &\cite{Abazov:2004sa}\\
\babar &Exclusive                  &99--00 &$1.082\pm 0.026\pm 0.012$          &\cite{Aubert:2001uw}\\
\belle &Exclusive                  &00--03 &$1.066\pm 0.008\pm 0.008$          &\cite{Abe:2004mz}\\
\hline
Average&                           &       & \hfagRTAUBU & \\   
\hline\hline
\multicolumn{5}{l}{$^a$ \footnotesize The combined SLD result quoted
	   in~\cite{Abe:1997ys} is $1.01 \pm 0.07 \pm 0.06$.}
\end{tabular}
\end{center}
\end{table}

The averaging of $\tau(\Bu)$, $\tau(\Bd)$ and $\tau(\Bu)/\tau(\Bd)$
measurements is summarized\footnote{
We do not include the old unpublished measurements of Refs.~\cite{CDFnote7514:2005,CDFnote7386:2005}.}
in \Tablesss{lifebd}{lifebu}{liferatioBuBd}.
For $\tau(\Bu)/\tau(\Bd)$ we averaged only the measurements of this
quantity provided by experiments rather than using all available
knowledge, which would have included, for example, $\tau(\Bu)$ and
$\tau(\Bd)$ measurements which did not contribute to any of the ratio
measurements.

The following sources of correlated (within experiment/machine)
systematic uncertainties have been considered:
\begin{itemize}
\item for SLC/LEP measurements -- \particle{D^{**}} branching ratio uncertainties~\cite{Abbaneo:2000ej_mod,*Abbaneo:2001bv_mod_cont},
momentum estimation of \b mesons from \particle{Z^0} decays
(\b-quark fragmentation parameter $\langle X_E \rangle = 0.702 \pm 0.008$~\cite{Abbaneo:2000ej_mod,*Abbaneo:2001bv_mod_cont}),
\Bs and \b baryon lifetimes (see \Secss{taubs}{taulb}),
and \b-hadron fractions at high energy (see \Table{fractions}); 
\item for \babar measurements -- alignment, $z$ scale, PEP-II boost,
sample composition (where applicable);
\item for \dzero and CDF Run~II measurements -- alignment (separately
within each experiment).
\end{itemize}
The resultant averages are:
\begin{eqnarray}
\tau(\Bd) & = & \hfagTAUBD \,, \\
\tau(\Bu) & = & \hfagTAUBU \,, \\
\tau(\Bu)/\tau(\Bd) & = & \hfagRTAUBU \,.
\end{eqnarray}
%
%
%

\mysubsubsection{\Bs lifetimes}
\labs{taubs}

Like neutral kaons, neutral \B mesons contain
short- and long-lived components, since the
light (L) and heavy (H)
eigenstates, $\B_{\rm L}$ and $\B_{\rm H}$, differ not only
in their masses, but also in their total decay widths,  
with a decay width difference defined as 
$\DG = \Gamma_{\rm L} - \Gamma_{\rm H}$. 
Neglecting \CP violation in $\B-\Bbar$ mixing, 
which is expected to be very small~\cite{Lenz:2011ti,*Lenz:2006hd,Beneke:1998sy},
the mass eigenstates are also \CP eigenstates,
with the light $\B_{\rm L}$ state being \CP-even 
and the heavy $\B_{\rm H}$ state being \CP-odd. 
While the decay width difference \DGd can be neglected in the \Bd system, 
the \Bs system exhibits a significant
value of \DGs: the sign of \DGs is known 
to be positive~\cite{Aaij:2012eq}, {\em i.e.}
the heavy eigenstates lives longer than the light eigenstate. 
Specific measurements of \DGs and 
$\Gs = (\Gamma_{\rm L} + \Gamma_{\rm H})/2$ are explained
and averaged in \Sec{DGs}, but the results for
$1/\Gamma_{\rm L}$, $1/\Gamma_{\rm H}$ and
the mean \Bs lifetime, defined as $\tau(\Bs) = 1/\Gs$,
are also quoted at the end of this section. 

Many \Bs lifetime analyses, in particular the early 
ones performed before the non-zero value of \DGs was 
firmly established, ignore \DGs and fit the proper time 
distribution of a sample of \Bs candidates 
reconstructed in a certain final state $f$
with a model assuming a single exponential function 
for the signal. We denote such {\rm effective lifetime}
measurements as $\tau_{\rm single}(\Bs\to f)$; 
their true values may lie {\em a priori} anywhere
between $1/\Gamma_{\rm L} = 1/(\Gs+\DGs/2)$ and
$1/\Gamma_{\rm H}= 1/(\Gs-\DGs/2)$, 
depending on the proportion of $\B_{\rm L}$ and $\B_{\rm H}$
in the final state $f$. \Table{lifebs}
summarizes the effective 
lifetime measurements.

Averaging measurements of $\tau_{\rm single}(\Bs\to f)$
over several final states $f$ will yield a result 
corresponding to an ill-defined observable
when the proportions of $\B_{\rm L}$ and $\B_{\rm H}$ differ. 
Therefore,
the effective \Bs lifetime measurements are broken down into
several categories and averaged separately.

\begin{table}[tb]
\caption{Measurements of the effective \Bs lifetimes obtained from single exponential fits,
without attempting to separate the \CP-even and \CP-odd components.}
\labt{lifebs}
\begin{center}
\begin{tabular}{lccrcl} \hline
Experiment & Final state $f$           & \multicolumn{2}{c}{Data set} & $\tau_{\rm single}(\Bs\to f)$ (ps) & Ref. \\
\hline \hline
ALEPH  & \particle{D_s \ell}  & 91--95 & & $1.54^{+0.14}_{-0.13}\pm 0.04$   & \cite{Buskulic:1996ei}          \\
CDF1   & \particle{D_s \ell}  & 92--96 & & $1.36\pm 0.09 ^{+0.06}_{-0.05}$  & \cite{Abe:1998cj}           \\
DELPHI & \particle{D_s \ell}  & 91--95 & & $1.42^{+0.14}_{-0.13}\pm 0.03$   & \cite{Abreu:2000sh}          \\
OPAL   & \particle{D_s \ell}  & 90--95 & & $1.50^{+0.16}_{-0.15}\pm 0.04$   & \cite{Ackerstaff:1997qi}  \\
\dzero & \particle{D_s \mu}  & 02--04 & & $1.398 \pm 0.044 ^{+0.028}_{-0.025}   $   & \cite{Abazov:2006cb}       \\ 
CDF2   & \particle{D_s \pi (X)} 
                              & 02--06 & 1.3 fb$^{-1}$ & $1.518 \pm 0.041 \pm 0.027     $   & \cite{Aaltonen:2011qsa,*Aaltonen:2011qsa_cont} \\ 
\multicolumn{4}{l}{Average of above 6 flavour-specific measurements} &  \hfagTAUBSSLnounit & \\  
\hline
ALEPH  & \particle{D_s h}     & 91--95 & & $1.47\pm 0.14\pm 0.08$           & \cite{Barate:1997ua}          \\
DELPHI & \particle{D_s h}     & 91--95 & & $1.53^{+0.16}_{-0.15}\pm 0.07$   & \cite{Abreu:2000ev} \\
OPAL   & \particle{D_s} incl. & 90--95 & & $1.72^{+0.20+0.18}_{-0.19-0.17}$ & \cite{Ackerstaff:1997ne}          \\ 
\hline
\multicolumn{4}{l}{Average of above 9 \particle{D_s} measurements} &  \hfagTAUBSnounit & \\ 
\hline\hline
CDF1     & \particle{\jpsi\phi} & 92--95 &  & $1.34^{+0.23}_{-0.19}    \pm 0.05$ & \cite{Abe:1997bd} \\
\dzero   & \particle{\jpsi\phi} & 02--04 &  & $1.444^{+0.098}_{-0.090} \pm 0.02$ & \cite{Abazov:2004ce}  \\
ATLAS & \particle{\jpsi\phi} & 2010 & 40 pb$^{-1}$ & $1.41 \pm0.08 \pm0.05$ & \cite{ATLAS-CONF-2011-092}$^p$ \\
LHCb  & \particle{\jpsi\phi} & 2010 & 36 pb$^{-1}$ & $1.447 \pm0.064 \pm 0.056$ & \cite{LHCb-CONF-2011-001}$^p$ \\
\hline 
\multicolumn{4}{l}{Average of above 4 \particle{\jpsi \phi} measurements} &  \hfagTAUBSJFnounit & \\ 
\hline\hline
ALEPH    & \particle{D_s^{(*)+}D_s^{(*)-}} & 91--95 & 4M \particle{Z\to q\bar{q}} & $1.27 \pm 0.33 \pm 0.08$ & \cite{Barate:2000kd} \\
\hline
LHCb    & \particle{K^+K^-}   &  10 & 0.037 fb$^{-1}$ & $1.440 \pm 0.096 \pm 0.009$ & \cite{Aaij:2011kn} \\
LHCb    & \particle{K^+K^-}   &  11 & 1.0 fb$^{-1}$ & $1.468 \pm 0.046 \pm 0.006$ & \cite{LHCb-CONF-2012-001}$^p$ \\
\hline \multicolumn{4}{l}{Average of above 2 $K^+K^-$ measurements} &  \hfagTAUBSSHORTnounit & \\ \hline \hline
CDF2     & \particle{\jpsi f_0(980)} & 02--08 & 3.8 fb$^{-1}$ & $1.70^{+0.12}_{-0.11} \pm 0.03$ & \cite{Aaltonen:2011nk} \\
\hline \multicolumn{4}{l}{Average of above 1 $\jpsi f_0(980)$ measurement} &  \hfagTAUBSLONGnounit & \\ \hline \hline
\hline
\multicolumn{5}{l}{$^p$ \footnotesize Preliminary.}
\end{tabular}
\end{center}
\end{table}

\begin{itemize}
\item 
{\bf\em Flavour-specific decays}, such as semileptonic
$\Bs \to \particle{D_s^- \ell^+ \nu}$
or $\Bs\to \particle {D_s^- \pi^+}$, 
have equal 
fractions of $\B_{\rm L}$ and $\B_{\rm H}$ at time
zero. 
If the resulting superposition of two exponential distributions
is fitted with a single exponential function, 
one obtains a measure of the so-called {\em flavour-specific lifetime}~\cite{Hartkorn:1999ga}:
\begin{equation}
\tau_{\rm single}(\Bs\to \mbox{flavour specific}) = \frac{1}{\Gs}
\frac{{1+\left(\frac{\DGs}{2\Gs}\right)^2}}{{1-\left(\frac{\DGs}{2\Gs}\right)^2}
}.
\labe{fslife}
\end{equation}
The average of all flavour-specific 
\Bs lifetime measurements\footnote{
An old unpublished measurement~\cite{CDFnote7757:2005} is not included.}
is
\begin{equation}
\tau_{\rm single}(\Bs\to \mbox{flavour specific}) = \hfagTAUBSSL \,.
\labe{tau_fs}
\end{equation}

\item
{\bf\em \boldmath $\Bs\to D_s^{\mp} X$ decays}
include flavour-specific decays but also decays 
with a less known mixture of light and heavy components. 
The corresponding effective lifetime average,
\begin{equation}
\tau_{\rm single}(\Bs\to D_s^{\mp} X) = \hfagTAUBS \,,
\end{equation}
can still be a useful input
for analyses examining an inclusive $D_s$ sample.
The following correlated systematic errors were considered:
average \B lifetime used in backgrounds,
\Bs decay multiplicity, and branching ratios used to determine 
backgrounds (\eg\ \BR{B\to D_s D}).
A knowledge of the multiplicity of \Bs decays is important for
measurements that partially reconstruct the final state such as 
\particle{\B\to D_s \mbox{$X$}} (where $X$ is not a lepton). 
The boost deduced from Monte Carlo simulation depends on the multiplicity used.
Since this is not well known, the multiplicity in the simulation is
varied and this range of values observed is taken to be a systematic.
Similarly not all the branching ratios for the potential background
processes are measured. Where they are available, the PDG values are
used for the error estimate. Where no measurements are available
estimates can usually be made by using measured branching ratios of
related processes and using some reasonable extrapolation.

\item
{\bf\em 
{\boldmath $\Bs \to \jpsi\phi$ \unboldmath}decays}
contain a well-defined mixture of \CP-even and \CP-odd states
There are no known correlations
between the existing 
\particle{\Bs\to \jpsi\phi}
effective lifetime measurements; these are combined  
into the average\footnote{
An old unpublished measurement~\cite{CDFnote8524:2007,*CDFnote8524:2007_cont} is not included.}
\begin{equation}
\tau_{\rm single}(\Bs\to \jpsi \phi) = \hfagTAUBSJF \,.
\end{equation}
A caveat is that different experimental acceptances
may lead to different admixtures of the 
\CP-even and \CP-odd states, and simple fits to a single
exponential may result in inherently different 
values of $\tau_{\rm single}(\Bs\to \jpsi \phi)$.
Analyses that separate the \CP-even and \CP-odd components in
this decay through a full angular study, outlined in \Sec{DGs},
provide directly measurements of $1/\Gs$ and $\DGs$ (see \Table{phisDGsGs}).

\item
{\bf\em Decays to (almost) pure \boldmath\CP-even eigenstates} have also 
been measured, in the modes $\Bs \to D_s^{(*)+}D_s^{(*)-}$ by ALEPH~\cite{Barate:2000kd}, 
$\Bs \to K^+ K^-$ by LHCb~\cite{Aaij:2011kn,LHCb-CONF-2012-001}%
\footnote{An old unpublished measurement of the $\Bs \to K^+ K^-$
effective lifetime by CDF~\cite{Tonelli:2006np} is no longer considered.}, 
and $\Bs \to \jpsi f_0(980)$ by CDF~\cite{Aaltonen:2011nk}.
The $\Bs \to D_s^{(*)+}D_s^{(*)-}$ decays
are expected to be mostly \CP-even, 
but a small \CP-odd component is most probably present. 
The decays $\Bs \to K^+ K^-$ and $\Bs \to \jpsi f_0(980)$ 
have \CP-even and \CP-odd final states, respectively; if these 
decays are dominated by a single weak phase and if \CP violation 
can be neglected, then $\tau_{\rm single}(\Bs \to K^+ K^-) \sim 1/\Gamma_{\rm L}$ 
and  $\tau_{\rm single}(\Bs \to \jpsi f_0(980)) \sim 1/\Gamma_{\rm H}$ 
(see \Eqss{tau_KK_approx}{tau_Jpsif0_approx} for approximate relations in presence of
\CP violation in the mixing). The averages for these two effective lifetimes are 
\begin{eqnarray}
\tau_{\rm single}(\Bs \to K^+ K^-) & = & \hfagTAUBSSHORT \,,
\labe{tau_KK}
\\
\tau_{\rm single}(\Bs \to \jpsi f_0(980)) & = & \hfagTAUBSLONG \,.
\labe{tau_Jpsif0}
\end{eqnarray}

\end{itemize}

As described in \Sec{DGs}, 
the effective lietime averages of \Eqsss{tau_fs}{tau_KK}{tau_Jpsif0}
are used as ingredients to improve the 
determination of $1/\Gs$ and \DGs obtained from the full angular analyses
of $\Bs\to \jpsi\phi$ decays. 
The resulting world averages for the \Bs lifetimes are
\begin{eqnarray}
\frac{1}{\Gamma_{\rm L}} = \frac{1}{\Gs+\DGs/2} & = & \hfagTAUBSLCON \,, \\
\frac{1}{\Gamma_{\rm H}} = \frac{1}{\Gs-\DGs/2} & = & \hfagTAUBSHCON \,, \\
\tau(\Bs) = \frac{1}{\Gs} = \frac{2}{\Gamma_{\rm L}+\Gamma_{\rm H}} & = & \hfagTAUBSMEANCON \,.
\labe{oneoverGs}
\end{eqnarray}

\mysubsubsection{\Bc lifetime}
\labs{taubc}

Early measurements of the \Bc meson lifetime,
from CDF~\cite{Abe:1998wi,CDFnote9294:2008,*Abulencia:2006zu_mod_cont} and \dzero~\cite{Abazov:2008rba},
use the semileptonic decay mode \particle{\Bc \to \jpsi \ell} and are based on a 
simultaneous fit to the mass and lifetime using the vertex formed
with the leptons from the decay of the \particle{\jpsi} and
the third lepton. Correction factors
to estimate the boost due to the missing neutrino are used.
In the analysis of the CDF Run~I data~\cite{Abe:1998wi},
a mass value of 
$6.40 \pm 0.39 \pm 0.13$~GeV/$c^2$ 
is found by fitting
to the tri-lepton invariant mass spectrum. 
In the CDF and \dzero Run~II results~\cite{CDFnote9294:2008,*Abulencia:2006zu_mod_cont,Abazov:2008rba}, 
the \Bc mass is assumed to be 
$6285.7 \pm 5.3 \pm 1.2$~MeV/$c^2$, taken from a 
CDF result~\cite{Abulencia:2005usa}. 
These mass measurements
are consistent within uncertainties, and also consistent with the
most recent precision determination from CDF of 
$6275.6 \pm 2.9 \pm 2.5$~MeV/$c^2$~\cite{Aaltonen:2007gv}.
Correlated systematic errors include the impact
of the uncertainty of the \Bc $p_T$ spectrum on the correction
factors, the level of feed-down from $\psi(2S)$, 
Monte-Carlo modeling of the decay model varying from phase space
to the ISGW model, and mass variations.

The most recent determination of the \Bc lifetime, from CDF2~\cite{CDFnote10533:2011}, is based on fully reconstructed 
$\Bc \to J/\psi \pi$ decays and does not suffer from a missing neutrino. 
All the measurements are summarized in 
\Table{lifebc} and the world average is determined to be
\begin{equation}
\tau(\Bc) = \hfagTAUBC \,.
\end{equation}

\begin{table}[tb]
\caption{Measurements of the \Bc lifetime.}
\labt{lifebc}
\begin{center}
\begin{tabular}{lccrcl} \hline
Experiment & Method                    & \multicolumn{2}{c}{Data set}  & $\tau(\Bc)$ (ps)
      & Ref.\\   \hline
CDF1       & \particle{\jpsi \ell} & 92--95 & 0.11 fb$^{-1}$ & $0.46^{+0.18}_{-0.16} \pm
 0.03$   & \cite{Abe:1998wi}  \\ 
CDF2       & \particle{\jpsi \ell} & 02--06 & 1.0 fb$^{-1}$ & $0.475^{+0.053}_{-0.049} \pm 0.018$   & \cite{CDFnote9294:2008,*Abulencia:2006zu_mod_cont}$^p$ \\
 \dzero & \particle{\jpsi \mu} & 02--06 & 1.3 fb$^{-1}$  & $0.448^{+0.038}_{-0.036} \pm 0.032$
   & \cite{Abazov:2008rba}  \\
CDF2       & \particle{\jpsi \pi} & & 6.7 fb$^{-1}$ & $0.452 \pm 0.048 \pm 0.027$  & \cite{CDFnote10533:2011}$^p$ \\
\hline
  \multicolumn{2}{l}{Average} & &  &  \hfagTAUBCnounit
                 &    \\   \hline
\multicolumn{5}{l}{$^p$ \footnotesize Preliminary.}
\end{tabular}
\end{center}
\end{table}

\mysubsubsection{\Lb and \b-baryon lifetimes}
\labs{taulb}

The first measurements of \b-baryon lifetimes
originate from two classes of partially reconstructed decays.
In the first class, decays with an exclusively 
reconstructed \Lc baryon
and a lepton of opposite charge are used. These products are
more likely to occur in the decay of \Lb baryons.
In the second class, more inclusive final states with a baryon
(\particle{p}, \particle{\bar{p}}, $\Lambda$, or $\bar{\Lambda}$) 
and a lepton have been used, and these final states can generally
arise from any \b baryon.  With the large \b-hadron samples available
at the Tevatron, the most precise measurements of \b-baryons now
come from fully reconstructed exclusive decays.

The following sources of correlated systematic uncertainties have 
been considered:
experimental time resolution within a given experiment, \b-quark
fragmentation distribution into weakly decaying \b baryons,
\Lb polarization, decay model,
and evaluation of the \b-baryon purity in the selected event samples.
In computing the averages
the central values of the masses are scaled to 
$M(\Lb) = 5620 \pm 2\MeVcc$~\cite{Acosta:2005mq} and
$M(\mbox{\b-baryon}) = 5670 \pm 100\MeVcc$.

For the semi-inclusive lifetime measurements, 
the meaning of decay model
systematic uncertainties
and the correlation of these uncertainties between measurements
are not always clear.
Uncertainties related to the decay model are dominated by
assumptions on the fraction of $n$-body semileptonic decays.
To be conservative it is assumed
that these are 100\%  correlated whenever given as an error.
DELPHI varies the fraction of 4-body decays from 0.0 to 0.3. 
In computing the average, the DELPHI
result is corrected to a value of  $0.2 \pm 0.2$ for this fraction.

Furthermore, in computing the average,
the semileptonic decay results from LEP are corrected for a polarization of 
$-0.45^{+0.19}_{-0.17}$~\cite{Abbaneo:2000ej_mod,*Abbaneo:2001bv_mod_cont} and  a 
\Lb fragmentation parameter
$\langle X_E \rangle =0.70\pm 0.03$~\cite{Buskulic:1995mf}.




\begin{table}[!t]
\caption{Measurements of the \b-baryon lifetimes.
}
\labt{lifelb}
\begin{center}
\begin{tabular}{lcccl} 
\hline
Experiment&Method                &Data set& Lifetime (ps) & Ref. \\\hline
ALEPH  &$\Lc\ell$             & 91--95 &$1.18^{+0.13}_{-0.12} \pm 0.03$ & \cite{Barate:1997if}$^a$\\
ALEPH  &$\Lambda\ell^-\ell^+$ & 91--95 &$1.30^{+0.26}_{-0.21} \pm 0.04$ & \cite{Barate:1997if}$^a$\\
DELPHI &$\Lc\ell$             & 91--94 &$1.11^{+0.19}_{-0.18} \pm 0.05$ & \cite{Abreu:1999hu}$^b$\\
OPAL   &$\Lc\ell$, $\Lambda\ell^-\ell^+$ 
                                 & 90--95 & $1.29^{+0.24}_{-0.22} \pm 0.06$ & \cite{Ackerstaff:1997qi}\\ 
CDF1   &$\Lc\ell$             & 91--95 &$1.32 \pm 0.15        \pm 0.07$ & \cite{Abe:1996df}\\
CDF2   &$\Lc\pi$              & 02--06 &$1.401 \pm 0.046 \pm 0.035$ & \cite{Aaltonen:2009zn} \\
CDF2   &$\jpsi \Lambda$      & 02--09 &$1.537 \pm 0.045 \pm 0.014$ & \cite{Aaltonen:2010pj,*Abulencia:2006dr_mod_cont}\\
\dzero &$\Lc\mu$              & 02--06 &$1.290^{+0.119+0.087}_{-0.110-0.091}$ & \cite{Abazov_mod:2007tha} \\
\dzero &$\jpsi \Lambda$      & 02--11 &$1.303 \pm 0.075 \pm 0.035$ & \cite{Abazov:2012iy,*Abazov:2007sf_mod_cont,*Abazov:2004bn_mod_cont} \\
LHCb   &$\jpsi \Lambda$      & 2010   &$1.353 \pm 0.108 \pm 0.035$ & \cite{LHCb-CONF-2011-001}$^p$ \\
\hline
\multicolumn{3}{l}{Average of above 10: \hfill \Lb lifetime $=$} & \hfagTAULBnounit & \\
\hline
ALEPH  &$\Lambda\ell$         & 91--95 &$1.20 \pm 0.08 \pm 0.06$ & \cite{Barate:1997if}\\
DELPHI &$\Lambda\ell\pi$ vtx  & 91--94 &$1.16 \pm 0.20 \pm 0.08$        & \cite{Abreu:1999hu}$^b$\\
DELPHI &$\Lambda\mu$ i.p.     & 91--94 &$1.10^{+0.19}_{-0.17} \pm 0.09$ & \cite{Abreu:1996nt}$^b$ \\
DELPHI &\particle{p\ell}      & 91--94 &$1.19 \pm 0.14 \pm 0.07$        & \cite{Abreu:1999hu}$^b$\\
OPAL   &$\Lambda\ell$ i.p.    & 90--94 &$1.21^{+0.15}_{-0.13} \pm 0.10$ & \cite{Akers:1995ui}$^c$  \\
OPAL   &$\Lambda\ell$ vtx     & 90--94 &$1.15 \pm 0.12 \pm 0.06$        & \cite{Akers:1995ui}$^c$ \\ 
\hline
\multicolumn{3}{l}{Average of above 16: \hfill mean \b-baryon lifetime $=$} & \hfagTAUBBnounit & \\  
\hline\hline
CDF2   &$\jpsi \Xi^-$        & 02--09 &$1.56 ^{+0.27}_{-0.25} \pm 0.02$ & \cite{Aaltonen:2009ny} \\
\hline
\multicolumn{3}{l}{Average of above 1: \hfill \Xibd lifetime $=$} & \hfagTAUXBDnounit & \\
\hline
ALEPH  &$\Xi\ell$             & 90--95 &$1.35^{+0.37+0.15}_{-0.28-0.17}$ & \cite{Buskulic:1996sm}\\
DELPHI &$\Xi\ell$             & 91--93 &$1.5 ^{+0.7}_{-0.4} \pm 0.3$     & \cite{Abreu:1995kt}$^d$ \\
DELPHI &$\Xi\ell$             & 92--95 &$1.45 ^{+0.55}_{-0.43} \pm 0.13$     & \cite{Abdallah:2005cw}$^d$ \\
\hline
\multicolumn{3}{l}{Average of above 4: \hfill mean \Xib lifetime $=$} & \hfagTAUXBnounit & \\
\hline\hline
CDF2   &$\jpsi \Omega^-$     & 02--09 &$1.13 ^{+0.53}_{-0.40} \pm 0.02$ & \cite{Aaltonen:2009ny} \\
\hline
\multicolumn{3}{l}{Average of above 1: \hfill \Omegab lifetime $=$} & \hfagTAUOBnounit & \\
\hline
\multicolumn{5}{l}{$^a$ \footnotesize The combined ALEPH result quoted 
in \cite{Barate:1997if} is $1.21 \pm 0.11$ ps.} \\[-0.5ex]
\multicolumn{5}{l}{$^b$ \footnotesize The combined DELPHI result quoted 
in \cite{Abreu:1999hu} is $1.14 \pm 0.08 \pm 0.04$ ps.} \\[-0.5ex]
\multicolumn{5}{l}{$^c$ \footnotesize The combined OPAL result quoted 
in \cite{Akers:1995ui} is $1.16 \pm 0.11 \pm 0.06$ ps.} \\[-0.5ex]
\multicolumn{5}{l}{$^d$ \footnotesize The combined DELPHI result quoted 
in \cite{Abdallah:2005cw} is $1.48 ^{+0.40}_{-0.31} \pm 0.12$ ps.}
\\[-0.5ex] \multicolumn{5}{l}{$^p$ \footnotesize Preliminary.}
\end{tabular}
\end{center}
\end{table}

Inputs to the averages are given in \Table{lifelb}.
The CDF $\Lambda_b \to \jpsi \Lambda$
lifetime result~\cite{Aaltonen:2010pj,*Abulencia:2006dr_mod_cont} is
$\hfagNSIGMATAULBCDFTWO\,\sigma$
larger than the world average computed excluding this result. 
It is nonetheless combined with the rest 
without adjustment of input errors.
The world average lifetime of \b baryons is then
\begin{equation}
\langle\tau(\mbox{\b-baryon})\rangle = \hfagTAUBB \,.
\end{equation}
Keeping only \particle{\Lambda^{\pm}_c \ell^{\mp}}, 
$\Lambda \ell^- \ell^+$, and fully exclusive
final states, as representative of
the \Lb baryon, the following lifetime is obtained:
\begin{equation}
\tau(\Lb) = \hfagTAULB \,. 
\end{equation}

Averaging the measurements based on the
$\Xi^{\mp} \ell^{\mp}$~\cite{Buskulic:1996sm,Abdallah:2005cw,Abreu:1995kt} 
and $\jpsi\Xi^{\mp}$~\cite{Aaltonen:2009ny}
final states gives
a lifetime value for a sample of events
containing $\Xib^0$ and $\Xib^-$ baryons:
\begin{equation}
\langle\tau(\Xib)\rangle = \hfagTAUXB \,.
\end{equation}
First measurements of fully reconstructed 
$\Xibd \to \jpsi\Xi^-$ and $\Omegab \to \jpsi\Omega^-$
baryons yield~\cite{Aaltonen:2009ny}
\begin{eqnarray}
\tau(\Xibd) &=& \hfagTAUXBD \,, \\
\tau(\Omegab) &=& \hfagTAUOB \,. 
\end{eqnarray}

\mysubsubsection{Summary and comparison with theoretical predictions}
\labs{lifesummary}

Averages of lifetimes of specific \b-hadron species are collected
in \Table{sumlife}.
\begin{table}[t]
\caption{Summary of lifetimes of different \b-hadron species.}
\labt{sumlife}
\begin{center}
\begin{tabular}{lc} \hline
\b-hadron species & Measured lifetime \\ \hline
\Bu                         & \hfagTAUBU   \\
\Bd                         & \hfagTAUBD   \\
\Bs ($1/\Gs$)               & \hfagTAUBSMEANC \\
\Bc                         & \hfagTAUBC   \\ 
\Lb                         & \hfagTAULB   \\
\Xib mixture                & \hfagTAUXB   \\
\b-baryon mixture           & \hfagTAUBB   \\
\b-hadron mixture           & \hfagTAUB    \\
\hline
\end{tabular}
\end{center}
\caption{Measured ratios of \b-hadron lifetimes relative to
the \Bd lifetime and ranges predicted
by theory~\cite{Tarantino:2003qw,*Gabbiani:2003pq,Gabbiani:2004tp}.}
\labt{liferatio}
%
%
%
\begin{center}
\begin{tabular}{lcc} \hline
Lifetime ratio & Measured value & Predicted range \\ \hline
$\tau(\Bu)/\tau(\Bd)$ & \hfagRTAUBU & 1.04 -- 1.08 \\
$\tau(\Bs)/\tau(\Bd)$ & \hfagRTAUBSMEANC & 0.99 -- 1.01 \\
$\tau(\Lb)/\tau(\Bd)$ & \hfagRTAULB & 0.86 -- 0.95    \\
$\tau(\mbox{\b-baryon})/\tau(\Bd)$  & \hfagRTAUBB & 0.86 -- 0.95 \\
\hline
\end{tabular}
\end{center}
\end{table}
As described in \Sec{lifetimes},
Heavy Quark Effective Theory
can be employed to explain the hierarchy of
$\tau(\Bc) \ll \tau(\Lb) < \tau(\Bs) \approx \tau(\Bd) < \tau(\Bu)$,
and used to predict the ratios between lifetimes.
Typical predictions are compared to the measured 
lifetime ratios in \Table{liferatio}.
The prediction of the ratio between the \Bu and \Bd lifetimes,
$1.06 \pm 0.02$~\cite{Tarantino:2003qw,*Gabbiani:2003pq}, 
is in good agreement with experiment. 

The total widths of the \Bs and \Bd mesons
are expected to be very close and differ by at most 
1\%~\cite{Beneke:1996gn,*Keum:1998fd,Gabbiani:2004tp}.
This prediction is consistent with the
experimental ratio $\tau(\Bs)/\tau(\Bd)=\Gd/\Gs$,
which is smaller than 1 by 
\hfagONEMINUSRTAUBSMEANCpercent. 

The ratio $\tau(\Lb)/\tau(\Bd)$ has particularly
been the source of theoretical
scrutiny since earlier calculations using Heavy Quark Effective Theory%
~\cite{Shifman:1986mx,*Chay:1990da,*Bigi:1992su,*Bigi:1992su_erratum,Voloshin:1999pz,*Guberina:1999bw,*Neubert:1996we,*Bigi:1997fj}
predicted a value larger than 0.90, almost $2\,\sigma$ 
above the world average at the time. 
Many predictions cluster around a most likely central value
of 0.94~\cite{Uraltsev:1996ta,*Pirjol:1998ur,*Colangelo:1996ta,*DiPierro:1999tb}.
More recent calculations
of this ratio that include higher-order effects predict a
lower ratio between the
\Lb and \Bd lifetimes~\cite{Tarantino:2003qw,*Gabbiani:2003pq,Gabbiani:2004tp}
and reduce this difference.
References~\cite{Tarantino:2003qw,*Gabbiani:2003pq,Gabbiani:2004tp} present probability density functions
of their predictions with variation of theoretical inputs, and the
indicated ranges in \Table{liferatio}
are the RMS of the distributions from the most probable values, and for 
$\tau(\Lb)/\tau(\Bd)$, also encompass the earlier theoretical predictions%
~\cite{Shifman:1986mx,*Chay:1990da,*Bigi:1992su,*Bigi:1992su_erratum,Voloshin:1999pz,*Guberina:1999bw,*Neubert:1996we,*Bigi:1997fj,Uraltsev:1996ta,*Pirjol:1998ur,*Colangelo:1996ta,*DiPierro:1999tb}.
Note that in contrast to the $B$ mesons, complete NLO QCD
corrections and
fully reliable lattice
determinations of the matrix elements for $\Lb$ are not
yet available.
As already mentioned, the CDF measurement of the $\Lambda_b$ lifetime
in the exclusive decay mode $\jpsi \Lambda$~\cite{Aaltonen:2010pj,*Abulencia:2006dr_mod_cont} is significantly 
higher than the world average before inclusion, with a ratio
to the $\tau(\Bd)$ world average of 
$\tau(\Lb)/\tau(\Bd) = 1.012 \pm 0.031$, 
%
resulting in continued interest in lifetimes of $b$ baryons.



\mysubsection{Neutral \B-meson mixing}
\labs{mixing}

The $\Bd-\Bdbar$ and $\Bs-\Bsbar$ systems
both exhibit the phenomenon of particle-antiparticle mixing. For each of them, 
there are two mass eigenstates which are linear combinations of the two flavour states,
\B and $\bar{\B}$. 
The heaviest (lightest) of the these mass states is denoted
$\B_{\rm H}$ ($\B_{\rm L}$),
with mass $m_{\rm H}$ ($m_{\rm L}$)
and total decay width $\Gamma_{\rm H}$ ($\Gamma_{\rm L}$). We define
\begin{eqnarray}
\Delta m = m_{\rm H} - m_{\rm L} \,, &~~~~&  x = \Delta m/\Gamma \,, \labe{dm} \\
\Delta \Gamma \, = \Gamma_{\rm L} - \Gamma_{\rm H} \,, ~ &~~~~&  y= \Delta\Gamma/(2\Gamma) \,, \labe{dg}
\end{eqnarray}
where 
$\Gamma = (\Gamma_{\rm H} + \Gamma_{\rm L})/2 =1/\bar{\tau}(\B)$ 
is the average decay width.
$\Delta m$ is positive by definition, and 
$\Delta \Gamma$ is expected to be positive within
the Standard Model.\footnote{For reason of symmetry in 
\Eqss{dm}{dg}, $\Delta \Gamma$ is sometimes defined with 
the opposite sign. The definition adopted here, \ie\
\Eq{dg}, is the one used by most experimentalists and many
phenomenologists in \B physics.}

There are four different time-dependent probabilities describing the 
case of a neutral \B meson produced 
as a flavour state and decaying to a flavour-specific final state.
If \CPT is conserved (which  
will be assumed throughout), they can be written as 
\begin{equation}
\left\{
\begin{array}{rcl}
{\cal P}(\B\to\B) & = &  \frac{e^{-\Gamma t}}{2} 
\left[ \cosh\!\left(\frac{\Delta\Gamma}{2}t\right) + \cos\!\left(\Delta m t\right)\right]  \\
{\cal P}(\B\to\bar{\B}) & = &  \frac{e^{-\Gamma t}}{2} 
\left[ \cosh\!\left(\frac{\Delta\Gamma}{2}t\right) - \cos\!\left(\Delta m t\right)\right] 
\left|\frac{q}{p}\right|^2 \\
{\cal P}(\bar{\B}\to\B) & = &  \frac{e^{-\Gamma t}}{2} 
\left[ \cosh\!\left(\frac{\Delta\Gamma}{2}t\right) - \cos\!\left(\Delta m t\right)\right] 
\left|\frac{p}{q}\right|^2 \\
{\cal P}(\bar{\B}\to\bar{\B}) & = &  \frac{e^{-\Gamma t}}{2} 
\left[ \cosh\!\left(\frac{\Delta\Gamma}{2}t\right) + \cos\!\left(\Delta m t\right)\right] 
\end{array} \right. \,,
\labe{oscillations}
\end{equation}
where $t$ is the proper time of the system (\ie\ the time interval between the production 
and the decay in the rest frame of the \B meson). 
At the \B factories, only the proper-time difference $\Delta t$ between the decays
of the two neutral \B mesons from the \Ups can be determined, but, 
because the two \B mesons evolve coherently (keeping opposite flavours as long as
none of them has decayed), the 
above formulae remain valid 
if $t$ is replaced with $\Delta t$ and the production flavour is replaced by the flavour 
at the time of the decay of the accompanying \B meson in a flavour-specific state.
As can be seen in the above expressions,
the mixing probabilities 
depend on three mixing observables:
$\Delta m$, $\Delta\Gamma$,
and $|q/p|^2$ which signals \CP violation in the mixing if $|q/p|^2 \ne 1$.

In the next sections we review in turn the experimental knowledge
on the \Bd decay-width and mass differences, 
the \Bs decay-width and mass differences,  
\CP violation in \Bd and \Bs mixing, and mixing-induced \CP violation in \Bs decays. 

\mysubsubsection{\Bd mixing parameters \DGd and \dmd}
\labs{DGd} \labs{dmd}

\begin{table}
\caption{Time-dependent measurements included in the \dmd average.
The results obtained from multi-dimensional fits involving also 
the \Bd (and \Bu) lifetimes
as free parameter(s)~\cite{Aubert:2002sh,Aubert:2005kf,Abe:2004mz} 
have been converted into one-dimensional measurements of \dmd.
All the measurements have then been adjusted to a common set of physics
parameters before being combined.}
\labt{dmd}
\begin{center}
\begin{tabular}{@{}rc@{}cc@{}c@{}cc@{}c@{}c@{}}
\hline
Experiment & \multicolumn{2}{c}{Method} & \multicolumn{3}{l}{\dmd in\invps}   
                                        & \multicolumn{3}{l}{\dmd in\invps}     \\
and Ref.   &  rec. & tag                & \multicolumn{3}{l}{before adjustment} 
                                        & \multicolumn{3}{l}{after adjustment} \\
\hline
 ALEPH~\cite{Buskulic:1996qt}  & \particle{ \ell  } & \particle{ \Qjet  } & $  0.404 $ & $ \pm  0.045 $ & $ \pm  0.027 $ & & & \\
 ALEPH~\cite{Buskulic:1996qt}  & \particle{ \ell  } & \particle{ \ell  } & $  0.452 $ & $ \pm  0.039 $ & $ \pm  0.044 $ & & & \\
 ALEPH~\cite{Buskulic:1996qt}  & \multicolumn{2}{c}{above two combined} & $  0.422 $ & $ \pm  0.032 $ & $ \pm  0.026 $ & $  0.442 $ & $ \pm  0.032 $ & $ ^{+  0.020 }_{-  0.019 } $ \\
 ALEPH~\cite{Buskulic:1996qt}  & \particle{ D^*  } & \particle{ \ell,\Qjet  } & $  0.482 $ & $ \pm  0.044 $ & $ \pm  0.024 $ & $  0.482 $ & $ \pm  0.044 $ & $ \pm  0.024 $ \\
 DELPHI~\cite{Abreu:1997xq}  & \particle{ \ell  } & \particle{ \Qjet  } & $  0.493 $ & $ \pm  0.042 $ & $ \pm  0.027 $ & $  0.503 $ & $ \pm  0.042 $ & $ \pm  0.024 $ \\
 DELPHI~\cite{Abreu:1997xq}  & \particle{ \pi^*\ell  } & \particle{ \Qjet  } & $  0.499 $ & $ \pm  0.053 $ & $ \pm  0.015 $ & $  0.501 $ & $ \pm  0.053 $ & $ \pm  0.015 $ \\
 DELPHI~\cite{Abreu:1997xq}  & \particle{ \ell  } & \particle{ \ell  } & $  0.480 $ & $ \pm  0.040 $ & $ \pm  0.051 $ & $  0.497 $ & $ \pm  0.040 $ & $ ^{+  0.042 }_{-  0.041 } $ \\
 DELPHI~\cite{Abreu:1997xq}  & \particle{ D^*  } & \particle{ \Qjet  } & $  0.523 $ & $ \pm  0.072 $ & $ \pm  0.043 $ & $  0.518 $ & $ \pm  0.072 $ & $ \pm  0.043 $ \\
 DELPHI~\cite{Abdallah:2002mr}  & \particle{ \mbox{vtx}  } & \particle{ \mbox{comb}  } & $  0.531 $ & $ \pm  0.025 $ & $ \pm  0.007 $ & $  0.527 $ & $ \pm  0.025 $ & $ \pm  0.006 $ \\
 L3~\cite{Acciarri:1998pq}  & \particle{ \ell  } & \particle{ \ell  } & $  0.458 $ & $ \pm  0.046 $ & $ \pm  0.032 $ & $  0.467 $ & $ \pm  0.046 $ & $ \pm  0.028 $ \\
 L3~\cite{Acciarri:1998pq}  & \particle{ \ell  } & \particle{ \Qjet  } & $  0.427 $ & $ \pm  0.044 $ & $ \pm  0.044 $ & $  0.439 $ & $ \pm  0.044 $ & $ \pm  0.042 $ \\
 L3~\cite{Acciarri:1998pq}  & \particle{ \ell  } & \particle{ \ell\mbox{(IP)}  } & $  0.462 $ & $ \pm  0.063 $ & $ \pm  0.053 $ & $  0.473 $ & $ \pm  0.063 $ & $ ^{+  0.045 }_{-  0.044 } $ \\
 OPAL~\cite{Ackerstaff:1997iw}  & \particle{ \ell  } & \particle{ \ell  } & $  0.430 $ & $ \pm  0.043 $ & $ ^{+  0.028 }_{-  0.030 } $ & $  0.466 $ & $ \pm  0.043 $ & $ ^{+  0.017 }_{-  0.016 } $ \\
 OPAL~\cite{Ackerstaff:1997vd}  & \particle{ \ell  } & \particle{ \Qjet  } & $  0.444 $ & $ \pm  0.029 $ & $ ^{+  0.020 }_{-  0.017 } $ & $  0.475 $ & $ \pm  0.029 $ & $ ^{+  0.014 }_{-  0.013 } $ \\
 OPAL~\cite{Alexander:1996id}  & \particle{ D^*\ell  } & \particle{ \Qjet  } & $  0.539 $ & $ \pm  0.060 $ & $ \pm  0.024 $ & $  0.544 $ & $ \pm  0.060 $ & $ \pm  0.023 $ \\
 OPAL~\cite{Alexander:1996id}  & \particle{ D^*  } & \particle{ \ell  } & $  0.567 $ & $ \pm  0.089 $ & $ ^{+  0.029 }_{-  0.023 } $ & $  0.572 $ & $ \pm  0.089 $ & $ ^{+  0.028 }_{-  0.022 } $ \\
 OPAL~\cite{Abbiendi:2000ec}  & \particle{ \pi^*\ell  } & \particle{ \Qjet  } & $  0.497 $ & $ \pm  0.024 $ & $ \pm  0.025 $ & $  0.496 $ & $ \pm  0.024 $ & $ \pm  0.025 $ \\
 CDF1~\cite{Abe:1997qf,*Abe:1998sq_mod_cont}  & \particle{ D\ell  } & \particle{ \mbox{SST}  } & $  0.471 $ & $ ^{+  0.078 }_{-  0.068 } $ & $ ^{+  0.033 }_{-  0.034 } $ & $  0.470 $ & $ ^{+  0.078 }_{-  0.068 } $ & $ ^{+  0.033 }_{-  0.034 } $ \\
 CDF1~\cite{Abe:1999pv}  & \particle{ \mu  } & \particle{ \mu  } & $  0.503 $ & $ \pm  0.064 $ & $ \pm  0.071 $ & $  0.515 $ & $ \pm  0.064 $ & $ \pm  0.070 $ \\
 CDF1~\cite{Abe:1999ds}  & \particle{ \ell  } & \particle{ \ell,\Qjet  } & $  0.500 $ & $ \pm  0.052 $ & $ \pm  0.043 $ & $  0.545 $ & $ \pm  0.052 $ & $ \pm  0.036 $ \\
 CDF1~\cite{Affolder:1999cn}  & \particle{ D^*\ell  } & \particle{ \ell  } & $  0.516 $ & $ \pm  0.099 $ & $ ^{+  0.029 }_{-  0.035 } $ & $  0.523 $ & $ \pm  0.099 $ & $ ^{+  0.028 }_{-  0.035 } $ \\
 \dzero~\cite{Abazov:2006qp}  & \particle{ D^{(*)}\mu  } & \particle{ \mbox{OST}  } & $  0.506 $ & $ \pm  0.020 $ & $ \pm  0.016 $ & $  0.506 $ & $ \pm  0.020 $ & $ \pm  0.016 $ \\
 \babar~\cite{Aubert:2001te,*Aubert:2002rg_cont}  & \particle{ \Bd  } & \particle{ \ell,K,\mbox{NN}  } & $  0.516 $ & $ \pm  0.016 $ & $ \pm  0.010 $ & $  0.521 $ & $ \pm  0.016 $ & $ \pm  0.008 $ \\
 \babar~\cite{Aubert:2001tf}  & \particle{ \ell  } & \particle{ \ell  } & $  0.493 $ & $ \pm  0.012 $ & $ \pm  0.009 $ & $  0.487 $ & $ \pm  0.012 $ & $ \pm  0.006 $ \\
 \babar~\cite{Aubert:2005kf}  & \particle{ D^*\ell\nu\mbox{(part)}  } & \particle{ \ell  } & $  0.511 $ & $ \pm  0.007 $ & $ \pm  0.007 $ & $  0.512 $ & $ \pm  0.007 $ & $ \pm  0.007 $ \\
 \babar~\cite{Aubert:2002sh}  & \particle{ D^*\ell\nu  } & \particle{ \ell,K,\mbox{NN}  } & $  0.492 $ & $ \pm  0.018 $ & $ \pm  0.014 $ & $  0.493 $ & $ \pm  0.018 $ & $ \pm  0.013 $ \\
 \belle~\cite{Zheng:2002jv}  & \particle{ D^*\pi\mbox{(part)}  } & \particle{ \ell  } & $  0.509 $ & $ \pm  0.017 $ & $ \pm  0.020 $ & $  0.513 $ & $ \pm  0.017 $ & $ \pm  0.019 $ \\
 \belle~\cite{Hastings:2002ff}  & \particle{ \ell  } & \particle{ \ell  } & $  0.503 $ & $ \pm  0.008 $ & $ \pm  0.010 $ & $  0.506 $ & $ \pm  0.008 $ & $ \pm  0.008 $ \\
 \belle~\cite{Abe:2004mz}  & \particle{ \Bd,D^*\ell\nu  } & \particle{ \mbox{comb}  } & $  0.511 $ & $ \pm  0.005 $ & $ \pm  0.006 $ & $  0.513 $ & $ \pm  0.005 $ & $ \pm  0.006 $ \\
 LHCb~\cite{LHCb-CONF-2011-010,*LHCb-CONF-2011-010_published}  & \particle{ \Bd  } & \particle{ \mbox{OST}  } & $  0.499 $ & $ \pm  0.032 $ & $ \pm  0.003 $ & $  0.499 $ & $ \pm  0.032 $ & $ \pm  0.003 $ \\
 \hline \\[-2.0ex]
 \multicolumn{6}{l}{World average (all above measurements included):} & $  0.507 $ & $ \pm  0.003 $ & $ \pm  0.003 $ \\

\\[-2.0ex]
\multicolumn{6}{l}{~~~ -- ALEPH, DELPHI, L3, OPAL and CDF1 only:}
     & \hfagDMDHval & \hfagDMDHsta & \hfagDMDHsys \\
\multicolumn{6}{l}{~~~ -- Above measurements of \babar and \belle only:}
     & \hfagDMDBval & \hfagDMDBsta & \hfagDMDBsys \\
\hline
\end{tabular}
\end{center}
\end{table}

Many time-dependent \Bd--\Bdbar oscillation analyses have been performed by the 
ALEPH, \babar, \belle, CDF, \dzero, DELPHI, L3 and OPAL collaborations. 
The corresponding measurements of \dmd are summarized in 
\Table{dmd},
where only the most recent results
are listed (\ie\ measurements superseded by more recent ones are omitted)\footnote{
Two old unpublished CDF2 measurements~\cite{CDFnote8235:2006,CDFnote7920:2005}
are also omitted from our averages, \Table{dmd} and \Fig{dmd}.}.
Although a variety of different techniques have been used, the 
individual \dmd
results obtained at high-energy colliders have remarkably similar precision.
Their average is compatible with the recent and more precise measurements 
from the asymmetric \B factories.
The systematic uncertainties are not negligible; 
they are often dominated by sample composition, mistag probability,
or \b-hadron lifetime contributions.
Before being combined, the measurements are adjusted on the basis of a 
common set of input values, including the averages of the 
\b-hadron fractions and lifetimes given in this report 
(see \Secss{fractions}{lifetimes}).
Some measurements are statistically correlated. 
Systematic correlations arise both from common physics sources 
(fractions, lifetimes, branching ratios of \b hadrons), and from purely 
experimental or algorithmic effects (efficiency, resolution, flavour tagging, 
background description). Combining all published measurements
listed in \Table{dmd}
and accounting for all identified correlations
as described in \Ref{Abbaneo:2000ej_mod,*Abbaneo:2001bv_mod_cont} yields $\dmd = \hfagDMDWfull$.

On the other hand, ARGUS and CLEO have published 
measurements of the time-integrated mixing probability 
\chid~\cite{Albrecht:1992yd,*Albrecht:1993gr_cont,Bartelt:1993cf,Behrens:2000qu}, 
which average to $\chid =\hfagCHIDU$.
Following \Ref{Behrens:2000qu}, 
the width difference \DGd could 
in principle be extracted from the
measured value of $\Gd=1/\tau(\Bd)$ and the above averages for 
\dmd and \chid 
(provided that \DGd has a negligible impact on 
the \dmd $\tau(\Bd)$ analyses that have assumed $\DGd=0$), 
using the relation
\begin{equation}
\chid = \frac{\xd^2+\yd^2}{2(\xd^2+1)} ~~~ \mbox{with} ~~ \xd=\frac{\dmd}{\Gd} 
~~~ \mbox{and} ~~ \yd=\frac{\DGd}{2\Gd} \,.
\labe{chid_definition}
\end{equation}
However, direct time-dependent studies provide much stronger constraints: 
$|\DGd|/\Gd < 18\%$ at \CL{95} from DELPHI~\cite{Abdallah:2002mr},
and $-6.8\% < {\rm sign}({\rm Re} \lambda_{\CP}) \DGGd < 8.4\%$
at \CL{90} from \babar~\cite{Aubert:2003hd,*Aubert:2004xga_mod_cont},
where $\lambda_{\CP} = (q/p)_{\particle{d}} (\bar{A}_{\CP}/A_{\CP})$
is defined for a \CP-even final state 
(the sensitivity to the overall sign of 
${\rm sign}({\rm Re} \lambda_{\CP}) \DGGd$ comes
from the use of \Bd decays to \CP final states).
Recently Belle has measured ${\rm sign}({\rm Re} \lambda_{\CP}) = 0.017 \pm 0.018 \pm 0.011$~\cite{Higuchi:2012kx}.
A combination of these three results (after adjusting the DELPHI and \babar ones to  
$1/\Gd=\tau(\Bd)=\hfagTAUBD$) yields
\begin{equation}
{\rm sign}({\rm Re} \lambda_{\CP}) \DGGd  = \hfagSDGDGD \,.
\end{equation}
The sign of ${\rm Re} \lambda_{\CP}$ is not measured,
but expected to be positive from the global fits
of the Unitarity Triangle within the Standard Model~\cite{Charles:2011va_mod,*Bona:2006ah}.

Assuming $\DGd=0$ 
and using $1/\Gd=\tau(\Bd)=\hfagTAUBD$,
the \dmd and \chid results are combined through \Eq{chid_definition} 
to yield the 
world average
\begin{equation} 
\dmd = \hfagDMDWU \,,
\labe{dmd}
\end{equation} 
or, equivalently,
\begin{equation} 
\xd= \hfagXDWU ~~~ \mbox{and} ~~~ \chid=\hfagCHIDWU \,.  
\labe{chid}
\end{equation}
\Figure{dmd} compares the \dmd values obtained by the different experiments.

\begin{figure}
\begin{center}
\epsfig{figure=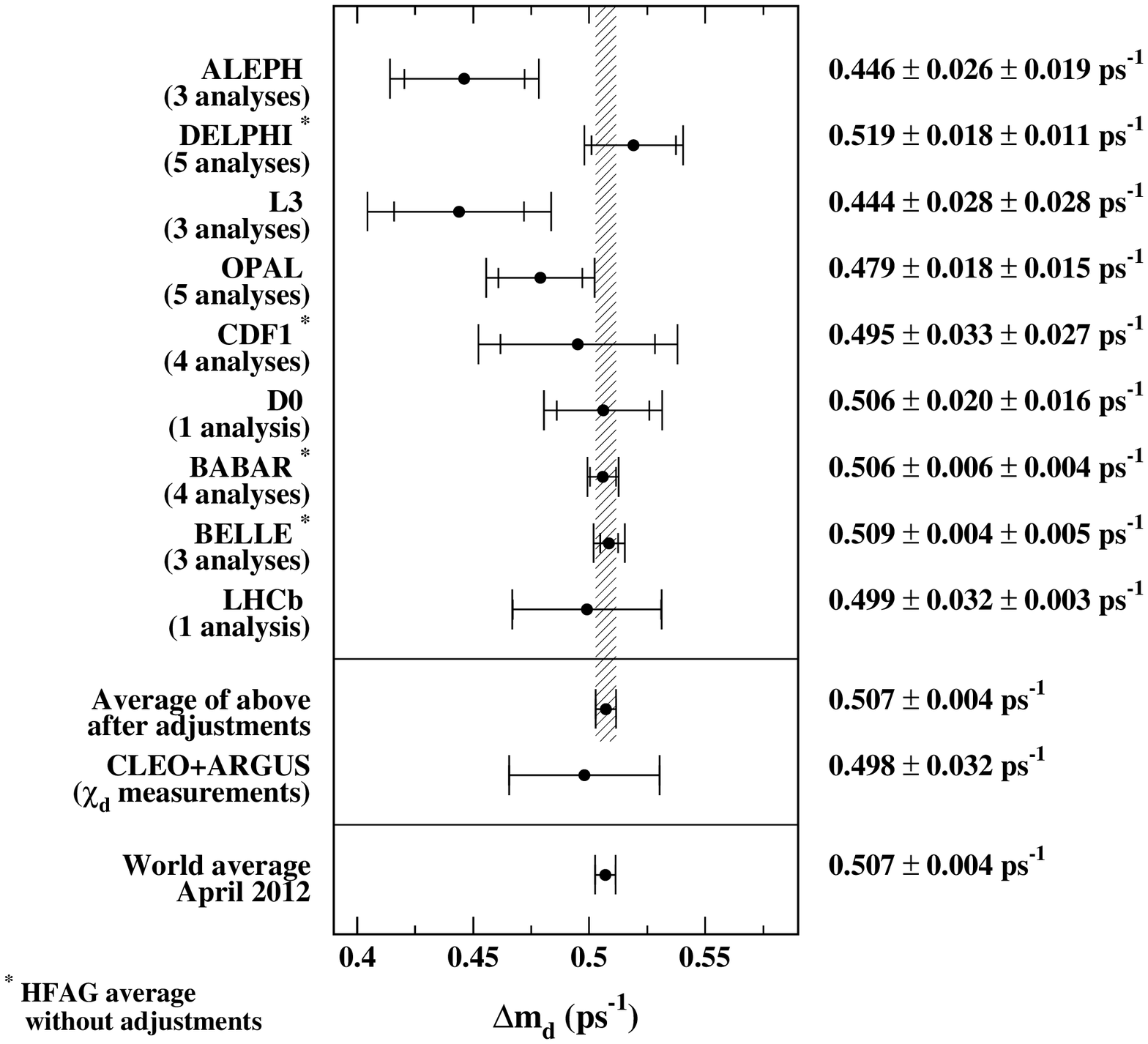,width=\textwidth}
\caption{The \Bd--\Bdbar oscillation frequency \dmd as measured by the different experiments. 
The averages quoted for ALEPH, L3 and OPAL are taken from the original publications, while the 
ones for DELPHI, CDF, \babar, and \belle have been computed from the individual results 
listed in \Table{dmd} without performing any adjustments. The time-integrated measurements 
of \chid from the symmetric \B factory experiments ARGUS and CLEO have been converted 
to a \dmd value using $\tau(\Bd)=\hfagTAUBD$. The two global averages have been obtained 
after adjustments of all the individual \dmd results of \Table{dmd} (see text).}
\labf{dmd}
\end{center}
\end{figure}

The \Bd mixing averages given in \Eqss{dmd}{chid}
and the \b-hadron fractions of \Table{fractions} have been obtained in a fully 
consistent way, taking into account the fact that the fractions are computed using 
the \chid value of \Eq{chid} and that many individual measurements of \dmd
at high energy depend on the assumed values for the \b-hadron fractions.
Furthermore, this set of averages is consistent with the lifetime averages 
of \Sec{lifetimes}.

\begin{table}
\caption{Simultaneous measurements of \dmd and $\tau(\Bd)$, and their average.
The \belle analysis also 
measures $\tau(\Bu)$ at the same time, but it is converted here into a two-dimensional measurement 
of \dmd and $\tau(\Bd)$, for an assumed value of $\tau(\Bu)$. 
The first quoted error on the measurements is statistical
and the second one systematic; in the case of adjusted measurements, the 
latter includes a contribution obtained from the variation of $\tau(\Bu)$ or 
$\tau(\Bu)/\tau(\Bd)$ in the indicated range. Units are\invps\ for \dmd
and\unit{ps} for lifetimes. 
The three different values of $\rho(\dmd,\tau(\Bd))$ correspond 
to the statistical, systematic and total correlation coefficients
between the adjusted measurements of \dmd and $\tau(\Bd)$.}
\labt{dmd2D}
\begin{center}
\begin{tabular}{@{}r@{~}c@{}c@{}c@{~}c@{}c@{}c@{~}c@{}c@{}c@{\hspace{0ex}}c@{}}
\hline
Exp.\ \& Ref.
& \multicolumn{3}{c}{Measured \dmd}   
& \multicolumn{3}{c}{Measured $\tau(\Bd)$}   
& \multicolumn{3}{c}{Measured $\tau(\Bu)$}   
&  Assumed $\tau(\Bu)$ \\
\hline
\babar \cite{Aubert:2002sh}  
      & $0.492$ & $\pm 0.018$ & $\pm 0.013$ 
      & $1.523$ & $\pm 0.024$ & $\pm 0.022$ 
      & \multicolumn{3}{c}{---}
      & $(1.083\pm 0.017)\tau(\Bd)$ \\  
\babar \cite{Aubert:2005kf}  
      & $0.511$ & $\pm 0.007$ & $^{+0.007}_{-0.006}$ 
      & $1.504$ & $\pm 0.013$ & $^{+0.018}_{-0.013}$
      & \multicolumn{3}{c}{---}
      & $1.671\pm 0.018$ \\  
\belle \cite{Abe:2004mz}  
      & $0.511$ & $\pm 0.005$ & $\pm 0.006$
      & $1.534$ & $\pm 0.008$ & $\pm 0.010$
      & $1.635$ & $\pm 0.011$ & $\pm 0.011$
      & --- \\  
\cline{2-10}
& \multicolumn{3}{c}{Adjusted \dmd}   
& \multicolumn{3}{c}{Adjusted $\tau(\Bd)$}   
& \multicolumn{3}{c}{$\rho(\dmd,\Bd)$} 
&  Assumed $\tau(\Bu)$ \\
\cline{2-10}
\babar \cite{Aubert:2002sh}  
      & $0.492$ & $\pm 0.018$ & $\pm 0.013$  
      & $1.523$ & $\pm 0.024$ & $\pm 0.022$  
      & $-0.22$ & $+0.71$ & $+0.16$ 
      & $(\hfagRTAUBUval$$\hfagRTAUBUerr)\tau(\Bd)$ \\  
\babar \cite{Aubert:2005kf} 
      & $0.512$ & $\pm 0.007$ & $\pm 0.007$  
      & $1.506$ & $\pm 0.013$ & $\pm 0.018$ 
      & $+0.01$ & $-0.85$ & $-0.48$ 
      & $\hfagTAUBUval$$\hfagTAUBUerr$ \\  
\belle \cite{Abe:2004mz}  
      & $0.511$ & $\pm 0.005$ & $\pm 0.006$ 
      & $1.535$ & $\pm 0.008$ & $\pm 0.011$ 
      & $-0.27$ & $-0.14$ & $-0.19$ 
      & $\hfagTAUBUval$$\hfagTAUBUerr$ \\  
\hline
\multicolumn{1}{l}{Average} 
      & \hfagDMDTWODval   & \hfagDMDTWODsta   & \hfagDMDTWODsys
      & \hfagTAUBDTWODval & \hfagTAUBDTWODsta & \hfagTAUBDTWODsys
      & \hfagRHOstaDMDTAUBD & \hfagRHOsysDMDTAUBD & \hfagRHODMDTAUBD 
      & $\hfagTAUBUval$$\hfagTAUBUerr$ \\  
\hline 
\end{tabular}
\end{center}
\end{table}
\begin{figure}
\begin{center}
\vspace{-0.5cm}
\epsfig{figure=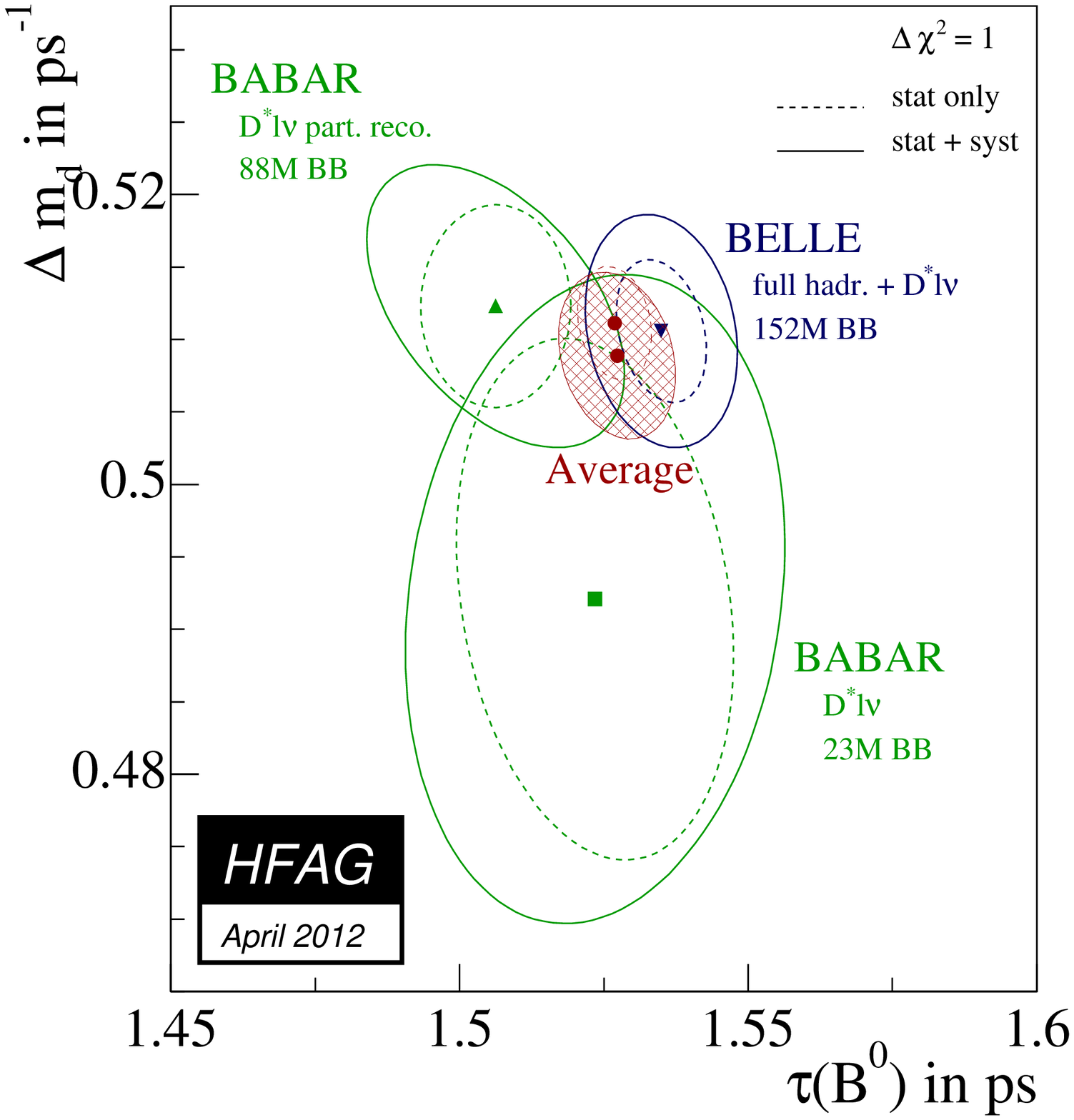,width=0.6\textwidth}
\vspace{-0.5cm}
\caption{Simultaneous measurements of
\dmd and $\tau(\Bd)$~\cite{Aubert:2002sh,Aubert:2005kf,Abe:2004mz}, 
after adjustment to a common set of parameters (see text). 
Statistical and total uncertainties are represented as dashed and
solid contours respectively.
The average of the three measurements
is indicated by a hatched ellipse.}
\labf{dmd2D}
\end{center}
\end{figure}

It should be noted that the most recent (and precise) analyses at the 
asymmetric \B factories measure \dmd
as a result of a multi-dimensional fit. 
Two \babar analyses~\cite{Aubert:2002sh,Aubert:2005kf},  
based on fully and partially reconstructed $\Bd \to D^*\ell\nu$ decays
respectively, 
extract simultaneously \dmd and $\tau(\Bd)$
while the latest \belle analysis~\cite{Abe:2004mz},  
based on fully reconstructed hadronic \Bd decays and $\Bd \to D^*\ell\nu$ decays, 
extracts simultaneously \dmd, $\tau(\Bd)$ and $\tau(\Bu)$.
The measurements of \dmd and $\tau(\Bd)$ of these three analyses 
are displayed in \Table{dmd2D} and in \Fig{dmd2D}. Their two-dimensional average, 
taking into account all statistical and systematic correlations, and expressed
at $\tau(\Bu)=\hfagTAUBU$, is
\begin{equation}
\left.
\begin{array}{r@{}l}
\dmd = \hfagDMDTWODnounit & \invps \\
\tau(\Bd) = \hfagTAUBDTWODnounit & \ps
\end{array}
\right\}
~\mbox{with a total correlation of \hfagRHODMDTAUBD.}
\end{equation}

\mysubsubsection{\Bs mixing parameters \DGs and \dms}
\labs{DGs} \labs{dms}

%
%
%


Definitions and an introduction to \DGs have been given in \Sec{taubs}.
Neglecting \CP violation, the mass eigenstates are
also \CP eigenstates, with the short-lived state being
\CP-even and the long-lived state being \CP-odd.

The best sensitivity to \DGs is currently achieved 
by the recent time-dependent measurements
of the $\hbox{\Bs} \to \jpsi \phi$ decay rates performed at
CDF~\cite{CDFnote10778:2012,*CDFnote10778:2012_cont,CDF:2011af,*Aaltonen:2007he_mod,*Aaltonen:2007gf_mod},
\dzero~\cite{Abazov:2011ry,*Abazov_mod:2008fj,*Abazov:2007tx_mod_cont}
and LHCb~\cite{LHCbnote002:2012,*LHCbnote002:2012_cont,LHCb:2011aa},
where the \CP-even and \CP-odd
amplitudes are statistically separated through a full angular analysis
(see last two columns of \Table{phisDGsGs}). 
In particular LHCb obtained the first observation of a non-zero 
value of $\DGs$~\cite{LHCbnote002:2012,*LHCbnote002:2012_cont}.
These studies use both untagged and tagged \Bs\ candidates and 
are optimized for the measurement of the \CP-violating 
phase \phiccbars, defined later in \Sec{phasebs}.
Recently the LHCb collaboration analyzed the $\Bs \to \jpsi K^+K^-$
decay, considering that the $K^+K^-$ system can be in a $P$-wave or $S$-wave state, 
and measured the dependence of the strong phase difference between the 
$P$-wave and $S$-wave amplitudes as a function of the $K^+K^-$ invariant
mass~\cite{Aaij:2012eq}. 
This allowed, for the first time, the unambiguous determination of the sign of 
$\DGs$, which was found to be positive at the $4.7\,\sigma$ level and the
following averages present only the $\DGs > 0$ solutions.

The combined fit procedure used to extract simultaneously \DGs\ and \phiccbars
is described in \Sec{phasebs}. 
The results, displayed as the red contours labelled ``$\Bs \to \jpsi\phi$ measurements'' in the 
plots of \Fig{DGs}, are given in the first column of numbers of \Table{tabtauLH}.
In those averages, the correlation between \DGs and \Gs has been neglected. 

\begin{figure}
\begin{center}
\epsfig{figure=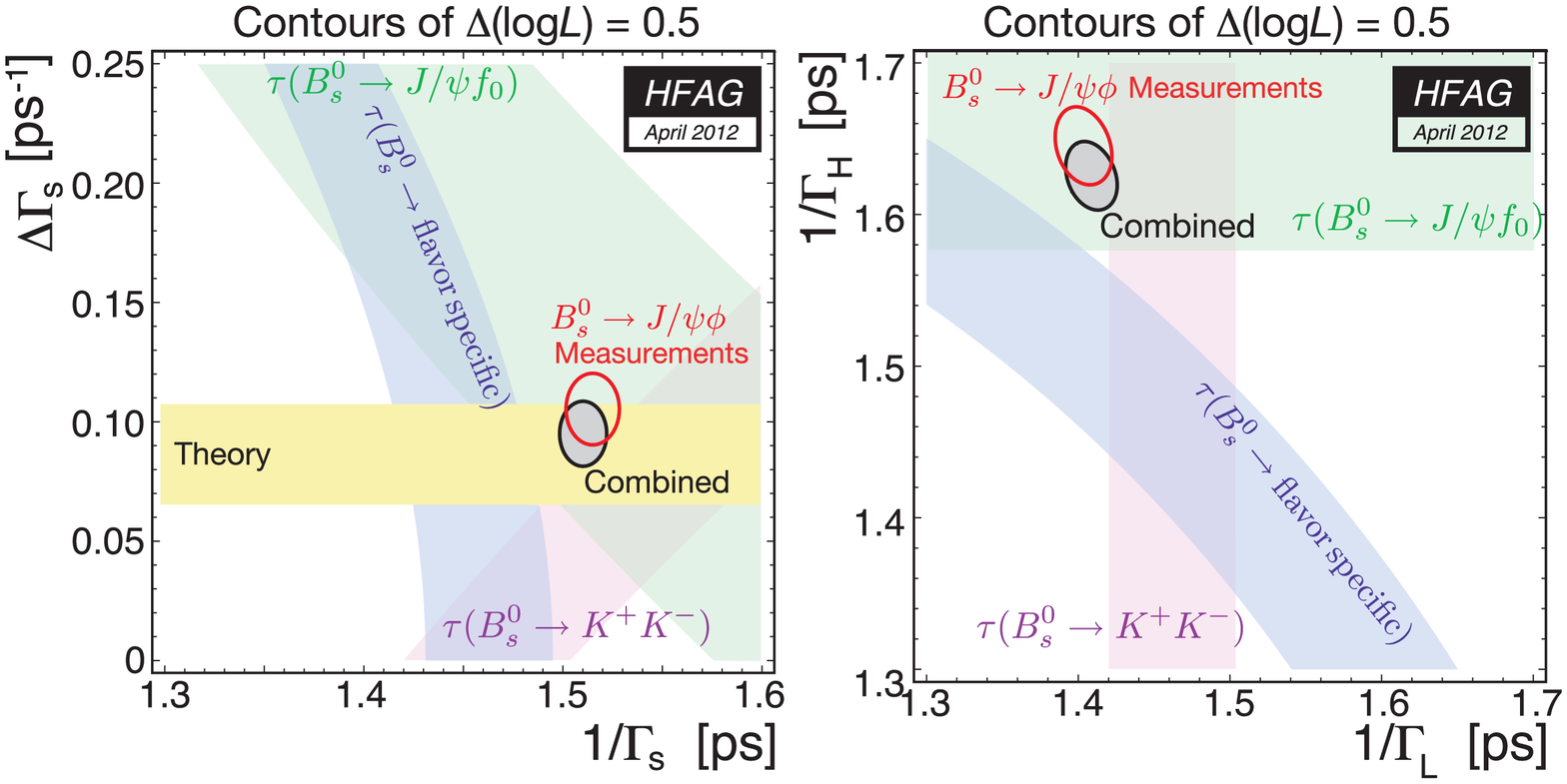,width=0.99\textwidth}
\caption{Contours of $\Delta \ln L = 0.5$ (39\% CL for the enclosed 2D regions, 68\% CL for the bands)
shown in the $(1/\Gs,\,\DGs)$ plane on the left
and in the $(1/\Gamma_{\rm L},\,1/\Gamma_{\rm H})$ plane on the right. 
The average of all the $\Bs \to \jpsi\phi$ results is shown as the red contour,
and the constraints given by the effective lifetime measurements of \Bs\ to flavour-specific final states, 
$\Bs \to \jpsi f_0(980)$ and $\Bs \to K^+K^-$ are shown as the blue, green and purple bands, 
respectively. The average taking all constraints into account is shown as the gray-filled contour.
The yellow band is a theory prediction
$\DGs = 0.087 \pm 0.021~\hbox{ps}^{-1}$~\cite{Lenz:2011ti,*Lenz:2006hd}
that assumes no new physics in \Bs\ mixing.}
\labf{DGs}
\end{center}
\end{figure}

\begin{table}
\caption{Averages of \DGs, $1/\Gs$ and related quantities, obtained from
$\Bs\to\jpsi\phi$ alone (first column), adding the constraints from the effective
lifetime measured in $\Bs\to\ K^+ K^-$ and $\Bs\to\jpsi f_0(980)$ (second column),
and adding the constraint from the average flavour-specific lifetime (third column,
recommended world averages).}
\labt{tabtauLH}
\begin{center}
\begin{tabular}{l|c|c|c}
\hline
                   & $\jpsi\phi$
                   & $\jpsi\phi, K^+K^-, \jpsi f_0$
                   & $\jpsi\phi, K^+K^-, \jpsi f_0, D_s^-\ell^+, D_s^-\pi^+$ \\
\hline
\DGs               & \hfagDGS       &  \hfagDGSCO       &  \hfagDGSCON       \\
$1/\Gs$            & \hfagTAUBSMEAN &  \hfagTAUBSMEANCO &  \hfagTAUBSMEANCON \\
$1/\Gamma_{\rm L}$ & \hfagTAUBSL    &  \hfagTAUBSLCO    &  \hfagTAUBSLCON    \\
$1/\Gamma_{\rm H}$ & \hfagTAUBSH    &  \hfagTAUBSHCO    &  \hfagTAUBSHCON    \\
\DGs/\Gs           & \hfagDGSGS     &  \hfagDGSGSCO     &  \hfagDGSGSCON     \\
\hline
\end{tabular}
\end{center}
\end{table}


An alternative approach, which is directly sensitive to first order in 
$\DGs/\Gs$, 
is to determine the effective lifetime of untagged \Bs\ candidates
decaying to 
\CP eigenstates; measurements exist for
$\Bs \to K^+K^-$~\cite{Aaij:2011kn,LHCb-CONF-2012-001}%
\footnote{An old unpublished measurement of the $\Bs \to K^+ K^-$
effective lifetime by CDF~\cite{Tonelli:2006np} is no longer considered.},
and $\Bs \to \jpsi f_0(980)$~\cite{Aaltonen:2011nk}.
The precise extraction of $1/\Gs$ and $\DGs$
from such measurements, discussed in detail in \Ref{Fleischer:2011cw}, 
requires additional information 
in the form of theoretical assumptions or
external inputs on weak phases and hadronic parameters. 
If $f$ designates a final state in which both \Bs and \Bsbar can decay,
the ratio of the effective \Bs lifetime decaying to $f$ relative to the mean
\Bs lifetime is~\cite{Fleischer:2011cw}%
\footnote{The definition of $A_f^{\DG}$ given in \Eq{ADG} has the sign opposite to that given in \Ref{Fleischer:2011cw}.}
\begin{equation}
  \frac{\tau_{\rm single}(\Bs \to f)}{\tau(\Bs)} = \frac{1}{1-y_s^2} \left[ \frac{1 - 2A_f^{\DG} y_s + y_s^2}{1 - A_f^{\DG} y_s}\right ] \,,
\labe{tauf_fleisch}
\end{equation}
where
\begin{equation}
A_f^{\DG} = -\frac{2 \Re(\lambda_f)} {1+|\lambda_f|^2} \,.
\labe{ADG}
\end{equation}
To include the measurements of the effective $\Bs \to K^+K^-$ and 
$\Bs \to \jpsi f_0(980)$ lifetimes 
as constraints in the \DGs fit,
we neglect sub-leading penguin contributions and possible direct \CP violation. 
Explicitly, in \Eq{ADG}, we set
$A_{KK}^{\DG} = \cos \phiccbars$
and $A_{\jpsi f_0}^{\DG} = -\cos \phiccbars$.
Given the small value of $\phiccbars$, we have, to first order in $y_s$:
\begin{eqnarray}
\tau_{\rm single}(\Bs \to K^+K^-)
& \approx & \frac{1}{\Gamma_{\rm L}} \left(1 + \frac{(\phiccbars)^2 y_s}{2} \right) \,,
\labe{tau_KK_approx}
\\
\tau_{\rm single}(\Bs \to \jpsi f_0(980))
& \approx & \frac{1}{\Gamma_{\rm H}} \left(1 - \frac{(\phiccbars)^2 y_s}{2} \right) \,.
\labe{tau_Jpsif0_approx}
\end{eqnarray}
The numerical inputs are taken from \Eqss{tau_KK}{tau_Jpsif0} 
and the resulting averages, combined with the $\Bs\to\jpsi\phi$ information,
are indicated in the second column of numbers of \Table{tabtauLH}. 

Information on \DGs can also be obtained from the study of the
proper time distribution of untagged samples
of flavour-specific \Bs decays~\cite{Hartkorn:1999ga}.
In the case of flavour-specific \Bs\ decays where the flavour,
\ie\ \Bs or \Bsbar, at the time of decay can be determined by
the decay products. In such decays,
\eg\ semileptonic \Bs decays, there is
an equal mix of the heavy and light mass eigenstates at time zero.
The proper time distribution is then a superposition 
of two exponential functions with decay constants
$\Gamma_{\rm L,H} = \Gs \pm \DGs/2$.
This provides sensitivity to both $1/\Gs$ and 
$(\DGs/\Gs)^2$. Ignoring \DGs and fitting for 
a single exponential leads to an estimate of \Gs with a 
relative bias proportional to $(\DGs/\Gs)^2$, as shown in \Eq{fslife}. 
Including the constraint from the world-average flavour-specific \Bs 
lifetime, given in \Eq{tau_fs}, leads to the results shown in the last column 
of \Table{tabtauLH}.
These world averages are displayed as the gray contours labelled ``Combined'' in the
plots of \Fig{DGs}. The average for the decay-width difference,
\begin{equation}
\DGs = \hfagDGSCON ~~~~\mbox{and} ~~~~~ \DGs/\Gs = \hfagDGSGSCON \,, 
\labe{DGs_DGsGs}
\end{equation}
is in good agreement with the Standard Model prediction 
$\DGs = 0.087 \pm 0.021~\hbox{ps}^{-1}$~\cite{Lenz:2011ti,*Lenz:2006hd}.

Independent estimates of $\DGs/\Gs$ obtained from measurements of the 
$\Bs \to D_s^{(*)+} D_s^{(*)-}$ branching fraction~\cite{Barate:2000kd,Esen:2010jq_mod,Abazov:2008ig,*Abazov:2007rb_mod_cont,Abulencia:2007zz}
have not been used\footnote{
Our average is ${\cal B} = \hfagBRDSDS$, from which one would get 
$\DGs/\Gs \sim 2{\cal B}/(1-{\cal B}) = \hfagDGSGSBRDSDS$.},
since they are based on the questionable~\cite{Lenz:2011ti,*Lenz:2006hd}
assumption that these decays account for all \CP-even final states.
The results of early lifetime analyses attempting
to measure $\DGs/\Gs$~\cite{Acciarri:1998uv,Abreu:2000sh,Abreu:2000ev,Abe:1997bd}
have not been used either. 

\comment{



Numerical results of the combination of the CDF2 and \dzero inputs
of \Table{dgammat} are:
\begin{eqnarray}
\DGGs &=& \hfagDGSGS \,, \\
\DGs &=& \hfagDGS \,, \\
\bar{\tau}(\Bs) = 1/\Gs &=& \hfagTAUBSMEAN \,, \\
1/\Gamma_{\rm L} = \tau_{\rm short} &=& \hfagTAUBSL \,, \\
1/\Gamma_{\rm H} = \tau_{\rm long}  &=& \hfagTAUBSH \,. 
\end{eqnarray}

Flavour-specific lifetime measurements are of an equal mix
of \CP-even and \CP-odd states at time zero, and  
if a single exponential function is used in the likelihood
lifetime fit of such a sample~\cite{Hartkorn:1999ga}, 
\begin{equation}
\tau(\Bs)_{\rm fs} = \frac{1}{\Gs}
\frac{{1+\left(\frac{\DGs}{2\Gs}\right)^2}}{{1-\left(\frac{\DGs}{2\Gs}\right)^2}
} \,.
\labe{fslife_const}
\end{equation}
Using the world average flavour-specific 
lifetime of \Eq{tau_fs} in \Sec{taubs}
the one-sigma blue bands shown in \Fig{DGs} are obtained. 
Higher-order corrections were checked to be negligible in the
combination.

When the flavour-specific lifetime measurements 
are combined with the 
CDF2 and \dzero measurements of \Table{dgammat}, the solid-outline
shaded
regions of \Fig{DGs} are obtained, with numerical results:
\begin{eqnarray}
\DGGs &=& \hfagDGSGSCON \,, \labe{DGGs_ave} \\
\DGs &=& \hfagDGSCON \,, \\
\bar{\tau}(\Bs) = 1/\Gs &=& \hfagTAUBSMEANC \,, \labe{oneoverGs} \\
1/\Gamma_{\rm L} = \tau_{\rm short} &=& \hfagTAUBSLCON \,, \\
1/\Gamma_{\rm H} = \tau_{\rm long}  &=& \hfagTAUBSHCON \,. 
\end{eqnarray}
These results can
be compared with the theoretical prediction of 
$\DGs = 0.096 \pm 0.039\invps$
(or $\DGs = 0.088 \pm 0.017\invps$ if there is no new physics in
\dms)~\cite{Lenz:2011ti,*Lenz:2006hd,Beneke:1998sy}.

Measurements of $\BR{B^0_s \to D_s^{(*)+} D_s^{(*)-}}$ can 
also be sensitive to \DGs.
The decay $\Bs \to D_s^{+} D_s^{-}$ is into
a final state that is purely \CP even. 
Under various theoretical assumptions~\cite{Aleksan:1993qp,Dunietz:2000cr}, the
inclusive decay into this plus the excited states
$\Bs \to D_s^{(*)+} D_s^{(*)-}$ is also \CP even
to within 5\%, and 
$\Bs \to D_s^{(*)+} D_s^{(*)-}$ saturates
$\Gs^{\CP \thinspace {\rm even}}$.
Under these assumptions, for no \CP violation, we have: 
\begin{equation}
\DGGs \approx
\frac{2 \BR{\Bs \to D_s^{(*)+} D_s^{(*)-}}}
{1 - \BR{\Bs \to D_s^{(*)+} D_s^{(*)-}}} \,.
\labe{dGsBr}
\end{equation}
However, there are concerns~\cite{Nierste_private:2006} 
that the assumptions needed
for the above are overly restrictive and that the inclusive branching
ratio may be \CP even to only 30\%.
In the application of the constraint as a Gaussian penalty
function, the theoretical uncertainty is dealt with in two ways:
the fraction of the \CP-odd component of the decay~\cite{Dunietz:2000cr} 
is taken
to be a uniform distribution ranging from 0 to 0.05 and
convoluted in the Gaussian, and the fractional uncertainty on the
average measured value is increased in quadrature by 
30\%.

\begin{table}
\caption{Measurements of $\BR{\Bs \to D_s^{(*)+} D_s^{(*)-}}$.}
\labt{dGsBr}
\begin{center}
\begin{tabular}{l|c|c|c}
\hline
Experiment & Method & Value & Ref.  \\
\hline
ALEPH         & $\phi$-$\phi$ correlations              
           & $0.115 \pm 0.050^{+0.095}_{-0.045}$  & \cite{Barate:2000kd}$^a$     \\
\dzero        & $D_s \to \phi \pi$, $D_s \to \phi \mu \nu$            
           & $0.035 \pm 0.010 \pm 0.011$  & \cite{Abazov:2008ig,*Abazov:2007rb_mod_cont}$^{~}$ \\
\belle      & full reco.\ in 6 excl.\ $D_s$ modes 
            & $0.0685 ^{+0.0153}_{-0.0130} {}^{+0.0179}_{-0.0180}$ & \cite{Esen:2010jq_mod}$^{~}$ \\
	 \hline
\multicolumn{2}{l}{Average of above 3} &   \hfagBRDSDS  &   \\
      \hline
\multicolumn{4}{l}{
$^a$ \footnotesize The value quoted in this table is half of 
$\BR{\Bs{\rm(short)} \to D_s^{(*)+} D_s^{(*)-}}$
given in \Ref{Barate:2000kd}.} \\[-0.5ex]
\multicolumn{4}{l}{$^{~}$ \footnotesize Before averaging, it has been adjusted the latest values
of \fBs at LEP and \BR{\Ds \to \phi X}.} 
\end{tabular}
\end{center}
\end{table}

Measurements for the branching fraction for this
decay channel are shown in \Table{dGsBr}.
Using their average value of \hfagBRDSDS with \Eq{dGsBr} yields
\begin{equation}
\DGGs = \hfagDGSGSBRDSDS \,,
\end{equation}
consistent with the value given in \Eq{DGGs_ave}. 

As described in \Sec{taubs}
and \Eq{tau_CPeven}, the average of the lifetime
measurements with \Bs $\to K^+ K^-$ and
$\Bs \to D_s^{(*)} D_s^{(*)}$ decays
can be used to measure the lifetime
of the \CP-even (or ``light" mass) eigenstate
$\tau(\Bs \to \CP\mbox{-even}) = \tau_L = 1/\Gamma_{\rm L} =
\hfagTAUBSSHORT$. These decays are assumed to be 100\% \CP even, with
a 5\% theoretical uncertainty on this assumption added in quadrature
for the combination.


CDF has also measured the exclusive branching fraction 
$\BR{B^0_s \to D^+_s D^-_s} = 
(9.4^{+4.4}_{-4.2}) \times 10^{-3}$~\cite{Abulencia:2007zz}, and
they use this to set a lower bound of
$\DGs^{\CP}/\Gs \geq 0.012$ at \CL{95} (since
on its own it does not saturate the \CP-even states).

} 


The strength of \Bs mixing is known to be large since more than 20 years. 
Indeed the time-integrated measurements of \chibar (see \Sec{chibar}),
when compared to our knowledge
of \chid and the \b-hadron fractions, indicated that 
\chis should be close to its maximal possible value of $1/2$.
Many searches of the time dependence of this mixing 
were performed by ALEPH~\cite{Heister:2002gk},
CDF (Run~I)~\cite{Abe:1998qj},
DELPHI~\cite{Abreu:2000sh,Abreu:2000ev,Abdallah:2002mr,Abdallah:2003we},
OPAL~\cite{Abbiendi:1999gm,Abbiendi:2000bh} and
SLD~\cite{Abe:2002ua,Abe:2002wfa,Abe:2000gp},
but did not have enough statistical power
and proper time resolution to resolve 
the small period of the \Bs\ oscillations.

\Bs oscillations have been observed for the first time in 2006
by the CDF collaboration~\cite{Abulencia:2006ze,*Abulencia:2006mq_mod_cont},
based on samples of flavour-tagged hadronic and semileptonic \Bs decays
(in flavour-specific final states), partially or fully reconstructed in 
$1\invfb$ of data collected during Tevatron's Run~II. 
This was shortly followed by an independent evidence obtained by the \dzero collaboration
with $2.4\invfb$ of
data~\cite{D0note5618:2008,*D0note5474:2007,*D0note5254:2006,*Abazov:2006dm_mod_cont}.
Recently the LHCb collaboration obtained the most precise results using fully reconstructed 
$\Bs \to D_s^-\pi^+$ and $\Bs \to D_s^-\pi^+\pi^-\pi^+$ decays at the 
LHC~\cite{Aaij:2011qx,LHCb-CONF-2011-050}. The measurements of \dms are summarized in \Table{dms}. 

\begin{table}[t]
\caption{Measurements of \dms.}
\labt{dms}
\begin{center}
\begin{tabular}{l@{}c@{}crl@{\,}l@{\,}ll} \hline
Experiment & Method           & \multicolumn{2}{c}{Data set} & \multicolumn{3}{c}{\dms (\!\!\invps)} & Ref. \\
\hline
CDF2   & \particle{D_s^{(*)-} \ell^+ \nu}, \particle{D_s^{(*)-} \pi^+}, \particle{D_s^{-} \rho^+}
       & & 1 \invfb & $17.77$ & $\pm 0.10$ & $\pm 0.07~$
       & \cite{Abulencia:2006ze,*Abulencia:2006mq_mod_cont} \\
\dzero & \particle{D_s^- \ell^+ X}, \particle{D_s^- \pi^+ X}
       &  & 2.4 \invfb & $18.53$ & $\pm 0.93$ & $\pm 0.30~$ 
       & \cite{D0note5618:2008,*D0note5474:2007,*D0note5254:2006,*Abazov:2006dm_mod_cont}$^u$ \\
LHCb   & \particle{D_s^- \pi^+}, \particle{D_s^- \pi^+\pi^-\pi^+}
       & 2010 & 0.034 \invfb & $17.63$ & $\pm 0.11$ & $\pm 0.02~$   
       & \cite{Aaij:2011qx} \\
LHCb   & \particle{D_s^- \pi^+}
       & 2011 & 0.34 \invfb & $17.725$ & $\pm 0.041$ & $\pm 0.026$ 
       & \cite{LHCb-CONF-2011-050}$^p$  \\
\hline
\multicolumn{4}{l}{Average of CDF and LHCb measurements} & $\hfagDMSval$ & $\hfagDMSsta$ & $\hfagDMSsys$ & \\  
\hline
\multicolumn{5}{l}{$^u$ \footnotesize Unpublished. ~~~ $^p$ \footnotesize Preliminary.} 
\end{tabular}
\end{center}
\end{table}

\begin{figure}
\begin{center}
\epsfig{figure=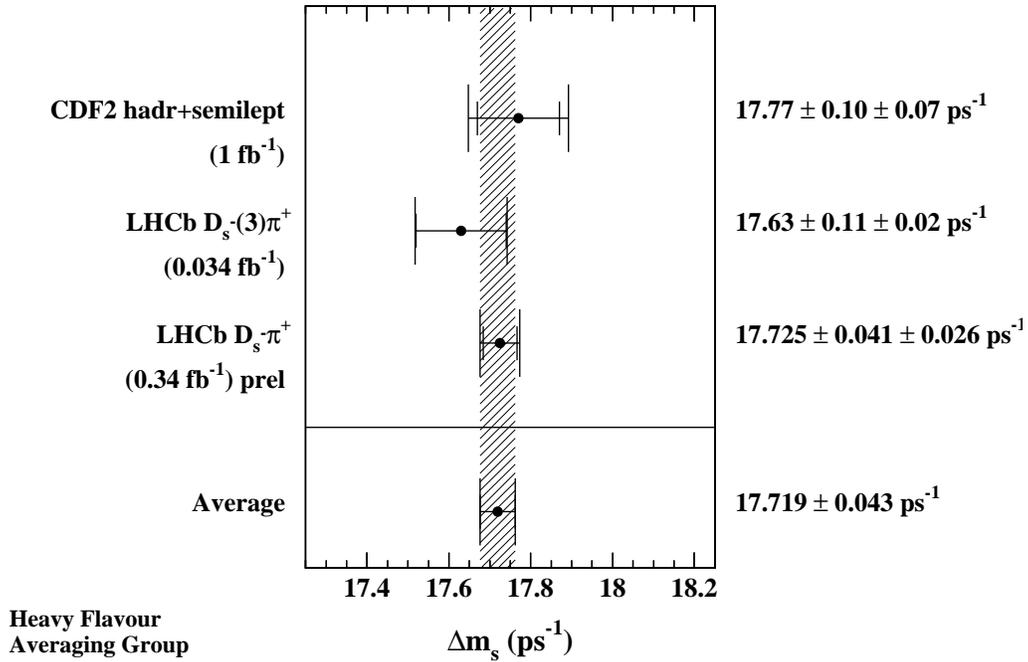,width=0.8\textwidth}
\caption{Published and recent preliminary measurements of \dms, together with their average.} 
\labf{dms}
\end{center}
\end{figure}

A simple average of the CDF and LHCb results\footnote{We do not
include the old unpublished
\dzero~\cite{D0note5618:2008,*D0note5474:2007,*D0note5254:2006,*Abazov:2006dm_mod_cont}
result in the average.},
taking into account the correlated systematic uncertainties between the two 
LHCb measurements, yields 
\begin{equation}
\dms = \hfagDMSfull = \hfagDMS \labe{dms}
\end{equation}
and is illustrated in \Figure{dms}.
Multiplying this result with the 
mean \Bs lifetime of \Eq{oneoverGs}, $1/\Gs=\hfagTAUBSMEANCON$,
yields
\begin{equation}
\xs = \frac{\dms}{\Gs} = \hfagXS \,. \labe{xs}
\end{equation}
With $2\ys = \DGGs=\hfagDGSGSCON$ 
(see \Eq{DGs_DGsGs})
and under the assumption of no \CP violation in \Bs mixing,
this corresponds to
\begin{equation}
\chis = \frac{\xs^2+\ys^2}{2(\xs^2+1)} = \hfagCHIS \,. \labe{chis}
\end{equation}
The ratio of the \Bd and \Bs oscillation frequencies, 
obtained from \Eqss{dmd}{dms}, 
\begin{equation}
\frac{\dmd}{\dms} = \hfagRATIODMDDMS \,, \labe{dmd_over_dms}
\end{equation}
can be used to extract the following ratio of CKM matrix elements, 
\begin{equation}
\left|\frac{V_{td}}{V_{ts}}\right| =
\xi \sqrt{\frac{\dmd}{\dms}\frac{m(\Bs)}{m(\Bd)}} = 
\hfagVTDVTSfull \,, \labe{Vtd_over_Vts}
\end{equation}
where the first quoted error is from experimental uncertainties 
(with the masses $m(\Bs)$ and $m(\Bd)$ taken from \Ref{PDG_2012}),
and where the second quoted error is from theoretical uncertainties 
in the estimation of the SU(3) flavour-symmetry breaking factor
$\xi 
= \hfagXI$
obtained from lattice QCD calculations~\cite{Laiho:2009eu_mod,*Evans:2008zzg_mod,*Gamiz:2009ku,*Albertus:2010nm}.

\mysubsubsection{\CP violation in \Bd and \Bs mixing}
\labs{qpd} \labs{qps}


Evidence for \CP violation in \Bd mixing
has been searched for,
both with flavour-specific and inclusive \Bd decays, 
in samples where the initial 
flavour state is tagged. In the case of semileptonic 
(or other flavour-specific) decays, 
where the final state tag is 
also available, the following asymmetry
\begin{equation} 
\ASLd = \frac{
N(\hbox{\Bdbar}(t) \to \ell^+      \nu_{\ell} X) -
N(\hbox{\Bd}(t)    \to \ell^- \bar{\nu}_{\ell} X) }{
N(\hbox{\Bdbar}(t) \to \ell^+      \nu_{\ell} X) +
N(\hbox{\Bd}(t)    \to \ell^- \bar{\nu}_{\ell} X) } 
= \frac{|p/q|_{\particle{d}}^2 - |q/p|_{\particle{d}}^2}%
{|p/q|_{\particle{d}}^2 + |q/p|_{\particle{d}}^2}
\labe{ASL}
\end{equation} 
has been measured, either in time-integrated analyses at 
CLEO~\cite{Bartelt:1993cf,Behrens:2000qu,Jaffe:2001hz},
CDF~\cite{Abe:1996zt,CDFnote9015:2007} and \dzero~\cite{Abazov:2011yk,*Abazov:2010hv_mod_cont,*Abazov:2010hj_mod_cont,*Abazov:2011yk_cont},
or in time-dependent analyses at 
OPAL~\cite{Ackerstaff:1997vd}, ALEPH~\cite{Barate:2000uk}, 
\babar~\cite{Aubert:2003hd,*Aubert:2004xga_mod_cont,Aubert:2006nf,*Aubert:2002mn_mod_cont,Aubert:2006sa}
and \belle~\cite{Nakano:2005jb}.
In the inclusive case, also investigated and published
at ALEPH~\cite{Barate:2000uk} and OPAL~\cite{Abbiendi:1998av},
no final state tag is used, and the asymmetry~\cite{Beneke:1996hv,*Dunietz:1998av}
\begin{equation} 
\frac{
N(\hbox{\Bd}(t) \to {\rm all}) -
N(\hbox{\Bdbar}(t) \to {\rm all}) }{
N(\hbox{\Bd}(t) \to {\rm all}) +
N(\hbox{\Bdbar}(t) \to {\rm all}) } 
\simeq
\ASLd \left[ \frac{\dmd}{2\Gd} \sin(\dmd \,t) - 
\sin^2\left(\frac{\dmd \,t}{2}\right)\right] 
\labe{ASLincl}
\end{equation} 
must be measured as a function of the proper time to extract information 
on \CP violation.
\Table{qoverp} summarized the different measurements: 
in all cases asymmetries compatible with zero have been found,  
with a precision limited by the available statistics. 

\begin{table}
\caption{Measurements\protect\footnotemark$^,$\protect\footnotemark\addtocounter{footnote}{0}
of \CP violation in \Bd mixing and their average
in terms of both \ASLd and $|q/p|_{\particle{d}}$.
The individual results are listed as quoted in the original publications, 
or converted\addtocounter{footnote}{2}\protect\footnotemark\addtocounter{footnote}{-5}
to an \ASLd value.
When two errors are quoted, the first one is statistical and the 
second one systematic. The last group of results from OPAL and ALEPH 
assume no \CP violation in \Bs mixing.}
\labt{qoverp}
\begin{center}
\begin{tabular}{@{}rcl@{$\,\pm$}l@{$\pm$}ll@{$\,\pm$}l@{$\pm$}l@{}}
\hline
Exp.\ \& Ref. & Method & \multicolumn{3}{c}{Measured \ASLd} 
                       & \multicolumn{3}{c}{Measured $|q/p|_{\particle{d}}$} \\
\hline
CLEO   \cite{Behrens:2000qu} & partial hadronic rec. 
                             & $+0.017$ & 0.070 & 0.014 
                             & \multicolumn{3}{c}{} \\
CLEO   \cite{Jaffe:2001hz}   & dileptons 
                             & $+0.013$ & 0.050 & 0.005 
                             & \multicolumn{3}{c}{} \\
CLEO   \cite{Jaffe:2001hz}   & average of above two 
                             & $+0.014$ & 0.041 & 0.006 
                             & \multicolumn{3}{c}{} \\
\babar \cite{Aubert:2003hd,*Aubert:2004xga_mod_cont}   & full hadronic rec. 
                             & \multicolumn{3}{c}{}  
                             & 1.029 & 0.013 & 0.011  \\
\babar \cite{Aubert:2006nf,*Aubert:2002mn_mod_cont}  & dileptons
                             & \multicolumn{3}{c}{}
                             & 0.9992 & 0.0027 & 0.0019 \\ 
\belle \cite{Nakano:2005jb}  & dileptons 
                             & $-0.0011$ & 0.0079 & 0.0085 
                             & 1.0005 & 0.0040 & 0.0043 \\
\multicolumn{2}{l}{Average of above 6 \B factory results} & \multicolumn{3}{l}{\hfagASLDB\ (tot)} 
                             & \multicolumn{3}{l}{\hfagQPDB\  (tot)} \\ 
\hline
\dzero  \cite{Abazov:2011yk,*Abazov:2010hv_mod_cont,*Abazov:2010hj_mod_cont,*Abazov:2011yk_cont}  & dimuons  
                             & $-0.0012$ & \multicolumn{2}{@{\hspace{0.26em}}l}{0.0052 (tot)}
                             & \multicolumn{3}{c}{} \\
\multicolumn{2}{l}{Average of above 7 direct measurements} & \multicolumn{3}{l}{\hfagASLDW\ (tot)} 
                             & \multicolumn{3}{l}{\hfagQPDW\  (tot)} \\ 
\hline
OPAL   \cite{Ackerstaff:1997vd}   & leptons     
                             & $+0.008$ & 0.028 & 0.012 
                             & \multicolumn{3}{c}{} \\
OPAL   \cite{Abbiendi:1998av}   & inclusive (\Eq{ASLincl}) 
                             & $+0.005$ & 0.055 & 0.013 
                             & \multicolumn{3}{c}{} \\
ALEPH  \cite{Barate:2000uk}       & leptons 
                             & $-0.037$ & 0.032 & 0.007 
                             & \multicolumn{3}{c}{} \\
ALEPH  \cite{Barate:2000uk}       & inclusive (\Eq{ASLincl}) 
                             & $+0.016$ & 0.034 & 0.009 
                             & \multicolumn{3}{c}{} \\
ALEPH  \cite{Barate:2000uk}       & average of above two 
                             & $-0.013$ & \multicolumn{2}{@{\hspace{0.26em}}l}{0.026 (tot)} 
                             & \multicolumn{3}{c}{} \\
\multicolumn{2}{l}{Average of above 12 results} & \multicolumn{3}{l}{\hfagASLDA\ (tot)} 
                             & \multicolumn{3}{l}{\hfagQPDA\  (tot)} \\ 
\hline
\multicolumn{5}{l}{Best fit value from 2D combination of} \\
\multicolumn{2}{l}{\ASLd and \ASLs results (see \Eq{ASLD})} & \multicolumn{3}{l}{\hfagASLD\ (tot)} 
                             & \multicolumn{3}{l}{\hfagQPD\  (tot)} \\ 
\hline
\end{tabular}
\end{center}
\end{table}

A simple average of all measurements performed at 
\B factories~\cite{Behrens:2000qu,Jaffe:2001hz,Aubert:2003hd,*Aubert:2004xga_mod_cont,Aubert:2006nf,*Aubert:2002mn_mod_cont,Nakano:2005jb}%
\footnote{An old unpublished measurement by \babar~\cite{Aubert:2006sa} 
in no longer included in our averages, nor in \Table{qoverp}.}
yields 
\begin{equation}
\ASLd = \hfagASLDB  ~~~ \Longleftrightarrow ~~~ |q/p|_{\particle{d}} = \hfagQPDB \,,
\labe{ASLDB}
\end{equation}
where the relation between \ASLd and $|q/p|_{\particle{d}}$ is given in \Eq{ASL}.
The latest dimuon \dzero analysis~\cite{Abazov:2011yk,*Abazov:2010hv_mod_cont,*Abazov:2010hj_mod_cont,*Abazov:2011yk_cont}
separates the \Bd and \Bs contributions by exploiting the dependence on the muon impact parameter cut; combining the 
\ASLd result quoted by \dzero with the above \B factory average yields
$\ASLd = \hfagASLDW$. 

All the other analyses performed at high energy, either at LEP or at the Tevatron,
did not separate the contributions from the \Bd and \Bs mesons.
Under the assumption of no \CP violation in \Bs mixing, a number of 
these analyses~\cite{Abazov:2006qw,Ackerstaff:1997vd,Barate:2000uk,Abbiendi:1998av}
quote a measurement of $\ASLd$ or $|q/p|_{\particle{d}}$ for the \Bd meson. Including also 
these results%
\footnote{A low-statistics result published by CDF using the Run I data~\cite{Abe:1996zt} and 
an unpublished result by CDF using Run II data~\cite{CDFnote9015:2007} 
are not included in our averages, nor in \Table{qoverp}.}
in the previous average 
leads to 
$\ASLd = \hfagASLDA$ 
under the assumption $\ASLs =0$. The latter assumption makes sense within the Standard Model, 
since \ASLs is predicted to be much smaller than \ASLd~\cite{Lenz:2011ti,*Lenz:2006hd}, but may not be suitable
in presence of New Physics. 


The following constraints on a combination of  \ASLd and \ASLs
(or equivalently $|q/p|_{\particle{d}}$ and $|q/p|_{\particle{s}}$)
have been obtained by the Tevatron 
experiments, using inclusive semileptonic decays of \b hadrons:
\begin{eqnarray}
\frac{1}{4}\left(f'_{\particle{d}} \,\chid \ASLd +
                 f'_{\particle{s}} \,\chis \ASLs \right) &=& 
+0.0015 \pm 0.0038 \mbox{(stat)} \pm 0.0020 \mbox{(syst)}
~~~~ \mbox{CDF1~\cite{Abe:1996zt}} \,, ~
\labe{CDF_ASLDS} \\
\ASLb = \frac{f'_{\particle{d}}Z_{\particle{d}} \ASLd + f'_{\particle{s}}Z_{\particle{s}} \ASLs}%
{f'_{\particle{d}}Z_{\particle{d}} + f'_{\particle{s}}Z_{\particle{s}}} &=&
 -0.00787 \pm 0.00172 \mbox{(stat)} \pm 0.00093 \mbox{(syst)}
~~~ \mbox{\dzero~\cite{Abazov:2011yk,*Abazov:2010hv_mod_cont,*Abazov:2010hj_mod_cont,*Abazov:2011yk_cont}} \,, ~
\labe{Dzero_ASLDS}
\end{eqnarray}
where\footnote{In \Ref{Abazov:2007zj}, the \dzero result
$\frac{1}{4}\left(\ASLd +
\ASLs \frac{f'_\particle{s}\chis}{f'_\particle{d}\chid} \right) =
-0.0023 \pm 0.0011 \mbox{(stat)} \pm 0.0008 \mbox{(syst)}$~\cite{Abazov:2006qw}
(now superseded by that of \Ref{Abazov:2011yk,*Abazov:2010hv_mod_cont,*Abazov:2010hj_mod_cont,*Abazov:2011yk_cont})
was reinterpreted by replacing $\chi_{\particle{s}}/\chi_{\particle{d}}$
with $Z_{\particle{s}}/Z_{\particle{d}}$.
For simplicity, and since this has anyway a negligible numerical effect on our
combined result of \Fig{ASLs}, 
we follow the same interpretation and set
$\chi_{\particle{q}}=Z_{\particle{q}}/2$ in \Eq{CDF_ASLDS}.
We also set $f'_{\particle{q}}=f_{\particle{q}}$.}
$Z_{\particle{q}} = 1/(1-y_{\particle{q}}^2)-1/(1+x_{\particle{q}}^2)
= 2 \chi_{\particle{q}}/(1-y_{\particle{q}}^2)$, $q=d,s$.
While the CDF measurement is compatible with no \CP violation%
\footnote{A more precise result from CDF2, 
$ \ASLb = +0.0080 \pm 0.0090 \mbox{(stat)} \pm 0.0068 \mbox{(syst)}$~\cite{CDFnote9015:2007},
is also compatible with no \CP violation, but since it is unpublished since 2007 
we no longer include it in our averages, nor in \Fig{ASLs}.},
the more precise \dzero result of \Eq{Dzero_ASLDS}, obtained by measuring
the charge asymmetry of like-sign dimuons, differs by 3.9 standard
deviations from the Standard Model prediction of
$\ASLb({\mathrm{SM}}) = (-2.8^{+0.5}_{-0.6}) \times 10^{-4}$%
~\cite{Abazov:2011yk,*Abazov:2010hv_mod_cont,*Abazov:2010hj_mod_cont,*Abazov:2011yk_cont,Lenz:2011ti,*Lenz:2006hd}.

Using the average $\ASLd = \hfagASLDB$ of \Eq{ASLDB},
obtained from results at \B factories, the averages 
of the Tevatron $b$-hadron fractions and their correlations listed in \Table{fractions},
and the averages of the mixing parameters 
presented in this chapter,
the two results of \Eqss{CDF_ASLDS}{Dzero_ASLDS}
are turned into the 
measurements of \ASLs displayed in the top part of \Fig{ASLs}.
\begin{figure}[t]
\begin{center}
\epsfig{figure=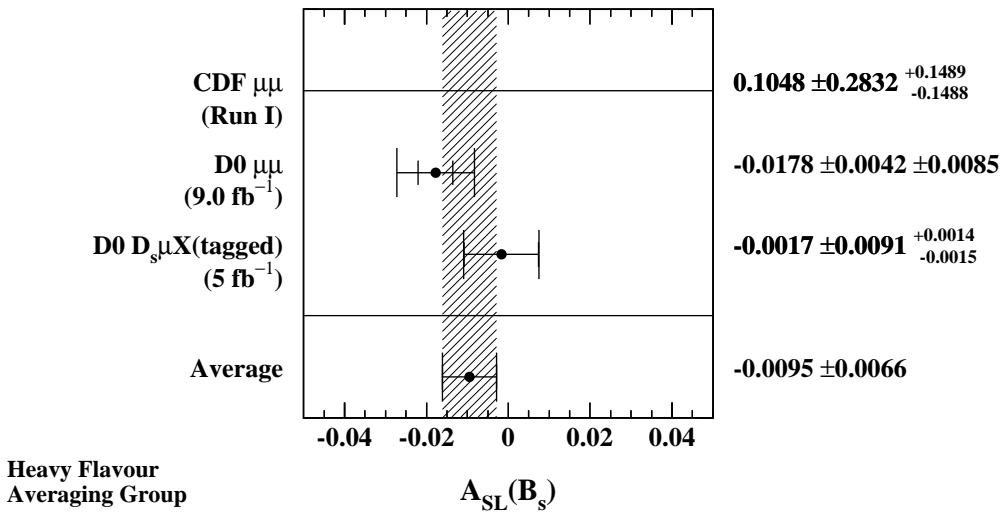,width=0.8\textwidth}
\caption{Measurements of \ASLs,
derived from CDF~\cite{Abe:1996zt}\addtocounter{footnote}{-1}\protect\footnotemark\ and
\dzero~\cite{Abazov:2011yk,*Abazov:2010hv_mod_cont,*Abazov:2010hj_mod_cont,*Abazov:2011yk_cont,Abazov:2009wg,*Abazov:2007nw_mod_cont}
analyses and adjusted to the \B factory 
average of \ASLd, the Tevatron averages of the \b-hadron fractions, and the latest averages of the mixing parameters. 
The combined value of \ASLs is also shown.}
\labf{ASLs}
\end{center}
\end{figure}
Taking into account the uncertainties
in $f'_{\particle{d}}, f'_{\particle{s}}, Z_{\particle{d}},$ and
$Z_{\particle{s}}$, the value derived from the \dzero result does not show evidence
of \CP violation in the \Bs system.
In addition, the third line of \Fig{ASLs} shows a direct determination of \ASLs
obtained by \dzero by measuring the charge asymmetry of
tagged $\Bs \to D_s \mu X$ decays~\cite{Abazov:2009wg,*Abazov:2007nw_mod_cont}.
The three results of \Fig{ASLs} are
combined to yield
$\ASLs = \hfagASLSWval\hfagASLSWsta\mbox{(stat)}\hfagASLSWsys\mbox{(syst)} = \hfagASLSW$
or, equivalently through \Eq{ASL},
$|q/p|_{\particle{s}} = \hfagQPSWval\hfagQPSWsta\mbox{(stat)}\hfagQPSWsys\mbox{(syst)} = \hfagQPSW$.
The quoted systematic errors include experimental systematics as well as the correlated dependence on external 
parameters. 

In the latest update of the \dzero like-sign dimuon analysis, 
the dependence of the charge asymmetry is investigated for the first time 
as a function of the muon impact parameters, allowing the separation of the 
\Bd and \Bs contributions to the result of \Eq{Dzero_ASLDS}. Using 
the mixing parameters and the LEP $b$-hadron fractions 
of \Ref{Asner:2010qj}, the \dzero collaboration
extracts~\cite{Abazov:2011yk,*Abazov:2010hv_mod_cont,*Abazov:2010hj_mod_cont,*Abazov:2011yk_cont}
values for \ASLd and \ASLs and their correlation coefficient, 
as shown in the first line of \Table{ASLs_ASLd}.
However, the individual 
contributions to the total quoted errors from this analysis and from the
external inputs are not given, so the adjustment of these results to different
or more recent values of the external inputs cannot (easily) be done. 
Using a two-dimensional fit, these values are combined with the 
\B factory average of \Eq{ASLDB} and with the result from the tagged
$\Bs \to D_s \mu X$ analysis~\cite{Abazov:2009wg,*Abazov:2007nw_mod_cont},
assumed to be independent and also shown in \Table{ASLs_ASLd}.
The result, shown graphically in \Fig{ASLs_ASLd}, is 
\begin{table}
\caption{Direct measurements of \CP violation in \Bs and \Bd mixing, together 
with their two-dimensional average. Only total errors are quoted.}
\labt{ASLs_ASLd}
\begin{center}
\begin{tabular}{ccccc}
\hline
Exp.\ \& Ref.\ & Method & Measured \ASLs & Measured \ASLd & $\rho(\ASLs,\ASLd)$ \\
\hline
\dzero  \cite{Abazov:2011yk,*Abazov:2010hv_mod_cont,*Abazov:2010hj_mod_cont,*Abazov:2011yk_cont}  & dimuons  
       & $-0.0181 \pm 0.0106$ 
       & $-0.0012 \pm 0.0052$ 
       & $-0.799$ \\          
\dzero  \cite{Abazov:2009wg,*Abazov:2007nw_mod_cont}  & tagged $\Bs \to D_s \mu X$ 
       & $-0.0017 \pm 0.0092$ 
       & & \\
\multicolumn{2}{l}{\B factory average of \Eq{ASLDB}} 
       & & \hfagASLDB & \\ 
\hline
\multicolumn{2}{l}{Average of all above}
       & \hfagASLS & \hfagASLD & $\hfagRHOASLSASLD$ \\ 
\hline
\end{tabular}
\end{center}
\end{table}
\begin{figure}
\begin{center}
\epsfig{figure=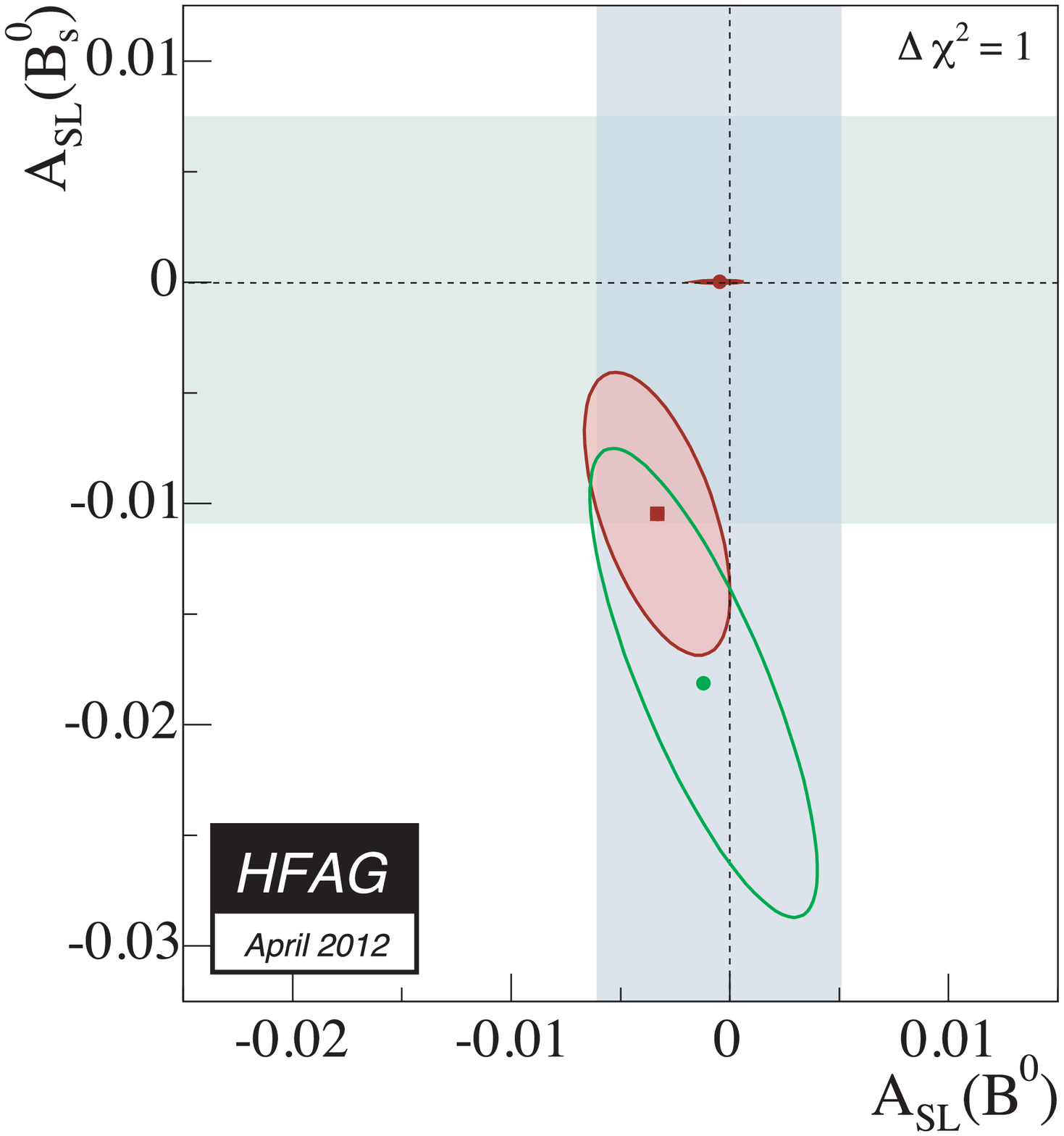,width=0.6\textwidth}
\end{center}
\vspace{-5mm}
\caption{
Direct measurements of \ASLs and \ASLd listed in \Table{ASLs_ASLd}
(\B-factory average as the vertical blue shaded band,
\dzero measurements as the horizontal green shaded band and as the green ellipse),
together with their two-dimensional average (red shaded ellipse).
The red point close to $(0,0)$ is the Standard Model prediction
of \Ref{Lenz:2011ti,*Lenz:2006hd} with error bars multiplied by 10.}
\labf{ASLs_ASLd}
\end{figure}
\begin{eqnarray}
\ASLd & = & \hfagASLD ~~~ \Longleftrightarrow ~~~ |q/p|_{\particle{d}} = \hfagQPD \,,
\labe{ASLD}
\\
\ASLs & = & \hfagASLS ~~~ \Longleftrightarrow ~~~ |q/p|_{\particle{s}} = \hfagQPS \,,
\labe{ASLS}
\\
\rho(\ASLd , \ASLs) & = & \hfagRHOASLSASLD \,.
\labe{rhoASLDASLS}
\end{eqnarray}

The average of \Fig{ASLs} 
ignores the impact parameter study of \dzero and is adjusted to the 
$b$-hadron fractions at the Tevatron. The average of \Eq{ASLS} ignores the CDF1 
result (which has a very large uncertainty anyway)
and is adjusted to the $b$-hadron fractions at LEP.
We choose the results of \Eqsss{ASLD}{ASLS}{rhoASLDASLS}
as our final averages\footnote{Early analyses and (perhaps hence) the PDG use the complex
parameter $\epsilon_{\B} = (p-q)/(p+q)$; if \CP violation in the mixing in small,
$\ASLd \cong 4 {\rm Re}(\epsilon_{\B})/(1+|\epsilon_{\B}|^2)$ and the averages of
\Eqss{ASLDB}{ASLD} 
correspond to ${\rm Re}(\epsilon_{\B})/(1+|\epsilon_{\B}|^2)=\hfagREBDB$ 
and $\hfagREBD$, respectively.},
since they better incorporate the available published data. 

The above averages are compatible with no \CP violation in \Bd and \Bs mixing. 
They are also compatible with the very small predictions of the Standard Model, 
${\ASLd}^{\rm SM} = -(4.1\pm 0.6)\times 10^{-4}$ and 
${\ASLs}^{\rm SM} = +(1.9\pm 0.3)\times 10^{-5}$~\cite{Lenz:2011ti,*Lenz:2006hd}.
However, given the current size of the experimental uncertainties, there is still 
a large room for a possible New Physics contribution, especially in the \Bs system. 
In this respect, the deviation of the \dzero dimuon
asymmetry~\cite{Abazov:2011yk,*Abazov:2010hv_mod_cont,*Abazov:2010hj_mod_cont,*Abazov:2011yk_cont}
from expectation has generated a lot of excitement,
and new experimental data (in particular from LHCb) is awaited eagerly. 

At the more fundamental level, \CP violation in \Bs
mixing\footnote{Of course, a similar formalism exists for the \Bd system; for 
simplicity we omit here the subscript $s$ for $\phi_{12}$, $M_{12}$ and $\Gamma_{12}$.}
is caused by the weak phase difference 
\begin{equation}
\phi_{12} = \arg \left[ -{M_{12}}/{\Gamma_{12}} \right], 
\end{equation}
where $M_{12}$ and $\Gamma_{12}$ are the off-diagonal
elements of the mass and decay matrices of the $\Bs-\Bsbar$ system.
This is related to the observed decay-width difference through the relation
\begin{equation}
\DGs = 2|\Gamma_{12}|\cos\phi_{12}+
{\cal O} \left( \left|\frac{\Gamma_{12}}{M_{12}}\right|^2 \right) \,,
\end{equation}
where quadratic (or higher-order) terms in the small quantity
$|\Gamma_{12}/M_{12}| \sim {\cal O}(m_b^2/m_t^2)$ can be neglected. 
The SM prediction for this phase is tiny,
$\phi_{12}^{\rm SM} = 0.0038\pm0.0010$~\cite{Lenz:2011ti,*Lenz:2006hd}; however,
new physics in \Bs mixing could change this observed phase to
\begin{equation}
\phi_{12} = \phi_{12}^{\rm SM} + \phi_{12}^{\rm NP} \,.
\labe{phi12NP}
\end{equation}
The \Bs semileptonic asymmetry can be expressed as~\cite{Beneke:2003az}
\begin{equation}
\ASLs = 
\Im \left(\frac{\Gamma_{12}}{M_{12}} \right) +
{\cal O} \left( \left|\frac{\Gamma_{12}}{M_{12}}\right|^2 \right) =
\frac{\DGs}{\dms}\tan\phi_{12} +
{\cal O} \left( \left|\frac{\Gamma_{12}}{M_{12}}\right|^2 \right) \,.
\labe{ASLS_tanphi12}
\end{equation}
Using this relation, the current knowledge of \ASLs, \DGs and \dms, 
given in \Eqsss{ASLS}{DGs_DGsGs}{dms} respectively, yield a very first
experimental determination of $ \phi_{12}$,
\begin{equation}
\tan\phi_{12} = \ASLs \frac{\dms}{\DGs} = -1.9 \pm 1.2 \,,
\comment{ 
from math import *
asls = -0.010460 ; easls = 0.006400
dms = 17.719032  ; edms = 0.042701
dgs = 0.0951919  ; edgs = 0.01362480381803716 
tanphi12 = asls*dms/dgs
etanphi12 = tanphi12*sqrt((easls/asls)**2+(edms/dms)**2+(edgs/dgs)**2)
print tanphi12, etanphi12
} 
\end{equation}
which only represents a very weak constraint at present.


\mysubsubsection{Mixing-induced \CP violation in \Bs decays}

%
%
%

\labs{phasebs}

\CP violation induced by $\Bs-\Bsbar$ mixing
has been a field of 
very active study and fast experimental progress 
in the past couple of years.
Similarly to what has happened at the \B factories 
a decade ago, when the \Bd mixing-induced phase $2\beta$
was measured, the Tevatron and LHC experiments are 
now obtaining point estimates
of the \Bs mixing-induced phase \phiccbars.
This \CP-violating phase is defined as 
the weak phase difference between
the $\Bs-\Bsbar$ mixing amplitude
and the $b \to c\bar{c}s$ decay amplitude.

The golden mode for such studies is 
$\Bs \to \jpsi\phi$, followed by $\jpsi \to \mu^+\mu-$ and 
$\phi\to K^+K^-$, for which a full angular 
analysis of the decay products is performed to 
separate statistically the \CP-even and \CP-odd
contributions in the final state. As already mentioned in 
\Sec{DGs},
CDF~\cite{CDFnote10778:2012,*CDFnote10778:2012_cont,CDF:2011af,*Aaltonen:2007he_mod,*Aaltonen:2007gf_mod},
\dzero~\cite{Abazov:2011ry,*Abazov_mod:2008fj,*Abazov:2007tx_mod_cont}
and LHCb~\cite{LHCbnote002:2012,*LHCbnote002:2012_cont,LHCb:2011aa} 
have used both untagged and tagged $\Bs \to \jpsi\phi$ events 
for the measurement of \phiccbars.
In addition, 
the newly observed \CP-odd decay mode $\Bs \to \jpsi f_0(980)$, 
$f_0(980)\to \pi^+\pi^-$ has also been analyzed by LHCb~\cite{LHCb:2011ab}, 
without the need for an angular analysis; this analysis was 
(superseded and) extended to the three-body decay mode
$\Bs \to \jpsi \pi^+\pi^-$~\cite{LHCb:Jpsipipi}, 
which has been shown to be almost \CP pure with a \CP-odd fraction 
larger than 0.977 at 95\% CL~\cite{:2012cy}.

All these analyses provide 
two mirror solutions related by the transformation 
$(\DGs, \phi_s) \to (-\DGs, \pi-\phi_s)$. However, a recent 
LHCb analysis of $\Bs \to \jpsi K^+K^-$ resolved this ambiguity and 
ruled out the solution with negative \DGs~\cite{Aaij:2012eq}.
Therefore, in what follows we only consider the solution with $\DGs > 0$.


\begin{table}
\caption{Direct experimental measurements of \phiccbars, \DGs and \Gs using
$\Bs\to \jpsi\phi$ and $\Bs\to\jpsi\pi\pi$ decays.
Only the solution with $\DGs > 0$ is shown, since the two-fold ambiguity has been
resolved in \Ref{Aaij:2012eq}. The first error is due to 
statistics, the second one to systematics. The last line gives our average.}
\labt{phisDGsGs}
\begin{center}
\begin{tabular}{@{}l@{\,}l@{\,}l@{\,}|@{\,}l@{\,}|@{\,}l@{\,}|@{\,}l@{}} 
\hline
Exp.\ & Mode & Ref.\ & \multicolumn{1}{c@{\,}|@{\,}}{\phiccbars}
                     & \multicolumn{1}{c@{\,}|@{\,}}{\DGs (\!\!\invps)}
                     & \multicolumn{1}{c@{}}{\Gs (\!\!\invps)} \\ 
\hline
CDF    & $\jpsi\phi$ & \cite{CDFnote10778:2012,*CDFnote10778:2012_cont}$^p$
       & $[-0.60,\, 0.12]$, 68\% CL & $0.068\pm0.026\pm0.007$ & $0.654\pm0.008\pm0.004$ \\
\dzero & $\jpsi\phi$ & \cite{Abazov:2011ry,*Abazov_mod:2008fj,*Abazov:2007tx_mod_cont}
       & $-0.55^{+0.38}_{-0.36}$ & $0.163^{+0.065}_{-0.064}$ & $0.693^{+0.018}_{-0.017}$  \\
LHCb   & $\jpsi\phi$ & \cite{LHCbnote002:2012,*LHCbnote002:2012_cont}$^{a,p}$
       & $-0.001\pm0.101\pm0.027$ & $0.116\pm0.018\pm0.006$ & $0.6580\pm0.0054\pm0.0066$  \\
LHCb   & $\jpsi\pi\pi$ & \cite{LHCb:Jpsipipi}$^a$ 
       & $-0.019 ^{+0.173 +0.004}_{-0.174 -0.003}$ & --- & --- \\
\hline
\multicolumn{3}{@{}l@{\,}|@{\,}}{Combined} & \hfagPHISCOMB & \hfagDGSCOMBnounit & \hfagGSnounit \\
\hline
\multicolumn{6}{l}{$^a$ \footnotesize The combined LHCb result quoted in \cite{LHCbnote002:2012,*LHCbnote002:2012_cont}
is $\phiccbars = -0.002\pm0.083\pm0.027$.} \\[-0.5ex]
\multicolumn{6}{l}{$^p$ {\footnotesize Preliminary.}}
\end{tabular}
\end{center}
\end{table}

\begin{figure}
\begin{center}
\epsfig{figure=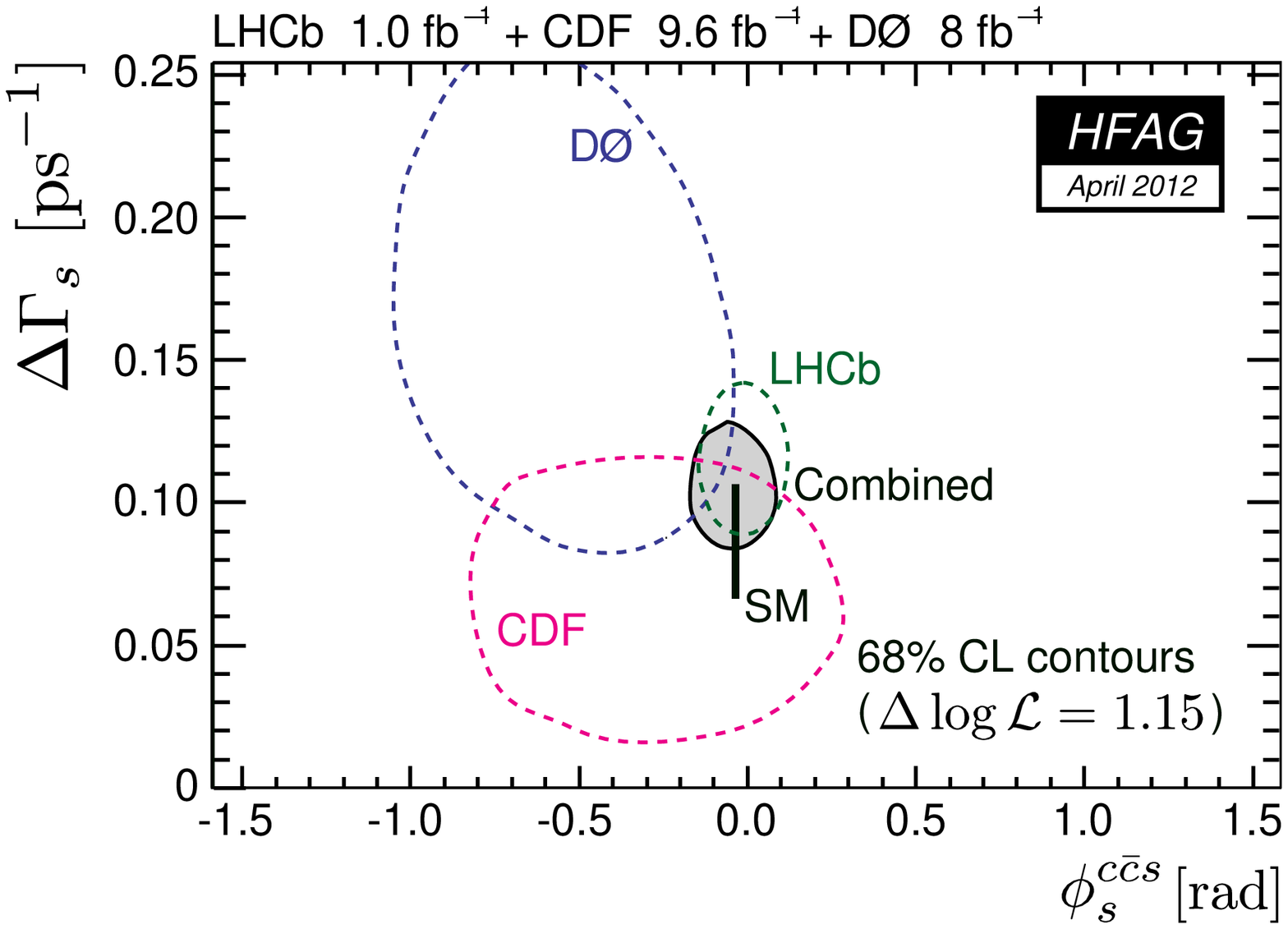,width=0.49\textwidth}
\epsfig{figure=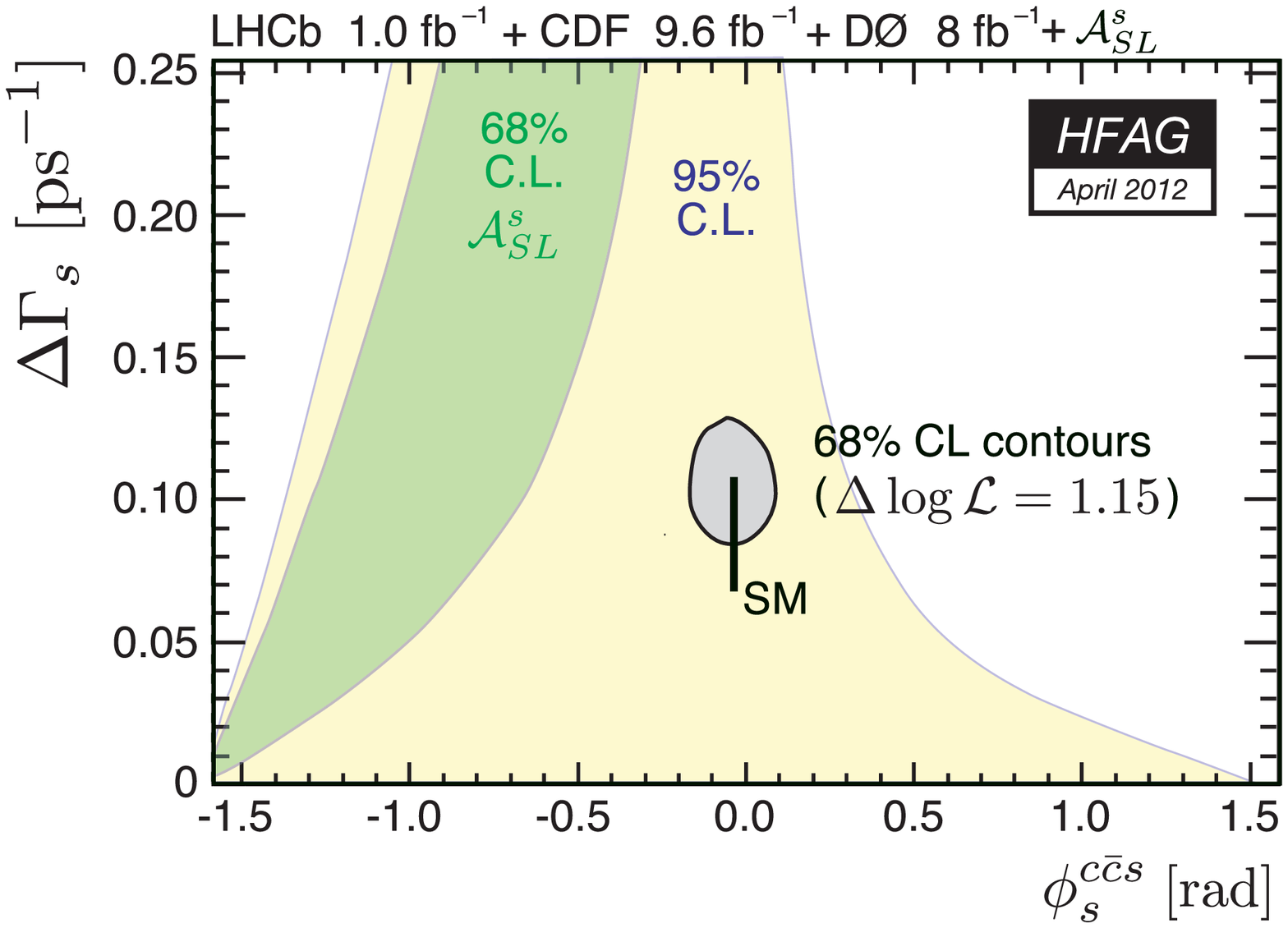,width=0.49\textwidth}
\caption{
Left: 68\% CL regions in \Bs width difference \DGs and weak phase \phiccbars
obtained from individual and combined CDF~\cite{CDFnote10778:2012,*CDFnote10778:2012_cont},
\dzero~\cite{Abazov:2011ry,*Abazov_mod:2008fj,*Abazov:2007tx_mod_cont} and
LHCb~\cite{LHCbnote002:2012,*LHCbnote002:2012_cont,LHCb:Jpsipipi} likelihoods of 
$\Bs\to \jpsi\phi$ and $\Bs\to\jpsi\pi\pi$~\cite{LHCb:Jpsipipi} samples.
Right: same combined contour compared with the 68\% CL (green) and 95\% CL (yellow)
regions allowed by the measurements of \ASLs and \dms.
The expectation within the Standard Model~\cite{Charles:2011va_mod,Lenz:2011ti,*Lenz:2006hd}
is shown as the black rectangle.}
\labf{DGs_phase}
\end{center}
\end{figure}

We perform a combination of the CDF~\cite{CDFnote10778:2012,*CDFnote10778:2012_cont},
\dzero~\cite{Abazov:2011ry,*Abazov_mod:2008fj,*Abazov:2007tx_mod_cont}
and LHCb~\cite{LHCbnote002:2012,*LHCbnote002:2012_cont,LHCb:Jpsipipi}
results summarized in \Table{phisDGsGs}.
This is done by adding the two-dimensional log profile-likelihood scans of
\DGs and \phiccbars from the three $\Bs\to\jpsi\phi$ analyses and 
a one-dimensional log profile-likelihood of \phiccbars
from the $\Bs\to\jpsi\pi^+\pi^-$ analysis, where in each case
the $-$log-likelihood is minimized with respect to all other
parameters, including \Gs.
Since the $\Bs\to\jpsi\phi$ two-dimensional scan provided by LHCb
in~\Ref{LHCbnote002:2012,*LHCbnote002:2012_cont}
contains only statistical uncertainty, on each (\DGs, \phiccbars) point,
we decrease the log-likelihood by the quantity
\begin{equation}
\Delta\log{\cal L}^{\rm new}  - \Delta\log{\cal L}^{\rm old} =
\frac{(\phiccbars-\phi_{s-{\rm min}}^{c\bar{c}s})^2 \sigma_{\phi-\rm syst}^2} 
{2 \sigma_{\phi-\rm stat}^2 (\sigma_{\phi-\rm stat}^2 + \sigma_{\phi-\rm syst}^2)} +  
\frac{ (\DGs - \DG_{s-{\rm min}})^2 \sigma_{\DG-{\rm syst}}^2} 
{2 \sigma_{\DG-\rm stat}^2 (\sigma_{\DG-\rm stat}^2 + \sigma_{\DG-\rm syst}^2)}\,, 
\end{equation}
where $\phi_{s-{\rm min}}$ and $\DG_{s-{\rm min}}$ are the values of \phiccbars
and \DGs at the minimum of the likelihood, and $\sigma_{\phi-\rm stat}$
($\sigma_{\DG-\rm stat}$) and $\sigma_{\phi-\rm syst}$ ($\sigma_{\DG-\rm syst}$)
the statistical and systematic uncertainties on \phiccbars (\DGs).  
This assumes that the systematic uncertainties are Gaussian
and independent of \DGs and \phiccbars. 
Both the \dzero and CDF log profile-likelihood scans are corrected
for coverage and include systematic uncertainties.
We obtain the individual and combined contours shown in \Fig{DGs_phase} (left).
Profiling the likelihood in each of the \DGs and $\phi_s$ dimensions,
we find, as summarized in \Table{phisDGsGs}:  
\begin{eqnarray}
\DGs &=& \hfagDGSCOMB \,, \\    
\phiccbars &=& \hfagPHISCOMB \,.
\labe{phis}
\end{eqnarray}

In the Standard Model and ignoring sub-leading penguin contributions, 
\phiccbars is expected to be equal to $-2\beta_s$, 
where
$\beta_s = \arg\left[-\left(V_{ts}V^*_{tb}\right)/\left(V_{cs}V^*_{cb}\right)\right]$ 
is a phase analogous to the angle $\beta$ of the usual CKM
unitarity triangle (aside from a sign change). 
An indirect determination via global fits to experimental data
gives~\cite{Charles:2011va_mod}
\begin{equation}
(\phiccbars)^{\rm SM} = -2\beta_s = -0.0363^{+0.0016}_{-0.0015} \,.
\end{equation}
The average value of \phiccbars from \Eq{phis} is consistent with this
Standard Model expectation.

New physics could contribute \phiccbars. Assuming that new physics only 
enters in $M_{12}$ (rather than in $\Gamma_{12}$),
one can write~\cite{Lenz:2011ti,*Lenz:2006hd}
\begin{equation}
\phiccbars = -  2\beta_s + \phi_{12}^{\rm NP} \,,
\end{equation}
where the new physics phase $\phi_{12}^{\rm NP}$ is the same as that appearing in \Eq{phi12NP}.
In this case
\begin{equation}
\phi_{12} = 
\phi_{12}^{\rm SM} +2\beta_s + \phiccbars 
\end{equation}
and \Eq{ASLS_tanphi12} then provides a relation between \DGs and \phiccbars, 
based on the measured values of \ASLs and \dms (\Eqss{ASLS}{dms}) 
as well as the expectations
for $\phi_{12}^{\rm SM}$ and $-2\beta_s$.
The allowed region in the (\DGs, \phiccbars) plane is shown in 
\Fig{DGs_phase} (right), where it is compared both with the
direct measurement of \DGs and \phiccbars,
and with the Standard Model expectations. 
No inconsistency is observed between all these data.


%


\comment{

For non-zero $|\Gamma_{12}|$, analysis of the time-dependent
decay \particle{\Bs \to \jpsi\phi} can measure
the weak phase.  Including information on the \Bs flavour at production
time via flavour tagging improves precision and also resolves the 
sign ambiguity on the weak phase angle for a given \DGs.
Both CDF~\cite{CDF:2011af,*Aaltonen:2007he_mod,*Aaltonen:2007gf_mod} 
and \dzero~\cite{Abazov:2011ry,*Abazov_mod:2008fj,*Abazov:2007tx_mod_cont} have performed 
such analyses and measure the same observed phase that we denote
$\phi_s^{\jpsi \phi} = -2\beta_s^{\jpsi \phi}$ to reflect
the different conventions of the experiments.

Under the assumption of non-zero $\phi_s^{\jpsi\phi}$, 
in addition to the result listed
in \Table{dgammat}, 
the \dzero collaboration~\cite{Abazov:2011ry,*Abazov_mod:2008fj,*Abazov:2007tx_mod_cont}  has also made simultaneous
fits allowing $\phi_s^{\jpsi\phi}$ to float while weakly 
constraining the strong phases, $\delta_i$ to find: 
\begin{eqnarray}
\DGs &=& +0.19 \pm 0.07 ^{+0.02}_{-0.01}~{\mathrm{ps}}^{-1}\,,  \\ 
\bar{\tau}(\Bs) &= &1/\Gs = 1.52 \pm 0.06~{\mathrm{ps}}\,,  \\
\phi_s^{\jpsi\phi} &=& -0.57 ^{+0.24+0.07}_{-0.30-0.02} \,. 
\end{eqnarray}
If the SM value of $\phi_s^{\jpsi\phi} = -0.04$ is assumed, a probability of 
6.6\% to obtain a value of $\phi_s^{\jpsi\phi}$ lower than $-0.57$ is
found.

The CDF 
analysis~\cite{CDF:2011af,*Aaltonen:2007he_mod,*Aaltonen:2007gf_mod} 
reports confidence regions
in the two-dimensional space of $2\beta_s^{\jpsi\phi}$ and \DGs.
They present a Feldman-Cousins confidence interval of $2\beta_s^{\jpsi\phi}$
where \DGs is treated as a nuisance parameter:
\begin{equation}
2\beta_s^{\jpsi\phi} = -\phi_s^{\jpsi\phi} \in [0.56,2.58]~{\mathrm{at~68\%~CL}}.
\end{equation}
Only a confidence range is quoted and a  point 
estimate is not given since biases were observed in the analysis.
Assuming the SM predictions for $2\beta_s$ and \DGs, they find
that the probability of a deviation as large as the level of the 
observed data is 7\%.
Note that CDF has very recently made a preliminary update~\cite{CDFnote10778:2012,*CDFnote10778:2012_cont}
to their
\particle{\Bs \to \jpsi\phi} analysis to an
integrated luminosity of 5.2~fb$^{-1}$ indicating a best-fit
confidence interval of:
\begin{equation}
2\beta_s^{\jpsi\phi} = -\phi_s^{\jpsi\phi} 
\in [0.04,1.04] \cup [2.16,3.10]~{\mathrm{at~68\%~CL}},
\end{equation}
where the probability
of a larger deviation from the SM prediction is 44\% or $0.8\,\sigma$.
However, this new result has not yet been used in the combinations
below.

Given the consistency of these two measurements of the weak phase,
as well as their
deviations from the SM, there is interest in combining the results and
using in global fits, \eg\ see \Ref{Bona:2008jn}.
To allow a combination on equal footing, the \dzero collaboration
has redone their fits~\cite{D0web:2009} 
allowing  strong phase values, $\delta_i$, to float
as in the CDF analysis.
Ensemble studies to test confidence level coverage were performed 
by both collaborations and used to adjust likelihood
values to correspond to the usual Gaussian confidence levels. 
Two-dimensional likelihoods were 
combined~\cite{CDFnote9787:2009,D0Note5928:2009}
with the result shown in 
\Fig{DGs_phase}(a).  
After the combination, consistency  
of the best fit values for $\phi_s^{\jpsi\phi} = -2\beta_s^{\jpsi\phi}$ with
SM predictions is at the level of $\hfagNSIGMASM\,\sigma$, with numerical results
for the two solutions given below.
Despite possible biases in the CDF input, point estimates are still
presented and the confidence level regions are straight projections
onto the \DGs or phase angle axes.
\begin{eqnarray}
\DGs &=& \hfagDGSCOMB \,, \\
\phi_s^{\jpsi\phi} = -2\beta_s^{\jpsi\phi} &=& \hfagPHISCOMB \,.
\end{eqnarray}

A comparison between
the above sum of the
CDF and \dzero likelihoods 
and the world average \Bs semileptonic asymmetry of
\Eq{ASLS} through~\cite{Beneke:2003az}:
\begin{equation}
\ASLs = 
\frac{|\Gamma^{12}_s|}{|M^{12}_s|}\sin\phi_s = \frac{\DGs}{\dms}\tan\phi_s
\end{equation}
is also made and shown in 
\Fig{DGs_phase}(a).
Consistency between the two is observed, and the value
of \ASLs is applied as a constraint
resulting in the
confidence level regions 
shown in \Fig{DGs_phase}(b)
including the region delineated by new physics traced by 
the relation of \Eq{new_phys_phase}. Numerical results for the 
two solutions are:
\begin{eqnarray}
\DGs &=& \hfagDGSCOMBCON \,, \\
\phi_s^{\jpsi\phi} = -2\beta_s^{\jpsi\phi} &=& \hfagPHISCOMBCON \,.
\end{eqnarray}
with a consistency
of the best fit values with
SM predictions of $2\beta_s$ at the level of $\hfagNSIGMASMCON\,\sigma$.

}

\clearpage
\mysection{Measurements related to Unitarity Triangle angles
}
\label{sec:cp_uta}

The charge of the ``$\CP(t)$ and Unitarity Triangle angles'' group
is to provide averages of measurements 
from time-dependent asymmetry analyses,
and other quantities that are related 
to the angles of the Unitarity Triangle (UT).
In cases where considerable theoretical input is required to 
extract the fundamental quantities, no attempt is made to do so at 
this stage. However, straightforward interpretations of the averages 
are given, where possible.

In Sec.~\ref{sec:cp_uta:introduction} 
a brief introduction to the relevant phenomenology is given.
In Sec.~\ref{sec:cp_uta:notations}
an attempt is made to clarify the various different notations in use.
In Sec.~\ref{sec:cp_uta:common_inputs}
the common inputs to which experimental results are rescaled in the
averaging procedure are listed. 
We also briefly introduce the treatment of experimental errors. 
In the remainder of this section,
the experimental results and their averages are given,
divided into subsections based on the underlying quark-level decays.

\mysubsection{Introduction
}
\label{sec:cp_uta:introduction}

The Standard Model Cabibbo-Kobayashi-Maskawa (CKM) quark mixing matrix $\VCKM$ 
must be unitary. A $3 \times 3$ unitary matrix has four free parameters,\footnote{
  In the general case there are nine free parameters,
  but five of these are absorbed into unobservable quark phases.}
and these are conventionally written by the product
of three (complex) rotation matrices~\cite{Chau:1984fp}, 
where the rotations are characterised by the Euler angles 
$\theta_{12}$, $\theta_{13}$ and $\theta_{23}$, which are the mixing angles
between the generations, and one overall phase $\delta$,
\begin{equation}
\label{eq:ckmPdg}
\VCKM =
        \left(
          \begin{array}{ccc}
            V_{ud} & V_{us} & V_{ub} \\
            V_{cd} & V_{cs} & V_{cb} \\
            V_{td} & V_{ts} & V_{tb} \\
          \end{array}
        \right)
        =
        \left(
        \begin{array}{ccc}
        c_{12}c_{13}    
                &    s_{12}c_{13}   
                        &   s_{13}e^{-i\delta}  \\
        -s_{12}c_{23}-c_{12}s_{23}s_{13}e^{i\delta} 
                &  c_{12}c_{23}-s_{12}s_{23}s_{13}e^{i\delta} 
                        & s_{23}c_{13} \\
        s_{12}s_{23}-c_{12}c_{23}s_{13}e^{i\delta}  
                &  -c_{12}s_{23}-s_{12}c_{23}s_{13}e^{i\delta} 
                        & c_{23}c_{13} 
        \end{array}
        \right)
\end{equation}
where $c_{ij}=\cos\theta_{ij}$, $s_{ij}=\sin\theta_{ij}$ for 
$i<j=1,2,3$. 

Following the observation of a hierarchy between the different
matrix elements, the Wolfenstein parametrisation~\cite{Wolfenstein:1983yz}
is an expansion of $\VCKM$ in terms of the four real parameters $\lambda$
(the expansion parameter), $A$, $\rho$ and $\eta$. Defining to 
all orders in $\lambda$~\cite{Buras:1994ec}
\begin{eqnarray}
  \label{eq:burasdef}
  s_{12}             &\equiv& \lambda\,,\nonumber \\ 
  s_{23}             &\equiv& A\lambda^2\,, \\
  s_{13}e^{-i\delta} &\equiv& A\lambda^3(\rho -i\eta)\,,\nonumber
\end{eqnarray}
and inserting these into the representation of Eq.~(\ref{eq:ckmPdg}), 
unitarity of the CKM matrix is achieved to all orders.
A Taylor expansion of $\VCKM$ leads to the familiar approximation
\begin{equation}
  \label{eq:cp_uta:ckm}
  \VCKM
  = 
  \left(
    \begin{array}{ccc}
      1 - \lambda^2/2 & \lambda & A \lambda^3 ( \rho - i \eta ) \\
      - \lambda & 1 - \lambda^2/2 & A \lambda^2 \\
      A \lambda^3 ( 1 - \rho - i \eta ) & - A \lambda^2 & 1 \\
    \end{array}
  \right) + {\cal O}\left( \lambda^4 \right) \, .
\end{equation}
At order $\lambda^{5}$, the obtained CKM matrix in this extended
Wolfenstein parametrisation is:
{\small
  \begin{equation}
    \label{eq:cp_uta:ckm_lambda5}
    \VCKM
    =
    \left(
      \begin{array}{ccc}
        1 - \frac{1}{2}\lambda^{2} - \frac{1}{8}\lambda^4 &
        \lambda &
        A \lambda^{3} (\rho - i \eta) \\
        - \lambda + \frac{1}{2} A^2 \lambda^5 \left[ 1 - 2 (\rho + i \eta) \right] &
        1 - \frac{1}{2}\lambda^{2} - \frac{1}{8}\lambda^4 (1+4A^2) &
        A \lambda^{2} \\
        A \lambda^{3} \left[ 1 - (1-\frac{1}{2}\lambda^2)(\rho + i \eta) \right] &
        -A \lambda^{2} + \frac{1}{2}A\lambda^4 \left[ 1 - 2(\rho + i \eta) \right] &
        1 - \frac{1}{2}A^2 \lambda^4
      \end{array} 
    \right) + {\cal O}\left( \lambda^{6} \right)\,.
  \end{equation}
}
The non-zero imaginary part of the CKM matrix,
which is the origin of $\CP$ violation in the Standard Model,
is encapsulated in a non-zero value of $\eta$.



The unitarity relation $\VCKM^\dagger\VCKM = {\mathit 1}$
results in a total of nine expressions,
that can be written as
$\sum_{i=u,c,t} V^*_{ij}V_{ik} = \delta_{jk}$,
where $\delta_{jk}$ is the Kronecker symbol.
Of the off-diagonal expressions ($j \neq k$),
three can be transformed into the other three 
leaving six relations, in which three complex numbers sum to zero,
which therefore can be expressed as triangles in the complex plane.
More details about unitarity triangles can be found in~\cite{Jarlskog:1985ht,Jarlskog:2005uq,Bjorken:2005rm,Harrison:2009bz,Frampton:2010ii,Frampton:2010uq}.

\begin{figure}[t]
  \begin{center}
    \resizebox{0.55\textwidth}{!}{\includegraphics{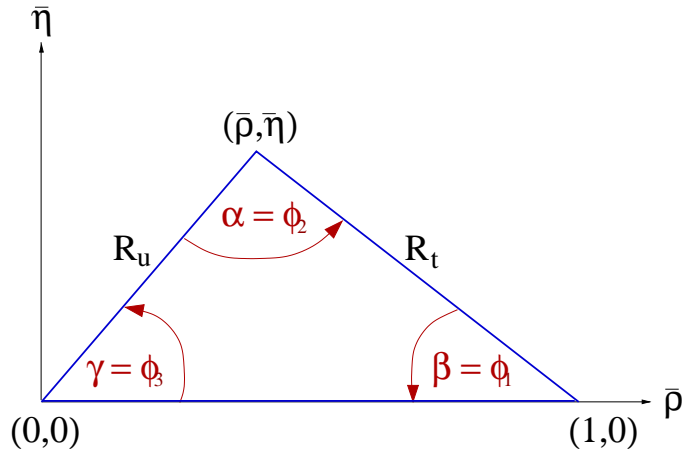}}
    \caption{The Unitarity Triangle.}
    \label{fig:cp_uta:ut}
  \end{center}
\end{figure}

One of these relations,
\begin{equation}
  \label{eq:cp_uta:ut}
  V_{ud}V^*_{ub} + V_{cd}V^*_{cb} + V_{td}V^*_{tb} = 0\,,
\end{equation}
is of particular importance to the $\B$ system, 
being specifically related to flavour changing 
neutral current $b \to d$ transitions.
The three terms in Eq.~(\ref{eq:cp_uta:ut}) are of the same order 
(${\cal O}\left( \lambda^3 \right)$),
and this relation is commonly known as the Unitarity Triangle.
For presentational purposes,
it is convenient to rescale the triangle by $(V_{cd}V^*_{cb})^{-1}$,
as shown in Fig.~\ref{fig:cp_uta:ut}.

Two popular naming conventions for the UT angles exist in the literature:
\begin{equation}
  \label{eq:cp_uta:abc}
  \alpha  \equiv  \phi_2  = 
  \arg\left[ - \frac{V_{td}V_{tb}^*}{V_{ud}V_{ub}^*} \right]\,,
  \hspace{0.5cm}
  \beta   \equiv   \phi_1 =  
  \arg\left[ - \frac{V_{cd}V_{cb}^*}{V_{td}V_{tb}^*} \right]\,,
  \hspace{0.5cm}
  \gamma  \equiv   \phi_3  =  
  \arg\left[ - \frac{V_{ud}V_{ub}^*}{V_{cd}V_{cb}^*} \right]\,.
\end{equation}
In this document the $\left( \alpha, \beta, \gamma \right)$ set is used.\footnote{
  The relevant unitarity triangle for the $\Bs$ system is obtained 
  by replacing $d \leftrightarrow s$ in Eq.~\ref{eq:cp_uta:ut}.
  Definitions of the set of angles $( \alpha_s, \beta_s, \gamma_s )$ 
  can be obtained using equivalent relations to those of Eq.~\ref{eq:cp_uta:abc},
  for example $\beta_s = \arg\left[ - (V_{cs}V_{cb}^*) / (V_{ts}V_{tb}^*) \right]$.
  This definition gives a value of $\beta_s$ that is negative in the Standard Model,
  so that the sign is often flipped in the literature.
}
The sides $R_u$ and $R_t$ of the Unitarity Triangle 
(the third side being normalised to unity) 
are given by
\begin{equation}
  \label{eq:ru_rt}
  R_u =
  \left|\frac{V_{ud}V_{ub}^*}{V_{cd}V_{cb}^*} \right|
  = \sqrt{\rhobar^2+\etabar^2} \,,
  \hspace{0.5cm}
  R_t = 
  \left|\frac{V_{td}V_{tb}^*}{V_{cd}V_{cb}^*}\right| 
  = \sqrt{(1-\rhobar)^2+\etabar^2} \,.
\end{equation} 
where $\rhobar$ and $\etabar$ define 
the apex of the Unitarity Triangle~\cite{Buras:1994ec} 
\begin{equation}
  \label{eq:rhoetabar}
  \rhobar + i\etabar
  \equiv -\frac{V_{ud}V_{ub}^*}{V_{cd}V_{cb}^*}
  \equiv 1 + \frac{V_{td}V_{tb}^*}{V_{cd}V_{cb}^*}
  = \frac{\sqrt{1-\lambda^2}\,(\rho + i \eta)}{\sqrt{1-A^2\lambda^4}+\sqrt{1-\lambda^2}A^2\lambda^4(\rho+i\eta)} \, .
\end{equation}
The exact relation between $\left( \rho, \eta \right)$ and 
$\left( \rhobar, \etabar \right)$ is
\begin{equation}
  \label{eq:rhoetabarinv}
  \rho + i\eta \;=\; 
  \frac{ 
    \sqrt{ 1-A^2\lambda^4 }(\rhobar+i\etabar) 
  }{
    \sqrt{ 1-\lambda^2 } \left[ 1-A^2\lambda^4(\rhobar+i\etabar) \right]
  } \, .
\end{equation}

By expanding in powers of $\lambda$, several useful approximate expressions
can be obtained, including
\begin{equation}
  \label{eq:rhoeta_approx}
  \rhobar = \rho (1 - \frac{1}{2}\lambda^{2}) + {\cal O}(\lambda^4) \, ,
  \hspace{0.5cm}
  \etabar = \eta (1 - \frac{1}{2}\lambda^{2}) + {\cal O}(\lambda^4) \, ,
  \hspace{0.5cm}
  V_{td} = A \lambda^{3} (1-\rhobar -i\etabar) + {\cal O}(\lambda^6) \, .
\end{equation}

\mysubsection{Notations
}
\label{sec:cp_uta:notations}

Several different notations for $\CP$ violation parameters
are commonly used.
This section reviews those found in the experimental literature,
in the hope of reducing the potential for confusion, 
and to define the frame that is used for the averages.

In some cases, when $\B$ mesons decay into 
multibody final states via broad resonances ($\rho$, $\Kstar$, \etc),
the experimental analyses ignore the effects of interference 
between the overlapping structures.
This is referred to as the quasi-two-body (Q2B) approximation
in the following.

\mysubsubsection{$\CP$ asymmetries
}
\label{sec:cp_uta:notations:pra}

The $\CP$ asymmetry is defined as the difference between the rate 
involving a $b$ quark and that involving a $\bar b$ quark, divided 
by the sum. For example, the partial rate (or charge) asymmetry for 
a charged $\B$ decay would be given as 
\begin{equation}
  \label{eq:cp_uta:pra}
  \Acp_{f} \;\equiv\; 
  \frac{\Gamma(\Bm \to f)-\Gamma(\Bp \to \bar{f})}{\Gamma(\Bm \to f)+\Gamma(\Bp \to \bar{f})}.
\end{equation}

\mysubsubsection{Time-dependent \CP asymmetries in decays to $\CP$ eigenstates
}
\label{sec:cp_uta:notations:cp_eigenstate}

If the amplitudes for $\Bz$ and $\Bzb$ to decay to a final state $f$, 
which is a $\CP$ eigenstate with eigenvalue $\etacpf$,
are given by $\Af$ and $\Abarf$, respectively, 
then the decay distributions for neutral $\B$ mesons, 
with known flavour at time $\Delta t =0$,
are given by
\begin{eqnarray}
  \label{eq:cp_uta:td_cp_asp1}
  \Gamma_{\Bzb \to f} (\Delta t) & = &
  \frac{e^{-| \Delta t | / \tau(\Bz)}}{4\tau(\Bz)}
  \left[ 
    1 +
    \frac{2\, \Im(\lambda_f)}{1 + |\lambda_f|^2} \sin(\Delta m \Delta t) -
    \frac{1 - |\lambda_f|^2}{1 + |\lambda_f|^2} \cos(\Delta m \Delta t)
  \right], \\
  \label{eq:cp_uta:td_cp_asp2}
  \Gamma_{\Bz \to f} (\Delta t) & = &
  \frac{e^{-| \Delta t | / \tau(\Bz)}}{4\tau(\Bz)}
  \left[ 
    1 -
    \frac{2\, \Im(\lambda_f)}{1 + |\lambda_f|^2} \sin(\Delta m \Delta t) +
    \frac{1 - |\lambda_f|^2}{1 + |\lambda_f|^2} \cos(\Delta m \Delta t)
  \right].
\end{eqnarray}
Here $\lambda_f = \frac{q}{p} \frac{\Abarf}{\Af}$ 
contains terms related to $\Bz$\textendash$\Bzb$ mixing and to the decay amplitude
(the eigenstates of the effective Hamiltonian in the $\BzBzb$ system 
are $\left| B_\pm \right> = p \left| \Bz \right> \pm q \left| \Bzb \right>$).
This formulation assumes $\CPT$ invariance, 
and neglects possible lifetime differences 
(between the eigenstates of the effective Hamiltonian;
see Section~\ref{sec:mixing} where the mass difference $\Delta m$ is also defined)
in the neutral $\B$ meson system.
The case where non-zero lifetime differences are taken into account is 
discussed in Section~\ref{sec:cp_uta:notations:Bs}.
Note that the notation and normalisation used here is that which is relevant for the $e^+e^-$ $B$ factory experiments.
At hadron collider experiments, the flavour tagging is done at production ($\Delta t = t = 0$), and therefore $t$ is usually used in place of $\Delta t$.
Moreover, since negative values of $t$ are not allowed, the normalisation is such that 
$\int_0^{+\infty} \left( 
\Gamma_{\Bzb \to f} (t) + \Gamma_{\Bzb \to f} (t) \right) dt = 1$,
rather than 
$\int_{-\infty}^{+\infty} \left( 
\Gamma_{\Bzb \to f} (\Delta t) + \Gamma_{\Bzb \to f} (\Delta t) \right) d(\Delta t) = 1$,
as in Eqs.~\ref{eq:cp_uta:td_cp_asp1} and~\ref{eq:cp_uta:td_cp_asp2}.

The time-dependent $\CP$ asymmetry,
again defined as the difference between the rate 
involving a $b$ quark and that involving a $\bar b$ quark,
is then given by
\begin{equation}
  \label{eq:cp_uta:td_cp_asp}
  \Acp_{f} \left(\Delta t\right) \; \equiv \;
  \frac{
    \Gamma_{\Bzb \to f} (\Delta t) - \Gamma_{\Bz \to f} (\Delta t)
  }{
    \Gamma_{\Bzb \to f} (\Delta t) + \Gamma_{\Bz \to f} (\Delta t)
  } \; = \;
  \frac{2\, \Im(\lambda_f)}{1 + |\lambda_f|^2} \sin(\Delta m \Delta t) -
  \frac{1 - |\lambda_f|^2}{1 + |\lambda_f|^2} \cos(\Delta m \Delta t).
\end{equation}

While the coefficient of the $\sin(\Delta m \Delta t)$ term in 
Eq.~(\ref{eq:cp_uta:td_cp_asp}) is everywhere\footnote
{
  Occasionally one also finds Eq.~(\ref{eq:cp_uta:td_cp_asp}) written as
  $\Acp_{f} \left(\Delta t\right) = 
  {\cal A}^{\rm mix}_f \sin(\Delta m \Delta t) + {\cal A}^{\rm dir}_f \cos(\Delta m \Delta t)$,
  or similar.
} denoted $S_f$:
\begin{equation}
  \label{eq:cp_uta:s_def}
  S_f \;\equiv\; \frac{2\, \Im(\lambda_f)}{1 + \left|\lambda_f\right|^2},
\end{equation}
different notations are in use for the
coefficient of the $\cos(\Delta m \Delta t)$ term:
\begin{equation}
  \label{eq:cp_uta:c_def}
  C_f \;\equiv\; - A_f \;\equiv\; \frac{1 - \left|\lambda_f\right|^2}{1 + \left|\lambda_f\right|^2}.
\end{equation}
The $C$ notation is used by the \babar\  collaboration 
(see \eg~\cite{Aubert:2001sp}), 
and also in this document.
The $A$ notation is used by the \belle\ collaboration
(see \eg~\cite{Abe:2001xe}).

Neglecting effects due to $\CP$ violation in mixing 
(by taking $|q/p| = 1$),
if the decay amplitude contains terms with 
a single weak (\ie, $\CP$ violating) phase
then $\left|\lambda_f\right| = 1$ and one finds
$S_f = -\etacpf \sin(\phi_{\rm mix} + \phi_{\rm dec})$, $C_f = 0$,
where $\phi_{\rm mix}=\arg(q/p)$ and $\phi_{\rm dec}=\arg(\Abarf/\Af)$.
Note that the $\Bz$--$\Bzb$ mixing phase $\phi_{\rm mix}\approx2\beta$
in the Standard Model (in the usual phase convention)~\cite{Carter:1980tk,Bigi:1981qs}. 

If amplitudes with different weak phases contribute to the decay, 
no clean interpretation of $S_f$ is possible without further input. 
If the decay amplitudes have in addition different $\CP$ conserving strong phases, then $\left| \lambda_f \right| \neq 1$ and additional input is required for interpretation.
The coefficient of the cosine term becomes non-zero,
indicating direct $\CP$ violation.
The sign of $A_f$ as defined above is consistent with that of $\Acp_{f}$ in 
Eq.~(\ref{eq:cp_uta:pra}).

Due to the fact that $\sin(\Delta m \Delta t)$ and $\cos(\Delta m \Delta t)$ are respectively odd and even functions of $\Delta t$, only small correlations (that can be induced by backgrounds, for example) between $S_f$ and $C_f$ are expected a $B$ factory experiments, where the range of $\Delta t$ is $-\infty < \Delta t < +\infty$.
The situation is different for measurements at hadron collider experiments, where the range of the time variable is $0 < \Delta t < +\infty$, so that more sizable correlations can be expected.  
We include the correlations in the averages where available.

Frequently, we are interested in combining measurements 
governed by similar or identical short-distance physics,
but with different final states
(\eg, $\Bz \to \jpsi \KS$ and $\Bz \to \jpsi \KL$).
In this case, we remove the dependence on the $\CP$ eigenvalue 
of the final state by quoting $-\etacp S_f$.
In cases where the final state is not a $\CP$ eigenstate but has
an effective $\CP$ content (see below),
the reported $-\etacp S$ is corrected by the effective $\CP$.

\mysubsubsection{Time-dependent distributions with non-zero decay width difference}
\label{sec:cp_uta:notations:Bs}

A complete analysis of the time-dependent decay rates of 
neutral $B$ mesons must also take into account the lifetime difference
between the eigenstates of the effective Hamiltonian, 
denoted by $\Delta \Gamma$.
This is particularly important in the $B_s$ system,
since non-negligible values of $\Delta \Gamma_s$ have now been established
(see Section~\ref{sec:mixing} for the latest experimental constraints).
Neglecting $\CP$ violation in mixing,
the relevant replacements for 
Eqs.~\ref{eq:cp_uta:td_cp_asp1}~\&~\ref{eq:cp_uta:td_cp_asp2} 
are~\cite{Dunietz:2000cr}
\begin{equation}
  \label{eq:cp_uta:td_cp_bs_asp1}
  \begin{array}{lcr}
    \mc{2}{l}{
      \Gamma_{\Bsb \to f} (\Delta t) = 
      {\cal N} 
      \frac{e^{-| \Delta t | / \tau(\Bs)}}{4\tau(\Bs)}
      \Big[ 
      \cosh(\frac{\Delta \Gamma \Delta t}{2}) +
    } & \hspace{40mm} \\
    \hspace{40mm} &
    \mc{2}{r}{
      \frac{2\, \Im(\lambda_f)}{1 + |\lambda_f|^2} \sin(\Delta m \Delta t) -
      \frac{1 - |\lambda_f|^2}{1 + |\lambda_f|^2} \cos(\Delta m \Delta t) -
      \frac{2\, \Re(\lambda_f)}{1 + |\lambda_f|^2} \sinh(\frac{\Delta \Gamma \Delta t}{2})
      \Big],
    } \\
  \end{array}
\end{equation}
and
\begin{equation}
  \label{eq:cp_uta:td_cp_bs_asp2}
  \begin{array}{lcr}
    \mc{2}{l}{
      \Gamma_{\Bs \to f} (\Delta t) =
      {\cal N} 
      \frac{e^{-| \Delta t | / \tau(\Bs)}}{4\tau(\Bs)}
      \Big[ 
      \cosh(\frac{\Delta \Gamma \Delta t}{2}) -
    } & \hspace{40mm} \\
    \hspace{40mm} & 
    \mc{2}{r}{
      \frac{2\, \Im(\lambda_f)}{1 + |\lambda_f|^2} \sin(\Delta m \Delta t) +
      \frac{1 - |\lambda_f|^2}{1 + |\lambda_f|^2} \cos(\Delta m \Delta t) -
      \frac{2\, \Re(\lambda_f)}{1 + |\lambda_f|^2} \sinh(\frac{\Delta \Gamma \Delta t}{2})
      \Big]. 
    } \\
  \end{array}
\end{equation}

To be consistent with our earlier notation,\footnote{
  As ever, alternative and conflicting notations appear in the literature.
  One popular alternative notation for this parameter is 
  ${\cal A}_{\Delta \Gamma}$.
  Particular care must be taken over the signs.
}
we write here the coefficient of the $\sinh$ term as
\begin{equation}
  A^{\Delta \Gamma}_f = - \frac{2\, \Re(\lambda_f)}{1 + |\lambda_f|^2} \, .
\end{equation}
A complete, tagged, time-dependent analysis of \CP asymmetries in 
$B_s$ decays to a \CP eigenstate $f$ can thus obtain the parameters 
$S_f$, $C_f$ and $A^{\Delta \Gamma}_f$.
Note that, by definition, 
\begin{equation}
  \left( S_f \right)^2 + \left( C_f \right)^2 + \left( A^{\Delta \Gamma}_f \right)^2 = 1 \, ,
\end{equation}
and this constraint can be imposed or not in the fits.
Since these parameters have sensitivity to both
$\Im(\lambda_f)$ and $\Re(\lambda_f)$,
alternative choices of parametrisation, 
including those directly involving \CP violating phases (such as $\beta_s$), 
are possible.
These can also be adopted for vector-vector final states.

The {\it untagged} time-dependent decay rate is given by
\begin{equation}
  \Gamma_{\Bsb \to f} (\Delta t) + \Gamma_{\Bs \to f} (\Delta t)
  = 
  {\cal N} 
  \frac{e^{-| \Delta t | / \tau(\Bs)}}{2\tau(\Bs)}
  \Big[ 
  \cosh\left(\frac{\Delta \Gamma \Delta t}{2}\right) -
  \frac{2\, \Re(\lambda_f)}{1 + |\lambda_f|^2} \sinh\left(\frac{\Delta \Gamma \Delta t}{2}\right)
  \Big] \, .
\end{equation}
With the requirement
$\int_{-\infty}^{+\infty} \Gamma_{\Bsb \to f} (\Delta t) + \Gamma_{\Bs \to f} (\Delta t) d(\Delta t) = 1$,
the normalisation factor ${\cal N}$ 
is fixed to $1 - (\frac{\Delta \Gamma}{2\Gamma})^2$.
Note that an untagged time-dependent analysis can probe
$\lambda_f$, through $\Re(\lambda_f)$, when $\Delta \Gamma \neq 0$.
This is equivalent to determining the ``{\it effective lifetime}''~\cite{Fleischer:2011cw}, as discussed in Sec.~\ref{sec:taubs}.
The tagged analysis is, of course, more sensitive.

Other expressions can be similarly modified to take into account 
non-zero lifetime differences.
Note that when the final state contains 
a mixture of $\CP$-even and $\CP$-odd states
(as, for example, for vector-vector or multibody self-conjugate states),
that $\Re(\lambda_f)$ contains terms proportional to 
both the sine and cosine of the weak phase difference, 
albeit with rather different sensitivities.

\mysubsubsection{Time-dependent \CP asymmetries in decays to vector-vector final states
}
\label{sec:cp_uta:notations:vv}

Consider \B decays to states consisting of two spin-1 particles,
such as $\jpsi K^{*0}(\to\KS\piz)$, $D^{*+}D^{*-}$ and $\rho^+\rho^-$,
which are eigenstates of charge conjugation but not of parity.\footnote{
  \noindent
  This is not true of all vector-vector final states,
  \eg, $D^{*\pm}\rho^{\mp}$ is clearly not an eigenstate of 
  charge conjugation.
}
In fact, for such a system, there are three possible final states;
in the helicity basis these can be written $h_{-1}, h_0, h_{+1}$.
The $h_0$ state is an eigenstate of parity, and hence of $\CP$;
however, $\CP$ transforms $h_{+1} \leftrightarrow h_{-1}$ (up to 
an unobservable phase). In the transversity basis, these states 
are transformed into  $h_\parallel =  (h_{+1} + h_{-1})/2$ and 
$h_\perp = (h_{+1} - h_{-1})/2$.
In this basis all three states are $\CP$ eigenstates, 
and $h_\perp$ has the opposite $\CP$ to the others.

The amplitudes to these states are usually given by $A_{0,\perp,\parallel}$
(here we use a normalisation such that 
$| A_0 |^2 + | A_\perp |^2 + | A_\parallel |^2 = 1$).
Then the effective $\CP$ of the vector-vector state is known if 
$| A_\perp |^2$ is measured.
An alternative strategy is to measure just the longitudinally polarised 
component,  $| A_0 |^2$
(sometimes denoted by $f_{\rm long}$), 
which allows a limit to be set on the effective $\CP$ since
$| A_\perp |^2 \leq | A_\perp |^2 + | A_\parallel |^2 = 1 - | A_0 |^2$.
The most complete treatment for 
neutral $\B$ decays to vector-vector final states
is time-dependent angular analysis 
(also known as time-dependent transversity analysis).
In such an analysis, 
the interference between the $\CP$-even and $\CP$-odd states 
provides additional sensitivity to the weak and strong phases involved.

In most analyses of time-dependent \CP asymmetries in decays to 
vector-vector final states carried out to date,
an assumption has been made that each helicity (or transversity) amplitude
has the same weak phase.
This is a good approximation for decays that are dominated by 
amplitudes with a single weak phase, such $\Bz \to \jpsi K^{*0}$,
and is a reasonable approximation in any mode for which only 
very limited statistics are available.
However, for modes that have contributions from amplitudes with different 
weak phases, the relative size of these contributions can be different 
for each helicity (or transversity) amplitude,
and therefore the time-dependent \CP asymmetry parameters can also differ.
The most generic analysis, suitable for modes with sufficient statistics,
would allow for this effect;
an intermediate analysis can allow different parameters for the 
$\CP$-even and $\CP$-odd components.
Such an analysis has been carried out by \babar\ for the decay
$\Bz \to D^{*+}D^{*-}$~\cite{Aubert:2008ah}.

\mysubsubsection{Time-dependent asymmetries: self-conjugate multiparticle final states
}
\label{sec:cp_uta:notations:dalitz}

Amplitudes for neutral \B decays into 
self-conjugate multiparticle final states
such as $\pi^+\pi^-\pi^0$, $K^+K^-\KS$, $\pi^+\pi^-\KS$,
$\jpsi \pi^+\pi^-$ or $D\pi^0$ with $D \to \KS\pi^+\pi^-$
may be written in terms of \CP-even and \CP-odd amplitudes.
As above, the interference between these terms 
provides additional sensitivity to the weak and strong phases
involved in the decay,
and the time-dependence depends on both the sine and cosine
of the weak phase difference.
In order to perform unbinned maximum likelihood fits,
and thereby extract as much information as possible from the distributions,
it is necessary to select a model for the multiparticle decay,
and therefore the results acquire some model dependence
(binned, model independent methods are also possible,
though are not as statistically powerful).
The number of observables depends on the final state (and on the model used);
the key feature is that as long as there are regions where both
\CP-even and \CP-odd amplitudes contribute,
the interference terms will be sensitive to the cosine 
of the weak phase difference.
Therefore, these measurements allow distinction between multiple solutions
for, \eg, the four values of $\beta$ from the measurement of $\sin(2\beta)$.

We now consider the various notations which have been used 
in experimental studies of
time-dependent asymmetries in decays to self-conjugate multiparticle final states.

\mysubsubsubsection{$\Bz \to D^{(*)}h^0$ with $D \to \KS\pi^+\pi^-$
}
\label{sec:cp_uta:notations:dalitz:dh0}

The states $D\pi^0$, $D^*\pi^0$, $D\eta$, $D^*\eta$, $D\omega$
are collectively denoted $D^{(*)}h^0$.
When the $D$ decay model is fixed,
fits to the time-dependent decay distributions can be performed
to extract the weak phase difference.
However, it is experimentally advantageous to use the sine and cosine of 
this phase as fit parameters, since these behave as essentially 
independent parameters, with low correlations and (potentially)
rather different uncertainties.
A parameter representing direct $\CP$ violation in the $B$ decay 
can also be floated.  
For consistency with other analyses, this could be chosen to be $C_f$,
but could equally well be $\left| \lambda_f \right|$, or other possibilities.

\belle\ performed an analysis of these channels
with $\sin(2\phi_1)$ and $\cos(2\phi_1)$ as free parameters~\cite{Krokovny:2006sv}.
\babar\ have performed an analysis floating also $\left| \lambda_f \right|$~\cite{Aubert:2007rp}
(and, of course, replacing $\phi_1 \Leftrightarrow \beta$).

\mysubsubsubsection{$\Bz \to D^{*+}D^{*-}\KS$
}
\label{sec:cp_uta:notations:dalitz:dstardstarks}

The hadronic structure of the $\Bz \to D^{*+}D^{*-}\KS$ decay
is not sufficiently well understood to perform a full 
time-dependent Dalitz plot analysis.
Instead, following Browder {\it et al.}~\cite{Browder:1999ng},
\babar~\cite{Aubert:2006fh} divide the Dalitz plane in two:
$m(D^{*+}\KS)^2 > m(D^{*-}\KS)^2$ $(\eta_y = +1)$ and 
$m(D^{*+}\KS)^2 < m(D^{*-}\KS)^2$ $(\eta_y = -1)$;
and then fit to a decay time distribution with asymmetry given by
\begin{equation}
  \Acp_{f} \left(\Delta t\right) =
  \eta_y \frac{J_c}{J_0} \cos(\Delta m \Delta t) -  
  \left[ 
    \frac{2J_{s1}}{J_0} \sin(2\beta) + \eta_y \frac{2J_{s2}}{J_0} \cos(2\beta) 
  \right] \sin(\Delta m \Delta t) \, .
\end{equation}
A similar analysis has also been carried out by \belle~\cite{Dalseno:2007hx}.
The measured values are $\frac{J_c}{J_0}$, $\frac{2J_{s1}}{J_0} \sin(2\beta)$
and $\frac{2J_{s2}}{J_0} \cos(2\beta)$, 
where the parameters $J_0$, $J_c$, $J_{s1}$ and $J_{s2}$ are the integrals 
over the half Dalitz plane $m(D^{*+}\KS)^2 < m(D^{*-}\KS)^2$ 
of the functions $|a|^2 + |\bar{a}|^2$, $|a|^2 - |\bar{a}|^2$, 
$\Re(\bar{a}a^*)$ and $\Im(\bar{a}a^*)$ respectively, 
where $a$ and $\bar{a}$ are the decay amplitudes of 
$\Bz \to D^{*+}D^{*-}\KS$ and $\Bzb \to D^{*+}D^{*-}\KS$ respectively. 
The parameter $J_{s2}$ (and hence $J_{s2}/J_0$) is predicted to be positive;
with this assumption is it possible to determine the sign of $\cos(2\beta)$.

\mysubsubsubsection{$\Bz \to K^+K^-\Kz$
}
\label{sec:cp_uta:notations:dalitz:kkk0}

Studies of $\Bz \to K^+K^-\Kz$~\cite{Aubert:2007sd,Nakahama:2010nj,Lees:2012kx} 
and of the related decay 
$\Bp \to K^+K^-K^+$~\cite{Garmash:2004wa,Aubert:2006nu,Lees:2012kx},
show that the decay is dominated by a large nonresonant contribution
with significant components from the 
intermediate $K^+K^-$ resonances $\phi(1020)$, $f_0(980)$,
and other higher resonances,\footnote{
  The broad structure that peaks near 
  $m(K^+K^-) \sim 1550 \ {\rm MeV}/c^2$ and was denoted $X_0(1550)$ 
  is now believed to originate from interference effects.
}
as well a contribution from $\chi_{c0}$.

The full time-dependent Dalitz plot analysis allows 
the complex amplitudes of each contributing term to be determined from data,
including $\CP$ violation effects
(\ie\ allowing the complex amplitude for the $\Bz$ decay to be independent
from that for $\Bzb$ decay), although one amplitude must be fixed 
to give a reference point.
There are several choices for parametrisation of the complex amplitudes 
(\eg\ real and imaginary part, or magnitude and phase).
Similarly, there are various approaches to include $\CP$ violation effects.
Note that positive definite parameters such as magnitudes are
disfavoured in certain circumstances 
(they inevitably lead to biases for small values).
In order to compare results between analyses,
it is useful for each experiment to present results in terms of the 
parameters that can be measured in a Q2B analysis
(such as $\Acp_{f}$, $S_f$, $C_f$, 
$\sin(2\beta^{\rm eff})$, $\cos(2\beta^{\rm eff})$, \etc)

In the \babar\ analysis of $\Bz \to K^+K^-\Kz$~\cite{Lees:2012kx},
the complex amplitude for each resonant contribution is written as
\begin{equation}
  A_f = c_f ( 1 + b_f ) e^{i ( \phi_f + \delta_f )} 
  \ , \ \ \ \ 
  \bar{A}_f = c_f ( 1 - b_f ) e^{i ( \phi_f - \delta_f )} \, ,
\end{equation}
where $b_f$ and $\delta_f$ introduce $\CP$ violation in the magnitude 
and phase respectively.
Belle~\cite{Nakahama:2010nj} use the same parametrisation but with a different notation for the parameters.\footnote{
  $(c, b, \phi, \delta) \leftrightarrow (a, c, b, d)$.
}
[The weak phase in $B^0$--$\bar{B}^0$ mixing ($2\beta$) also appears 
in the full formula for the time-dependent decay distribution.]
The Q2B direct $\CP$ violation parameter is directly related to $b_f$
\begin{equation}
  \Acp_{f} = \frac{-2b_f}{1+b_f^2} \approx C_f \, ,
\end{equation}
and the mixing-induced $\CP$ violation parameter can be used to obtain
$\sin(2\beta^{\rm eff})$
\begin{equation}
  -\eta_f S_f \approx \frac{1-b_f^2}{1+b_f^2}\sin(2\beta^{\rm eff}_f) \, ,
\end{equation}
where the approximations are exact in the case that $\left| q/p \right| = 1$.

Both \babar~\cite{Lees:2012kx} and \belle~\cite{Nakahama:2010nj} present results for $c_f$ and $\phi_f$,
for each resonant contribution,
and in addition present results for $\Acp_{f}$ and $\beta^{\rm eff}_{f}$ for $\phi(1020) \Kz$, $f_0(980) \Kz$ and for the remainder of the contributions to the $K^+K^-\Kz$ Dalitz plot combined.\footnote{
  \babar also present results for the Q2B parameter $S_{f}$ for these channels.
}
The models used to describe the resonant structure of the Dalitz plot differ, however.  Both analyses suffer from multiple solutions, from which we select only one for averaging.

\mysubsubsubsection{$\Bz \to \pi^+\pi^-\KS$
}
\label{sec:cp_uta:notations:dalitz:pipik0}

Studies of $\Bz \to \pi^+\pi^-\KS$~\cite{Aubert:2009me,:2008wwa}
and of the related decay
$\Bp \to \pi^+\pi^-K^+$~\cite{Garmash:2004wa,Garmash:2005rv,Aubert:2005ce,Aubert:2008bj}
show that the decay is dominated by components from intermediate resonances 
in the $K\pi$ ($K^*(892)$, $K^*_0(1430)$) 
and $\pi\pi$ ($\rho(770)$, $f_0(980)$, $f_2(1270)$) spectra,
together with a poorly understood scalar structure that peaks near 
$m(\pi\pi) \sim 1300 \ {\rm MeV}/c^2$ and is denoted $f_X(1300)$
(that could be identified as either the $f_0(1370)$ or $f_0(1500)$),
and a large nonresonant component.
There is also a contribution from the $\chi_{c0}$ state.

The full time-dependent Dalitz plot analysis allows 
the complex amplitudes of each contributing term to be determined from data,
including $\CP$ violation effects.
In the \babar\ analysis~\cite{Aubert:2009me}, 
the magnitude and phase of each component (for both $\Bz$ and $\Bzb$ decays) 
are measured relative to $\Bz \to f_0(980)\KS$, using the following
parametrisation
\begin{equation}
  A_f = \left| A_f \right| e^{i\,{\rm arg}(A_f)}
  \ , \ \ \ \ 
  \bar{A}_f = \left| \bar{A}_f \right| e^{i\,{\rm arg}(\bar{A}_f)} \, .
\end{equation}
In the \belle\ analysis~\cite{:2008wwa}, the $\Bz \to K^{*+}\pi^-$ amplitude
is chosen as the reference, and the amplitudes are parametrised as 
\begin{equation}
  A_f = a_f ( 1 + c_f ) e^{i ( b_f + d_f )} 
  \ , \ \ \ \ 
  \bar{A}_f = a_f ( 1 - c_f ) e^{i ( b_f - d_f )} \, .
\end{equation}
In both cases, the results are translated into quasi-two-body parameters 
such as $2\beta^{\rm eff}_f$, $S_f$, $C_f$ for each \CP\ eigenstate $f$,
and direct \CP\ asymmetries for each flavour-specific state.
Relative phase differences between resonant terms are also extracted.

\mysubsubsubsection{$\Bz \to \pi^+\pi^-\pi^0$
}
\label{sec:cp_uta:notations:dalitz:pipipi0}

The $\Bz \to \pi^+\pi^-\pi^0$ decay is dominated by 
intermediate $\rho$ resonances.
Though it is possible, as above, 
to determine directly the complex amplitudes for each component,
an alternative approach~\cite{Snyder:1993mx,Quinn:2000by},
has been used by both \babar~\cite{Aubert:2007jn}
and \belle~\cite{Kusaka:2007dv,:2007mj}.
The amplitudes for $\Bz$ and $\Bzb$ to $\pi^+\pi^-\pi^0$ are written
\begin{equation}
  A_{3\pi} = f_+ A_+ + f_- A_- + f_0 A_0
  \ , \ \ \ 
  \bar{A}_{3\pi} = f_+ \bar{A}_+ + f_- \bar{A}_- + f_0 \bar{A}_0
\end{equation}
respectively.
$A_+$, $A_-$ and $A_0$
represent the complex decay amplitudes for 
$\Bz \to \rho^+\pi^-$, $\Bz \to \rho^-\pi^+$ and $\Bz \to \rho^0\pi^0$
while 
$\bar{A}_+$, $\bar{A}_-$ and $\bar{A}_0$
represent those for 
$\Bzb \to \rho^+\pi^-$, $\Bzb \to \rho^-\pi^+$ and $\Bzb \to \rho^0\pi^0$
respectively.
$f_+$, $f_-$ and $f_0$ incorporate kinematic and dynamical factors
and depend on the Dalitz plot coordinates.
The full time-dependent decay distribution can then be written 
in terms of 27 free parameters,
one for each coefficient of the form factor bilinears,
as listed in Table~\ref{tab:cp_uta:pipipi0:uandi}.
These parameters are often referred to as ``the $U$s and $I$s'',
and can be expressed in terms of 
$A_+$, $A_-$, $A_0$, $\bar{A}_+$, $\bar{A}_-$ and $\bar{A}_0$.
If the full set of parameters is determined,
together with their correlations,
other parameters, such as weak and strong phases,
direct $\CP$ violation parameters, \etc, 
can be subsequently extracted.
Note that one of the parameters (typically $U_+^+$)
is often fixed to unity to provide a reference point;
this does not affect the analysis.


\begin{table}[htb]
  \begin{center}
    \caption{
      Definitions of the $U$ and $I$ coefficients.
      Modified from~\cite{Aubert:2007jn}.
    }
    \label{tab:cp_uta:pipipi0:uandi}
    \setlength{\tabcolsep}{0.3pc}
    \begin{tabular}{l@{\extracolsep{5mm}}l}
      \hline
      Parameter   & Description \\
      \hline
      $U_+^+$          & Coefficient of $|f_+|^2$ \\
      $U_0^+$          & Coefficient of $|f_0|^2$ \\
      $U_-^+$          & Coefficient of $|f_-|^2$ \\
      [0.15cm]
      $U_0^-$          & Coefficient of $|f_0|^2\cos(\Delta m\Delta t)$ \\
      $U_-^-$          & Coefficient of $|f_-|^2\cos(\Delta m\Delta t)$ \\
      $U_+^-$          & Coefficient of $|f_+|^2\cos(\Delta m\Delta t)$ \\
      [0.15cm]
      $I_0$            & Coefficient of $|f_0|^2\sin(\Delta m\Delta t)$ \\
      $I_-$            & Coefficient of $|f_-|^2\sin(\Delta m\Delta t)$ \\
      $I_+$            & Coefficient of $|f_+|^2\sin(\Delta m\Delta t)$ \\
      [0.15cm]
      $U_{+-}^{+,\Im}$ & Coefficient of $\Im[f_+f_-^*]$ \\
      $U_{+-}^{+,\Re}$ & Coefficient of $\Re[f_+f_-^*]$ \\
      $U_{+-}^{-,\Im}$ & Coefficient of $\Im[f_+f_-^*]\cos(\Delta m\Delta t)$ \\
      $U_{+-}^{-,\Re}$ & Coefficient of $\Re[f_+f_-^*]\cos(\Delta m\Delta t)$ \\
      $I_{+-}^{\Im}$   & Coefficient of $\Im[f_+f_-^*]\sin(\Delta m\Delta t)$ \\
      $I_{+-}^{\Re}$   & Coefficient of $\Re[f_+f_-^*]\sin(\Delta m\Delta t)$ \\
      [0.15cm]
      $U_{+0}^{+,\Im}$ & Coefficient of $\Im[f_+f_0^*]$ \\
      $U_{+0}^{+,\Re}$ & Coefficient of $\Re[f_+f_0^*]$ \\
      $U_{+0}^{-,\Im}$ & Coefficient of $\Im[f_+f_0^*]\cos(\Delta m\Delta t)$ \\
      $U_{+0}^{-,\Re}$ & Coefficient of $\Re[f_+f_0^*]\cos(\Delta m\Delta t)$ \\
      $I_{+0}^{\Im}$   & Coefficient of $\Im[f_+f_0^*]\sin(\Delta m\Delta t)$ \\
      $I_{+0}^{\Re}$   & Coefficient of $\Re[f_+f_0^*]\sin(\Delta m\Delta t)$ \\
      [0.15cm]
      $U_{-0}^{+,\Im}$ & Coefficient of $\Im[f_-f_0^*]$ \\
      $U_{-0}^{+,\Re}$ & Coefficient of $\Re[f_-f_0^*]$ \\
      $U_{-0}^{-,\Im}$ & Coefficient of $\Im[f_-f_0^*]\cos(\Delta m\Delta t)$ \\
      $U_{-0}^{-,\Re}$ & Coefficient of $\Re[f_-f_0^*]\cos(\Delta m\Delta t)$ \\
      $I_{-0}^{\Im}$   & Coefficient of $\Im[f_-f_0^*]\sin(\Delta m\Delta t)$ \\
      $I_{-0}^{\Re}$   & Coefficient of $\Re[f_-f_0^*]\sin(\Delta m\Delta t)$ \\     
      \hline
    \end{tabular}
  \end{center}
\end{table}

\mysubsubsection{Time-dependent \CP asymmetries in decays to non-$\CP$ eigenstates
}
\label{sec:cp_uta:notations:non_cp}

Consider a non-$\CP$ eigenstate $f$, and its conjugate $\bar{f}$. 
For neutral $\B$ decays to these final states,
there are four amplitudes to consider:
those for $\Bz$ to decay to $f$ and $\bar{f}$
($\Af$ and $\Afbar$, respectively),
and the equivalents for $\Bzb$
($\Abarf$ and $\Abarfbar$).
If $\CP$ is conserved in the decay, then
$\Af = \Abarfbar$ and $\Afbar = \Abarf$.


The time-dependent decay distributions can be written in many different ways.
Here, we follow Sec.~\ref{sec:cp_uta:notations:cp_eigenstate}
and define $\lambda_f = \frac{q}{p}\frac{\Abarf}{\Af}$ and
$\lambda_{\bar f} = \frac{q}{p}\frac{\Abarfbar}{\Afbar}$.
The time-dependent \CP asymmetries then follow Eq.~(\ref{eq:cp_uta:td_cp_asp}):
\begin{eqnarray}
\label{eq:cp_uta:non-cp-obs}
  {\cal A}_f (\Delta t) \; \equiv \;
  \frac{
    \Gamma_{\Bzb \to f} (\Delta t) - \Gamma_{\Bz \to f} (\Delta t)
  }{
    \Gamma_{\Bzb \to f} (\Delta t) + \Gamma_{\Bz \to f} (\Delta t)
  } & = & S_f \sin(\Delta m \Delta t) - C_f \cos(\Delta m \Delta t), \\
  {\cal A}_{\bar{f}} (\Delta t) \; \equiv \;
  \frac{
    \Gamma_{\Bzb \to \bar{f}} (\Delta t) - \Gamma_{\Bz \to \bar{f}} (\Delta t)
  }{
    \Gamma_{\Bzb \to \bar{f}} (\Delta t) + \Gamma_{\Bz \to \bar{f}} (\Delta t)
  } & = & S_{\bar{f}} \sin(\Delta m \Delta t) - C_{\bar{f}} \cos(\Delta m \Delta t),
\end{eqnarray}
with the definitions of the parameters 
$C_f$, $S_f$, $C_{\bar{f}}$ and $S_{\bar{f}}$,
following Eqs.~(\ref{eq:cp_uta:s_def}) and~(\ref{eq:cp_uta:c_def}).

The time-dependent decay rates are given by
\begin{eqnarray}
  \Gamma_{\Bzb \to f} (\Delta t) & = &
  \frac{e^{-\left| \Delta t \right| / \tau(\Bz)}}{8\tau(\Bz)} 
  ( 1 + \Adirnoncp ) 
  \left\{ 
    1 + S_f \sin(\Delta m \Delta t) - C_f \cos(\Delta m \Delta t) 
  \right\},
  \\
  \Gamma_{\Bz \to f} (\Delta t) & = &
  \frac{e^{-\left| \Delta t \right| / \tau(\Bz)}}{8\tau(\Bz)} 
  ( 1 + \Adirnoncp ) 
  \left\{ 
    1 - S_f \sin(\Delta m \Delta t) + C_f \cos(\Delta m \Delta t) 
  \right\},
  \\
  \Gamma_{\Bzb \to \bar{f}} (\Delta t) & = &
  \frac{e^{-\left| \Delta t \right| / \tau(\Bz)}}{8\tau(\Bz)} 
  ( 1 - \Adirnoncp ) 
  \left\{ 
    1 + S_{\bar{f}} \sin(\Delta m \Delta t) - C_{\bar{f}} \cos(\Delta m \Delta t) 
  \right\},
  \\
  \Gamma_{\Bz \to \bar{f}} (\Delta t) & = &
    \frac{e^{-\left| \Delta t \right| / \tau(\Bz)}}{8\tau(\Bz)} 
  ( 1 - \Adirnoncp ) 
  \left\{ 
    1 - S_{\bar{f}} \sin(\Delta m \Delta t) + C_{\bar{f}} \cos(\Delta m \Delta t) 
  \right\},
\end{eqnarray}
where the time-independent parameter \Adirnoncp
represents an overall asymmetry in the production of the 
$f$ and $\bar{f}$ final states,\footnote{
  This parameter is often denoted ${\cal A}_f$ (or ${\cal A}_{\CP}$),
  but here we avoid this notation to prevent confusion with the
  time-dependent $\CP$ asymmetry.
}
\begin{equation}
  \Adirnoncp = 
  \frac{
    \left( 
      \left| \Af \right|^2 + \left| \Abarf \right|^2
    \right) - 
    \left( 
      \left| \Afbar \right|^2 + \left| \Abarfbar \right|^2
    \right)
  }{
    \left( 
      \left| \Af \right|^2 + \left| \Abarf \right|^2
    \right) +
    \left( 
      \left| \Afbar \right|^2 + \left| \Abarfbar \right|^2
    \right)
  }.
\end{equation}
Assuming $|q/p| = 1$,
the parameters $C_f$ and $C_{\bar{f}}$
can also be written in terms of the decay amplitudes as follows:
\begin{equation}
  C_f = 
  \frac{
    \left| \Af \right|^2 - \left| \Abarf \right|^2 
  }{
    \left| \Af \right|^2 + \left| \Abarf \right|^2
  }
  \hspace{5mm}
  {\rm and}
  \hspace{5mm}
  C_{\bar{f}} = 
  \frac{
    \left| \Afbar \right|^2 - \left| \Abarfbar \right|^2
  }{
    \left| \Afbar \right|^2 + \left| \Abarfbar \right|^2
  },
\end{equation}
giving asymmetries in the decay amplitudes of $\Bz$ and $\Bzb$
to the final states $f$ and $\bar{f}$ respectively.
In this notation, the direct $\CP$ invariance conditions are
$\Adirnoncp = 0$ and $C_f = - C_{\bar{f}}$.
Note that $C_f$ and $C_{\bar{f}}$ are typically non-zero;
\eg, for a flavour-specific final state, 
$\Abarf = \Afbar = 0$ ($\Af = \Abarfbar = 0$), they take the values
$C_f = - C_{\bar{f}} = 1$ ($C_f = - C_{\bar{f}} = -1$).

The coefficients of the sine terms
contain information about the weak phase. 
In the case that each decay amplitude contains only a single weak phase
(\ie, no direct $\CP$ violation),
these terms can be written
\begin{equation}
  S_f = 
  \frac{ 
    - 2 \left| \Af \right| \left| \Abarf \right| 
    \sin( \phi_{\rm mix} + \phi_{\rm dec} - \delta_f )
  }{
    \left| \Af \right|^2 + \left| \Abarf \right|^2
  } 
  \hspace{5mm}
  {\rm and}
  \hspace{5mm}
  S_{\bar{f}} = 
  \frac{
    - 2 \left| \Afbar \right| \left| \Abarfbar \right| 
    \sin( \phi_{\rm mix} + \phi_{\rm dec} + \delta_f )
  }{
    \left| \Afbar \right|^2 + \left| \Abarfbar \right|^2
  },
\end{equation}
where $\delta_f$ is the strong phase difference between the decay amplitudes.
If there is no $\CP$ violation, the condition $S_f = - S_{\bar{f}}$ holds.
If decay amplitudes with different weak and strong phases contribute,
no clean interpretation of $S_f$ and $S_{\bar{f}}$ is possible.

Since two of the $\CP$ invariance conditions are 
$C_f = - C_{\bar{f}}$ and $S_f = - S_{\bar{f}}$,
there is motivation for a rotation of the parameters:
\begin{equation}
\label{eq:cp_uta:non-cp-s_and_deltas}
  S_{f\bar{f}} = \frac{S_{f} + S_{\bar{f}}}{2},
  \hspace{4mm}
  \Delta S_{f\bar{f}} = \frac{S_{f} - S_{\bar{f}}}{2},
  \hspace{4mm}
  C_{f\bar{f}} = \frac{C_{f} + C_{\bar{f}}}{2},
  \hspace{4mm}
  \Delta C_{f\bar{f}} = \frac{C_{f} - C_{\bar{f}}}{2}.
\end{equation}
With these parameters, the $\CP$ invariance conditions become
$S_{f\bar{f}} = 0$ and $C_{f\bar{f}} = 0$. 
The parameter $\Delta C_{f\bar{f}}$ gives a measure of the ``flavour-specificity''
of the decay:
$\Delta C_{f\bar{f}}=\pm1$ corresponds to a completely flavour-specific decay,
in which no interference between decays with and without mixing can occur,
while $\Delta C_{f\bar{f}} = 0$ results in 
maximum sensitivity to mixing-induced $\CP$ violation.
The parameter $\Delta S_{f\bar{f}}$ is related to the strong phase difference 
between the decay amplitudes of $\Bz$ to $f$ and to $\bar f$. 
We note that the observables of Eq.~(\ref{eq:cp_uta:non-cp-s_and_deltas})
exhibit experimental correlations 
(typically of $\sim 20\%$, depending on the tagging purity, and other effects)
between $S_{f\bar{f}}$ and  $\Delta S_{f\bar{f}}$, 
and between $C_{f\bar{f}}$ and $\Delta C_{f\bar{f}}$. 
On the other hand, 
the final state specific observables of Eq.~(\ref{eq:cp_uta:non-cp-obs})
tend to have low correlations.

Alternatively, if we recall that the $\CP$ invariance
conditions at the decay amplitude level are
$\Af = \Abarfbar$ and $\Afbar = \Abarf$,
we are led to consider the parameters~\cite{Charles:2004jd}
\begin{equation}
  \label{eq:cp_uta:non-cp-directcp}
  {\cal A}_{f\bar{f}} = 
  \frac{
    \left| \Abarfbar \right|^2 - \left| \Af \right|^2 
  }{
    \left| \Abarfbar \right|^2 + \left| \Af \right|^2
  }
  \hspace{5mm}
  {\rm and}
  \hspace{5mm}
  {\cal A}_{\bar{f}f} = 
  \frac{
    \left| \Abarf \right|^2 - \left| \Afbar \right|^2
  }{
    \left| \Abarf \right|^2 + \left| \Afbar \right|^2
  }.
\end{equation}
These are sometimes considered more physically intuitive parameters
since they characterise direct $\CP$ violation 
in decays with particular topologies.
For example, in the case of $\Bz \to \rho^\pm\pi^\mp$
(choosing $f =  \rho^+\pi^-$ and $\bar{f} = \rho^-\pi^+$),
${\cal A}_{f\bar{f}}$ (also denoted ${\cal A}^{+-}_{\rho\pi}$)
parametrises direct $\CP$ violation
in decays in which the produced $\rho$ meson does not contain the 
spectator quark,
while ${\cal A}_{\bar{f}f}$ (also denoted ${\cal A}^{-+}_{\rho\pi}$)
parametrises direct $\CP$ violation 
in decays in which it does.
Note that we have again followed the sign convention that the asymmetry 
is the difference between the rate involving a $b$ quark and that
involving a $\bar{b}$ quark, \cf\ Eq.~(\ref{eq:cp_uta:pra}). 
Of course, these parameters are not independent of the 
other sets of parameters given above, and can be written
\begin{equation}
  {\cal A}_{f\bar{f}} =
  - \frac{
    \Adirnoncp + C_{f\bar{f}} + \Adirnoncp \Delta C_{f\bar{f}} 
  }{
    1 + \Delta C_{f\bar{f}} + \Adirnoncp C_{f\bar{f}} 
  }
  \hspace{5mm}
  {\rm and}
  \hspace{5mm}
  {\cal A}_{\bar{f}f} =
  \frac{
    - \Adirnoncp + C_{f\bar{f}} + \Adirnoncp \Delta C_{f\bar{f}} 
  }{
    - 1 + \Delta C_{f\bar{f}} + \Adirnoncp C_{f\bar{f}}  
  }.
\end{equation}
They usually exhibit strong correlations.

We now consider the various notations which have been used 
in experimental studies of
time-dependent $\CP$ asymmetries in decays to non-$\CP$ eigenstates.

\mysubsubsubsection{$\Bz \to D^{*\pm}D^\mp$
}
\label{sec:cp_uta:notations:non_cp:dstard}

The ($\Adirnoncp$, $C_f$, $S_f$, $C_{\bar{f}}$, $S_{\bar{f}}$),
set of parameters was used in early publications by both \babar~\cite{Aubert:2007pa} and \belle~\cite{Aushev:2004uc} (albeit with slightly different notations) in the $D^{*\pm}D^{\mp}$ system ($f = D^{*+}D^-$, $\bar{f} = D^{*-}D^+$).
In their most recent paper on this topic \belle~\cite{Rohrken:2012ta} instead used the parametrisation ($A_{D^*D}$, $S_{D^*D}$, $\Delta S_{D^*D}$, $C_{D^*D}$, $\Delta C_{D^*D}$), while \babar~\cite{Aubert:2008ah} give results in both sets of parameters.
We therefore use the ($A_{D^*D}$, $S_{D^*D}$, $\Delta S_{D^*D}$, $C_{D^*D}$, $\Delta C_{D^*D}$) set.

\mysubsubsubsection{$\Bz \to \rho^{\pm}\pi^\mp$
}
\label{sec:cp_uta:notations:non_cp:rhopi}

In the $\rho^\pm\pi^\mp$ system, the 
($\Adirnoncp$, $C_{f\bar{f}}$, $S_{f\bar{f}}$, $\Delta C_{f\bar{f}}$, 
$\Delta S_{f\bar{f}}$)
set of parameters has been used 
originally by \babar~\cite{Aubert:2003wr} and \belle~\cite{Wang:2004va}, 
in the Q2B approximation; 
the exact names\footnote{
  \babar\ has used the notations
  $A_{\CP}^{\rho\pi}$~\cite{Aubert:2003wr} and 
  ${\cal A}_{\rho\pi}$~\cite{Aubert:2007jn}
  in place of ${\cal A}_{\CP}^{\rho\pi}$.
}
used in this case are
$\left( 
  {\cal A}_{\CP}^{\rho\pi}, C_{\rho\pi}, S_{\rho\pi}, \Delta C_{\rho\pi}, \Delta S_{\rho\pi}
\right)$,
and these names are also used in this document.

Since $\rho^\pm\pi^\mp$ is reconstructed in the final state $\pi^+\pi^-\pi^0$,
the interference between the $\rho$ resonances
can provide additional information about the phases 
(see Sec.~\ref{sec:cp_uta:notations:dalitz}).
Both \babar~\cite{Aubert:2007jn} 
and \belle~\cite{Kusaka:2007dv,:2007mj}
have performed time-dependent Dalitz plot analyses, 
from which the weak phase $\alpha$ is directly extracted.
In such an analysis, the measured Q2B parameters are 
also naturally corrected for interference effects.
See Sec.~\ref{sec:cp_uta:notations:dalitz:pipipi0}.

\mysubsubsubsection{$\Bz \to D^{\pm}\pi^{\mp}, D^{*\pm}\pi^{\mp}, D^{\pm}\rho^{\mp}$
}
\label{sec:cp_uta:notations:non_cp:dstarpi}

Time-dependent $\CP$ analyses have also been performed for the
final states $D^{\pm}\pi^{\mp}$, $D^{*\pm}\pi^{\mp}$ and $D^{\pm}\rho^{\mp}$.
In these theoretically clean cases, no penguin contributions are possible,
so there is no direct $\CP$ violation.
Furthermore, due to the smallness of the ratio of the magnitudes of the 
suppressed ($b \to u$) and favoured ($b \to c$) amplitudes (denoted $R_f$),
to a very good approximation, $C_f = - C_{\bar{f}} = 1$
(using $f = D^{(*)-}h^+$, $\bar{f} = D^{(*)+}h^-$ $h = \pi,\rho$),
and the coefficients of the sine terms are given by
\begin{equation}
  S_f = - 2 R_f \sin( \phi_{\rm mix} + \phi_{\rm dec} - \delta_f )
  \hspace{5mm}
  {\rm and}
  \hspace{5mm}
  S_{\bar{f}} = - 2 R_f \sin( \phi_{\rm mix} + \phi_{\rm dec} + \delta_f ).
\end{equation}
Thus weak phase information can be cleanly obtained from measurements
of $S_f$ and $S_{\bar{f}}$, 
although external information on at least one of $R_f$ or $\delta_f$ is necessary.
(Note that $\phi_{\rm mix} + \phi_{\rm dec} = 2\beta + \gamma$ for all the decay modes 
in question, while $R_f$ and $\delta_f$ depend on the decay mode.)

Again, different notations have been used in the literature.
\babar~\cite{Aubert:2006tw,Aubert:2005yf}
defines the time-dependent probability function by
\begin{equation}
  f^\pm (\eta, \Delta t) = \frac{e^{-|\Delta t|/\tau}}{4\tau} 
  \left[  
    1 \mp S_\zeta \sin (\Delta m \Delta t) \mp \eta C_\zeta \cos(\Delta m \Delta t) 
  \right],
\end{equation} 
where the upper (lower) sign corresponds to 
the tagging meson being a $\Bz$ ($\Bzb$). 
[Note here that a tagging $\Bz$ ($\Bzb$) corresponds to $-S_\zeta$ ($+S_\zeta$).]
The parameters $\eta$ and $\zeta$ take the values $+1$ and $+$ ($-1$ and $-$) 
when the final state is, \eg, $D^-\pi^+$ ($D^+\pi^-$). 
However, in the fit, the substitutions $C_\zeta = 1$ and 
$S_\zeta = a \mp \eta b_i - \eta c_i$ are made.\footnote{
  The subscript $i$ denotes tagging category.
}
[Note that, neglecting $b$ terms, $S_+ = a - c$ and $S_- = a + c$, 
so that $a = (S_+ + S_-)/2$, $c = (S_- - S_+)/2$, in analogy to 
the parameters of Eq.~(\ref{eq:cp_uta:non-cp-s_and_deltas}).] 
The subscript $i$ denotes the tagging category. 
These are motivated by the possibility of 
$\CP$ violation on the tag side~\cite{Long:2003wq}, 
which is absent for semileptonic $\B$ decays (mostly lepton tags). 
The parameter $a$ is not affected by tag side $\CP$ violation. 
The parameter $b$ only depends on tag side $\CP$ violation parameters 
and is not directly useful for determining UT angles.
A clean interpretation of the $c$ parameter is only possible for 
lepton-tagged events,
so the \babar\ measurements report $c$ measured with those events only.

The parameters used by \belle\ in the analysis using 
partially reconstructed $\B$ decays~\cite{Bahinipati:2011yq}, 
are similar to the $S_\zeta$ parameters defined above. 
However, in the \belle\ convention, 
a tagging $\Bz$ corresponds to a $+$ sign in front of the sine coefficient; 
furthermore the correspondence between the super/subscript 
and the final state is opposite, so that $S_\pm$ (\babar) = $- S^\mp$ (\belle). 
In this analysis, only lepton tags are used, 
so there is no effect from tag side $\CP$ violation. 
In the \belle\ analysis using 
fully reconstructed $\B$ decays~\cite{Ronga:2006hv}, 
this effect is measured and taken into account using $\Dstar l \nu$ decays; 
in neither \belle\ analysis are the $a$, $b$ and $c$ parameters used. 
In the latter case, the measured parameters are 
$2 R_{D^{(*)}\pi} \sin( 2\phi_1 + \phi_3 \pm \delta_{D^{(*)}\pi} )$; 
the definition is such that 
$S^\pm$ (\belle) = $- 2 R_{\Dstar \pi} \sin( 2\phi_1 + \phi_3 \pm \delta_{\Dstar \pi} )$. 
However, the definition includes an 
angular momentum factor $(-1)^L$~\cite{Fleischer:2003yb}, 
and so for the results in the $D\pi$ system, 
there is an additional factor of $-1$ in the conversion.

Explicitly, the conversion then reads as given in 
Table~\ref{tab:cp_uta:notations:non_cp:dstarpi}, 
where we have neglected the $b_i$ terms used by \babar
(which are zero in the absence of tag side $\CP$ violation).
For the averages in this document,
we use the $a$ and $c$ parameters,
and give the explicit translations used in 
Table~\ref{tab:cp_uta:notations:non_cp:dstarpi2}.
It is to be fervently hoped that the experiments will
converge on a common notation in future.

\begin{table}
  \begin{center} 
    \caption{
      Conversion between the various notations used for 
      $\CP$ violation parameters in the 
      $D^{\pm}\pi^{\mp}$, $D^{*\pm}\pi^{\mp}$ and $D^{\pm}\rho^{\mp}$ systems.
      The $b_i$ terms used by \babar\ have been neglected.
      Recall that $\left( \alpha, \beta, \gamma \right) = \left( \phi_2, \phi_1, \phi_3 \right)$.
    }
    \vspace{0.2cm}
    \setlength{\tabcolsep}{0.0pc}
    \begin{tabular*}{\textwidth}{@{\extracolsep{\fill}}cccc} \hline 
      & \babar\ & \belle\ partial rec. & \belle\ full rec. \\
      \hline
      $S_{D^+\pi^-}$    & $- S_- = - (a + c_i)$ &  N/A  &
      $2 R_{D\pi} \sin( 2\phi_1 + \phi_3 + \delta_{D\pi} )$ \\
      $S_{D^-\pi^+}$    & $- S_+ = - (a - c_i)$ &  N/A  &
      $2 R_{D\pi} \sin( 2\phi_1 + \phi_3 - \delta_{D\pi} )$ \\
      $S_{D^{*+}\pi^-}$ & $- S_- = - (a + c_i)$ & $S^+$ &   
      $- 2 R_{\Dstar \pi} \sin( 2\phi_1 + \phi_3 + \delta_{\Dstar \pi} )$ \\
      $S_{D^{*-}\pi^+}$ & $- S_+ = - (a - c_i)$ & $S^-$ &
      $- 2 R_{\Dstar \pi} \sin( 2\phi_1 + \phi_3 - \delta_{\Dstar \pi} )$ \\
      $S_{D^+\rho^-}$    & $- S_- = - (a + c_i)$ &  N/A  &  N/A  \\
      $S_{D^-\rho^+}$    & $- S_+ = - (a - c_i)$ &  N/A  &  N/A  \\
      \hline 
    \end{tabular*}
    \label{tab:cp_uta:notations:non_cp:dstarpi}
  \end{center}
\end{table}
   
\begin{table}
  \begin{center} 
    \caption{
      Translations used to convert the parameters measured by \belle
      to the parameters used for averaging in this document.
      The angular momentum factor $L$ is $-1$ for $\Dstar\pi$ and $+1$ for $D\pi$.
      Recall that $\left( \alpha, \beta, \gamma \right) = \left( \phi_2, \phi_1, \phi_3 \right)$.
    }
    \vspace{0.2cm}
    \setlength{\tabcolsep}{0.0pc}
    \begin{tabular*}{\textwidth}{@{\extracolsep{\fill}}ccc} \hline 
        & $\Dstar\pi$ partial rec. & $D^{(*)}\pi$ full rec. \\
        \hline
        $a$ & $- (S^+ + S^-)$ &
        $\frac{1}{2} (-1)^{L+1}
        \left(
          2 R_{D^{(*)}\pi} \sin( 2\phi_1 + \phi_3 + \delta_{D^{(*)}\pi} ) + 
          2 R_{D^{(*)}\pi} \sin( 2\phi_1 + \phi_3 - \delta_{D^{(*)}\pi} )
        \right)$ \\
        $c$ & $- (S^+ - S^-)$ & 
        $\frac{1}{2} (-1)^{L+1}
        \left(
          2 R_{D^{(*)}\pi} \sin( 2\phi_1 + \phi_3 + \delta_{D^{(*)}\pi} ) -
          2 R_{D^{(*)}\pi} \sin( 2\phi_1 + \phi_3 - \delta_{D^{(*)}\pi} )
        \right)$ \\
        \hline 
      \end{tabular*}
    \label{tab:cp_uta:notations:non_cp:dstarpi2}
  \end{center}
\end{table}

\mysubsubsubsection{Time-dependent asymmetries in radiative $\B$ decays
}
\label{sec:cp_uta:notations:non_cp:radiative}

As a special case of decays to non-$\CP$ eigenstates,
let us consider radiative $\B$ decays.
Here, the emitted photon has a distinct helicity,
which is in principle observable, but in practise is not usually measured.
Thus the measured time-dependent decay rates 
are given by~\cite{Atwood:1997zr,Atwood:2004jj}
\begin{eqnarray}
  \Gamma_{\Bzb \to X \gamma} (\Delta t) & = &
  \Gamma_{\Bzb \to X \gamma_L} (\Delta t) + \Gamma_{\Bzb \to X \gamma_R} (\Delta t) \\ \nonumber
  & = &
  \frac{e^{-\left| \Delta t \right| / \tau(\Bz)}}{4\tau(\Bz)} 
  \left\{ 
    1 + 
    \left( S_L + S_R \right) \sin(\Delta m \Delta t) - 
    \left( C_L + C_R \right) \cos(\Delta m \Delta t) 
  \right\},
  \\
  \Gamma_{\Bz \to X \gamma} (\Delta t) & = & 
  \Gamma_{\Bz \to X \gamma_L} (\Delta t) + \Gamma_{\Bz \to X \gamma_R} (\Delta t) \\ \nonumber 
  & = &
  \frac{e^{-\left| \Delta t \right| / \tau(\Bz)}}{4\tau(\Bz)} 
  \left\{ 
    1 - 
    \left( S_L + S_R \right) \sin(\Delta m \Delta t) + 
    \left( C_L + C_R \right) \cos(\Delta m \Delta t) 
  \right\},
\end{eqnarray}
where in place of the subscripts $f$ and $\bar{f}$ we have used $L$ and $R$
to indicate the photon helicity.
In order for interference between decays with and without $\Bz$-$\Bzb$ mixing
to occur, the $X$ system must not be flavour-specific,
\eg, in case of $\Bz \to K^{*0}\gamma$, the final state must be $\KS \pi^0 \gamma$.
The sign of the sine term depends on the $C$ eigenvalue of the $X$ system.
At leading order, the photons from 
$b \to q \gamma$ ($\bar{b} \to \bar{q} \gamma$) are predominantly
left (right) polarised, with corrections of order of $m_q/m_b$,
thus interference effects are suppressed.
Higher order effects can lead to corrections of order 
$\Lambda_{\rm QCD}/m_b$~\cite{Grinstein:2004uu,Grinstein:2005nu},
though explicit calculations indicate such corrections are small
for exclusive final states~\cite{Matsumori:2005ax,Ball:2006cva}.
The predicted smallness of the $S$ terms in the Standard Model
results in sensitivity to new physics contributions.

\mysubsubsection{Asymmetries in $\B \to \DorDstar K^{(*)}$ decays
}
\label{sec:cp_uta:notations:cus}

$\CP$ asymmetries in $\B \to \DorDstar K^{(*)}$ decays are sensitive to $\gamma$.
The neutral $D^{(*)}$ meson produced 
is an admixture of $\DorDstarz$ (produced by a $b \to c$ transition) and 
$\DorDstarzb$ (produced by a colour-suppressed $b \to u$ transition) states.
If the final state is chosen so that both $\DorDstarz$ and $\DorDstarzb$ 
can contribute, the two amplitudes interfere,
and the resulting observables are sensitive to $\gamma$, 
the relative weak phase between 
the two $\B$ decay amplitudes~\cite{Bigi:1988ym}.
Various methods have been proposed to exploit this interference,
including those where the neutral $D$ meson is reconstructed 
as a $\CP$ eigenstate (GLW)~\cite{Gronau:1990ra,Gronau:1991dp},
in a suppressed final state (ADS)~\cite{Atwood:1996ci,Atwood:2000ck},
or in a self-conjugate three-body final state, 
such as $\KS \pi^+\pi^-$ (Dalitz)~\cite{Giri:2003ty,Poluektov:2004mf}.
It should be emphasised that while each method 
differs in the choice of $D$ decay,
they are all sensitive to the same parameters of the $B$ decay,
and can be considered as variations of the same technique.

Consider the case of $\Bmp \to D \Kmp$,
with $D$ decaying to a final state $f$,
which is accessible to both $\Dz$ and $\Dzb$.
We can write the decay rates for $\Bm$ and $\Bp$ ($\Gamma_\mp$), 
the charge averaged rate ($\Gamma = (\Gamma_- + \Gamma_+)/2$)
and the charge asymmetry 
(${\cal A} = (\Gamma_- - \Gamma_+)/(\Gamma_- + \Gamma_+)$, see Eq.~(\ref{eq:cp_uta:pra})) as 
\begin{eqnarray}
  \label{eq:cp_uta:dk:rate_def}
  \Gamma_\mp  & \propto & 
  r_B^2 + r_D^2 + 2 r_B r_D \cos \left( \delta_B + \delta_D \mp \gamma \right), \\
  \label{eq:cp_uta:dk:av_rate_def}
  \Gamma & \propto &  
  r_B^2 + r_D^2 + 2 r_B r_D \cos \left( \delta_B + \delta_D \right) \cos \left( \gamma \right), \\
  \label{eq:cp_uta:dk:acp_def}
  {\cal A} & = & 
  \frac{
    2 r_B r_D \sin \left( \delta_B + \delta_D \right) \sin \left( \gamma \right)
  }{
    r_B^2 + r_D^2 + 2 r_B r_D \cos \left( \delta_B + \delta_D \right) \cos \left( \gamma \right),  
  }
\end{eqnarray}
where the ratio of $\B$ decay amplitudes\footnote{
  Note that here we use the notation $r_B$ to denote the ratio
  of $\B$ decay amplitudes, 
  whereas in Sec.~\ref{sec:cp_uta:notations:non_cp:dstarpi} 
  we used, \eg, $R_{D\pi}$, for a rather similar quantity.
  The reason is that here we need to be concerned also with 
  $D$ decay amplitudes,
  and so it is convenient to use the subscript to denote the decaying particle.
  Hopefully, using $r$ in place of $R$ will help reduce potential confusion.
} 
is usually defined to be less than one,
\begin{equation}
  \label{eq:cp_uta:dk:rb_def}
  r_B = 
  \frac{
    \left| A\left( \Bm \to \Dzb K^- \right) \right|
  }{
    \left| A\left( \Bm \to \Dz  K^- \right) \right|
  },
\end{equation}
and the ratio of $D$ decay amplitudes is correspondingly defined by
\begin{equation}
  \label{eq:cp_uta:dk:rd_def}
  r_D = 
  \frac{
    \left| A\left( \Dz  \to f \right) \right|
  }{
    \left| A\left( \Dzb \to f \right) \right|
  }.
\end{equation}
The strong phase differences between the $\B$ and $D$ decay amplitudes 
are given by $\delta_B$ and $\delta_D$, respectively.
The values of $r_D$ and $\delta_D$ depend on the final state $f$:
for the GLW analysis, $r_D = 1$ and $\delta_D$ is trivial (either zero or $\pi$),
in the Dalitz plot analysis $r_D$ and $\delta_D$ vary across the Dalitz plot,
and depend on the $D$ decay model used,
for the ADS analysis, the values of $r_D$ and $\delta_D$ are not trivial.

Note that, for given values of $r_B$ and $r_D$, 
the maximum size of ${\cal A}$ (at $\sin \left( \delta_B + \delta_D \right) = 1$)
is $2 r_B r_D \sin \left( \gamma \right) / \left( r_B^2 + r_D^2 \right)$.
Thus even for $D$ decay modes with small $r_D$, 
large asymmetries, and hence sensitivity to $\gamma$, 
may occur for $B$ decay modes with similar values of $r_B$.
For this reason, the ADS analysis of the decay $B^\mp \to D \pi^\mp$ 
is also of interest.

In the GLW analysis, the measured quantities are the 
partial rate asymmetry, and the charge averaged rate,
which are measured both for $\CP$-even and $\CP$-odd $D$ decays.
The former is defined as 
\begin{equation}
  \label{eq:cp_uta:dk:glw-rdef}
  R_{\CP} = 
  \frac{2 \, \Gamma \left( \Bm \to D_{\CP} \Km  \right)}
  {\Gamma\left( \Bm \to \Dz \Km \right)} \, .
\end{equation}
It is experimentally convenient to measure $R_{\CP}$ using a double ratio,
\begin{equation}
  \label{eq:cp_uta:dk:double_ratio}
  R_{\CP} = 
  \frac{
    \Gamma\left( \Bm \to D_{\CP} \Km  \right) \, / \, \Gamma\left( \Bm \to \Dz \Km \right)
  }{
    \Gamma\left( \Bm \to D_{\CP} \pim \right) \, / \, \Gamma\left( \Bm \to \Dz \pim \right)
  }
\end{equation}
that is normalised both to the rate for the favoured $\Dz \to \Km\pip$ decay, 
and to the equivalent quantities for $\Bm \to D\pim$ decays
(charge conjugate modes are implicitly included in 
Eq.~(\ref{eq:cp_uta:dk:glw-rdef}) and~(\ref{eq:cp_uta:dk:double_ratio})).
In this way the constant of proportionality drops out of 
Eq.~(\ref{eq:cp_uta:dk:av_rate_def}).
Eq.~(\ref{eq:cp_uta:dk:double_ratio}) is exact in the limit that the
contribution of the $b \to u$ decay amplitude to $\Bm \to D \pim$ vanishes and
when the flavour-specific rates $\Gamma\left( \Bm \to \Dz h^- \right)$ ($h =
\pi,K$) are determined using appropriately flavour-specific $D$ decays.
In reality, the decay $D \to K\pi$ is invariable used, leading to a small source of systematic uncertainty.
The direct \CP\ asymmetry is defined as
\begin{equation}
  \label{eq:cp_uta:dk:glw-adef}
  A_{\CP} = \frac{
    \Gamma\left(\Bm\to D_{\CP}\Km\right) - \Gamma\left(\Bp\to D_{\CP}\Kp\right)
  }{
    \Gamma\left(\Bm\to D_{\CP}\Km\right) + \Gamma\left(\Bp\to D_{\CP}\Kp\right)
  } \, .
\end{equation}

For the ADS analysis, using a suppressed $D \to f$ decay,
the measured quantities are again the partial rate asymmetry, 
and the charge averaged rate.
In this case it is sufficient to measure the rate in a single ratio
(normalised to the favoured $D \to \bar{f}$ decay)
since detection systematics cancel naturally;
the observed quantity is then
\begin{equation}
  \label{eq:cp_uta:dk:r_ads}
  R_{\rm ADS} = 
  \frac{
    \Gamma \left( \Bm \to \left[ f \right]_D \Km \right)
  }{
    \Gamma \left( \Bm \to \left[ \bar{f} \right]_D \Km \right)
  } \, ,
\end{equation}
where inclusion of charge conjugate modes is implied.
The direct \CP\ asymmetry is defined as
\begin{equation}
  \label{eq:cp_uta:dk:a_ads}
  A_{\rm ADS} = 
  \frac{
    \Gamma\left(\Bm\to\left[f\right]_D\Km\right)-
    \Gamma\left(\Bp\to\left[f\right]_D\Kp\right)
  }{
    \Gamma\left(\Bm\to\left[f\right]_D\Km\right)+
    \Gamma\left(\Bp\to\left[f\right]_D\Kp\right)
  } \, .
\end{equation}
Since the uncertainty of $A_{\rm ADS}$ depends on the central value of $R_{\rm ADS}$, for some statistical treatments it is preferable to use an alternative pair of parameters (as discussed in Refs.~\cite{Bondar:2004bi,Rama:2010bw})
\begin{equation}
  R_- = \frac{
    \Gamma \left( \Bm \to \left[ f \right]_D \Km \right)
  }{
    \Gamma \left( \Bm \to \left[ \bar{f} \right]_D \Km \right)
  } \, 
  \hspace{5mm}
  R_+ = \frac{
    \Gamma \left( \Bp \to \left[ \bar{f} \right]_D \Kp \right)
  }{
    \Gamma \left( \Bp \to \left[ f \right]_D \Kp \right)
  } \, ,
\end{equation}
where there is no inclusion of charge conjugated processes.
We use the $(R_{\rm ADS}, A_{\rm ADS})$ set in our compilation.

In the ADS analysis, there are an additional two unknowns ($r_D$ and $\delta_D$)
compared to the GLW case.  
However, the value of $r_D$ can be measured using 
decays of $D$ mesons of known flavour, and $\delta_D$ can be measured from interference effects in decays of quantum-correlated $D\bar{D}$ pairs produced at the $\psi(3770)$ resonance.
In fact, the most precise information on both $r_D$ and $\delta_D$ currently comes from global fits on charm mixing parameters, as discussed in Sec.~\ref{sec:charm:mixcpv}.

In the Dalitz plot analysis,
once a model is assumed for the $D$ decay, 
which gives the values of $r_D$ and $\delta_D$ across the Dalitz plot,
it is possible to perform a simultaneous fit to the $B^+$ and $B^-$ samples 
and directly extract $\gamma$, $r_B$ and $\delta_B$.
However, the uncertainties on the phases depend approximately inversely on $r_B$.
Furthermore, $r_B$ is positive definite (and small), 
and therefore tends to be overestimated,
which can lead to an underestimation of the uncertainty.
Some statistical treatment is necessary to correct for this bias.
An alternative approach is to extract from the data the ``Cartesian''
variables
\begin{equation}
  \left( x_\pm, y_\pm \right) = 
  \left( \Re(r_B e^{i(\delta_B\pm\gamma)}), \Im(r_B e^{i(\delta_B\pm\gamma)}) \right) = 
  \left( r_B \cos(\delta_B\pm\gamma), r_B \sin(\delta_B\pm\gamma) \right).
\end{equation}
These are (a) approximately statistically uncorrelated 
and (b) almost Gaussian.
The pairs of variables $\left( x_\pm, y_\pm \right)$ can be extracted
from independent fits of the $B^\pm$ data samples.
Use of these variables makes the combination of results much simpler.

However, if the Dalitz plot is effectively dominated by one $CP$ state,
there will be additional sensitivity to $\gamma$ in the numbers of events
in the $B^\pm$ data samples.
This can be taken into account in various ways.
One possibility is to extract GLW-like variables 
in addition to the $\left( x_\pm, y_\pm \right)$ parameters.
An alternative proceeds by defining $z_\pm = x_\pm + i y_\pm$
and $x_0 = - \int \Re \left[ f(s_1,s_2)f^*(s_2,s_1) \right] ds_1ds_2$,
where $s_1, s_2$ are the coordinates of invariant mass squared that
define the Dalitz plot and $f$ is the complex amplitude for $D$ decay
as a function of the Dalitz plot coordinates.\footnote{
  The $x_0$ parameter is closely related to the $c_i$ parameters of 
  the model dependent Dalitz plot analysis~\cite{Giri:2003ty,Bondar:2005ki,Bondar:2008hh},
  and the coherence factor of inclusive ADS-type analyses~\cite{Atwood:2003mj},
  integrated over the entire Dalitz plot.
}
The fitted parameters ($\rho^\pm, \theta^\pm$) are then defined by
\begin{equation}
  \rho^\pm e^{i \theta^\pm} = z_\pm - x_0 \, .
\end{equation}
Note that the yields of $B^\pm$ decays are proportional 
to $1 + (\rho^\pm)^2 - (x_0)^2$. 
This choice of variables has been used by \babar\ in the analysis of
$\Bmp \to D\Kmp$ with $D \to \pi^+\pi^-\pi^0$~\cite{Aubert:2007ii};
for this $D$ decay, $x_0 = 0.850$.

The relations between the measured quantities and the
underlying parameters are summarised in Table~\ref{tab:cp_uta:notations:dk}.
Note carefully that the hadronic factors $r_B$ and $\delta_B$ 
are different, in general, for each $\B$ decay mode.

\begin{table}[htb]
  \begin{center} 
    \caption{
      Summary of relations between measured and physical parameters 
      in GLW, ADS and Dalitz analyses of $\B \to \DorDstar K^{(*)}$.
    }
    \vspace{0.2cm}
    \setlength{\tabcolsep}{1.0pc}
    \begin{tabular}{cc} \hline 
      \mc{2}{c}{GLW analysis} \\
      $R_{\CP\pm}$ & $1 + r_B^2 \pm 2 r_B \cos \left( \delta_B \right) \cos \left( \gamma \right)$ \\
      $A_{\CP\pm}$ & $\pm 2 r_B \sin \left( \delta_B \right) \sin \left( \gamma \right) / R_{\CP\pm}$ \\
      \hline
      \mc{2}{c}{ADS analysis} \\
      $R_{\rm ADS}$ & $r_B^2 + r_D^2 + 2 r_B r_D \cos \left( \delta_B + \delta_D \right) \cos \left( \gamma \right)$ \\
      $A_{\rm ADS}$ & $2 r_B r_D \sin \left( \delta_B + \delta_D \right) \sin \left( \gamma \right) / R_{\rm ADS}$ \\
      \hline
      \mc{2}{c}{Dalitz analysis ($D \to \KS \pi^+\pi^-$)} \\
      $x_\pm$ & $r_B \cos(\delta_B\pm\gamma)$ \\
      $y_\pm$ & $r_B \sin(\delta_B\pm\gamma)$ \\
      \hline
      \mc{2}{c}{Dalitz analysis ($D \to \pi^+\pi^-\pi^0$)} \\
      $\rho^\pm$ & $|z_\pm - x_0|$ \\
      $\theta^\pm$ & $\tan^{-1}(\Im(z_\pm)/(\Re(z_\pm) - x_0))$ \\
      \hline
    \end{tabular}
    \label{tab:cp_uta:notations:dk}
  \end{center}
\end{table}

Results from model-dependent Dalitz plot fits tend to suffer from significant uncertainties due to the choice of model to describe hadronic effects.
This can be obviated by a model-independent analysis, in which the Dalitz plot is binned~\cite{Giri:2003ty,Bondar:2005ki,Bondar:2008hh}.  It is then necessary to gain information on effective parameters which describe the average strong phase difference between a certain bin and its conjugate (found by reflecting in the symmetry axis of the Dalitz plot\footnote{Here we restrict the discussion to three-body self conjugate final states such as $K_S^0\pip\pim$ and $K_S^0\Kp\Km$, though it can be extended to other modes, including four-body final states.}).
Such information can be obtained from interference effects in decays of quantum-correlated $D\bar{D}$ pairs produced at the $\psi(3770)$ resonance.

\mysubsection{Common inputs and error treatment
}
\label{sec:cp_uta:common_inputs}

The common inputs used for rescaling are listed in 
Table~\ref{tab:cp_uta:common_inputs}.
The $\Bz$ lifetime ($\tau(\Bz)$), mixing parameter ($\Delta m_d$) and relative width difference ($\Delta\Gamma_d / \Gamma_d$)
averages are provided by the HFAG Lifetimes and Oscillations 
subgroup (Sec.~\ref{sec:life_mix}).
The fraction of the perpendicularly polarised component 
($\left| A_{\perp} \right|^2$) in $\B \to \jpsi \Kstar(892)$ decays,
which determines the $\CP$ composition in these decays, 
is averaged from results by 
\babar~\cite{Aubert:2007hz}, \belle~\cite{Itoh:2005ks}, CDF~\cite{Acosta:2004gt}, D0~\cite{Abazov:2008jz} and LHCb~\cite{LHCb-CONF-2011-002}.
See also HFAG $B$ to Charm Decay Parameters subgroup (Sec.~\ref{sec:b2c}).

At present, we only rescale to a common set of input parameters
for modes with reasonably small statistical errors
($b \to c\bar{c}s$ transitions).
Correlated systematic errors are taken into account
in these modes as well.
For all other modes, the effect of such a procedure is 
currently negligible.

\begin{table}[htb]
  \begin{center}
    \caption{
      Common inputs used in calculating the averages.
    }
    \vspace{0.2cm}
    \setlength{\tabcolsep}{1.0pc}
    \begin{tabular}{cc} \hline 
      $\tau(\Bz)$ $({\rm ps})$  & $1.519 \pm 0.007$  \\
      $\Delta m_d$ $({\rm ps}^{-1})$ & $0.507 \pm 0.004$ \\
      $\Delta\Gamma_d / \Gamma_d$ & $0.015 \pm 0.018$ \\
      $\left| A_{\perp} \right|^2 (\jpsi \Kstar)$ & $0.213 \pm 0.008$ \\
      \hline
    \end{tabular}
    \label{tab:cp_uta:common_inputs}
  \end{center}
\end{table}

As explained in Sec.~\ref{sec:intro},
we do not apply a rescaling factor on the error of an average
that has $\chi^2/\dof > 1$ 
(unlike the procedure currently used by the PDG~\cite{PDG_2010}).
We provide a confidence level of the fit so that
one can know the consistency of the measurements included in the average,
and attach comments in case some care needs to be taken in the interpretation.
Note that, in general, results obtained from data samples with low statistics
will exhibit some non-Gaussian behaviour.
We average measurements with asymmetric errors 
using the PDG~\cite{PDG_2010} prescription.
In cases where several measurements are correlated
(\eg\ $S_f$ and $C_f$ in measurements of time-dependent $\CP$ violation
in $B$ decays to a particular $\CP$ eigenstate)
we take these into account in the averaging procedure
if the uncertainties are sufficiently Gaussian.
For measurements where one error is given, 
it represents the total error, 
where statistical and systematic uncertainties have been added in quadrature.
If two errors are given, the first is statistical and the second systematic.
If more than two errors are given,
the origin of the additional uncertainty will be explained in the text.

\clearpage
\mysubsection{Time-dependent asymmetries in $b \to c\bar{c}s$ transitions
}
\label{sec:cp_uta:ccs}

\mysubsubsection{Time-dependent $\CP$ asymmetries in $b \to c\bar{c}s$ decays to $\CP$ eigenstates
}
\label{sec:cp_uta:ccs:cp_eigen}

\begin{figure}[htb]
  \begin{center}
    \resizebox{0.51\textwidth}{!}{
      \includegraphics{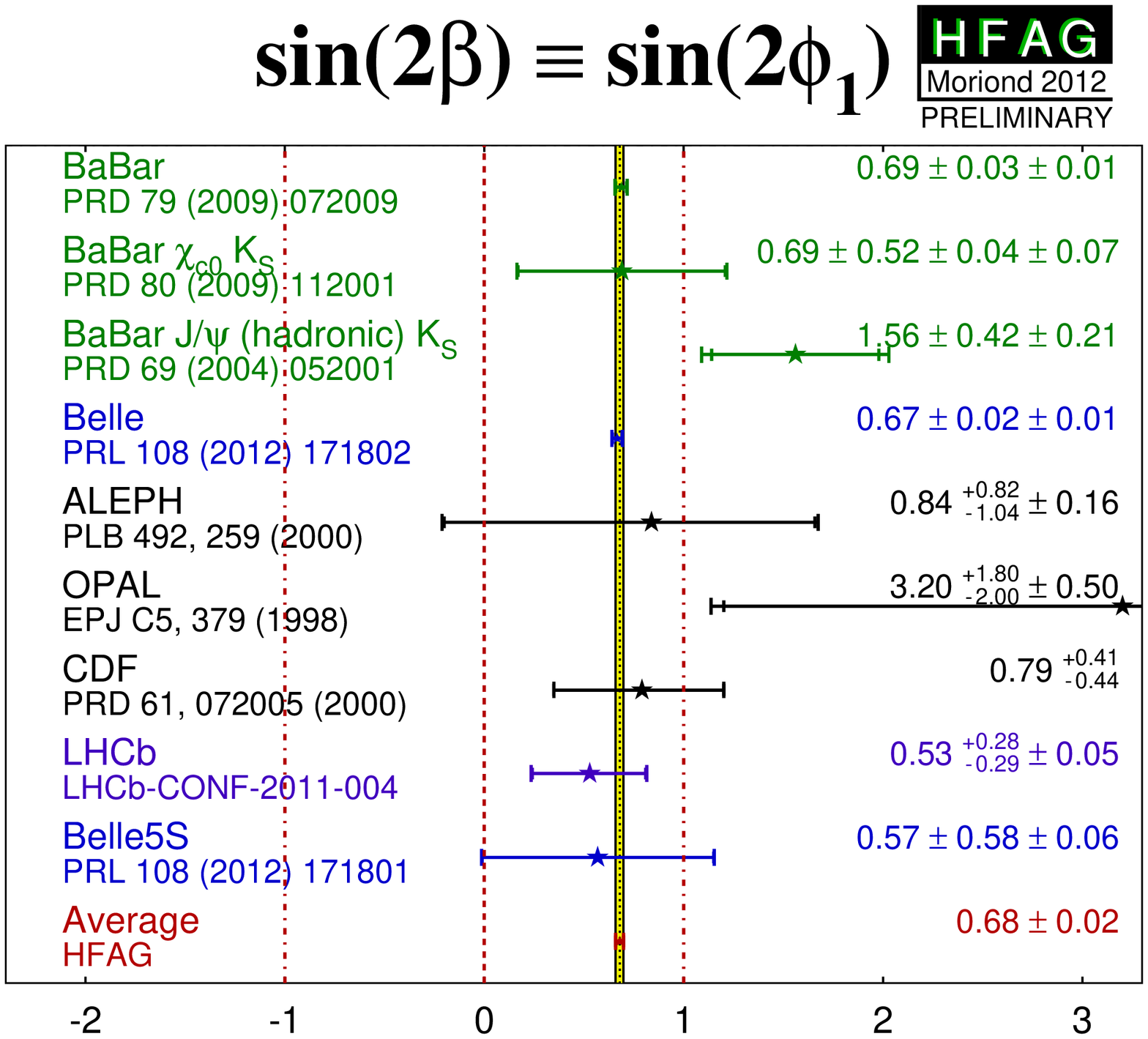}
    }
    \hfill
    \resizebox{0.48\textwidth}{!}{
      \includegraphics{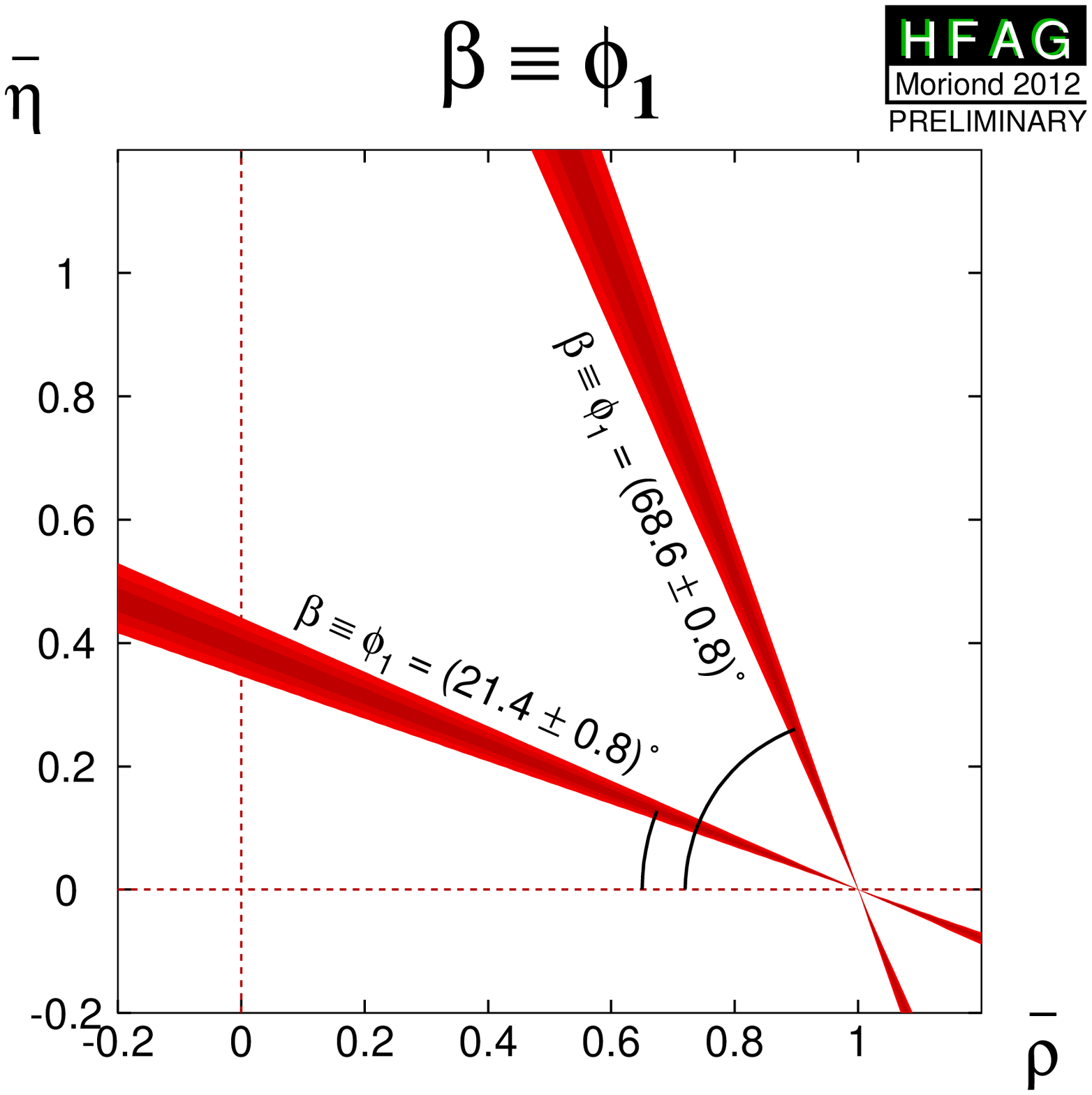}
    }
  \end{center}
  \vspace{-0.5cm}
  \caption{
    (Left) Average of measurements of $S_{b \to c\bar c s}$.
    (Right) Constraints on the $(\rhobar,\etabar)$ plane,
    obtained from the average of $-\etacp S_{b \to c\bar c s}$ 
    and Eq.~\ref{eq:cp_uta:sin2beta}.
  }
  \label{fig:cp_uta:ccs}
\end{figure}

In the Standard Model, the time-dependent parameters for
$b \to c\bar c s$ transitions are predicted to be: 
$S_{b \to c\bar c s} = - \etacp \sin(2\beta)$,
$C_{b \to c\bar c s} = 0$ to very good accuracy.
The averages for $-\etacp S_{b \to c\bar c s}$ and $C_{b \to c\bar c s}$
are provided in Table~\ref{tab:cp_uta:ccs}.
The averages for $-\etacp S_{b \to c\bar c s}$ 
are shown in Fig.~\ref{fig:cp_uta:ccs}.

Both \babar\  and \belle\ have used the $\etacp = -1$ modes
$\jpsi \KS$, $\psi(2S) \KS$, $\chi_{c1} \KS$ and $\eta_c \KS$, 
as well as $\jpsi \KL$, which has $\etacp = +1$
and $\jpsi K^{*0}(892)$, which is found to have $\etacp$ close to $+1$
based on the measurement of $\left| A_\perp \right|$ 
(see Sec.~\ref{sec:cp_uta:common_inputs}).
The most recent \belle\ result does not use $\eta_c \KS$ or $\jpsi K^{*0}(892)$.
ALEPH, OPAL, CDF and LHCb have used only the $\jpsi \KS$ final state.
\babar\ have also determined the \CP-violation parameters of the
$\Bz\to\chi_{c0} \KS$ decay from the time-dependent Dalitz plot analysis of
$\Bz \to \pi^+\pi^-\KS$ (see subsection~\ref{sec:cp_uta:qqs:dp}).
In addition, \belle\ have performed a measurement with data accumulated at the $\Upsilon(5S)$ resonance, using the $\jpsi\KS$ final state -- this involves a different flavour tagging method compared to the measurements performed with data accumulated at the $\Upsilon(4S)$ resonance.
A breakdown of results in each charmonium-kaon final state is given in 
Table~\ref{tab:cp_uta:ccs-BF}.

\begin{table}[htb]
	\begin{center}
		\caption{
                        $S_{b \to c\bar c s}$ and $C_{b \to c\bar c s}$.
                }
		\vspace{0.2cm}
		\setlength{\tabcolsep}{0.0pc}
		\begin{tabular*}{\textwidth}{@{\extracolsep{\fill}}lrccc} \hline
      \mc{2}{l}{Experiment} & Sample size & $- \etacp S_{b \to c\bar c s}$ & $C_{b \to c\bar c s}$ \\
      \hline
	\babar & \cite{:2009yr} & $N(B\bar{B})$ = 465M & $0.687 \pm 0.028 \pm 0.012$ & $0.024 \pm 0.020 \pm 0.016$ \\
	\babar\ $\chi_{c0} \KS$ & \cite{Aubert:2009me} & $N(B\bar{B})$ = 383M & $0.69 \pm 0.52 \pm 0.04 \pm 0.07$ & $-0.29 \,^{+0.53}_{-0.44} \pm 0.03 \pm 0.05$ \\
	\babar\ $J/\psi \KS$ ($^{*}$) & \cite{Aubert:2003xn} & $N(B\bar{B})$ = 88M & $1.56 \pm 0.42 \pm 0.21$ &  \textendash{} \\
	\belle & \cite{Adachi:2012et} & $N(B\bar{B})$ = 722M & $0.667 \pm 0.023 \pm 0.012$ & $-0.006 \pm 0.016 \pm 0.012$ \\
	\mc{3}{l}{\bf \boldmath $\B$ factory average} & $0.679 \pm 0.020$ & $0.005 \pm 0.017$ \\
	\mc{3}{l}{\small Confidence level} & {\small $0.28$} & {\small $0.47$} \\
        \hline
        ALEPH & \cite{Barate:2000tf} & \textendash{} & $0.84 \, ^{+0.82}_{-1.04} \pm 0.16$ &  \textendash{} \\
        OPAL  & \cite{Ackerstaff:1998xz} & \textendash{} & $3.2 \, ^{+1.8}_{-2.0} \pm 0.5$ &  \textendash{} \\
        CDF   & \cite{Affolder:1999gg} & \textendash{} & $0.79 \, ^{+0.41}_{-0.44}$ &  \textendash{} \\
	LHCb & \cite{LHCb-CONF-2011-004} & $0.035\ {\rm fb}^{-1}$ & $0.53 \,^{+0.28}_{-0.29} \pm 0.05$ &  \textendash{} \\
	Belle $\Upsilon(5S)$ & \cite{Sato:2012hu} & $121\ {\rm fb}^{-1}$ & $0.57 \pm 0.58 \pm 0.06$ &  \textendash{} \\
        \mc{3}{l}{\bf Average} & $0.679 \pm 0.020$ & $0.005 \pm 0.017$ \\
		\hline
		\end{tabular*}
                \label{tab:cp_uta:ccs}
        \end{center}
$^{*}$ {\small This result uses "{\it hadronic and previously unused muonic decays of the $J/\psi$}". We neglect a small possible correlation of this result with the main \babar\ result~\cite{:2009yr} that could be caused by reprocessing of the data.}
\end{table}

\begin{table}[htb]
	\begin{center}
		\caption{
                        Breakdown of $B$ factory results on $S_{b \to c\bar c s}$ and $C_{b \to c\bar c s}$.
                }
		\vspace{0.2cm}
		\setlength{\tabcolsep}{0.0pc}
		\begin{tabular*}{\textwidth}{@{\extracolsep{\fill}}lrccc} \hline
        \mc{2}{l}{Mode} & $N(B\bar{B})$ & $- \etacp S_{b \to c\bar c s}$ & $C_{b \to c\bar c s}$ \\
        \hline
        \mc{5}{c}{\babar} \\
        $J/\psi \KS$ & \cite{:2009yr} & 465M & $0.657 \pm 0.036 \pm 0.012$ & $0.026 \pm 0.025 \pm 0.016$ \\
        $J/\psi \KL$ & \cite{:2009yr} & 465M & $0.694 \pm 0.061 \pm 0.031$ & $-0.033 \pm 0.050 \pm 0.027$ \\
        {\bf \boldmath $J/\psi K^0$} & \cite{:2009yr} & 465M & $0.666 \pm 0.031 \pm 0.013$ & $0.016 \pm 0.023 \pm 0.018$ \\
        $\psi(2S) \KS$ & \cite{:2009yr} & 465M & $0.897 \pm 0.100 \pm 0.036$ & $0.089 \pm 0.076 \pm 0.020$ \\
        $\chi_{c1} \KS$ & \cite{:2009yr} & 465M & $0.614 \pm 0.160 \pm 0.040$ & $0.129 \pm 0.109 \pm 0.025$ \\
        $\eta_c \KS$ & \cite{:2009yr} & 465M & $0.925 \pm 0.160 \pm 0.057$ & $0.080 \pm 0.124 \pm 0.029$ \\
        $\jpsi K^{*0}(892)$ & \cite{:2009yr} & 465M & $0.601 \pm 0.239 \pm 0.087$ & $0.025 \pm 0.083 \pm 0.054$ \\
        {\bf All} & \cite{:2009yr} & 465M & $0.687 \pm 0.028 \pm 0.012$ & $0.024 \pm 0.020 \pm 0.016$ \\
	\hline
	\mc{5}{c}{\bf \belle} \\
        $J/\psi \KS$ & \cite{Adachi:2012et} & 722M & $0.670 \pm 0.029 \pm 0.013$ & $0.015 \pm 0.021 \,^{+0.023}_{-0.045}$ \\
        $J/\psi \KL$ & \cite{Adachi:2012et} & 722M & $0.642 \pm 0.047 \pm 0.021$ & $-0.019 \pm 0.026 \,^{+0.041}_{-0.017}$ \\
	$\psi(2{\rm S}) \KS$ & \cite{Adachi:2012et} & 722M & $0.738 \pm 0.079 \pm 0.036$ & $-0.104 \pm 0.055 \,^{+0.027}_{-0.047}$ \\
	$\chi_{c1} \KS$ & \cite{Adachi:2012et} & 722M & $0.640 \pm 0.117 \pm 0.040$ & $0.017 \pm 0.083 \,^{+0.026}_{-0.046}$ \\
        {\bf All} & \cite{Adachi:2012et} & 722M & $0.667 \pm 0.023 \pm 0.012$ & $-0.006 \pm 0.016 \pm 0.012$ \\
	\hline
	\mc{5}{c}{\bf Averages} \\
        \mc{3}{l}{$J/\psi \KS$} & $0.665 \pm 0.024$ & $0.024 \pm 0.026$ \\
        \mc{3}{l}{$J/\psi \KL$} & $0.663 \pm 0.041$ & $-0.023 \pm 0.030$ \\
        \mc{3}{l}{$\psi(2{\rm S}) \KS$} & $0.807 \pm 0.067$ & $-0.009 \pm 0.055$ \\
        \mc{3}{l}{$\chi_{c1} \KS$} & $0.632 \pm 0.099$ & $0.066 \pm 0.074$ \\
		\hline
		\end{tabular*}
                \label{tab:cp_uta:ccs-BF}
        \end{center}
\end{table}

It should be noted that, while the uncertainty in the average for 
$-\etacp S_{b \to c\bar c s}$ is still limited by statistics,
that for $C_{b \to c\bar c s}$ is close to being dominated by systematics.
This occurs due to the possible effect of tag side interference on the
$C_{b \to c\bar c s}$ measurement, an effect which is correlated between
the different experiments.
Understanding of this effect may continue to improve in future,
allowing the uncertainty to reduce.

From the average for $-\etacp S_{b \to c\bar c s}$ above, 
we obtain the following solutions for $\beta$
(in $\left[ 0, \pi \right]$):
\begin{equation}
  \beta = \left( 21.4 \pm 0.8 \right)^\circ
  \hspace{5mm}
  {\rm or}
  \hspace{5mm}
  \beta = \left( 68.6 \pm 0.8 \right)^\circ
  \label{eq:cp_uta:sin2beta}
\end{equation}
In radians, these values are 
$\beta = \left( 0.375 \pm 0.014 \right)$, 
$\beta = \left( 1.197 \pm 0.014 \right)$.

This result gives a precise constraint on the $(\rhobar,\etabar)$ plane,
as shown in Fig.~\ref{fig:cp_uta:ccs}.
The measurement is in remarkable agreement with other constraints from 
$\CP$ conserving quantities, 
and with $\CP$ violation in the kaon system, in the form of the parameter $\epsilon_K$.
Such comparisons have been performed by various phenomenological groups,
such as CKMfitter~\cite{Charles:2004jd} 
and UTFit~\cite{Bona:2005vz}.

\mysubsubsection{Time-dependent transversity analysis of $\Bz \to J/\psi K^{*0}$
}
\label{sec:cp_uta:ccs:vv}

$\B$ meson decays to the vector-vector final state $J/\psi K^{*0}$
are also mediated by the $b \to c \bar c s$ transition.
When a final state which is not flavour-specific ($K^{*0} \to \KS \pi^0$) is used,
a time-dependent transversity analysis can be performed 
allowing sensitivity to both 
$\sin(2\beta)$ and $\cos(2\beta)$~\cite{Dunietz:1990cj}.
Such analyses have been performed by both $\B$ factory experiments.
In principle, the strong phases between the transversity amplitudes
are not uniquely determined by such an analysis, 
leading to a discrete ambiguity in the sign of $\cos(2\beta)$.
The \babar\ collaboration resolves 
this ambiguity using the known variation~\cite{Aston:1987ir}
of the P-wave phase (fast) relative to the S-wave phase (slow) 
with the invariant mass of the $K\pi$ system 
in the vicinity of the $K^*(892)$ resonance. 
The result is in agreement with the prediction from 
$s$ quark helicity conservation,
and corresponds to Solution II defined by Suzuki~\cite{Suzuki:2001za}.
We use this phase convention for the averages given in 
Table~\ref{tab:cp_uta:ccs:psi_kstar}.

\begin{table}[htb]
	\begin{center}
		\caption{
			Averages from $\Bz \to J/\psi K^{*0}$ transversity analyses.
		}
		\vspace{0.2cm}
		\setlength{\tabcolsep}{0.0pc}
		\begin{tabular*}{\textwidth}{@{\extracolsep{\fill}}lrcccc} \hline
		\mc{2}{l}{Experiment} & $N(B\bar{B})$ & $\sin 2\beta$ & $\cos 2\beta$ & Correlation \\
		\hline
	\babar & \cite{Aubert:2004cp} & 88M & $-0.10 \pm 0.57 \pm 0.14$ & $3.32 ^{+0.76}_{-0.96} \pm 0.27$ & $-0.37$ \\
	\belle & \cite{Itoh:2005ks} & 275M & $0.24 \pm 0.31 \pm 0.05$ & $0.56 \pm 0.79 \pm 0.11$ & $0.22$ \\
	\mc{3}{l}{\bf Average} & $0.16 \pm 0.28$ & $1.64 \pm 0.62$ &  \hspace{-8mm} {\small uncorrelated averages}  \\
        \mc{3}{l}{\small Confidence level} & {\small $0.61~(0.5\sigma)$} & {\small $0.03~(2.2\sigma)$} & \\
		\hline
		\end{tabular*}
		\label{tab:cp_uta:ccs:psi_kstar}
	\end{center}
\end{table}

At present the results are dominated by 
large and non-Gaussian statistical errors,
and exhibit significant correlations.
We perform uncorrelated averages, 
the interpretation of which has to be done with the greatest care. 
Nonetheless, it is clear that $\cos(2\beta)>0$ is preferred 
by the experimental data in $J/\psi \Kstar$.
[\babar~\cite{Aubert:2004cp} 
find a confidence level for $\cos(2\beta)>0$ of $89\%$.]

\mysubsubsection{Time-dependent $\CP$ asymmetries in $\Bz \to \Dstarp \Dstarm \KS$ decays
}
\label{sec:cp_uta:ccs:DstarDstarKs}

\begin{figure}[htb]
  \begin{center}
    \begin{tabular}{c@{\hspace{-1mm}}c@{\hspace{-1mm}}c}
      \resizebox{0.32\textwidth}{!}{
        \includegraphics{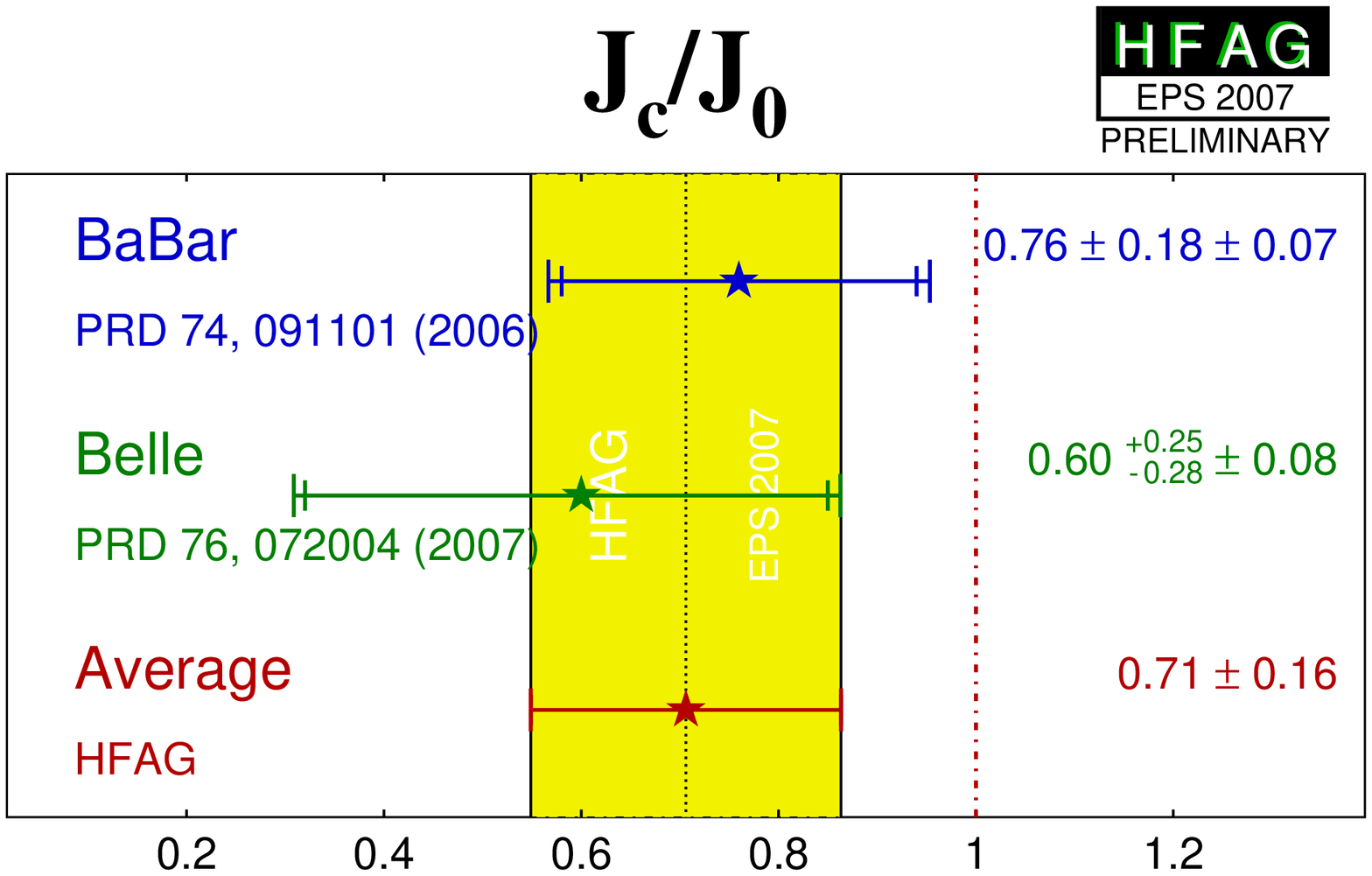}
      }
      &
      \resizebox{0.32\textwidth}{!}{
        \includegraphics{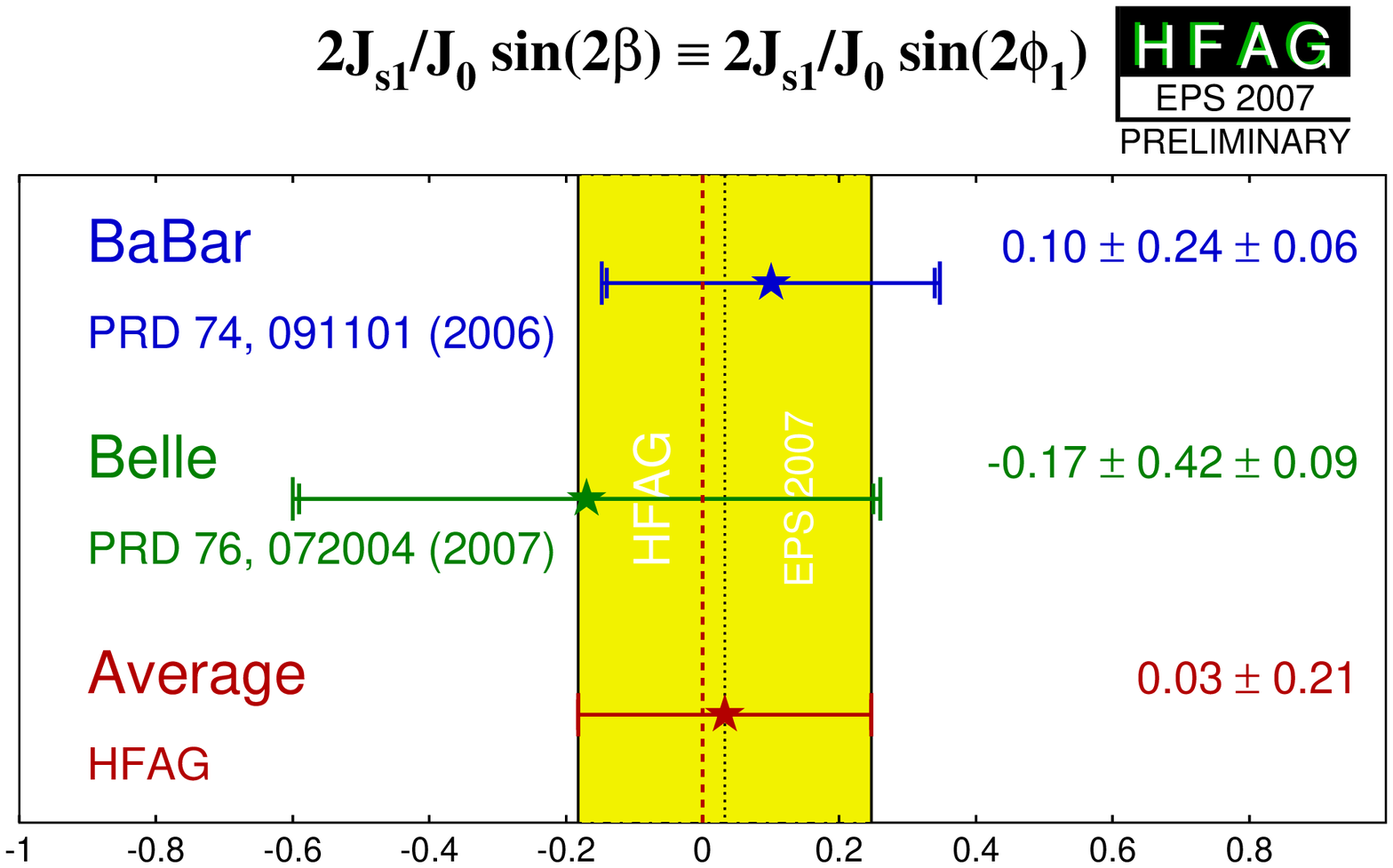}
      }
      &
      \resizebox{0.32\textwidth}{!}{
        \includegraphics{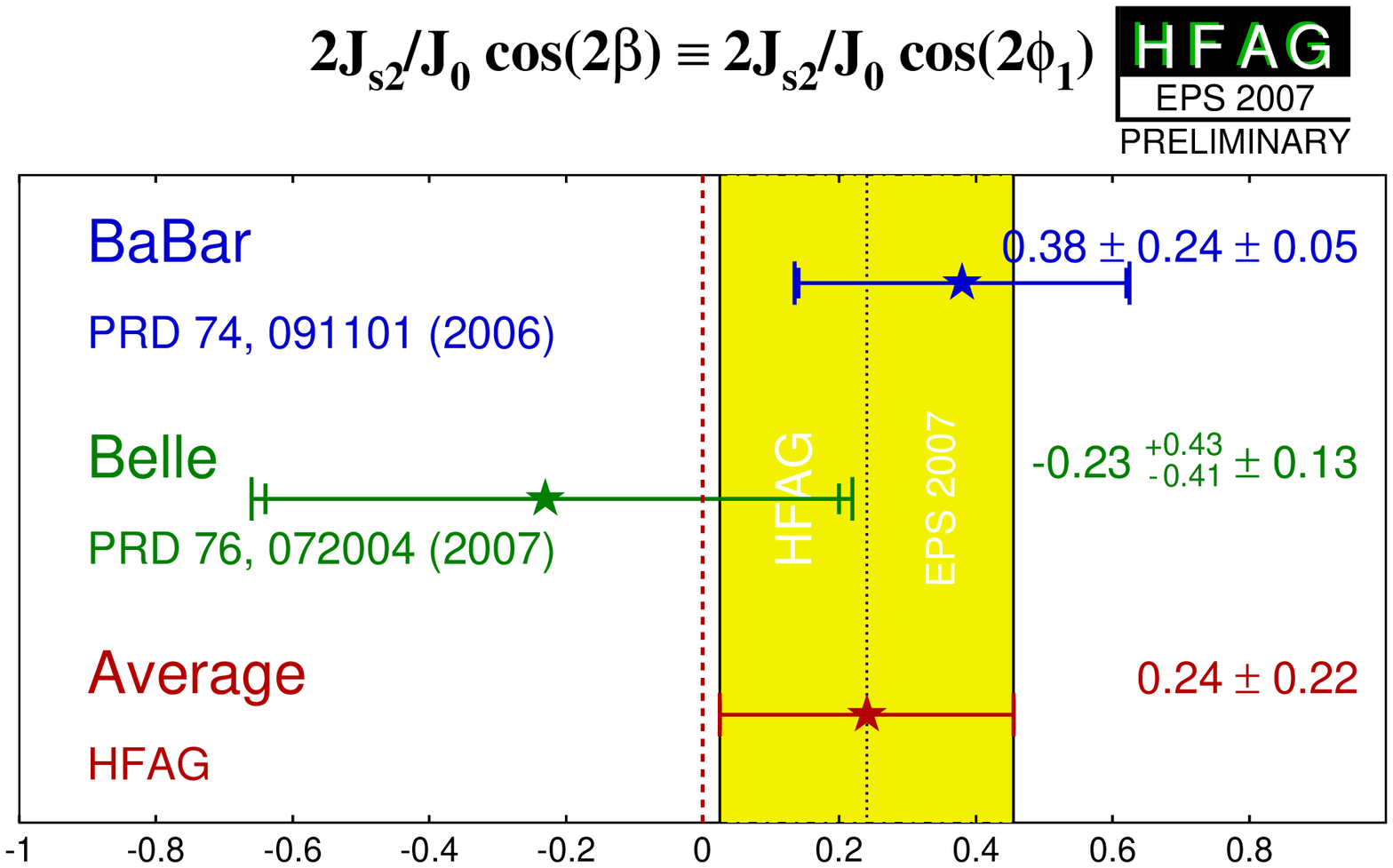}
      }
    \end{tabular}
  \end{center}
  \vspace{-0.8cm}
  \caption{
    Averages of 
    (left) $(J_c/J_0)$, (middle) $(2J_{s1}/J_0) \sin(2\beta)$ and 
    (right) $(2J_{s2}/J_0) \cos(2\beta)$
    from time-dependent analyses of $\Bz \to \Dstarp \Dstarm \KS$ decays.
  }
  \label{fig:cp_uta:ccs:dstardstarks}
\end{figure}

Both \babar~\cite{Aubert:2006fh} and \belle~\cite{Dalseno:2007hx} have performed
time-dependent analyses of the $\Bz \to \Dstarp \Dstarm \KS$ decay,
to obtain information on the sign of $\cos(2\beta)$.
More information can be found in 
Sec.~\ref{sec:cp_uta:notations:dalitz:dstardstarks}.
The results are shown in Table~\ref{tab:cp_uta:ccs:dstardstarks}, 
and Fig.~\ref{fig:cp_uta:ccs:dstardstarks}.

\begin{table}[htb]
	\begin{center}
		\caption{
                        Results from time-dependent analysis of $\Bz \to \Dstarp \Dstarm \KS$.
		}
		\vspace{0.2cm}
		\setlength{\tabcolsep}{0.0pc}
		\begin{tabular*}{\textwidth}{@{\extracolsep{\fill}}lrcccc} \hline
                \mc{2}{l}{Experiment} & $N(B\bar{B})$ & $\frac{J_c}{J_0}$ & $\frac{2J_{s1}}{J_0} \sin(2\beta)$ &  $\frac{2J_{s2}}{J_0} \cos(2\beta)$ \\
		\hline
	\babar & \cite{Aubert:2006fh} & 230M & $0.76 \pm 0.18 \pm 0.07$ & $0.10 \pm 0.24 \pm 0.06$ & $0.38 \pm 0.24 \pm 0.05$ \\
	\belle & \cite{Dalseno:2007hx} & 449M & $0.60 \,^{+0.25}_{-0.28} \pm 0.08$ & $-0.17 \pm 0.42 \pm 0.09$ & $-0.23 \,^{+0.43}_{-0.41} \pm 0.13$ \\
	\mc{3}{l}{\bf Average} & $0.71 \pm 0.16$ & $0.03 \pm 0.21$ & $0.24 \pm 0.22$ \\
	\mc{3}{l}{\small Confidence level} & {\small $0.63~(0.5\sigma)$} & {\small $0.59~(0.5\sigma)$} & {\small $0.23~(1.2\sigma)$} \\
		\hline
		\end{tabular*}
		\label{tab:cp_uta:ccs:dstardstarks}
	\end{center}
\end{table}

From the above result and the assumption that $J_{s2}>0$, 
\babar\ infer that $\cos(2\beta)>0$ at the $94\%$ confidence level~\cite{Aubert:2006fh}.

\mysubsubsection{Time-dependent analysis of $\Bs \to J/\psi \phi$
}
\label{sec:cp_uta:ccs:jpsiphi}

As described in Sec.~\ref{sec:cp_uta:notations:Bs},
time-dependent analysis of $\Bs \to J/\psi \phi$ probes the 
$\CP$ violating phase of $\Bs$--$\Bsb$ oscillations, $\phi_s$.
Within the Standard Model, this parameter is predicted to be small.\footnote{
   We make the approximation $\phi_s = ¡Ý2 \beta_s$, 
   where $\phi_s \equiv \arg\left[ -M_{12}/\Gamma_{12} \right]$ 
   and $2\beta_s \equiv 2 \arg\left[ -(V_{ts}V_{tb}^*)/(V_{cs}V_{cb}^*) \right]$
   (see Section~\ref{sec:cp_uta:introduction}). 
   This is a reasonable approximation since, 
   although the equality does not hold in the Standard Model~\cite{Lenz:2011ti,*Lenz:2006hd}, 
   both are much smaller than the current experimental resolution, 
   whereas new physics contributions add a phase $\phi_{\rm NP}$ to $\phi_s$
   and subtract the same phase from $2\beta_s$, 
   so that the approximation remains valid.
}
The combination of results is performed by the HFAG Lifetimes and Oscillations group, see Sec.~\ref{sec:life_mix}.

\clearpage
\mysubsection{Time-dependent $\CP$ asymmetries in colour-suppressed $b \to c\bar{u}d$ transitions
}
\label{sec:cp_uta:cud_beta}

Decays of $\B$ mesons to final states such as $D\pi^0$ are 
governed by $b \to c\bar{u}d$ transitions. 
If the final state is a $\CP$ eigenstate, \eg\ $D_{\CP}\pi^0$, 
the usual time-dependence formulae are recovered, 
with the sine coefficient sensitive to $\sin(2\beta)$. 
Since there is no penguin contribution to these decays, 
there is even less associated theoretical uncertainty 
than for $b \to c\bar{c}s$ decays like $\B \to \jpsi \KS$.
Such measurements therefore allow to test the Standard Model prediction
that the $\CP$ violation parameters in $b \to c\bar{u}d$ transitions
are the same as those in $b \to c\bar{c}s$~\cite{Grossman:1996ke}.

Note that there is an additional contribution from CKM suppressed
$b \to u \bar{c} d$ decays.
The effect of this contribution is small, and can be taken into 
account in the analysis~\cite{Fleischer:2003ai,Fleischer:2003aj}.

Results of such an analysis are available from \babar~\cite{Aubert:2007mn}.
The decays $\Bz \to D\pi^0$, $\Bz \to D\eta$, $\Bz \to D\omega$,
$\Bz \to D^*\pi^0$ and $\Bz \to D^*\eta$ are used.
The daughter decay $D^* \to D\pi^0$ is used.
The $\CP$-even $D$ decay to $K^+K^-$ is used for all decay modes,
with the $\CP$-odd $D$ decay to $\KS\omega$ also used in $\Bz \to D^{(*)}\pi^0$
and the additional $\CP$-odd $D$ decay to $\KS\pi^0$ 
also used in $\Bz \to D\omega$.
Results are presented separately for $\CP$-even and $\CP$-odd 
$D^{(*)}$ decays (denoted $D^{(*)}_+ h^0$ and $D^{(*)}_- h^0$ respectively),
and for both combined, with the different $\CP$ factors accounted for
(denoted $D^{(*)}_{CP} h^0$).
The results are summarised in Table~\ref{tab:cp_uta:cud_cp_beta}.

\begin{table}[htb]
	\begin{center}
		\caption{
			Results from analyses of $\Bz \to D^{(*)}h^0$, $D \to CP$ eigenstates decays.
		}
		\vspace{0.2cm}
		\setlength{\tabcolsep}{0.0pc}
		\begin{tabular*}{\textwidth}{@{\extracolsep{\fill}}lrcccc} \hline
	\mc{2}{l}{Experiment} & $N(B\bar{B})$ & $S_{CP}$ & $C_{CP}$ & Correlation \\
	\hline
        \mc{6}{c}{$D^{(*)}_+ h^0$}  \\
	\babar & \cite{Aubert:2007mn} & 383M & $-0.65 \pm 0.26 \pm 0.06$ & $-0.33 \pm 0.19 \pm 0.04$ & $0.04$ \\
	\hline

        \mc{6}{c}{$D^{(*)}_- h^0$} \\
	\babar & \cite{Aubert:2007mn} & 383M & $-0.46 \pm 0.46 \pm 0.13$ & $-0.03 \pm 0.28 \pm 0.07$ & $-0.14$ \\
	\hline

        \mc{6}{c}{$D^{(*)}_{CP} h^0$} \\
	\babar & \cite{Aubert:2007mn} & 383M & $-0.56 \pm 0.23 \pm 0.05$ & $-0.23 \pm 0.16 \pm 0.04$ & $-0.02$ \\
	\hline
		\end{tabular*}
		\label{tab:cp_uta:cud_cp_beta}
	\end{center}
\end{table}

When multibody $D$ decays, such as $D \to \KS\pi^+\pi^-$ are used, 
a time-dependent analysis of the Dalitz plot of the neutral $D$ decay 
allows a direct determination of the weak phase: $2\beta$. 
(Equivalently, both $\sin(2\beta)$ and $\cos(2\beta)$ can be measured.)
This information allows to resolve the ambiguity in the 
measurement of $2\beta$ from $\sin(2\beta)$~\cite{Bondar:2005gk}.

\begin{figure}[htb]
  \begin{center}
    \begin{tabular}{cc}
      \resizebox{0.46\textwidth}{!}{
        \includegraphics{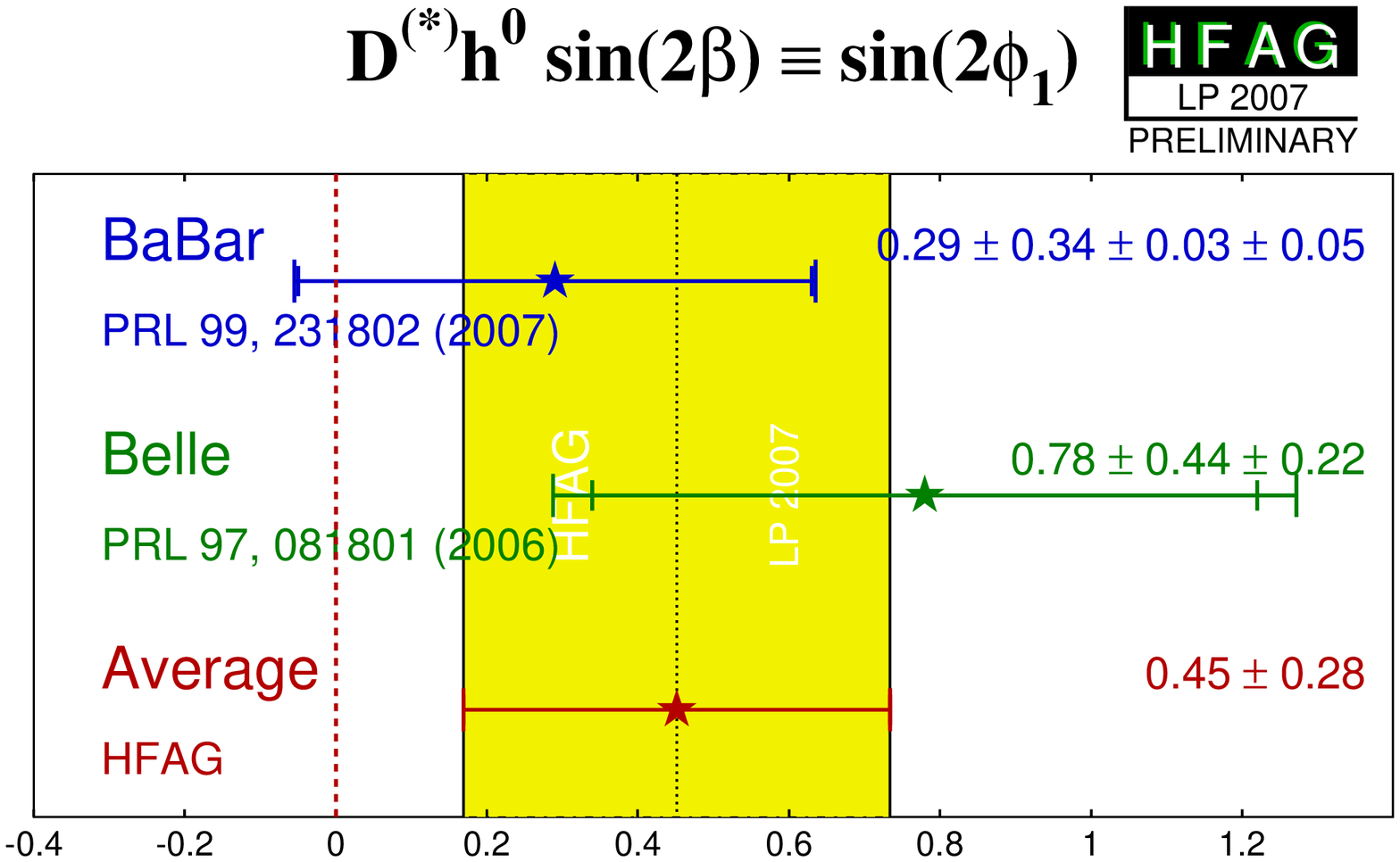}
      }
      &
      \resizebox{0.46\textwidth}{!}{
        \includegraphics{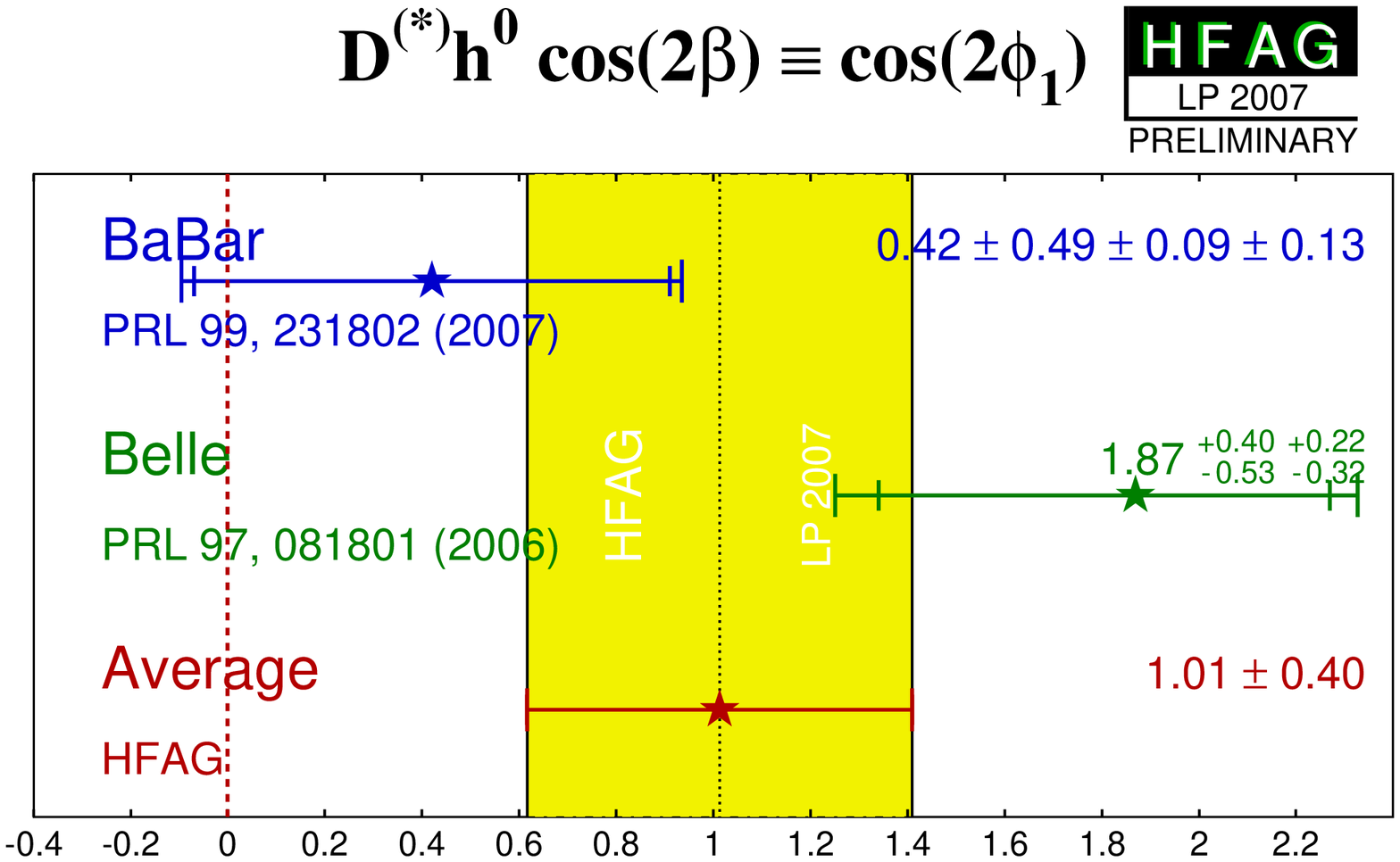}
      }
    \end{tabular}
  \end{center}
  \vspace{-0.8cm}
  \caption{
    Averages of 
    (left) $\sin(2\beta)$ and (right) $\cos(2\beta)$
    measured in colour-suppressed $b \to c\bar{u}d$ transitions.
  }
  \label{fig:cp_uta:cud_beta}
\end{figure}

Results of such analyses are available from both 
\belle~\cite{Krokovny:2006sv} and \babar~\cite{Aubert:2007rp}.
The decays $\B \to D\pi^0$, $\B \to D\eta$, $\B \to D\omega$, 
$\B \to D^*\pi^0$ and $\B \to D^*\eta$ are used. 
[This collection of states is denoted by $D^{(*)}h^0$.]
The daughter decays are $D^* \to D\pi^0$ and $D \to \KS\pi^+\pi^-$.
The results are shown in Table~\ref{tab:cp_uta:cud_beta},
and Fig.~\ref{fig:cp_uta:cud_beta}.
Note that \babar\ quote uncertainties due to the $D$ decay model 
separately from other systematic errors, while \belle\ do not.

\begin{table}[htb]
	\begin{center}
		\caption{
			Averages from $\Bz \to D^{(*)}h^0$, $D \to K_S\pi^+\pi^-$ analyses.
		}
		\vspace{0.2cm}
		\setlength{\tabcolsep}{0.0pc}
    \resizebox{\textwidth}{!}{
      		\begin{tabular*}{\textwidth}{@{\extracolsep{\fill}}lrcccc} \hline
	\mc{2}{l}{Experiment} & $N(B\bar{B})$ & $\sin 2\beta$ & $\cos 2\beta$ & $|\lambda|$ \\
		\hline
	\babar & \cite{Aubert:2007rp} & 383M & $0.29 \pm 0.34 \pm 0.03 \pm 0.05$ & $0.42 \pm 0.49 \pm 0.09 \pm 0.13$ & $1.01 \pm 0.08 \pm 0.02$ \\
	\belle & \cite{Krokovny:2006sv} & 386M & $0.78 \pm 0.44 \pm 0.22$ & $1.87 \,^{+0.40}_{-0.53} \,^{+0.22}_{-0.32}$ & \textendash{} \\
	\mc{3}{l}{\bf Average} & $0.45 \pm 0.28$ & $1.01 \pm 0.40$ & $1.01 \pm 0.08$ \\
	\mc{3}{l}{\small Confidence level} & {\small $0.59~(0.5\sigma)$} & {\small $0.12~(1.6\sigma)$} & \textendash{} \\
		\hline
		\end{tabular*}
    }
		\label{tab:cp_uta:cud_beta}
	\end{center}
\end{table}

Again, it is clear that the data prefer $\cos(2\beta)>0$.
Indeed, \belle~\cite{Krokovny:2006sv} 
determine the sign of $\cos(2\phi_1)$ to be positive at $98.3\%$ confidence level,
while \babar~\cite{Aubert:2007rp} 
favour the solution of $\beta$ with $\cos(2\beta)>0$ at $87\%$ confidence level.
Note, however, that the Belle measurement has strongly non-Gaussian behaviour. 
Therefore, we perform uncorrelated averages, 
from which any interpretation has to be done with the greatest care. 

\clearpage
\mysubsection{Time-dependent $\CP$ asymmetries in charmless $b \to q\bar{q}s$ transitions
}
\label{sec:cp_uta:qqs}

The flavour changing neutral current $b \to s$ penguin
can be mediated by any up-type quark in the loop, 
and hence the amplitude can be written as
\begin{equation}
  \label{eq:cp_uta:b_to_s}
  \begin{array}{ccccc}
    A_{b \to s} & = & 
    \mc{3}{l}{F_u V_{ub}V^*_{us} + F_c V_{cb}V^*_{cs} + F_t V_{tb}V^*_{ts}} \\
    & = & (F_u - F_c) V_{ub}V^*_{us} & + & (F_t - F_c) V_{tb}V^*_{ts} \\
    & = & {\cal O}(\lambda^4) & + & {\cal O}(\lambda^2) \\
  \end{array}
\end{equation}
using the unitarity of the CKM matrix.
Therefore, in the Standard Model, 
this amplitude is dominated by $V_{tb}V^*_{ts}$, 
and to within a few degrees 
($\delta\beta \lesssim 2^\circ$ for $\beta \approx 20^\circ$) 
the time-dependent parameters can be written\footnote
{
  The presence of a small (${\cal O}(\lambda^2)$) weak phase in 
  the dominant amplitude of the $s$ penguin decays introduces 
  a phase shift given by
  $S_{b \to q\bar q s} = -\eta\sin(2\beta)\cdot(1 + \Delta)$. 
  Using the CKMfitter results for the Wolfenstein 
  parameters~\cite{Charles:2004jd}, one finds: 
  $\Delta \simeq 0.033$, which corresponds to a shift of 
  $2\beta$ of $+2.1$ degrees. 
  Nonperturbative contributions can alter this result.
}
$S_{b \to q\bar q s} \approx - \etacp \sin(2\beta)$,
$C_{b \to q\bar q s} \approx 0$,
assuming $b \to s$ penguin contributions only ($q = u,d,s$).

Due to the large virtual mass scales occurring in the penguin loops,
additional diagrams from physics beyond the Standard Model,
with heavy particles in the loops, may contribute.
In general, these contributions will affect the values of 
$S_{b \to q\bar q s}$ and $C_{b \to q\bar q s}$.
A discrepancy between the values of 
$S_{b \to c\bar c s}$ and $S_{b \to q\bar q s}$
can therefore provide a clean indication of new physics~\cite{Grossman:1996ke,Fleischer:1996bv,London:1997zk,Ciuchini:1997zp}.

However, there is an additional consideration to take into account.
The above argument assumes only the $b \to s$ penguin contributes
to the $b \to q\bar q s$ transition.
For $q = s$ this is a good assumption, which neglects only rescattering effects.
However, for $q = u$ there is a colour-suppressed $b \to u$ tree diagram
(of order ${\cal O}(\lambda^4)$), 
which has a different weak (and possibly strong) phase.
In the case $q = d$, any light neutral meson that is formed from $d \bar{d}$ 
also has a $u \bar{u}$ component, and so again there is ``tree pollution''. 
The \Bz decays to $\piz\KS$, $\rhoz\KS$ and $\omega\KS$ belong to this category.
The mesons $\phi$, $f_0$ and $\etapr$ are expected to have predominant
$s\bar{s}$ parts, which reduces the relative size of the possible tree
pollution. 
If the inclusive decay $\Bz\to\Kp\Km\Kz$ (excluding $\phi\Kz$) is dominated by
a nonresonant three-body transition, 
an OZI-rule suppressed tree-level diagram can occur 
through insertion of an $s\sbar$ pair. 
The corresponding penguin-type transition 
proceeds via insertion of a $u\ubar$ pair, which is expected
to be favoured over the $s\sbar$ insertion by fragmentation models.
Neglecting rescattering, the final state $\Kz\Kzb\Kz$ 
(reconstructed as $\KS\KS\KS$) has no tree pollution~\cite{Gershon:2004tk}.
Various estimates, using different theoretical approaches,
of the values of $\Delta S = S_{b \to q\bar q s} - S_{b \to c\bar c s}$
exist in the literature~\cite{Grossman:2003qp,Gronau:2003ep,Gronau:2003kx,Gronau:2004hp,Cheng:2005bg,Gronau:2005gz,Buchalla:2005us,Beneke:2005pu,Engelhard:2005hu,Cheng:2005ug,Engelhard:2005ky,Gronau:2006qh,Silvestrini:2007yf,Dutta:2008xw}.
In general, there is agreement that the modes
$\phi\Kz$, $\etapr\Kz$ and $\Kz\Kzb\Kz$ are the cleanest,
with values of $\left| \Delta S \right|$ at or below the few percent level 
($\Delta S$ is usually positive).

\mysubsubsection{Time-dependent $\CP$ asymmetries: $b \to q\bar{q}s$ decays to $\CP$ eigenstates
}
\label{sec:cp_uta:qqs:cp_eigen}

\begin{figure}[htb]
  \begin{center}
    \resizebox{0.45\textwidth}{!}{
      \includegraphics{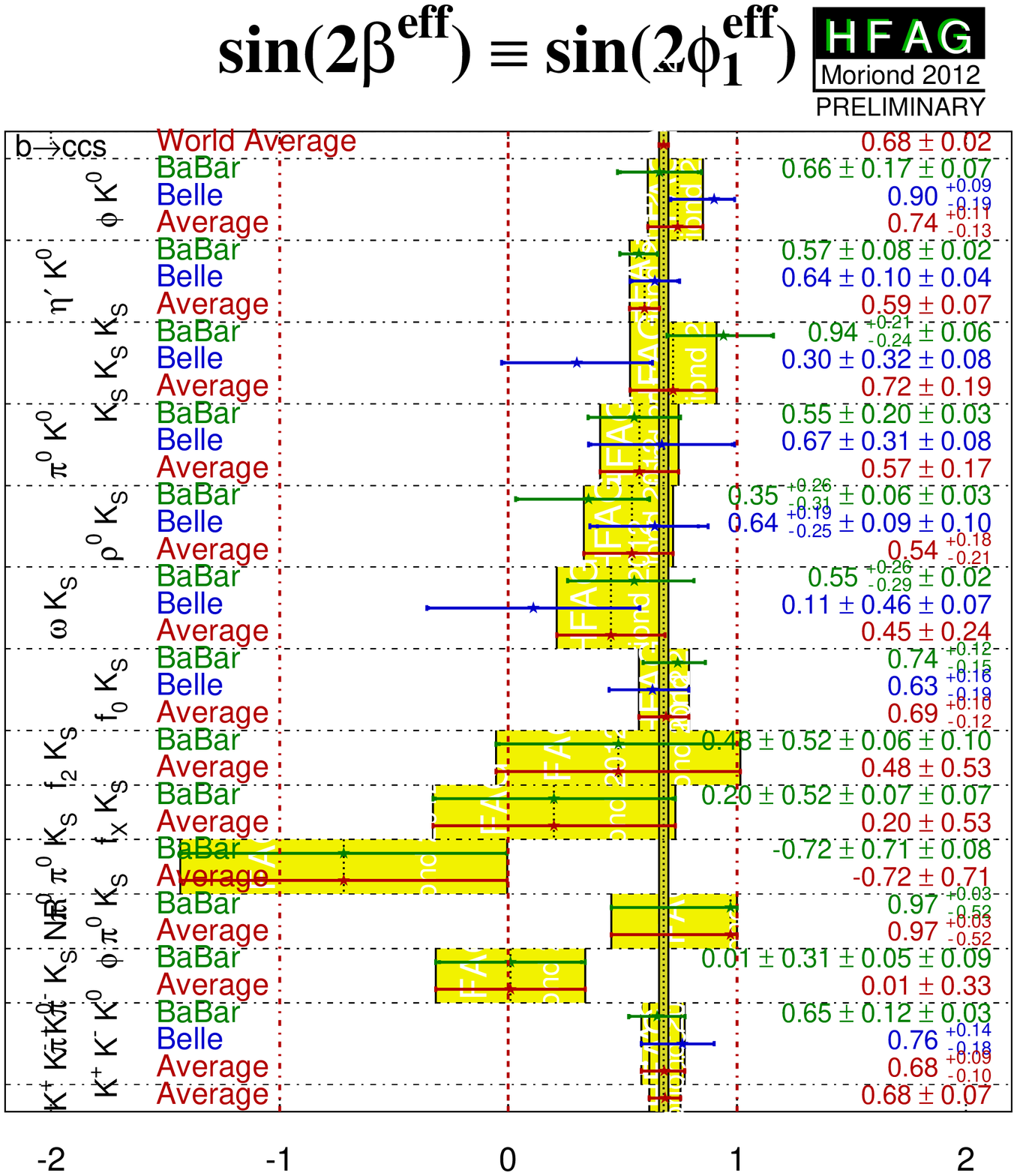}
    }
    \hfill
    \resizebox{0.45\textwidth}{!}{
      \includegraphics{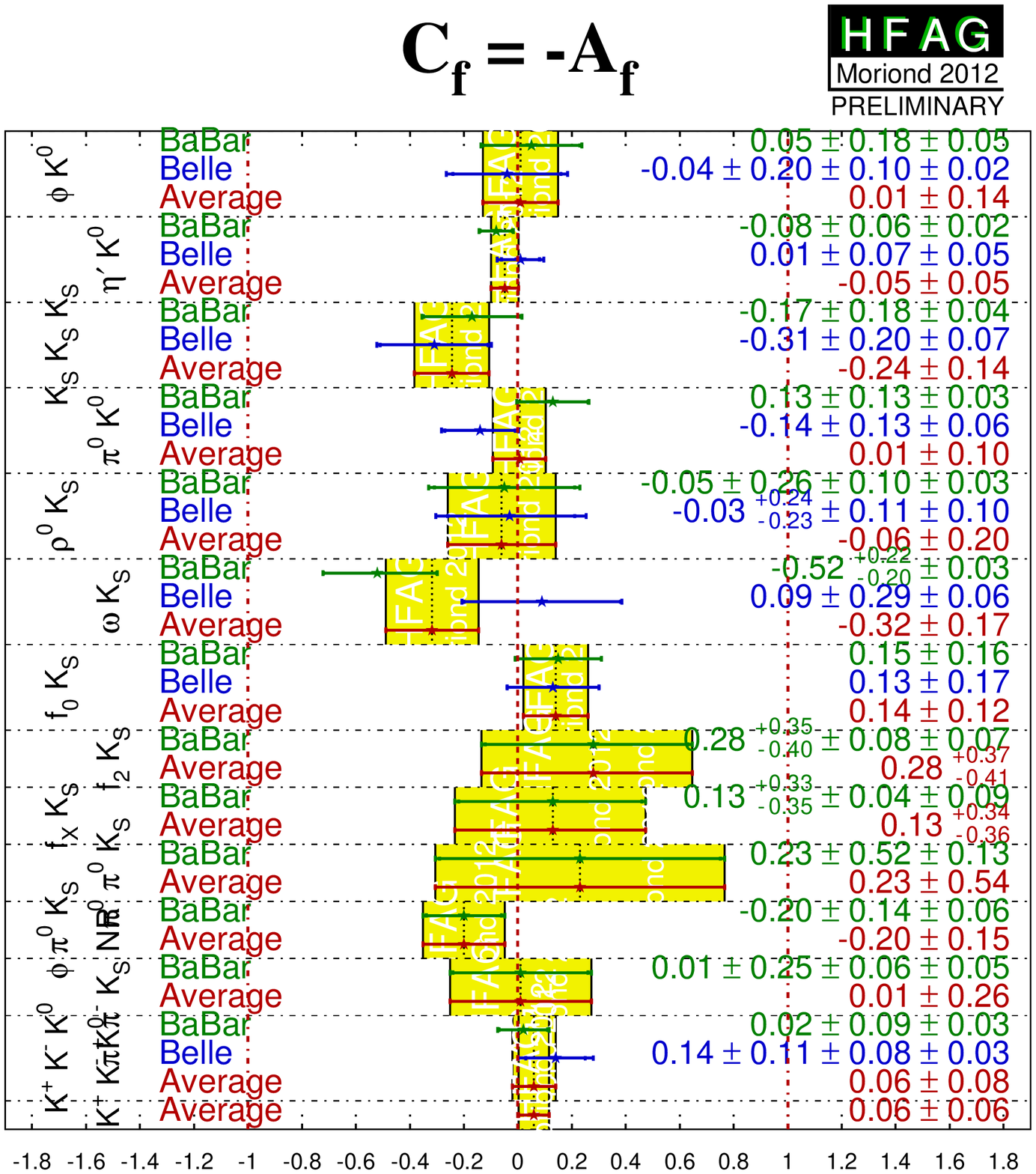}
    }
    \\
    \resizebox{0.45\textwidth}{!}{
      \includegraphics{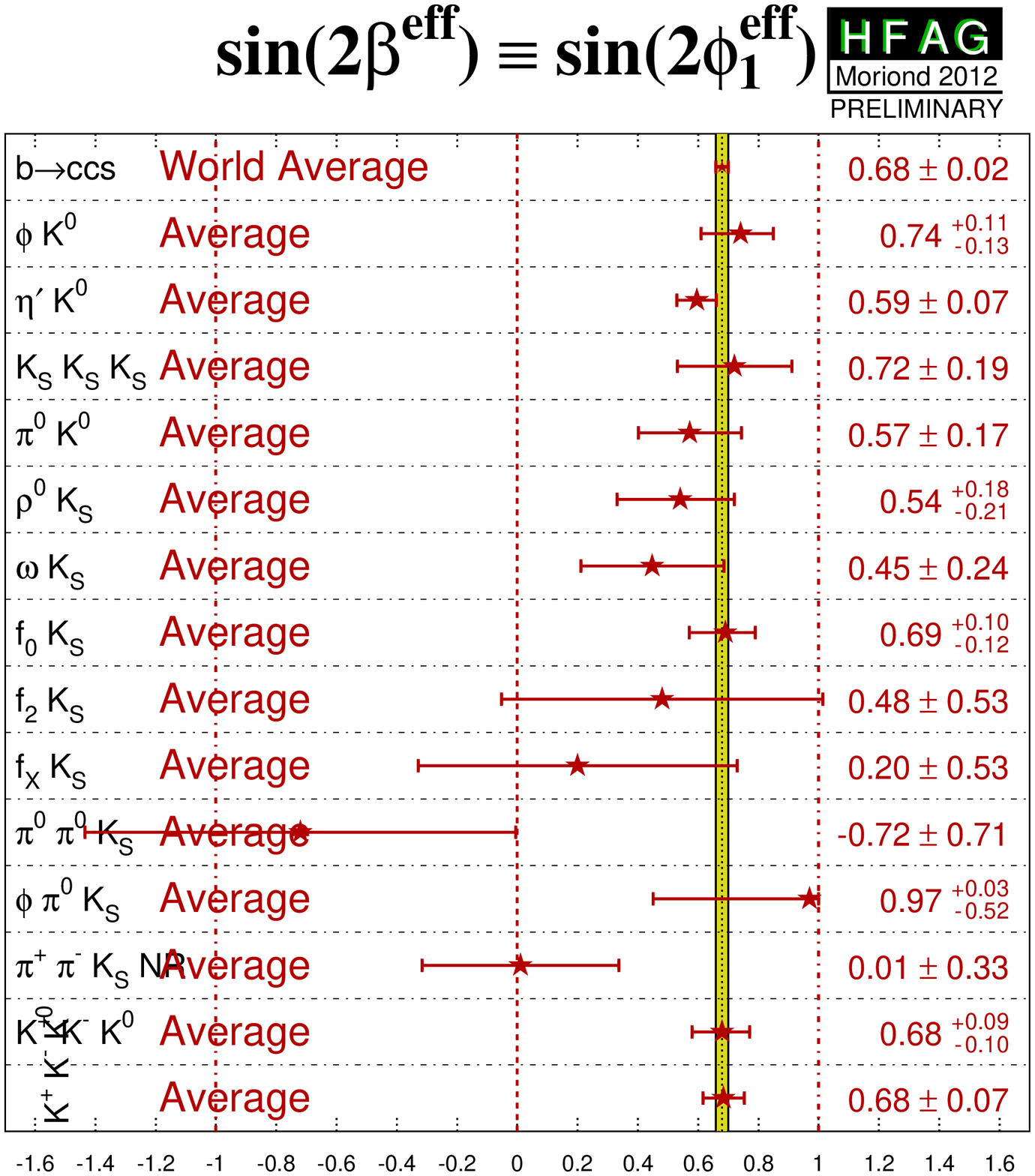}
    }
    \hfill
    \resizebox{0.45\textwidth}{!}{
      \includegraphics{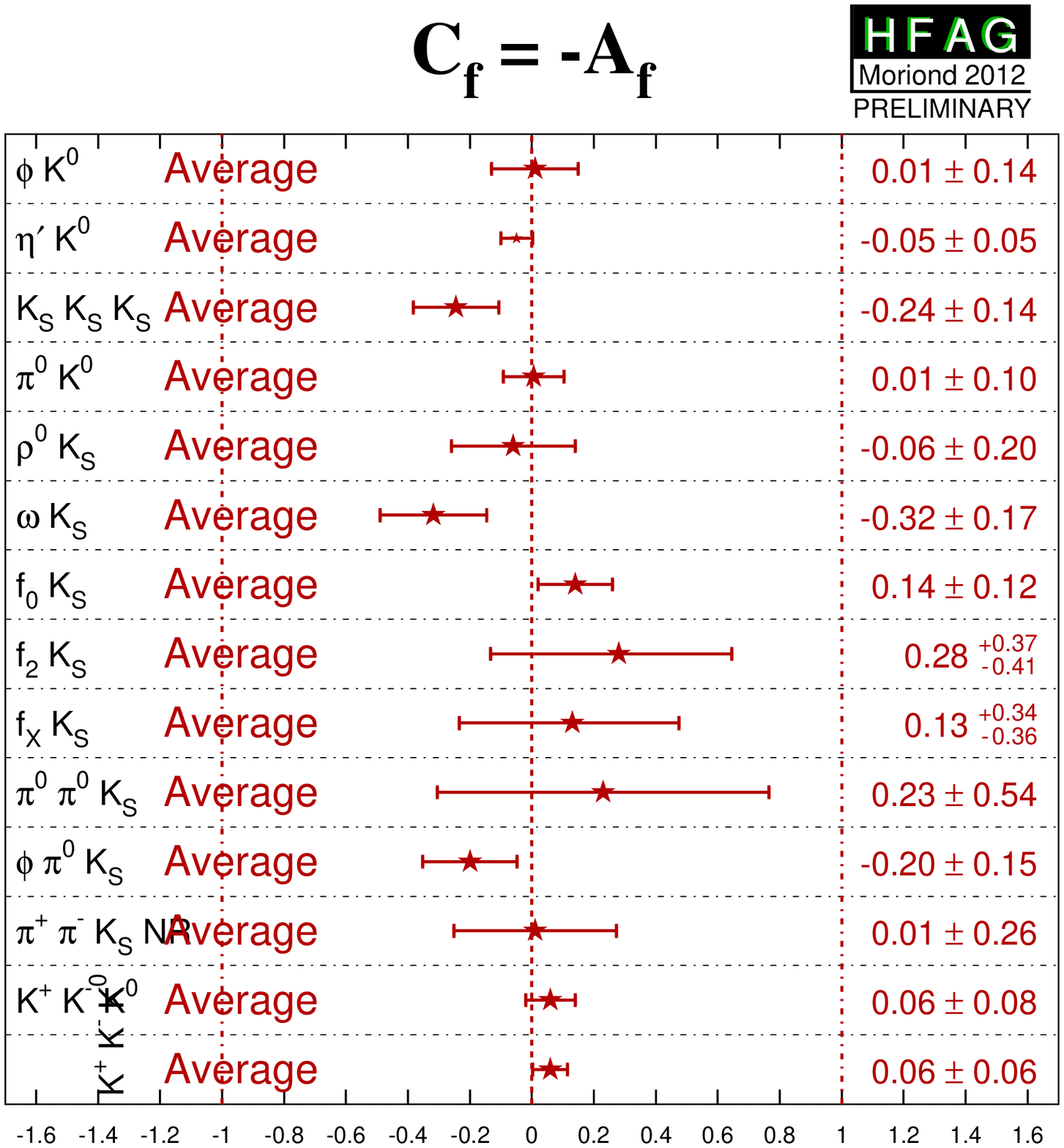}
    }
  \end{center}
  \vspace{-0.8cm}
  \caption{
    (Top)
    Averages of 
    (left) $-\etacp S_{b \to q\bar q s}$ and (right) $C_{b \to q\bar q s}$.
    The $-\etacp S_{b \to q\bar q s}$ figure compares the results to 
    the world average 
    for $-\etacp S_{b \to c\bar c s}$ (see Section~\ref{sec:cp_uta:ccs:cp_eigen}).
    (Bottom) Same, but only averages for each mode are shown.
    More figures are available from the HFAG web pages.
  }
  \label{fig:cp_uta:qqs}
\end{figure}

\begin{figure}[htb]
  \begin{center}
    \resizebox{0.33\textwidth}{!}{
      \includegraphics{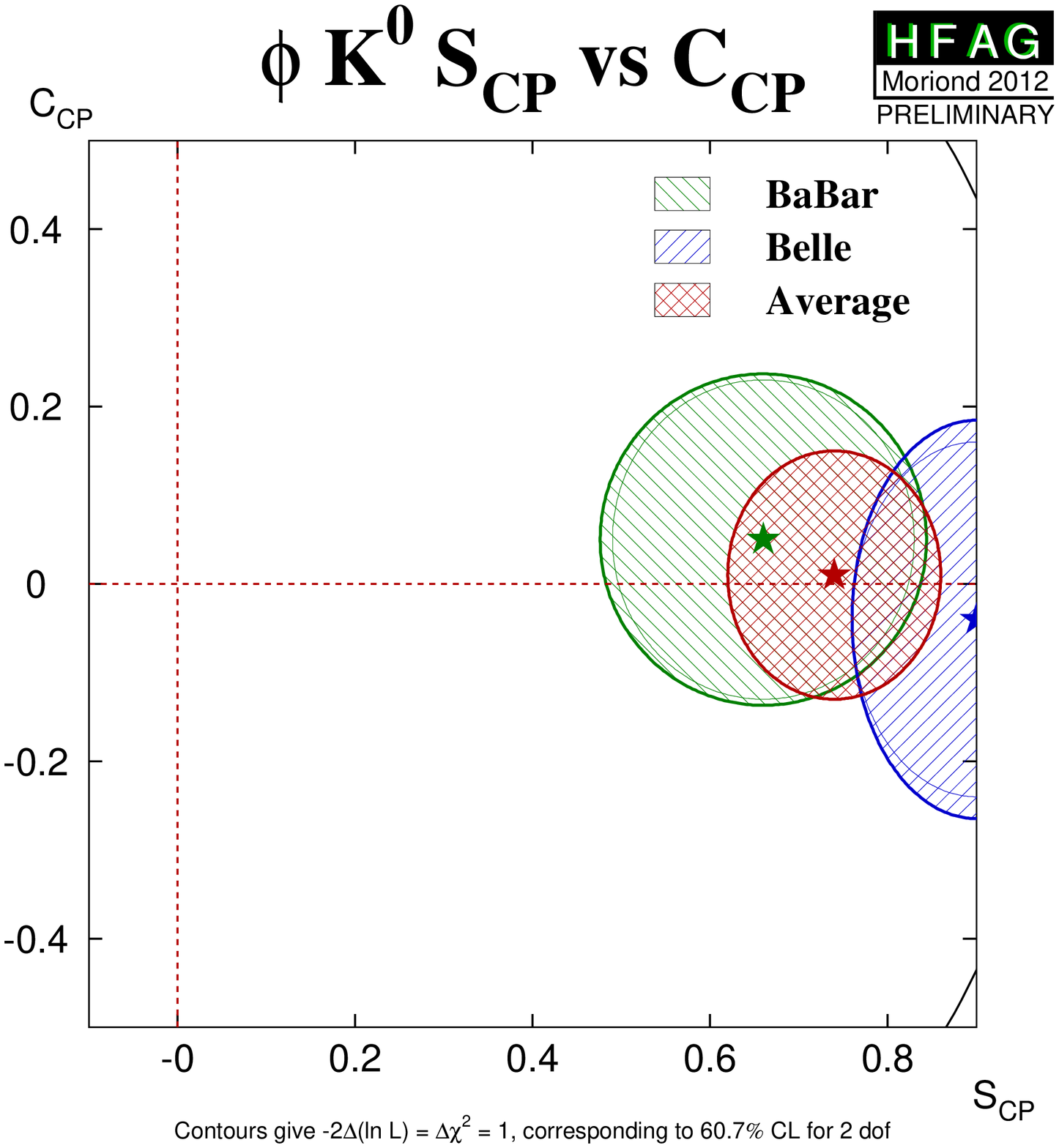}
    }
    \hspace{0.08\textwidth}
    \resizebox{0.33\textwidth}{!}{
      \includegraphics{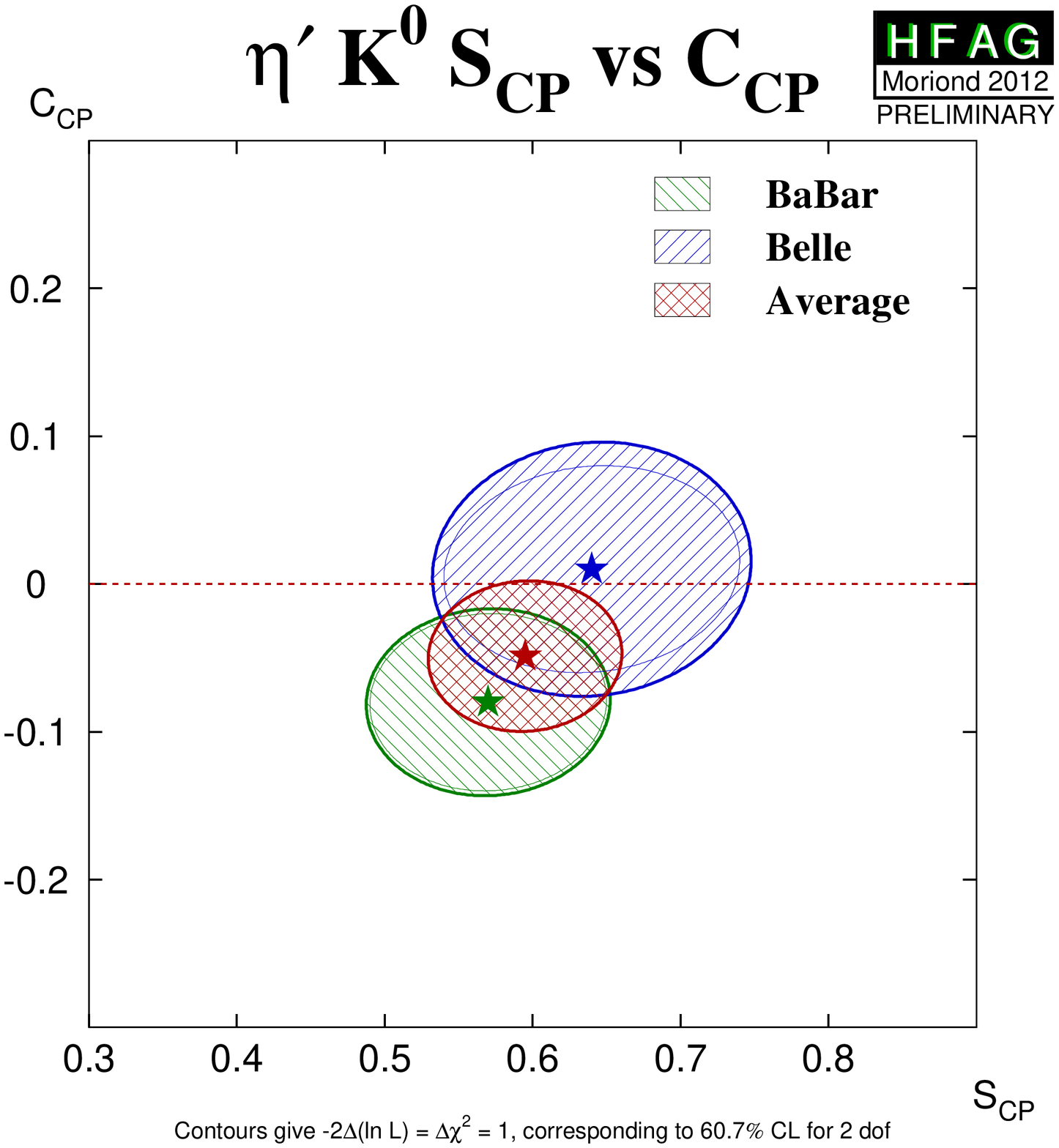}
    }
    \\
    \resizebox{0.33\textwidth}{!}{
      \includegraphics{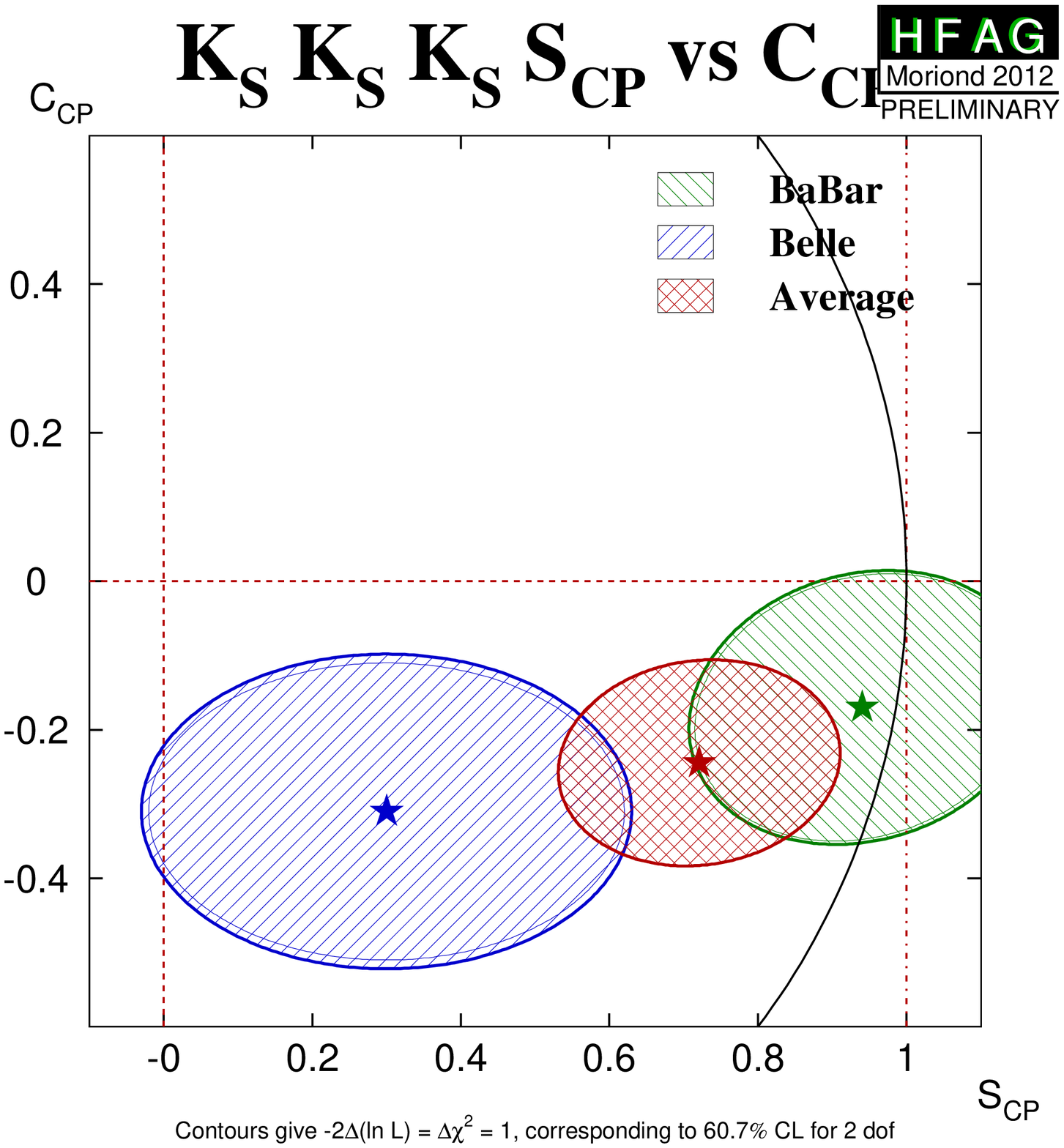}
    }
    \hspace{0.08\textwidth}
    \resizebox{0.33\textwidth}{!}{
      \includegraphics{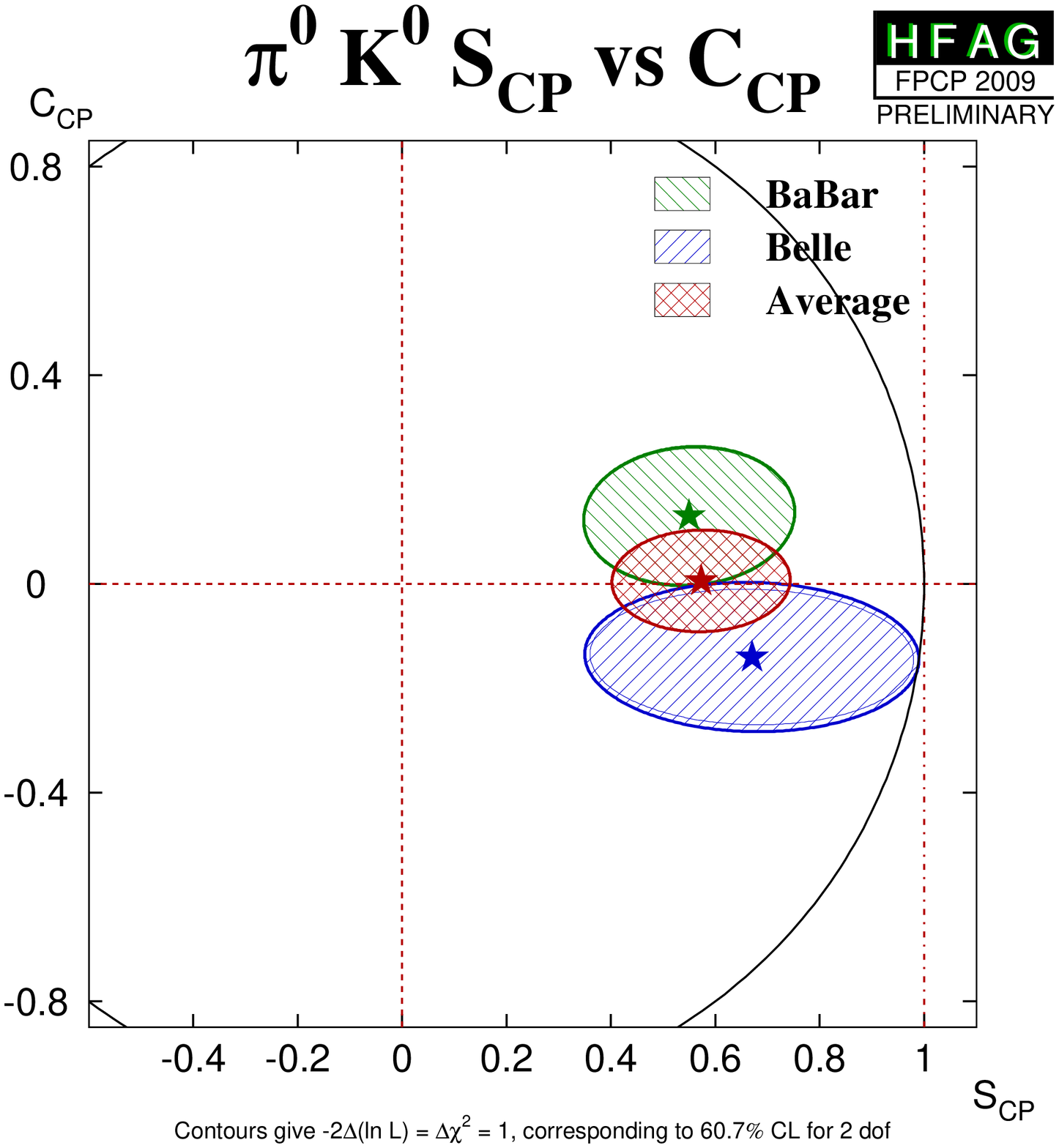}
    }
  \end{center}
  \vspace{-0.8cm}
  \caption{
    Averages of four $b \to q\bar q s$ dominated channels,
    for which correlated averages are performed,
    in the $S_{\CP}$ \vs\ $C_{\CP}$ plane,
    where $S_{\CP}$ has been corrected by the $\CP$ eigenvalue to give
    $\sin(2\beta^{\rm eff})$.
    (Top left) $\Bz \to \phi\Kz$,
    (top right) $\Bz \to \eta^\prime\Kz$,
    (bottom left) $\Bz \to \KS\KS\KS$,
    (bottom right) $\Bz \to \pi^0\KS$.
    More figures are available from the HFAG web pages.
  }
  \label{fig:cp_uta:qqs_SvsC}
\end{figure}

\begin{figure}[htb]
  \begin{center}
    \resizebox{0.66\textwidth}{!}{
      \includegraphics{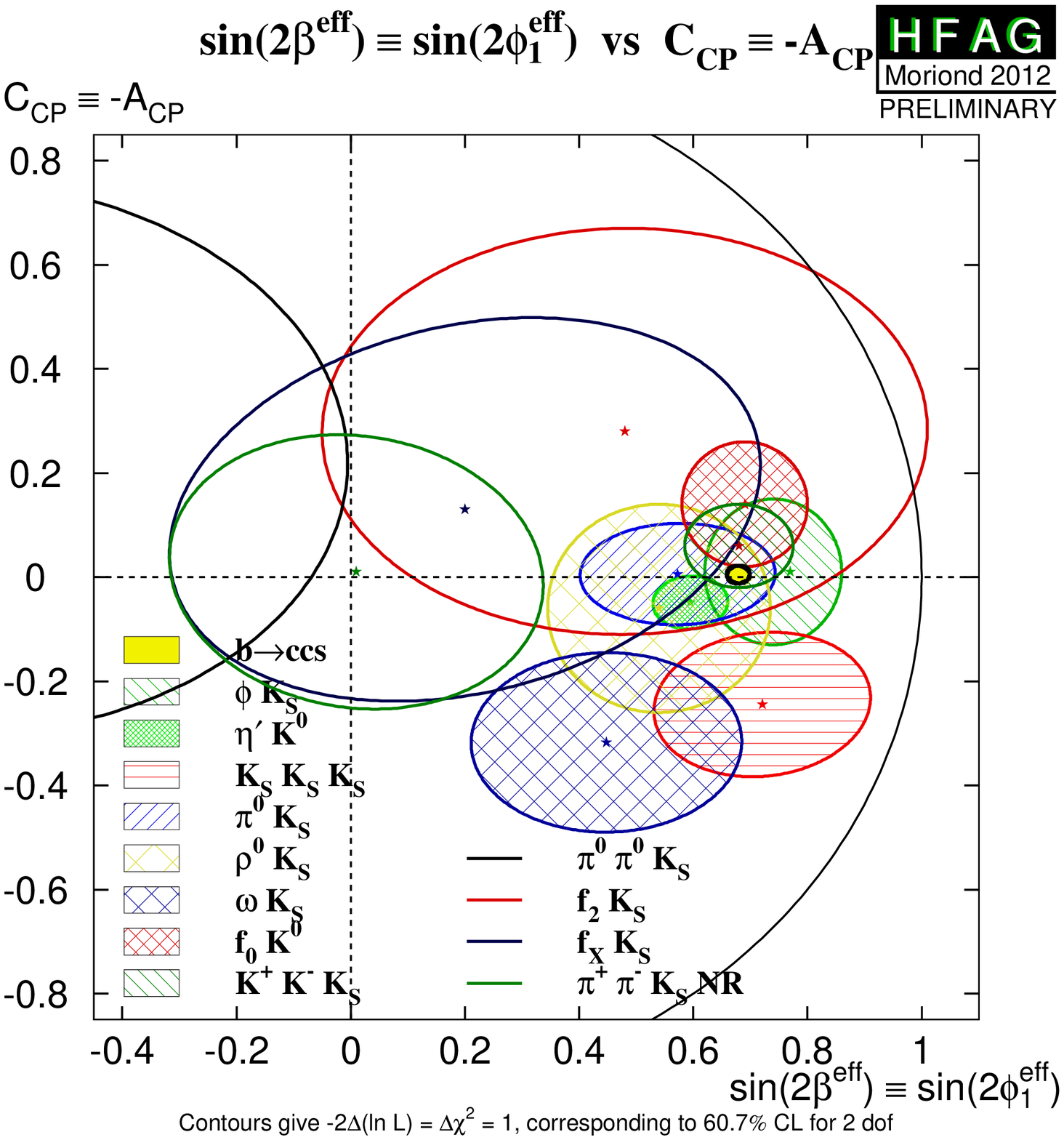}
    }
  \end{center}
  \vspace{-0.8cm}
  \caption{
    Compilation of constraints in the 
    $-\etacp S_{b \to q\bar q s}$ \vs\ $C_{b \to q\bar q s}$ plane.
  }
  \label{fig:cp_uta:qqs_SvsC-all}
\end{figure}

The averages for $-\etacp S_{b \to q\bar q s}$ and $C_{b \to q\bar q s}$
can be found in Table~\ref{tab:cp_uta:qqs},
and are shown in Figs.~\ref{fig:cp_uta:qqs},~\ref{fig:cp_uta:qqs_SvsC} 
and~\ref{fig:cp_uta:qqs_SvsC-all}.
Results from both \babar\  and \belle\ are averaged for the modes
$\etapr\Kz$ ($\Kz$ indicates that both $\KS$ and $\KL$ are used)
$\KS\KS\KS$, $\pi^0 \KS$\footnote{
  \belle~\cite{Fujikawa:2008pk} include the $\pi^0\KL$ final state in order to
  improve the constraint on the direct \CP\ violation parameter; these events
  cannot be used for time-dependent analysis.
} and $\omega\KS$.
Results on $\phi\KS$ and $\Kp\Km\KS$ (implicitly excluding $\phi\KS$ and $f_0\KS$) are taken from time-dependent Dalitz plot analyses of $\Kp\Km\KS$;
results on $\rhoz\KS$, $f_2\KS$, $f_{\rm X}\KS$ and $\pip\pim\KS$ nonresonant are taken from time-dependent Dalitz plot analyses of $\pip\pim\KS$ (see
subsection~\ref{sec:cp_uta:qqs:dp}).
The results on $f_0\KS$ are from combinations of both Dalitz plot analyses.
\babar\ also has presented results with the final states
$\pi^0\pi^0\KS$,\footnote{
  We do not include a preliminary result from \belle~\cite{:2007xd}, which
  remains unpublished after more than two years.
}
and $\phi \KS \pi^0$. 

Of these final states,
$\phi\KS$, $\etapr\KS$, $\pi^0 \KS$, $\rho^0\KS$, $\omega\KS$ and $f_0\KL$
have $\CP$ eigenvalue $\etacp = -1$, 
while $\phi\KL$, $\etapr\KL$, $\KS\KS\KS$, $f_0 \KS$, $f_2 \KS$, 
$f_{\rm X} \KS$,\footnote{ 
  The $f_{\rm X}$ is assumed to be spin even.
} $\pi^0\pi^0\KS$ and $\pi^+ \pi^- \KS$ nonresonant have $\etacp = +1$.
The final state $K^+K^-\KS$ (with $\phi\KS$ and $f_0\KS$ implicitly excluded)
is not a $\CP$ eigenstate, but the \CP-content can be absorbed in the amplitude analysis to allow the determination of a single effective $S$ parameter.
(In earlier analyses of the $K^+K^-\Kz$ final state,
its $\CP$ composition was determined using an isospin argument~\cite{Abe:2006gy}
and a moments analysis~\cite{Aubert:2005ja}.)

\begin{table}[!htb]
	\begin{center}
		\caption{
      Averages of $-\etacp S_{b \to q\bar q s}$ and $C_{b \to q\bar q s}$.
		}
		\vspace{0.2cm}
		\setlength{\tabcolsep}{0.0pc}
		\begin{tabular*}{\textwidth}{@{\extracolsep{\fill}}lrccc@{\hspace{-3pt}}c} \hline
        \mc{2}{l}{Experiment} & $N(B\bar{B})$ & $- \etacp S_{b \to q\bar q s}$ & $C_{b \to q\bar q s}$ & Correlation \\
	\hline
      \mc{6}{c}{$\phi \Kz$} \\
	\babar & \cite{Lees:2012kx} & 470M & $0.66 \pm 0.17 \pm 0.07$ & $0.05 \pm 0.18 \pm 0.05$ & \textendash{} \\
	\belle & \cite{Nakahama:2010nj} & 657M & $0.90 \,^{+0.09}_{-0.19}$ & $-0.04 \pm 0.20 \pm 0.10 \pm 0.02$ & \textendash{} \\
	\mc{3}{l}{\bf Average} & $0.74 \,^{+0.11}_{-0.13}$ & $0.01 \pm 0.14$ & {\small uncorrelated averages} \\
		\hline

      \mc{6}{c}{$\etapr \Kz$} \\
	\babar & \cite{:2008se} & 467M & $0.57 \pm 0.08 \pm 0.02$ & $-0.08 \pm 0.06 \pm 0.02$ & $0.03$ \\
	\belle & \cite{Chen:2006nk} & 535M & $0.64 \pm 0.10 \pm 0.04$ & $0.01 \pm 0.07 \pm 0.05$ & $0.09$ \\
	\mc{3}{l}{\bf Average} & $0.59 \pm 0.07$ & $-0.05 \pm 0.05$ & $0.04$ \\
	\mc{3}{l}{\small Confidence level} & \mc{2}{c}{\small $0.63~(0.5\sigma)$} & \\
		\hline

      \mc{6}{c}{$\KS\KS\KS$} \\
	\babar & \cite{Lees:2011nf} & 468M & $0.94 \,^{+0.21}_{-0.24} \pm 0.06$ & $-0.17 \pm 0.18 \pm 0.04$ & $0.16$ \\
	\belle & \cite{Chen:2006nk} & 535M & $0.30 \pm 0.32 \pm 0.08$ & $-0.31 \pm 0.20 \pm 0.07$ & \textendash{} \\
	\mc{3}{l}{\bf Average} & $0.72 \pm 0.19$ & $-0.24 \pm 0.14$ & $0.09$ \\
	\mc{3}{l}{\small Confidence level} & \mc{2}{c}{\small $0.26~(1.1\sigma)$} & \\
		\hline

      \mc{6}{c}{$\pi^0 K^0$} \\
	\babar & \cite{:2008se} & 467M & $0.55 \pm 0.20 \pm 0.03$ & $0.13 \pm 0.13 \pm 0.03$ & $0.06$ \\
	\belle & \cite{Fujikawa:2008pk} & 657M & $0.67 \pm 0.31 \pm 0.08$ & $-0.14 \pm 0.13 \pm 0.06$ & $-0.04$ \\
	\mc{3}{l}{\bf Average} & $0.57 \pm 0.17$ & $0.01 \pm 0.10$ & $0.02$ \\
	\mc{3}{l}{\small Confidence level} & \mc{2}{c}{\small $0.37~(0.9\sigma)$} & \\
		\hline

		\hline
      \mc{6}{c}{$\rho^0 \KS$} \\
	\babar & \cite{Aubert:2009me} & 383M & $0.35 \,^{+0.26}_{-0.31} \pm 0.06 \pm 0.03$ & $-0.05 \pm 0.26 \pm 0.10 \pm 0.03$ & \textendash{} \\
	\belle & \cite{:2008wwa} & 657M & $0.64 \,^{+0.19}_{-0.25} \pm 0.09 \pm 0.10$ & $-0.03 \,^{+0.24}_{-0.23} \pm 0.11 \pm 0.10$ & \textendash{} \\
	\mc{3}{l}{\bf Average} & $0.54 \,^{+0.18}_{-0.21}$ & $-0.06 \pm 0.20$ & {\small uncorrelated averages} \\
		\hline

      \mc{6}{c}{$\omega \KS$} \\
	\babar & \cite{:2008se} & 467M & $0.55 \,^{+0.26}_{-0.29} \pm 0.02$ & $-0.52 \,^{+0.22}_{-0.20} \pm 0.03$ & $0.03$ \\
	\belle & \cite{Abe:2006gy} & 535M & $0.11 \pm 0.46 \pm 0.07$ & $0.09 \pm 0.29 \pm 0.06$ & $-0.04$ \\
	\mc{3}{l}{\bf Average} & $0.45 \pm 0.24$ & $-0.32 \pm 0.17$ & $0.01$ \\
	\mc{3}{l}{\small Confidence level} & \mc{2}{c}{\small $0.18~(1.3\sigma)$} & \\
		\hline

      \mc{6}{c}{$f_0 \Kz$} \\
	\babar & \cite{Lees:2012kx,Aubert:2009me} & \textendash{} & $0.74 \,^{+0.12}_{-0.15}$ & $0.15 \pm 0.16$ & \textendash{} \\
	\belle & \cite{Nakahama:2010nj,:2008wwa} & \textendash{} & $0.63 \,^{+0.16}_{-0.19}$ & $0.13 \pm 0.17$ & \textendash{} \\
	\mc{3}{l}{\bf Average} & $0.69 \,^{+0.10}_{-0.12}$ & $0.14 \pm 0.12$ & {\small uncorrelated averages} \\
		\hline

      \mc{6}{c}{$f_2 \KS$} \\
	\babar & \cite{Aubert:2009me} & 383M & $0.48 \pm 0.52 \pm 0.06 \pm 0.10$ & $0.28 \,^{+0.35}_{-0.40} \pm 0.08 \pm 0.07$ & \textendash{} \\
		\hline

      \mc{6}{c}{$f_{\rm X} \KS$} \\
	\babar & \cite{Aubert:2009me} & 383M & $0.20 \pm 0.52 \pm 0.07 \pm 0.07$ & $0.13 \,^{+0.33}_{-0.35} \pm 0.04 \pm 0.09$ & \textendash{} \\
		\hline

 		\end{tabular*}
		\label{tab:cp_uta:qqs}
	\end{center}
\end{table}

\begin{table}[!htb]
	\begin{center}
		\caption{
      Averages of $-\etacp S_{b \to q\bar q s}$ and $C_{b \to q\bar q s}$ (continued).
		}
		\vspace{0.2cm}
		\setlength{\tabcolsep}{0.0pc}
		\begin{tabular*}{\textwidth}{@{\extracolsep{\fill}}lrccc@{\hspace{-3pt}}c} \hline
        \mc{2}{l}{Experiment} & $N(B\bar{B})$ & $- \etacp S_{b \to q\bar q s}$ & $C_{b \to q\bar q s}$ & Correlation \\
	\hline
      \mc{6}{c}{$\pi^0 \pi^0 \KS$} \\
	\babar & \cite{Aubert:2007ub} & 227M & $-0.72 \pm 0.71 \pm 0.08$ & $0.23 \pm 0.52 \pm 0.13$ & $-0.02$ \\
		\hline

      \mc{6}{c}{$\phi \KS \pi^0$} \\
	\babar & \cite{Aubert:2008zza} & 465M & $0.97 \,^{+0.03}_{-0.52}$ & $-0.20 \pm 0.14 \pm 0.06$ & \textendash{} \\
 		\hline

      \mc{6}{c}{$\pi^+ \pi^- \KS$ nonresonant} \\
	\babar & \cite{Aubert:2009me} & 383M & $0.01 \pm 0.31 \pm 0.05 \pm 0.09$ & $0.01 \pm 0.25 \pm 0.06 \pm 0.05$ & \textendash{} \\
 		\hline

      \mc{6}{c}{$K^+K^- \Kz$} \\
	\babar & \cite{Lees:2012kx} & 470M & $0.65 \pm 0.12 \pm 0.03$ & $0.02 \pm 0.09 \pm 0.03$ & \textendash{} \\
	\belle & \cite{Nakahama:2010nj} & 657M & $0.76 \,^{+0.14}_{-0.18}$ & $0.14 \pm 0.11 \pm 0.08 \pm 0.03$ & \textendash{} \\
	\mc{3}{l}{\bf Average} & $0.68 \,^{+0.09}_{-0.10}$ & $0.06 \pm 0.08$ & {\small uncorrelated averages} \\
		\hline



		\hline
		\end{tabular*}
		\label{tab:cp_uta:qqs2}
	\end{center}
\end{table}

The final state $\phi \KS \pi^0$ is also not a \CP-eigenstate but its
\CP-composition can be determined from an angular analysis.
Since the angular parameters are common to the $\Bz\to\phi \KS \pi^0$ and
$\Bz\to \phi \Kp\pim$ decays (because only $K\pi$ resonance contribute),
\babar\ perform a simultaneous analysis of the two final
states~\cite{Aubert:2008zza} (see subsection~\ref{sec:cp_uta:qqs:vv}).

It must be noted that Q2B parameters extracted from Dalitz plot analyses 
are constrained to lie within the physical boundary ($S_{\CP}^2 + C_{\CP}^2 < 1$)
and consequently the obtained errors are highly non-Gaussian when
the central value is close to the boundary.  
This is particularly evident in the \babar\ results for 
$\Bz \to f_0\Kz$ with $f_0 \to \pi^+\pi^-$~\cite{Aubert:2009me}.
These results must be treated with extreme caution.

As explained above,
each of the modes listed in Table~\ref{tab:cp_uta:qqs} has
different uncertainties within the Standard Model,
and so each may have a different value of $-\etacp S_{b \to q\bar q s}$.
Therefore, there is no strong motivation to make a combined average
over the different modes.
We refer to such an average as a ``na\"\i ve $s$-penguin average.''
It is na\"\i ve not only because of the neglect of the theoretical uncertainty,
but also since possible correlations of systematic effects 
between different modes are neglected.
In spite of these caveats, there remains substantial interest 
in the value of this quantity,
and therefore it is given here:
$\langle -\etacp S_{b \to q\bar q s} \rangle = 0.64 \pm 0.03$,
with confidence level $0.74~(0.3\sigma)$.
This value is in agreement with the average 
$-\etacp S_{b \to c\bar c s}$ given in Sec.~\ref{sec:cp_uta:ccs:cp_eigen}.
%
(The average for $C_{b \to q\bar q s}$ is 
$\langle C_{b \to q\bar q s} \rangle = -0.01 \pm 0.03$
with confidence level $0.74~(0.3\sigma)$.)
%
We emphasise again that we do not advocate the use of these averages,
and that the values should be treated with {\it extreme caution}, if at all.

From Table~\ref{tab:cp_uta:qqs} it may be noted 
that the averages for $-\etacp S_{b \to q\bar q s}$ in 
$\phi\KS$, $\etapr \Kz$, $f_0\KS$ and $\Kp\Km\KS$
are all now more than $5\sigma$ away from zero, 
so that $\CP$ violation in these modes can be considered well established.
There is no evidence (above $2\sigma$) for direct $\CP$ violation 
in any $b \to q \bar q s$ mode.

\mysubsubsection{Time-dependent Dalitz plot analyses: $\Bz \to K^+K^-\Kz$ and $\Bz \to \pi^+\pi^-\KS$}
\label{sec:cp_uta:qqs:dp}

\begin{figure}[htb]
  \begin{center}
    \resizebox{0.45\textwidth}{!}{
      \includegraphics{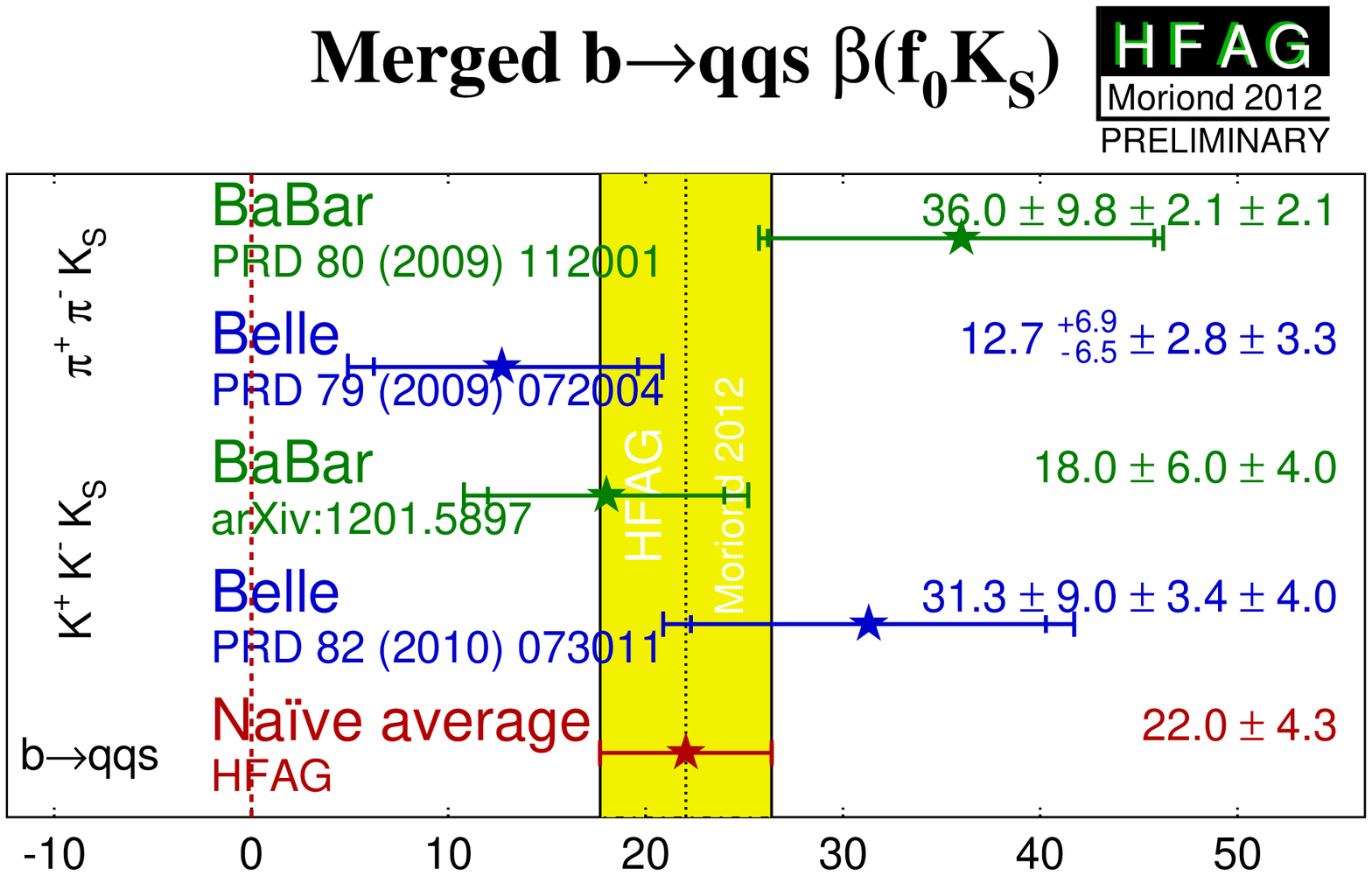}
    }
    \hfill
    \resizebox{0.45\textwidth}{!}{
      \includegraphics{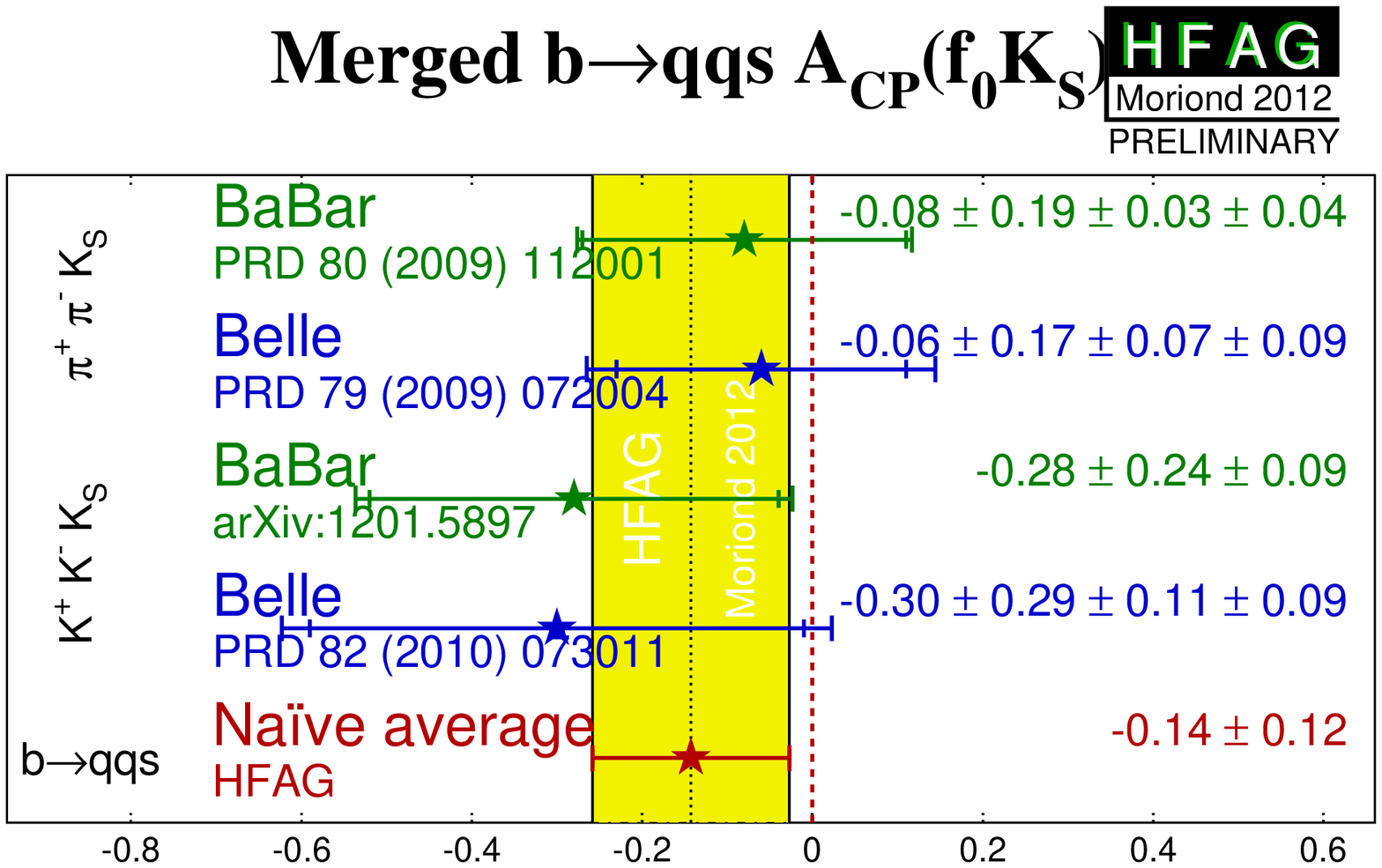}
    }
  \end{center}
  \vspace{-0.8cm}
  \caption{
    (Top)
    Averages of 
    (left) $\beta^{\rm eff} \equiv \phi_1^{\rm eff}$ and (right) $A_{CP}$
    for the $\Bz\to f_0\KS$ decay including measurements from Dalitz plot analyses of both $\Bz\to K^+K^-\KS$ and $\Bz\to \pi^+\pi^-\KS$.
  }
  \label{fig:cp_uta:qqs:f0KS}
\end{figure}

As mentioned in Sec.~\ref{sec:cp_uta:notations:dalitz:kkk0} and above,
both \babar\ and \belle\ have performed time-dependent Dalitz plot analysis of
$\Bz \to K^+K^-\Kz$ and $\Bz \to \pi^+\pi^-\KS$ decays.
The results are summarised in Tabs.~\ref{tab:cp_uta:kkk0_tddp} 
and~\ref{tab:cp_uta:pipik0_tddp}.
Averages for the $\Bz\to f_0 \KS$ decay, which contributes to both Dalitz
plots, are shown in Fig.~\ref{fig:cp_uta:qqs:f0KS}.
Results are presented in terms of the effective weak phase (from mixing and
decay) difference $\beta^{\rm eff}$ and the direct $\CP$ violation parameter
$\Acp$ ($\Acp = -C$) for each of the resonant contributions.
Note that Dalitz plot analyses, including all those included in these
averages, often suffer from ambiguous solutions -- we quote the results
corresponding to those presented as solution 1 in all cases.
Results on flavour specific amplitudes that may contribute to these Dalitz
plots (such as $K^{*+}\pi^-$) are averaged by the HFAG Rare Decays subgroup 
(Sec.~\ref{sec:rare}).

\begin{sidewaystable}
  \begin{center}
    \caption{
      Results from time-dependent Dalitz plot analysis of 
      the $\Bz \to K^+K^-\Kz$ decay.
      Correlations (not shown) are taken into account in the average.
    }
    \vspace{0.2cm}
    \setlength{\tabcolsep}{0.0pc}
    \resizebox{\textwidth}{!}{
      \begin{tabular}{l@{\hspace{2mm}}r@{\hspace{2mm}}c@{\hspace{2mm}}|@{\hspace{2mm}}c@{\hspace{2mm}}c@{\hspace{2mm}}|@{\hspace{2mm}}c@{\hspace{2mm}}c|@{\hspace{2mm}}c@{\hspace{2mm}}c} 
        \hline 
        \mc{2}{l}{Experiment} & $N(B\bar{B})$ &
        \mc{2}{c}{$\phi\KS$} & \mc{2}{c}{$f_0\KS$} & \mc{2}{c}{$K^+K^-\KS$} \\
        & & & $\beta^{\rm eff}\,(^\circ)$ & $\Acp$ & $\beta^{\rm eff}\,(^\circ)$ & $\Acp$ & $\beta^{\rm eff}\,(^\circ)$ & $\Acp$ \\
	\babar & \cite{Lees:2012kx} & 470M & $21 \pm 6 \pm 2$ & $-0.05 \pm 0.18 \pm 0.05$ & $18 \pm 6 \pm 4$ & $-0.28 \pm 0.24 \pm 0.09$ & $20.3 \pm 4.3 \pm 1.2$ & $-0.02 \pm 0.09 \pm 0.03$ \\
	\belle & \cite{Nakahama:2010nj} & 657M & $32.2 \pm 9.0 \pm 2.6 \pm 1.4$ & $0.04 \pm 0.20 \pm 0.10 \pm 0.02$ & $31.3 \pm 9.0 \pm 3.4 \pm 4.0$ & $-0.30 \pm 0.29 \pm 0.11 \pm 0.09$ & $24.9 \pm 6.4 \pm 2.1 \pm 2.5$ & $-0.14 \pm 0.11 \pm 0.08 \pm 0.03$ \\
	\mc{2}{l}{\bf Average} & & $24 \pm 5$ & $-0.01 \pm 0.14$ & $22 \pm 6$ & $-0.29 \pm 0.20$ & $21.6 \pm 3.7$ & $-0.06 \pm 0.08$ \\
	\mc{3}{l}{\small Confidence level} & \mc{6}{c}{\small $0.93~(0.1\sigma)$} \\
        \hline
      \end{tabular}
    }
    
    \label{tab:cp_uta:kkk0_tddp}
  \end{center}
\end{sidewaystable}

\begin{sidewaystable}
  \begin{center}
    \caption{
      Results from time-dependent Dalitz plot analysis of 
      the $\Bz \to \pi^+\pi^-\KS$ decay.
      Correlations (not shown) are taken into account in the average.
    }
    \vspace{0.2cm}
    \setlength{\tabcolsep}{0.0pc}
    \resizebox{\textwidth}{!}{
      \begin{tabular}{l@{\hspace{2mm}}r@{\hspace{2mm}}c@{\hspace{2mm}}|@{\hspace{2mm}}c@{\hspace{2mm}}c|@{\hspace{2mm}}c@{\hspace{2mm}}c} 
        \hline 
        \mc{2}{l}{Experiment} & $N(B\bar{B})$ & 
        \mc{2}{c}{$\rho^0\KS$} & \mc{2}{c}{$f_0\KS$} \\
        & & & $\beta^{\rm eff}$ & $\Acp$ & $\beta^{\rm eff}$ & $\Acp$ \\
        \hline
        \babar & \cite{Aubert:2009me} & 383M & $(10.2 \pm 8.9 \pm 3.0 \pm 1.9)^\circ$ & $0.05 \pm 0.26 \pm 0.10 \pm 0.03$ & $(36.0 \pm 9.8 \pm 2.1 \pm 2.1)^\circ$ & $-0.08 \pm 0.19 \pm 0.03 \pm 0.04$ \\
        \belle & \cite{:2008wwa} & 657M & $(20.0 \,^{+8.6}_{-8.5} \pm 3.2 \pm 3.5)^\circ$ & $0.03 \,^{+0.23}_{-0.24} \pm 0.11 \pm 0.10$ & $(12.7 \,^{+6.9}_{-6.5} \pm 2.8 \pm 3.3)^\circ$ & $-0.06 \pm 0.17 \pm 0.07 \pm 0.09$ \\
        \hline
        \mc{2}{l}{\bf Average} & & $16.4 \pm 6.8$ & $0.06 \pm 0.20$ & $20.6 \pm 6.2$ & $-0.07 \pm 0.14$  \\
        \mc{3}{l}{\small Confidence level} & \mc{4}{c}{\small $0.39~(0.9\sigma)$} \\
        \hline
      \end{tabular}
    }

    \vspace{2ex}

    \setlength{\tabcolsep}{0.0pc}
    \resizebox{\textwidth}{!}{
      \begin{tabular}{l@{\hspace{2mm}}r@{\hspace{2mm}}c@{\hspace{2mm}}|@{\hspace{2mm}}c@{\hspace{2mm}}c|@{\hspace{2mm}}c@{\hspace{2mm}}c} 
        \hline 
        \mc{2}{l}{Experiment} & $N(B\bar{B})$ & 
        \mc{2}{c}{$f_2\KS$} & \mc{2}{c}{$f_{\rm X}\KS$} \\
        & & & $\beta^{\rm eff}$ & $\Acp$ & $\beta^{\rm eff}$ & $\Acp$ \\
        \babar & \cite{Aubert:2009me} & 383M & $(14.9 \pm 17.9 \pm 3.1 \pm 5.2)^\circ$ & $-0.28 \,^{+0.40}_{-0.35} \pm 0.08 \pm 0.07$ & $(5.8 \pm 15.2 \pm 2.2 \pm 2.3)^\circ$ & $-0.13 \,^{+0.35}_{-0.33} \pm 0.04 \pm 0.09$ \\
        \hline
      \end{tabular}
    }

    \vspace{2ex}

    \setlength{\tabcolsep}{0.0pc}
    \resizebox{\textwidth}{!}{
      \begin{tabular}{l@{\hspace{2mm}}r@{\hspace{2mm}}c@{\hspace{2mm}}|@{\hspace{2mm}}c@{\hspace{2mm}}c|@{\hspace{2mm}}c@{\hspace{2mm}}c} 
        \hline 
        \mc{2}{l}{Experiment} & $N(B\bar{B})$ & 
        \mc{2}{c}{$\Bz \to \pi^+\pi^-\KS$ nonresonant} & \mc{2}{c}{$\chi_{c0}\KS$} \\
        & & & $\beta^{\rm eff}$ & $\Acp$ & $\beta^{\rm eff}$ & $\Acp$ \\
        \babar & \cite{Aubert:2009me} & 383M & $(0.4 \pm 8.8 \pm 1.9 \pm 3.8)^\circ$ & $-0.01 \pm 0.25 \pm 0.06 \pm 0.05$ & $(23.2 \pm 22.4 \pm 2.3 \pm 4.2)^\circ$ & $0.29 \,^{+0.44}_{-0.53} \pm 0.03 \pm 0.05$ \\
        \hline
      \end{tabular}
    }

    \label{tab:cp_uta:pipik0_tddp}
  \end{center}
\end{sidewaystable}


\mysubsubsection{Time-dependent analyses of $\Bz \to \phi \KS \pi^0$}
\label{sec:cp_uta:qqs:vv}

The final state in the decay $\Bz \to \phi \KS \pi^0$ is a mixture of \CP-even
and \CP-odd amplitudes. However, since only $\phi K^{*0}$ resonant states
contribute (in particular, $\phi K^{*0}(892)$, $\phi K^{*0}_0(1430)$ and $\phi
K^{*0}_2(1430)$ are seen), the composition can be determined from the analysis
of $B \to \phi K^+ \pi^-$, assuming only that the ratio of branching fractions
${\cal B}(K^{*0} \to \KS \pi^0)/{\cal B}(K^{*0} \to K^+ \pi^-)$ is the same
for each exited kaon state. 

\babar~\cite{Aubert:2008zza} have performed a simultaneous analysis of 
$\Bz \to \phi \KS \pi^0$ and $\Bz \to \phi K^+ \pi^-$ that is time-dependent
for the former mode and time-integrated for the latter. Such an analysis
allows, in principle, all parameters of the $\Bz \to \phi K^{*0}$ system to be
determined, including mixing-induced \CP violation effects. The latter is
determined to be $\Delta\phi_{00} = 0.28 \pm 0.42 \pm 0.04$, where
$\Delta\phi_{00}$ is half the weak phase difference between $\Bz$ and $\Bzb$
decays to $\phi K^{*0}_0(1430)$. As discussed above, this can also be
presented in terms of the quasi-two-body parameter $\sin(2\beta^{\rm eff}_{00}) =
\sin(2\beta+2\Delta\phi_{00}) = 0.97 \,^{+0.03}_{-0.52}$. The highly asymmetric
uncertainty arises due to the conversion from the phase to the sine of the
phase, and the proximity of the physical boundary. 

Similar $\sin(2\beta^{\rm eff})$ parameters can be defined for each of the
helicity amplitudes for both $\phi K^{*0}(892)$ and $\phi
K^{*0}_2(1430)$. However, the relative phases between these decays are
constrained due to the nature of the simultaneous analysis of $\Bz \to \phi
\KS \pi^0$ and $\Bz \to \phi K^+ \pi^-$, and therefore these measurements are
highly correlated. Instead of quoting all these results, \babar provide an
illustration of their measurements with the following differences: 
\begin{eqnarray}
  \sin(2\beta - 2\Delta\delta_{01}) - \sin(2\beta) & = & -0.42\,^{+0.26}_{-0.34} \, \\
  \sin(2\beta - 2\Delta\phi_{\parallel1}) - \sin(2\beta) & = & -0.32\,^{+0.22}_{-0.30} \, \\
  \sin(2\beta - 2\Delta\phi_{\perp1}) - \sin(2\beta) & = & -0.30\,^{+0.23}_{-0.32} \, \\
  \sin(2\beta - 2\Delta\phi_{\perp1}) - \sin(2\beta - 2\Delta\phi_{\parallel1})
  & = & 0.02 \pm 0.23 \, \\
  \sin(2\beta - 2\Delta\delta_{02}) - \sin(2\beta) & = & -0.10\,^{+0.18}_{-0.29} \,
\end{eqnarray}
where the first subscript indicates the helicity amplitude and the second
indicates the spin of the kaon resonance. For the complete definitions of the
$\Delta\delta$ and $\Delta\phi$ parameters, please refer to the \babar\ paper~\cite{Aubert:2008zza}.

Direct \CP violation parameters for each of the contributing helicity
amplitudes can also be measured. Again, these are determined from a
simultaneous fit of $\Bz \to \phi \KS \pi^0$ and $\Bz \to \phi K^+ \pi^-$,
with the precision being dominated by the statistics of the latter
mode. Direct \CP violation measurements are tabulated by HFAG - Rare Decays 
(Sec.~\ref{sec:rare}). 

\mysubsubsection{Time-dependent \CP asymmetries in $\Bs \to \Kp\Km$}
\label{sec:cp_uta:qqs:BstoKK}

The decay $\Bs \to \Kp\Km$ involves a $b \to u\bar{u}s$ transition, and hence has both penguin and tree contributions. Both mixing-induced and direct \CP violation effects may arise, and additional input is needed to disentangle the contributions and determine $\gamma$ and $\beta_s^{\rm eff}$. For example, the observables in $\Bd \to \pip\pim$ can be related using U-spin, as proposed by Fleischer~\cite{Fleischer:1999pa}.

The observables are $A_{\rm mix} = S_{\CP}$, $A_{\rm dir} = -C_{\CP}$, and $A_{\Delta\Gamma}$. They can all be treated as free parameters, but are physically constrained to satisfy $A_{\rm mix}^2 + A_{\rm dir}^2 + A_{\Delta\Gamma}^2 = 1$. Note that the untagged decay distribution, from which an ``effective lifetime'' can be measured, retains sensitivity to $A_{\Delta\Gamma}$. Averages of effective lifetimes are performed by the HFAG Lifetimes and Oscillations group, see Sec.~\ref{sec:life_mix}.

The observables in $\Bs \to \Kp\Km$ have been measured by LHCb, who impose the constraint mentioned above to eliminate $A_{\rm \Delta\Gamma}$. 

\begin{table}[!htb]
	\begin{center}
		\caption{
      Results from time-dependent analysis of the $\Bs \to K^{+} K^{-}$ decay.
		}
		\vspace{0.2cm}
		\setlength{\tabcolsep}{0.0pc}
		\begin{tabular*}{\textwidth}{@{\extracolsep{\fill}}lrcccc} \hline
	\mc{2}{l}{Experiment} & Sample size & $A_{\rm mix}$ & $A_{\rm dir}$ & Correlation \\
	\hline
	LHCb & \cite{LHCb-CONF-2012-007} & 0.7 ${\rm fb}^{-1}$ & $0.17 \pm 0.18 \pm 0.05$ & $0.02 \pm 0.18 \pm 0.04$ & $-0.10$ \\
	\hline
		\end{tabular*}
		\label{tab:cp_uta:BstoKK}
	\end{center}
\end{table}

\clearpage
\mysubsection{Time-dependent $\CP$ asymmetries in $b \to c\bar{c}d$ transitions
}
\label{sec:cp_uta:ccd}

The transition $b \to c\bar c d$ can occur via either a $b \to c$ tree
or a $b \to d$ penguin amplitude.  
Similarly to Eq.~(\ref{eq:cp_uta:b_to_s}), the amplitude for 
the $b \to d$ penguin can be written
\begin{equation}
  \label{eq:cp_uta:b_to_d}
  \begin{array}{ccccc}
    A_{b \to d} & = & 
    \mc{3}{l}{F_u V_{ub}V^*_{ud} + F_c V_{cb}V^*_{cd} + F_t V_{tb}V^*_{td}} \\
    & = & (F_u - F_c) V_{ub}V^*_{ud} & + & (F_t - F_c) V_{tb}V^*_{td} \\
    & = & {\cal O}(\lambda^3) & + & {\cal O}(\lambda^3). \\
  \end{array}
\end{equation}
From this it can be seen that the $b \to d$ penguin amplitude 
contains terms with different weak phases at the same order of
CKM suppression.

In the above, we have followed Eq.~(\ref{eq:cp_uta:b_to_s}) 
by eliminating the $F_c$ term using unitarity.
However, we could equally well write
\begin{equation}
  \label{eq:cp_uta:b_to_d_alt}
  \begin{array}{ccccc}
    A_{b \to d} 
    & = & (F_u - F_t) V_{ub}V^*_{ud} & + & (F_c - F_t) V_{cb}V^*_{cd}, \\
    & = & (F_c - F_u) V_{cb}V^*_{cd} & + & (F_t - F_u) V_{tb}V^*_{td}. \\
  \end{array}
\end{equation}
Since the $b \to c\bar{c}d$ tree amplitude 
has the weak phase of $V_{cb}V^*_{cd}$,
either of the above expressions allow the penguin to be decomposed into 
parts with weak phases the same and different to the tree amplitude
(the relative weak phase can be chosen to be either $\beta$ or $\gamma$).
However, if the tree amplitude dominates,
there is little sensitivity to any phase 
other than that from $\Bz$\textendash$\Bzb$ mixing.

\begin{figure}[htb]
  \begin{center}
    \begin{tabular}{cc}
      \resizebox{0.46\textwidth}{!}{
        \includegraphics{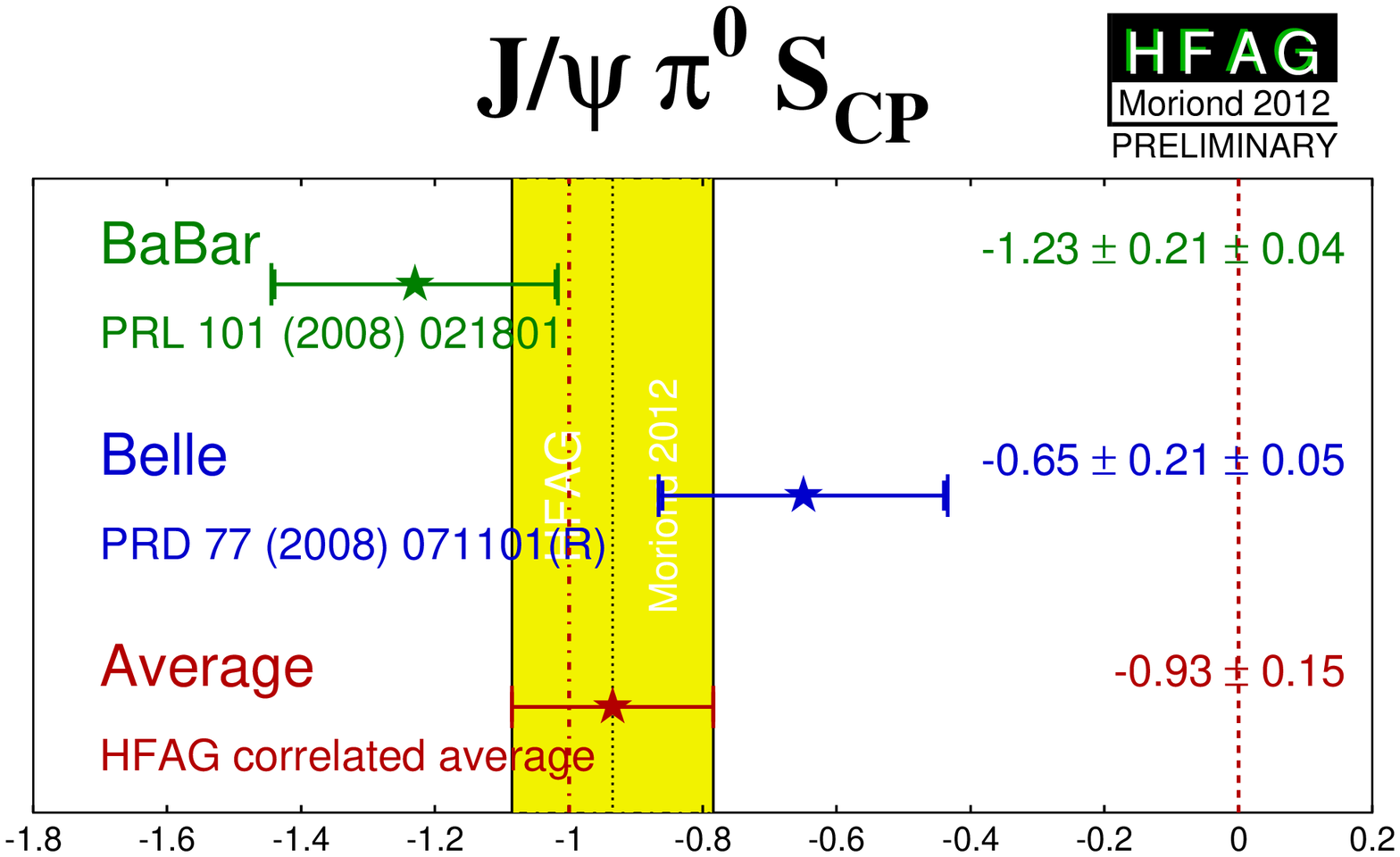}
      }
      &
      \resizebox{0.46\textwidth}{!}{
        \includegraphics{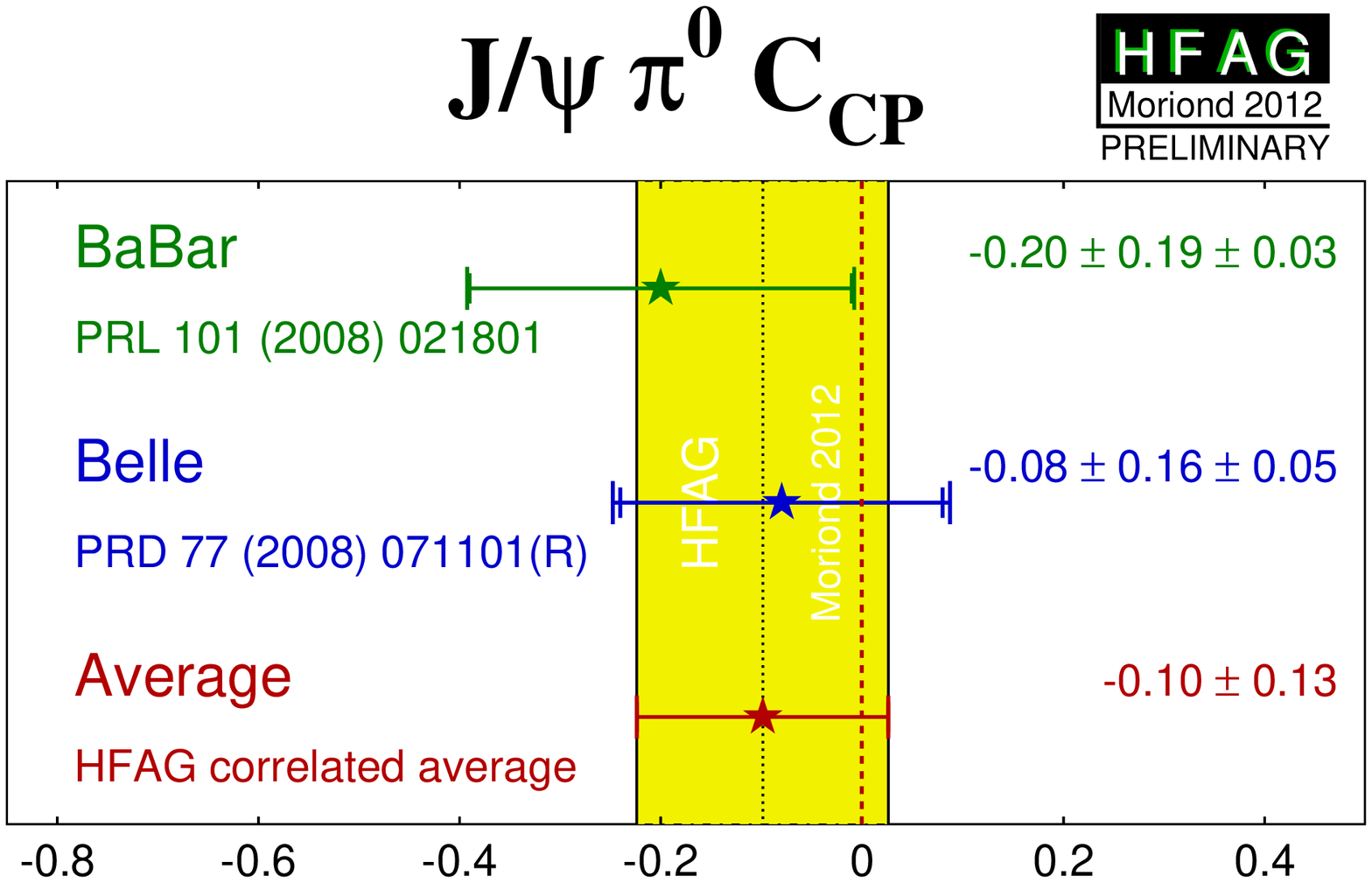}
      }
    \end{tabular}
  \end{center}
  \vspace{-0.8cm}
  \caption{
    Averages of 
    (left) $S_{b \to c\bar c d}$ and (right) $C_{b \to c\bar c d}$ 
    for the mode $\Bz \to J/ \psi \pi^0$.
  }
  \label{fig:cp_uta:ccd:psipi0}
\end{figure}

\begin{figure}[htb]
  \begin{center}
    \begin{tabular}{cc}
      \resizebox{0.46\textwidth}{!}{
        \includegraphics{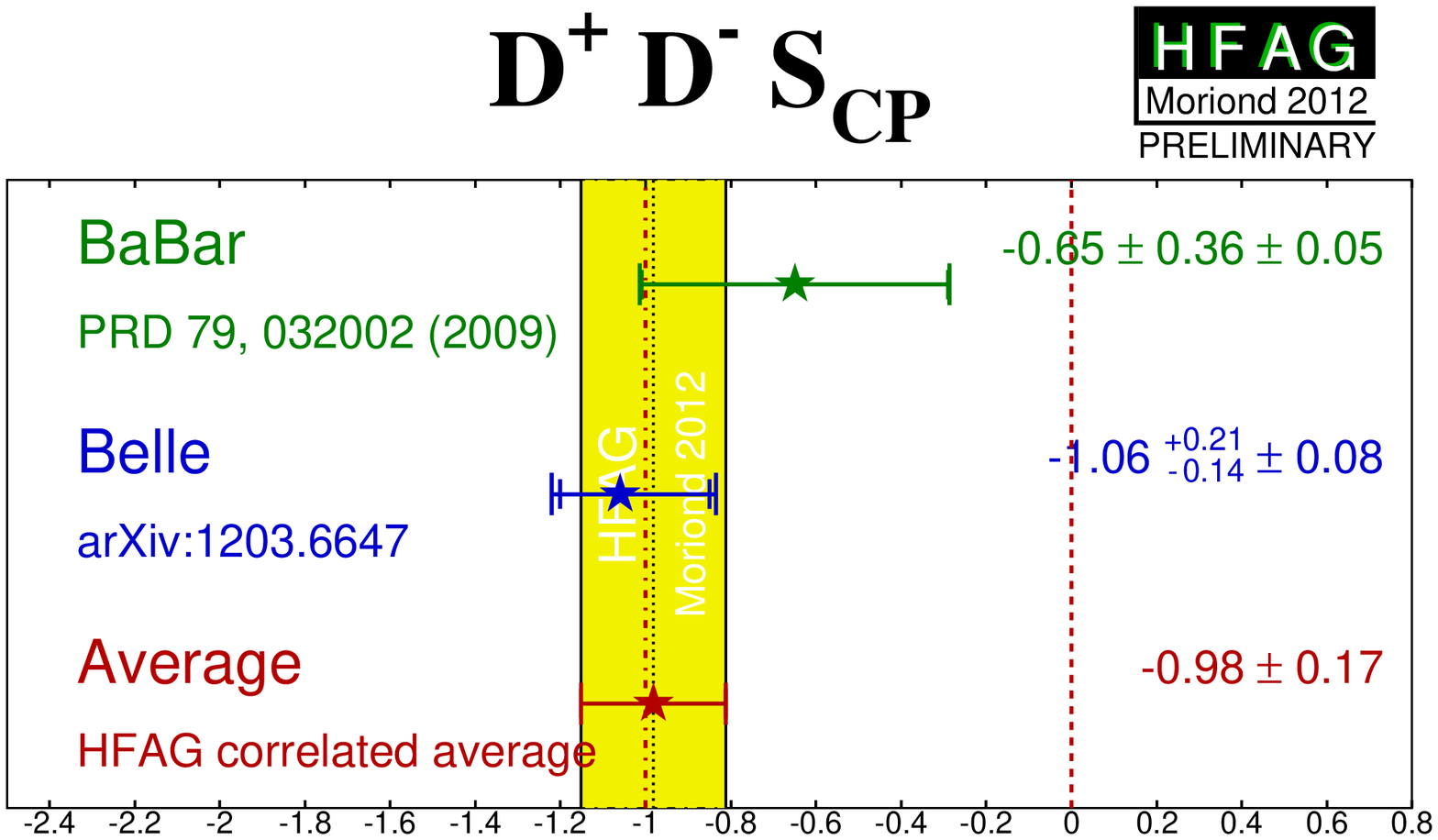}
      }
      &
      \resizebox{0.46\textwidth}{!}{
        \includegraphics{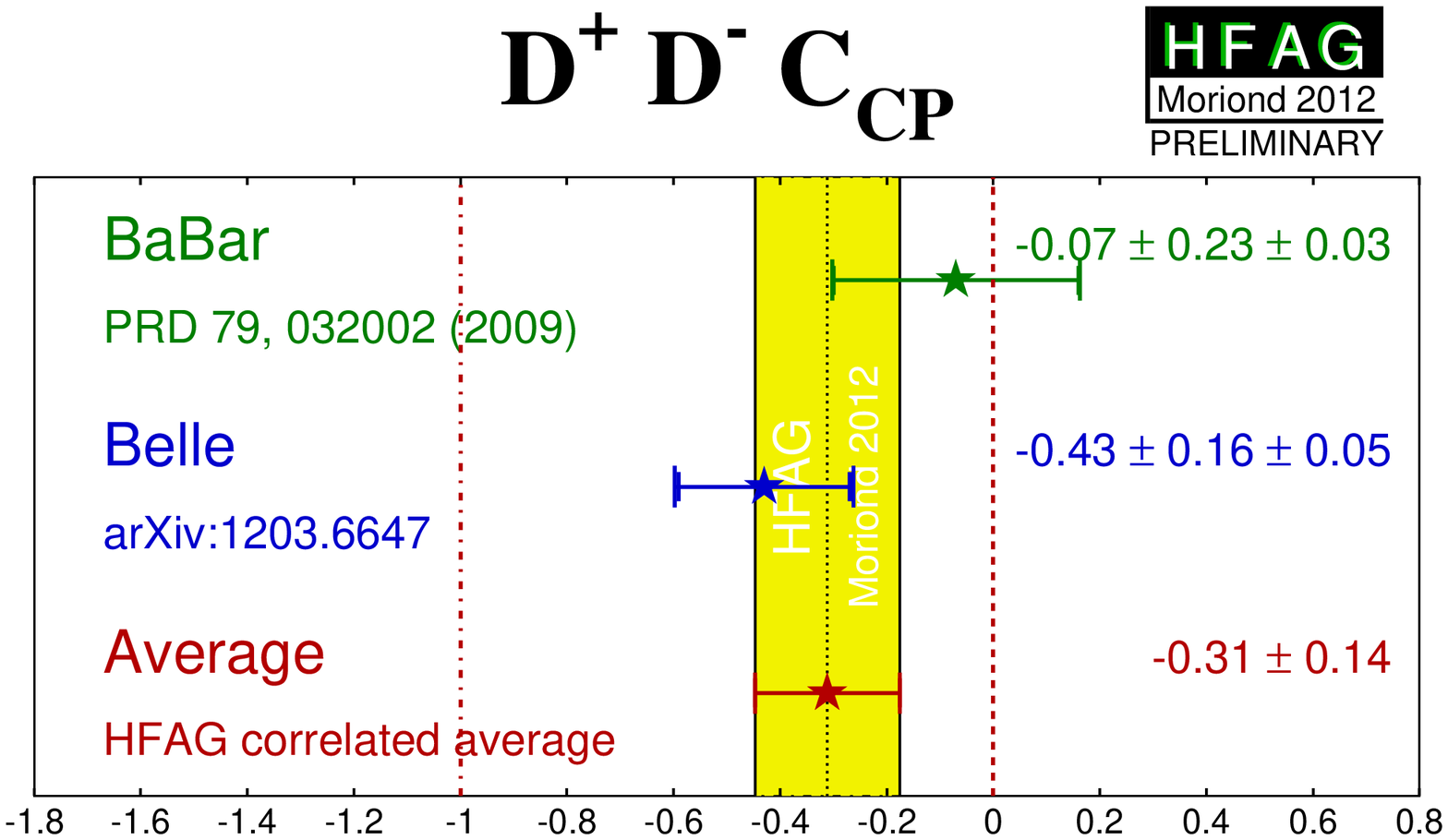}
      }
    \end{tabular}
  \end{center}
  \vspace{-0.8cm}
  \caption{
    Averages of 
    (left) $S_{b \to c\bar c d}$ and (right) $C_{b \to c\bar c d}$ 
    for the mode $\Bz \to D^+D^-$.
  }
  \label{fig:cp_uta:ccd:dd}
\end{figure}

\begin{figure}[htb]
  \begin{center}
    \begin{tabular}{cc}
      \resizebox{0.46\textwidth}{!}{
        \includegraphics{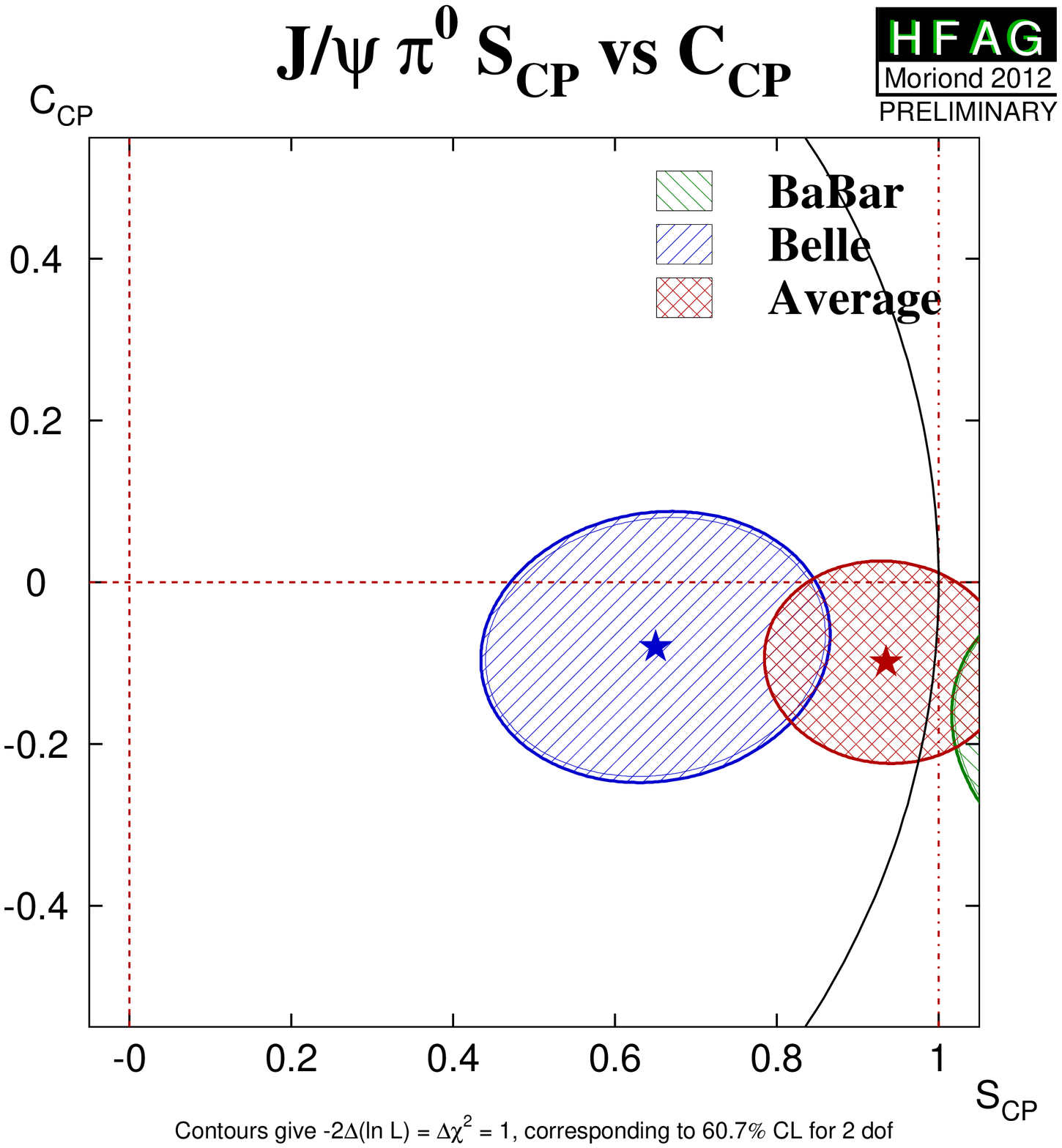}
      }
      &
      \resizebox{0.46\textwidth}{!}{
        \includegraphics{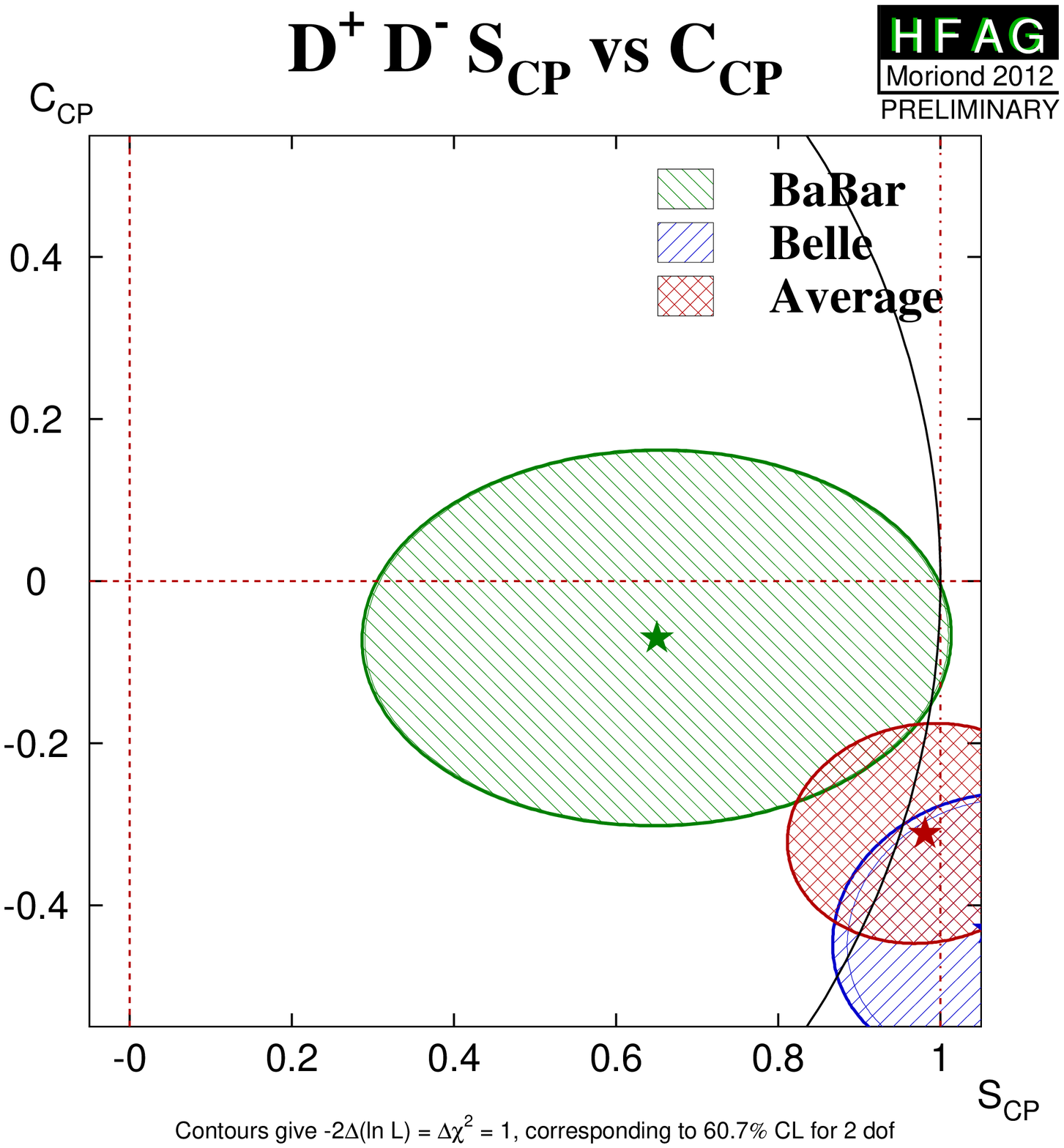}
      }
    \end{tabular}
  \end{center}
  \vspace{-0.8cm}
  \caption{
    Averages of two $b \to c\bar c d$ dominated channels,
    for which correlated averages are performed,
    in the $S_{\CP}$ \vs\ $C_{\CP}$ plane.
    (Left) $\Bz \to J/ \psi \pi^0$ and (right) $\Bz \to D^+D^-$.
  }
  \label{fig:cp_uta:ccd_SvsC}
\end{figure}

The $b \to c\bar{c}d$ transitions can be investigated with studies 
of various different final states. 
Results are available from both \babar\  and \belle\ 
using the final states $\jpsi \pi^0$, $D^+D^-$, 
$D^{*+}D^{*-}$ and $D^{*\pm}D^{\mp}$,
the averages of these results are given in Tables~\ref{tab:cp_uta:ccd1} and~\ref{tab:cp_uta:ccd2}.
The results using the $\CP$ eigenstate ($\etacp = +1$) modes
$\jpsi \pi^0$ and $D^+D^-$
are shown in Fig.~\ref{fig:cp_uta:ccd:psipi0} and 
Fig.~\ref{fig:cp_uta:ccd:dd} respectively,
with two-dimensional constraints shown in Fig.~\ref{fig:cp_uta:ccd_SvsC}.

\begin{figure}[htb]
  \begin{center}
    \begin{tabular}{cc}
      \resizebox{0.46\textwidth}{!}{
        \includegraphics{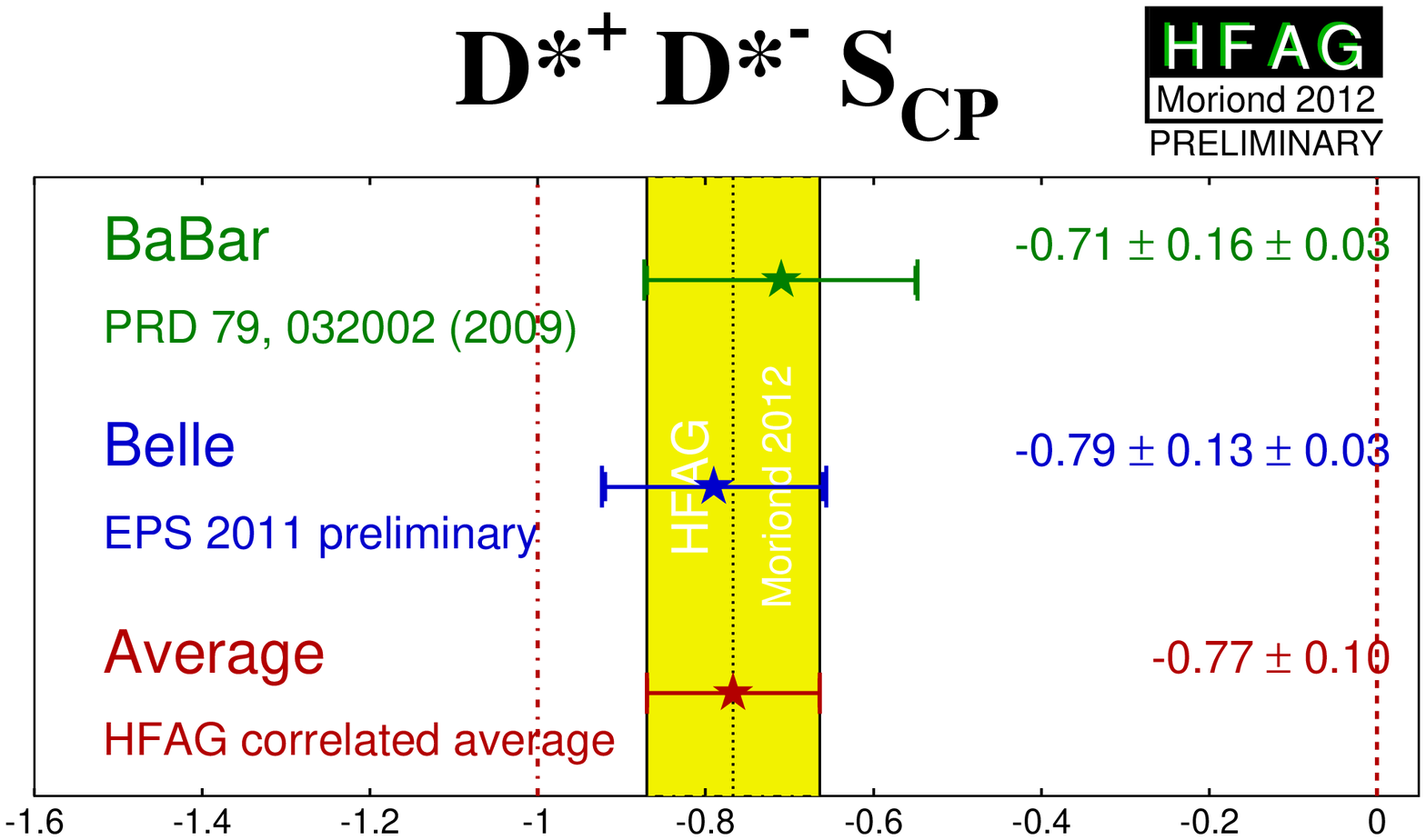}
      }
      &
      \resizebox{0.46\textwidth}{!}{
        \includegraphics{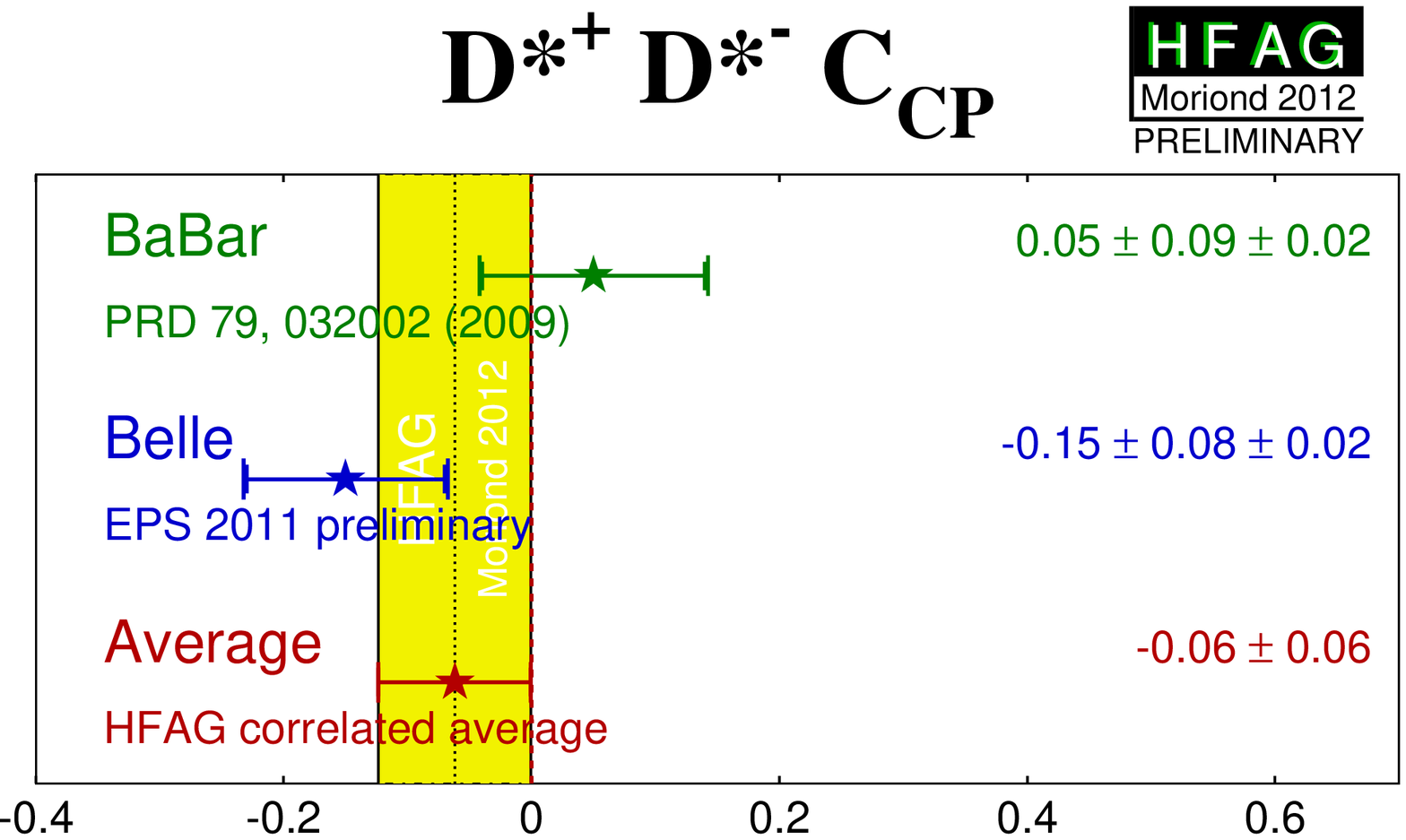}
      }
    \end{tabular}
  \end{center}
  \vspace{-0.8cm}
  \caption{
    Averages of 
    (left) $S_{b \to c\bar c d}$ and (right) $C_{b \to c\bar c d}$ 
    for the mode $\Bz \to D^{*+}D^{*-}$.
  }
  \label{fig:cp_uta:ccd:dstardstar}
\end{figure}

The vector-vector mode $D^{*+}D^{*-}$ 
is found to be dominated by the $\CP$-even longitudinally polarised component;
\babar\ measures a $\CP$-odd fraction of 
$0.158 \pm 0.028 \pm 0.006$~\cite{Aubert:2008ah} while
\belle\ measures a $\CP$-odd fraction of 
$0.125 \pm 0.043 \pm 0.023$~\cite{:2009za}.
These values, listed as $R_\perp$, are included in the averages which ensures
the correlations to be taken into account.\footnote{
  Note that the \babar\ value given in Table~\ref{tab:cp_uta:ccd2} differs from
  that given above, since that in the table is not corrected for efficiency.
}
\babar\ have also performed an additional fit in which the 
$\CP$-even and $\CP$-odd components are allowed to have different 
$\CP$ violation parameters $S$ and $C$.  
These results are included in Table~\ref{tab:cp_uta:ccd2}.
Results using $D^{*+}D^{*-}$ are shown in Fig.~\ref{fig:cp_uta:ccd:dstardstar}.


As discussed in Sec.~\ref{sec:cp_uta:notations:non_cp}, the most recent papers on the non-$\CP$ eigenstate mode $D^{*\pm}D^{\mp}$ use the ($A$, $S$, $\Delta S$, $C$, $\Delta C$) set of parameters, and we therefore perform the averages with this choice.

\begin{table}[htb]
	\begin{center}
		\caption{
     Averages for the $b \to c\bar{c}d$ modes,
     $\Bz \to J/\psi \pi^{0}$ and $D^+D^-$.
		}
		\vspace{0.2cm}
		\setlength{\tabcolsep}{0.0pc}
		\begin{tabular*}{\textwidth}{@{\extracolsep{\fill}}lrcccc} \hline
	\mc{2}{l}{Experiment} & $N(B\bar{B})$ & $S_{CP}$ & $C_{CP}$ & Correlation \\
	\hline
        \mc{6}{c}{$J/\psi \pi^{0}$} \\
	\babar & \cite{Aubert:2008bs} & 466M & $-1.23 \pm 0.21 \pm 0.04$ & $-0.20 \pm 0.19 \pm 0.03$ & $0.20$ \\
	\belle & \cite{:2007wd} & 535M & $-0.65 \pm 0.21 \pm 0.05$ & $-0.08 \pm 0.16 \pm 0.05$ & $-0.10$ \\
	\mc{3}{l}{\bf Average} & $-0.93 \pm 0.15$ & $-0.10 \pm 0.13$ & $0.04$ \\
	\mc{3}{l}{\small Confidence level} & \mc{2}{c}{\small $0.15~(1.4\sigma)$} & \\
		\hline

        \mc{6}{c}{$D^{+} D^{-}$} \\
	\babar & \cite{Aubert:2008ah} & 467M & $-0.65 \pm 0.36 \pm 0.05$ & $-0.07 \pm 0.23 \pm 0.03$ & $-0.01$ \\
	\belle & \cite{Rohrken:2012ta} & 772M & $-1.06 \,^{+0.21}_{-0.14} \pm 0.08$ & $-0.43 \pm 0.16 \pm 0.05$ & $-0.12$ \\
	\mc{3}{l}{\bf Average} & $-0.98 \pm 0.17$ & $-0.31 \pm 0.14$ & $-0.08$ \\
	\mc{3}{l}{\small Confidence level} & \mc{2}{c}{\small $0.26~(1.1\sigma)$} & \\
		\hline
 		\end{tabular*}
 		\label{tab:cp_uta:ccd1}
 	\end{center}
 \end{table}

\begin{sidewaystable}
 	\begin{center}
 		\caption{
      Averages for the $b \to c\bar{c}d$ modes,
      $D^{*+} D^{*-}$ and $D^{*\pm}D^\mp$.
 		}

 		\begin{tabular*}{\textwidth}{@{\extracolsep{\fill}}lrcccc} \hline
 		\mc{2}{l}{Experiment} & $N(B\bar{B})$ & $S_{CP}$ & $C_{CP}$ & $R_\perp$ \\
 		\hline
        \mc{6}{c}{$D^{*+} D^{*-}$} \\
	\babar & \cite{Aubert:2008ah} & 467M & $-0.71 \pm 0.16 \pm 0.03$ & $0.05 \pm 0.09 \pm 0.02$ & $0.17 \pm 0.03$ \\
	\belle & \cite{belle:dstardstar:prelim} & 772M & $-0.79 \pm 0.13 \pm 0.03$ & $-0.15 \pm 0.08 \pm 0.02$ & $0.14 \pm 0.02 \pm 0.01$ \\
	\mc{3}{l}{\bf Average} & $-0.77 \pm 0.10$ & $-0.06 \pm 0.06$ & $0.15 \pm 0.02$ \\
	\mc{3}{l}{\small Confidence level} & \mc{3}{c}{\small $0.31~(1.0\sigma)$} \\
		\hline
		\end{tabular*}

                \vspace{2ex}

    \resizebox{\textwidth}{!}{
		\begin{tabular}{@{\extracolsep{2mm}}lrcccccc} \hline
	\mc{2}{l}{Experiment} & $N(B\bar{B})$ & $S_{CP+}$ & $C_{CP+}$ & $S_{CP-}$ & $C_{CP-}$ & $R_\perp$ \\
	\hline
        \mc{7}{c}{$D^{*+} D^{*-}$} \\
	\babar & \cite{Aubert:2008ah} & 467M & $-0.76 \pm 0.16 \pm 0.04$ & $0.02 \pm 0.12 \pm 0.02$ & $-1.81 \pm 0.71 \pm 0.16$ & $0.41 \pm 0.50 \pm 0.08$ & $0.15 \pm 0.03$ \\
		\hline
		\end{tabular}
    }

                \vspace{2ex}

    \resizebox{\textwidth}{!}{
		\begin{tabular}{@{\extracolsep{2mm}}lrcccccc} \hline
	\mc{2}{l}{Experiment} & $N(B\bar{B})$ & $S$ & $C$ & $\Delta S$ & $\Delta C$ & ${\cal A}$ \\
        \hline
        \mc{8}{c}{$D^{*\pm} D^{\mp}$} \\
	\babar & \cite{Aubert:2008ah} & 467M & $-0.68 \pm 0.15 \pm 0.04$ & $0.04 \pm 0.12 \pm 0.03$ & $0.05 \pm 0.15 \pm 0.02$ & $0.04 \pm 0.12 \pm 0.03$ & $0.01 \pm 0.05 \pm 0.01$ \\
	\belle & \cite{Rohrken:2012ta} & 772M & $-0.78 \pm 0.15 \pm 0.05$ & $-0.01 \pm 0.11 \pm 0.04$ & $-0.13 \pm 0.15 \pm 0.04$ & $0.12 \pm 0.11 \pm 0.03$ & $0.06 \pm 0.05 \pm 0.02$ \\
	\mc{3}{l}{\bf Average} & $-0.73 \pm 0.11$ & $0.01 \pm 0.09$ & $-0.04 \pm 0.11$ & $0.08 \pm 0.08$ & $0.03 \pm 0.04$ \\
	\mc{3}{l}{\small Confidence level} & {\small $0.65~(0.5\sigma)$} & {\small $0.77~(0.3\sigma)$} & {\small $0.41~(0.8\sigma)$} & {\small $0.63~(0.5\sigma)$} & {\small $0.48~(0.7\sigma)$} \\
        \hline
                \end{tabular}
    }
		\label{tab:cp_uta:ccd2}
	\end{center}
\end{sidewaystable}

In the absence of the penguin contribution (tree dominance),
the time-dependent parameters would be given by
$S_{b \to c\bar c d} = - \etacp \sin(2\beta)$,
$C_{b \to c\bar c d} = 0$,
$S_{+-} = \sin(2\beta + \delta)$,
$S_{-+} = \sin(2\beta - \delta)$,
$C_{+-} = - C_{-+}$ and 
${\cal A} = 0$,
where $\delta$ is the strong phase difference between the 
$D^{*+}D^-$ and $D^{*-}D^+$ decay amplitudes.
In the presence of the penguin contribution,
there is no clean interpretation in terms of CKM parameters,
however
direct $\CP$ violation may be observed as any of
$C_{b \to c\bar c d} \neq 0$, $C_{+-} \neq - C_{-+}$ or $A_{+-} \neq 0$.

\begin{figure}[htb]
  \begin{center}
    \begin{tabular}{cc}
      \resizebox{0.46\textwidth}{!}{
        \includegraphics{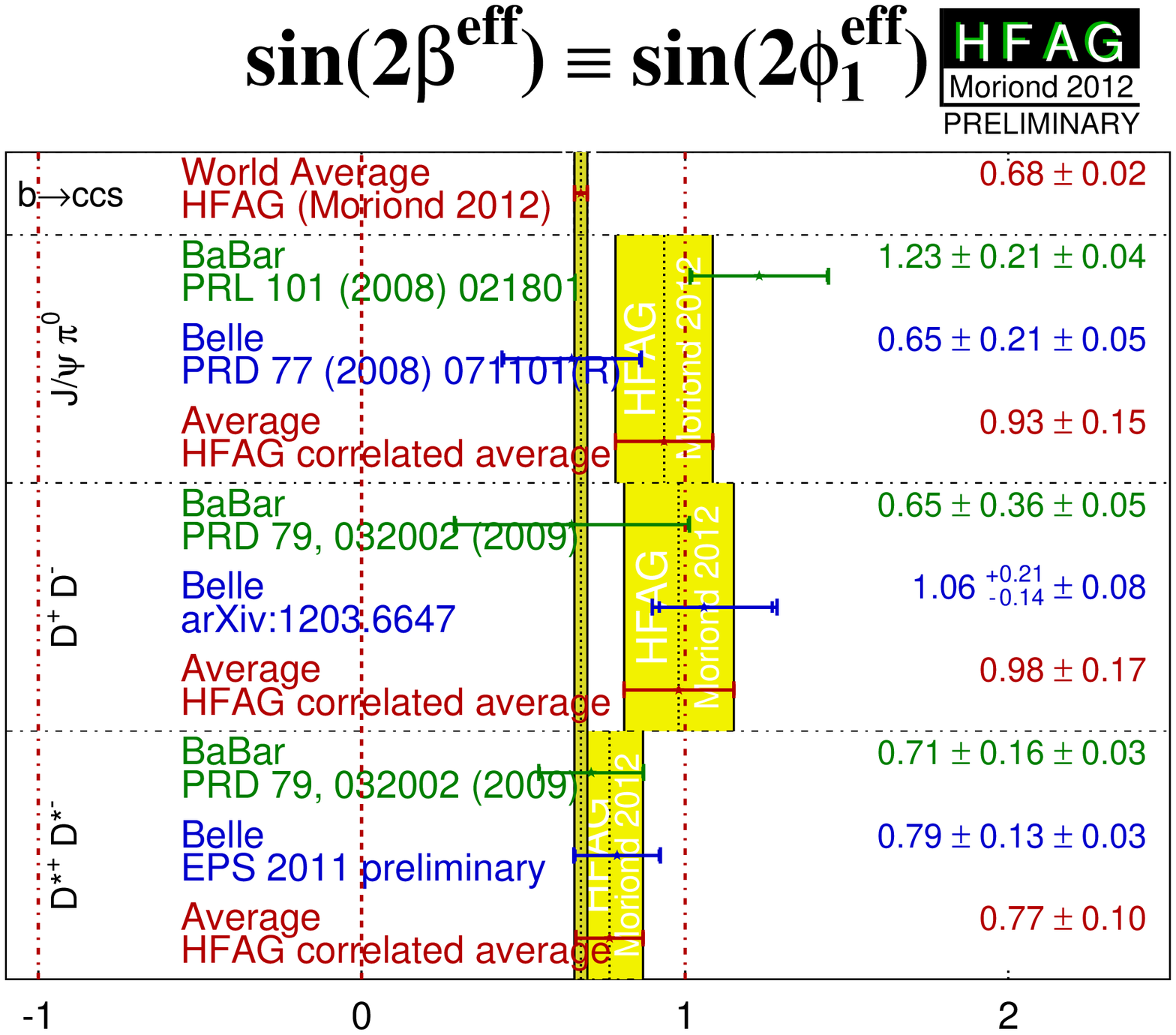}
      }
      &
      \resizebox{0.46\textwidth}{!}{
        \includegraphics{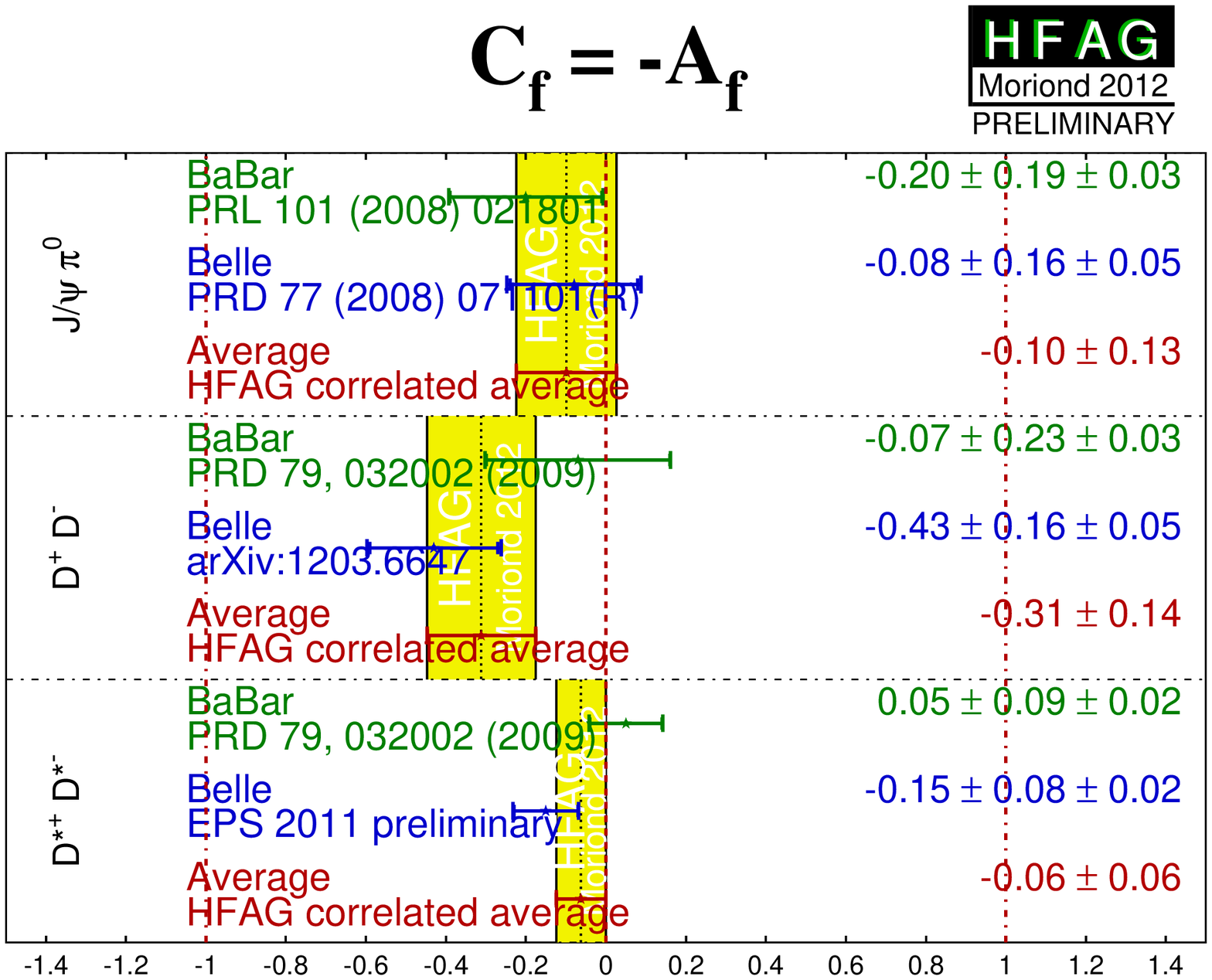}
      }
    \end{tabular}
  \end{center}
  \vspace{-0.8cm}
  \caption{
    Averages of 
    (left) $-\etacp S_{b \to c\bar c d}$ and (right) $C_{b \to c\bar c d}$.
    The $-\etacp S_{b \to q\bar q s}$ figure compares the results to 
    the world average 
    for $-\etacp S_{b \to c\bar c s}$ (see Section~\ref{sec:cp_uta:ccs:cp_eigen}).
  }
  \label{fig:cp_uta:ccd}
\end{figure}

\begin{figure}[htb]
  \begin{center}
    \resizebox{0.66\textwidth}{!}{
      \includegraphics{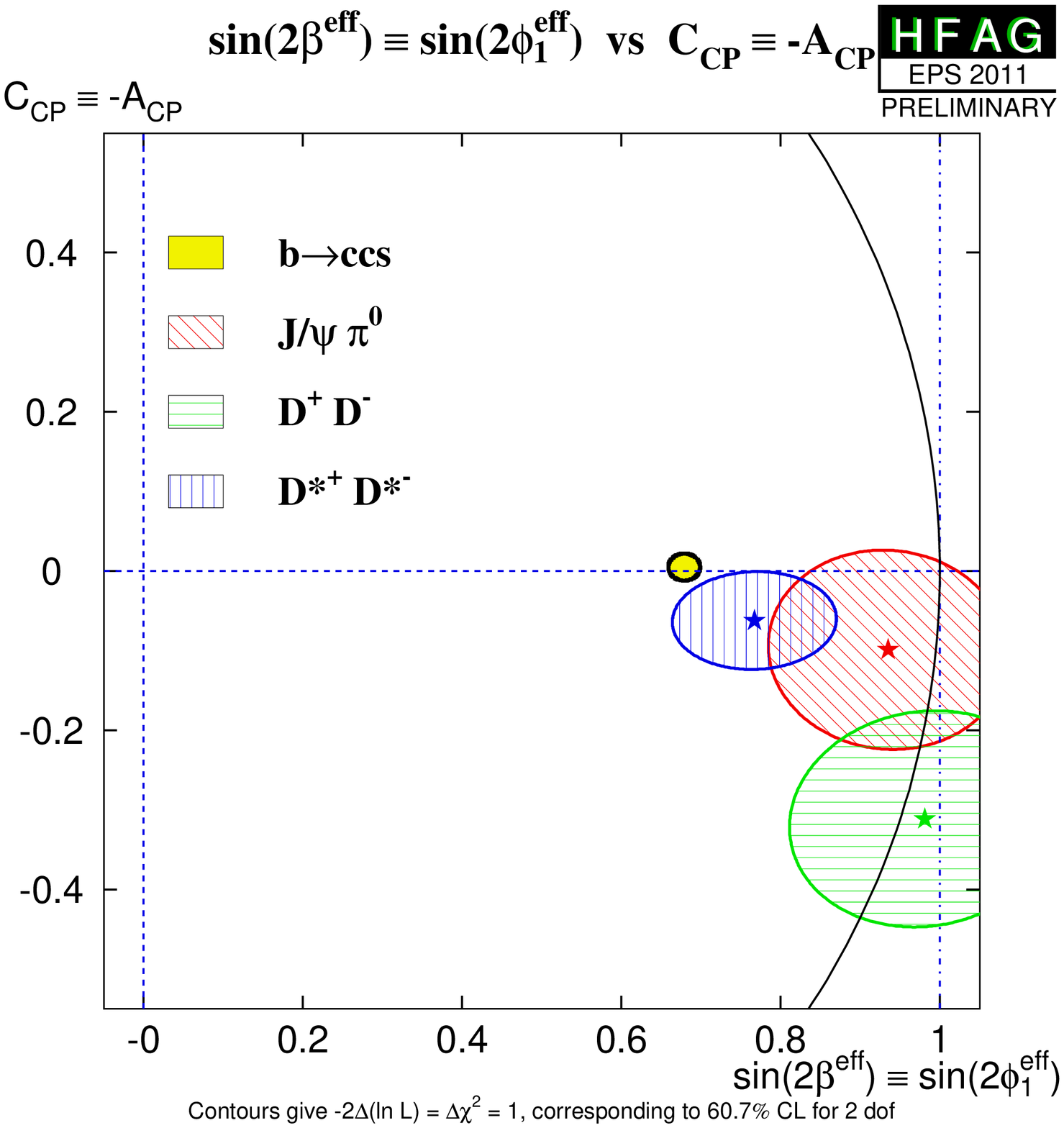}
    }
  \end{center}
  \vspace{-0.8cm}
  \caption{
    Compilation of constraints in the 
    $-\etacp S_{b \to c\bar c d}$ \vs\ $C_{b \to c\bar c d}$ plane.
  }
  \label{fig:cp_uta:ccd_SvsC-all}
\end{figure}

The averages for the $b \to c\bar c d$ modes 
are shown in Figs.~\ref{fig:cp_uta:ccd} and~\ref{fig:cp_uta:ccd_SvsC-all}.
Results are consistent with tree dominance,
and with the Standard Model,
though the \belle\ results in $\Bz \to D^+D^-$~\cite{Fratina:2007zk}
show an indication of direct $\CP$ violation,
and hence a non-zero penguin contribution.
The average of $S_{b \to c\bar c d}$ in both $J/\psi \pi^{0}$ and
$D^{*+}D^{*-}$ final states is more than $5\sigma$ from zero, corresponding to
observations of \CP violation in these decay channels.,
That in the $D^+D^-$ final state is more than $3\sigma$ from zero;
however, due to the large uncertainty and possible non-Gaussian effects,
any strong conclusion should be deferred.


\clearpage
\mysubsection{Time-dependent $\CP$ asymmetries in $b \to q\bar{q}d$ transitions
}
\label{sec:cp_uta:qqd}

Decays such as $\Bz\to\KS\KS$ are pure $b \to q\bar{q}d$ penguin transitions.
As shown in Eq.~\ref{eq:cp_uta:b_to_d},
this diagram has different contributing weak phases,
and therefore the observables are sensitive to the difference 
(which can be chosen to be either $\beta$ or $\gamma$).
Note that if the contribution with the top quark in the loop dominates,
the weak phase from the decay amplitudes should cancel that from mixing,
so that no $\CP$ violation (neither mixing-induced nor direct) occurs.
Non-zero contributions from loops with intermediate up and charm quarks
can result in both types of effect 
(as usual, a strong phase difference is required for direct $\CP$ violation
to occur).

\begin{figure}[htb]
  \begin{center}
    \begin{tabular}{cc}
      \resizebox{0.46\textwidth}{!}{
        \includegraphics{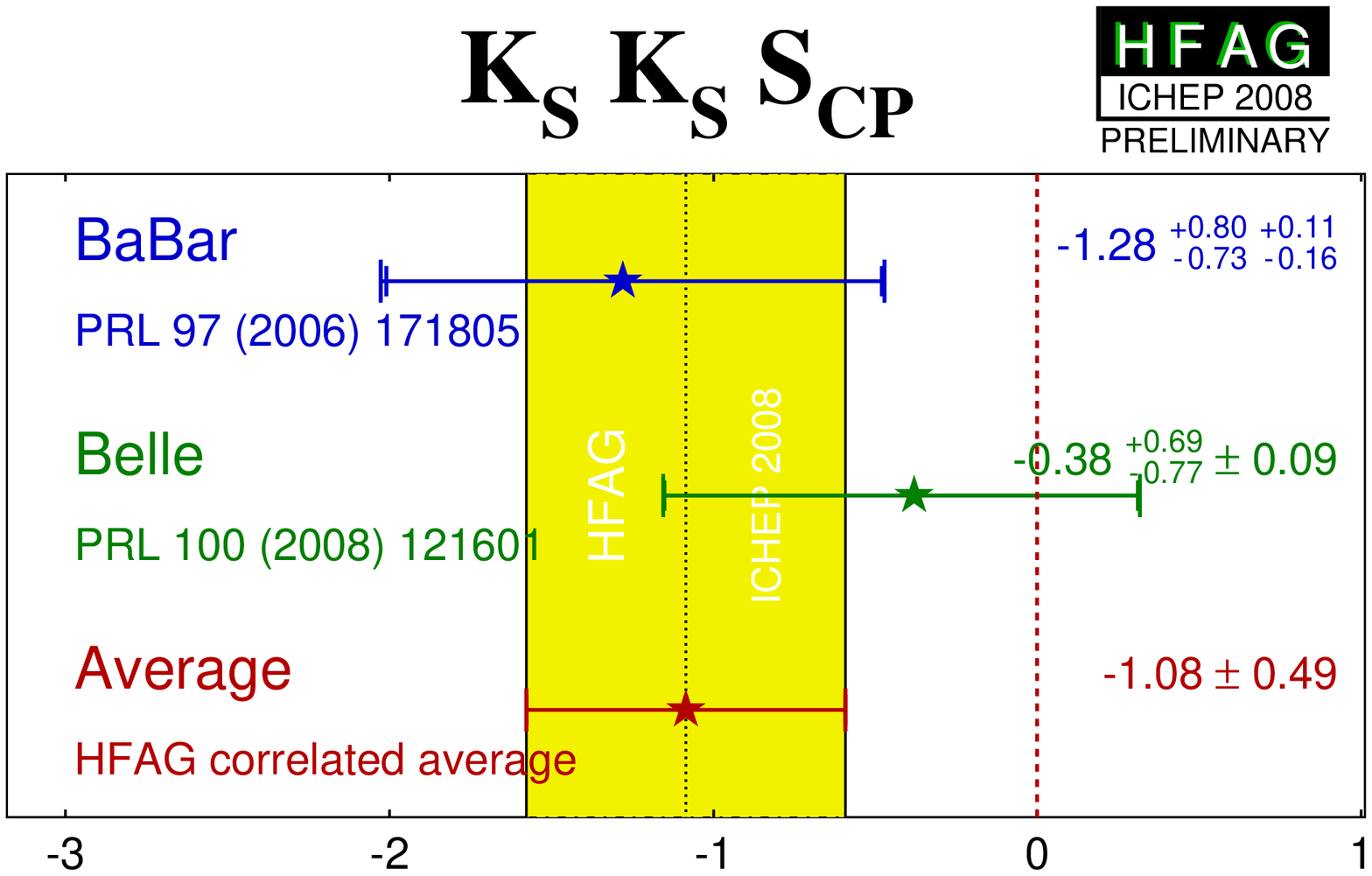}
      }
      &
      \resizebox{0.46\textwidth}{!}{
        \includegraphics{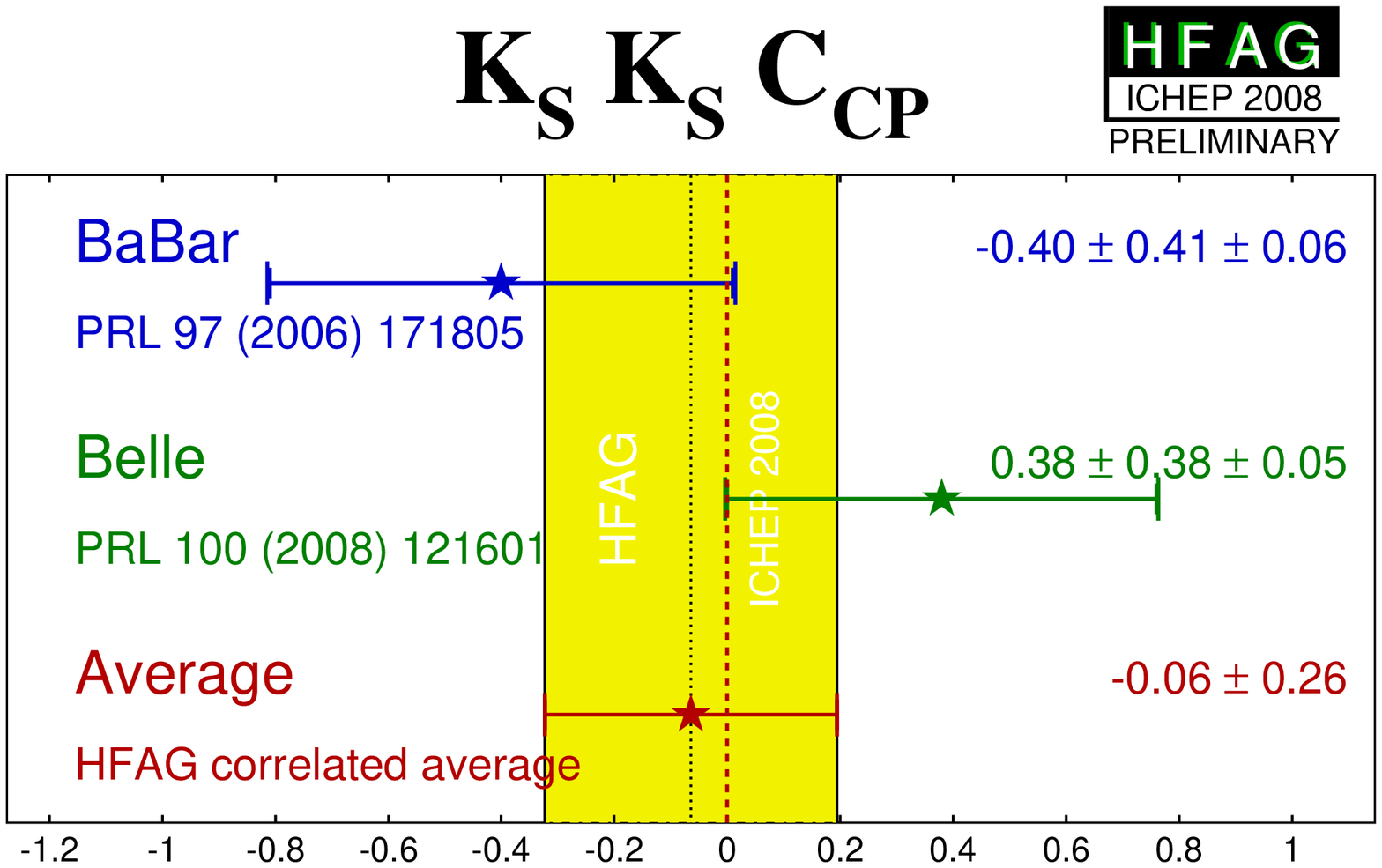}
      }
    \end{tabular}
  \end{center}
  \vspace{-0.8cm}
  \caption{
    Averages of 
    (left) $S_{b \to q\bar q d}$ and (right) $C_{b \to q\bar q d}$ 
    for the mode $\Bz \to \KS\KS$.
  }
  \label{fig:cp_uta:qqd:ksks}
\end{figure}

Both \babar~\cite{Aubert:2006gm} and \belle~\cite{Nakahama:2007dg}
have performed time-dependent analyses of $\Bz\to\KS\KS$.
The results are shown in Table~\ref{tab:cp_uta:qqd}
and Fig.~\ref{fig:cp_uta:qqd:ksks}.

\begin{table}[htb]
	\begin{center}
		\caption{
			Results for $\Bz \to \KS\KS$.
		}
		\vspace{0.2cm}
		\setlength{\tabcolsep}{0.0pc}
		\begin{tabular*}{\textwidth}{@{\extracolsep{\fill}}lrcccc} \hline
	\mc{2}{l}{Experiment} & $N(B\bar{B})$ & $S_{CP}$ & $C_{CP}$ & Correlation \\
	\hline
	\babar & \cite{Aubert:2006gm} & 350M & $-1.28 \,^{+0.80}_{-0.73} \,^{+0.11}_{-0.16}$ & $-0.40 \pm 0.41 \pm 0.06$ & $-0.32$ \\
	\belle & \cite{Nakahama:2007dg} & 657M & $-0.38 \,^{+0.69}_{-0.77} \pm 0.09$ & $0.38 \pm 0.38 \pm 0.05$ & $0.48$ \\
	\hline
	\mc{3}{l}{\bf Average} & $-1.08 \pm 0.49$ & $-0.06 \pm 0.26$ & $0.14$ \\
	\mc{3}{l}{\small Confidence level} & \mc{2}{c}{\small $0.29~(1.1\sigma)$} & \\
		\hline
		\end{tabular*}
		\label{tab:cp_uta:qqd}
	\end{center}
\end{table}

\clearpage
\mysubsection{Time-dependent asymmetries in $b \to s\gamma$ transitions
}
\label{sec:cp_uta:bsg}

The radiative decays $b \to s\gamma$ produce photons 
which are highly polarised in the Standard Model.
The decays $\Bz \to F \gamma$ and $\Bzb \to F \gamma$ 
produce photons with opposite helicities, 
and since the polarisation is, in principle, observable,
these final states cannot interfere.
The finite mass of the $s$ quark introduces small corrections
to the limit of maximum polarisation,
but any large mixing induced $\CP$ violation would be a signal for new physics.
Since a single weak phase dominates the $b \to s \gamma$ transition in the 
Standard Model, the cosine term is also expected to be small.

Atwood {\it et al.}~\cite{Atwood:2004jj} have shown that 
an inclusive analysis with respect to $\KS\pi^0\gamma$ can be performed,
since the properties of the decay amplitudes 
are independent of the angular momentum of the $\KS\pi^0$ system. 
However, if non-dipole operators contribute significantly to the amplitudes, 
then the Standard Model mixing-induced $\CP$ violation could be larger 
than the na\"\i ve expectation 
$S \simeq -2 (m_s/m_b) \sin \left(2\beta\right)$~\cite{Grinstein:2004uu,Grinstein:2005nu}.
In this case, 
the $\CP$ parameters may vary over the $\KS\pi^0\gamma$ Dalitz plot, 
for example as a function of the $\KS\pi^0$ invariant mass.
Explicit calculations indicate such corrections are small
for exclusive final states~\cite{Matsumori:2005ax,Ball:2006cva}.

With the above in mind, 
we quote two averages: one for $K^*(892)$ candidates only, 
and the other one for the inclusive $\KS\pi^0\gamma$ decay (including the $K^*(892)$).
If the Standard Model dipole operator is dominant, 
both should give the same quantities 
(the latter naturally with smaller statistical error). 
If not, care needs to be taken in interpretation of the inclusive parameters, 
while the results on the $K^*(892)$ resonance remain relatively clean.
Results from \babar~\cite{Aubert:2008gy} and \belle~\cite{Ushiroda:2006fi} are
used for both averages; both experiments use the invariant mass range 
$0.60 \ {\rm GeV}/c^2 < M_{\KS\pi^0} < 1.80 \ {\rm GeV}/c^2$
in the inclusive analysis.
In addition to the $\KS\pi^0\gamma$ decay, \babar\ have presented results
using $\KS\eta\gamma$~\cite{Aubert:2008js}, and \belle\ have presented results
using $\KS\rho\gamma$~\cite{Li:2008qma} and $\KS\phi\gamma$~\cite{Sahoo:2011zd}.

\begin{table}[htb]
	\begin{center}
		\caption{
      Averages for $b \to s \gamma$ modes.
		}
		\vspace{0.2cm}
		\setlength{\tabcolsep}{0.0pc}
		\begin{tabular*}{\textwidth}{@{\extracolsep{\fill}}lrcccc} \hline
	\mc{2}{l}{Experiment} & $N(B\bar{B})$ & $S_{CP} (b \to s \gamma)$ & $C_{CP} (b \to s \gamma)$ & Correlation \\
        \hline
        \mc{6}{c}{$\Kstar(892)\gamma$} \\
	\babar & \cite{Aubert:2008gy} & 467M & $-0.03 \pm 0.29 \pm 0.03$ & $-0.14 \pm 0.16 \pm 0.03$ & $0.05$ \\
	\belle & \cite{Ushiroda:2006fi} & 535M & $-0.32 \,^{+0.36}_{-0.33} \pm 0.05$ & $0.20 \pm 0.24 \pm 0.05$ & $0.08$ \\
	\mc{3}{l}{\bf Average} & $-0.16 \pm 0.22$ & $-0.04 \pm 0.14$ & $0.06$ \\
	\mc{3}{l}{\small Confidence level} & \mc{2}{c}{\small $0.40~(0.9\sigma)$} & \\
		\hline

        \mc{6}{c}{$\KS \pi^0 \gamma$ (including $\Kstar(892)\gamma$)} \\
	\babar & \cite{Aubert:2008gy} & 467M & $-0.17 \pm 0.26 \pm 0.03$ & $-0.19 \pm 0.14 \pm 0.03$ & $0.04$ \\
	\belle & \cite{Ushiroda:2006fi} & 535M & $-0.10 \pm 0.31 \pm 0.07$ & $0.20 \pm 0.20 \pm 0.06$ & $0.08$ \\
	\mc{3}{l}{\bf Average} & $-0.15 \pm 0.20$ & $-0.07 \pm 0.12$ & $0.05$ \\
        \mc{3}{l}{\small Confidence level} & \mc{2}{c}{\small $0.30~(1.0\sigma)$} & \\

		\hline

        \mc{6}{c}{$\KS \eta \gamma$} \\
	\babar & \cite{Aubert:2008js} & 465M & $-0.18 \,^{+0.49}_{-0.46} \pm 0.12$ & $-0.32 \,^{+0.40}_{-0.39} \pm 0.07$ & $-0.17$ \\
	\hline

        \mc{6}{c}{$\KS \rho^0 \gamma$} \\
	\belle & \cite{Li:2008qma} & 657M & $0.11 \pm 0.33 \,^{+0.05}_{-0.09}$ & $-0.05 \pm 0.18 \pm 0.06$ & $0.04$ \\
	\hline

        \mc{6}{c}{$\KS \phi \gamma$} \\
	\belle & \cite{Sahoo:2011zd} & 772M & $0.74 \,^{+0.72}_{-1.05} \,^{+0.10}_{-0.24}$ & $-0.35 \pm 0.58 \,^{+0.10}_{-0.23}$ & \textendash{} \\
	\hline
		\end{tabular*}
		\label{tab:cp_uta:bsg}
	\end{center}
\end{table}

\begin{figure}[htb]
  \begin{center}
    \begin{tabular}{cc}
      \resizebox{0.46\textwidth}{!}{
        \includegraphics{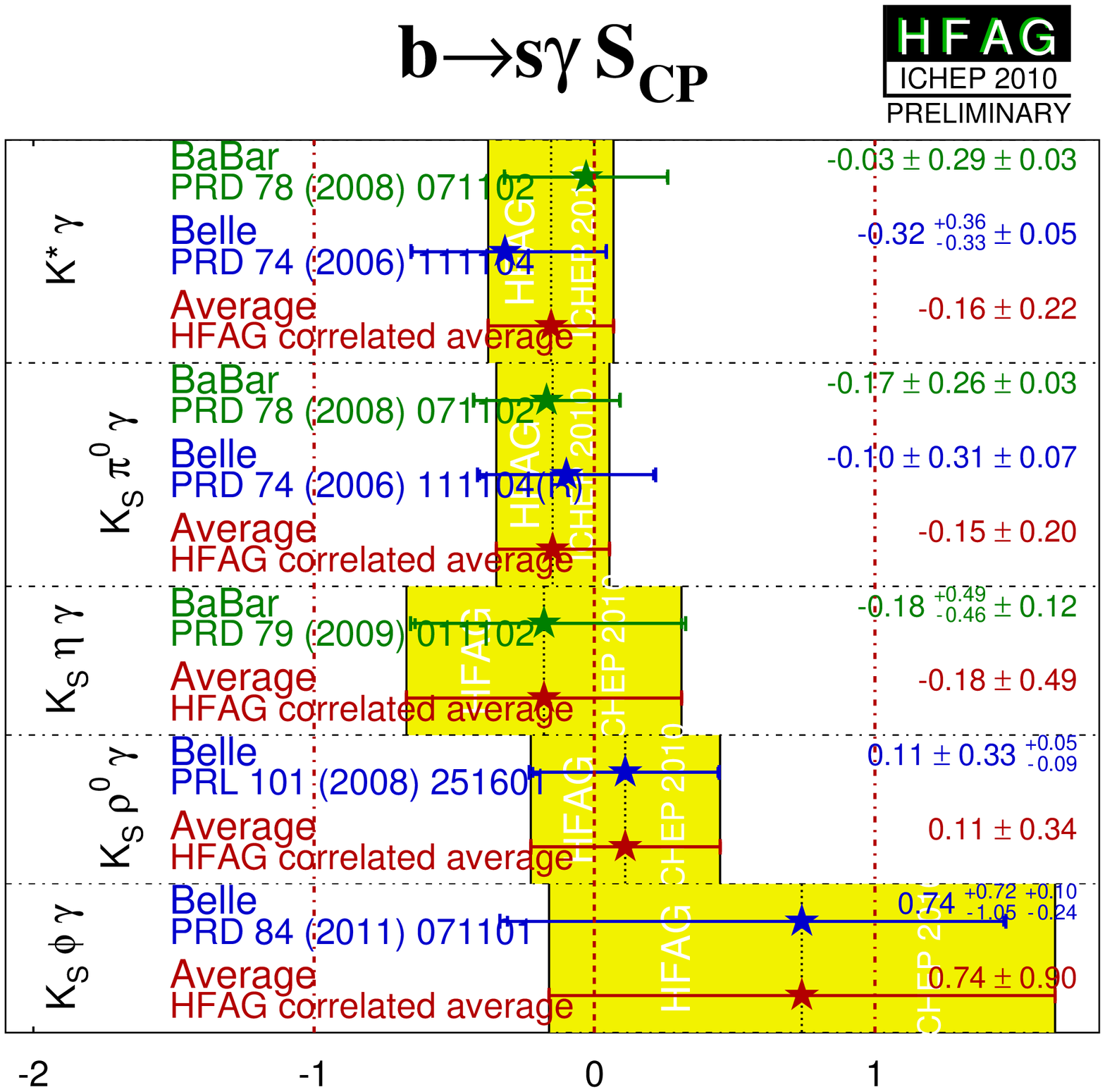}
      }
      &
      \resizebox{0.46\textwidth}{!}{
        \includegraphics{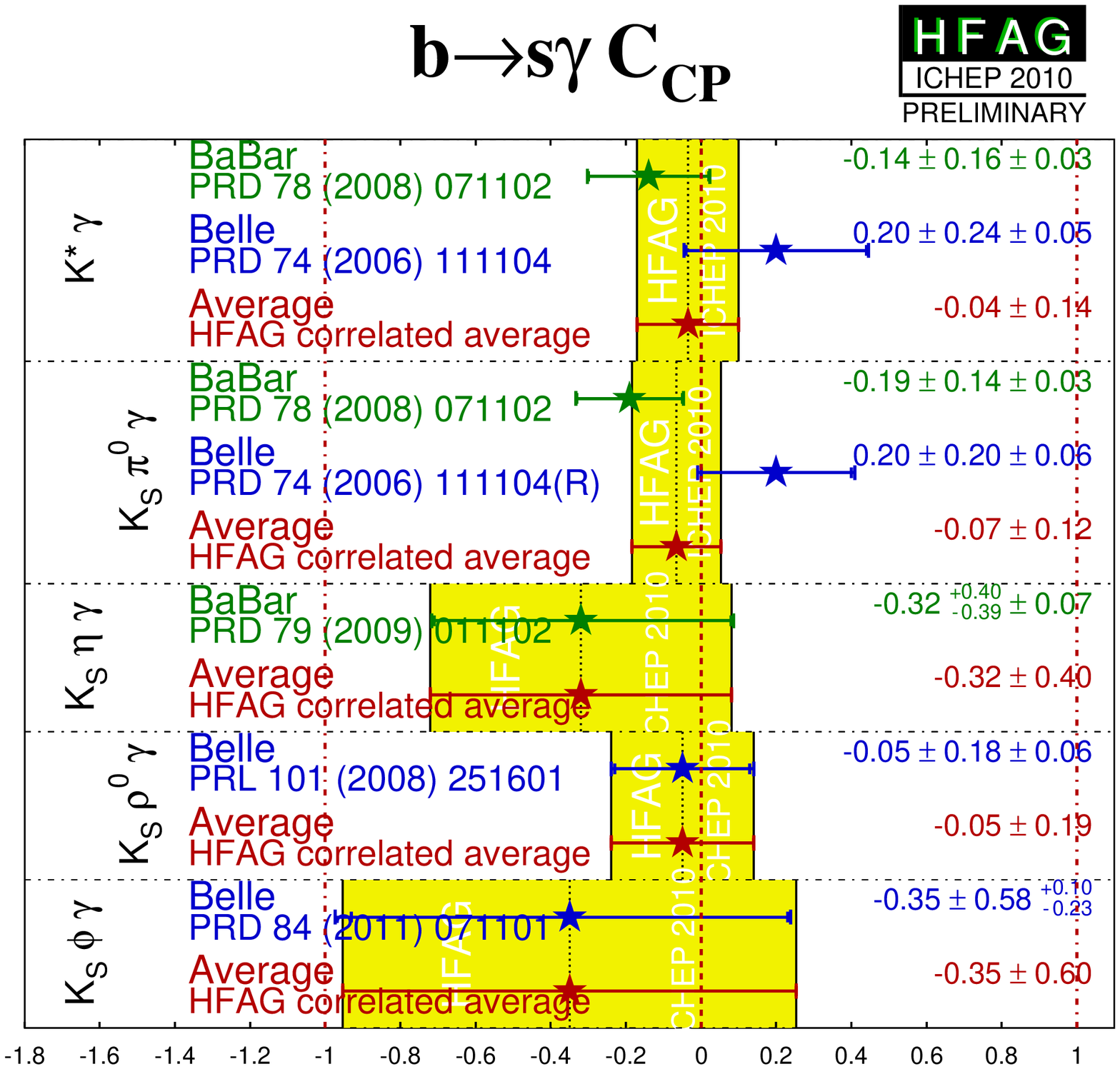}
      }
    \end{tabular}
  \end{center}
  \vspace{-0.8cm}
  \caption{
    Averages of (left) $S_{b \to s \gamma}$ and (right) $C_{b \to s \gamma}$.
    Recall that the data for $K^*\gamma$ is a subset of that for $\KS\pi^0\gamma$.
  }
  \label{fig:cp_uta:bsg}
\end{figure}

\begin{figure}[htb]
  \begin{center}
    \begin{tabular}{cc}
      \resizebox{0.46\textwidth}{!}{
        \includegraphics{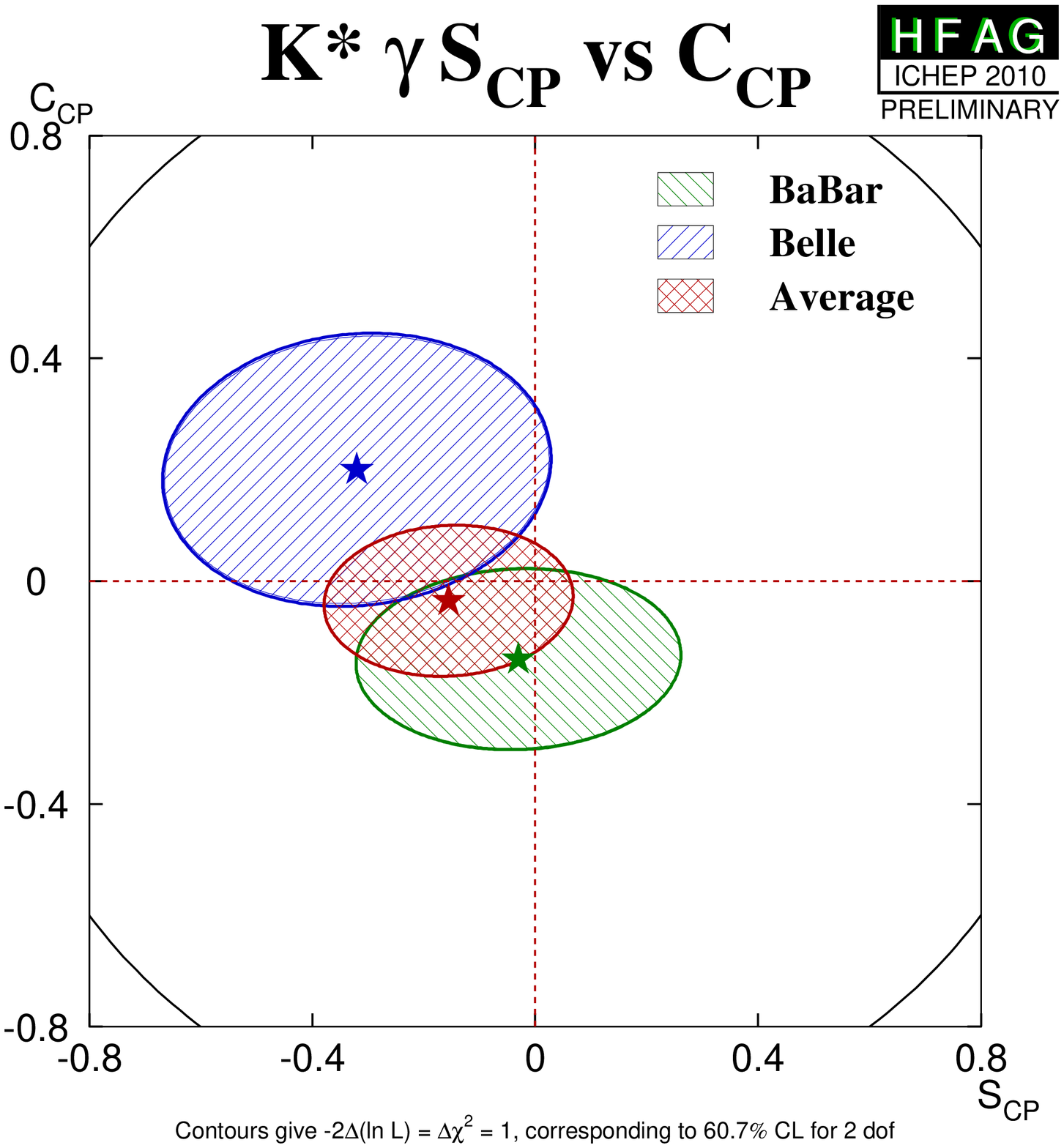}
      }
      &
      \resizebox{0.46\textwidth}{!}{
        \includegraphics{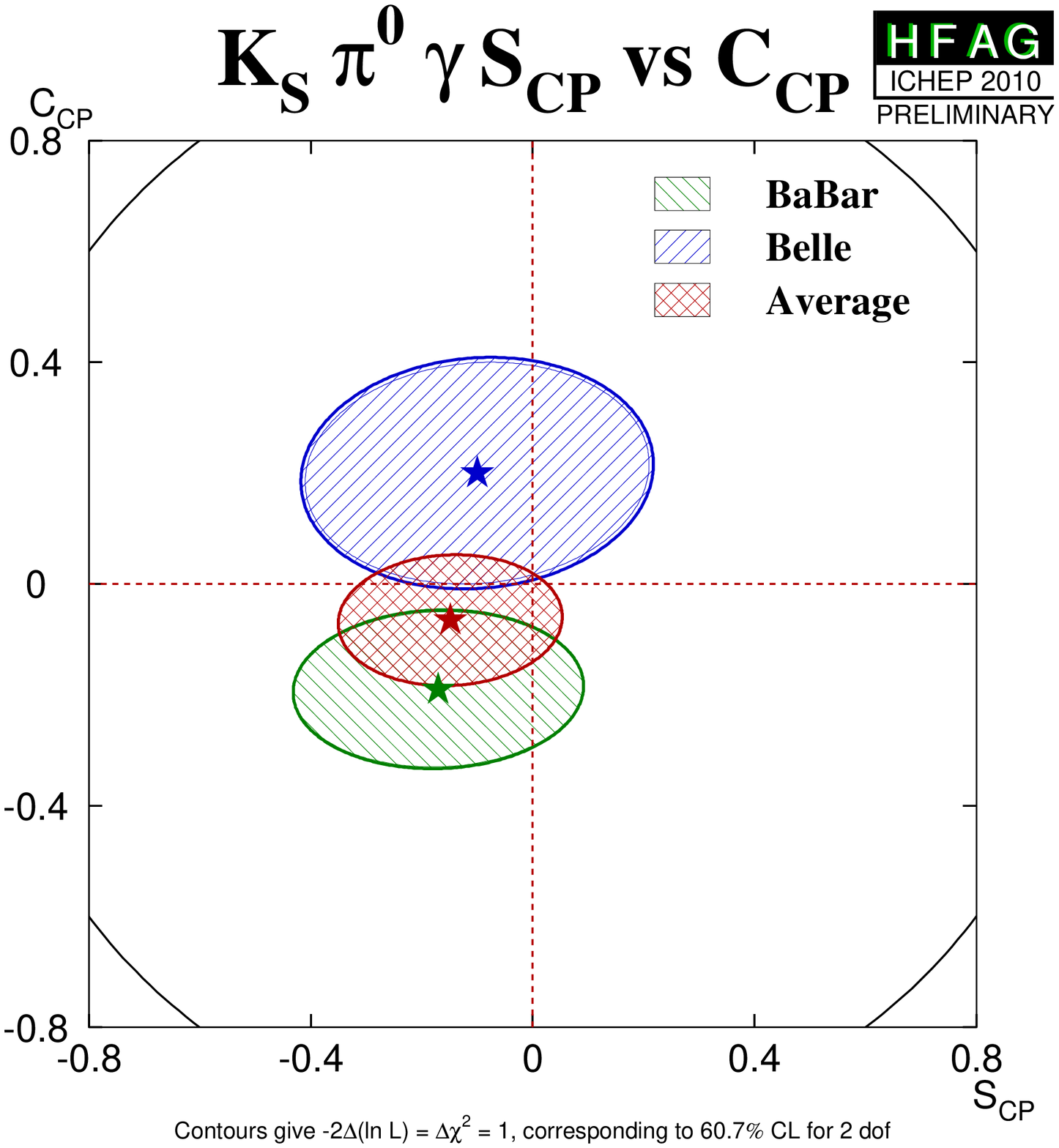}
      }
    \end{tabular}
  \end{center}
  \vspace{-0.8cm}
  \caption{
    Averages of $b \to s\gamma$ dominated channels,
    for which correlated averages are performed,
    in the $S_{\CP}$ \vs\ $C_{\CP}$ plane.
    (Left) $\Bz \to K^*\gamma$ and 
    (right) $\Bz \to \KS\pi^0\gamma$ (including $K^*\gamma$).
  }
  \label{fig:cp_uta:bsg_SvsC}
\end{figure}

The results are shown in Table~\ref{tab:cp_uta:bsg},
and in Figs.~\ref{fig:cp_uta:bsg} and~~\ref{fig:cp_uta:bsg_SvsC}.
No significant $\CP$ violation results are seen;
the results are consistent with the Standard Model
and with other measurements in the $b \to s\gamma$ system (see Sec.~\ref{sec:rare}).

\mysubsection{Time-dependent asymmetries in $b \to d\gamma$ transitions
}
\label{sec:cp_uta:bdg}

The formalism for the radiative decays $b \to d\gamma$ is much the same
as that for $b \to s\gamma$ discussed above.
Assuming dominance of the top quark in the loop,
the weak phase in decay should cancel with that from mixing,
so that the mixing-induced \CP\ violation parameter $S_{\CP}$ 
should be very small.
Corrections due to the finite light quark mass are smaller
compared to $b \to s\gamma$, since $m_d < m_s$,
and although QCD corrections may still play a role,
they cannot significantly affect the prediction $S_{b \to d \gamma} \simeq 0$.
Large direction \CP\ violation effects could, however, be seen through
a non-zero value of $C_{b \to d \gamma}$, 
since the top loop is not the only contribution.

Results using the mode $\Bz \to \rho^0\gamma$ are available from 
\belle\ and are shown in Table~\ref{tab:cp_uta:bdg}.

\begin{table}[htb]
	\begin{center}
		\caption{
			Averages for $\Bz \to \rho^{0} \gamma$.
		}
		\vspace{0.2cm}
		\setlength{\tabcolsep}{0.0pc}
		\begin{tabular*}{\textwidth}{@{\extracolsep{\fill}}lrcccc} \hline
	\mc{2}{l}{Experiment} & $N(B\bar{B})$ & $S_{CP}$ & $C_{CP}$ & Correlation \\
	\hline
	\belle & \cite{:2007jf} & 657M & $-0.83 \pm 0.65 \pm 0.18$ & $0.44 \pm 0.49 \pm 0.14$ & $-0.08$ \\
		\hline
		\end{tabular*}
		\label{tab:cp_uta:bdg}
	\end{center}
\end{table}

\clearpage
\mysubsection{Time-dependent $\CP$ asymmetries in $b \to u\bar{u}d$ transitions
}
\label{sec:cp_uta:uud}

The $b \to u \bar u d$ transition can be mediated by either 
a $b \to u$ tree amplitude or a $b \to d$ penguin amplitude.
These transitions can be investigated using 
the time dependence of $\Bz$ decays to final states containing light mesons.
Results are available from both \babar\ and \belle\ for the 
$\CP$ eigenstate ($\etacp = +1$) $\pi^+\pi^-$ final state
and for the vector-vector final state $\rho^+\rho^-$,
which is found to be dominated by the $\CP$-even
longitudinally polarised component
(\babar\ measure $f_{\rm long} = 
0.992 \pm 0.024 \, ^{+0.026}_{-0.013}$~\cite{Aubert:2007nua}
while \belle\ measure $f_{\rm long} = 
0.941 \, ^{+0.034}_{-0.040} \pm 0.030$~\cite{Somov:2006sg}).
\babar\ have also performed a time-dependent analysis of the 
vector-vector final state $\rho^0\rho^0$~\cite{:2008iha},
in which they measure  $f_{\rm long} = 0.70 \pm 0.14 \pm 0.05$;
\belle\ measures a smaller branching fraction than \babar\ for
$\Bz\to\rho^0\rho^0$~\cite{:2008et} with corresponding signal yields too small
to perform time-dependent or angular analyses.
\babar\ have furthermore performed a time-dependent analysis of the 
$\Bz \to a_1^\pm \pi^\mp$ decay~\cite{Aubert:2006gb}; further experimental
input for the extraction of $\alpha$ from this channel is reported in a later
publication~\cite{:2009ii}.

\begin{figure}[htb]
  \begin{center}
    \begin{tabular}{cc}
      \resizebox{0.46\textwidth}{!}{
        \includegraphics{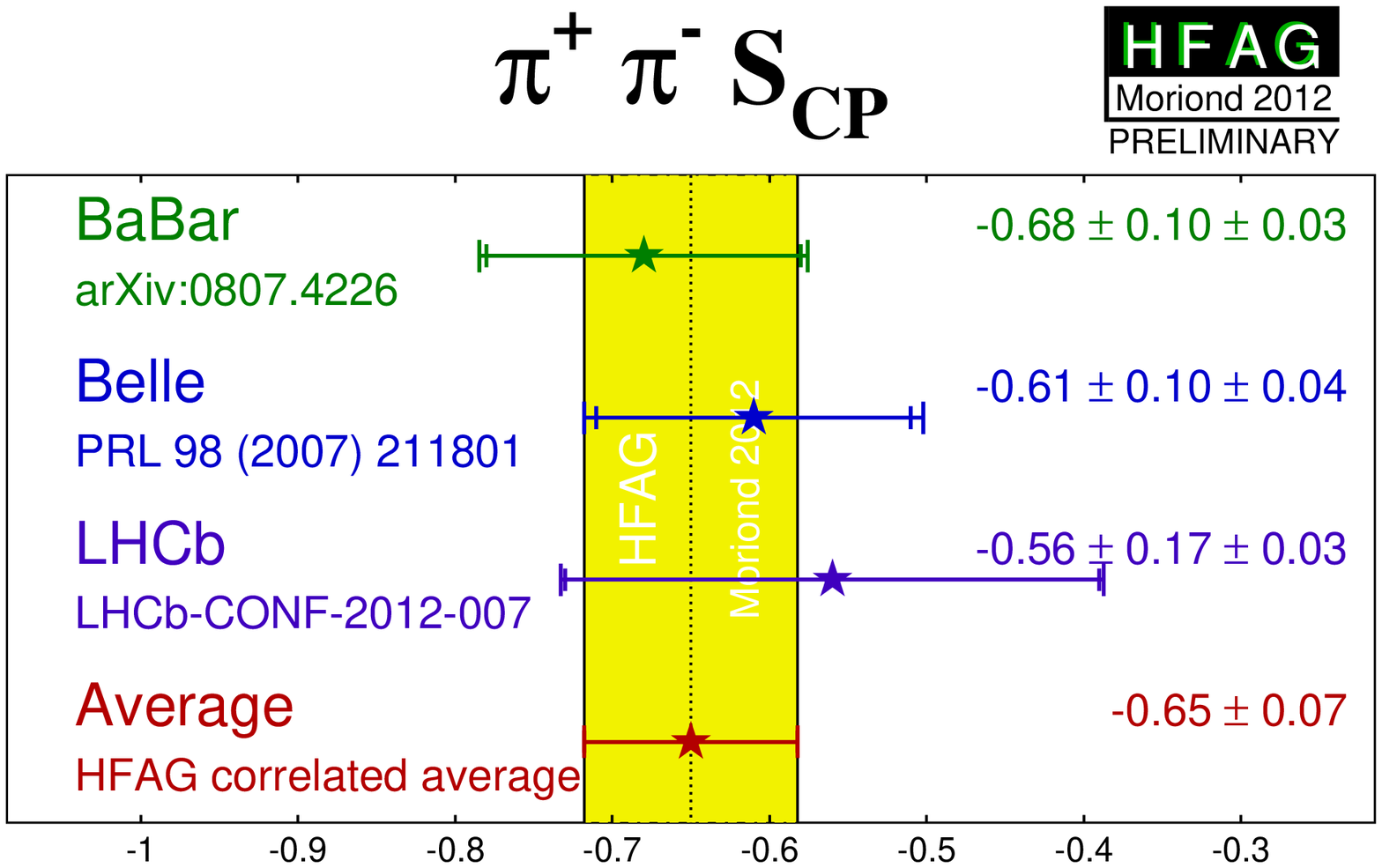}
      }
      &
      \resizebox{0.46\textwidth}{!}{
        \includegraphics{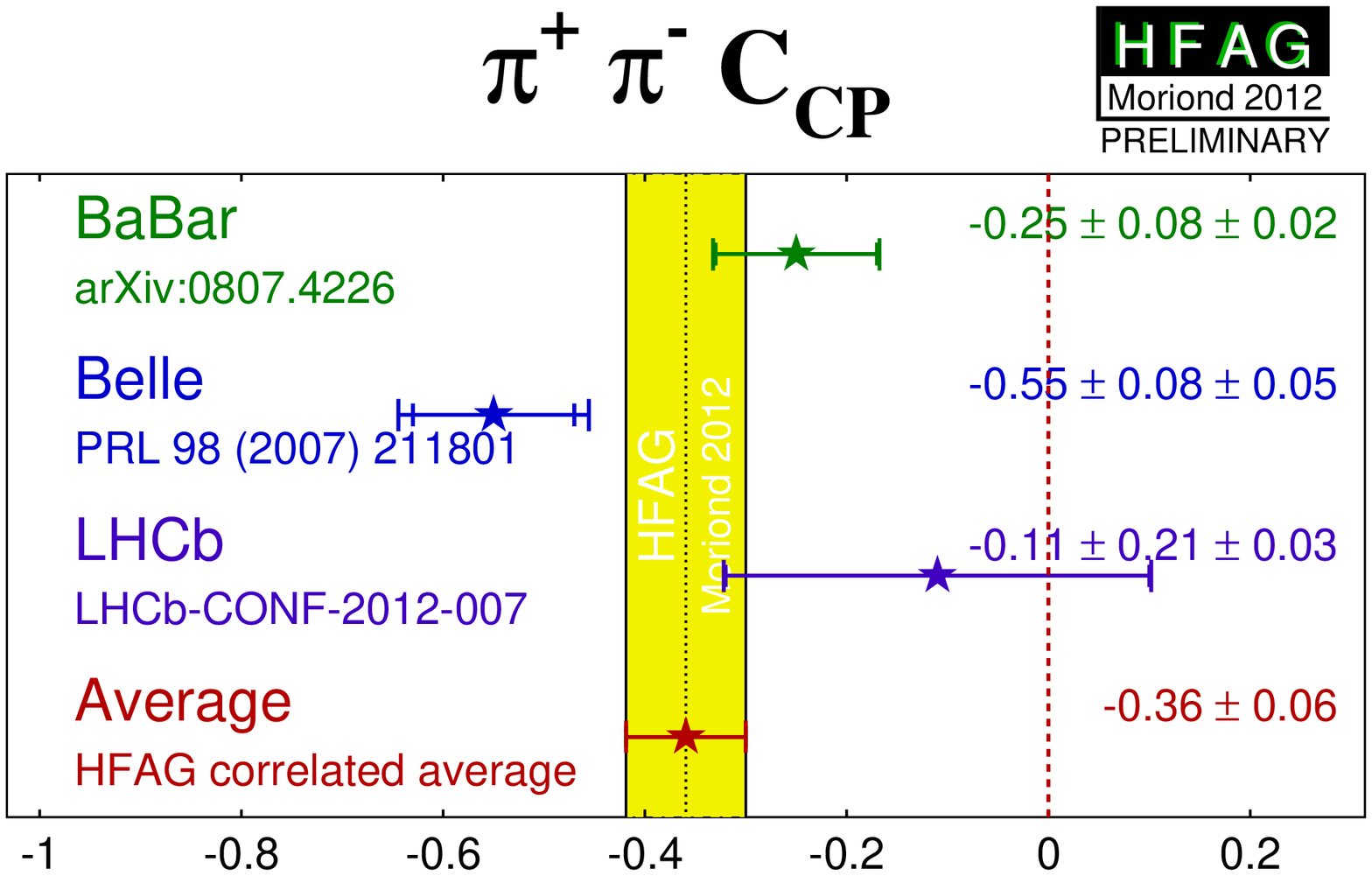}
      }
    \end{tabular}
  \end{center}
  \vspace{-0.8cm}
  \caption{
    Averages of (left) $S_{b \to u\bar u d}$ and (right) $C_{b \to u\bar u d}$
    for the mode $\Bz \to \pi^+\pi^-$.
  }
  \label{fig:cp_uta:uud:pipi}
\end{figure}

\begin{figure}[htb]
  \begin{center}
    \begin{tabular}{cc}
      \resizebox{0.46\textwidth}{!}{
        \includegraphics{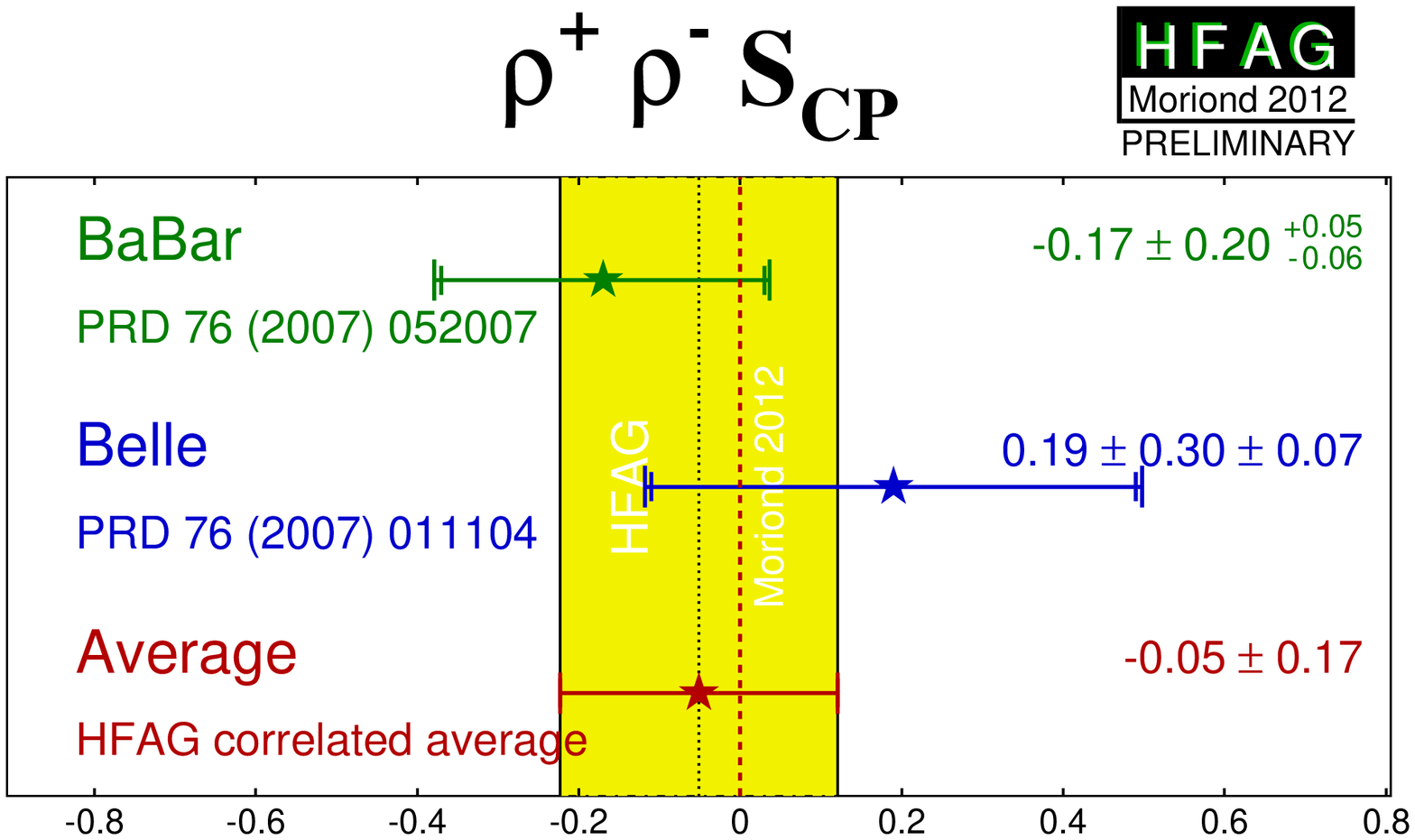}
      }
      &
      \resizebox{0.46\textwidth}{!}{
        \includegraphics{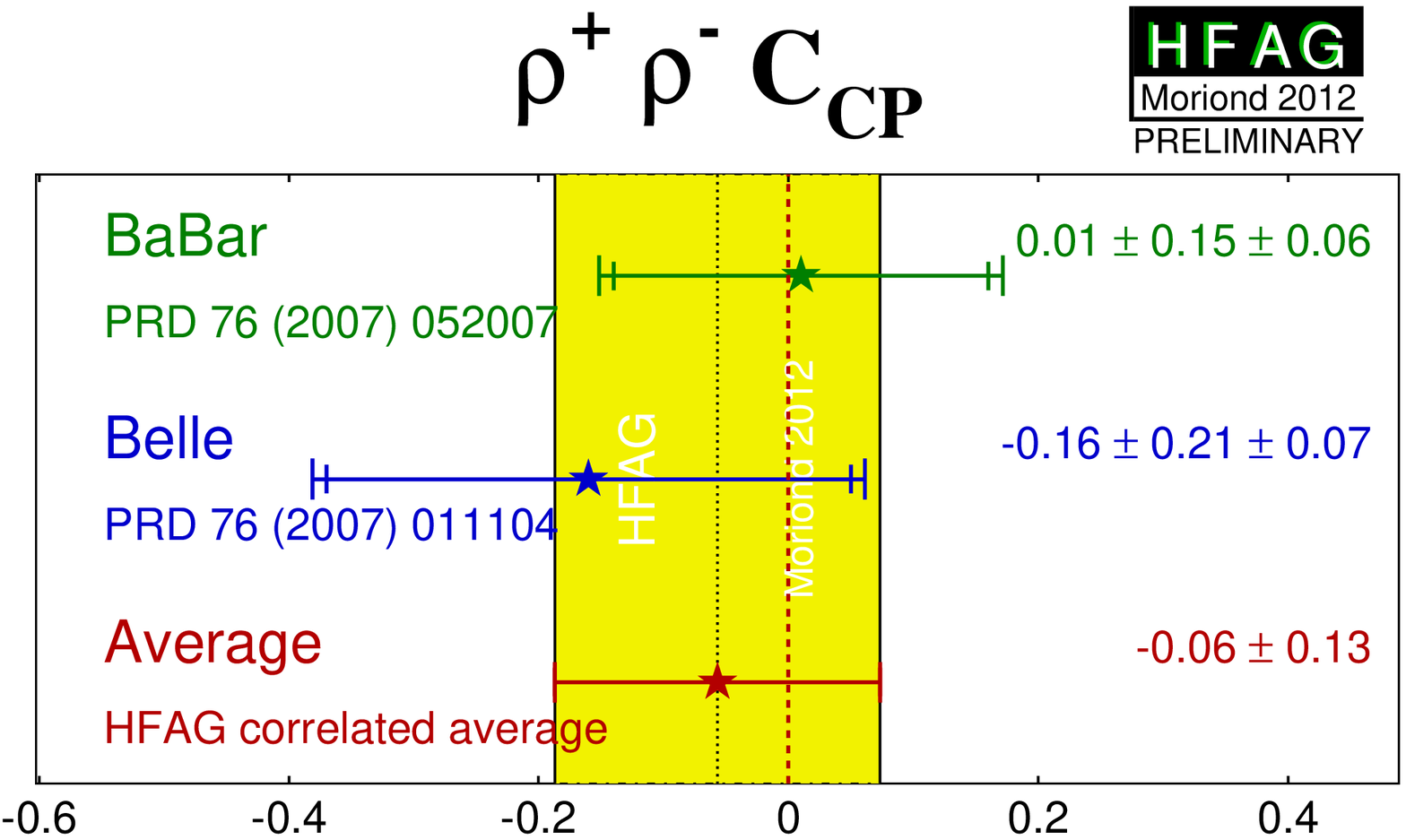}
      }
    \end{tabular}
  \end{center}
  \vspace{-0.8cm}
  \caption{
    Averages of (left) $S_{b \to u\bar u d}$ and (right) $C_{b \to u\bar u d}$
    for the mode $\Bz \to \rho^+\rho^-$.
  }
  \label{fig:cp_uta:uud:rhorho}
\end{figure}

\begin{figure}[htb]
  \begin{center}
    \begin{tabular}{cc}
      \resizebox{0.46\textwidth}{!}{
        \includegraphics{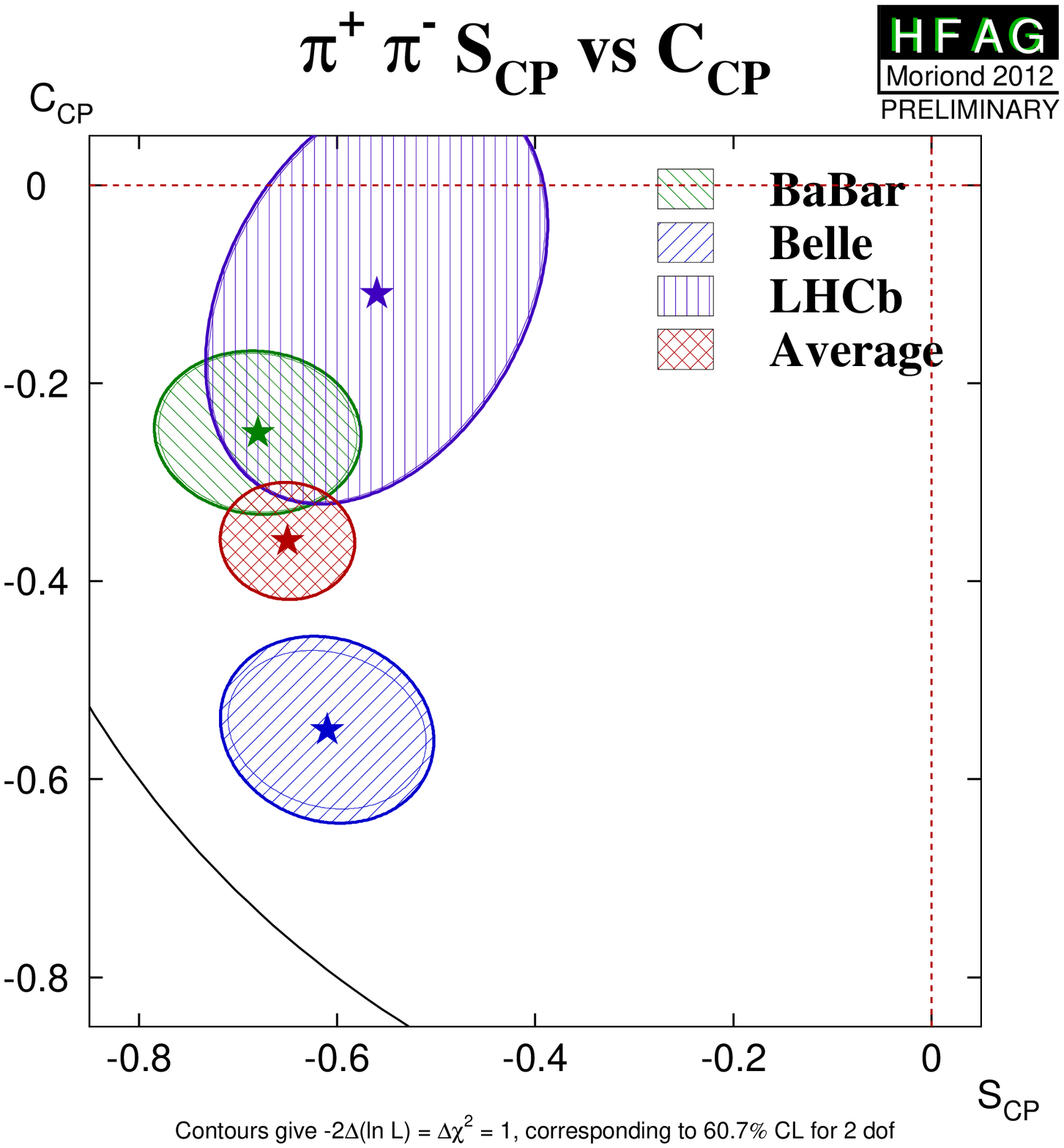}
      }      
      &
      \resizebox{0.46\textwidth}{!}{
        \includegraphics{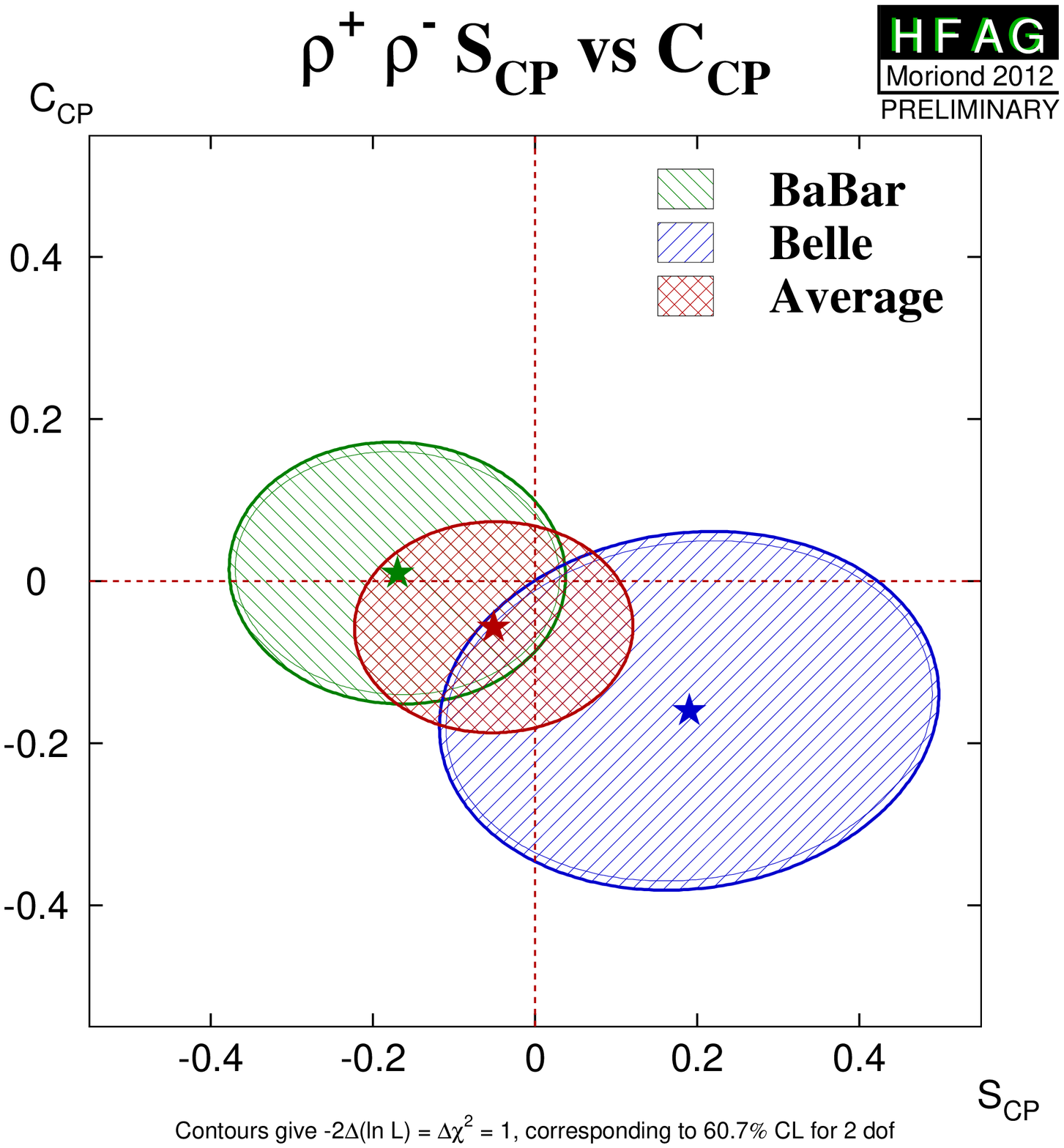}
      }
    \end{tabular}
  \end{center}
  \vspace{-0.8cm}
  \caption{
    Averages of $b \to u\bar u d$ dominated channels,
    for which correlated averages are performed,
    in the $S_{\CP}$ \vs\ $C_{\CP}$ plane.
    (Left) $\Bz \to \pi^+\pi^-$ and (right) $\Bz \to \rho^+\rho^-$.
  }
  \label{fig:cp_uta:uud_SvsC}
\end{figure}

Results, and averages, of time-dependent \CP-violation parameters in 
$b \to u \bar u d$ transitions are listed in Table~\ref{tab:cp_uta:uud}.
The averages for $\pi^+\pi^-$ are shown in Fig.~\ref{fig:cp_uta:uud:pipi},
and those for $\rho^+\rho^-$ are shown in Fig.~\ref{fig:cp_uta:uud:rhorho},
with the averages in the $S_{\CP}$ \vs\ $C_{\CP}$ plane 
shown in Fig.~\ref{fig:cp_uta:uud_SvsC}.

\begin{sidewaystable}
	\begin{center}
		\caption{
      Averages for $b \to u \bar u d$ modes.
		}
		\vspace{0.2cm}
		\setlength{\tabcolsep}{0.0pc}
		\begin{tabular*}{\textwidth}{@{\extracolsep{\fill}}lrcccc} \hline
	\mc{2}{l}{Experiment} & Sample size & $S_{CP}$ & $C_{CP}$ & Correlation \\
	\hline
      \mc{6}{c}{$\pi^{+} \pi^{-}$} \\
	\babar & \cite{Aubert:2008sb} & $N(B\bar{B}) =$ 467M & $-0.68 \pm 0.10 \pm 0.03$ & $-0.25 \pm 0.08 \pm 0.02$ & $-0.06$ \\
	\belle & \cite{Ishino:2006if} & $N(B\bar{B}) =$ 535M & $-0.61 \pm 0.10 \pm 0.04$ & $-0.55 \pm 0.08 \pm 0.05$ & $-0.15$ \\
	LHCb & \cite{LHCb-CONF-2012-007} & 0.7 ${\rm fb}^{-1}$ & $-0.56 \pm 0.17 \pm 0.03$ & $-0.11 \pm 0.21 \pm 0.03$ & $0.34$ \\
	\mc{3}{l}{\bf Average} & $-0.65 \pm 0.07$ & $-0.36 \pm 0.06$ & $-0.03$ \\
	\mc{3}{l}{\small Confidence level} & \mc{2}{c}{\small $0.12~(1.6\sigma)$} & \\
		\hline

      \mc{6}{c}{$\rho^{+} \rho^{-}$} \\
	\babar & \cite{Aubert:2007nua} & $N(B\bar{B}) =$ 387M & $-0.17 \pm 0.20 \,^{+0.05}_{-0.06}$ & $0.01 \pm 0.15 \pm 0.06$ & $-0.04$ \\
	\belle & \cite{Abe:2007ez} & $N(B\bar{B}) =$ 535M & $0.19 \pm 0.30 \pm 0.07$ & $-0.16 \pm 0.21 \pm 0.07$ & $0.10$ \\
	\mc{3}{l}{\bf Average} & $-0.05 \pm 0.17$ & $-0.06 \pm 0.13$ & $0.01$ \\
	\mc{3}{l}{\small Confidence level} & \mc{2}{c}{\small $0.50~(0.7\sigma)$} & \\
		\hline

      \mc{6}{c}{$\rho^{0} \rho^{0}$} \\
	\babar & \cite{:2008iha} & $N(B\bar{B}) =$ 465M & $0.3 \pm 0.7 \pm 0.2$ & $0.2 \pm 0.8 \pm 0.3$ & $-0.04$ \\
 		\hline
 		\end{tabular*}

                \vspace{2ex}

    \resizebox{\textwidth}{!}{
 		\begin{tabular}{@{\extracolsep{2mm}}lrcccccc} \hline
 		\mc{2}{l}{Experiment} & $N(B\bar{B})$ & $A_{CP}$ & $C$ & $S$ & $\Delta C$ & $\Delta S$ \\
 		\hline
      \mc{8}{c}{$a_1^{\pm} \pi^{\mp}$} \\
	\babar & \cite{Aubert:2006gb} & 384M & $-0.07 \pm 0.07 \pm 0.02$ & $-0.10 \pm 0.15 \pm 0.09$ & $0.37 \pm 0.21 \pm 0.07$ & $0.26 \pm 0.15 \pm 0.07$ & $-0.14 \pm 0.21 \pm 0.06$ \\
 	\hline
		\end{tabular}
              }
		\label{tab:cp_uta:uud}
	\end{center}
\end{sidewaystable}

If the penguin contribution is negligible, 
the time-dependent parameters for $\Bz \to \pi^+\pi^-$ 
and $\Bz \to \rho^+\rho^-$ are given by
$S_{b \to u\bar u d} = \etacp \sin(2\alpha)$ and
$C_{b \to u\bar u d} = 0$.
In the presence of the penguin contribution, 
direct $\CP$ violation may arise, 
and there is no straightforward interpretation 
of $S_{b \to u\bar u d}$ and $C_{b \to u\bar u d}$.
An isospin analysis~\cite{Gronau:1990ka} 
can be used to disentangle the contributions and extract $\alpha$.

\begin{figure}[htb]
  \begin{center}
    \begin{tabular}{cc}
      \resizebox{0.46\textwidth}{!}{
        \includegraphics{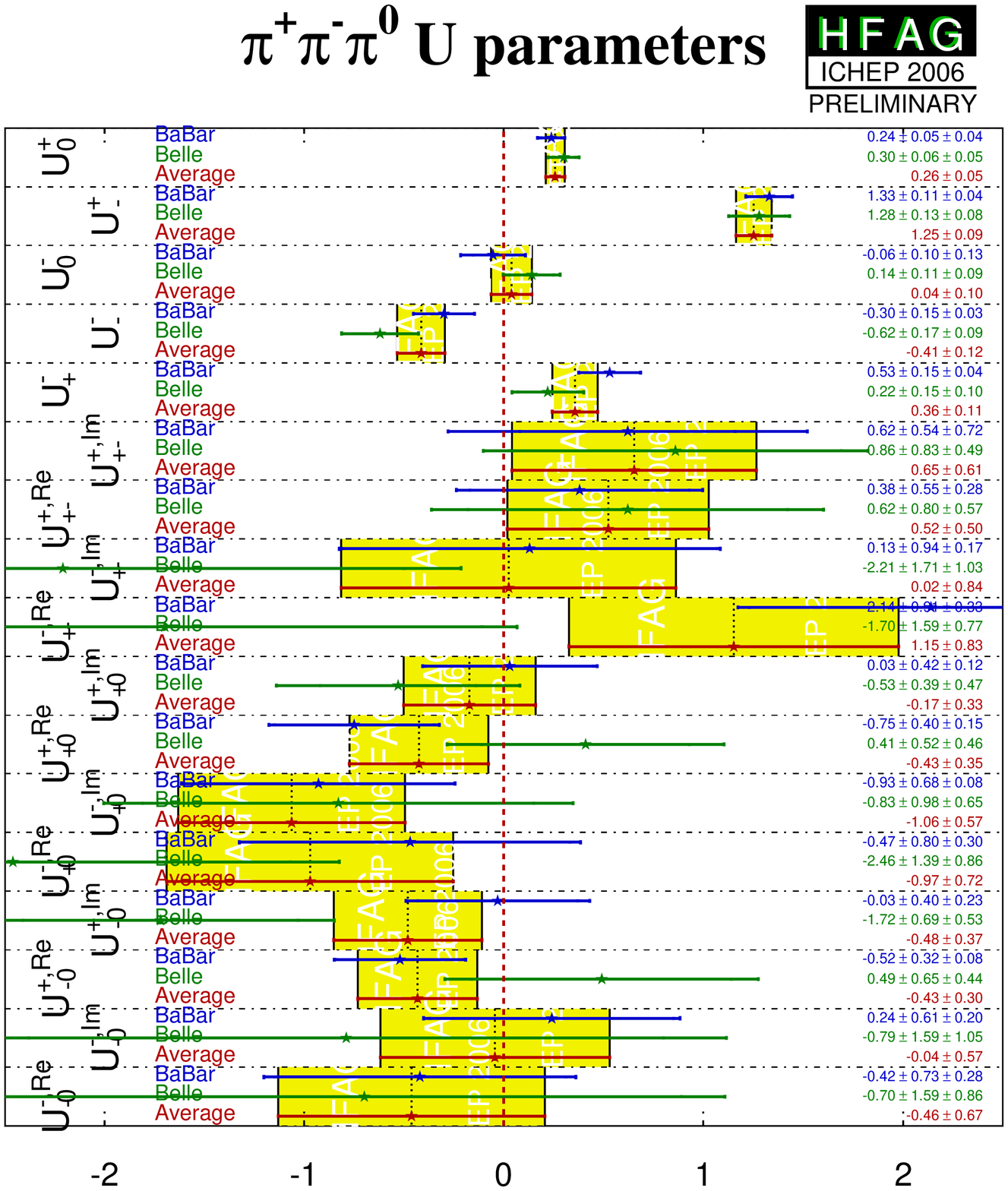}
      }
      &
      \resizebox{0.46\textwidth}{!}{
        \includegraphics{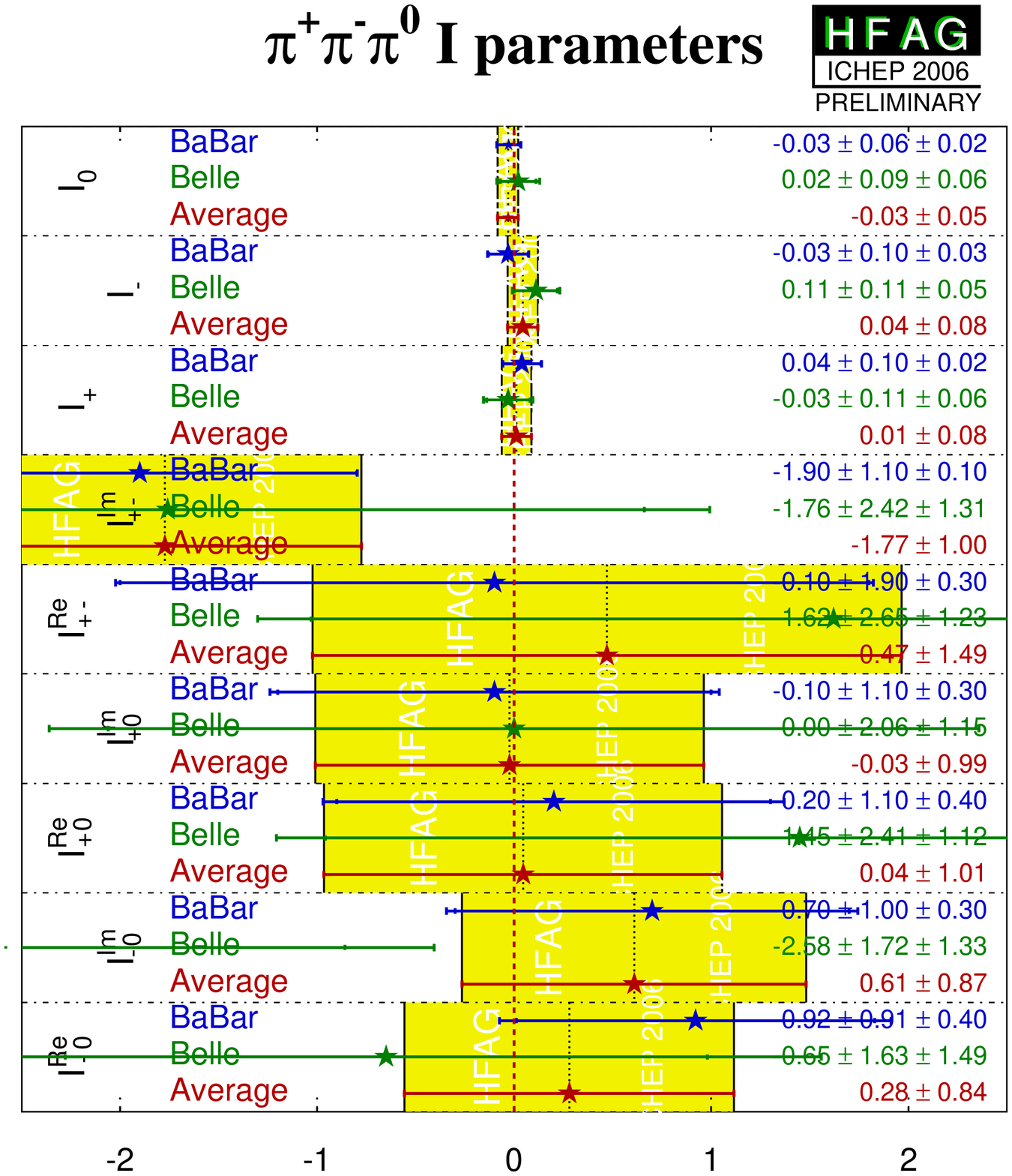}
      }
    \end{tabular}
  \end{center}
  \vspace{-0.8cm}
  \caption{
    Summary of the $U$ and $I$ parameters measured in the 
    time-dependent $\Bz \to \pi^+\pi^-\pi^0$ Dalitz plot analysis.
  }
  \label{fig:cp_uta:uud:uandi}
\end{figure}

For the non-$\CP$ eigenstate $\rho^{\pm}\pi^{\mp}$, 
both \babar~\cite{Aubert:2007jn} 
and \belle~\cite{Kusaka:2007dv,:2007mj} have performed 
time-dependent Dalitz plot (DP) analyses
of the $\pi^+\pi^-\pi^0$ final state~\cite{Snyder:1993mx};
such analyses allow direct measurements of the phases.
Both experiments have measured the $U$ and $I$ parameters discussed in 
Sec.~\ref{sec:cp_uta:notations:dalitz:pipipi0} and defined in 
Table~\ref{tab:cp_uta:pipipi0:uandi}.
We have performed a full correlated average of these parameters,
the results of which are summarised in Fig.~\ref{fig:cp_uta:uud:uandi}.

\begin{figure}[htb]
  \begin{center}
    \begin{tabular}{cc}
      \resizebox{0.46\textwidth}{!}{
        \includegraphics{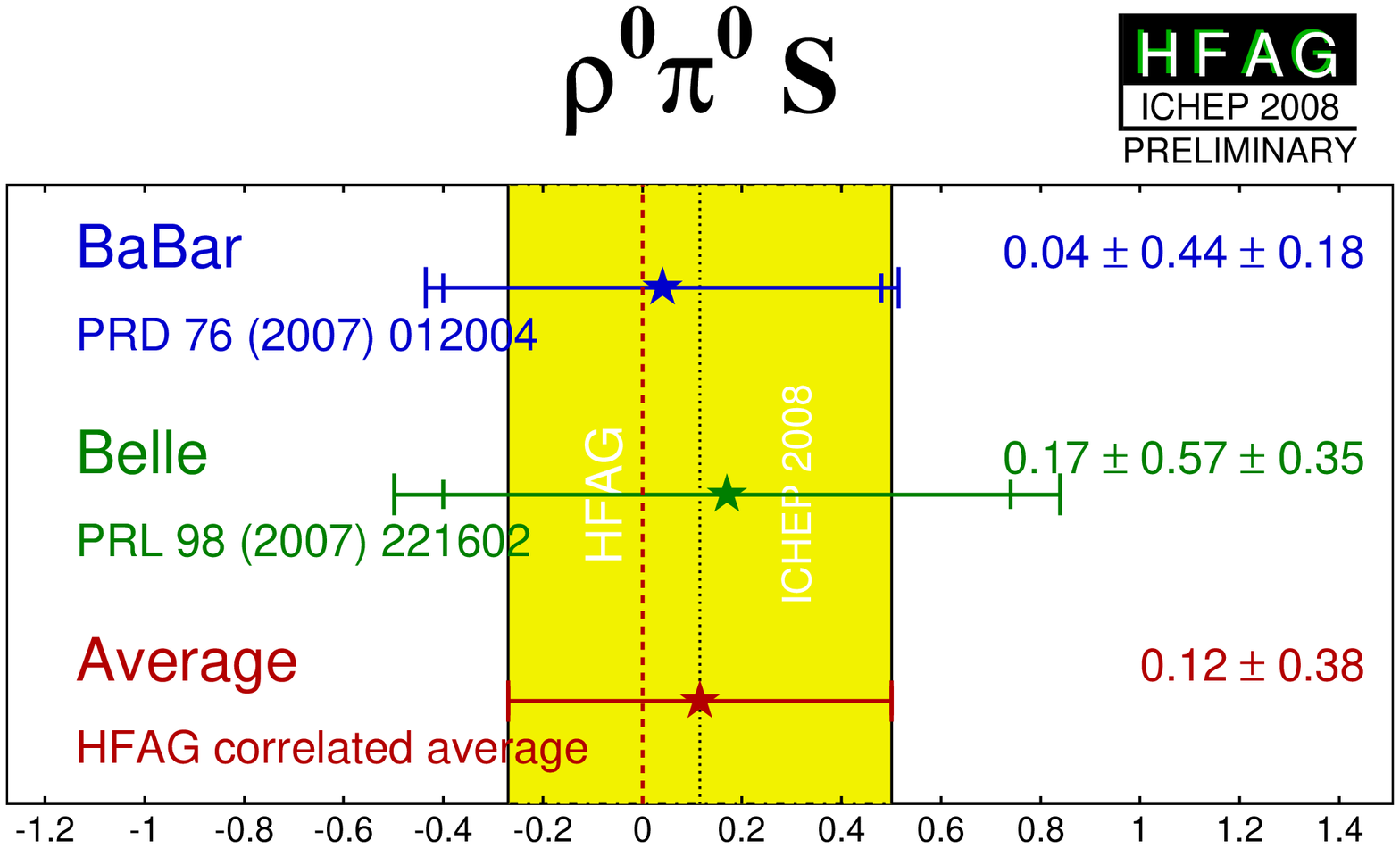}
      }
      &
      \resizebox{0.46\textwidth}{!}{
        \includegraphics{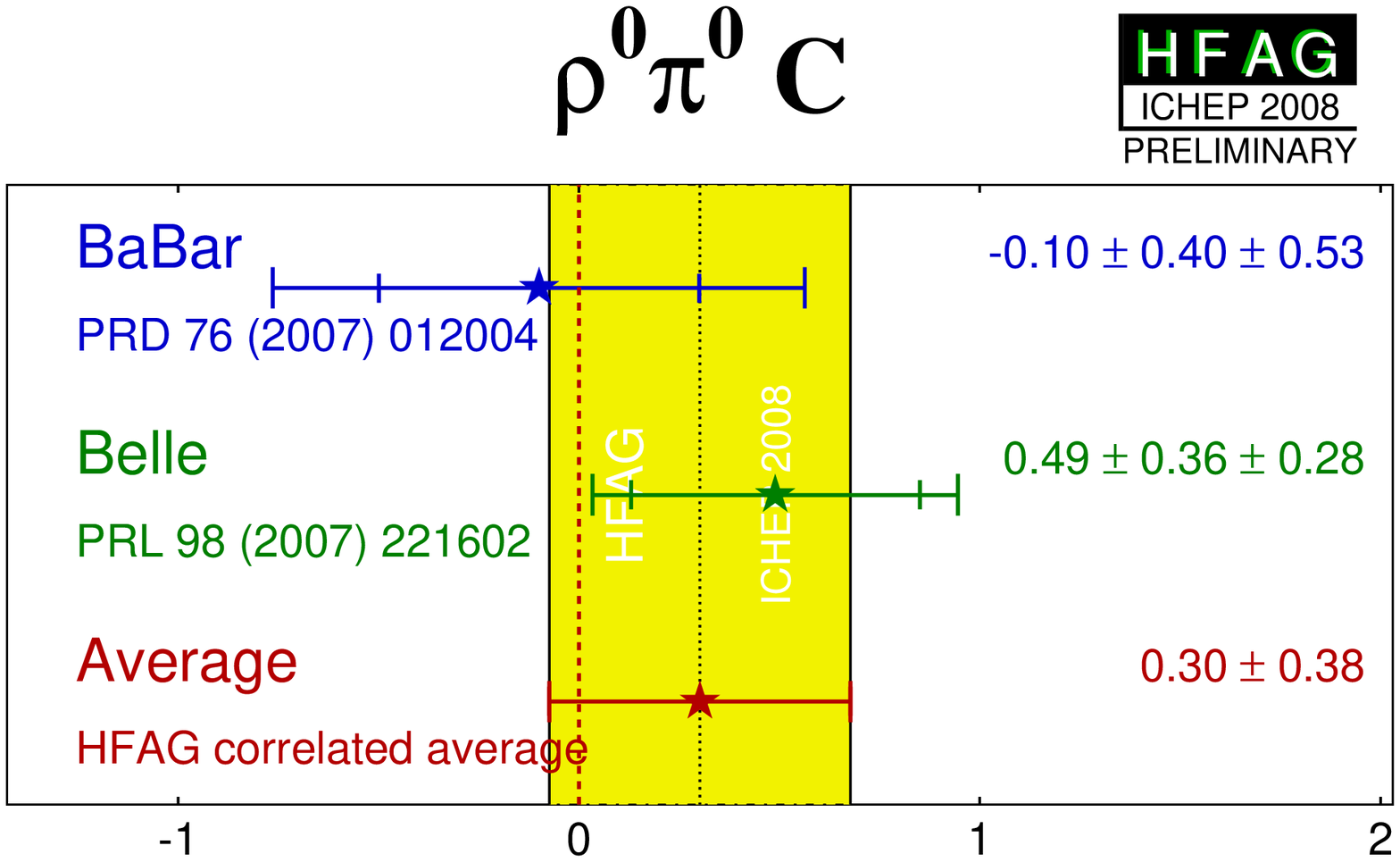}
      }
    \end{tabular}
  \end{center}
  \vspace{-0.8cm}
  \caption{
    Averages of (left) $S_{b \to u\bar u d}$ and (right) $C_{b \to u\bar u d}$
    for the mode $\Bz \to \rho^0\pi^0$.
  }
  \label{fig:cp_uta:uud:rho0pi0}
\end{figure}

\begin{figure}[htb]
  \begin{center}
    \resizebox{0.46\textwidth}{!}{
      \includegraphics{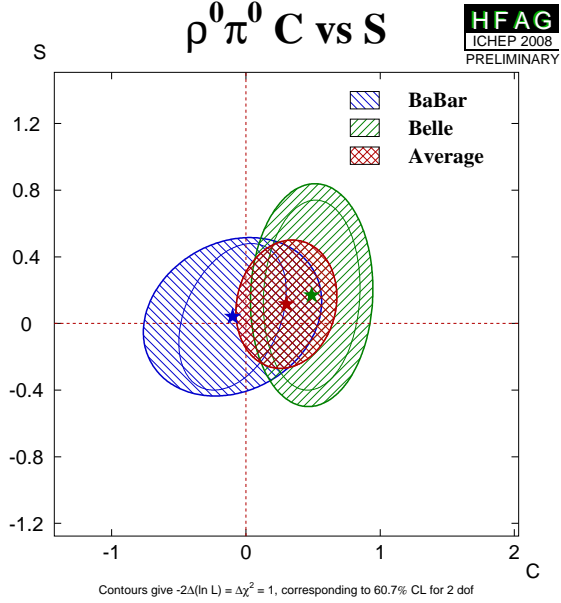}
    }      
  \end{center}
  \vspace{-0.8cm}
  \caption{
    Averages of $b \to u\bar u d$ dominated channels,
    for the mode $\Bz \to \rho^0\pi^0$
    in the $S_{\CP}$ \vs\ $C_{\CP}$ plane.
  }
  \label{fig:cp_uta:uud:rho0pi0_SvsC}
\end{figure}

Both experiments have also extracted the Q2B parameters.
We have performed a full correlated average of these parameters,
which is equivalent to determining the values from the 
averaged $U$ and $I$ parameters.
The results are shown in Table.~\ref{tab:cp_uta:uud:rhopi_q2b}.
Averages of the $\Bz \to \rho^0\pi^0$ Q2B parameters are shown in 
Figs.~\ref{fig:cp_uta:uud:rho0pi0} and~\ref{fig:cp_uta:uud:rho0pi0_SvsC}.

\begin{sidewaystable}
	\begin{center}
		\caption{
                  Averages of quasi-two-body parameters extracted
                  from time-dependent Dalitz plot analysis of 
                  $\Bz \to \pi^+\pi^-\pi^0$.
		}
		\vspace{0.2cm}
		\setlength{\tabcolsep}{0.0pc}
    \resizebox{\textwidth}{!}{
		\begin{tabular}{@{\extracolsep{2mm}}lrcccccc} \hline
		\mc{2}{l}{Experiment} & $N(B\bar{B})$ & ${\cal A}_{CP}^{\rho\pi}$ & $C_{\rho\pi}$ & $S_{\rho\pi}$ & $\Delta C_{\rho\pi}$ & $\Delta S_{\rho\pi}$ \\
	\hline
	\babar & \cite{Aubert:2007jn} & 375M & $-0.14 \pm 0.05 \pm 0.02$ & $0.15 \pm 0.09 \pm 0.05$ & $-0.03 \pm 0.11 \pm 0.04$ & $0.39 \pm 0.09 \pm 0.09$ & $-0.01 \pm 0.14 \pm 0.06$ \\
	\belle & \cite{Kusaka:2007dv,:2007mj} & 449M & $-0.12 \pm 0.05 \pm 0.04$ & $-0.13 \pm 0.09 \pm 0.05$ & $0.06 \pm 0.13 \pm 0.05$ & $0.36 \pm 0.10 \pm 0.05$ & $-0.08 \pm 0.13 \pm 0.05$ \\
	\mc{3}{l}{\bf Average} & $-0.13 \pm 0.04$ & $0.01 \pm 0.07$ & $0.01 \pm 0.09$ & $0.37 \pm 0.08$ & $-0.04 \pm 0.10$ \\
	\mc{3}{l}{\small Confidence level} & \mc{5}{c}{\small $0.52~(0.6\sigma)$} \\
        \hline
		\end{tabular}
              }

                \vspace{2ex}

		\begin{tabular*}{\textwidth}{@{\extracolsep{\fill}}lrcccc} \hline
		\mc{2}{l}{Experiment} & $N(B\bar{B})$ & ${\cal A}^{-+}_{\rho\pi}$ & ${\cal A}^{+-}_{\rho\pi}$ & Correlation \\
		\hline
	\babar & \cite{Aubert:2007jn} & 375M & $-0.37 \,^{+0.16}_{-0.10} \pm 0.09$ & $0.03 \pm 0.07 \pm 0.04$ & $0.62$ \\
	\belle & \cite{Kusaka:2007dv,:2007mj} & 449M & $0.08 \pm 0.16 \pm 0.11$ & $0.21 \pm 0.08 \pm 0.04$ & $0.47$ \\
	\mc{3}{l}{\bf Average} & $-0.18 \pm 0.12$ & $0.11 \pm 0.06$ & $0.40$ \\
        \mc{3}{l}{\small Confidence level} & \mc{2}{c}{\small $0.14~(1.5\sigma)$} \\
		\hline
		\end{tabular*}

                \vspace{2ex}

		\begin{tabular*}{\textwidth}{@{\extracolsep{\fill}}lrcccc} \hline
		\mc{2}{l}{Experiment} & $N(B\bar{B})$ & $C_{\rho^0\pi^0}$ & $S_{\rho^0\pi^0}$ & Correlation \\
		\hline
	\babar & \cite{Aubert:2007jn} & 375M & $-0.10 \pm 0.40 \pm 0.53$ & $0.04 \pm 0.44 \pm 0.18$ & $0.35$ \\
	\belle & \cite{Kusaka:2007dv,:2007mj} & 449M & $0.49 \pm 0.36 \pm 0.28$ & $0.17 \pm 0.57 \pm 0.35$ & $0.08$ \\
	\mc{3}{l}{\bf Average} & $0.30 \pm 0.38$ & $0.12 \pm 0.38$ & $0.12$ \\
	\mc{3}{l}{\small Confidence level} & \mc{2}{c}{\small $0.76~(0.3\sigma)$} \\
		\hline
		\end{tabular*}
		\label{tab:cp_uta:uud:rhopi_q2b}
	\end{center}
\end{sidewaystable}

\begin{figure}[htb]
  \begin{center}
    \resizebox{0.46\textwidth}{!}{
      \includegraphics{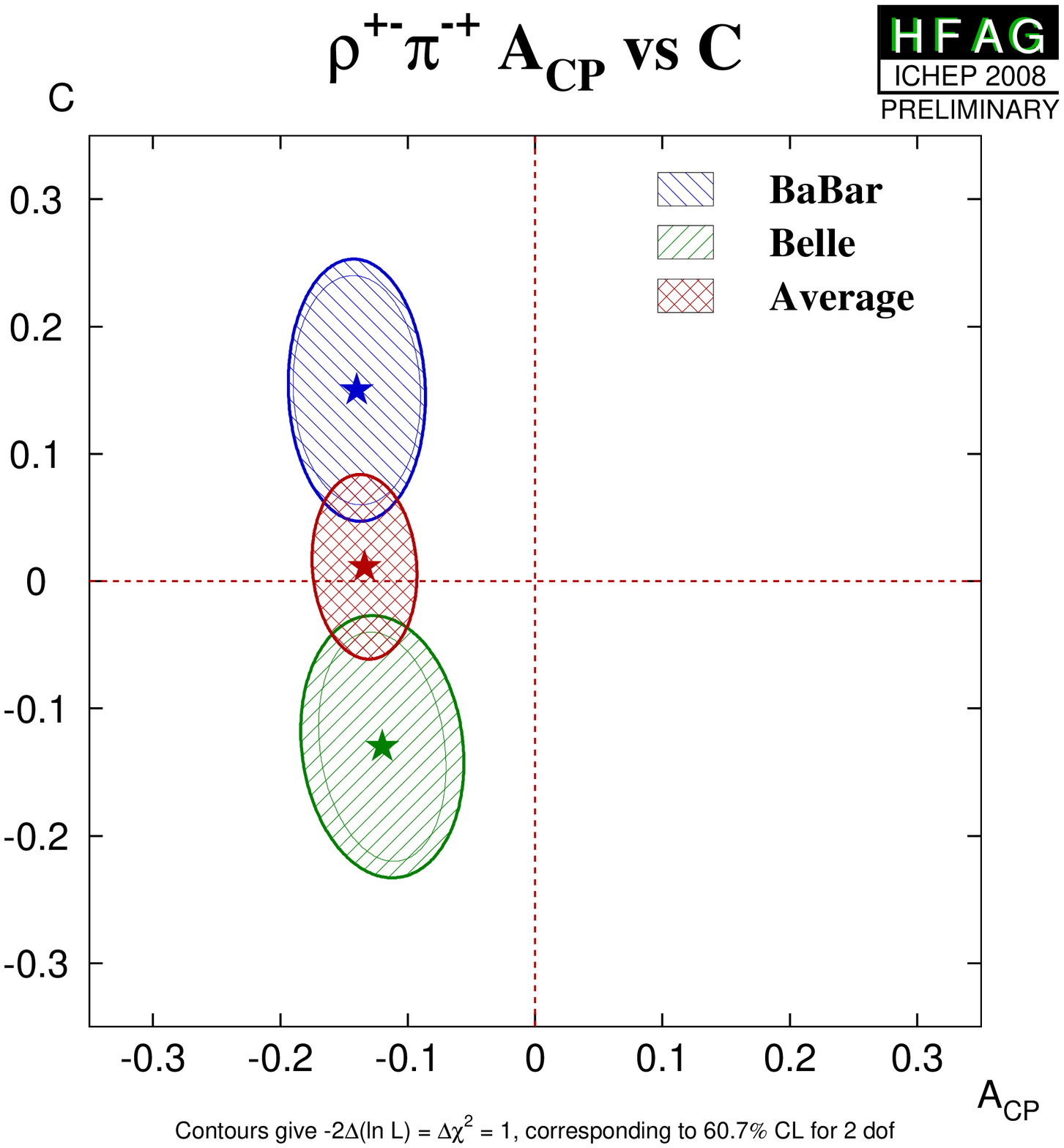}
    }
    \hfill
    \resizebox{0.46\textwidth}{!}{
      \includegraphics{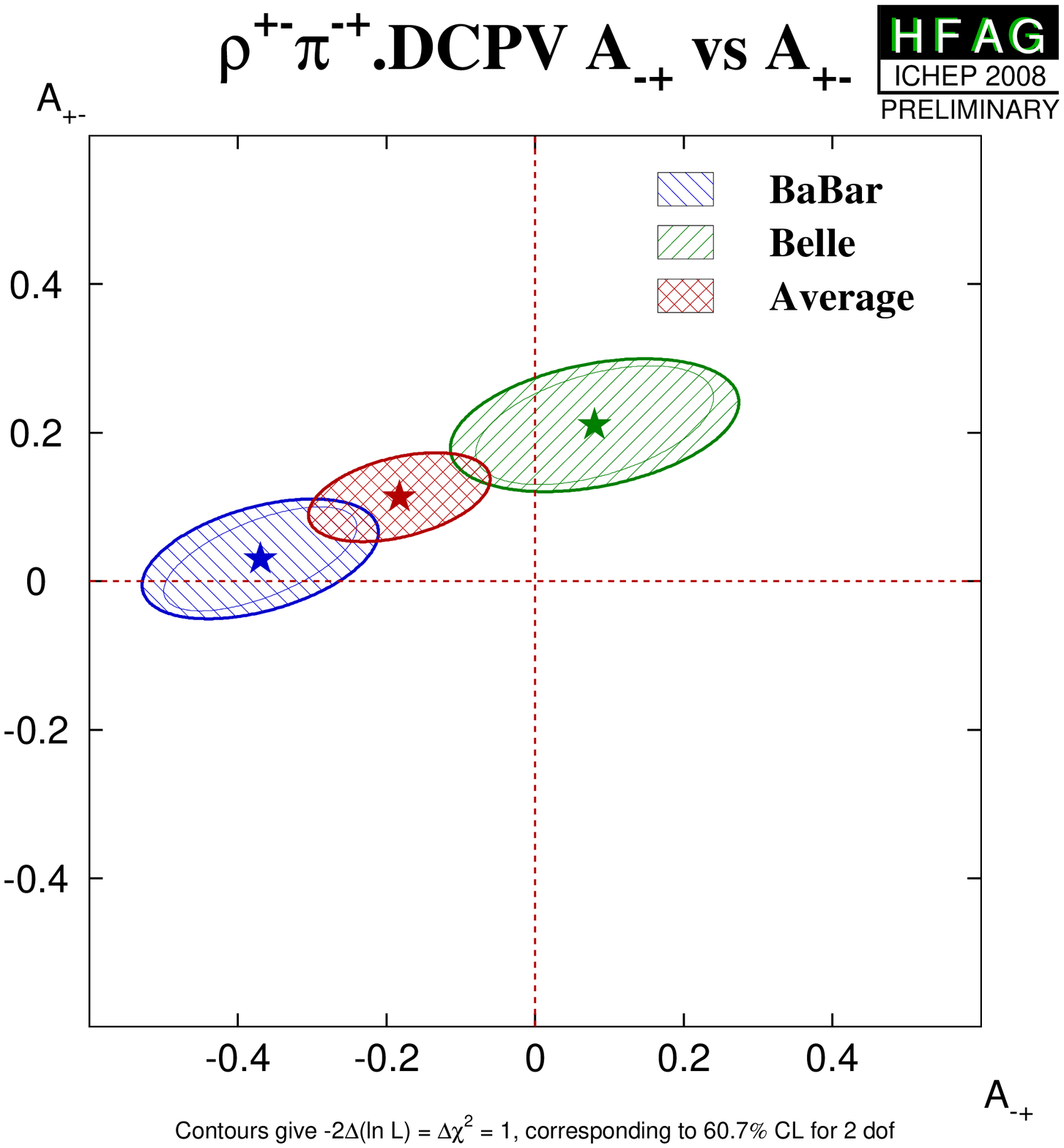}
    }
  \end{center}
  \vspace{-0.8cm}
  \caption{
    Direct $\CP$ violation in $\Bz\to\rho^\pm\pi^\mp$.
    (Left) ${\cal A}^{\rho\pi}_{\CP}$ \vs\ $C_{\rho\pi}$ space,
    (right) ${\cal A}^{-+}_{\rho\pi}$ \vs\ ${\cal A}^{+-}_{\rho\pi}$ space.
  }
  \label{fig:cp_uta:uud:rhopi-dircp}
\end{figure}

With the notation described in Sec.~\ref{sec:cp_uta:notations}
(Eq.~(\ref{eq:cp_uta:non-cp-s_and_deltas})), 
the time-dependent parameters for the Q2B $\Bz \to \rho^\pm\pi^\mp$ analysis are,
neglecting penguin contributions, given by
\begin{equation}
  S_{\rho\pi} = 
  \sqrt{1 - \left(\frac{\Delta C}{2}\right)^2}\sin(2\alpha)\cos(\delta)
  \ , \ \ \ 
  \Delta S_{\rho\pi} = 
  \sqrt{1 - \left(\frac{\Delta C}{2}\right)^2}\cos(2\alpha)\sin(\delta)
\end{equation} 
and $C_{\rho\pi} = {\cal A}_{\CP}^{\rho\pi} = 0$,
where $\delta=\arg(A_{-+}A^*_{+-})$ is the strong phase difference 
between the $\rho^-\pi^+$ and $\rho^+\pi^-$ decay amplitudes.
In the presence of the penguin contribution, there is no straightforward 
interpretation of the Q2B observables in the $\Bz \to \rho^\pm\pi^\mp$ system
in terms of CKM parameters.
However direct $\CP$ violation may arise,
resulting in either or both of $C_{\rho\pi} \neq 0$ and ${\cal A}_{\CP}^{\rho\pi} \neq 0$.
Equivalently,
direct $\CP$ violation may be seen by either of
the decay-type-specific observables ${\cal A}^{+-}_{\rho\pi}$ 
and ${\cal A}^{-+}_{\rho\pi}$, defined in Eq.~(\ref{eq:cp_uta:non-cp-directcp}), 
deviating from zero.
Results and averages for these parameters
are also given in Table~\ref{tab:cp_uta:uud:rhopi_q2b}.
Averages of the direct $\CP$ violation effect in $\Bz \to \rho^\pm\pi^\mp$
are shown in Fig.~\ref{fig:cp_uta:uud:rhopi-dircp},
both in 
${\cal A}^{\rho\pi}_{\CP}$ \vs\ $C_{\rho\pi}$ space and in 
${\cal A}^{-+}_{\rho\pi}$ \vs\ ${\cal A}^{+-}_{\rho\pi}$ space.

Some difference is seen between the 
\babar\ and \belle\ measurements in the $\pi^+\pi^-$ system.
The confidence level of the average is $0.034$,
which corresponds to a $2.1\sigma$ discrepancy.  Since there is no
evidence of systematic problems in either analysis,
we do not rescale the errors of the averages.
The averages for $S_{b \to u\bar u d}$ and $C_{b \to u\bar u d}$ 
in $\Bz \to \pi^+\pi^-$ are both more than $5\sigma$ away from zero,
suggesting that both mixing-induced and direct $\CP$ violation 
are well-established in this channel.
Nonetheless, due to the possible discrepancy mentioned above,
a slightly cautious interpretation should be made 
with regard to the significance of direct $\CP$ violation.

In $\Bz \to \rho^\pm\pi^\mp$, however,
both experiments see an indication of direct $\CP$ violation in the 
${\cal A}^{\rho\pi}_{\CP}$ parameter 
(as seen in Fig.~\ref{fig:cp_uta:uud:rhopi-dircp}).
The average is more than $3\sigma$ from zero,
providing evidence of direct $\CP$ violation in this channel.

\vspace{3ex}

\noindent
\underline{\large Constraints on $\alpha$}

The precision of the measured $\CP$ violation parameters in
$b \to u\bar{u}d$ transitions allows 
constraints to be set on the UT angle $\alpha$. 
Constraints have been obtained with various methods:
\begin{itemize}\setlength{\itemsep}{0.5ex}
\item 
  Both \babar~\cite{Aubert:2007hh}
  and  \belle~\cite{Ishino:2006if} have performed 
  isospin analyses in the $\pi\pi$ system.
  \belle\ exclude $9^\circ < \phi_2 < 81^\circ$ at the $95.4\%$  C.L. while
  \babar\ give a confidence level interpretation for $\alpha$,
  exclude the range $23^\circ < \alpha < 67^\circ$ at the $90\%$  C.L.
  In both cases, only solutions in $0^\circ$--$180^\circ$ are considered.

\item
  Both experiments have also performed isospin analyses in the $\rho\rho$
  system. 
  The most recent result from \babar\ is given in an update of the
  measurements of the $B^+\to\rho^+\rho^0$ decay~\cite{Aubert:2009it}, and
  sets the constraint $\alpha = \left( 92.4 \,^{+6.0}_{-6.5}\right)^\circ$.
  The most recent result from \belle\ is given in an update of the
  search for the $\Bz \to \rho^0\rho^0$ decay and sets the constraint
  $\phi_2 = \left( 91.7 \pm 14.9 \right)^\circ$~\cite{:2008et}.

\item
  The time-dependent Dalitz plot analysis of the $\Bz \to \pi^+\pi^-\pi^0$
  decay allows a determination of $\alpha$ without input from any other 
  channels.
  \babar~\cite{Aubert:2007jn} obtain the constraint 
  $75^\circ < \alpha < 152^\circ$ at $68\%$ C.L.
  \belle~\cite{Kusaka:2007dv,:2007mj} have performed a similar analysis,
  and in addition have included information from the SU(2) partners of 
  $B \to \rho\pi$, which can be used to constrain $\alpha$
  via an isospin pentagon relation~\cite{Lipkin:1991st}. 
  With this analysis,
  \belle\ obtain the tighter constraint $\phi_2 = (83 \, ^{+12}_{-23})^\circ$
  (where the errors correspond to $1\sigma$, \ie\ $68.3\%$ confidence level).

\item 
  The results from \babar\ on $\Bz \to a_1^\pm \pi^\mp$~\cite{Aubert:2006gb} can be
  combined with results from modes related by isospin~\cite{Gronau:2005kw}
  leading to the following constraint: 
  $\alpha = \left( 79 \pm 7 \pm 11 \right)^\circ$~\cite{:2009ii}.

\item 
  Each experiment has obtained a value of $\alpha$ from combining its 
  results in the different $b \to u \bar{u} d$ modes 
  (with some input also from HFAG).
  These values have appeared in talks, but not in publications,
  and are not listed here.

\item 
  The CKMfitter~\cite{Charles:2004jd} and 
  UTFit~\cite{Bona:2005vz} groups use the measurements 
  from \belle\ and \babar\ given above
  with other branching fractions and \CP asymmetries in 
  $\B\to\pi\pi$, $\rho\pi$ and $\rho\rho$ modes, 
  to perform isospin analyses for each system, 
  and to make combined constraints on $\alpha$.
\end{itemize}

Note that methods based on isospin symmetry make extensive use of 
measurements of branching fractions and direct $\CP$ asymmetries,
as averaged by the HFAG Rare Decays subgroup (Sec.~\ref{sec:rare}).
Note also that each method suffers from discrete ambiguities in the solutions.
The model assumption in the $\Bz \to \pi^+\pi^-\pi^0$ analysis 
allows to resolve some of the multiple solutions, 
and results in a single preferred value for $\alpha$ in $\left[ 0, \pi \right]$.
All the above measurements correspond to the choice
that is in agreement with the global CKM fit.

At present we make no attempt to provide an HFAG average for $\alpha$.
More details on procedures to calculate a best fit value for $\alpha$ 
can be found in Refs.~\cite{Charles:2004jd,Bona:2005vz}.

\clearpage
\mysubsection{Time-dependent $\CP$ asymmetries in $b \to c\bar{u}d / u\bar{c}d$ transitions
}
\label{sec:cp_uta:cud}

Non-$\CP$ eigenstates such as $D^\pm\pi^\mp$, $D^{*\pm}\pi^\mp$ and $D^\pm\rho^\mp$ can be produced 
in decays of $\Bz$ mesons either via Cabibbo favoured ($b \to c$) or
doubly Cabibbo suppressed ($b \to u$) tree amplitudes. 
Since no penguin contribution is possible,
these modes are theoretically clean.
The ratio of the magnitudes of the suppressed and favoured amplitudes, $R$,
is sufficiently small (predicted to be about $0.02$),
that terms of ${\cal O}(R^2)$ can be neglected, 
and the sine terms give sensitivity to the combination of UT angles $2\beta+\gamma$.

\begin{figure}[htb]
  \begin{center}
    \begin{tabular}{cc}
      \resizebox{0.46\textwidth}{!}{
        \includegraphics{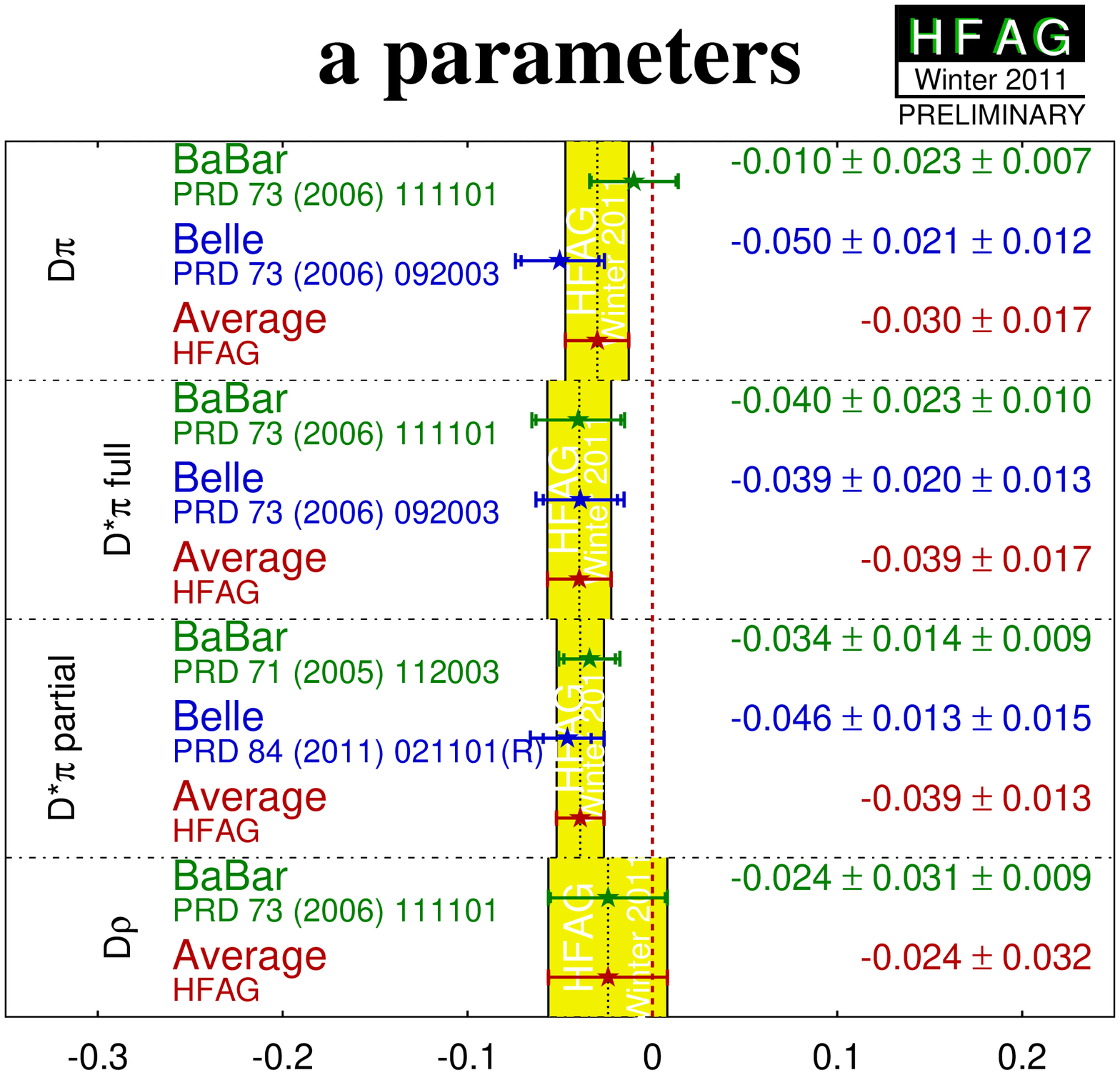}
      }
      &
      \resizebox{0.46\textwidth}{!}{
        \includegraphics{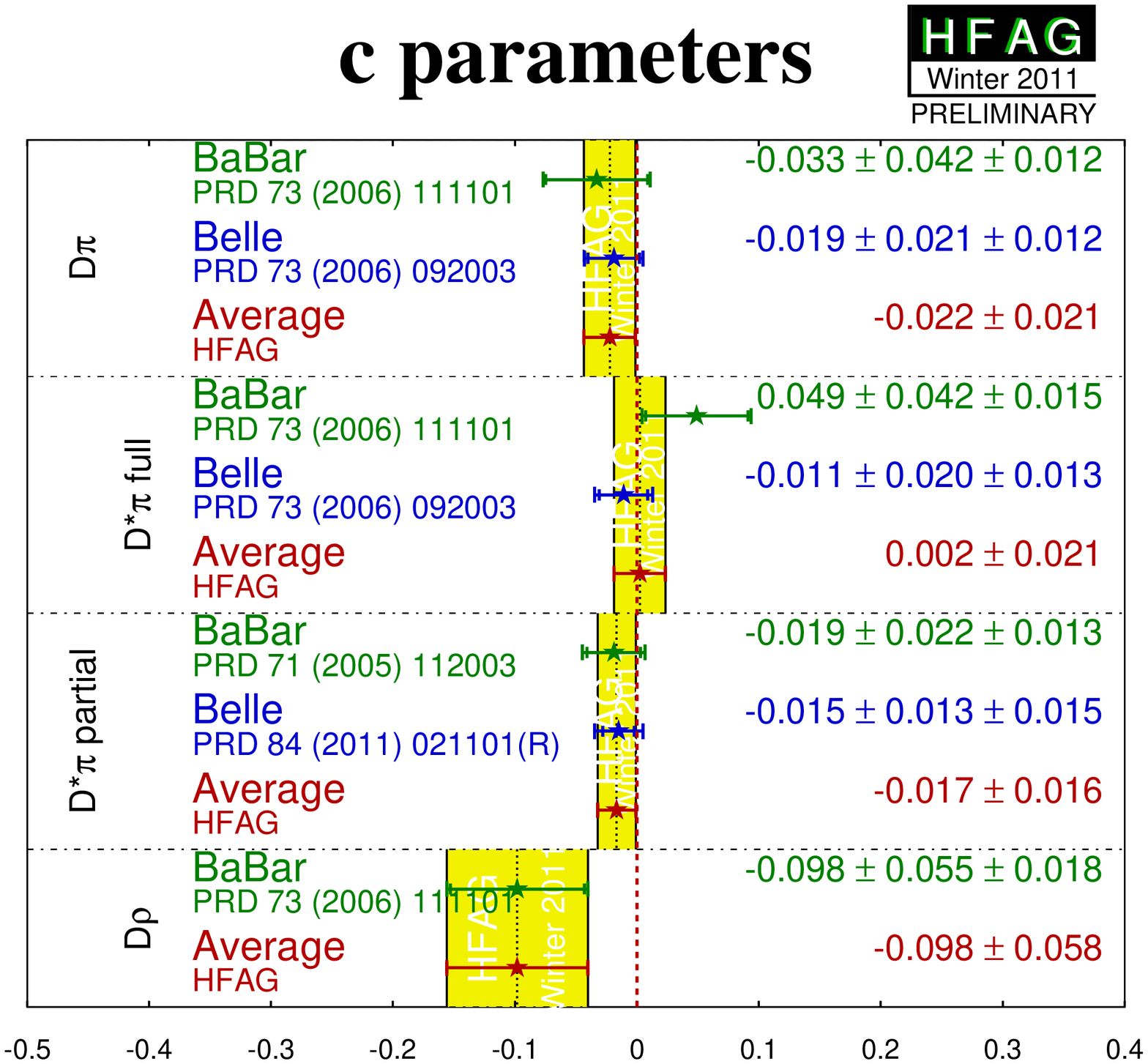}
      }
    \end{tabular}
  \end{center}
  \vspace{-0.8cm}
  \caption{
    Averages for $b \to c\bar{u}d / u\bar{c}d$ modes.
  }
  \label{fig:cp_uta:cud}
\end{figure}

\begin{figure}[htb]
  \begin{center}
    \begin{tabular}{cc}
      \resizebox{0.46\textwidth}{!}{
        \includegraphics{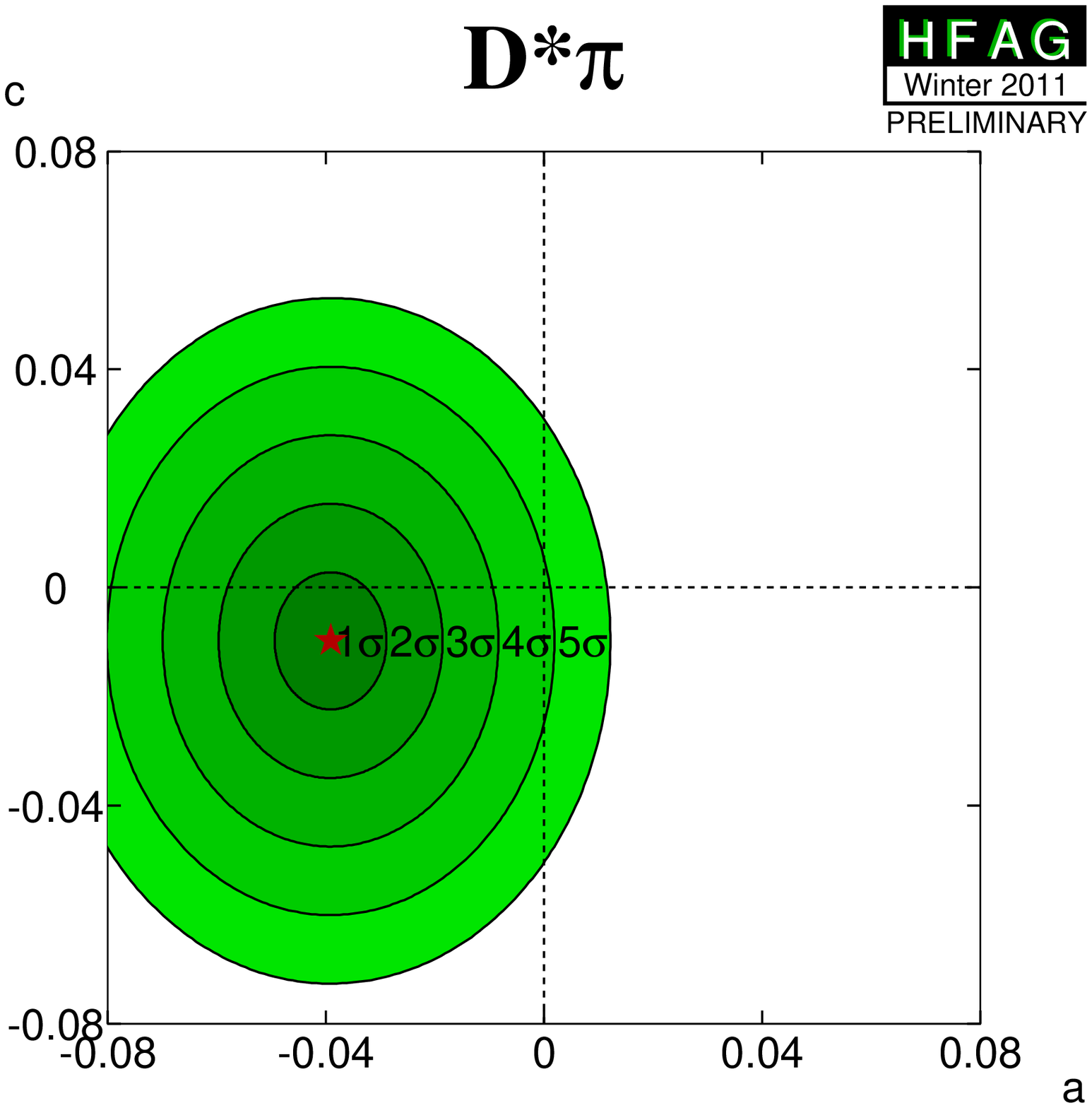}
      }
      &
      \resizebox{0.46\textwidth}{!}{
        \includegraphics{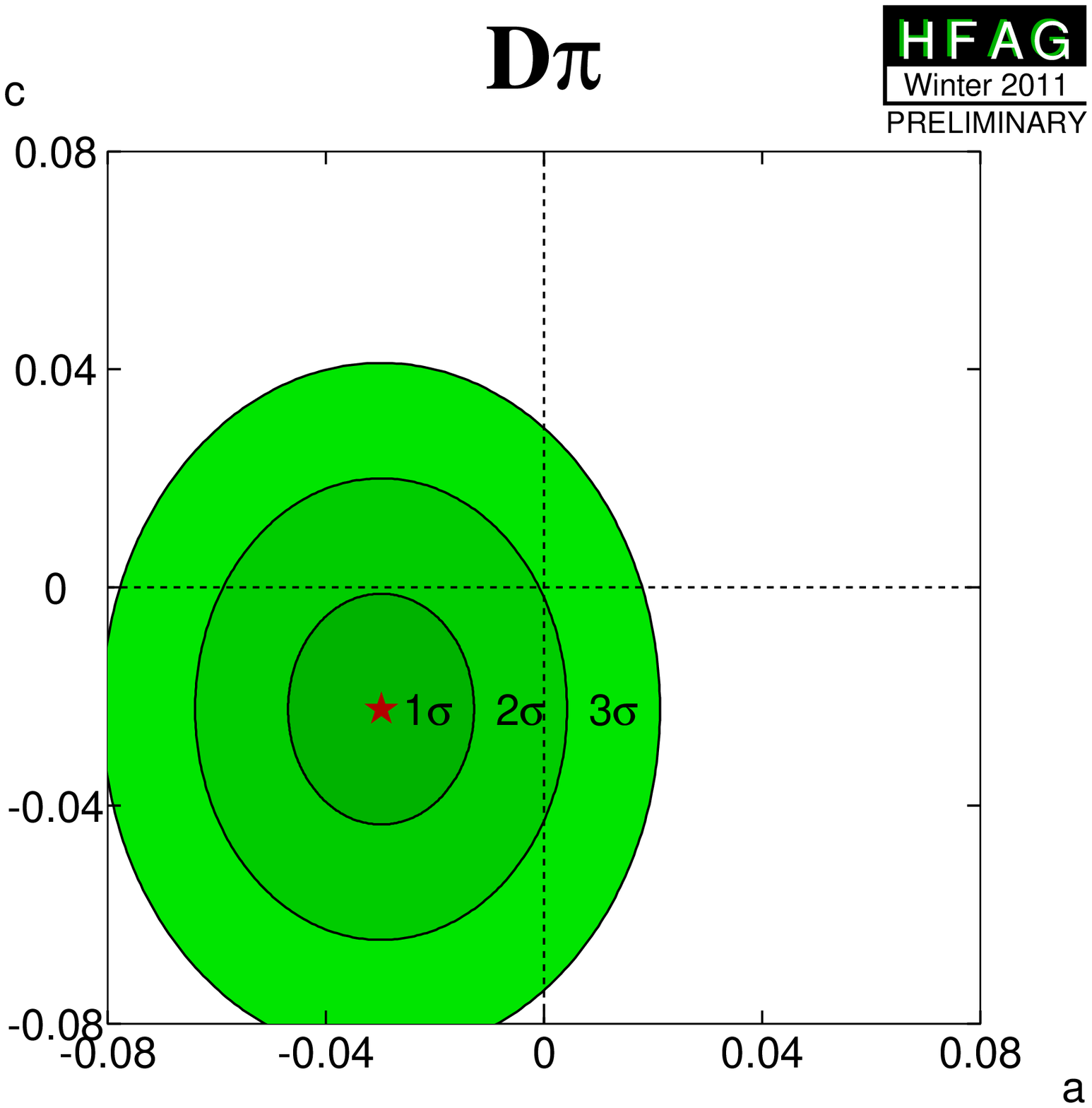}
      } \\
      \resizebox{0.46\textwidth}{!}{
        \includegraphics{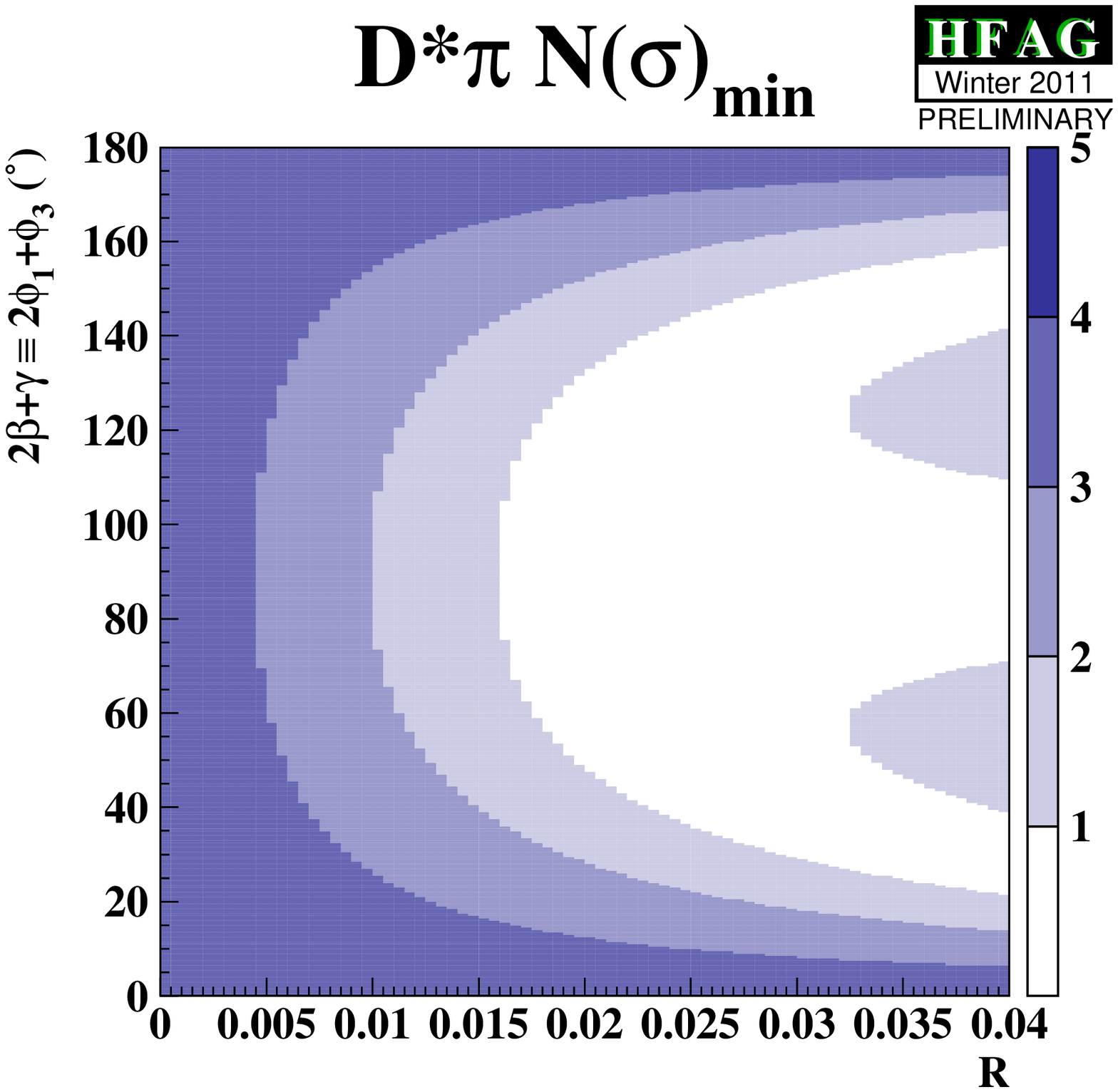}
      }
      &
      \resizebox{0.46\textwidth}{!}{
        \includegraphics{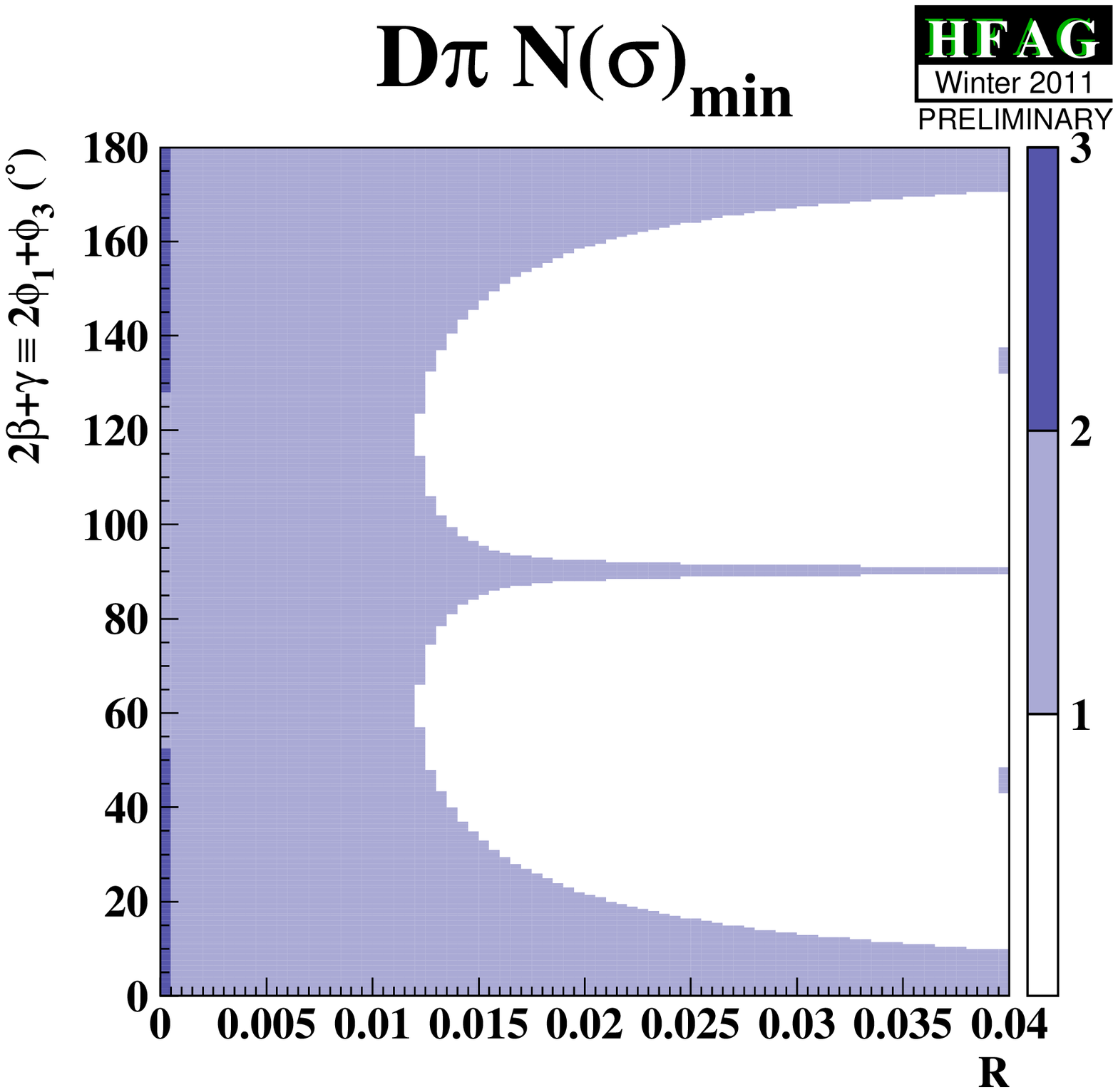}
      }          
    \end{tabular}
  \end{center}
  \vspace{-0.8cm}
  \caption{
    Results from $b \to c\bar{u}d / u\bar{c}d$ modes.
    (Top) Constraints in $c$ {\it vs.} $a$ space.
    (Bottom) Constraints in $2\beta+\gamma$ {\it vs.} $R$ space.
    (Left) $D^*\pi$ and (right) $D\pi$ modes.
  }
  \label{fig:cp_uta:cud_constraints}
\end{figure}

As described in Sec.~\ref{sec:cp_uta:notations:non_cp:dstarpi},
the averages are given in terms of parameters $a$ and $c$.
$\CP$ violation would appear as $a \neq 0$.
Results are available from both \babar\ and \belle\ in the modes
$D^\pm\pi^\mp$ and $D^{*\pm}\pi^\mp$; for the latter mode both experiments 
have used both full and partial reconstruction techniques.
Results are also available from \babar\ using $D^\pm\rho^\mp$.
These results, and their averages, are listed in Table~\ref{tab:cp_uta:cud},
and are shown in Fig.~\ref{fig:cp_uta:cud}.
The constraints in $c$ \vs\ $a$ space for the $D\pi$ and $D^*\pi$ modes
are shown in Fig.~\ref{fig:cp_uta:cud_constraints}.
It is notable that the average value of $a$ from $D^*\pi$ is more than
$3\sigma$ from zero, providing evidence of $\CP$ violation in this channel.

\begin{table}[htb]
	\begin{center}
		\caption{
      Averages for $b \to c\bar{u}d / u\bar{c}d$ modes.
                }
                \vspace{0.2cm}
                \setlength{\tabcolsep}{0.0pc}
                \begin{tabular*}{\textwidth}{@{\extracolsep{\fill}}lrccc} \hline 
	\mc{2}{l}{Experiment} & $N(B\bar{B})$ & $a$ & $c$ \\
	\hline
      \mc{5}{c}{$D^{\pm}\pi^{\mp}$} \\
	\babar (full rec.) & \cite{Aubert:2006tw} & 232M & $-0.010 \pm 0.023 \pm 0.007$ & $-0.033 \pm 0.042 \pm 0.012$ \\
	\belle (full rec.) & \cite{Ronga:2006hv} & 386M & $-0.050 \pm 0.021 \pm 0.012$ & $-0.019 \pm 0.021 \pm 0.012$ \\
        \mc{3}{l}{\bf Average} & $ -0.030 \pm 0.017$ & $ -0.022 \pm 0.021 $ \\
        \mc{3}{l}{\small Confidence level} & {\small $0.24~(1.2\sigma)$} & {\small $0.78~(0.3\sigma)$} \\
        \hline
      \mc{5}{c}{$D^{*\pm}\pi^{\mp}$} \\
      \babar (full rec.) & \cite{Aubert:2006tw} & 232M & $-0.040 \pm 0.023 \pm 0.010$ & $0.049 \pm 0.042 \pm 0.015$ \\
      \babar (partial rec.)  & \cite{Aubert:2005yf} & 232M & $-0.034 \pm 0.014 \pm 0.009$ & $-0.019 \pm 0.022 \pm 0.013$ \\
      \belle (full rec.) & \cite{Ronga:2006hv} & 386M & $-0.039 \pm 0.020 \pm 0.013$ & $-0.011 \pm 0.020 \pm 0.013$ \\
      \belle (partial rec.) & \cite{Bahinipati:2011yq} & 657M & $-0.046 \pm 0.013 \pm 0.015$ & $-0.015 \pm 0.013 \pm 0.015$ \\
	\mc{3}{l}{\bf Average} & $-0.039 \pm 0.013$ & $-0.017 \pm 0.016$ \\
      \mc{3}{l}{\small Confidence level} & {\small $0.97~(0.03\sigma)$} & {\small $0.59~(0.6\sigma)$} \\
      \hline
      \mc{5}{c}{$D^{\pm}\rho^{\mp}$} \\
      \babar (full rec.) & \cite{Aubert:2006tw} & 232M & $-0.024 \pm 0.031 \pm 0.009$ & $-0.098 \pm 0.055 \pm 0.018$ \\
      \hline 
    \end{tabular*}
    \label{tab:cp_uta:cud}
  \end{center}
\end{table}

For each of $D\pi$, $D^*\pi$ and $D\rho$, 
there are two measurements ($a$ and $c$, or $S^+$ and $S^-$) 
which depend on three unknowns ($R$, $\delta$ and $2\beta+\gamma$), 
of which two are different for each decay mode. 
Therefore, there is not enough information to solve directly for $2\beta+\gamma$. 
However, for each choice of $R$ and $2\beta+\gamma$, 
one can find the value of $\delta$ that allows $a$ and $c$ to be closest 
to their measured values, 
and calculate the distance in terms of numbers of standard deviations.
(We currently neglect experimental correlations in this analysis.) 
These values of $N(\sigma)_{\rm min}$ can then be plotted 
as a function of $R$ and $2\beta+\gamma$
(and can trivially be converted to confidence levels). 
These plots are given for the $D\pi$ and $D^*\pi$ modes 
in Figure~\ref{fig:cp_uta:cud_constraints}; 
the uncertainties in the $D\rho$ mode are currently too large 
to give any meaningful constraint.

The constraints can be tightened if one is willing 
to use theoretical input on the values of $R$ and/or $\delta$. 
One popular choice is the use of SU(3) symmetry to obtain 
$R$ by relating the suppressed decay mode to $\B$ decays 
involving $D_s$ mesons. 
More details can be found 
in Refs.~\cite{Charles:2004jd,Bona:2005vz}.

\mysubsection{Time-dependent $\CP$ asymmetries in $b \to c\bar{u}s / u\bar{c}s$ transitions
}
\label{sec:cp_uta:cus-td}

Time-dependent analyses of transitions such as $\Bz \to D^\pm \KS \pi^\mp$ can
be used to probe $\sin(2\beta+\gamma)$ in a similar way to that discussed
above (Sec.~\ref{sec:cp_uta:cud}). Since the final state contains three
particles, a Dalitz plot analysis is necessary to maximise the
sensitivity. \babar~\cite{Aubert:2007qe} have carried out such an
analysis. They obtain $2\beta+\gamma = \left( 83 \pm 53 \pm 20 \right)^\circ$
(with an ambiguity $2\beta+\gamma \leftrightarrow 2\beta+\gamma+\pi$) assuming
the ratio of the $b \to u$ and $b \to c$ amplitude to be constant across the
Dalitz plot at 0.3.

\clearpage
\mysubsection{Rates and asymmetries in $\Bmp \to \DorDstar K^{(*)\mp}$ decays
}
\label{sec:cp_uta:cus}

As explained in Sec.~\ref{sec:cp_uta:notations:cus},
rates and asymmetries in $\Bmp \to \DorDstar K^{(*)\mp}$ decays
are sensitive to $\gamma$.
Various methods using different $\DorDstar$ final states exist.

\mysubsubsection{$D$ decays to $\CP$ eigenstates
}
\label{sec:cp_uta:cus:glw}

\begin{figure}[htb]
  \begin{center}
    \begin{tabular}{cc}
      \resizebox{0.46\textwidth}{!}{
        \includegraphics{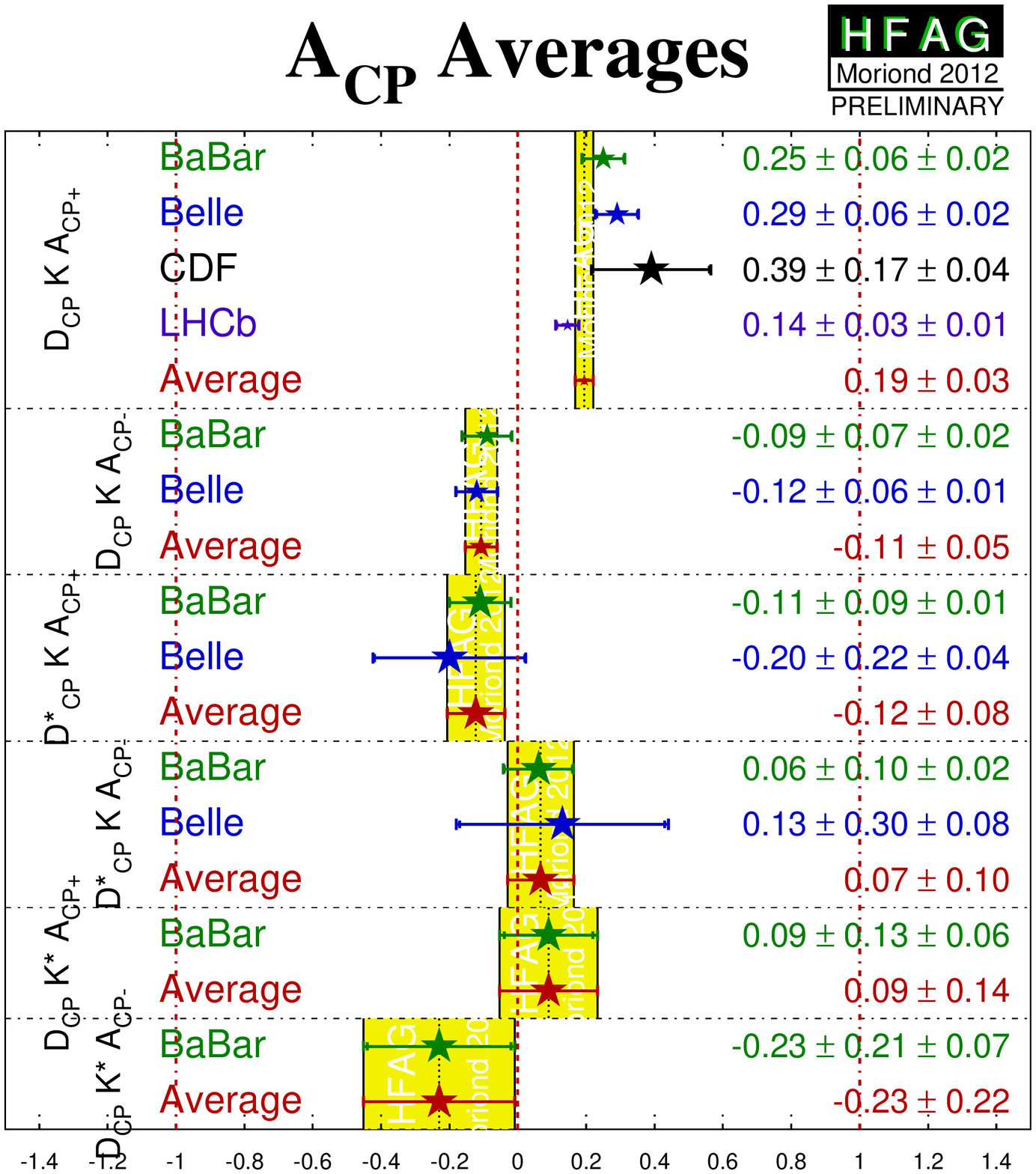}
      }
      &
      \resizebox{0.46\textwidth}{!}{
        \includegraphics{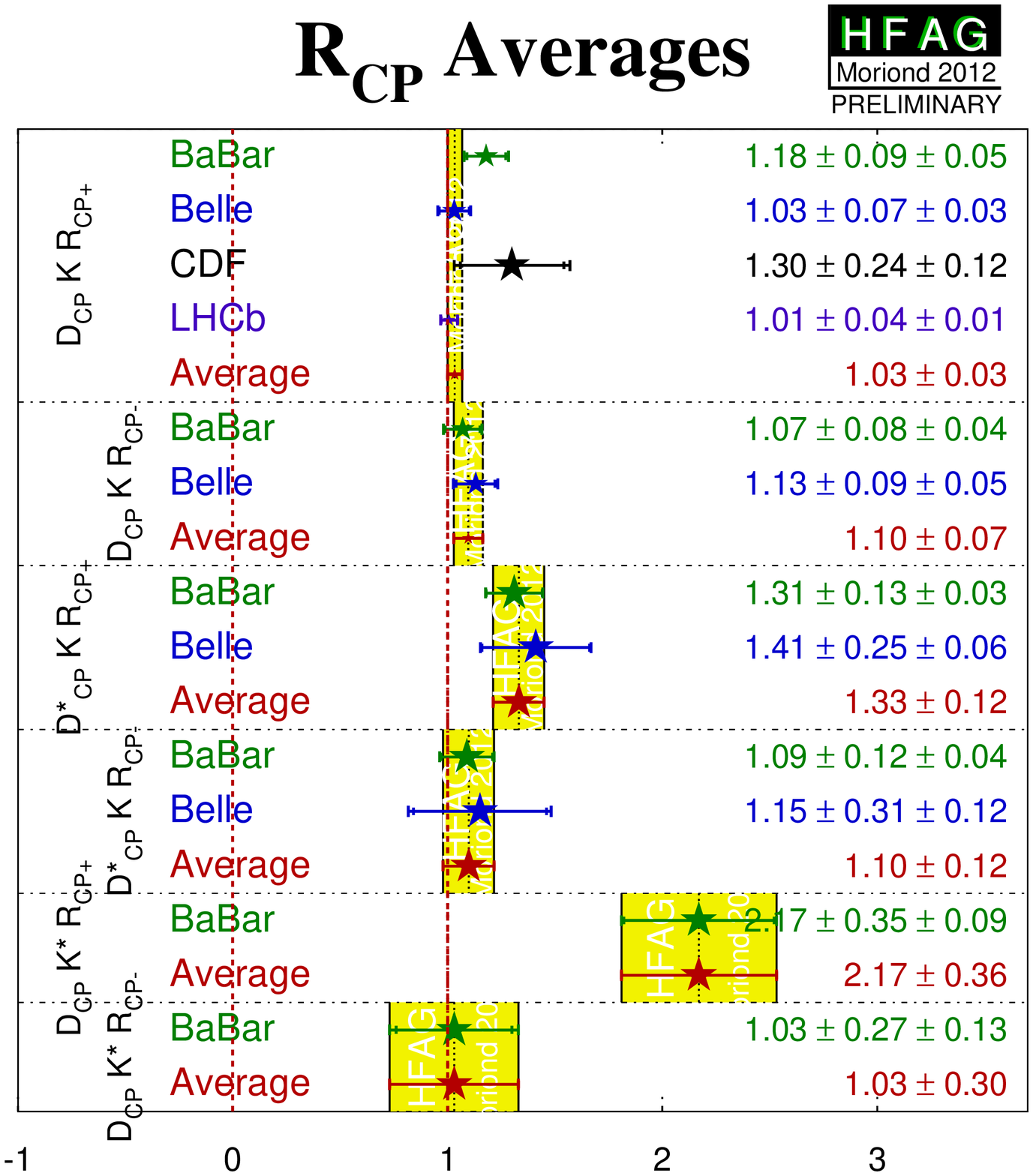}
      }
    \end{tabular}
 \end{center}
  \vspace{-0.8cm}
  \caption{
    Averages of $A_{\CP}$ and $R_{\CP}$ from GLW analyses.
  }
  \label{fig:cp_uta:cus:glw}
\end{figure}

Results are available from both \babar\ and \belle\ on GLW analyses in the
decay modes $\Bmp \to D\Kmp$, $\Bmp \to \Dstar\Kmp$ and 
$\Bmp \to D\Kstarmp$.\footnote{
  We do not include a preliminary result from \belle~\cite{Abe:2003rg}, which
  remains unpublished after more than two years.
}
Both experiments use the $\CP$-even $D$ decay final states $K^+K^-$ and
$\pi^+\pi^-$ in all three modes; both experiments generally use the \CP-odd
decay modes $\KS\pi^0$, $\KS\omega$ and $\KS\phi$, though care is taken to
avoid statistical overlap with the $\KS K^+K^-$ sample used for Dalitz plot
analysis (see Sec.~\ref{sec:cp_uta:cus:dalitz}), 
and asymmetric systematic errors are assigned due to $\CP$-even pollution
under the $\KS\omega$ and $\KS\phi$ signals.
Both experiments also use the $\Dstar \to D\pi^0$ decay, 
which gives $\CP(\Dstar) = \CP(D)$;
\babar\ in addition use the $\Dstar \to D\gamma$ decays, 
which gives $\CP(\Dstar) = -\CP(D)$.
In addition, results from CDF and LHCb are available in the
decay mode $\Bmp \to D\Kmp$, 
for $\CP$-even final states ($K^+K^-$ and $\pipi$) only.
The results and averages are given in Table~\ref{tab:cp_uta:cus:glw}
and shown in Fig.~\ref{fig:cp_uta:cus:glw}.

\begin{table}[htb]
	\begin{center}
		\caption{
                        Averages from GLW analyses of $b \to c\bar{u}s / u\bar{c}s$ modes.
                }
                \vspace{0.2cm}
    \resizebox{\textwidth}{!}{
      \setlength{\tabcolsep}{0.0pc}
      \begin{tabular}{@{\extracolsep{2mm}}lrccccc} \hline 
        \mc{2}{l}{Experiment} & Sample size & $A_{\CP+}$ & $A_{\CP-}$ & $R_{\CP+}$ & $R_{\CP-}$ \\
        \hline
        \mc{7}{c}{$D_{\CP} K^-$} \\
	\babar & \cite{delAmoSanchez:2010ji} & $N(B\bar{B}) =$ 467M & $0.25 \pm 0.06 \pm 0.02$ & $-0.09 \pm 0.07 \pm 0.02$ & $1.18 \pm 0.09 \pm 0.05$ & $1.07 \pm 0.08 \pm 0.04$ \\
	\belle & \cite{belle:glwads:prelim} & $N(B\bar{B}) =$ 772M & $0.29 \pm 0.06 \pm 0.02$ & $-0.12 \pm 0.06 \pm 0.01$ & $1.03 \pm 0.07 \pm 0.03$ & $1.13 \pm 0.09 \pm 0.05$ \\
	CDF & \cite{Aaltonen:2009hz} & 1 ${\rm fb}^{-1}$ & $0.39 \pm 0.17 \pm 0.04$ & \textendash{} & $1.30 \pm 0.24 \pm 0.12$ &  \textendash{} \\
	LHCb & \cite{Aaij:2012kz} & 1 ${\rm fb}^{-1}$ & $0.14 \pm 0.03 \pm 0.01$ &  \textendash{} & $1.01 \pm 0.04 \pm 0.01$ &  \textendash{} \\
	\mc{3}{l}{\bf Average} & $0.19 \pm 0.03$ & $-0.11 \pm 0.05$ & $1.03 \pm 0.03$ & $1.10 \pm 0.07$ \\
	\mc{3}{l}{\small Confidence level} & {\small $0.09~(1.7\sigma)$} & {\small $0.75~(0.3\sigma)$} & {\small $0.33~(1.0\sigma)$} & {\small $0.66~(0.4\sigma)$} \\
		\hline

        \mc{7}{c}{$\Dstar_{\CP} K^-$} \\
	\babar & \cite{:2008jd} & $N(B\bar{B}) =$ 383M & $-0.11 \pm 0.09 \pm 0.01$ & $0.06 \pm 0.10 \pm 0.02$ & $1.31 \pm 0.13 \pm 0.03$ & $1.09 \pm 0.12 \pm 0.04$ \\
	\belle & \cite{Abe:2006hc} & $N(B\bar{B}) =$ 275M & $-0.20 \pm 0.22 \pm 0.04$ & $0.13 \pm 0.30 \pm 0.08$ & $1.41 \pm 0.25 \pm 0.06$ & $1.15 \pm 0.31 \pm 0.12$ \\
	\mc{3}{l}{\bf Average} & $-0.12 \pm 0.08$ & $0.07 \pm 0.10$ & $1.33 \pm 0.12$ & $1.10 \pm 0.12$ \\
	\mc{3}{l}{\small Confidence level} & {\small $0.71~(0.4\sigma)$} & {\small $0.83~(0.2\sigma)$} & {\small $0.73~(0.4\sigma)$} & {\small $0.87~(0.2\sigma)$} \\
		\hline

        \mc{7}{c}{$D_{\CP} K^{*-}$} \\
	\babar & \cite{Aubert:2009yw} & $N(B\bar{B}) =$ 379M & $0.09 \pm 0.13 \pm 0.06$ & $-0.23 \pm 0.21 \pm 0.07$ & $2.17 \pm 0.35 \pm 0.09$ & $1.03 \pm 0.27 \pm 0.13$ \\
		\hline
      \end{tabular}
    }
    \label{tab:cp_uta:cus:glw}
	\end{center}
\end{table}

\mysubsubsection{$D$ decays to suppressed final states}
\label{sec:cp_uta:cus:ads}

\begin{figure}[htb]
  \begin{center}
    \begin{tabular}{cc}
      \resizebox{0.46\textwidth}{!}{
        \includegraphics{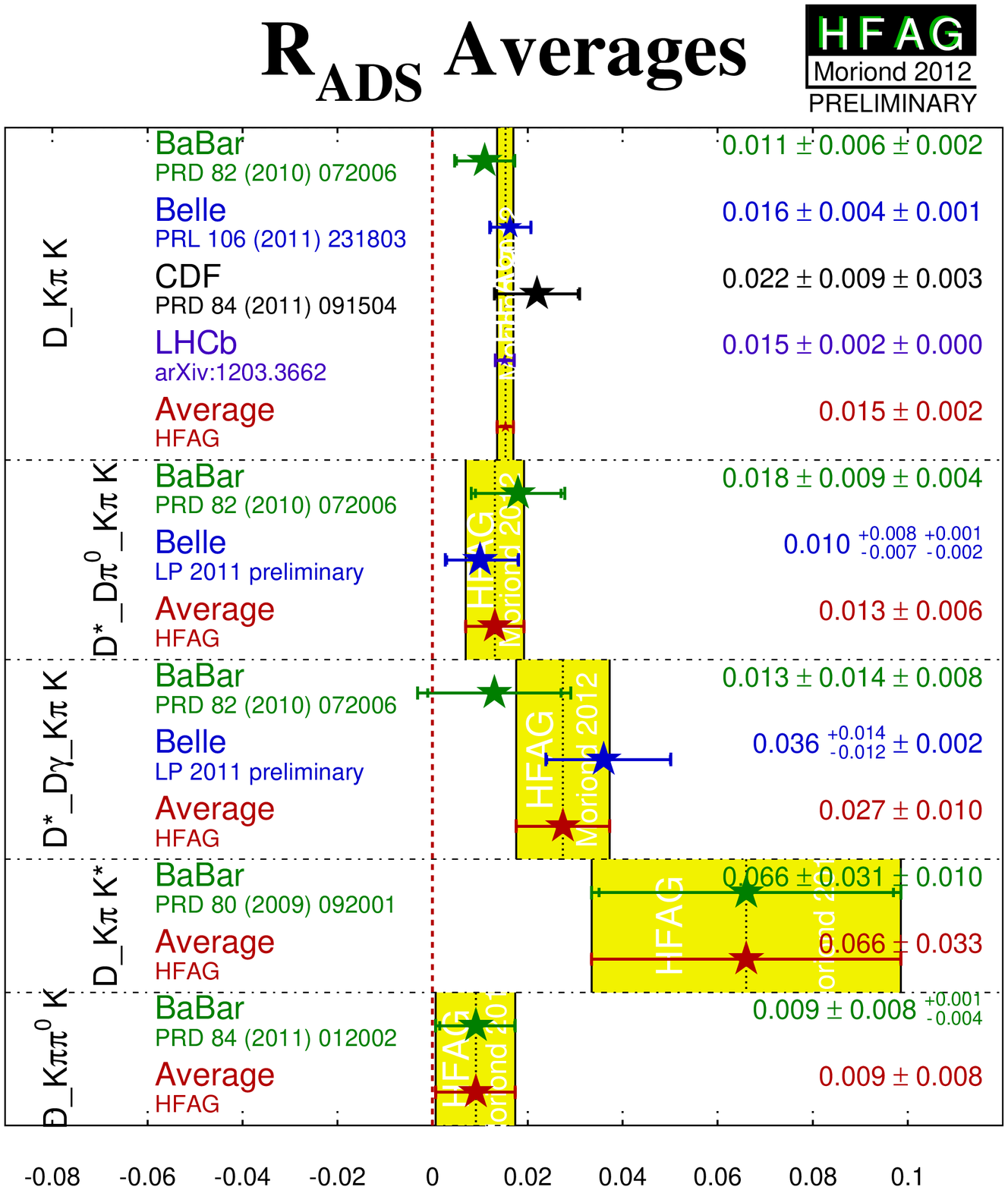}
      }
      &
      \resizebox{0.46\textwidth}{!}{
        \includegraphics{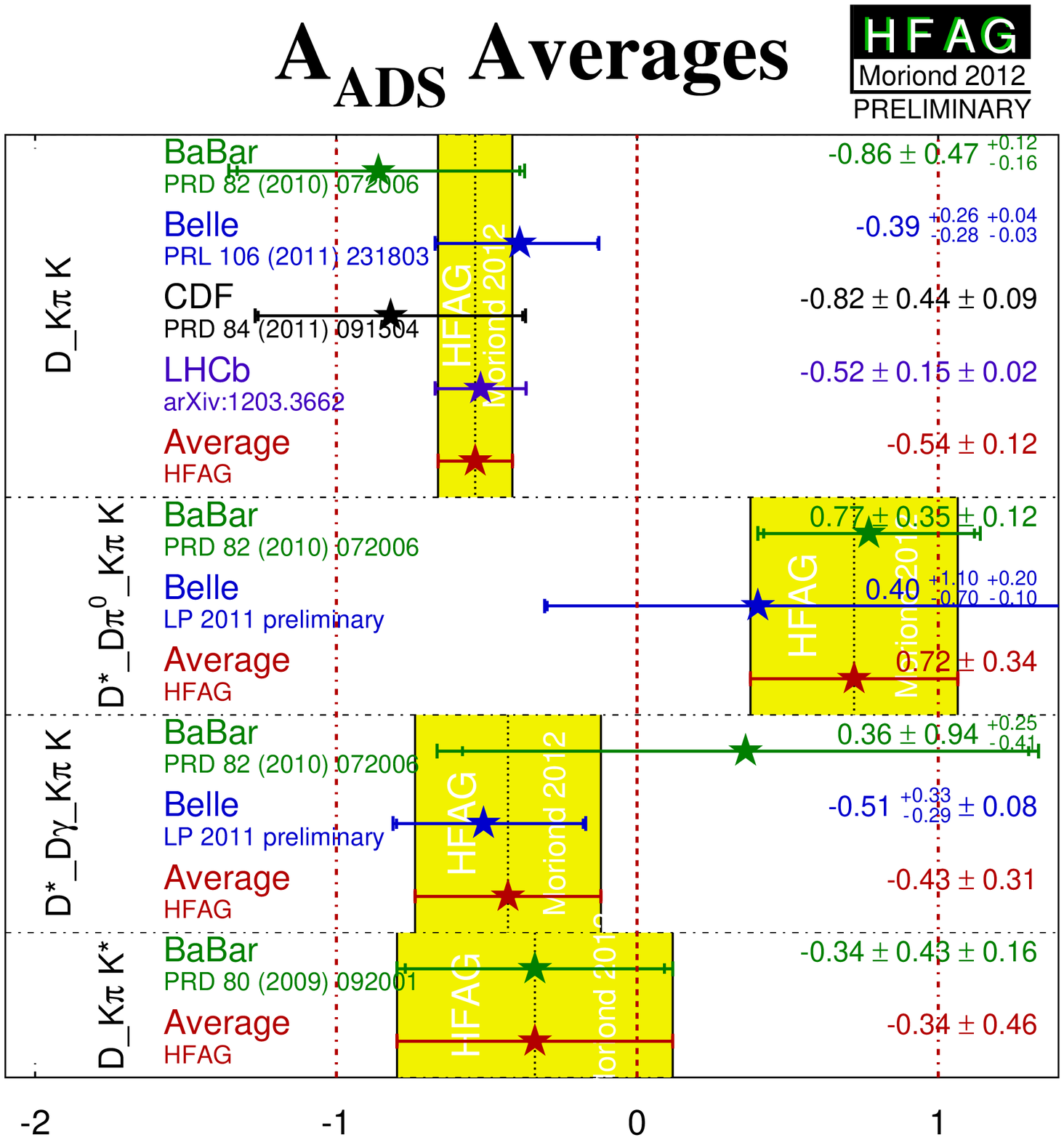}
      }
    \end{tabular}
  \end{center}
  \vspace{-0.8cm}
  \caption{
    Averages of $R_{\rm ADS}$ and $A_{\rm ADS}$ for $B \to D^{(*)}K^{(*)}$ decays.
  }
  \label{fig:cp_uta:cus:ads}
\end{figure}

\begin{figure}[htb]
  \begin{center}
    \begin{tabular}{cc}
      \resizebox{0.46\textwidth}{!}{
        \includegraphics{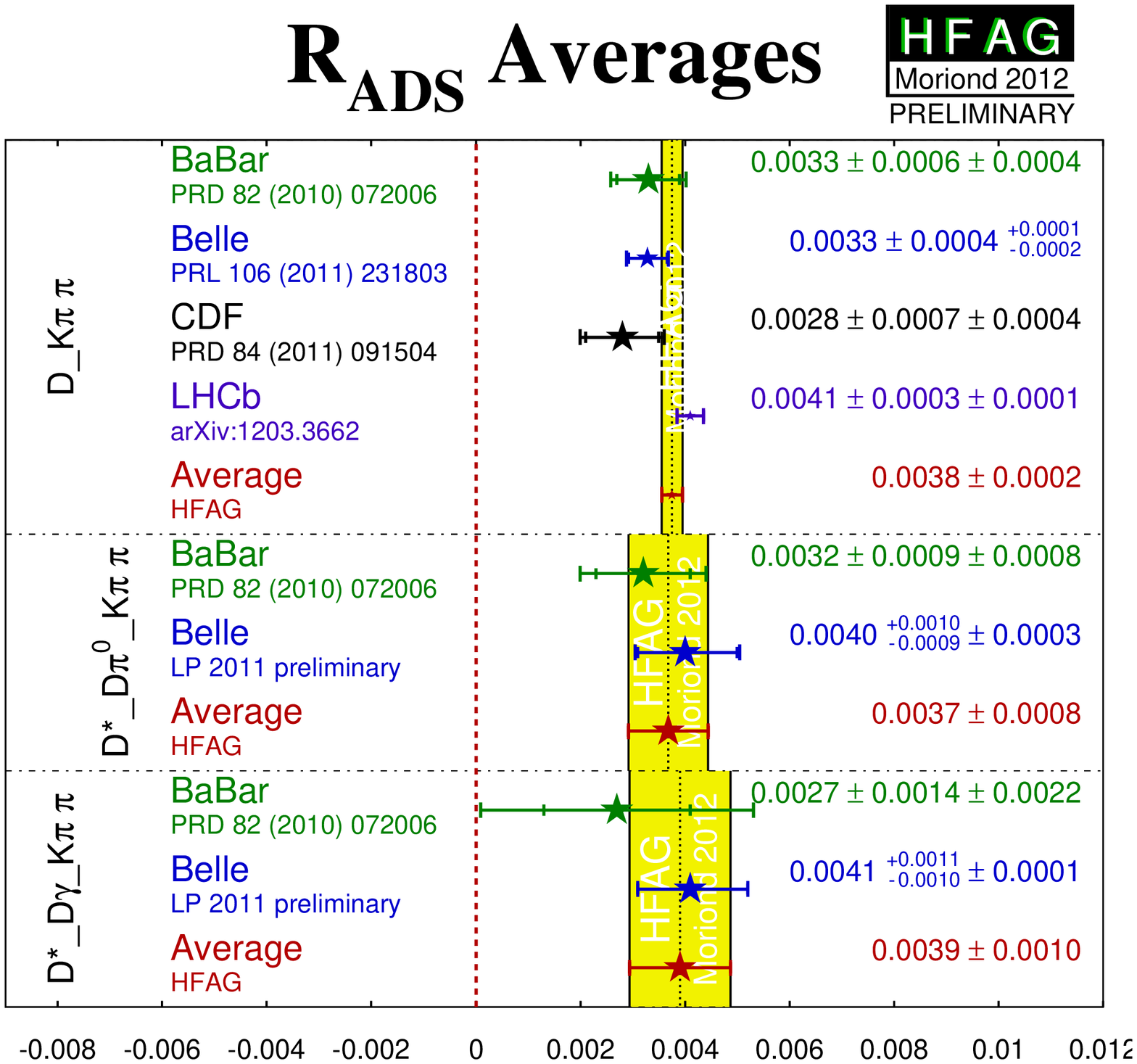}
      }
      &
      \resizebox{0.46\textwidth}{!}{
        \includegraphics{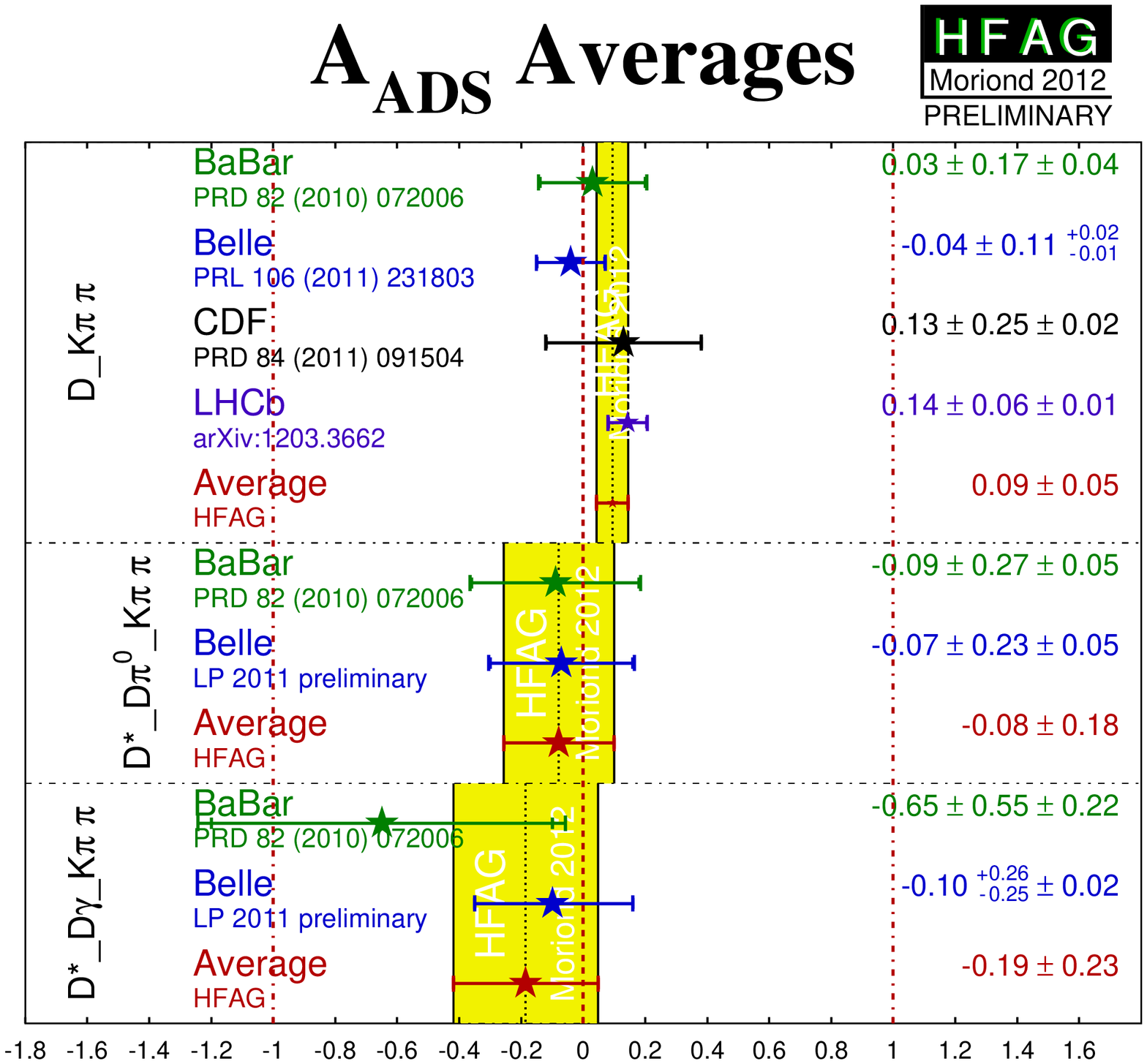}
      }
    \end{tabular}
  \end{center}
  \vspace{-0.8cm}
  \caption{
    Averages of $R_{\rm ADS}$ and $A_{\rm ADS}$ for $B \to D^{(*)}\pi$ decays.
  }
  \label{fig:cp_uta:cus:ads-Dpi}
\end{figure}

For ADS analysis, both \babar\ and \belle\ have studied the modes 
$\Bmp \to D\Kmp$ and $\Bmp \to D\pi^\mp$. \babar\ has also analysed the 
$\Bmp \to \Dstar\Kmp$ and $\Bmp \to D\Kstarmp$ modes.
There is an effective shift of $\pi$ in the strong phase difference between
the cases that the $\Dstar$ is reconstructed as $D\pi^0$ and
$D\gamma$~\cite{Bondar:2004bi}, therefore these modes are studied separately.
$\Kstarmp$ is reconstructed as $\KS\pi^\mp$.
In all cases the suppressed decay $D \to K^+\pi^-$ has been used.
\babar\ also has results using $\Bmp \to D\Kmp$ with $D \to K^+\pi^-\pi^0$.
The results and averages are given in Table~\ref{tab:cp_uta:cus:ads}
and shown in Figs.~\ref{fig:cp_uta:cus:ads} and~\ref{fig:cp_uta:cus:ads-Dpi}.

\begin{table}[htb]
	\begin{center}
		\caption{
      Averages from ADS analyses of $b \to c\bar{u}s / u\bar{c}s$ and 
      $b \to c\bar{u}d / u\bar{c}d$ modes.
                }
                \vspace{0.2cm}
                \setlength{\tabcolsep}{0.0pc}
                \begin{tabular*}{\textwidth}{@{\extracolsep{\fill}}lrccc} \hline 
        \mc{2}{l}{Experiment} & Sample size & $A_{\rm ADS}$ & $R_{\rm ADS}$ \\
	\hline
        \mc{5}{c}{$D K^-$, $D \to K^+\pi^-$} \\
	\babar & \cite{delAmoSanchez:2010dz} & $N(B\bar{B}) =$ 467M & $-0.86 \pm 0.47 \,^{+0.12}_{-0.16}$ & $0.011 \pm 0.006 \pm 0.002$ \\
	\belle & \cite{Belle:2011ac} & $N(B\bar{B}) =$ 772M & $-0.39 \,^{+0.26}_{-0.28} \,^{+0.04}_{-0.03}$ & $0.0163 \,^{+0.0044}_{-0.0041} \,^{+0.0007}_{-0.0013}$ \\
	CDF & \cite{Aaltonen:2011uu} & 7 ${\rm fb}^{-1}$ & $-0.82 \pm 0.44 \pm 0.09$ & $0.0220 \pm 0.0086 \pm 0.0026$ \\
	LHCb & \cite{Aaij:2012kz} & 1 ${\rm fb}^{-1}$ & $-0.52 \pm 0.15 \pm 0.02$ & $0.0152 \pm 0.0020 \pm 0.0004$ \\
	\mc{3}{l}{\bf Average} & $-0.54 \pm 0.12$ & $0.0153 \pm 0.0017$ \\
	\mc{3}{l}{\small Confidence level} & {\small $0.77~(0.3\sigma)$} & {\small $0.78~(0.3\sigma)$} \\
		\hline

        \mc{2}{l}{Experiment} & $N(B\bar{B})$ & $A_{\rm ADS}$ & $R_{\rm ADS}$ \\
        \mc{5}{c}{$\Dstar K^-$, $\Dstar \to D\pi^0$, $D \to K^+\pi^-$} \\
	\babar & \cite{delAmoSanchez:2010dz} & 467M & $0.77 \pm 0.35 \pm 0.12$ & $0.018 \pm 0.009 \pm 0.004$ \\
	\belle & \cite{belle:glwads:prelim} & 772M & $0.4 \,^{+1.1}_{-0.7} \,^{+0.2}_{-0.1}$ & $0.010 \,^{+0.008}_{-0.007} \,^{+0.001}_{-0.002}$ \\
	\mc{3}{l}{\bf Average} & $0.72 \pm 0.34$ & $0.013 \pm 0.006$ \\
	\mc{3}{l}{\small Confidence level} & {\small $0.71~(0.4\sigma)$} & {\small $0.52~(0.6\sigma)$} \\
 	\hline

        \mc{5}{c}{$\Dstar K^-$, $\Dstar \to D\gamma$, $D \to K^+\pi^-$} \\
	\babar & \cite{delAmoSanchez:2010dz} & 467M & $0.36 \pm 0.94 \,^{+0.25}_{-0.41}$ & $0.013 \pm 0.014 \pm 0.008$ \\
	\belle & \cite{belle:glwads:prelim} & 772M & $-0.51 \,^{+0.33}_{-0.29} \pm 0.08$ & $0.036 \,^{+0.014}_{-0.012} \pm 0.002$ \\
	\mc{3}{l}{\bf Average} & $-0.43 \pm 0.31$ & $0.027 \pm 0.010$ \\
	\mc{3}{l}{\small Confidence level} & {\small $0.42~(0.8\sigma)$} & {\small $0.26~(1.1\sigma)$} \\
		\hline

        \mc{5}{c}{$D K^{*-}$, $D \to K^+\pi^-$, $K^{*-} \to \KS \pi^-$} \\
	\babar & \cite{Aubert:2009yw} & 379M & $-0.34 \pm 0.43 \pm 0.16$ & $0.066 \pm 0.031 \pm 0.010$ \\
        \hline

        \mc{5}{c}{$D K^{-}$, $D \to K^+\pi^-\pi^0$} \\
	\babar & \cite{Lees:2011up} & 474M & $0.0091 \,^{+0.0082}_{-0.0076} \,^{+0.0014}_{-0.0037}$ \\
        \hline 

        \vspace{1ex} \\

        \hline
 	\mc{2}{l}{Experiment} & Sample size & $R_{\rm ADS}$ & Correlation \\
       \mc{5}{c}{$D \pi^-$, $D \to K^+\pi^-$} \\
	\babar & \cite{delAmoSanchez:2010dz} & $N(B\bar{B}) =$ 467M & $0.03 \pm 0.17 \pm 0.04$ & $0.0033 \pm 0.0006 \pm 0.0004$ \\
	\belle & \cite{Belle:2011ac} & $N(B\bar{B}) =$ 772M & $-0.04 \pm 0.11 \,^{+0.02}_{-0.01}$ & $0.00328 \,^{+0.00038}_{-0.00036} \,^{+0.00012}_{-0.00018}$ \\
	CDF & \cite{Aaltonen:2011uu} & 7 ${\rm fb}^{-1}$ & $0.13 \pm 0.25 \pm 0.02$ & $0.00280 \pm 0.00070 \pm 0.00040$ \\
	LHCb & \cite{Aaij:2012kz} & 1 ${\rm fb}^{-1}$ & $0.14300 \pm 0.06200 \pm 0.01100$ & $0.00410 \pm 0.00025 \pm 0.00005$ \\
	\mc{3}{l}{\bf Average} & $0.09 \pm 0.05$ & $0.00375 \pm 0.00020$ \\
	\mc{3}{l}{\small Confidence level} & {\small $0.53~(0.6\sigma)$} & {\small $0.17~(1.4\sigma)$} \\
       \hline 
 	\mc{2}{l}{Experiment} & $N(B\bar{B})$ & $R_{\rm ADS}$ & Correlation \\
       \mc{5}{c}{$\Dstar \pi^-$, $\Dstar \to D\pi^0$, $D \to K^+\pi^-$} \\
	\babar & \cite{delAmoSanchez:2010dz} & 467M & $-0.09 \pm 0.27 \pm 0.05$ & $0.0032 \pm 0.0009 \pm 0.0008$ \\
	\belle & \cite{belle:glwads:prelim} & 772M & $-0.07 \pm 0.23 \pm 0.05$ & $0.0040 \,^{+0.0010}_{-0.0009} \pm 0.0003$ \\
	\mc{3}{l}{\bf Average} & $-0.08 \pm 0.18$ & $0.0037 \pm 0.0008$ \\
	\mc{3}{l}{\small Confidence level} & {\small $0.96~(0.1\sigma)$} & {\small $0.61~(0.5\sigma)$} \\
       \hline 
       \mc{5}{c}{$\Dstar \pi^-$, $\Dstar \to D\gamma$, $D \to K^+\pi^-$} \\
	\babar & \cite{delAmoSanchez:2010dz} & 467M & $-0.65 \pm 0.55 \pm 0.22$ & $0.0027 \pm 0.0014 \pm 0.0022$ \\
	\belle & \cite{belle:glwads:prelim} & 772M & $-0.10 \,^{+0.26}_{-0.25} \pm 0.02$ & $0.0041 \,^{+0.0011}_{-0.0010} \pm 0.0001$ \\
	\mc{3}{l}{\bf Average} & $-0.19 \pm 0.23$ & $0.0039 \pm 0.0010$ \\
	\mc{3}{l}{\small Confidence level} & {\small $0.39~(0.9\sigma)$} & {\small $0.62~(0.5\sigma)$} \\
        \hline
 		\end{tabular*}
                \label{tab:cp_uta:cus:ads}
 	\end{center}
 \end{table}

\babar~\cite{:2009au} have also presented results on a similar analysis with
self-tagging neutral $B$ decays: $\Bz \to DK^{*0}$ with $D \to K^-\pi^+$, 
$D \to K^-\pi^+\pi^0$ and $D \to K^-\pi^+\pi^+\pi^-$ 
(all with $K^{*0} \to K^+\pi^-$). 
Effects due to the natural width of the $K^{*0}$ are
handled using the parametrisation suggested by Gronau~\cite{Gronau:2002mu}. 

The following 95\% C.L. limits are set:
\begin{equation}
  R_{\rm ADS}(K\pi) < 0.244 \hspace{5mm}
  R_{\rm ADS}(K\pi\pi^0) < 0.181 \hspace{5mm}
  R_{\rm ADS}(K\pi\pi\pi) < 0.391 \, .
\end{equation}

Combining the results and using additional input from
CLEOc~\cite{Asner:2008ft,Lowery:2009id} a limit on the ratio between the 
$b \to u$ and $b \to c$ amplitudes of $r_s \in \left[ 0.07,0.41 \right]$ 
at 95\% C.L. limit is set.

\belle~\cite{:2012uh} have obtained the constraint
\begin{equation}
  R_{\rm ADS}(K\pi) < 0.16 \, .
\end{equation}

\mysubsubsection{$D$ decays to multiparticle self-conjugate final states}
\label{sec:cp_uta:cus:dalitz}

For the Dalitz plot analysis, both 
\babar~\cite{Aubert:2008bd} and
\belle~\cite{Poluektov:2010wz,Poluektov:2006ia} have studied the modes 
$\Bmp \to D\Kmp$, $\Bmp \to \Dstar\Kmp$ and $\Bmp \to D\Kstarmp$.
For $\Bmp \to \Dstar\Kmp$,
both experiments have used both $\Dstar$ decay modes, $\Dstar \to D\pi^0$ and
$\Dstar \to D\gamma$, taking the effective shift in the strong phase
difference into account. 
In all cases the decay $D \to \KS\pi^+\pi^-$ has been used.
\babar\ also used the decay $D \to \KS K^+K^-$ .
\babar\ has also performed an analysis of $\Bmp \to D\Kmp$ with 
$D \to \pi^+\pi^-\pi^0$~\cite{Aubert:2007ii}.
Results and averages are given in Table~\ref{tab:cp_uta:cus:dalitz}.
The third error on each measurement is due to $D$ decay model uncertainty.

The parameters measured in the analyses are explained in
Sec.~\ref{sec:cp_uta:notations:cus}.
Both \babar\ and \belle\ have measured the ``Cartesian''
$(x_\pm,y_\pm)$ variables,
and perform frequentist statistical procedures,
to convert these into measurements of $\gamma$, $r_B$ and $\delta_B$.
In the $\Bmp \to D\Kmp$ with $D \to \pi^+\pi^-\pi^0$ analysis,
the parameters $(\rho^{\pm}, \theta^\pm)$ are used instead.

Both experiments reconstruct $\Kstarmp$ as $\KS\pi^\mp$,
but the treatment of possible nonresonant $\KS\pi^\mp$ differs:
\belle\ assign an additional model uncertainty,
while \babar\ use a parametrisation suggested by Gronau~\cite{Gronau:2002mu}.
The parameters $r_B$ and $\delta_B$ are replaced with 
effective parameters $\kappa r_s$ and $\delta_s$;
no attempt is made to extract the true hadronic parameters 
of the $\Bmp \to D\Kstarmp$ decay.

We perform averages using the following procedure, which is based on a set of
(more or less) reasonable, though imperfect, assumptions. 

\begin{itemize}\setlength{\itemsep}{0.5ex}
\item 
  It is assumed that effects due to the different $D$ decay models 
  used by the two experiments are negligible. 
  Therefore, we do not rescale the results to a common model.
\item 
  It is further assumed that the model uncertainty is $100\%$ 
  correlated between experiments, 
  and therefore this source of error is not used in the averaging procedure.
  (This approximation is significantly less valid now that the \babar\ results
  include $D \to \KS K^+K^-$ decays in addition to $D \to \KS\pi^+\pi^-$.)
\item 
  We include in the average the effect of correlations 
  within each experiments set of measurements.
\item 
  At present it is unclear how to assign an average model uncertainty. 
  We have not attempted to do so. 
  Our average includes only statistical and systematic error. 
  An unknown amount of model uncertainty should be added to the final error.
\item 
  We follow the suggestion of Gronau~\cite{Gronau:2002mu} 
  in making the $DK^*$ averages. 
  Explicitly, we assume that the selection of $K^{*\pm} \to \KS\pi^\pm$
  is the same in both experiments 
  (so that $\kappa$, $r_s$ and $\delta_s$ are the same), 
  and drop the additional source of model uncertainty 
  assigned by Belle due to possible nonresonant decays.
\item 
  We do not consider common systematic errors, 
  other than the $D$ decay model. 
\end{itemize}

\begin{sidewaystable}
	\begin{center}
		\caption{
      Averages from Dalitz plot analyses of $b \to c\bar{u}s / u\bar{c}s$ modes.
      Note that the uncertainities assigned to the averages do not include model errors.	
		}
		\vspace{0.2cm}
		\setlength{\tabcolsep}{0.0pc}
    \resizebox{\textwidth}{!}{
		\begin{tabular}{@{\extracolsep{2mm}}lrccccc} \hline
	\mc{2}{l}{Experiment} & $N(B\bar{B})$ & $x_+$ & $y_+$ & $x_-$ & $y_-$ \\
	\hline
        \mc{7}{c}{$D K^-$, $D \to \KS \pi^+\pi^-$} \\
	\babar & \cite{delAmoSanchez:2010rq} & 468M & $-0.103 \pm 0.037 \pm 0.006 \pm 0.007$ & $-0.021 \pm 0.048 \pm 0.004 \pm 0.009$ & $0.060 \pm 0.039 \pm 0.007 \pm 0.006$ & $0.062 \pm 0.045 \pm 0.004 \pm 0.006$ \\
	\belle & \cite{Poluektov:2010wz} & 657M & $-0.107 \pm 0.043 \pm 0.011 \pm 0.055$ & $-0.067 \pm 0.059 \pm 0.018 \pm 0.063$ & $0.105 \pm 0.047 \pm 0.011 \pm 0.064$ & $0.177 \pm 0.060 \pm 0.018 \pm 0.054$ \\
	\mc{3}{l}{\bf Average} & $-0.104 \pm 0.029$ & $-0.038 \pm 0.038$ & $0.085 \pm 0.030$ & $0.105 \pm 0.036$ \\
        \mc{3}{l}{\small Confidence level} &  \mc{4}{c}{\small $0.47~(0.7\sigma)$} \\
 		\hline

                \mc{7}{c}{$\Dstar K^-$, $\Dstar \to D\pi^0$ or $D\gamma$, $D \to \KS \pi^+\pi^-$} \\
	\babar & \cite{delAmoSanchez:2010rq} & 468M & $0.147 \pm 0.053 \pm 0.017 \pm 0.003$ & $-0.032 \pm 0.077 \pm 0.008 \pm 0.006$ & $-0.104 \pm 0.051 \pm 0.019 \pm 0.002$ & $-0.052 \pm 0.063 \pm 0.009 \pm 0.007$ \\
	\belle & \cite{Poluektov:2010wz} & 657M & $0.083 \pm 0.092 \pm 0.081$ & $0.157 \pm 0.109 \pm 0.063$ & $-0.036 \pm 0.127 \pm 0.090$ & $-0.249 \pm 0.118 \pm 0.049$ \\
	\mc{3}{l}{\bf Average} & $0.130 \pm 0.048$ & $0.031 \pm 0.063$ & $-0.090 \pm 0.050$ & $-0.099 \pm 0.056$ \\
        \mc{3}{l}{\small Confidence level} & \mc{4}{c}{\small $0.29~(1.1\sigma)$} \\
 		\hline

                \mc{7}{c}{$D K^{*-}$, $D \to \KS \pi^+\pi^-$} \\
	\babar & \cite{delAmoSanchez:2010rq} & 468M & $-0.151 \pm 0.083 \pm 0.029 \pm 0.006$ & $0.045 \pm 0.106 \pm 0.036 \pm 0.008$ & $0.075 \pm 0.096 \pm 0.029 \pm 0.007$ & $0.127 \pm 0.095 \pm 0.027 \pm 0.006$ \\
 	\belle & \cite{Poluektov:2006ia} & 386M & $-0.105 \,^{+0.177}_{-0.167} \pm 0.006 \pm 0.088$ & $-0.004 \,^{+0.164}_{-0.156} \pm 0.013 \pm 0.095$ & $-0.784 \,^{+0.249}_{-0.295} \pm 0.029 \pm 0.097$ & $-0.281 \,^{+0.440}_{-0.335} \pm 0.046 \pm 0.086$ \\
	\mc{3}{l}{\bf Average} & $-0.152 \pm 0.077$ & $0.024 \pm 0.091$ & $-0.043 \pm 0.094$ & $0.091 \pm 0.096$ \\
        \mc{3}{l}{\small Confidence level} & \mc{4}{c}{\small $0.011~(2.5\sigma)$} \\
 		\hline

                \vspace{1ex} \\

	\hline
	\mc{2}{l}{Experiment} & $N(B\bar{B})$ & $\rho^{+}$ & $\theta^+$ & $\rho^{-}$ & $\theta^-$ \\
	\hline
        \mc{7}{c}{$D K^-$, $D \to \pi^+\pi^-\pi^0$} \\
	\babar & \cite{Aubert:2007ii} & 324M & $0.75 \pm 0.11 \pm 0.04$ & $147 \pm 23 \pm 1$ & $0.72 \pm 0.11 \pm 0.04$ & $173 \pm 42 \pm 2$ \\
	\hline
		\end{tabular}
              }
		\label{tab:cp_uta:cus:dalitz}
	\end{center}
\end{sidewaystable}

\begin{figure}[htb]
  \begin{center}
    \resizebox{0.30\textwidth}{!}{
      \includegraphics{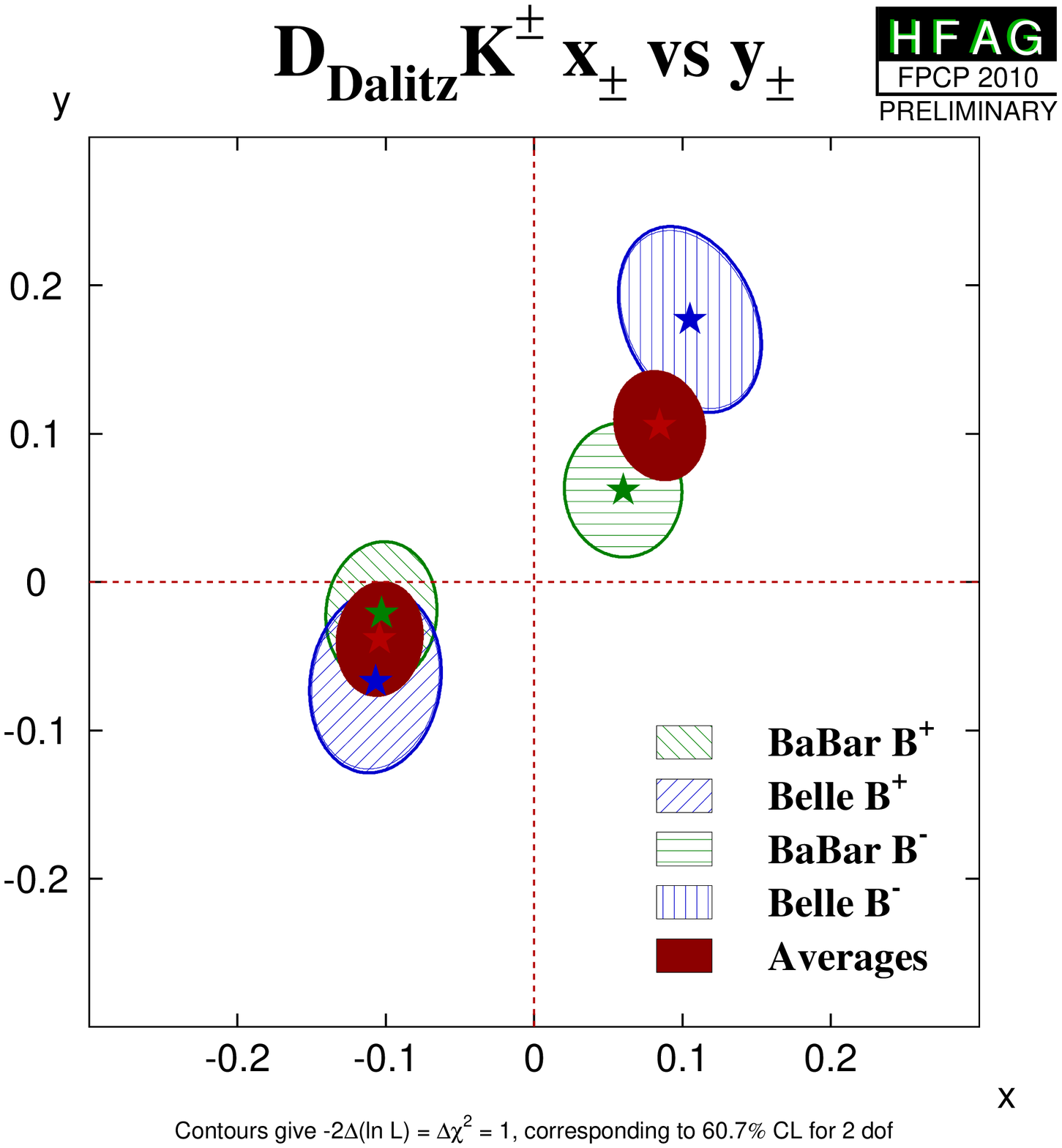}
    }
    \hfill
    \resizebox{0.30\textwidth}{!}{
      \includegraphics{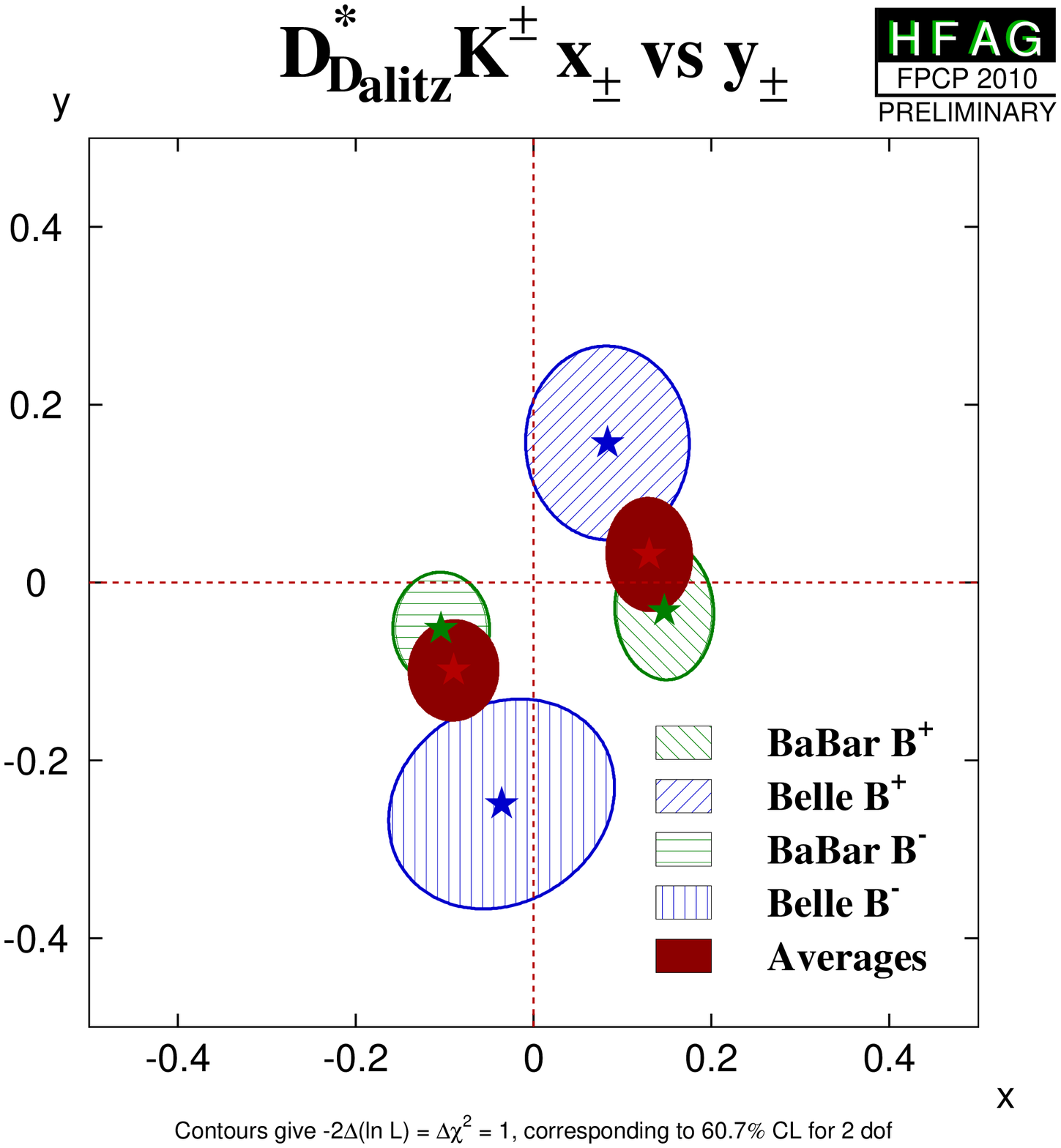}
    }
    \hfill
    \resizebox{0.30\textwidth}{!}{
      \includegraphics{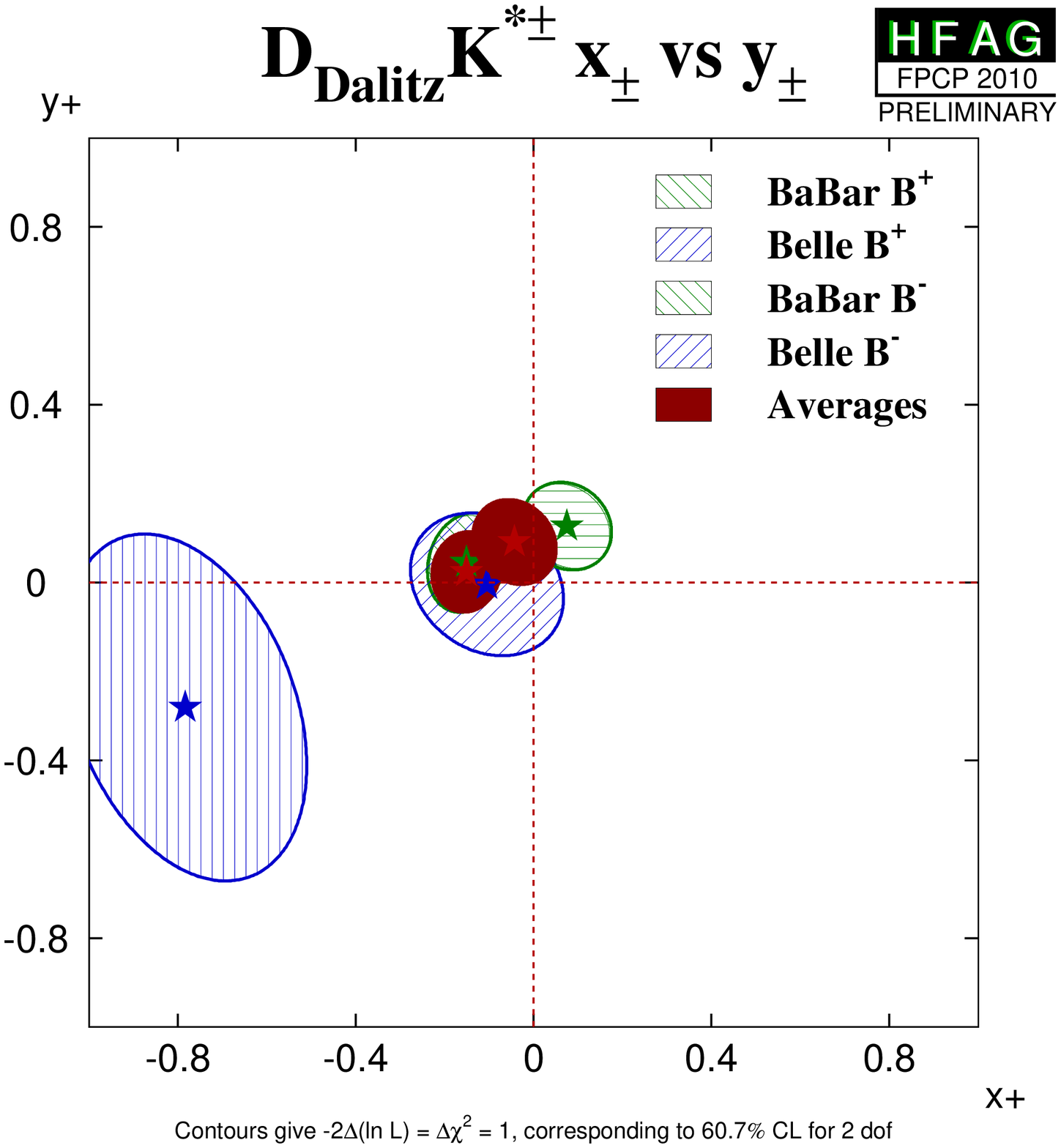}
    }
  \end{center}
  \vspace{-0.8cm}
  \caption{
    Contours in the $(x_\pm, y_\pm)$ from $\Bmp \to D^{(*)}K^{(*)\pm}$.
    (Left) $\Bmp \to D\Kmp$, 
    (middle) $\Bmp \to \Dstar\Kmp$,
    (right) $\Bmp \to D\Kstarmp$.
    Note that the uncertainties assigned to the averages given in these plots
    do not include model errors.        
  }
  \label{fig:cp_uta:cus:dalitz_2d}
\end{figure}

\begin{figure}[htb]
  \begin{center}
    \resizebox{0.40\textwidth}{!}{
      \includegraphics{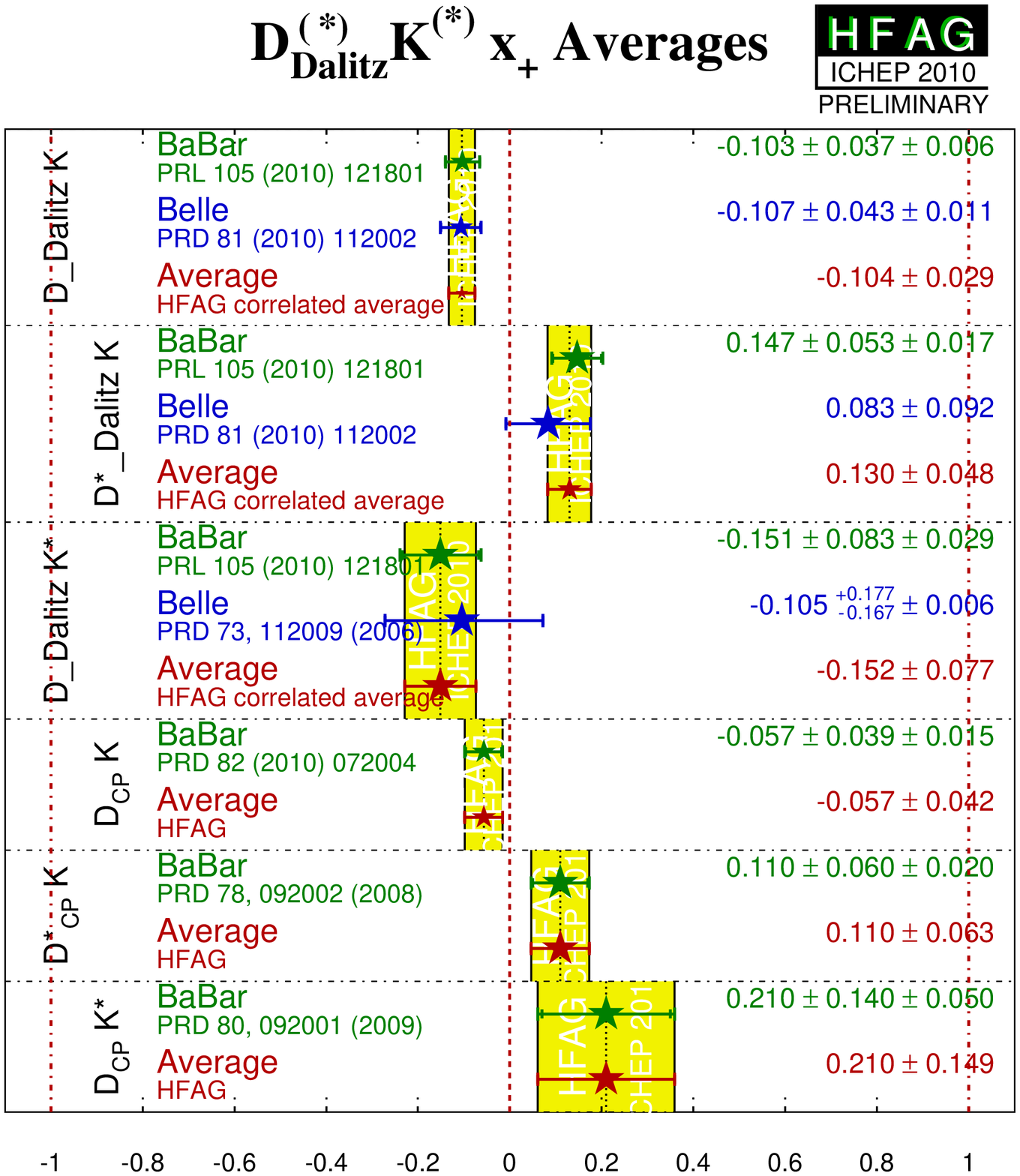}
    }
    \hspace{0.1\textwidth}
    \resizebox{0.40\textwidth}{!}{
      \includegraphics{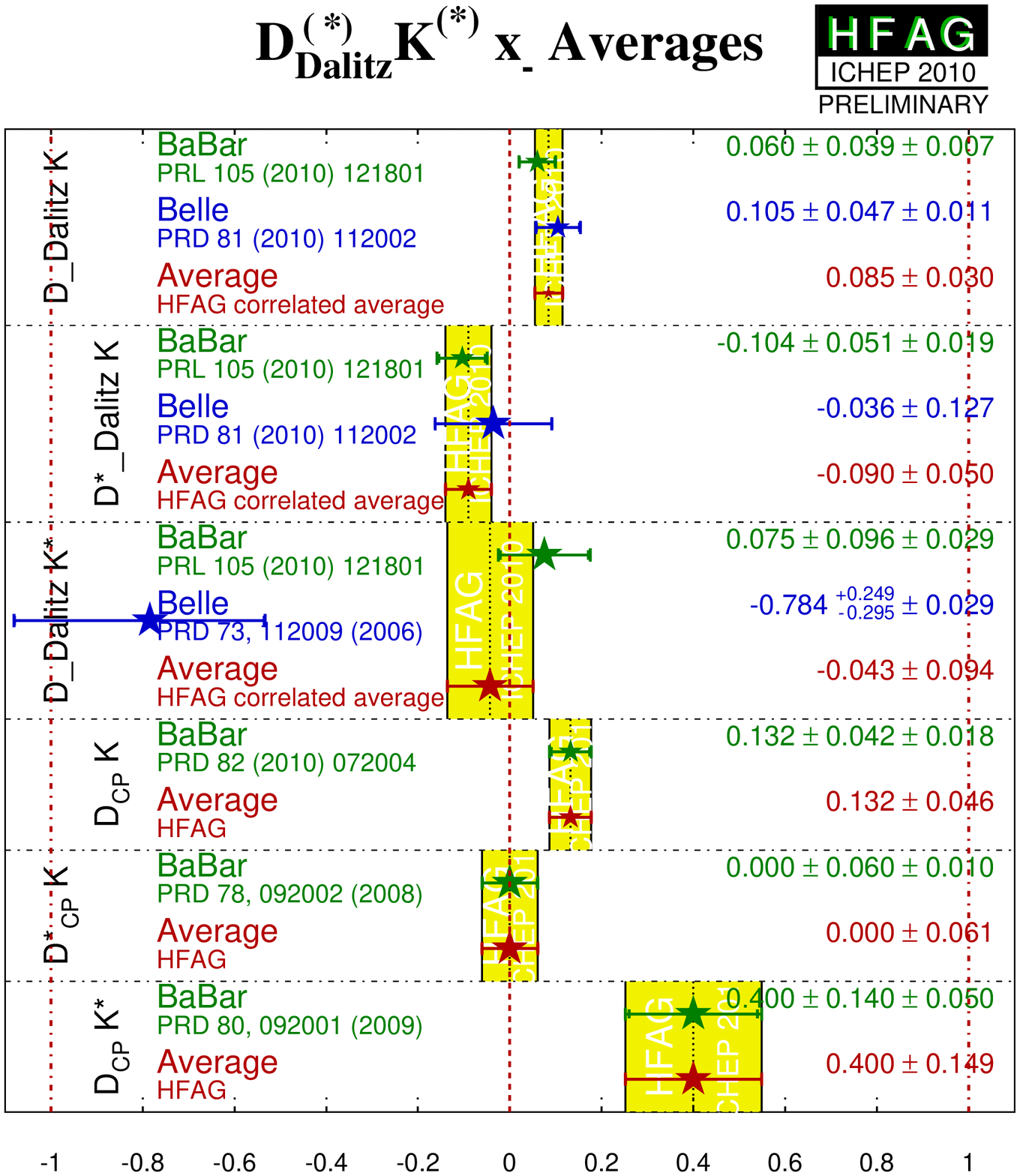}
    }
    \\
    \resizebox{0.40\textwidth}{!}{
      \includegraphics{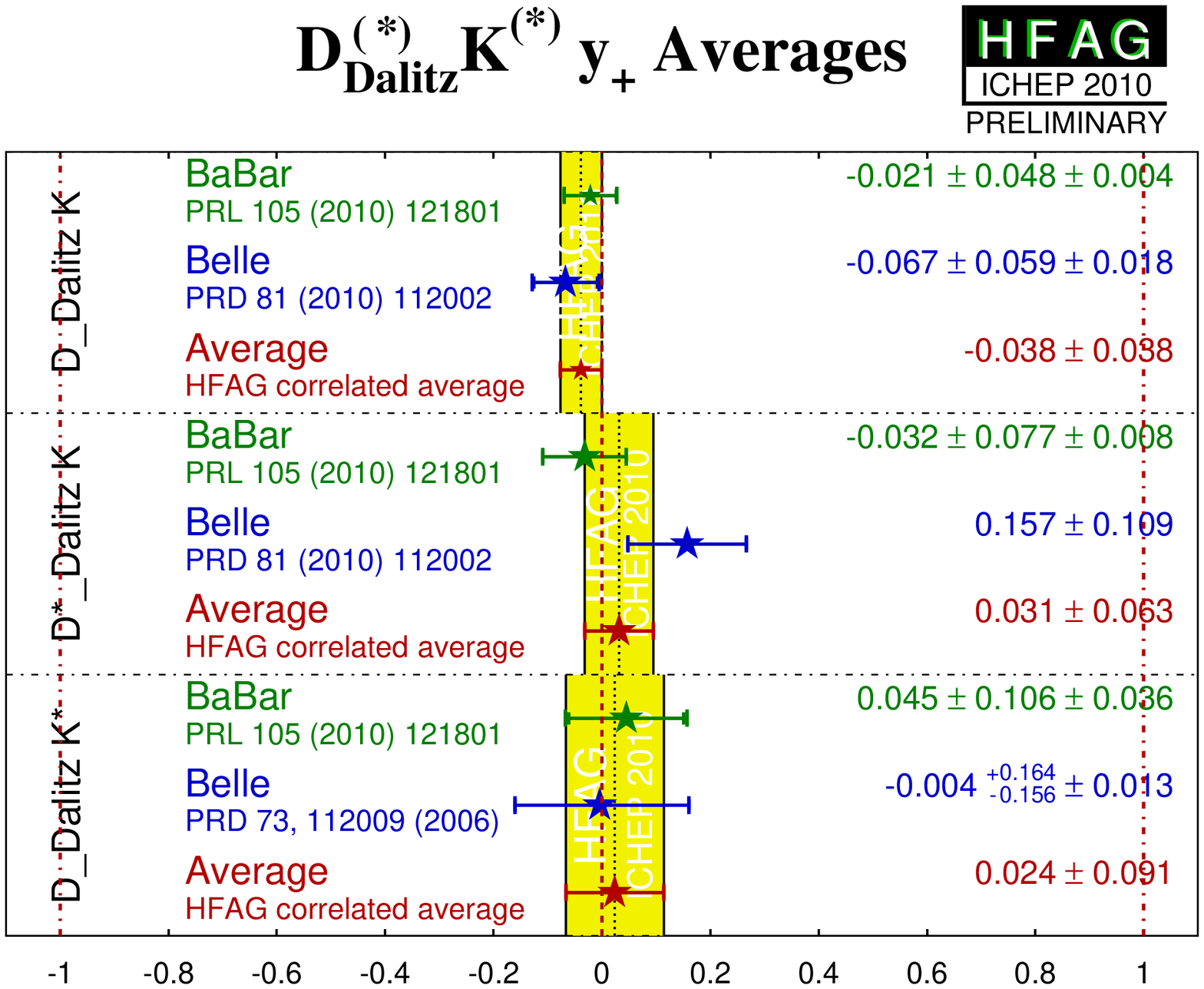}
    }
    \hspace{0.1\textwidth}
    \resizebox{0.40\textwidth}{!}{
      \includegraphics{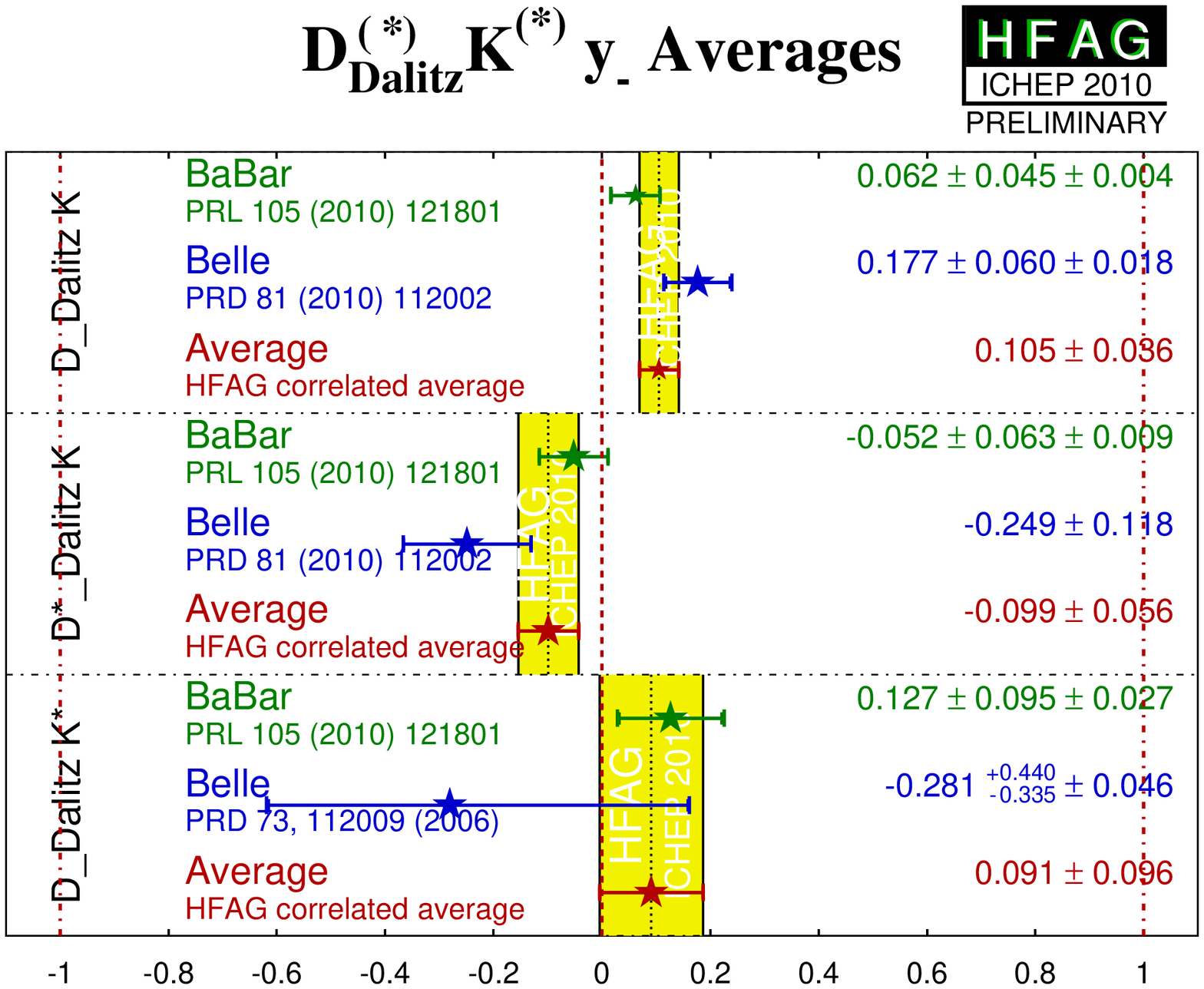}
    }
  \end{center}
  \vspace{-0.8cm}
  \caption{
    Averages of $(x_\pm, y_\pm)$ from $\Bpm \to D^{(*)}K^{(*)\pm}$.
    (Top left) $x_+$, (top right) $x_-$,
    (bottom left) $y_+$, (bottom right) $y_-$.
    The top plots include constraints on $x_{\pm}$ obtained from GLW analyses (see Sec.~\ref{sec:cp_uta:cus:glw}).
    Note that the uncertainties assigned to the averages given in these plots
    do not include model errors.        
  }
  \label{fig:cp_uta:cus:dalitz_1d}
\end{figure}

\vspace{3ex}

\noindent
\underline{\large Constraints on $\gamma$}

The measurements of $(x_\pm, y_\pm)$ can be used to obtain constraints on 
$\gamma$, as well as the hadronic parameters $r_B$ and $\delta_B$.
Both
\babar~\cite{delAmoSanchez:2010rq} and 
\belle~\cite{Poluektov:2010wz,Poluektov:2006ia} 
have done so using a frequentist procedure 
(there are some differences in the details of the techniques used).

\begin{itemize}\setlength{\itemsep}{0.5ex}

\item 
  \babar\ obtain $\gamma = (68 \,^{+15}_{-14} \pm 4 \pm 3)^\circ$
  from $D\Kpm$, $\Dstar\Kpm$ and $D\Kstarpm$

\item
  \belle\ obtain $\phi_3 = (78 \,^{+11}_{-12} \pm 4 \pm 9)^\circ$
  from $D\Kpm$ and $\Dstar\Kpm$

\item
  The experiments also obtain values for the hadronic parameters as detailed
  in Tab.~\ref{tab:cp_uta:rBdeltaB_summary}.

\item 
  Improved constraints can be achieved combining the information from
  $\Bpm \to D\Kpm$ analysis with different $D$ decay modes.
  The experiments have not yet published such results,
  and none are listed here.

\item 
  The CKMfitter~\cite{Charles:2004jd} and 
  UTFit~\cite{Bona:2005vz} groups use the measurements 
  from \belle\ and \babar\ given above
  to make combined constraints on $\gamma$.

\item 
  In the \babar\ analysis of $\Bmp \to D\Kmp$ with 
  $D \to \pi^+\pi^-\pi^0$~\cite{Aubert:2007ii},
  a constraint of $-30^\circ < \gamma < 76^\circ$ is obtained 
  at the 68\% confidence level.

\end{itemize}

\begin{table}
  \begin{center}
  \caption{
    Summary of constraints on hadronic parameters 
    in $\Bpm \to \DorDstar\KorKstarpm$ decays.
    Note the alternative parametrisation of the hadronic parameters used by
    \babar\ in the $D\Kstarpm$ mode.
  }
  \label{tab:cp_uta:rBdeltaB_summary}
  \begin{tabular}{lcc}
    \hline
    & $r_B$ & $\delta_B$ \\
    \hline
    \multicolumn{3}{c}{In $D\Kpm$} \\
    \babar & $0.096 \pm 0.029 \pm 0.005 \pm 0.004$ & $\delta_B (D\Kpm) = (119 \,^{+19}_{-20} \pm 3 \pm 3)^\circ$ \\
    \belle & $0.160 \,^{+0.040}_{-0.038} \pm 0.011 \,^{+0.05}_{-0.010}$ & 
    $(138 \,^{+13}_{-16} \pm 4 \pm 23)^\circ$ \\
    \hline
    \multicolumn{3}{c}{In $\Dstar\Kpm$} \\
    \babar & $0.133 \,^{+0.042}_{-0.039} \pm 0.014 \pm 0.003$ & $(-82 \pm 21 \pm 5 \pm 3)^\circ$ \\
    \belle & $0.196 \,^{+0.072}_{-0.069} \pm 0.012 \,^{+0.062}_{-0.012}$ &
    $(342 \,^{+19}_{-21} \pm 3 \pm 23)^\circ$ \\
    \hline
    \multicolumn{3}{c}{In $D\Kstarpm$} \\
    \babar & $\kappa r_S = 0.149 \,^{+0.066}_{-0.062} \pm 0.026 \pm 0.006$ &
    $\delta_S = (111 \pm 32 \pm 11 \pm 3)^\circ$ \\
    \belle & $0.56 \,^{+0.22}_{-0.16} \pm 0.04 \pm 0.08$ & 
    $(243 \,^{+20}_{-23} \pm 3 \pm 50)^\circ$ \\
    \hline
  \end{tabular}
  \end{center}
\end{table}

At present we make no attempt to provide an HFAG average for $\gamma$,
nor indeed for the hadronic parameters.
More details on procedures to calculate a best fit value for $\gamma$ 
can be found in Refs.~\cite{Charles:2004jd,Bona:2005vz}.

\babar~\cite{Aubert:2008yn} have also performed a similar Dalitz plot analysis
to that described above using the self-tagging neutral $B$ decay $\Bz \to
DK^{*0}$ (with $K^{*0} \to K^+\pi^-$). Effects due to the natural width of the
$K^{*0}$ are handled using the parametrisation suggested by
Gronau~\cite{Gronau:2002mu}.

\babar\ extract the three-dimensional likelihood for the parameters 
$\left( \gamma, \delta_S, r_S \right)$ and, combining with a separately
measured PDF for $r_S$ (using a Bayesian technique), obtain bounds on each of
the three parameters. 
\begin{equation}
  \gamma = (162 \pm 56)^\circ \hspace{5mm}
  \delta_S = (62 \pm 57)^\circ \hspace{5mm}
  r_S < 0.55  \, ,
\end{equation}
where the limit on $r_S$ is at 95\% probability.
Note that there is an ambiguity in the solutions 
$\left( \gamma, \delta_S \leftrightarrow \gamma+\pi, \delta_S+\pi \right)$.

\clearpage


\section{Semileptonic $B$ decays}
\label{sec:slbdecays}

Measurements of semileptonic $B$-meson decays are an important tool to
study the magnitude of the CKM matrix elements $|V_{cb}|$ and
$|V_{ub}|$, the Heavy Quark parameters (e.g. $b$ and $c$--quark masses),
QCD form factors, QCD dynamics, new physics, etc.

In the following, we provide averages of exclusive and inclusive
 branching fractions, the product of $|V_{cb}|$ and the form factor
 normalization ${\cal F}(1)$ and ${\cal G}(1)$ for $\bar{B} \to D^* \ell^-\bar{\nu}_{\ell}$ and
$\bar{B} \to D \ell^-\bar{\nu}_{\ell}$ decays, respectively, and $|V_{ub}|$ as determined from
inclusive and exclusive measurements of $\B\to X_u \ell \nul$ decays.
We will compute Heavy Quark parameters and extract QCD form factors
for $\bar{B} \to D^* \ell^-\bar{\nu}_{\ell}$ decays.
Throughout this section, charge conjugate states are implicitly included, 
unless otherwise indicated.

Brief descriptions of all parameters
and analyses (published or preliminary) relevant for the
determination of the combined results are given.  The descriptions are
based on the information available on the web page at\\
 \centerline{\tt http://www.slac.stanford.edu/xorg/hfag/semi/EndOfYear11}
A description of the technique employed for calculating averages
was presented in the previous update~\cite{Barberio:2008fa}. 
Asymmetric errors have been introduced in the current averages
for $\Bb\to X_u\ell\nub$ decays to take into account theoretical 
asymmetric errors. 






\subsection{Exclusive CKM-favored decays}
\label{slbdecays_b2cexcl}
This section contains the measurements of $\bar B\to
D^*\ell^-\bar\nu_\ell$ and $\bar B\to D\ell^-\bar\nu_\ell$, relevant
for the determination of $\vcb$ from exclusive decays. We then provide
averages for the inclusive branching fractions $\cbf(\bar B\to
D^{(*)}\pi \ell^-\bar\nu_\ell)$ and for $B$ semileptonic decays into
orbitally-excited $P$-wave charm mesons ($D^{**}$). As the $D^{**}$
branching fraction is poorly known, we report the averages for the products 
$\cbf(B^-\to D^{**}(D^{(*)}\pi)\ell^-\bar\nu_\ell)\times
\cbf(D^{**}\to D^{(*)}\pi)$.


\mysubsubsection{$\bar B\to D^*\ell^-\bar\nu_\ell$}
\label{slbdecays_dstarlnu}

In the parameterization of Caprini, Lellouch and Neubert
(CLN), the shape and normalization of the form factor ${\cal F}(w)$, which describes
the decay $\bar B\to D^*\ell^-\bar\nu_\ell$, can be described by four
quantities: ${\cal F}(1)\vcb$, $\rho^2$, $R_1(1)$ and
$R_2(1)$~\cite{CLN}. Our average and the determination of $\vcb$ are
based on this parameterization.

We use the measurements of these form factor parameters shown in
Table~\ref{tab:vcbf1} and rescale them to the latest values of the
input parameters (mainly branching fractions of charmed
mesons)~\cite{HFAG_sl:inputparams}. Most of the measurements in
Table~\ref{tab:vcbf1} are based exclusively on the decay $\bar B^0\to
D^{*+}\ell^-\bar\nu_\ell$. Some
measurements~\cite{Adam:2002uw,Aubert:2009_1} are sensitive also to
$B^-\to D^{*0}\ell^-\bar\nu_\ell$ and one
measurement~\cite{Aubert:2009_3} is based exclusively on the decay
$B^-\to D^{*0}\ell^-\bar\nu_\ell$. Our analysis thus assumes isospin
symmetry.
\begin{table}[!htb]
\caption{Measurements of $\bar B\to D^*\ell^-\bar\nu_\ell$ in the
  parameterization of Caprini, Lellouch and Neubert
  (CLN)~\cite{CLN}. The average is the result of a 4-dimensional fit
  to the rescaled measurements of ${\cal F}(1)\vcb$, $\rho^2$, $R_1(1)$ and
  $R_2(1)$. The $\chi^2$~value of the combination is 29.7 for 23
  degrees of freedom (CL=$15.7\%$). The total  correlation between the
  average ${\cal F}(1)\vcb$ and $\rho^2$ is 0.32.}
\begin{center}
\resizebox{0.99\textwidth}{!}{
\begin{tabular}{|l|c|c|}
  \hline
  Experiment
  & ${\cal F}(1)\vcb [10^{-3}]$ (rescaled)
  & $\rho^2$ (rescaled)\\
  & ${\cal F}(1)\vcb [10^{-3}]$ (published)
  & $\rho^2$ (published)\\
  \hline\hline
  ALEPH~\hfill\cite{Buskulic:1996yq}
  & $31.34\pm 1.80_{\rm stat}\pm 1.26_{\rm syst}$
  & $0.490\pm 0.227_{\rm stat}\pm 0.146_{\rm syst}$\\
  & $31.9\pm 1.8_{\rm stat}\pm 1.9_{\rm syst}$
  & $0.37\pm 0.26_{\rm stat}\pm 0.14_{\rm syst}$\\
  \hline
  CLEO~\hfill\cite{Adam:2002uw}
  & $40.00\pm 1.24_{\rm stat}\pm 1.62_{\rm syst}$
  & $1.366\pm 0.085_{\rm stat}\pm 0.088_{\rm syst}$\\
  & $43.1\pm 1.3_{\rm stat}\pm 1.8_{\rm syst}$
  & $1.61\pm 0.09_{\rm stat}\pm 0.21_{\rm syst}$\\
  \hline
  OPAL excl~\hfill\cite{Abbiendi:2000hk}
  & $36.57\pm 1.60_{\rm stat}\pm 1.48_{\rm syst}$
  & $1.234\pm 0.212_{\rm stat}\pm 0.145_{\rm syst}$\\
  & $36.8\pm 1.6_{\rm stat}\pm 2.0_{\rm syst}$
  & $1.31\pm 0.21_{\rm stat}\pm 0.16_{\rm syst}$\\
  \hline
  OPAL partial reco~\hfill\cite{Abbiendi:2000hk}
  & $37.20\pm 1.19_{\rm stat}\pm 2.35_{\rm syst}$
  & $1.149\pm 0.145_{\rm stat}\pm 0.295_{\rm syst}$\\
  & $37.5\pm 1.2_{\rm stat}\pm 2.5_{\rm syst}$
  & $1.12\pm 0.14_{\rm stat}\pm 0.29_{\rm syst}$\\
  \hline
  DELPHI partial reco~\hfill\cite{Abreu:2001ic}
  & $35.38\pm 1.40_{\rm stat}\pm 2.34_{\rm syst}$
  & $1.174\pm 0.126_{\rm stat} \pm 0.376_{\rm syst}$\\
  & $35.5\pm 1.4_{\rm stat}\ {}^{+2.3}_{-2.4}{}_{\rm syst}$
  & $1.34\pm 0.14_{\rm stat}\ {}^{+0.24}_{-0.22}{}_{\rm syst}$\\
  \hline
  DELPHI excl~\hfill\cite{Abdallah:2004rz}
  & $36.19\pm 1.70_{\rm stat}\pm 1.98_{\rm syst}$
  & $1.082\pm 0.142_{\rm stat} \pm 0.154_{\rm syst}$\\
  & $39.2\pm 1.8_{\rm stat}\pm 2.3_{\rm syst}$
  & $1.32\pm 0.15_{\rm stat}\pm 0.33_{\rm syst}$\\
  \hline
  \belle~\hfill\cite{Dungel:2010uk}
  & $34.73\pm 0.17_{\rm stat}\pm 1.02_{\rm syst}$
  & $1.214\pm 0.034_{\rm stat}\pm 0.008_{\rm syst}$\\
  & $34.6\pm 0.2_{\rm stat}\pm 1.0_{\rm syst}$
  & $1.214\pm 0.034_{\rm stat} \pm 0.009_{\rm syst}$\\
  \hline
  \babar\ excl~\hfill\cite{Aubert:2006mb}
  & $34.09\pm 0.30_{\rm stat}\pm 1.00_{\rm syst}$
  & $1.184\pm 0.048_{\rm stat}\pm 0.029_{\rm syst}$\\
  & $34.7\pm 0.3_{\rm stat}\pm 1.1_{\rm syst}$
  & $1.18\pm 0.05_{\rm stat}\pm 0.03_{\rm syst}$\\
  \hline
  \babar\ $D^{*0}$~\hfill\cite{Aubert:2009_3}
  & $35.14\pm 0.59_{\rm stat}\pm 1.33_{\rm syst}$
  & $1.126\pm 0.058_{\rm stat}\pm 0.055_{\rm syst}$\\
  & $35.9\pm 0.6_{\rm stat}\pm 1.4_{\rm syst}$
  & $1.16\pm 0.06_{\rm stat}\pm 0.08_{\rm syst}$\\
  \hline
  \babar\ global fit~\hfill\cite{Aubert:2009_1}
  & $35.83\pm 0.20_{\rm stat}\pm 1.10_{\rm syst}$
  & $1.194\pm 0.020_{\rm stat}\pm 0.061_{\rm syst}$\\
  & $35.7\pm 0.2_{\rm stat}\pm 1.2_{\rm syst}$
  & $1.21\pm 0.02_{\rm stat}\pm 0.07_{\rm syst}$\\
  \hline
  {\bf Average}
  & \mathversion{bold} $35.90\pm 0.11_{\rm stat}\pm 0.44_{\rm syst}$ &
  \mathversion{bold} $1.207\pm 0.015_{\rm stat}\pm 0.021_{\rm syst}$\\
  \hline 
\end{tabular}
}
\end{center}
\label{tab:vcbf1}
\end{table}

\begin{figure}[!ht]
  \begin{center}
  \unitlength 1.0cm 
  \begin{picture}(14.,8.0)
    \put(  8.0,
    -0.2){\includegraphics[width=8.0cm]{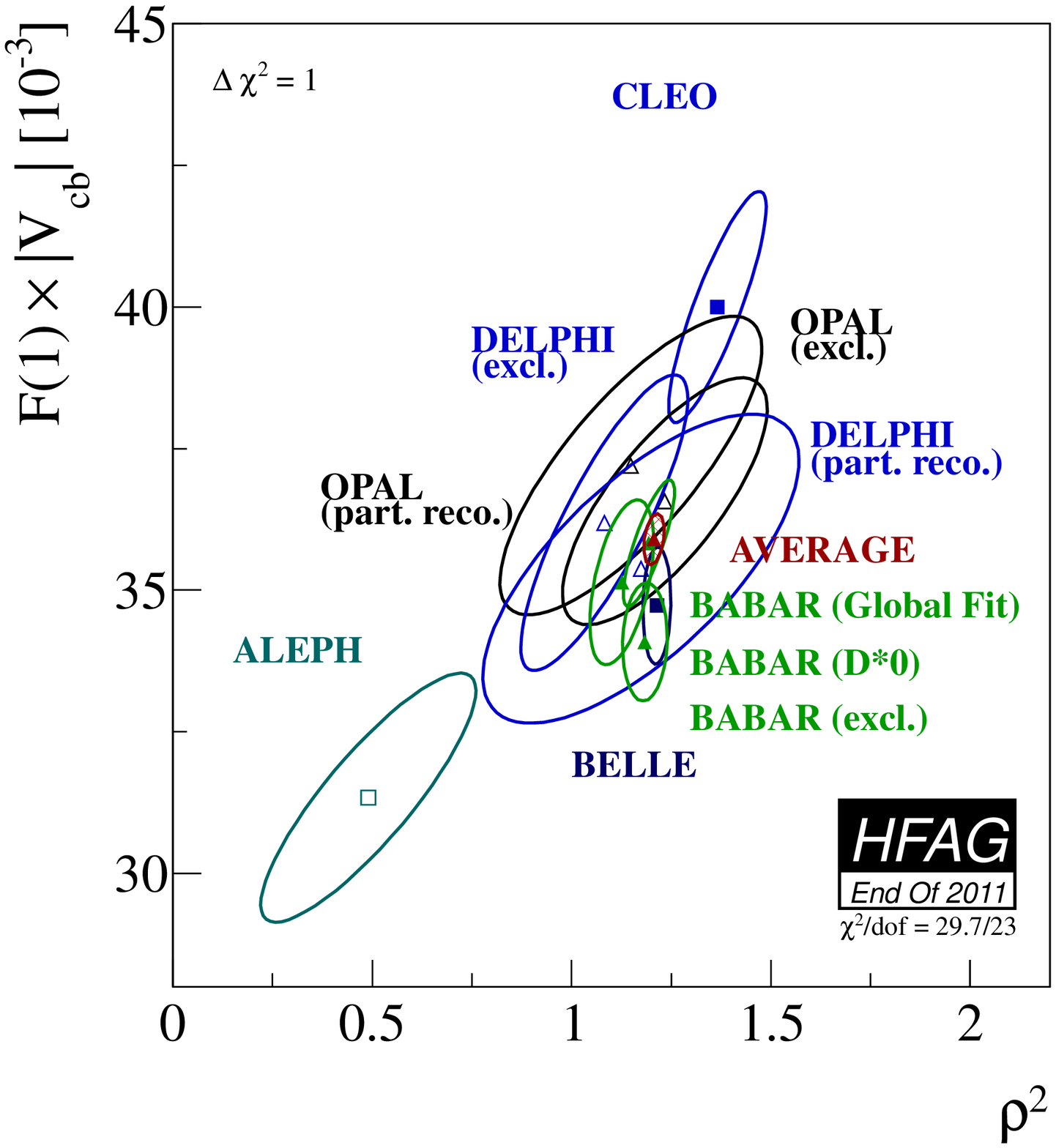}}
    \put( -0.5,
    0.0){\includegraphics[width=7.5cm]{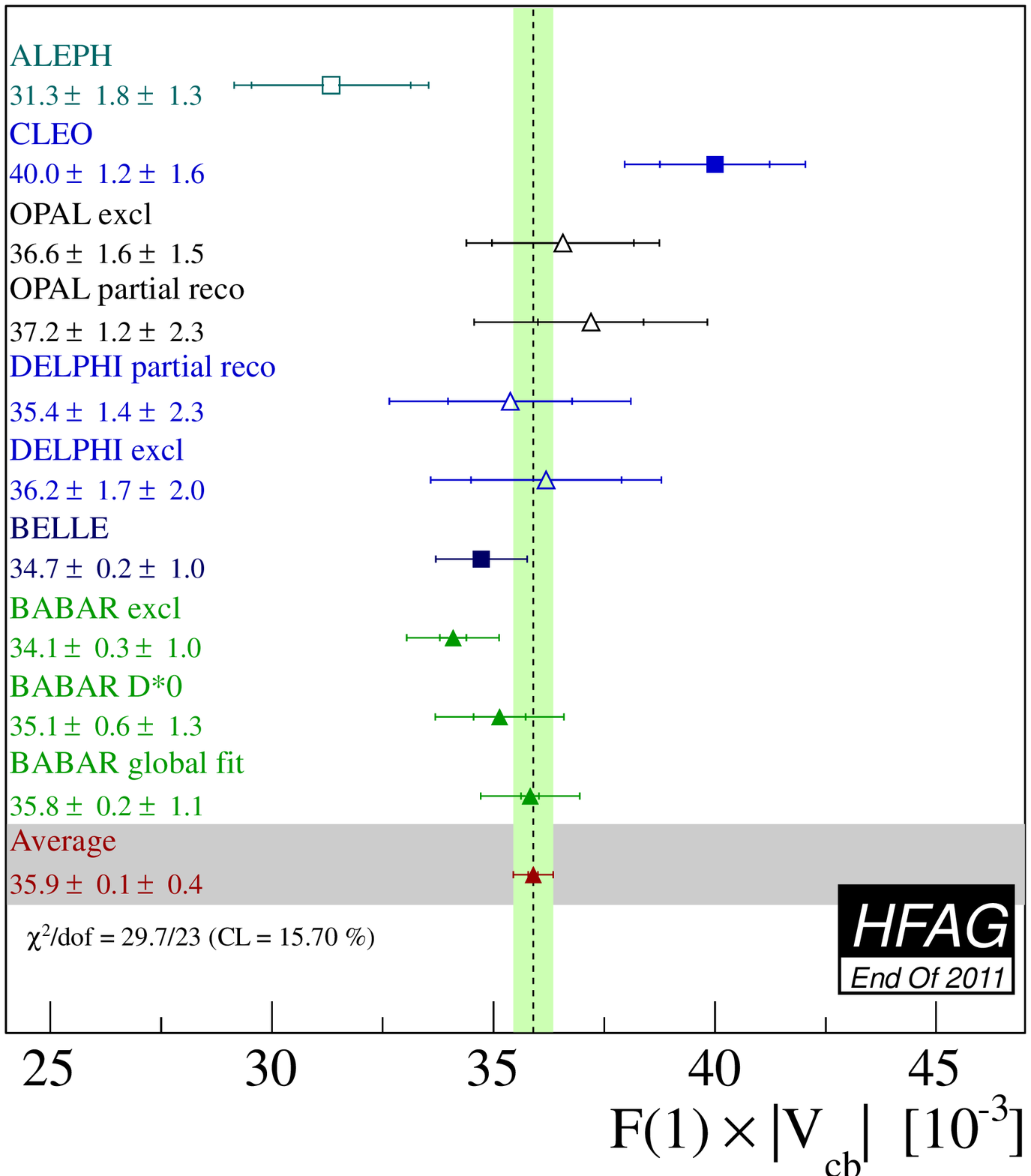}}
    \put(  5.5,  6.8){{\large\bf a)}}  
    \put( 14.4,  6.8){{\large\bf b)}}
  \end{picture}
  \caption{(a) Illustration of the ${\cal F}(1)\vcb$ average. (b)
    Illustration of the ${\cal F}(1)\vcb$ vs.\ $\rho^2$ average. The error
    ellipses correspond  to $\Delta\chi^2 = 1$
    (CL=39\%).} \label{fig:vcbf1}
  \end{center}
\end{figure}

In the next step, we perform a four-dimensional fit of the parameters
${\cal F}(1)\vcb$, $\rho^2$, $R_1(1)$ and $R_2(1)$ using the rescaled
measurements and taking into account correlated systematic
uncertainties. Only two measurements constrain all four
parameters~\cite{Dungel:2010uk,Aubert:2006mb}, the remaining
measurements constrain only the normalization ${\cal F}(1)\vcb$ and the
slope $\rho^2$. The result of the fit is
\begin{eqnarray}
  {\cal F}(1)\vcb & = & (35.90\pm 0.45)\times 10^{-3}~, \label{eq:vcbf1} \\
  \rho^2 & = & 1.207 \pm 0.026~,\\
  R_1(1) & = & 1.403\pm 0.033~, \label{eq:r1} \\
  R_2(1) & = & 0.854\pm 0.020~, \label{eq:r2}
\end{eqnarray}
and the correlation coefficients are
\begin{eqnarray}
  \rho_{{\cal F}(1)\vcb,\rho^2} & = & 0.324~,\\
  \rho_{{\cal F}(1)\vcb,R_1(1)} & = & -0.109~,\\
  \rho_{{\cal F}(1)\vcb,R_2(1)} & = & -0.063~,\\
  \rho_{\rho^2,R_1(1)} & = & 0.566~,\\
  \rho_{\rho^2,R_2(1)} & = & -0.807~,\\
  \rho_{R_1(1),R_2(1)} & = & -0.759~.\\
\end{eqnarray}
The uncertainties and correlations quoted here include both
statistical and systematic contributions. The $\chi^2$ of the fit is
29.7 for 23 degrees of freedom, which corresponds to a confidence
level of 15.7\%. An illustration of this fit result is given in
Fig.~\ref{fig:vcbf1}.

Using the form factor normalization ${\cal F}(1)$ of the latest LQCD
calculation~\cite{Bailey:2010gb},
\begin{equation}
  {\cal F}(1) = 0.908\pm 0.017~,
\end{equation}
we obtain the following determination of $\vcb$ from
Eq.~\ref{eq:vcbf1},
\begin{equation}
  \vcb = (39.54\pm 0.50_{\rm exp}\pm 0.74_{\rm th})\times
  10^{-3}~, \label{eq:vcbdstar}
\end{equation}
where the first uncertainty is experimental and the second stems from
the LQCD calculation.

\begin{figure}[!ht]
  \begin{center}
  \unitlength1.0cm 
  \begin{picture}(14.,8.0)  
    \put( -0.5, 0.0){\includegraphics[width=7.5cm]{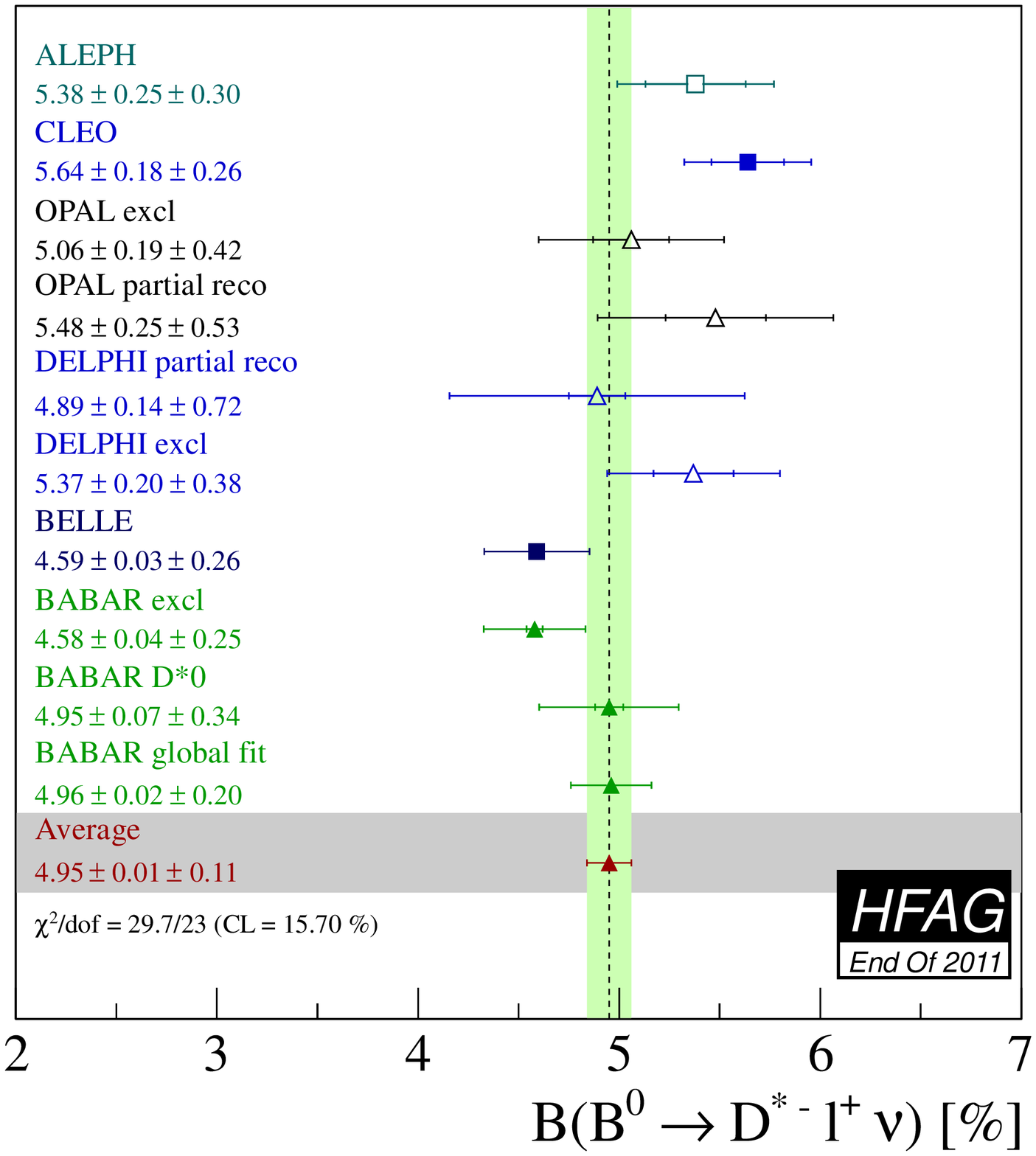}}
    \put(  8.0, 0.0){\includegraphics[width=7.5cm]{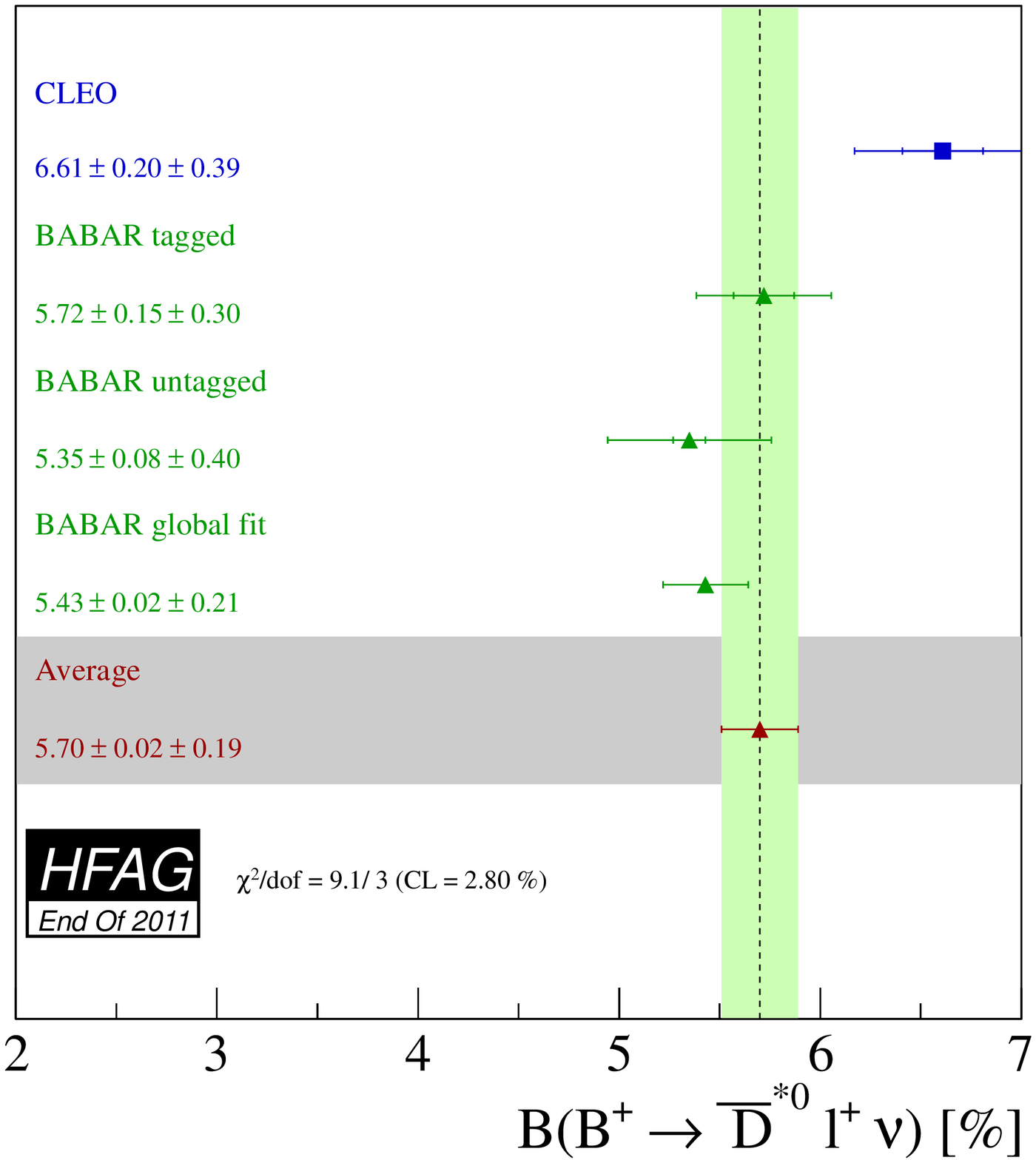}}
    \put(  5.5, 6.8){{\large\bf a)}}
    \put( 14.0, 6.8){{\large\bf b)}}
  \end{picture}
  \caption{(a) Average branching fractions of exclusive semileptonic
    $B$ decays $\bar B\to D^*\ell^-\bar\nu_\ell$: (a) $\BzbDstarlnu$
    (Table~\ref{tab:dstarlnu}) and (b) $B^-\to
    D^{*0}\ell^-\bar\nu_\ell$ (Table~\ref{tab:dstar0lnu}).} \label{fig:brdsl}
  \end{center}
\end{figure}

From each rescaled measurement in Table~\ref{tab:vcbf1}, we
calculate the $\bar B\to D^*\ell^-\bar\nu_\ell$ form factor ${\cal F}(w)$
and, by numerical integration, the branching ratio of the decay $\bar B^0\to
D^{*+}\ell^-\bar\nu_\ell$. For measurements which do not determine the
parameters $R_1(1)$ and $R_2(1)$ we assume the average
values~Eqs.~\ref{eq:r1} and \ref{eq:r2}. The results are quoted in
Table~\ref{tab:dstarlnu}. The branching ratio found for the average
values of ${\cal F}(1)\vcb$, $\rho^2$, $R_1(1)$ and $R_2(1)$ is
\begin{equation}
  \cbf(\BzbDstarlnu)=(4.95\pm 0.11)\%~.
\end{equation}
Again, this analysis assumes isospin symmetry although most of the
measurements included are sensitive only to $\BzbDstarlnu$. For an
independent analysis of the decay $B^-\to D^{*0}\ell^-\bar\nu_\ell$,
we have performed a simple 1-dimensional average of measurements
sensitive to the decay $B^-\to D^{*0}\ell^-\bar\nu_\ell$ only, which is
shown in Table~\ref{tab:dstar0lnu}. Fig.~\ref{fig:brdsl} illustrates
these two averages of $\bar B\to D^*\ell^-\bar\nu_\ell$.
\begin{table}[!htb]
\caption{$\BzbDstarlnu$ branching fractions calculated from the
  rescaled measurements in Table~\ref{tab:vcbf1}, asumming the CLN
  parameterization of the form factor~\cite{CLN}. The branching ratios
  published in Refs.~\cite{Aubert:2009_3,Aubert:2009_1} have been
  rescaled by the factor $\tau(B^0)/\tau(B^+)$. While the fit assumes
  isospin symmetry, most measurements included here use only the decay
  $\BzbDstarlnu$.}
\begin{center} 
\resizebox{0.99\textwidth}{!}{
\begin{tabular}{|l|c|c|}\hline
  Experiment & $\cbf(\BzbDstarlnu)$ [\%] (calculated) &
  $\cbf(\BzbDstarlnu)$ [\%] (published)\\
  \hline\hline
  ALEPH~\hfill\cite{Buskulic:1996yq}
  & $5.38\pm 0.25_{\rm stat} \pm 0.30_{\rm syst}$
  & $5.53\pm 0.26_{\rm stat} \pm 0.52_{\rm syst}$\\
  CLEO~\hfill\cite{Adam:2002uw}
  & $5.64\pm 0.18_{\rm stat} \pm 0.26_{\rm syst}$
  & $6.09\pm 0.19_{\rm stat} \pm 0.40_{\rm syst}$\\
  OPAL excl~\hfill\cite{Abbiendi:2000hk}
  & $5.06\pm 0.19_{\rm stat} \pm 0.42_{\rm syst}$
  & $5.11\pm 0.19_{\rm stat} \pm 0.49_{\rm syst}$\\
  OPAL partial reco~\hfill\cite{Abbiendi:2000hk}
  & $5.48\pm 0.25_{\rm stat} \pm 0.53_{\rm syst}$
  & $5.92\pm 0.27_{\rm stat} \pm 0.68_{\rm syst}$\\
  DELPHI partial reco~\hfill\cite{Abreu:2001ic}
  & $4.89\pm 0.14_{\rm stat} \pm 0.72_{\rm syst}$
  & $4.70\pm 0.13_{\rm stat} \ {}^{+0.36}_{-0.31}\ {}_{\rm syst}$\\
  DELPHI excl~\hfill\cite{Abdallah:2004rz}
  & $5.37\pm 0.20_{\rm stat} \pm 0.38_{\rm syst}$
  & $5.90\pm 0.22_{\rm stat} \pm 0.50_{\rm syst}$\\
  \belle~\hfill\cite{Dungel:2010uk}
  & $4.59\pm 0.03_{\rm stat} \pm 0.26_{\rm syst}$
  & $4.58\pm 0.03_{\rm stat} \pm 0.26_{\rm syst}$\\
  \babar\ excl~\hfill\cite{Aubert:2006mb}
  & $4.58\pm 0.04_{\rm stat}\pm 0.25_{\rm syst}$
  & $4.69\pm 0.04_{\rm stat} \pm 0.34_{\rm syst}$\\
  \babar\ $D^{*0}$~\hfill\cite{Aubert:2009_3}
  & $4.95\pm 0.07_{\rm stat}\pm 0.34_{\rm syst}$
  & $5.15\pm 0.07_{\rm stat} \pm 0.38_{\rm syst}$\\
  \babar\ global fit~\hfill\cite{Aubert:2009_1}
  & $4.96\pm 0.02_{\rm stat}\pm 0.20_{\rm syst}$
  & $5.00\pm 0.02_{\rm stat} \pm 0.19_{\rm syst}$\\
  \hline 
  {\bf Average} & \mathversion{bold}$4.95\pm 0.01_{\rm stat}\pm
  0.11_{\rm syst}$ & \mathversion{bold}$\chi^2/\dof = 29.7/23$ (CL=$15.7\%$)\\
  \hline 
\end{tabular}
}
\end{center}
\label{tab:dstarlnu}
\end{table}

\begin{table}[!htb]
\caption{Average of the $B^-\to D^{*0}\ell^-\bar\nu_\ell$ branching
  fraction measurements. This fit uses only measurements of $B^-\to
  D^{*0}\ell^-\bar\nu_\ell$.}
\begin{center}
\begin{tabular}{|l|c|c|}
  \hline
  Experiment & $\cbf(B^-\to D^{*0}\ell^-\bar\nu_\ell)$ [\%] (rescaled) &
  $\cbf(B^-\to D^{*0}\ell^-\bar\nu_\ell)$ [\%] (published)\\
  \hline \hline
  CLEO~\hfill\cite{Adam:2002uw}
  & $6.61\pm 0.20_{\rm stat}\pm 0.39_{\rm syst}$
  & $6.50\pm 0.20_{\rm stat}\pm 0.43_{\rm syst}$\\
  \babar tagged~\hfill\cite{Aubert:vcbExcl}
  & $5.72\pm 0.15_{\rm stat}\pm 0.30_{\rm syst}$
  & $5.83\pm 0.15_{\rm stat}\pm0.30_{\rm syst}$\\
  \babar~\hfill\cite{Aubert:2009_3}
  & $5.35\pm0.08_{\rm stat}\pm 0.40_{\rm syst}$
  & $5.56\pm0.08_{\rm stat} \pm0.41_{\rm syst}$\\
  \babar~\hfill\cite{Aubert:2009_1}
  & $5.43\pm 0.02_{\rm stat}\pm 0.21_{\rm syst}$
  & $5.40\pm 0.02_{\rm stat}\pm 0.21_{\rm syst}$\\
  \hline
  {\bf Average} & \mathversion{bold}$5.70\pm 0.02_{\rm stat}\pm
  0.19_{\rm syst}$ & \mathversion{bold}$\chi^2/\dof = 9.1/3$ (CL=$2.8\%$)\\
  \hline 
\end{tabular}
\end{center}
\label{tab:dstar0lnu}
\end{table}

\mysubsubsection{$\bar B\to D\ell^-\bar\nu_\ell$}
\label{slbdecays_dlnu}

Similarly, the average of $\bar B\to D\ell^-\bar\nu_\ell$ is also
based on the CLN parameterization~\cite{CLN}. The form factor ${\cal G}(w)$
of the decay is described by only two parameters: the normalization
${\cal G}(1)\vcb$ and the slope $\rho^2$.

We use the measurements of these two form factor parameters shown in
Table~\ref{tab:vcbg1} and correct them to match the latest values of the
input parameters~\cite{HFAG_sl:inputparams}. These measurements are
sensitive to both isospin states ($\BzbDplnu$ and $B^-\to
D^0\ell^-\bar\nu_\ell$). So, isospin symmetry is assumed in this
analysis.
\begin{table}[!htb]
\caption{Measurements of $\bar B\to D\ell^-\bar\nu_\ell$ in the
  parameterization of Caprini, Lellouch and Neubert
  (CLN)~\cite{CLN}. The average is the result of a 2-dimensional fit
  to the rescaled measurements of ${\cal G}(1)\vcb$ and $\rho^2$. The
  $\chi^2$~value of the combination is 0.5 for 8 degrees of freedom
  (CL=$100.0\%$). The total correlation between the average ${\cal G}(1)\vcb$
  and $\rho^2$ is 0.83.}
\begin{center}
\begin{tabular}{|l|c|c|}
  \hline
  Experiment
  & ${\cal G}(1)\vcb$ [10$^{-3}$] (rescaled)
  & $\rho^2$ (rescaled)\\
  & ${\cal G}(1)\vcb$ [10$^{-3}$] (published)
  & $\rho^2$ (published)\\
  \hline \hline
  ALEPH~\hfill\cite{Buskulic:1996yq}
  & $38.89\pm 11.80_{\rm stat}\pm 6.09_{\rm syst}$
  & $0.951\pm 0.980_{\rm stat}\pm 0.357_{\rm syst}$\\
  & $31.1\pm 9.9_{\rm stat}\pm 8.6_{\rm syst}$
  & $0.70\pm 0.98_{\rm stat}\pm 0.50_{\rm syst}$\\
  \hline
  CLEO~\hfill\cite{Bartelt:1998dq}
  & $44.90\pm 5.97_{\rm stat}\pm 3.30_{\rm syst}$
  & $1.270\pm 0.250_{\rm stat}\pm 0.140_{\rm syst}$\\
  & $44.8\pm 6.1_{\rm stat}\pm 3.7_{\rm syst}$
  & $1.30\pm 0.27_{\rm stat}\pm 0.14_{\rm syst}$\\
  \hline
  \belle~\hfill\cite{Abe:2001yf}
  & $40.84\pm 4.37_{\rm stat}\pm 5.17_{\rm syst}$
  & $1.120\pm 0.220_{\rm stat}\pm 0.140_{\rm syst}$\\
  & $41.1\pm 4.4_{\rm stat}\pm 5.1_{\rm syst}$
  & $1.12\pm 0.22_{\rm stat}\pm 0.14_{\rm syst}$\\
  \hline
  \babar global fit~\hfill\cite{Aubert:2009_1}
  & $43.42\pm 0.81_{\rm stat}\pm 2.08_{\rm syst}$
  & $1.204\pm 0.040_{\rm stat}\pm 0.057_{\rm syst}$\\
  & $43.1\pm 0.8_{\rm stat}\pm 2.3_{\rm syst}$
  & $1.20\pm 0.04_{\rm stat}\pm 0.07_{\rm syst}$\\
  \hline
  \babar tagged~\hfill\cite{Aubert:2009_2}
  & $42.45\pm 1.88_{\rm stat}\pm 1.05_{\rm syst}$
  & $1.180\pm 0.089_{\rm stat}\pm 0.051_{\rm syst}$\\
  & $42.3\pm 1.9_{\rm stat}\pm 1.0_{\rm syst}$
  & $1.20\pm 0.09_{\rm stat}\pm 0.04_{\rm syst}$\\
  \hline 
  {\bf Average }
  & \mathversion{bold}$42.64\pm 0.72_{\rm stat}\pm 1.35_{\rm syst}$
  & \mathversion{bold}$1.186\pm 0.036_{\rm stat}\pm 0.041_{\rm syst}$\\
  \hline 
\end{tabular}
\end{center}
\label{tab:vcbg1}
\end{table}

\begin{figure}[!ht]
  \begin{center}
  \unitlength1.0cm 
  \begin{picture}(14.,8.) 
    \put(  8.0, -0.2){\includegraphics[width=8.0cm]{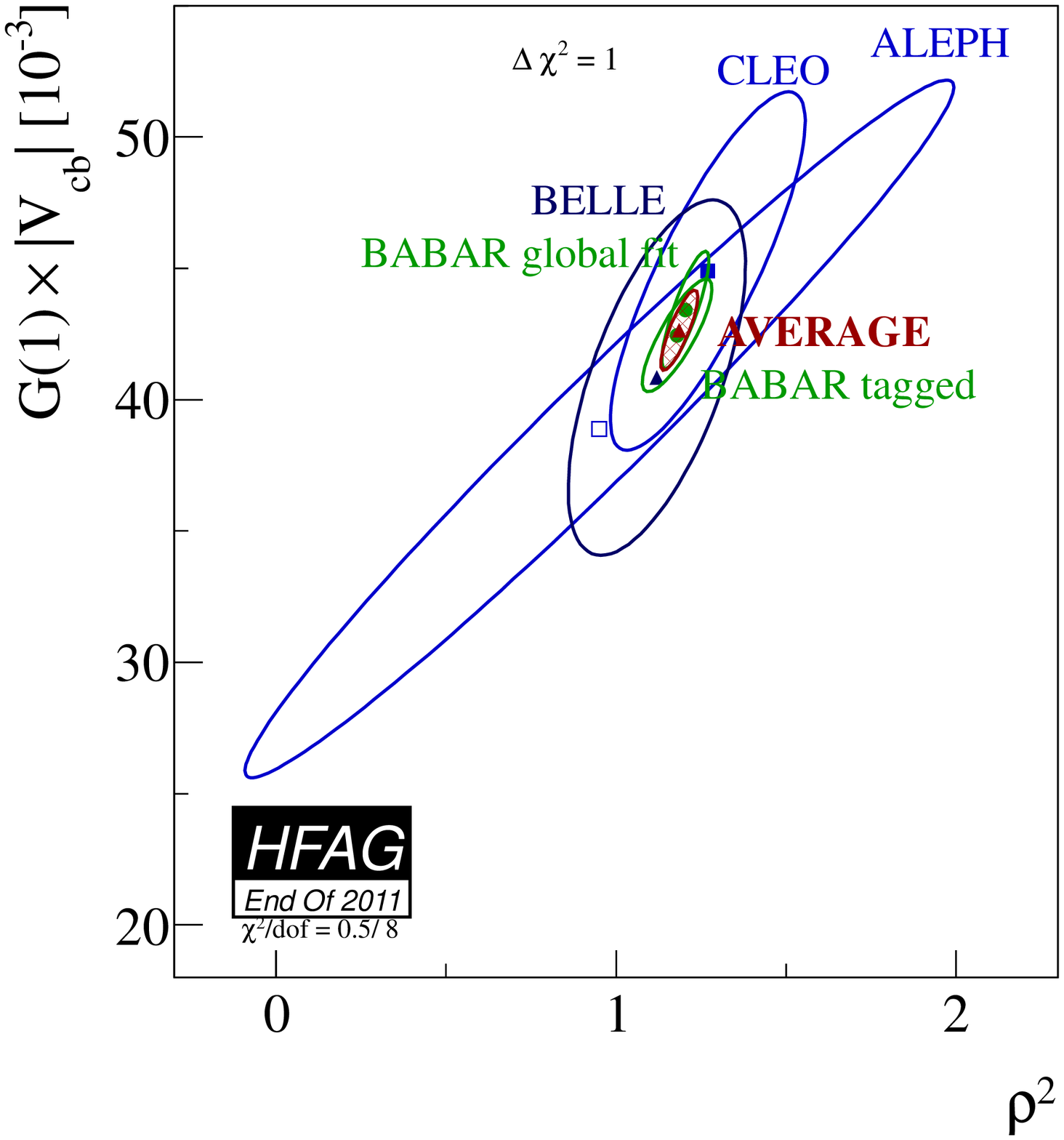}}
    \put( -0.5,  0.0){\includegraphics[width=7.5cm]{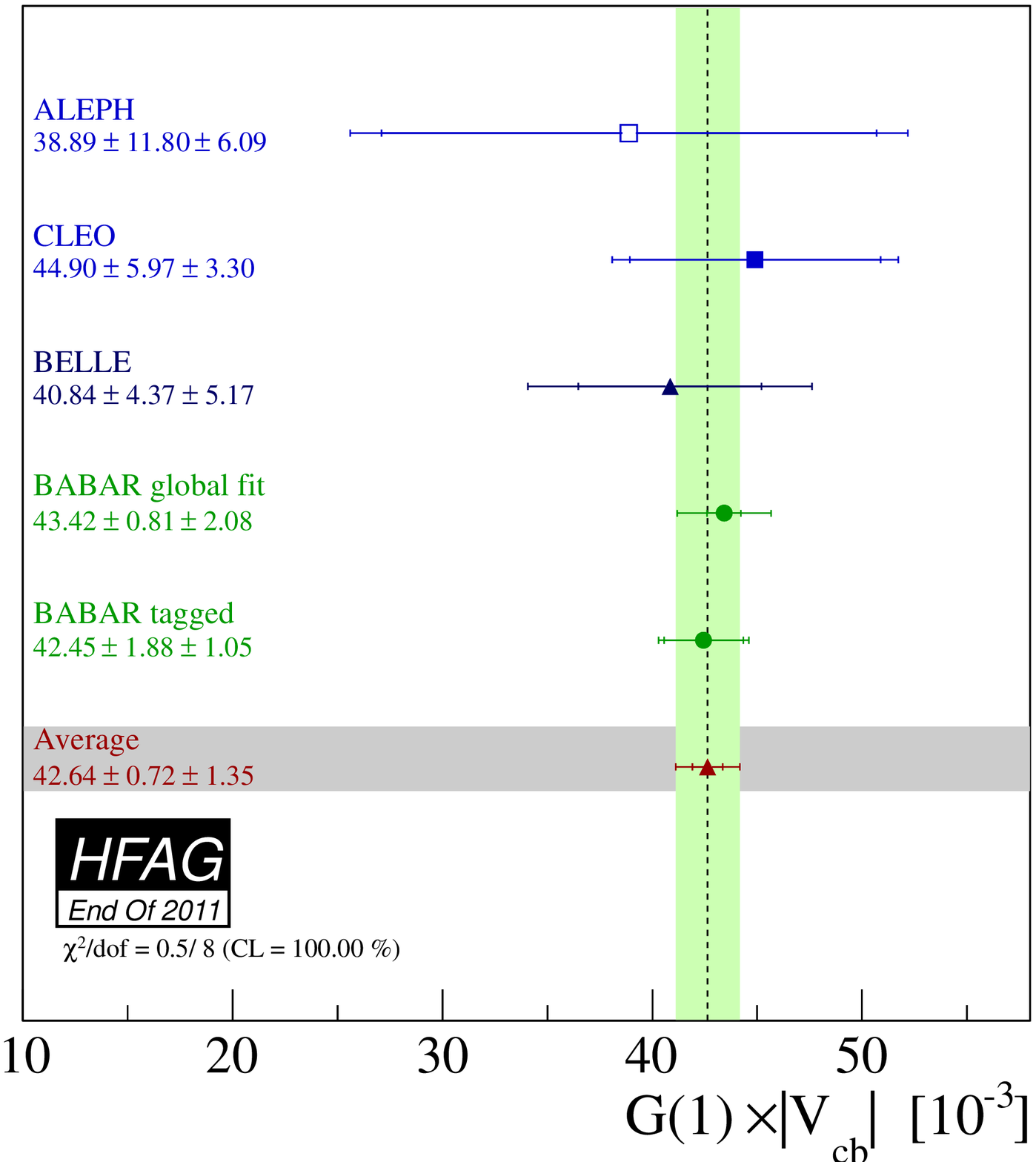}}
    \put(  5.5,  6.8){{\large\bf a)}}
    \put( 9.4,  6.8){{\large\bf b)}}
  \end{picture}
  \caption{(a) Illustration of the ${\cal G}(1)\vcb$ average. (b)
    Illustration of the ${\cal G}(1)\vcb$ vs.\ $\rho^2$ average. The error
    ellipses correspond  to $\Delta\chi^2 = 1$ (CL=39\%).}
  \label{fig:vcbg1}
  \end{center}
\end{figure}

The form factor parameters are extracted by a two-dimensional fit to
the rescaled measurements of ${\cal G}(1)\vcb$ and $\rho^2$ taking into
account correlated systematic uncertainties. The result of the fit
reads
\begin{eqnarray}
  {\cal G}(1)\vcb & = & (42.64\pm 1.53)\times 10^{-3}~, \label{eq:vcbg1} \\
  \rho^2 & = & 1.186 \pm 0.054~,
\end{eqnarray}
with a correlation of
\begin{equation}
  \rho_{{\cal G}(1)\vcb,\rho^2} = 0.829~.
\end{equation}
The uncertainties and the correlation coefficient include both
statistical and systematic contributions. The $\chi^2$ of the fit is
0.5 for 8 degrees of freedom, which corresponds to a confidence
level of 100.0\%. An illustration of this fit result is given in
Fig.~\ref{fig:vcbg1}.

The most recent result obtained for the form factor normalization
${\cal G}(1)$ in LQCD is~\cite{Okamoto:2004xg}
\begin{equation}
  {\cal G}(1) = 1.074\pm 0.024~,
\end{equation}
which can be used to turn Eq.~\ref{eq:vcbg1} into a determination of
$\vcb$,
\begin{equation}
  \vcb = (39.70\pm 1.42_{\rm exp}\pm 0.89_{\rm th})\times 10^{-3}~,
\end{equation}
where the first error is experimental and the second theoretical. This
number is in excellent agreement with $\vcb$ obtained from decays
$\bar B\to D^*\ell^-\bar\nu_\ell$, Eq.~\ref{eq:vcbdstar}.

\begin{figure}[!ht]
  \begin{center}
  \unitlength1.0cm 
  \begin{picture}(8.,8.0)  
  \put( -0.5,  0.0){\includegraphics[width=7.55cm]{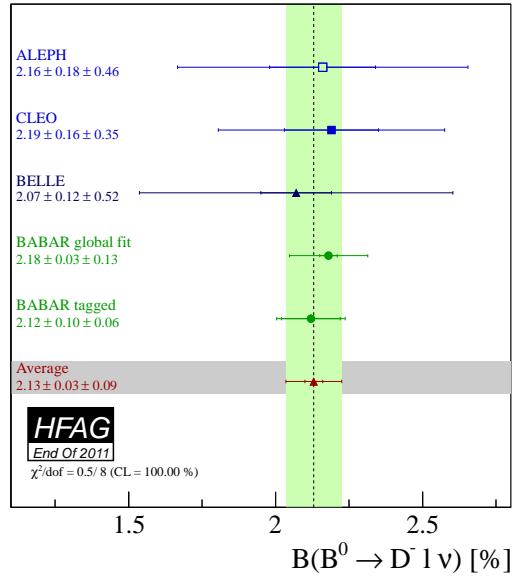}}
  \end{picture}
  \caption{Average branching fraction of exclusive semileptonic
    $B$~decays $\bar B^0\to D^+\ell^-\bar\nu_\ell$
    (Table~\ref{tab:dlnuIso}). The fit assumes isospin conservation.}
  \label{fig:brdlIso}
  \end{center}
\end{figure}

From each rescaled measurement in Table~\ref{tab:vcbg1}, we have
calculated the $\bar B\to D\ell^-\bar\nu_\ell$ form factor ${\cal G}(w)$
and, by numerical integration, the branching ratio of the decay
$\BzbDplnu$. The results are quoted in Table~\ref{tab:dlnuIso} and
illustrated in Fig.~\ref{fig:brdlIso}. The branching ratio found for
the average values of ${\cal G}(1)\vcb$ and $\rho^2$ is
\begin{equation}
  \cbf(\BzbDplnu)=(2.13\pm 0.09)\%~.
\end{equation}
This analysis assumes isospin symmetry.
\begin{table}[!htb]
\caption{$\bar B^0\to D^+\ell^-\bar\nu_\ell$ branching fractions
  calculated from the rescaled measurements in Table~\ref{tab:vcbg1},
  asumming the CLN parameterization of the form factor~\cite{CLN}. The
  fit assumes isospin symmetry.}
\begin{center}
\resizebox{0.99\textwidth}{!}{
\begin{tabular}{|l|c|c|}
  \hline
  Experiment
  & $\cbf(\bar B^0\to D^+\ell^-\bar\nu_\ell)$ [\%] (calculated)
  & $\cbf(\bar B^0\to D^+\ell^-\bar\nu_\ell)$ [\%] (published)\\
  \hline \hline
  ALEPH~\hfill\cite{Buskulic:1996yq}
  & $2.16\pm 0.18_{\rm stat}\pm 0.46_{\rm syst}$
  & $2.35\pm 0.20_{\rm stat}\pm 0.44_{\rm syst}$\\
  CLEO~\hfill\cite{Bartelt:1998dq}
  & $2.19\pm 0.16_{\rm stat}\pm 0.35_{\rm syst}$
  & $2.20\pm 0.16_{\rm stat}\pm 0.19_{\rm syst}$\\
  \belle~\hfill\cite{Abe:2001yf}
  & $2.07\pm 0.12_{\rm stat}\pm 0.52_{\rm syst}$
  & $2.13\pm 0.12_{\rm stat}\pm 0.39_{\rm syst}$\\
  \babar global fit~\hfill\cite{Aubert:2009_1}
  & $2.18\pm 0.03_{\rm stat}\pm 0.13_{\rm syst}$
  & $2.34\pm 0.03_{\rm stat}\pm 0.13_{\rm syst}$\\
  \babar tagged~\hfill\cite{Aubert:2009_2}
  & $2.12\pm 0.10_{\rm stat}\pm 0.06_{\rm syst}$
  & $2.23\pm 0.11_{\rm stat}\pm 0.11_{\rm syst}$\\
  \hline 
  {\bf Average}
  & \mathversion{bold}$2.13\pm 0.03_{\rm stat}\pm 0.09_{\rm syst}$
  & \mathversion{bold}$\chi^2/\dof = 0.5/8$ (CL=$100.0\%$)\\
  \hline 
\end{tabular}
}
\end{center}
\label{tab:dlnuIso}
\end{table}

We have also performed simple 1-dimensional averages of measurements
of $\BzbDplnu$ and $B^-\to D^0\ell^-\bar\nu_\ell$. These fits are
shown Tables~\ref{tab:dlnu} and \ref{tab:d0lnu}.
\begin{table}[!htb]
\caption{Average of the $\BzbDplnu$ branching fraction
  measurements. This fit uses only measurements of $\BzbDplnu$.}
\begin{center}
\begin{tabular}{|l|c|c|}
  \hline
  Experiment
  & $\cbf(\BzbDplnu)$ [\%] (rescaled)
  & $\cbf(\BzbDplnu)$ [\%] (published)\\
  \hline \hline
  ALEPH~\hfill\cite{Buskulic:1996yq}
  & $2.29\pm 0.18_{\rm stat}\pm 0.35_{\rm syst}$
  & $2.35\pm 0.20_{\rm stat}\pm 0.44_{\rm syst}$\\
  CLEO~\hfill\cite{Bartelt:1998dq}
  & $2.13\pm 0.13_{\rm stat}\pm 0.15_{\rm syst}$
  & $2.20\pm 0.16_{\rm stat}\pm 0.19_{\rm syst}$\\
  \belle~\hfill\cite{Abe:2001yf}
  & $2.10\pm 0.12_{\rm stat}\pm 0.39_{\rm syst}$
  & $2.13\pm 0.12_{\rm stat}\pm 0.39_{\rm syst}$\\
  \babar~\hfill\cite{Aubert:vcbExcl}
  & $2.21\pm 0.11_{\rm stat}\pm 0.12_{\rm syst}$
  & $2.21\pm 0.11_{\rm stat}\pm 0.12_{\rm syst}$\\
  \hline 
  {\bf Average}
  & \mathversion{bold}$2.18\pm 0.06_{\rm stat}\pm 0.10_{\rm syst}$
  & \mathversion{bold}$\chi^2/\dof = 0.2/3$ (CL=$97.4\%$)\\
  \hline 
\end{tabular}
\end{center}
\label{tab:dlnu}
\end{table}

\begin{table}[!htb]
\caption{Average of the $B^-\to D^0\ell^-\bar\nu_\ell$ branching fraction
  measurements. This fit uses only measurements of $B^-\to
  D^0\ell^-\bar\nu_\ell$.}
\begin{center}
\begin{tabular}{|l|c|c|}
  \hline
  Experiment
  & $\cbf(B^-\to D^0\ell^-\bar\nu_\ell)$ [\%] (rescaled)
  & $\cbf(B^-\to D^0\ell^-\bar\nu_\ell)$ [\%] (published)\\
  \hline \hline
  CLEO~\hfill\cite{Bartelt:1998dq}
  & $2.21\pm 0.13_{\rm stat}\pm 0.17_{\rm syst}$
  & $2.32\pm 0.17_{\rm stat}\pm 0.20_{\rm syst}$\\
  \babar~\hfill\cite{Aubert:vcbExcl}
  & $2.28\pm 0.09_{\rm stat}\pm 0.09_{\rm syst}$
  & $2.33\pm 0.09_{\rm stat}\pm 0.09_{\rm syst}$\\
  \hline
  {\bf Average}
  & \mathversion{bold}$2.26\pm 0.07_{\rm stat}\pm 0.08_{\rm syst}$
  & \mathversion{bold}$\chi^2/\dof = 0.1/1$ (CL=$76.0\%$)\\
  \hline
\end{tabular}
\end{center}
\label{tab:d0lnu}
\end{table}


\mysubsubsection{$\bar{B} \to D^{(*)}\pi \ell^-\bar{\nu}_{\ell}$}
\label{slbdecays_dpilnu}

The average inclusive branching fractions for $\bar{B} \to D^{*}\pi \ell^-\bar{\nu}_{\ell}$
decays, where no constrain is applied to the hadronic $D^{(*)}\pi$ system, 
are determined by the
combination of the results provided in Table~\ref{tab:dpilnu} for 
$\bar{B}^0 \to D^0 \pi^+ \ell^-\bar{\nu}_{\ell}$, $\bar{B}^0 \to D^{*0} \pi^+
\ell^-\bar{\nu}_{\ell}$, 
$B^- \to D^+ \pi^-
\ell^-\bar{\nu}_{\ell}$, and $B^- \to D^{*+} \pi^- \ell^-\bar{\nu}_{\ell}$.
The measurements included in the average 
are scaled to a consistent set of input
parameters and their errors~\cite{HFAG_sl:inputparams}.

For both the \babar\ and Belle results, the $B$ semileptonic signal yields are
 extracted from a fit to the missing mass squared in a sample of fully
 reconstructed \BB\ events. 
 
\begin{figure}[!ht]
 \begin{center}
  \unitlength1.0cm 
  \begin{picture}(14.,8.0)  
   \put( -0.5,  0.0){\includegraphics[width=7.55cm]{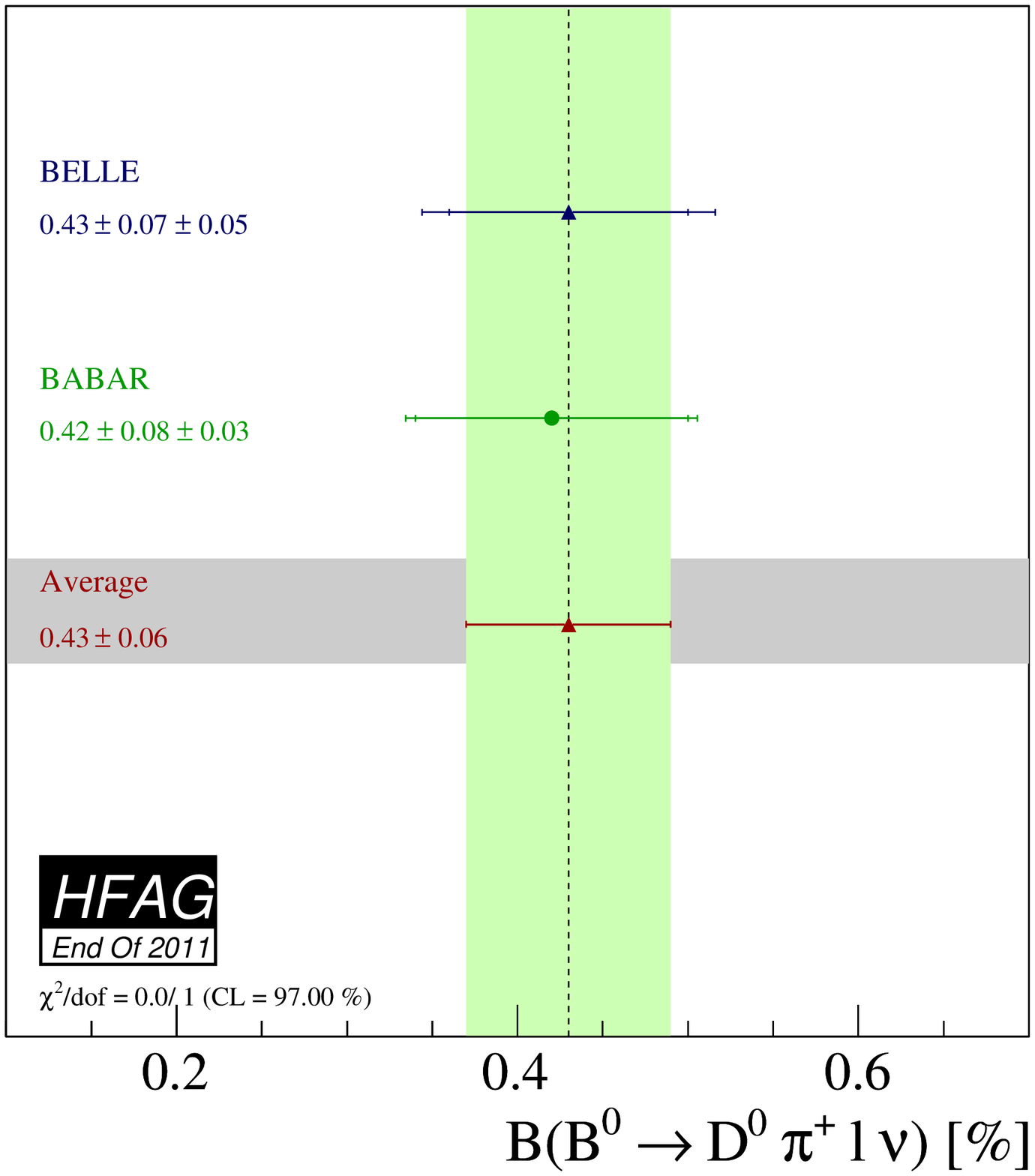}}
   \put(  8.0,  0.0){\includegraphics[width=7.8cm]{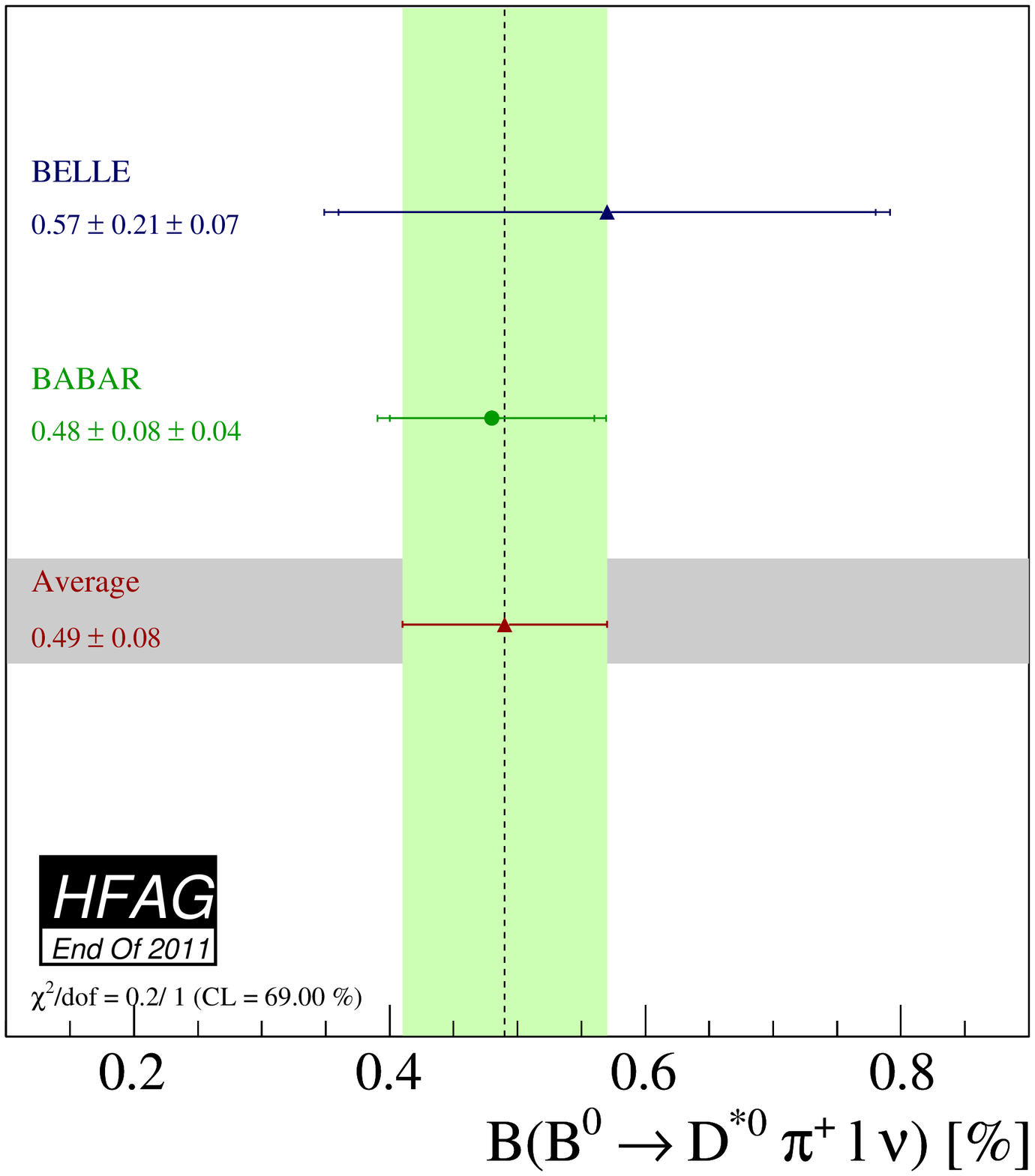}}
   \put(  5.5,  6.8){{\large\bf a)}}
   \put( 14.0,  6.8){{\large\bf b)}}
  \end{picture}
  \begin{picture}(14.,8.0)  
   \put( -0.5,  0.0){\includegraphics[width=7.55cm]{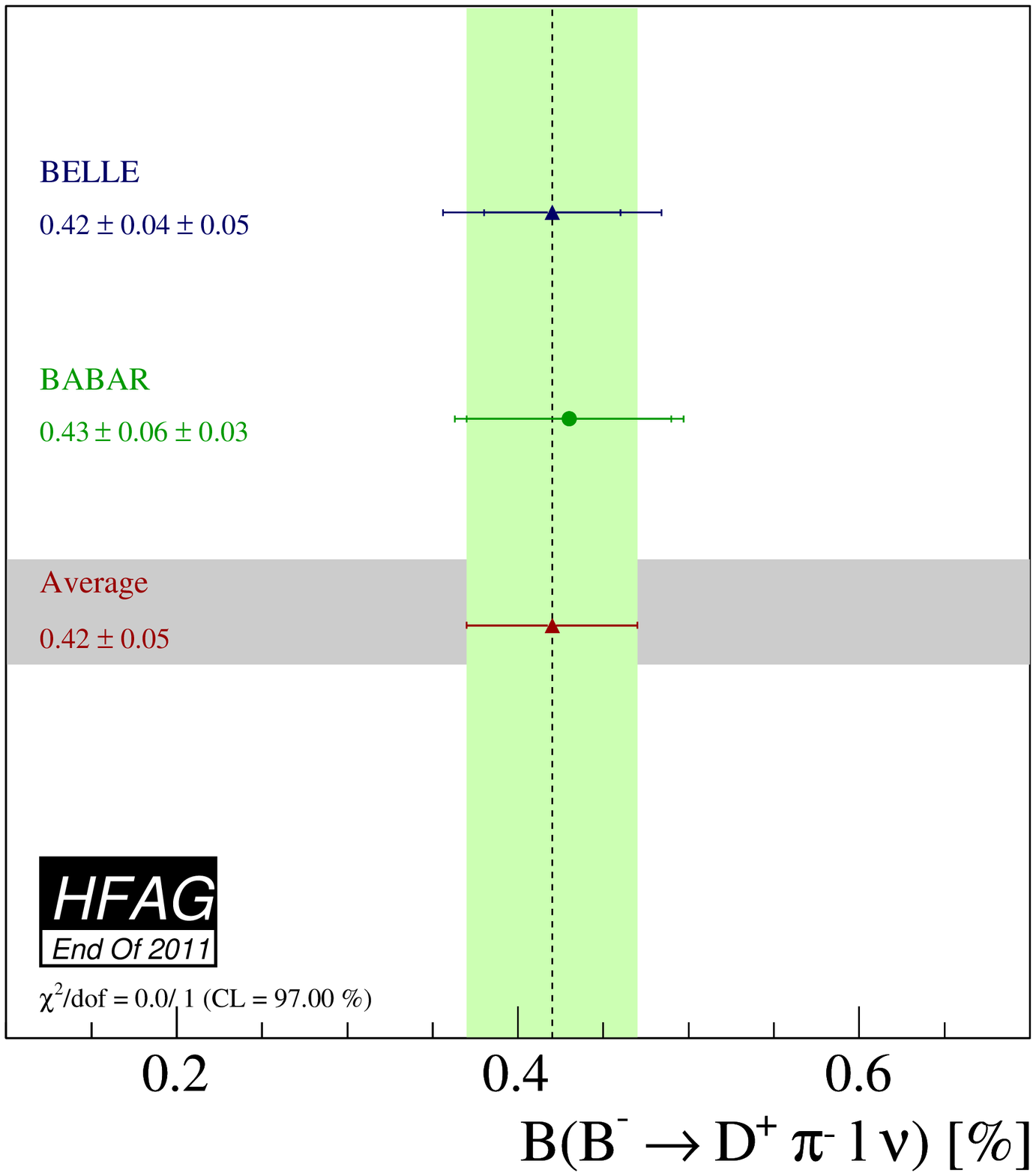}}
   \put(  8.0,  0.0){\includegraphics[width=7.8cm]{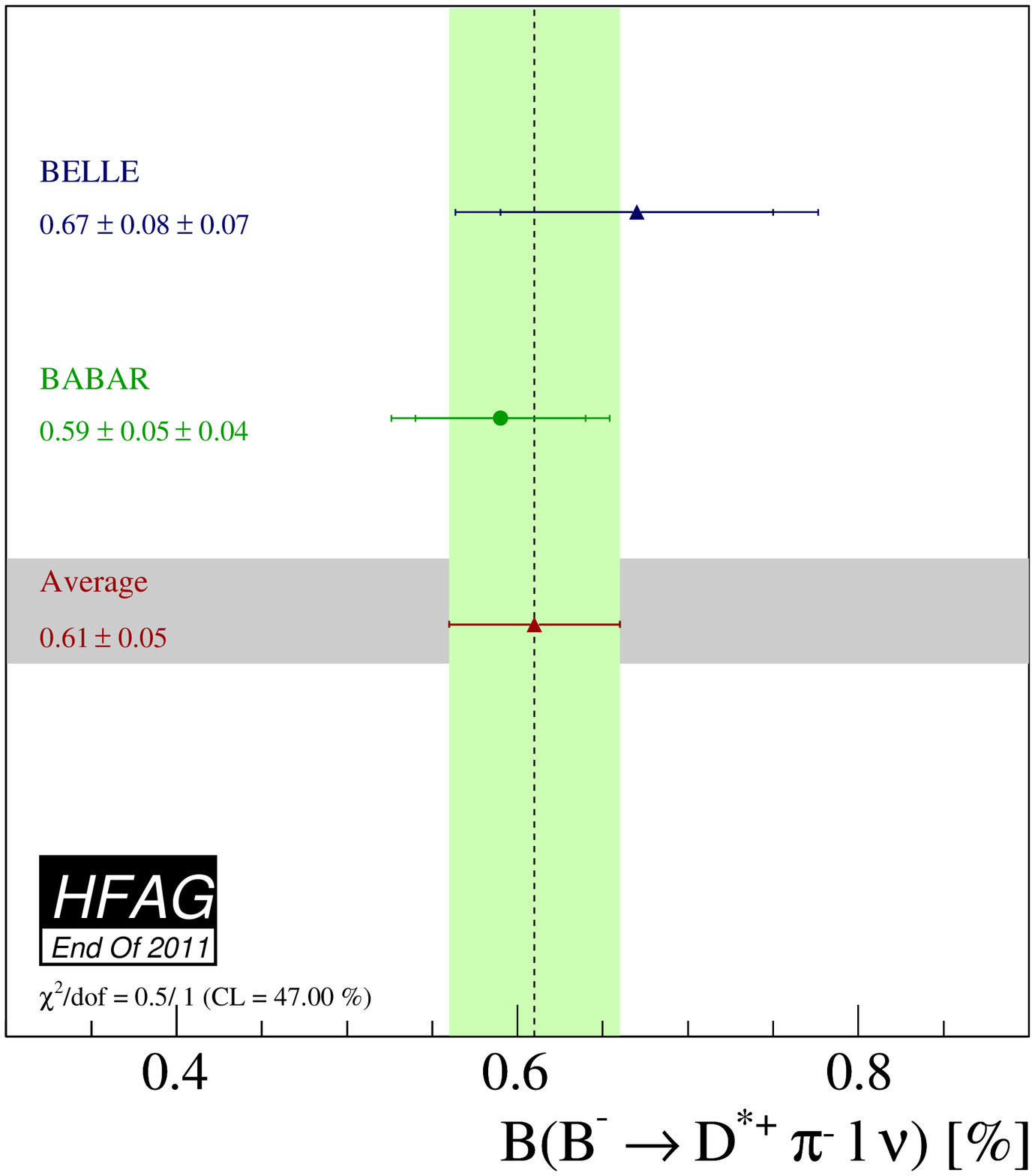}}
   \put(  5.5,  6.8){{\large\bf c)}}
   \put( 14.0,  6.8){{\large\bf d)}}
  \end{picture}
  \caption{Average branching fraction  of exclusive semileptonic $B$ decays
(a) $\bar{B}^0 \to D^0 \pi^+ \ell^-\bar{\nu}_{\ell}$, (b) $\bar{B}^0 \to D^{*0} \pi^+
\ell^-\bar{\nu}_{\ell}$, (c) $B^- \to D^+ \pi^-
\ell^-\bar{\nu}_{\ell}$, and (d) $B^- \to D^{*+} \pi^- \ell^-\bar{\nu}_{\ell}$.
The corresponding individual
  results are also shown.}
  \label{fig:brdpil}
 \end{center}
\end{figure}

Figure~\ref{fig:brdpil} illustrates the measurements and the
resulting average.

\begin{table}[!htb]
\caption{Average of the branching fraction $B \to D^{(*)} \pi^- \ell^-\bar{\nu}_{\ell}$ and individual
results.}
\begin{center}
\begin{tabular}{|l|c c|}\hline
Experiment                                 &$\cbf(B^- \to D^+ \pi^- \ell^-\bar{\nu}_{\ell}) [\%]$ (rescaled) &\\
\hline
\belle  ~\hfill\cite{Live:Dss}           &$0.42 \pm0.04_{\rm stat} \pm0.05_{\rm syst}$  & \\
\babar  ~\hfill\cite{Aubert:vcbExcl}       &$0.42 \pm0.06_{\rm stat} \pm0.03_{\rm syst}$ & \\
\hline 
{\bf Average}                              &\mathversion{bold}$0.42 \pm0.05$  
   &\mathversion{bold}$\chi^2/\dof = 0.001$ (CL=$97\%$)  \\
\hline\hline

Experiment                                 &$\cbf(B^- \to D^{*+} \pi^- \ell^-\bar{\nu}_{\ell}) [\%]$ (rescaled) & \\
\hline\hline 
\belle  ~\hfill\cite{Live:Dss}           &$0.67 \pm0.08_{\rm stat} \pm0.07_{\rm syst}$  
& \\
\babar  ~\hfill\cite{Aubert:vcbExcl}       &$0.59 \pm0.05_{\rm stat} \pm0.04_{\rm syst}$  
& \\
\hline 
{\bf Average}                              &\mathversion{bold}$0.61 \pm0.05$  
 & \mathversion{bold}$\chi^2/\dof = 0.15$ (CL=$69\%$)    \\
\hline \hline

Experiment                               &$\cbf(\bar{B}^0 \to D^0 \pi^+ \ell^-\bar{\nu}_{\ell}) [\%]$ (rescaled) & \\
\hline\hline 
\belle  ~\hfill\cite{Live:Dss}           &$0.43 \pm0.07_{\rm stat} \pm0.05_{\rm syst}$ &  \\
\babar  ~\hfill\cite{Aubert:vcbExcl}     &$0.43 \pm0.08_{\rm stat} \pm0.03_{\rm syst}$ &  \\
\hline 
{\bf Average}                              &\mathversion{bold}$0.43 \pm0.06$  
   &\mathversion{bold}$\chi^2/\dof = 0.002$ (CL=$97\%$)  \\
\hline\hline

Experiment                                 &$\cbf(\bar{B}^0 \to D^{*0} \pi^+\ell^-\bar{\nu}_{\ell}) [\%]$ (rescaled) & \\
\hline\hline 
\belle  ~\hfill\cite{Live:Dss}           &$0.57 \pm0.21_{\rm stat} \pm0.07_{\rm syst}$  & \\
\babar  ~\hfill\cite{Aubert:vcbExcl}       &$0.48 \pm0.08_{\rm stat} \pm0.04_{\rm syst}$ & \\ 
\hline 
{\bf Average}                              &\mathversion{bold}$0.49 \pm0.08$  
   &\mathversion{bold}$\chi^2/\dof = 0.15$ (CL=$69\%$)  \\
\hline\hline

\end{tabular}
\end{center}
\label{tab:dpilnu}
\end{table}

\mysubsubsection{$\bar{B} \to D^{**} \ell^-\bar{\nu}_{\ell}$}
\label{slbdecays_dsslnu}

The $D^{**}$ mesons contain one charm quark and one light quark with relative angular momentum $L=1$. According to Heavy Quark Symmetry (HQS)~\cite{Isgur:1991wq}, they form one doublet of states with angular momentum $j \equiv s_q + L= 3/2$  $\left[D_1(2420), D_2^*(2460)\right]$ and another doublet with $j=1/2$ $\left[D^*_0(2400), D_1'(2430)\right]$, where $s_q$ is the light quark spin. Parity and angular momentum conservation constrain the decays allowed for each state. The $D_1$ and $D_2^*$ states decay through a D-wave to $D^*\pi$ and $D^{(*)}\pi$, respectively, and have small decay widths, while the $D_0^*$ and $D_1'$ 
states decay through an S-wave to $D\pi$ and $D^*\pi$ and are very broad.
For the narrow states, the average 
are determined by the
combination of the results provided in Table~\ref{tab:dss1lnu} and \ref{tab:dss2lnu} for 
$\cbf(B^- \to D_1^0(D^{*+}\pi^-)\ell^-\bar{\nu}_{\ell})
\times \cbf(D_1^0 \to D^{*+}\pi^-)$ and $\cbf(B^- \to D_2^0(D^{*+}\pi^-)\ell^-\bar{\nu}_{\ell})
\times \cbf(D_2^0 \to D^{*+}\pi^-)$. 
For the broad states, the average 
are determined by the
combination of the results provided in Table~\ref{tab:dss1plnu} and \ref{tab:dss0lnu} for 
$\cbf(B^- \to D_1'^0(D^{*+}\pi^-)\ell^-\bar{\nu}_{\ell})
\times \cbf(D_1'^0 \to D^{*+}\pi^-)$ and $\cbf(B^- \to D_0^{*0}(D^{+}\pi^-)\ell^-\bar{\nu}_{\ell})
\times \cbf(D_0^{*0} \to D^{+}\pi^-)$. 
The measurements included in the average 
are scaled to a consistent set of input
parameters and their errors~\cite{HFAG_sl:inputparams}.  

\begin{figure}[!ht]
 \begin{center}
  \unitlength1.0cm 
  \begin{picture}(14.,8.0)  
   \put( -0.5,  0.0){\includegraphics[width=7.55cm]{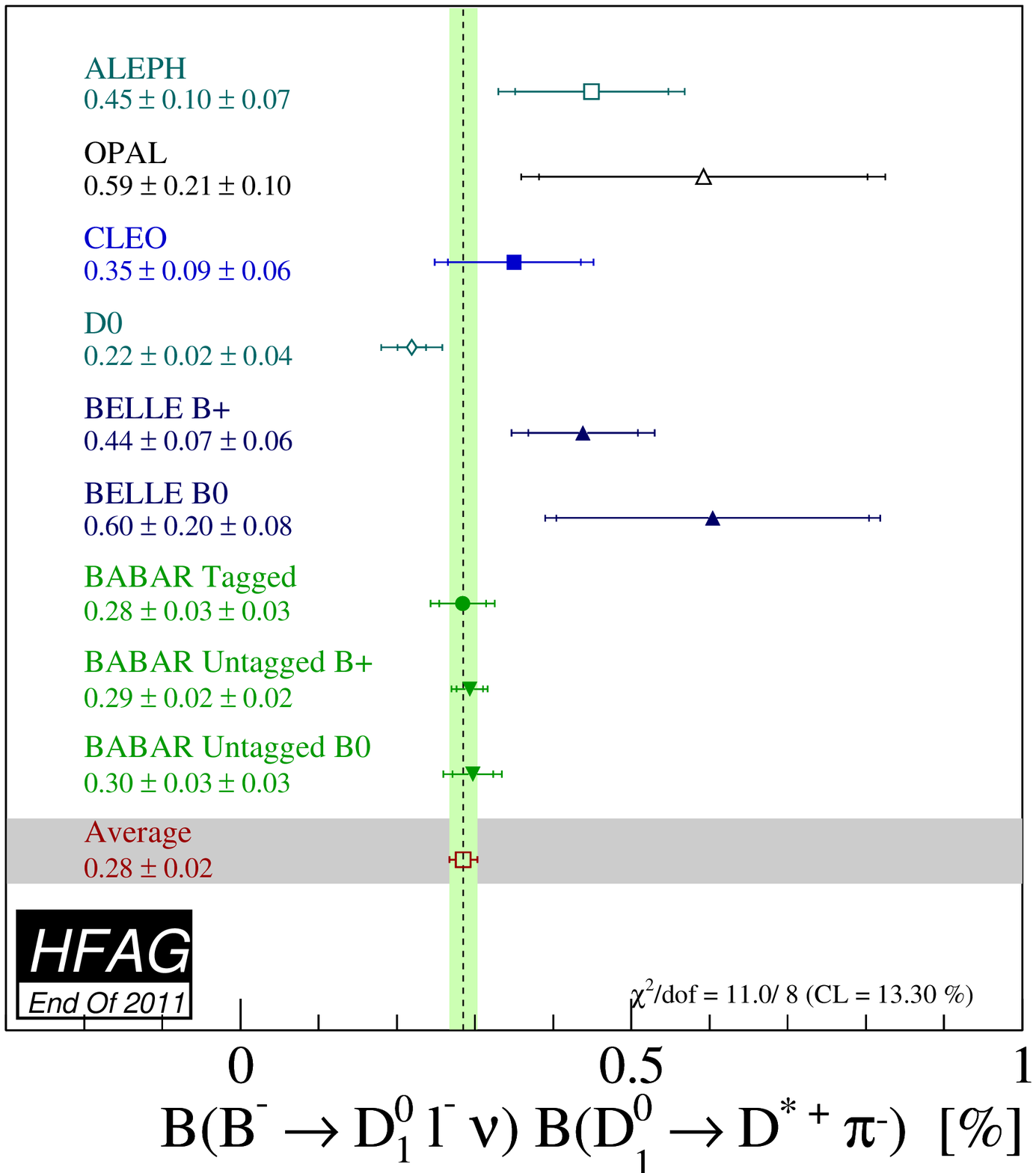}}
   \put(  8.0,  0.0){\includegraphics[width=7.8cm]{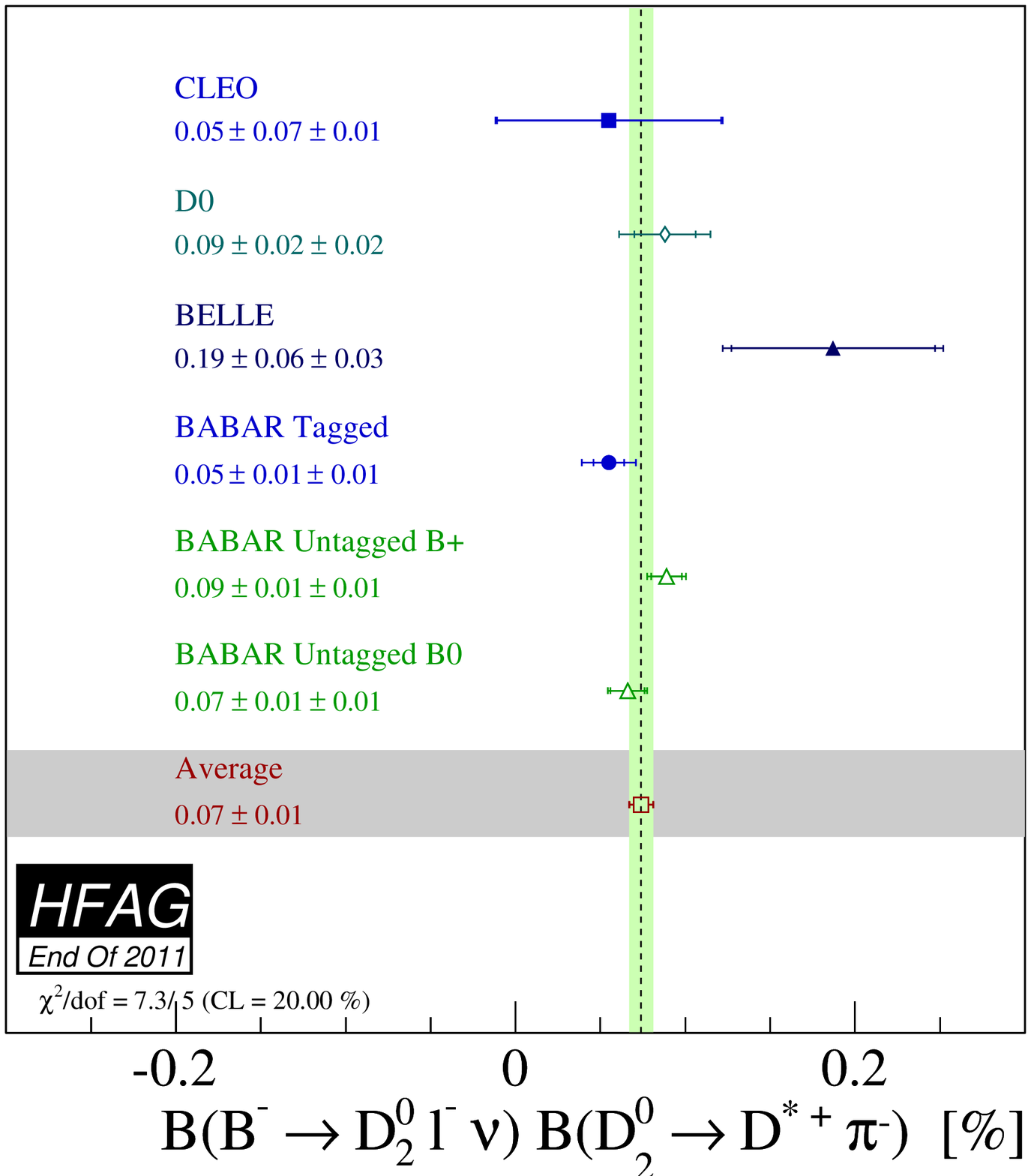}}
   \put(  5.5,  6.8){{\large\bf a)}}
   \put( 14.0,  6.8){{\large\bf b)}}
  \end{picture}
  \caption{Average of the product of branching fraction (a) 
  $\cbf(B^- \to D_1^0(D^{*+}\pi^-)\ell^-\bar{\nu}_{\ell})
\times \cbf(D_1^0 \to D^{*+}\pi^-)$ and (b) $\cbf(B^- \to D_2^0(D^{*+}\pi^-)\ell^-\bar{\nu}_{\ell})
\times \cbf(D_2^0 \to D^{*+}\pi^-)$
The corresponding individual
  results are also shown.}
  \label{fig:brdssl}
 \end{center}
\end{figure}

\begin{figure}[!ht]
 \begin{center}
  \unitlength1.0cm 
  \begin{picture}(14.,8.0)  
   \put( -0.5,  0.0){\includegraphics[width=7.55cm]{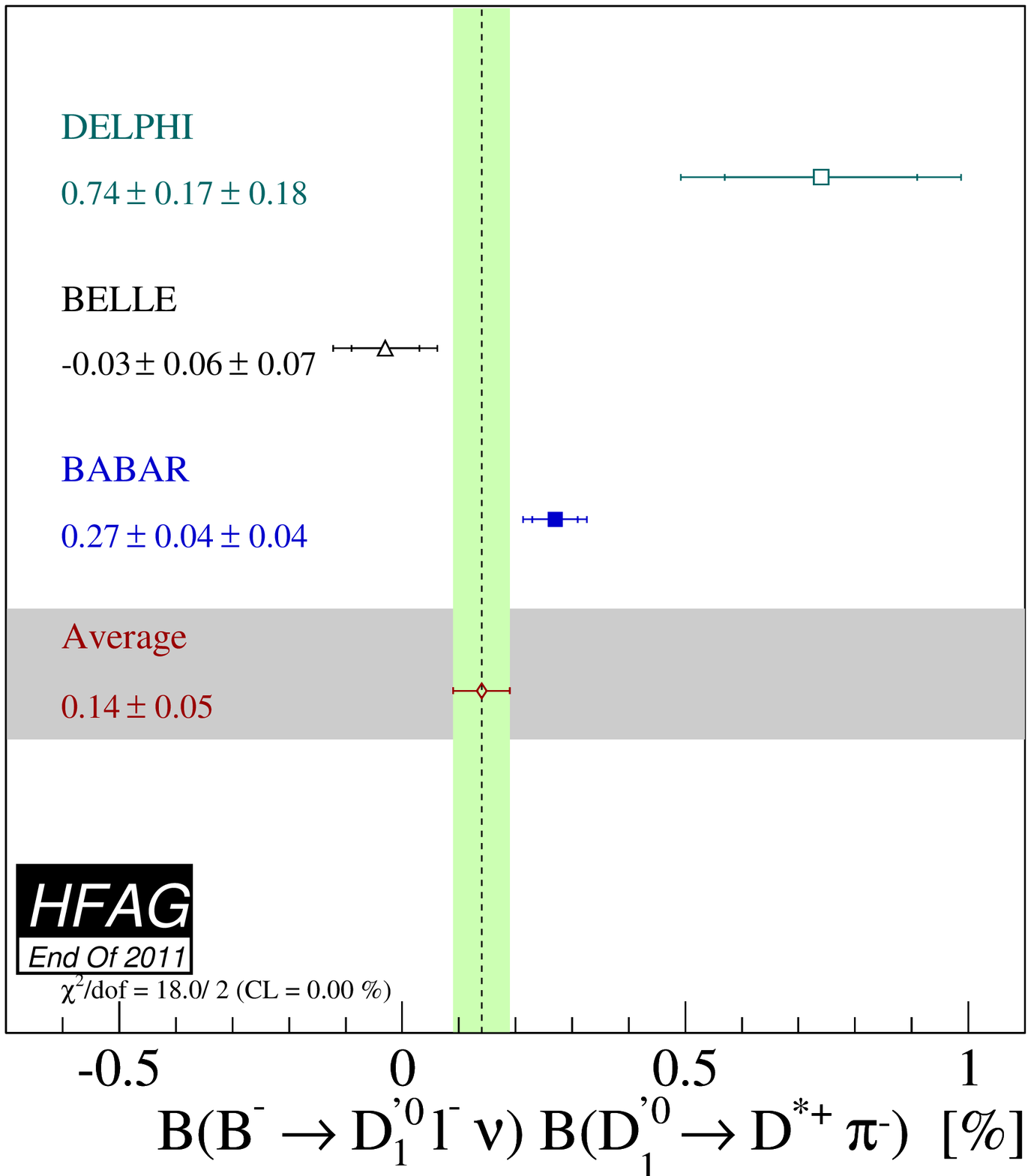}}
   \put(  8.0,  0.0){\includegraphics[width=7.8cm]{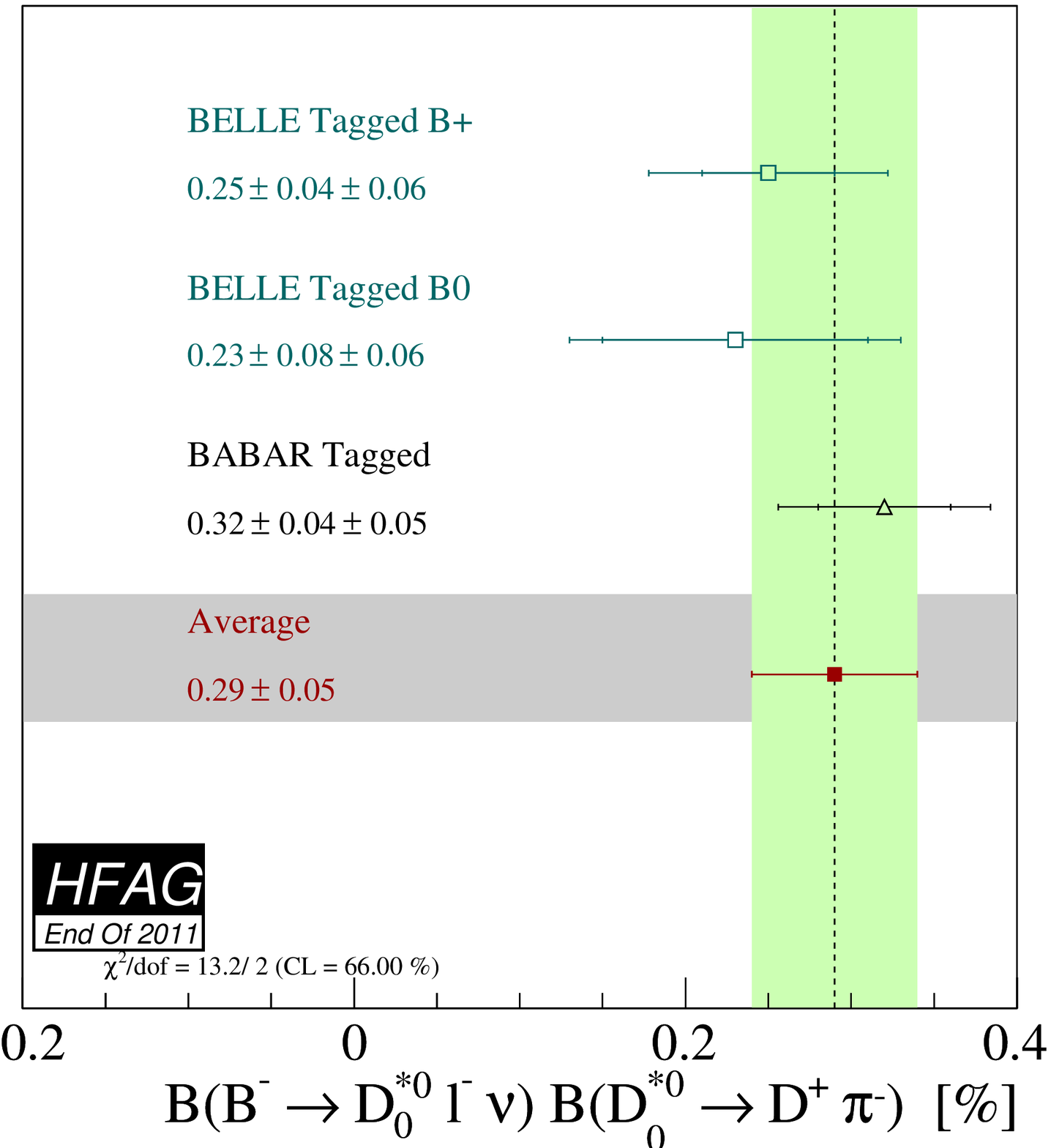}}
   \put(  5.5,  6.8){{\large\bf a)}}
   \put( 14.0,  6.8){{\large\bf b)}}
  \end{picture}
  \caption{Average of the product of branching fraction (a) 
  $\cbf(B^- \to D_1'^0(D^{*+}\pi^-)\ell^-\bar{\nu}_{\ell})
\times \cbf(D_1'^0 \to D^{*+}\pi^-)$ and (b) $\cbf(B^- \to D_0^{*0}(D^{*+}\pi^-)\ell^-\bar{\nu}_{\ell})
\times \cbf(D_0^{*0} \to D^{+}\pi^-)$
The corresponding individual
  results are also shown.}
  \label{fig:brdssl2}
 \end{center}
\end{figure}

For both the B-factory and the LEP and Tevatron results, the $B$ semileptonic 
signal yields are
 extracted from a fit to the invariant mass distribution of the $D^{(*)+}\pi^-$ system.
 Apart for the CLEO and Belle results, the other measurements 
 are for the final state $\bar{B} \to D_2(D^{*+}\pi^-)X \ell^- \bar{\nu}_{\ell}$. 
 We assume that no particle is left in the X system. 
Figure~\ref{fig:brdssl} and ~\ref{fig:brdssl2} illustrate the measurements and the
resulting average.

\begin{table}[!htb]
\caption{Average of the branching fraction $\cbf(B^- \to D_1^0(D^{*+}\pi^-)\ell^-\bar{\nu}_{\ell})
\times \cbf(D_1^0 \to D^{*+}\pi^-))$ and individual
results. }
\begin{center}
\resizebox{0.99\textwidth}{!}{
\begin{tabular}{|l|c|c|}\hline
Experiment                                 &$\cbf(B^- \to D_1^0(D^{*+}\pi^-)\ell^-\bar{\nu}_{\ell})
) [\%]$  &$\cbf(B^- \to D_1^0(D^{*+}\pi^-)\ell^-\bar{\nu}_{\ell})
) [\%]$  \\
                                                & (rescaled) & (published) \\

\hline\hline 
ALEPH ~\hfill\cite{Aleph:Dss}        &$0.45 \pm0.10_{\rm stat} \pm0.07_{\rm syst}$ 
 &$0.47 \pm0.098_{\rm stat} \pm0.074_{\rm syst}$ \\
OPAL  ~\hfill\cite{opal:Dss}         &$0.59 \pm0.21_{\rm stat} \pm0.10_{\rm syst}$  
&$0.698 \pm0.21_{\rm stat} \pm0.10_{\rm syst}$ \\
CLEO  ~\hfill\cite{cleo:Dss}         &$0.35 \pm0.09_{\rm stat} \pm0.06_{\rm syst}$ 
 &$0.373 \pm0.085_{\rm stat} \pm0.057_{\rm syst}$ \\
D0  ~\hfill\cite{D0:Dss}         &$0.22 \pm0.02_{\rm stat} \pm0.04_{\rm syst}$  
&$0.219 \pm0.018_{\rm stat} \pm0.035_{\rm syst}$ \\
\belle Tagged $B^-$ ~\hfill\cite{Live:Dss}           &$0.44 \pm0.07_{\rm stat} \pm0.06_{\rm syst}$  
&$0.42 \pm0.07_{\rm stat} \pm0.07_{\rm syst}$ \\
\belle Tagged $B^0$ ~\hfill\cite{Live:Dss}           &$0.60 \pm0.20_{\rm stat} \pm0.08_{\rm syst}$  
&$0.42 \pm0.07_{\rm stat} \pm0.07_{\rm syst}$ \\ 
\babar Tagged ~\hfill\cite{Aubert:2009_4}           &$0.28 \pm0.03_{\rm stat} \pm0.03_{\rm syst}$
&$0.29 \pm0.03_{\rm stat} \pm0.03_{\rm syst}$ \\
\babar Untagged $B^-$ ~\hfill\cite{Aubert:2009_5}           &$0.29 \pm0.02_{\rm stat} \pm0.02_{\rm syst}$
&$0.30 \pm0.02_{\rm stat} \pm0.02_{\rm syst}$ \\
\babar Untagged $B^0$ ~\hfill\cite{Aubert:2009_5}           &$0.30 \pm0.03_{\rm stat} \pm0.03_{\rm syst}$
&$0.30 \pm0.02_{\rm stat} \pm0.02_{\rm syst}$ \\
\hline
{\bf Average}                              &\mathversion{bold}$0.285 \pm0.018$ 
    &\mathversion{bold}$\chi^2/\dof = 11.0/8$ (CL=$13.3\%$)  \\
\hline 
\end{tabular}
}
\end{center}
\label{tab:dss1lnu}
\end{table}

\begin{table}[!htb]
\caption{Average of the branching fraction $\cbf(B^- \to D_2^0(D^{*+}\pi^-)\ell^-\bar{\nu}_{\ell})
\times \cbf(D_2^0 \to D^{*+}\pi^-))$ and individual
results. }
\begin{center}
\resizebox{0.99\textwidth}{!}{
\begin{tabular}{|l|c|c|}\hline
Experiment                                 &$\cbf(B^- \to D_2^0(D^{*+}\pi^-)\ell^-\bar{\nu}_{\ell})
) [\%]$  &$\cbf(B^- \to D_2^0(D^{*+}\pi^-)\ell^-\bar{\nu}_{\ell})
) [\%]$  \\
                                                & (rescaled) & (published) \\
\hline\hline 
CLEO  ~\hfill\cite{cleo:Dss}         &$0.055 \pm0.07_{\rm stat} \pm0.01_{\rm syst}$ 
 &$0.059 \pm0.066_{\rm stat} \pm0.011_{\rm syst}$ \\
D0  ~\hfill\cite{D0:Dss}         &$0.088 \pm0.018_{\rm stat} \pm0.020_{\rm syst}$  
&$0.088 \pm0.018_{\rm stat} \pm0.020_{\rm syst}$ \\
\belle  ~\hfill\cite{Live:Dss}           &$0.187 \pm0.060_{\rm stat} \pm0.025_{\rm syst}$  
&$0.18 \pm0.06_{\rm stat} \pm0.03_{\rm syst}$ \\
\babar Tagged ~\hfill\cite{Aubert:2009_4}           &$0.068 \pm0.009_{\rm stat} \pm0.016_{\rm syst}$
&$0.068 \pm0.009_{\rm stat} \pm0.016_{\rm syst}$ \\
\babar Untagged $B^-$ ~\hfill\cite{Aubert:2009_5}           &$0.089 \pm0.009_{\rm stat} \pm0.007_{\rm syst}$
&$0.087 \pm0.013_{\rm stat} \pm0.007_{\rm syst}$ \\
\babar Untagged $B^0$ ~\hfill\cite{Aubert:2009_5}           &$0.066 \pm0.010_{\rm stat} \pm0.006_{\rm syst}$
&$0.087 \pm0.013_{\rm stat} \pm0.007_{\rm syst}$ \\
\hline
{\bf Average}                              &\mathversion{bold}$0.074 \pm0.007$ 
    &\mathversion{bold}$\chi^2/\dof = 7.3/5$ (CL=$20\%$)  \\
\hline 
\end{tabular}
}
\end{center}
\label{tab:dss2lnu}
\end{table}

\begin{table}[!htb]
\caption{Average of the branching fraction $\cbf(B^- \to D_1^{'0}(D^{*+}\pi^-)\ell^-\bar{\nu}_{\ell})
\times \cbf(D_1^{'0} \to D^{*+}\pi^-))$ and individual
results. }
\begin{center}
\begin{tabular}{|l|c|c|}\hline
Experiment                                 &$\cbf(B^- \to D_1^{'0}(D^{*+}\pi^-)\ell^-\bar{\nu}_{\ell})
) [\%]$  &$\cbf(B^- \to D_1^{'0}(D^{*+}\pi^-)\ell^-\bar{\nu}_{\ell})
) [\%]$  \\
                                                & (rescaled) & (published) \\
\hline\hline 
DELPHI ~\hfill\cite{Abdallah:2005cx}        &$0.74 \pm0.17_{\rm stat} \pm0.18_{\rm syst}$ 
 &$0.83 \pm0.17_{\rm stat} \pm0.18_{\rm syst}$ \\
\belle  ~\hfill\cite{Live:Dss}           &$-0.03 \pm0.06_{\rm stat} \pm0.07_{\rm syst}$  
&$-0.03 \pm0.06_{\rm stat} \pm0.07_{\rm syst}$ \\
\babar  ~\hfill\cite{Aubert:2009_4}           &$0.27 \pm0.04_{\rm stat} \pm0.04_{\rm syst}$
&$0.27 \pm0.04_{\rm stat} \pm0.05_{\rm syst}$ \\
\hline
{\bf Average}                              &\mathversion{bold}$0.13 \pm0.04$ 
    &\mathversion{bold}$\chi^2/\dof = 18./2$ (CL=$0.001\%$)  \\
\hline 
\end{tabular}
\end{center}
\label{tab:dss1plnu}
\end{table}

\begin{table}[!htb]
\caption{Average of the branching fraction $\cbf(B^- \to D_0^{*0}(D^{+}\pi^-)\ell^-\bar{\nu}_{\ell})
\times \cbf(D_0^{*0} \to D^{+}\pi^-))$ and individual
results. }
\begin{center}
\begin{tabular}{|l|c|c|}\hline
Experiment                                 &$\cbf(B^- \to D_0^{*0}(D^{+}\pi^-)\ell^-\bar{\nu}_{\ell})
) [\%]$  &$\cbf(B^- \to D_0^{*0}(D^{+}\pi^-)\ell^-\bar{\nu}_{\ell})
) [\%]$ \\
						& (rescaled) & (published) \\
\hline\hline 
\belle Tagged $B^-$ ~\hfill\cite{Live:Dss}           &$0.25 \pm0.04_{\rm stat} \pm0.06_{\rm syst}$  
&$0.24 \pm0.04_{\rm stat} \pm0.06_{\rm syst}$ \\
\belle Tagged $B^0$ ~\hfill\cite{Live:Dss}           &$0.23 \pm0.08_{\rm stat} \pm0.06_{\rm syst}$  
&$0.24 \pm0.04_{\rm stat} \pm0.06_{\rm syst}$ \\
\babar Tagged ~\hfill\cite{Aubert:2009_4}            &$0.32 \pm0.04_{\rm stat} \pm0.05_{\rm syst}$
&$0.26 \pm0.05_{\rm stat} \pm0.04_{\rm syst}$ \\
\hline
{\bf Average}                              &\mathversion{bold}$0.29 \pm0.05$ 
    &\mathversion{bold}$\chi^2/\dof = 0.83/2$ (CL=$66\%$)  \\
\hline 
\end{tabular}
\end{center}
\label{tab:dss0lnu}
\end{table}

%
\subsection{Inclusive CKM-favored decays}
\label{slbdecays_b2cincl}

\subsubsection{Global analysis of $\bar B\to X_c\ell^-\bar\nu_\ell$}

The semileptonic width $\Gamma(\bar B\to X_c\ell^-\bar\nu_\ell)$ has
been calculated in the framework of the Operator Product
Expansion. The result is a double-expansion in $\Lambda_{\rm QCD}/m_b$
and $\alpha_s$, which depends on a number of non-perturbative
parameters. These parameters can be measured using other observables
in $\bar B\to X_c\ell^-\bar\nu_\ell$ decays, such as the moments of
the lepton energy and the hadronic mass spectrum.

Two independent sets of theoretical expressions, referred to as
kinetic~\cite{Benson:2003kp,Gambino:2004qm,Gambino:2011cq} and 1S
schemes~\cite{Bauer:2004ve} are available for this kind of
analysis. The non-perturbative parameters in the kinetic scheme
are: the quark masses $m_b$ and $m_c$, $\mu^2_\pi$ and
$\mu^2_G$ at $O(1/m^2_b)$, and $\rho^3_D$ and $\rho^3_{LS}$ at
$O(1/m^3_b)$. In the 1S scheme, the parameters are: $m_b$, $\lambda_1$
at $O(1/m^2_b)$, and $\rho_1$, $\tau_1$, $\tau_2$ and $\tau_3$ at
$O(1/m^3_b)$. Note that due to the different definitions, the results
for the quark masses cannot be compared directly between the two
schemes.

Our analysis uses all available measurements of moments in $\bar B\to
X_c\ell^-\bar\nu_\ell$, excluding only points with too high
correlation to avoid numerical issues. The list of included
measurements is given in
Table~\ref{tab:gf_input}. The only external input is the average
lifetime~$\tau_B$ of neutral and charged $B$~mesons, taken to be
$(1.582\pm 0.007)$~ps (Sect.~\ref{sec:life_mix}).
\begin{table}[!htb]
\caption{Experimental inputs used in the global analysis of $\bar B\to
  X_c\ell^-\bar\nu_\ell$. $n$ is the order of the moment, $c$ is the
  threshold value in GeV. In total, there are 29 measurements from
  \babar, 25 measurements from Belle and 12 from other
  experiments.} \label{tab:gf_input}
\begin{center}
\resizebox{0.99\textwidth}{!}{
\begin{tabular}{|l|l|l|l|}
  \hline
  Experiment
  & Hadron moments $\langle M^n_X\rangle$
  & Lepton moments $\langle E^n_\ell\rangle$
  & Photons moment $\langle E^n_\gamma\rangle$\\
  \hline \hline
  \babar & $n=2$, $c=0.9,1.1,1.3,1.5$ & $n=0$, $c=0.6,1.2,1.5$ & $n=1$,
  $c=1.9,2.0$\\
  & $n=4$, $c=0.8,1.0,1.2,1.4$ & $n=1$, $c=0.6,0.8,1.0,1.2,1.5$ & $n=2$,
  $c=1.9$~\cite{Aubert:2005cua,Aubert:2006gg}\\
  & $n=6$, $c=0.9,1.3$~\cite{Aubert:2009qda} & $n=2$, $c=0.6,1.0,1.5$
  & \\
  & & $n=3$, $c=0.8,1.2$~\cite{Aubert:2009qda,Aubert:2004td} & \\
  \hline
  Belle & $n=2$, $c=0.7,1.1,1.3,1.5$ & $n=0$, $c=0.6,1.0,1.4$ & $n=1$,
  $c=1.8,1.9$\\
  & $n=4$, $c=0.7,0.9,1.3$~\cite{Schwanda:2006nf} & $n=1$,
  $c=0.6,0.8,1.0,1.2,1.4$ & $n=2$, $c=1.8,2.0$~\cite{Limosani:2009qg}\\
  & & $n=2$, $c=0.6,1.0,1.4$ & \\
  & & $n=3$, $c=0.8,1.0, 1.2$~\cite{Urquijo:2006wd} & \\
  \hline
  CDF & $n=2$, $c=0.7$ & & \\
  & $n=4$, $c=0.7$~\cite{Acosta:2005qh} & & \\
  \hline
  CLEO & $n=2$, $c=1.0,1.5$ & & $n=1$, $c=2.0$~\cite{Chen:2001fja}\\
  & $n=4$, $c=1.0,1.5$~\cite{Csorna:2004kp} & & \\
  \hline
  DELPHI & $n=2$, $c=0.0$ & $n=1$, $c=0.0$ & \\
  & $n=4$, $c=0.0$~\cite{Abdallah:2005cx} & $n=2$, $c=0.0$ & \\
  & & $n=3$, $c=0.0$~\cite{Abdallah:2005cx} & \\
  \hline
\end{tabular}
}
\end{center}
\end{table}

Both in the kinetic and 1S scheme, the moments in $\bar B\to
X_c\ell^-\bar\nu_\ell$ are not sufficient to constrain the $b$-quark
mass precisely, which limits the precision of the determination of
$\vcb$. This limitation can be overcome:
\begin{itemize}
  \item by including the photon energy moments in $B\to X_s\gamma$
    into the fit, or
  \item by applying a precise constraint on the $c$-quark mass.
\end{itemize}
For the former, calculations of the $B\to X_s\gamma$~moments are
available both in the kinetic~\cite{Benson:2004sg} and the
1S~scheme~\cite{Bauer:2004ve}. For the latter, we use the $c$-quark
mass calculated in Ref.~\cite{Dehnadi:2011gc} in the kinetic scheme
analysis,
\begin{equation}
  m_c^{\overline{\rm MS}}(3~{\rm GeV})=(0.998\pm 0.029)~{\rm GeV}~.
\end{equation}

\subsubsection{Analysis in the kinetic scheme}
\label{globalfitsKinetic}

The fit relies on the calculations of the spectral moments in $\bar
B\to X_c\ell^-\bar\nu_\ell$~decays described in
Ref.~\cite{Gambino:2011cq}. The
photon energy moments are calculated in
Ref.~\cite{Benson:2004sg}. The theoretical uncertainties and
correlations are estimated as explained in
Ref.~\cite{Schwanda:2008kw}. Namely, we assume 100\% correlation
between calculations of the same moment at different threshold values
and no theory correlation between different moments. The fit
determines $\vcb$ and the 6 non-perturbative parameters mentioned
above.

\begin{figure}
\begin{center}
  \includegraphics[width=0.45\columnwidth]{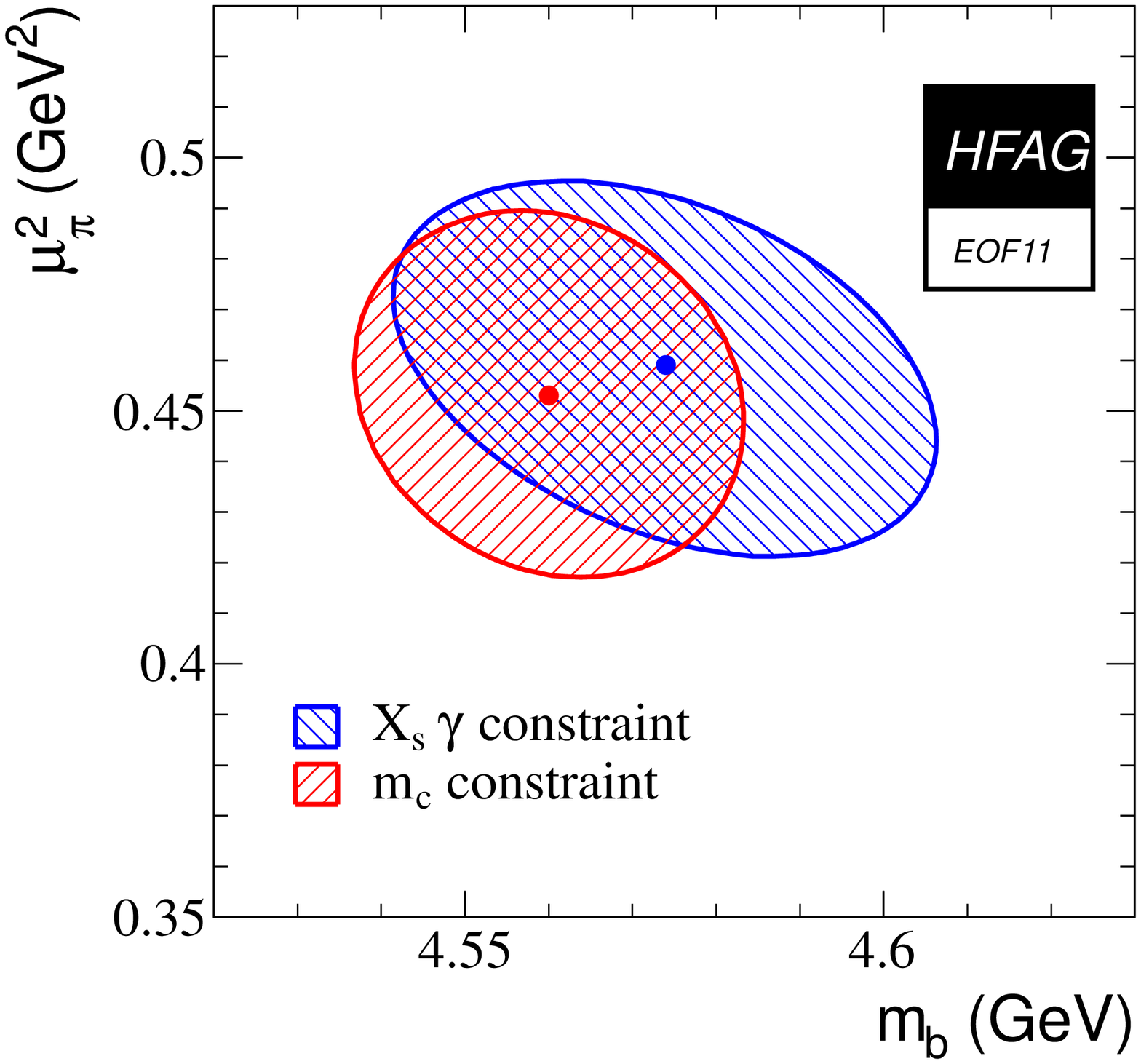}
  \includegraphics[width=0.45\columnwidth]{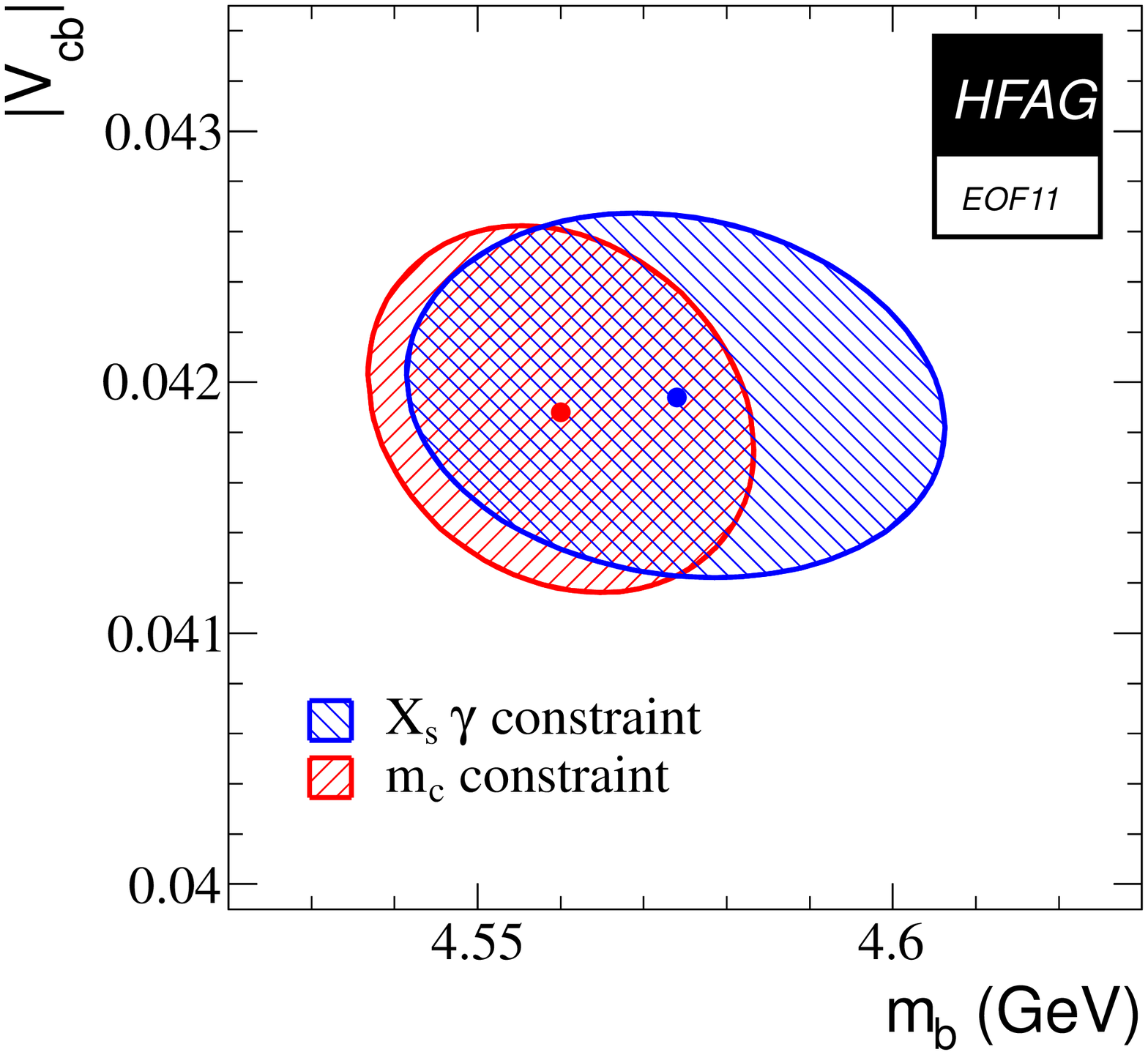}
\end{center}
\caption{$\Delta\chi^2=1$~contours of the fit result in the kinetic mass
  scheme.} \label{fig:gf_res_kin}
\end{figure}

The result of the fit using the $c$-quark mass constraint is
\begin{eqnarray}
  \vcb & = & (41.88\pm 0.73)\times 10^{-3}~, \\
  m_b^{\rm kin} & = & 4.560\pm 0.023~{\rm GeV}~, \\
  \mu^2_\pi & = & 0.453\pm 0.036~{\rm GeV^2}~,
\end{eqnarray}
with a $\chi^2$ of 33.4 for $55-7$ degrees of freedom. The detailed
result of the fit is given in Table~\ref{tab:gf_res_mc_kin}. This
result is also consistent with the fit using the $B\to
X_s\gamma$~constraint (Table~\ref{tab:gf_res_kin}). An illustration of
the fit is given in Fig.~\ref{fig:gf_res_kin}.
\begin{table}[!htb]
\caption{Fit result in the kinetic scheme, using a precise $c$-quark
  mass constraint. The error matrix of the fit ($\sigma_{\rm fit}$) contains
  experimental and theoretical contributions. The expression for
  calculating $\vcb$ has an additional uncertainty of 1.4\%
  ($\sigma_{\rm th}$). In the lower part of the table, the
  correlation matrix of the parameters is given.} \label{tab:gf_res_mc_kin}
\begin{center}
\resizebox{0.99\textwidth}{!}{
\begin{tabular}{|l|ccccccc|}
  \hline
  & \vcb\ [10$^{-3}$] & $m_b^{\rm kin}$ [GeV] &
  $m_c^{\overline{\rm MS}}$ [GeV] & $\mu^2_\pi$ [GeV$^2$]
  & $\rho^3_D$ [GeV$^3$] & $\mu^2_G$ [GeV$^2$] & $\rho^3_{LS}$ [GeV$^3$]\\
  \hline \hline
  value & 41.88 & \phantom{$-$}4.560 & \phantom{$-$}1.010 &
  \phantom{$-$}0.453 & \phantom{$-$}0.164 & \phantom{$-$}0.229 &
  $-$0.140\\
  $\sigma_{\rm fit}$ & 0.44 & \phantom{$-$}0.023 &
  \phantom{$-$}0.027 & \phantom{$-$}0.036 & \phantom{$-$}0.020 &
  \phantom{$-$}0.043 & \phantom{$-$}0.086\\
  $\sigma_{\rm th}$ & 0.59 & & & & & & \\
  \hline
  $|V_{cb}|$ & 1.000 & $-$0.164 & \phantom{$-$}0.137 &
  \phantom{$-$}0.089 & \phantom{$-$}0.328 & $-$0.324 &
  \phantom{$-$}0.146\\
  $m_b^{\rm kin}$ & & \phantom{$-$}1.000 & \phantom{$-$}0.745 &
  $-$0.117 & $-$0.177 & \phantom{$-$}0.128 & $-$0.179\\
  $m_c^{\overline{\rm MS}}$ & & & \phantom{$-$}1.000
  & $-$0.199 & $-$0.006 & $-$0.433 & \phantom{$-$}0.258\\
  $\mu^2_\pi$ & & & & \phantom{$-$}1.000 & \phantom{$-$}0.335 &
  $-$0.109 & $-$0.078\\
  $\rho^3_D$ & & & & & \phantom{$-$}1.000 & $-$0.308 & $-$0.238\\
  $\mu^2_G$ & & & & & & \phantom{$-$}1.000 & $-$0.323\\
  $\rho^3_{LS}$ & & & & & & & \phantom{$-$}1.000\\
  \hline
\end{tabular}
}
\end{center}
\end{table}
\begin{table}[!htb]
\caption{Fit result in the kinetic scheme for different
  constraints. Refer to the text for more
  details.} \label{tab:gf_res_kin}
\begin{center}
\begin{tabular}{|c|c|c|c|c|}
  \hline
  Constraint & $\vcb$ [10$^{-3}$] & $m_b^{\rm kin}$ [GeV] &
  $\mu^2_\pi$ [GeV$^2$] & $\chi^2/$d.o.f.\\
  \hline \hline
  $B\to X_s\gamma$ & $41.94\pm 0.43_{\rm fit}\pm 0.59_{\rm th}$ &
  $4.574\pm 0.032$ & $0.459\pm 0.037$ & $27.0/(66-7)$\\
  $m_c^{\overline{\rm MS}}(3~{\rm GeV})$ & $41.88\pm 0.44_{\rm fit}\pm
  0.59_{\rm th}$ & $4.560\pm 0.023$ & $0.453\pm 0.036$ &
  $33.4/(55-7)$\\
  \hline
\end{tabular}
\end{center}
\end{table}

The fit using the $c$-quark mass constraint yields a $\bar B\to
X_c\ell^-\bar\nu_\ell$ branching fraction of
\begin{equation}
  \cbf(\bar B\to X_c\ell^-\bar\nu_\ell)=(10.51\pm 0.13)\%~.
\end{equation}
Correcting for charmless semileptonic decays
(Sect.~\ref{slbdecays_b2uincl}), $\cbf(\bar B\to
X_u\ell^-\bar\nu_\ell)=(2.08\pm 0.30)\times 10^{-3}$, we obtain the
semileptonic branching fraction,
\begin{equation}
  \cbf(\bar B\to X\ell^-\bar\nu_\ell)=(10.72\pm 0.13)\%~.
\end{equation}

\subsubsection{Analysis in the 1S scheme}
\label{globalfits1S}

The fit relies on the calculations of the spectral moments described in
Ref.~\cite{Bauer:2004ve}. The theoretical uncertainties are estimated
as explained in Ref.~\cite{Schwanda:2008kw}. Only trivial theory
correlations, {\it i.e.}, between the same moment at the same
threshold are included in the analysis. The fit determines $\vcb$ and
the 6 non-perturbative parameters mentioned above.

The result of the fit using the $B\to X_s\gamma$ constraint is
\begin{eqnarray}
  \vcb & = & (41.96\pm 0.45)\times 10^{-3}~, \\
  m_b^{1S} & = & 4.691\pm 0.037~{\rm GeV}~, \\
  \lambda_1 & = & -0.362\pm 0.067~{\rm GeV^2}~,
\end{eqnarray}
with a $\chi^2$ of 23.0 for $66-7$ degrees of freedom. The detailed
result of the fit is given in Table~\ref{tab:gf_res_xsgamma_1s}. This
result is consistent with the fit using the $\bar B\to
X_c\ell^-\bar\nu_\ell$~data only (Table~\ref{tab:gf_res_1s}).
\begin{table}[!htb]
\caption{Fit result in the 1S scheme, using $B\to X_s\gamma$~moments
  as a constraint. In the lower part of the table, the correlation
  matrix of the parameters is given.} \label{tab:gf_res_xsgamma_1s}
\begin{center}
\begin{tabular}{|l|ccccccc|}
  \hline
  & $m_b^{1S}$ [GeV] & $\lambda_1$ [GeV$^2$] & $\rho_1$ [GeV$^3$] &
  $\tau_1$ [GeV$^3$] & $\tau_2$ [GeV$^3$] & $\tau_3$ [GeV$^3$] &
  $\vcb$ [10$^{-3}$]\\
  \hline \hline
  value & 4.691 & $-0.362$ & \phantom{$-$}0.043 &
  \phantom{$-$}0.161 & $-0.017$ & \phantom{$-$}0.213 &
  \phantom{$-$}41.96\\
  error & 0.037 & \phantom{$-$}0.067 & \phantom{$-$}0.048 &
  \phantom{$-$}0.122 & \phantom{$-$}0.062 & \phantom{$-$}0.102 &
  \phantom{$-$}0.45\\
  \hline
  $m_b^{1S}$ & 1.000 & \phantom{$-$}0.434 & \phantom{$-$}0.213 &
  $-0.058$ & $-0.629$ & $-0.019$ & $-0.215$\\
  $\lambda_1$ & & \phantom{$-$}1.000 & $-0.467$ & $-0.602$ & $-0.239$
  & $-0.547$ & $-0.403$\\
  $\rho_1$ & & & \phantom{$-$}1.000 & \phantom{$-$}0.129 & $-0.624$ &
  \phantom{$-$}0.494 & \phantom{$-$}0.286\\
  $\tau_1$ & & & & \phantom{$-$}1.000 & \phantom{$-$}0.062 & $-0.148$ &
  \phantom{$-$}0.194\\
  $\tau_2$ & & & & & \phantom{$-$}1.000 & $-0.009$ & $-0.145$\\
  $\tau_3$ & & & & & & \phantom{$-$}1.000 & \phantom{$-$}0.376\\
  $\vcb$ & & & & & & & \phantom{$-$}1.000\\
  \hline
\end{tabular}
\end{center}
\end{table}
\begin{table}[!htb]
\caption{Fit result in the 1S scheme for different data
  sets.} \label{tab:gf_res_1s}
\begin{center}
\begin{tabular}{|c|c|c|c|c|}
  \hline
  Data & $\vcb$ [10$^{-3}$] & $m_b^{1S}$ [GeV] &
  $\lambda_1$ [GeV$^2$] & $\chi^2/$d.o.f.\\
  \hline \hline
  $X_c\ell\nu$ and $X_s\gamma$ & $41.96\pm 0.45$ & $4.691\pm 0.037$ &
  $-0.362\pm 0.067$ & $23.0/(66-7)$\\
  $X_c\ell\nu$ only & $42.37\pm 0.65$ & $4.622\pm 0.085$ & $-0.412\pm
  0.084$ & $13.7/(55-7)$\\
  \hline
\end{tabular}
\end{center}
\end{table}

\subsection{Exclusive CKM-suppressed decays}
\label{slbdecays_b2uexcl}
In this section, we list results on exclusive charmless semileptonic branching fractions
and determinations of $\vub$ based on $\Bb\to\pi\ell\nub$ decays.
The measurements are based on two different event selections: tagged
events, in which case the second $B$ meson in the event is fully
reconstructed in either a hadronic decay (``$B_{reco}$'') or in a 
CKM-favored semileptonic decay (``SL''); and untagged events, in which case the selection infers the momentum
of the undetected neutrino based on measurements of the total 
momentum sum of detected particles and knowledge of the initial state.
We present averages for $\Bb\to\rho\ell\nub$ and $\Bb\to\omega\ell\nub$. Moreover, the average for the  branching fraction $\Bb\to\eta\ell\nub$ is presented for the first time. 

\begin{figure}[!ht]
 \begin{center}
  \unitlength1.0cm 
  \begin{picture}(14.,8.0)  
   \put(  8.0,  0.0){\includegraphics[width=8.0cm]{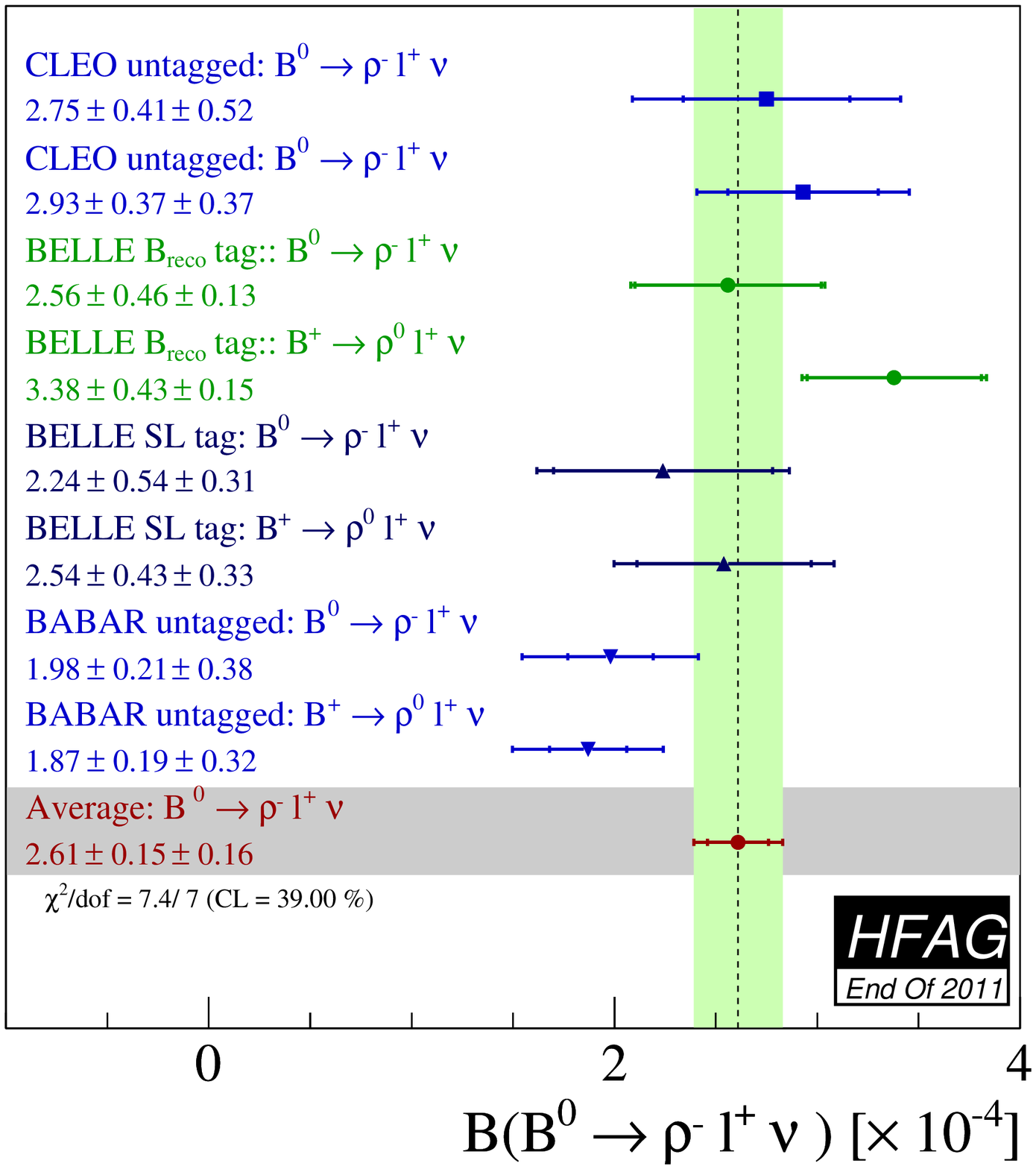}}
   \put( -0.5,  0.0){\includegraphics[width=8.0cm]{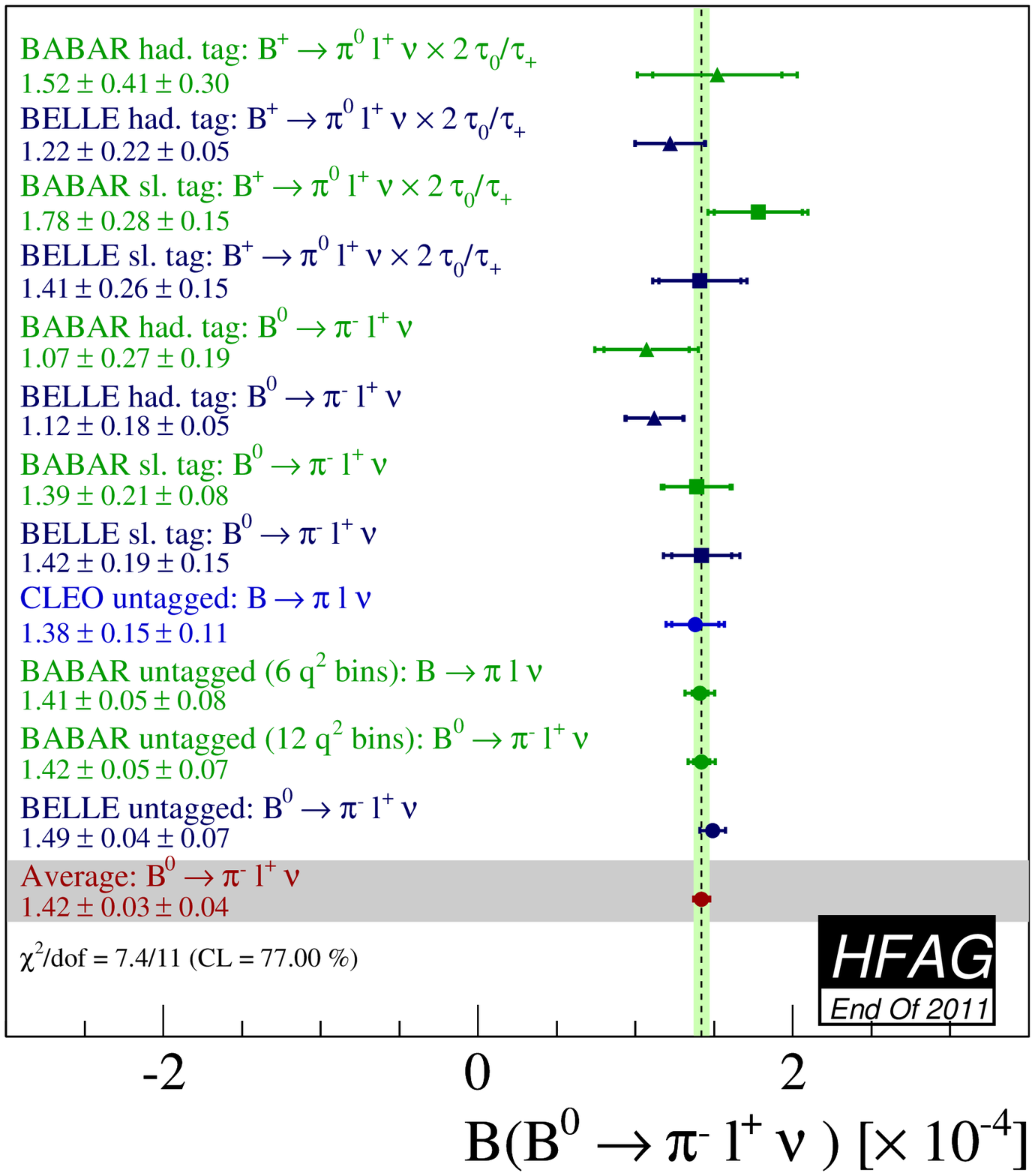}}
   \put(  5.5,  7.3){{\large\bf a)}}  
   \put( 14.4,  7.3){{\large\bf b)}}
   \end{picture} \caption{
(a) Summary of exclusive determinations of $\cbf(\Bb\to\pi
\ell\nub)$ and their average.
Measured branching fractions for $B\rightarrow \pi^0 l \nu$ have been
multiplied by $2\times \tau_{B^0}/\tau_{B^+}$ in accordance with
isospin symmetry. The labels ``$B_{reco}$'' and ``SL''
refer to type of $B$ decay tag used in a measurement. ``untagged'' refers to an untagged measurement.
(b) Summary of exclusive determinations of $\cbf(\Bb\to\rho\ell\nub)$ and their average.
}
\label{fig:xlnu}
\end{center}
\end{figure}

The results for the full and partial branching fraction for $\Bb\to\pi\ell\nub$ are given
in Table~\ref{tab:pilnubf} and shown in Figure~\ref{fig:xlnu} (a).   

When averaging these results, systematic uncertainties due to external
inputs, e.g., form factor shapes and background estimates from the
modeling of $\Bb\to X_c\ell\nub$ and $\Bb\to X_u\ell\nub$ decays, are
treated as fully correlated (in the sense of Eq.~\ref{eq:correlrho}).
Uncertainties due to experimental reconstruction effects are treated
as fully correlated among measurements from a given experiment.  Varying
the assumed dependence of the quoted errors on the measured value
for error sources where the dependence was not obvious had no significant impact.

\begin{sidewaystable}[!htb]
\begin{center}
\caption{\label{tab:pilnubf}
Summary of exclusive determinations of $\cbf(\Bb\to\pi
\ell\nub)$. The errors quoted
correspond to statistical and systematic uncertainties, respectively.
Measured branching fractions for $B\rightarrow \pi^0 l \nu$ have been
multiplied by $2\times \tau_{B^0}/\tau_{B^+}$ in accordance with
isospin symmetry. The labels ``$B_{reco}$'' and ``SL'' tags refer to
the type of $B$
decay tag used in a measurement, and ``untagged'' refers to an untagged measurement.}
\begin{small}
\begin{tabular}{|lcccc|}
\hline
& $\cbf [10^{-4}]$
& $\cbf(q^2<12\,\gev^2/c^2) [10^{-4}]$
& $\cbf(q^2<16\,\gev^2/c^2) [10^{-4}]$
& $\cbf(q^2>16\,\gev^2/c^2) [10^{-4}]$
\\
\hline\hline
CLEO $\pi^+,\pi^0$~\cite{Adam:2007pv}
& $1.38\pm 0.15\pm 0.11\ $ 
& $0.70\pm 0.12\pm 0.07$
& $0.97\pm 0.13\pm 0.09$
& $0.41\pm 0.08\pm 0.04$
\\ 
\babar $\pi^+,\pi^0$~\cite{delAmoSanchez:2010af}
& $1.41\pm 0.05\pm 0.08\ $
& $0.88\pm 0.04\pm 0.05$
& $1.10\pm 0.04\pm 0.06$
& $0.32\pm 0.02\pm 0.03$
\\  
\babar $\pi^+$~\cite{delAmoSanchez:2010zd}
& $1.42\pm 0.05\pm 0.07\ $
& $0.83\pm 0.03\pm 0.04$
& $1.09\pm 0.04\pm 0.05$
& $0.33\pm 0.03\pm 0.03$
\\  
Belle $\pi^+$~\cite{Ha:2010rf}
& $1.49\pm 0.04\pm 0.07\ $
& $0.83\pm 0.03\pm 0.04$
& $1.10\pm 0.03\pm 0.05$
& $0.40\pm 0.02\pm 0.02$
\\  
Belle SL $\pi^+$~\cite{Hokuue:2006nr}
& $1.42\pm 0.19\pm 0.15\ $
& $0.80\pm 0.14\pm 0.08$
& $1.04\pm 0.16\pm 0.11$
& $0.37\pm 0.10\pm 0.04$
\\ 
Belle SL $\pi^0$~\cite{Hokuue:2006nr}
& $1.41\pm 0.26\pm 0.15\ $
& $0.71\pm 0.17\pm 0.08$
& $1.04\pm 0.22\pm 0.12$
& $0.36\pm 0.15\pm 0.04$
\\ 
\babar SL $\pi^+$~\cite{Aubert:2008bf}
& $1.39\pm 0.21\pm 0.08\ $
& $0.77\pm 0.14\pm 0.05$
& $0.92\pm 0.16\pm 0.05$
& $0.46\pm 0.13\pm 0.03$
\\ 
\babar SL $\pi^0$~\cite{Aubert:2008bf}
& $1.78\pm 0.28\pm 0.15\ $
& $1.07\pm 0.20\pm 0.09$
& $1.34\pm 0.22\pm 0.11$
& $0.44\pm 0.17\pm 0.06$
\\ 
\babar $B_{reco}$ $\pi^+$~\cite{Aubert:2006ry}
& $1.07\pm 0.27\pm 0.19\ $
& $0.26\pm 0.15\pm 0.04$
& $0.42\pm 0.18\pm 0.06$
& $0.65\pm 0.20\pm 0.13$
\\ 
\babar $B_{reco}$ $\pi^0$~\cite{Aubert:2006ry}
& $1.52\pm 0.41 \pm0.30\ $
& $0.67\pm 0.30\pm 0.12$
& $1.04\pm 0.35\pm 0.18$
& $0.48\pm 0.22\pm 0.12$
\\ 
Belle $B_{reco}$ $\pi^+$~\cite{:2008kn}
& $1.12\pm 0.18\pm 0.05\ $
& $0.65\pm 0.14\pm 0.03$
& $0.85\pm 0.16\pm 0.04$
& $0.26\pm 0.08\pm 0.01$
\\ 
Belle $B_{reco}$ $\pi^0$~\cite{:2008kn}
& $1.22\pm 0.22\pm 0.05\ $
& $0.65\pm 0.19\pm 0.03$
& $0.80\pm 0.19\pm 0.03$
& $0.41\pm 0.11\pm 0.02$
\\  \hline
{\bf Average}
& \mathversion{bold}$1.42\pm 0.03\pm 0.04\ $
& \mathversion{bold}$0.81\pm 0.02\pm 0.03$
& \mathversion{bold}$1.05\pm 0.02\pm 0.03$
& \mathversion{bold}$0.37\pm 0.01\pm 0.02$
\\ 
\hline
\end{tabular}\\
\end{small}
\end{center}
\end{sidewaystable}

The determination of \vub\ from $\Bb\to\pi\ell\nub$ decays is
shown in Table~\ref{tab:pilnuvub}, and uses our averages for the partial branching
fractions given in Table~\ref{tab:pilnubf}. Two theoretical approaches are
used: unquenched Lattice QCD and QCD light-cone sum rules.
Lattice calculations of the form factors are limited to small hadron momenta, i.e.
large $q^2$, while calculations based on light-cone sum rules are restricted
to small $q^2$.

\begin{table}[hbtf]
\caption{\label{tab:pilnuvub}
Determinations of \vub\ based on the average partial
$\Bb\to\pi\ell\nub$ decay branching fractions stated in
Table~\ref{tab:pilnubf}. The first
uncertainty is experimental and the second is from theory.  The
full or partial branching fractions are used as indicated. 
Acronyms for the calculations refer to either the method (LCSR) or
the collaboration working on it (HPQCD, FNAL/MILC).
}
\begin{center}
\begin{tabular}{|lc|}
\hline
Method                                                     & $\Vub [10^{-3}]$ \\\hline\hline
LCSR~1,    $q^2<12\,\gev^2/c^2$~\cite{Khodjamirian:2011ub} & $3.40\pm 0.07 {}^{+0.37}_{-0.32}$ \\ \hline
LCSR~2,    $q^2<16\,\gev^2/c^2$~\cite{Ball:2004ye}         & $3.57\pm 0.06 {}^{+0.59}_{-0.39}$ \\ \hline
HPQCD,     $q^2>16\,\gev^2/c^2$~\cite{Dalgic:2006dt}       & $3.45\pm 0.09 {}^{+0.60}_{-0.39}$ \\  \hline
FNAL/MILC, $q^2>16\,\gev^2/c^2$~\cite{Bailey:2008wp}       & $3.30\pm 0.09 {}^{+0.37}_{-0.30}$ \\  \hline
\hline
\end{tabular}
\end{center}
\end{table}

\begin{figure}[!ht]
 \begin{center}
  \unitlength1.0cm 
  \begin{picture}(14.,8.0)  
   \put( -0.0,  0.0){\includegraphics[width=8.0cm]{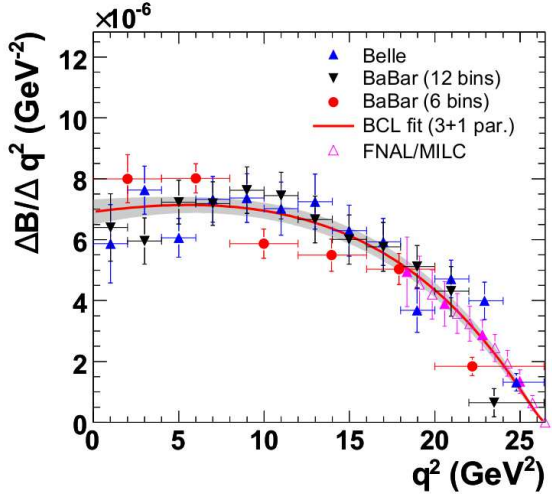}}
   \end{picture} \caption{
    Simultaneous fit of the untagged $\Bb\to\pi\ell\nub$ measurements from \babar and Belle and the
    FNAL/MILC LQCD calculations. This fit yields $\Vub = (3.23 \pm 0.30) \times 10^{-3}$.
}
\label{fig:vub_pilnu_simultaneous}
\end{center}
\end{figure}

An alternative method to determine \vub\ from $\Bb\to\pi\ell\nub$ decays that makes use
of the measurement over the full $q^2$ range is based on a simultaneous fit of the 
BCL (Bourrely, Caprini, Lellouch) form factor parameterization to the data and the LQCD predictions.
The result of the simultaneous fit to the three untagged measurements from \babar and Belle and the 
FNAL/MILC LQCD calculations is shown in Figure~\ref{fig:vub_pilnu_simultaneous}.
A value of $\Vub = (3.23 \pm 0.30) \times 10^{-3}$ is obtained.

The branching fractions for 
$\Bb\to \rho\ell\nub$ decays is computed based on the measurements in
Table~\ref{tab:rholnu} and is shown in Figure~\ref{fig:xlnu} (b). The determination of $\Vub$
from these other channels looks less promising than for
$\Bb\to\pi\ell\nub$ and at the moment it is not extracted.

\begin{table}[!htb]
\begin{center}
\caption{Summary of exclusive determinations of $\cbf(\Bb\to\rho
\ell\nub)$. The errors quoted
correspond to statistical and systematic uncertainties, respectively.}
\label{tab:rholnu}
\begin{small}
\begin{tabular}{|lc|}
\hline
& $\cbf [10^{-4}]$
\\
\hline\hline
CLEO $\rho^+$~\cite{Behrens:1999vv}
& $2.75\pm 0.41\pm 0.52\ $ 
\\ 
CLEO $\rho^+$~\cite{Adam:2007pv}
& $2.93\pm 0.37\pm 0.37\ $ 
\\ 
%
Belle $\rho^+$~\cite{:2008kn}
& $2.56\pm 0.46\pm 0.13\ $
\\
Belle $\rho^0$~\cite{:2008kn}
& $3.38\pm 0.43\pm 0.15\ $
\\
Belle $\rho^+$~\cite{Hokuue:2006nr}
& $2.24\pm 0.54\pm 0.31\ $
\\
Belle $\rho^0$~\cite{Hokuue:2006nr}
& $2.54\pm 0.43\pm 0.33\ $
\\
\babar $\rho^+$~\cite{delAmoSanchez:2010af}
& $1.98\pm 0.21\pm 0.38\ $
\\
\babar $\rho^0$~\cite{delAmoSanchez:2010af}
& $1.87\pm 0.19\pm 0.32\ $

\\  \hline
{\bf Average}
& \mathversion{bold}$2.61 \pm 0.15\pm 0.16 $
\\ 
\hline
\end{tabular}\\
\end{small}
\end{center}
\end{table}

\begin{figure}[!ht]
 \begin{center}
  \unitlength1.0cm 
  \begin{picture}(14.,8.0)  
   \put( -0.5,  0.0){\includegraphics[width=8.0cm]{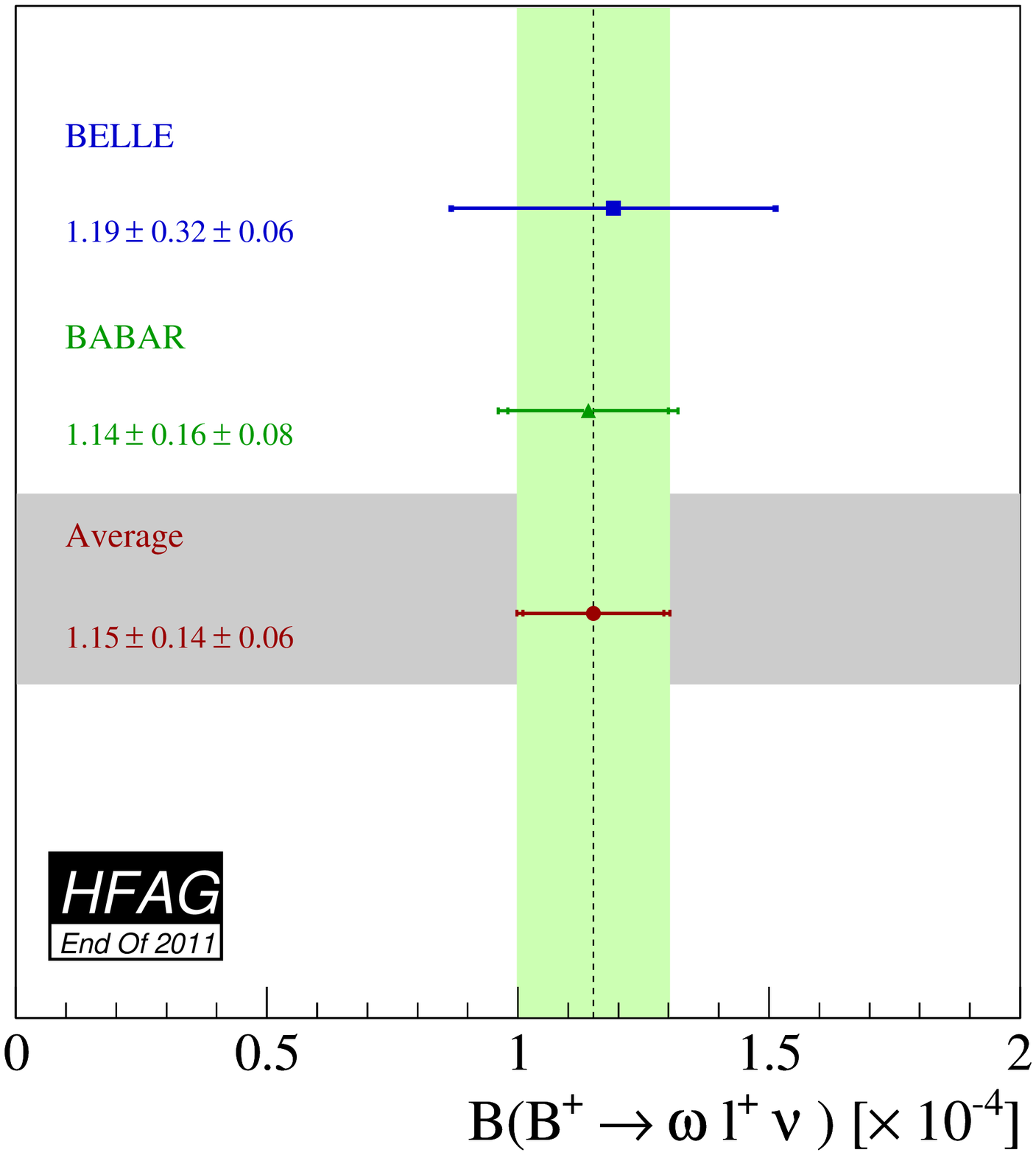}}
   \put(  5.5,  7.3){{\large\bf a)}}  
   \end{picture} \caption{
(a) Summary of exclusive determinations of $\cbf(\Bb\to\omega\ell\nub)$ and their average.
}
\label{fig:xulnu1}
\end{center}
\end{figure}

\begin{figure}[!ht]
 \begin{center}
  \unitlength1.0cm 
  \begin{picture}(14.,8.0)  
   \put( -0.5,  0.0){\includegraphics[width=8.0cm]{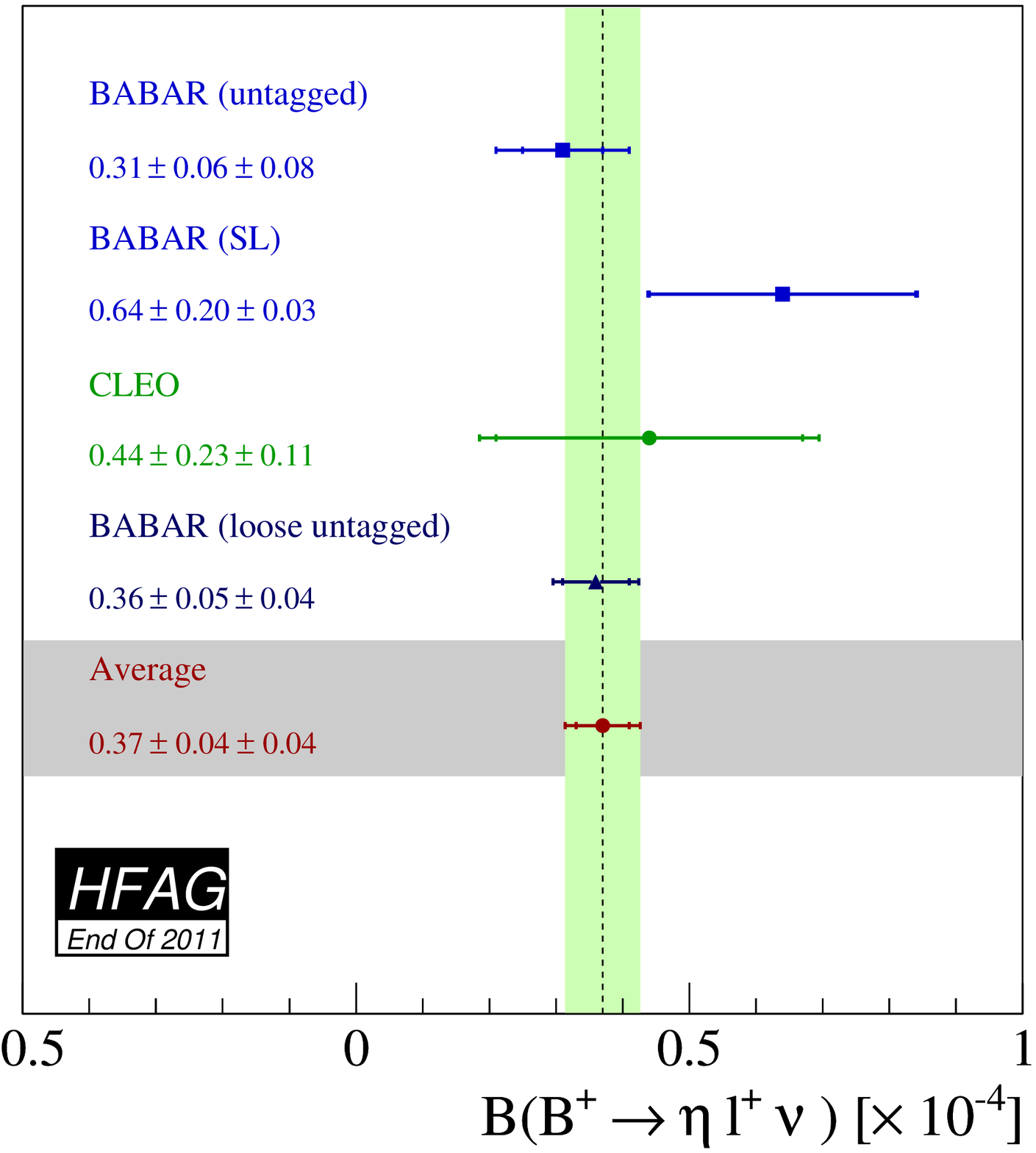}}
   \put( 8.0,  0.0){\includegraphics[width=8.0cm]{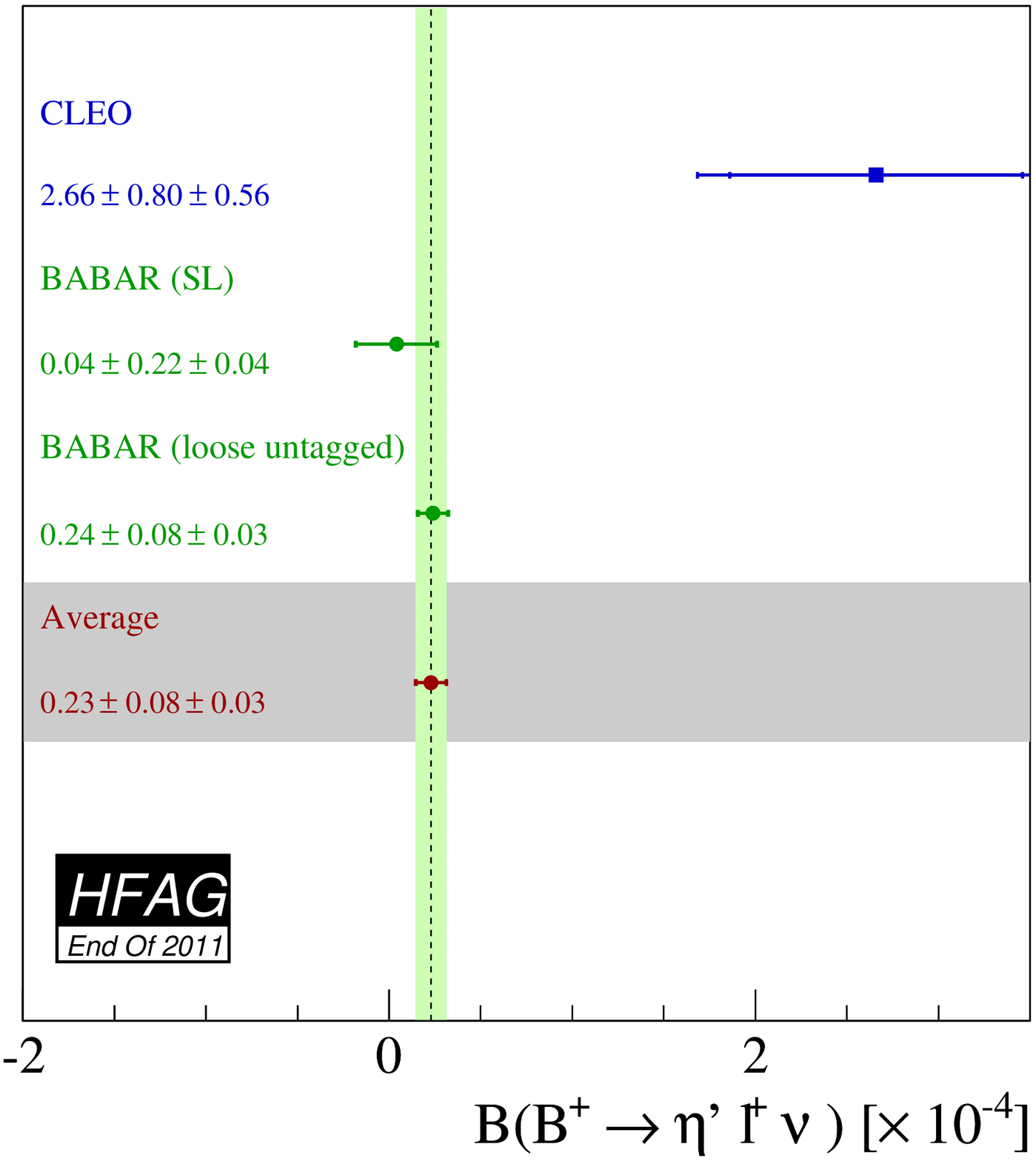}} 
   \put(  5.5,  7.3){{\large\bf a)}}     
   \put( 18.4,  7.3){{\large\bf b)}}
   
   \end{picture} \caption{
(a) Summary of exclusive determinations of $\cbf(\Bb\to\eta\ell\nub)$ and their average.
(b) Summary of exclusive determinations of $\cbf(\Bb\to\eta'\ell\nub)$ and their average.
}
\label{fig:xulnu2}
\end{center}
\end{figure}

We also report the branching fraction average for $\Bb\to\omega\ell\nub$, $\Bb\to\eta\ell\nub$ 
and $\Bb\to\eta'\ell\nub$. The measurements for $\Bb\to\omega\ell\nub$ are reported in Table~\ref{tab:omegalnu} 
and shown in Figure~\ref{fig:xulnu1}, while the ones for $\Bb\to\eta\ell\nub$ and  $\Bb\to\eta'\ell\nub$ are reported in Table~\ref{tab:etalnu} and~\ref{tab:etaprimelnu},  and are shown in Figure~\ref{fig:xulnu2}. 

\begin{table}[!htb]
\begin{center}
\caption{Summary of exclusive determinations of $\cbf(\Bb\to\omega
\ell\nub)$. The errors quoted
correspond to statistical and systematic uncertainties, respectively.}
\label{tab:omegalnu}
\begin{small}
\begin{tabular}{|lc|}
\hline
& $\cbf [10^{-4}]$
\\
\hline\hline
Belle $\omega$~\cite{:2008kn}
& $1.19\pm 0.32\pm 0.06\ $
\\
\babar $\omega$~\cite{Aubert:2008ct}
& $1.14\pm 0.16\pm 0.08\ $
\\  \hline
{\bf Average}
& \mathversion{bold}$1.15 \pm 014 \pm 0.06\ $
\\ 
\hline
\end{tabular}\\
\end{small}
\end{center}
\end{table}

\begin{table}[!htb]
\begin{center}
\caption{Summary of exclusive determinations of $\cbf(\Bb\to\eta
\ell\nub)$. The errors quoted
correspond to statistical and systematic uncertainties, respectively.}
\label{tab:etalnu}
\begin{small}
\begin{tabular}{|lc|}
\hline
& $\cbf [10^{-4}]$
\\
\hline\hline
CLEO $\eta$~\cite{Gray:2007pw}
& $0.44\pm 0.23\pm 0.11\ $
\\
BABAR $\eta$~\cite{Aubert:2008ct}
& $0.31\pm 0.06\pm 0.08\ $
\\ 
BABAR $\eta$~\cite{Aubert:2008bf}
& $0.64\pm 0.20\pm 0.03\ $
\\
BABAR $\eta$~\cite{delAmoSanchez:2010zd}
& $0.36\pm 0.05\pm 0.04\ $
\\  
 \hline
{\bf Average}
& \mathversion{bold}$0.37 \pm 0.04 \pm 0.04 $
\\ 
\hline
\end{tabular}\\
\end{small}
\end{center}
\end{table}

\begin{table}[!htb]
\begin{center}
\caption{Summary of exclusive determinations of $\cbf(\Bb\to\eta'
\ell\nub)$. The errors quoted
correspond to statistical and systematic uncertainties, respectively.}
\label{tab:etaprimelnu}
\begin{small}
\begin{tabular}{|lc|}
\hline
& $\cbf [10^{-4}]$
\\
\hline\hline
CLEO $\eta'$~\cite{Gray:2007pw}
& $2.66\pm 0.80\pm 0.56\ $
\\
BABAR $\eta'$~\cite{Aubert:2008bf}
& $0.04\pm 0.22\pm 0.04\ $
\\ 
BABAR $\eta'$~\cite{delAmoSanchez:2010zd}
& $0.24\pm 0.08\pm 0.03\ $
\\  
 \hline
{\bf Average}
& \mathversion{bold}$0.23 \pm 0.08 \pm 0.03 $
\\ 
\hline
\end{tabular}\\
\end{small}
\end{center}
\end{table}

%
\subsection{Inclusive CKM-suppressed decays}
\label{slbdecays_b2uincl}
The large background from $\B\to X_c\ell^+\nul$ decays is the chief
experimental limitation in determinations of $\vub$.  Cuts designed to
reject this background limit the acceptance for $\B\to X_u\ell^+\nul$
decays. The calculation of partial rates for these restricted
acceptances is more complicated and requires substantial theoretical machinery.
In this update, we use several theoretical calculations
to extract \vub. We do not advocate the use of one method over another.
The authors for the different calculations have provided 
codes to compute the partial rates in limited regions of phase space covered by the measurements. 
Latest results by Belle~\cite{ref:belle-multivariate} and \babar~\cite{ref:babar-finalupdate} 
explore bigger and bigger portions of phase space, with a consequent reduction of the theoretical 
uncertainties. 

For the averages we performed, the systematic errors associated with the
modeling of $\B\to X_c\ell^+\nul$ and $\B\to X_u\ell^+\nul$ decays and the theoretical
uncertainties are taken as fully correlated among all measurements.
Reconstruction-related uncertainties are taken as fully correlated within a given experiment.
We use all results published by \babar\ in~\cite{ref:babar-finalupdate}, since the 
statistical correlations are given. 
To make use of the theoretical calculations of Ref.~\cite{ref:BLL}, we restrict the
kinematic range in $M_X$ and $q^2$, thereby reducing the size of the data
sample significantly, but also the theoretical uncertainty, as stated by the
authors~\cite{ref:BLL}.
The dependence of the quoted error on the measured value for each source of error
is taken into account in the calculation of the averages.
Measurements of partial branching fractions for $\B\to X_u\ell^+\nul$
transitions from $\Upsilon(4S)$ decays, together with the corresponding accepted region, 
are given in Table~\ref{tab:BFbulnu}.  
The signal yields for all the measurements shown in Table~\ref{tab:BFbulnu}
are not rescaled to common input values of the $B$ meson lifetime (see Sect.~\ref{sec:life_mix})
and the semileptonic width~\cite{PDG_2008}.

It has been first suggested by Neubert~\cite{Neubert:1993um} and later detailed by Leibovich, 
Low, and Rothstein (LLR)~\cite{Leibovich:1999xf} and Lange, Neubert and Paz (LNP)~\cite{Lange:2005qn}, 
that the uncertainty of
the leading shape functions can be eliminated by comparing inclusive rates for
$\B\to X_u\ell^+\nul$ decays with the inclusive photon spectrum in $\B\to X_s\gamma$,
based on the assumption that the shape functions for transitions to light
quarks, $u$ or $s$, are the same to first order.
However, shape function uncertainties are only eliminated at the leading order
and they still enter via the signal models used for the determination of efficiency. 
For completeness, we provide a comparison of the results using 
calculations with reduced dependence on the shape function, as just
introduced, with our averages using different theoretical approaches.
Results are presented by \babar\ in Ref.\cite{Aubert:2006qi} using the LLR prescription. 
In another work (Ref.~\cite{Golubev:2007cs}), \vub\ was extracted from the 
endpoint spectrum of $\B\to X_u\ell^+\nul$ from \babar~\cite{ref:babar-endpoint}, 
using several theoretical approaches with reduced dependence on the shape function.
In both cases, the photon energy spectrum in the 
rest frame of the $B$-meson by \babar~\cite{Aubert:2005cua} has been used.

\begin{table}[!htb]
\caption{\label{tab:BFbulnu}
Summary of inclusive determinations of partial branching
fractions for $B\rightarrow X_u \ell^+ \nu_{\ell}$ decays.
The errors quoted on $\Delta\cbf$ correspond to
statistical and systematic uncertainties.
The $s_\mathrm{h}^{\mathrm{max}}$ variable is described in Refs.~\cite{ref:shmax,ref:babar-elq2}. }
\begin{center}
\begin{small}
\begin{tabular}{|llcl|}
\hline
Measurement & Accepted region &  $\Delta\cbf [10^{-4}]$ & Notes\\
\hline\hline
CLEO~\cite{ref:cleo-endpoint}
& $E_e>2.1\,\gev$ & $3.3\pm 0.2\pm 0.7$ &  \\ 
\babar~\cite{ref:babar-elq2}
& $E_e>2.0\,\gev$, $s_\mathrm{h}^{\mathrm{max}}<3.5\,\mathrm{GeV^2}$ & $4.4\pm 0.4\pm 0.4$ & \\
\babar~\cite{ref:babar-endpoint}
& $E_e>2.0\,\gev$  & $5.7\pm 0.4\pm 0.5$ & \\
Belle~\cite{ref:belle-endpoint}
& $E_e>1.9\,\gev$  & $8.5\pm 0.4\pm 1.5$ & \\
\babar~\cite{ref:babar-finalupdate}
& $M_X<1.7\,\gev/c^2, q^2>8\,\gev^2/c^2$ & $6.8\pm 0.6\pm 0.4$ & 
\\
Belle~\cite{ref:belle-mxq2Anneal}
& $M_X<1.7\,\gev/c^2, q^2>8\,\gev^2/c^2$ & $7.4\pm 0.9\pm 1.3$ & \\
Belle~\cite{ref:belle-mx}
& $M_X<1.7\,\gev/c^2, q^2>8\,\gev^2/c^2$ & $8.4\pm 0.8\pm 1.0$ & used only in BLL average\\
\babar~\cite{ref:babar-finalupdate}
& $P_+<0.66\,\gev$  & $9.8\pm 0.9\pm 0.8 $ & 
\\
\babar~\cite{ref:babar-finalupdate}
& $M_X<1.7\,\gev/c^2$ & $11.5\pm 1.0\pm 0.8 $ &
\\ 
\babar~\cite{ref:babar-finalupdate}
& $M_X<1.55\,\gev/c^2$ & $10.8\pm 0.8\pm 0.6 $ & 
\\ 
Belle~\cite{ref:belle-multivariate}
& $p^*_{\ell} > 1 \gev/c$ & $19.6\pm 1.7\pm 1.6$ & \\
\babar~\cite{ref:babar-finalupdate}
& ($M_X, q^2$) fit, $p^*_{\ell} > 1 \gev/c$  & $18.0\pm 1.3\pm 1.5$ & 
\\ 
\babar~\cite{ref:babar-finalupdate}
& $p^*_{\ell} > 1.3 \gev/c$  & $15.3\pm 1.3\pm 1.4$ & 
\\ \hline
\end{tabular}\\
\end{small}
\end{center}
\end{table}

\subsubsection{BLNP}
\begin{figure}
\begin{center}
\includegraphics[width=0.48\textwidth]{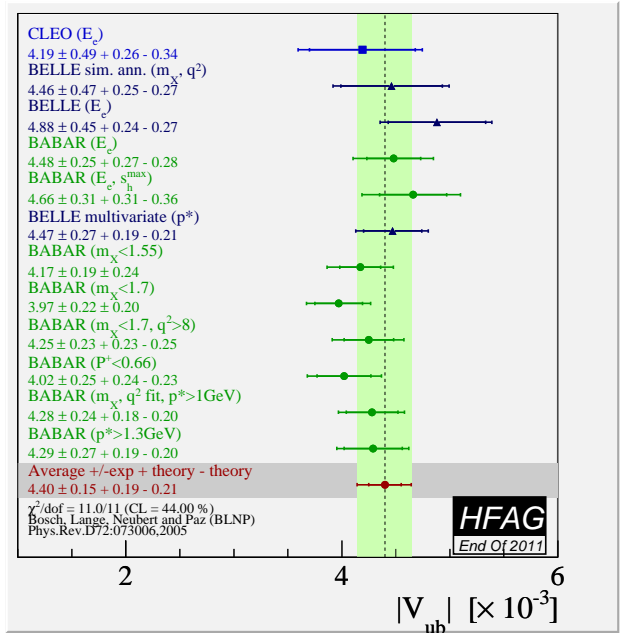}
\end{center}
\caption{Measurements of $\vub$ from inclusive semileptonic decays 
and their average based on the BLNP prescription.
``$E_e$'', ``$M_X$'', ``$(M_X,q^2)$'', ``$P^+$'', ``$p^*$ and ``($E_e,s^{max}_h$)'' indicate the 
distributions and cuts used for the measurement of the partial decay rates.}
\label{fig:BLNP}
\end{figure}

Bosch, Lange, Neubert and Paz (BLNP)~\cite{ref:BLNP,
  ref:Neubert-new-1,ref:Neubert-new-2,ref:Neubert-new-3}
provide theoretical expressions for the triple
differential decay rate for $B\to X_u \ell^+ \nul$ events, incorporating all known
contributions, whilst smoothly interpolating between the 
``shape-function region'' of large hadronic
energy and small invariant mass, and the ``OPE region'' in which all
hadronic kinematical variables scale with the $b$-quark mass. BLNP assign
uncertainties to the $b$-quark mass which enters through the leading shape function, 
to sub-leading shape function forms, to possible weak annihilation
contribution, and to matching scales. 
The BLNP calculation uses the shape function renormalization scheme; the heavy quark parameters determined  
from the global fit in the kinetic scheme, described in \ref{globalfitsKinetic}, were therefore 
translated into the shape function scheme by using a prescription by Neubert 
\cite{Neubert:2004sp,Neubert:2005nt}. The resulting parameters are 
$m_b(SF)=(4.588 \pm 0.023 \pm 0.011)$ GeV, 
$\mu_\pi^2(SF)=(0.189 \pm 0.041 ^{+0.020}_{-0.040})$ GeV$^2$, 
where the second uncertainty is due to the scheme translation. 
The extracted values of \vub\, for each measurement along with their average are given in
Table~\ref{tab:bulnu} and illustrated in Figure~\ref{fig:BLNP}. 
The total uncertainty is $^{+5.6}_{-5.9}\%$ and is due to:
statistics ($^{+2.1}_{-2.1}\%$),
detector ($^{+1.7}_{-1.8}\%$),
$B\to X_c \ell^+ \nul$ model ($^{+1.2}_{-1.2}\%$),
$B\to X_u \ell^+ \nul$ model ($^{+1.7}_{-1.6}\%$),
heavy quark parameters ($^{+2.3}_{-2.4}\%$),
SF functional form ($^{+0.3}_{-0.3}\%$),
sub-leading shape functions ($^{+0.5}_{-0.7}\%$),
BLNP theory: matching scales $\mu,\mu_i,\mu_h$ ($^{+3.7}_{-3.7}\%$), and
weak annihilation ($^{+0.0}_{-1.7}\%$).
The error on the HQE parameters ($b$-quark mass and $\mu_\pi^2)$ 
is the source of the largest uncertainty, while the
uncertainty assigned for the matching scales is a close second. The uncertainty due to 
weak annihilation has been assumed to be asymmetric, i.e. it only tends to decrease \vub.

\begin{table}[!htb]
\caption{\label{tab:bulnu}
Summary of input parameters used by the different theory calculations,
corresponding inclusive determinations of $\vub$ and their average.
The errors quoted on \vub\ correspond to
experimental and theoretical uncertainties, respectively.}
\begin{center}
\resizebox{0.99\textwidth}{!}{
\begin{tabular}{|lccccc|}
\hline
 & BLNP &DGE & GGOU & ADFR &BLL \\
\hline\hline
\multicolumn{6}{|c|}{Input parameters}\\ \hline
scheme & SF           & $\overline{MS}$ & kinetic &  $\overline{MS}$ & $1S$ \\ 
Ref.       & \cite{Neubert:2004sp,Neubert:2005nt} & Ref.~\cite{PDG_2010} & 
see Sect.~\ref{globalfitsKinetic}  & Ref.~\cite{PDG_2010} & Ref.~\cite{Barberio:2008fa} \\
$m_b$ (GeV)           & 4.588 $\pm$ 0.025 & 4.194 $\pm 0.043$ & 4.560 $\pm 0.023$ & 4.194 $\pm 0.043$ & 4.704 $\pm 0.029$ \\
$\mu_\pi^2$ (GeV$^2$) & 0.189 $^{+0.046}_{-0.057}$ & -                 & 0.453 $\pm 0.036$ & - &  - \\
\hline\hline
Ref. & \multicolumn{5}{c|}{$|V_{ub}|$ values}\\ 
\hline
$E_e$~\cite{ref:cleo-endpoint} &
$4.19\pm 0.49 ^{+0.26}_{-0.34}$ &
$3.82\pm 0.45 ^{+0.23}_{-0.26}$ &
$3.93\pm 0.46 ^{+0.22}_{-0.29}$ &
$3.43\pm 0.40 ^{+0.16}_{-0.17}$ &
- \\

$M_X, q^2$~\cite{ref:belle-mxq2Anneal}&
$4.46\pm 0.47 ^{+0.25}_{-0.27}$ &
$4.40\pm 0.46 ^{+0.19}_{-0.20}$ &
$4.37\pm 0.46 ^{+0.23}_{-0.26}$ &
$3.89\pm 0.41 ^{+0.17}_{-0.18}$ &
$4.68\pm 0.49 ^{+0.30}_{-0.30}$ \\

$E_e$~\cite{ref:belle-endpoint}&
$4.88\pm 0.45 ^{+0.24}_{-0.27}$ &
$4.79\pm 0.44 ^{+0.21}_{-0.24}$ &
$4.75\pm 0.44 ^{+0.17}_{-0.22}$ &
$4.48\pm 0.42 ^{+0.20}_{-0.20}$ &
-\\

$E_e$~\cite{ref:babar-endpoint}&
$4.48\pm 0.25 ^{+0.27}_{-0.28}$ &
$4.28\pm 0.24 ^{+0.22}_{-0.24}$ &
$4.29\pm 0.24 ^{+0.18}_{-0.24}$ &
$3.94\pm 0.22 ^{+0.19}_{-0.20}$ &
-\\

$E_e,s_\mathrm{h}^{\mathrm{max}}$~\cite{ref:babar-elq2}&
$4.66\pm 0.31 ^{+0.31}_{-0.36}$ &
$4.32\pm 0.29 ^{+0.24}_{-0.29}$ &
- &
$3.82\pm 0.26 ^{+0.17}_{-0.18}$ &
 \\ 

$p^*_{\ell}$~\cite{ref:belle-multivariate}&
$4.47\pm 0.27 ^{+0.19}_{-0.21}$ &
$4.60\pm 0.27 ^{+0.11}_{-0.13}$ &
$4.54\pm 0.27 ^{+0.10}_{-0.11}$ &
$4.48\pm 0.30 ^{+0.19}_{-0.19}$ &
- \\

$M_X$~\cite{ref:babar-finalupdate}&
$4.17\pm 0.19 ^{+0.24}_{-0.24}$ &
$4.40\pm 0.20 ^{+0.24}_{-0.19}$ &
$4.08\pm 0.19 ^{+0.20}_{-0.21}$ &
$3.81\pm 0.18 ^{+0.18}_{-0.20}$ &
- \\
$M_X$~\cite{ref:babar-finalupdate}&
$3.97\pm 0.22 ^{+0.20}_{-0.20}$ &
$4.16\pm 0.23 ^{+0.26}_{-0.22}$ &
$3.94\pm 0.22 ^{+0.16}_{-0.17}$ &
$3.73\pm 0.21 ^{+0.17}_{-0.18}$ &
- \\

$M_X,q^2$~\cite{ref:babar-finalupdate}&
$4.25\pm 0.23 ^{+0.23}_{-0.25}$  &
$4.19\pm 0.22 ^{+0.18}_{-0.19}$  &
$4.17\pm 0.22 ^{+0.22}_{-0.25}$  &
$3.74\pm 0.20 ^{+0.16}_{-0.17}$  &
$4.50\pm 0.24 ^{+0.29}_{-0.29}$ \\

$P_+$~\cite{ref:babar-finalupdate}&
$4.02\pm 0.25 ^{+0.24}_{-0.23}$  &
$4.10\pm 0.25 ^{+0.37}_{-0.28}$  &
$3.75\pm 0.23 ^{+0.30}_{-0.32}$  &
$3.56\pm 0.22 ^{+0.18}_{-0.19}$  &
- \\

$p^*_{\ell}$, $(M_X,q^2)$ fit~\cite{ref:babar-finalupdate}&
$4.28\pm 0.24 ^{+0.18}_{-0.20}$  &
$4.40\pm 0.24 ^{+0.12}_{-0.13}$  &
$4.35\pm 0.24 ^{+0.09}_{-0.10}$  &
$4.29\pm 0.24 ^{+0.18}_{-0.19}$  &
- \\

$p^*_{\ell}$~\cite{ref:babar-finalupdate}&
$4.29\pm 0.27 ^{+0.19}_{-0.20}$  &
$4.39\pm 0.27 ^{+0.15}_{-0.14}$  &
$4.33\pm 0.27 ^{+0.10}_{-0.11}$  &
$4.27\pm 0.26 ^{+0.18}_{-0.19}$  &
- \\

$M_X,q^2$~\cite{ref:belle-mx}&
- &
- &
- &
- &
$5.01\pm 0.39 ^{+0.32}_{-0.32}$ \\
\hline
Average &
$4.40\pm 0.15 ^{+0.19}_{-0.21}$ &
$4.45\pm 0.15 ^{+0.15}_{-0.16}$ &
$4.39\pm 0.15 ^{+0.12}_{-0.14}$ &
$4.03\pm 0.13 ^{+0.18}_{-0.12}$ &
$4.62\pm 0.20 ^{+0.29}_{-0.29}$ \\
\hline
\end{tabular}
}
\end{center}
\end{table}

\subsubsection{DGE}
\begin{figure}
\begin{center}
\includegraphics[width=0.48\textwidth]{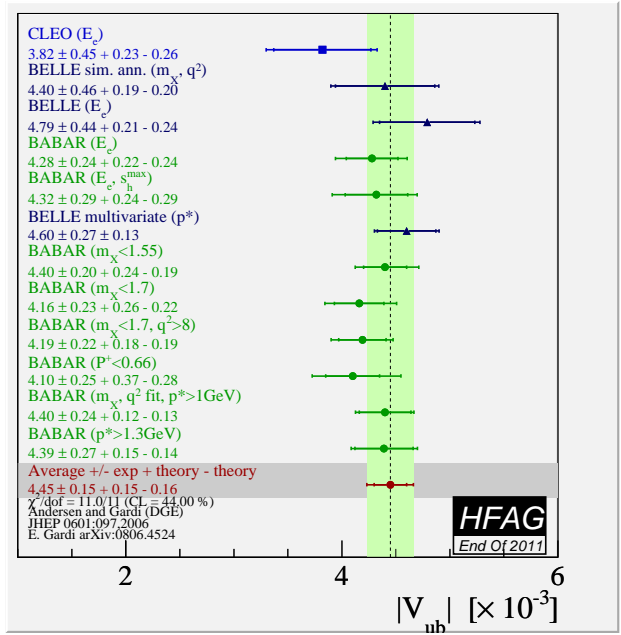}
\end{center}
\caption{Measurements of $\vub$ from inclusive semileptonic decays 
and their average based on the DGE prescription.
``$E_e$'', ``$M_X$'', ``$(M_X,q^2)$'' `$P^+$'', ``$p^*$ and ``($E_e,s^{max}_h$)'' indicate the 
analysis type and applied cut.}
\label{fig:DGE}
\end{figure}

J.R.~Andersen and E.~Gardi (Dressed Gluon Exponentiation, DGE)~\cite{ref:DGE} provide
a framework where the on-shell $b$-quark calculation, converted into hadronic variables, is
directly used as an approximation to the meson decay spectrum without
the use of a leading-power non-perturbative function (or, in other words,
a shape function). The on-shell mass of the $b$-quark within the $B$-meson ($m_b$) is
required as input. 
The DGE calculation uses the $\overline{MS}$ renormalization scheme; the heavy quark parameters determined  
from the global fit in the kinetic scheme, described in \ref{globalfitsKinetic}, were therefore 
translated into the $\overline{MS}$ scheme by using a calculation by Gardi, giving 
$m_b({\overline{MS}})=(4.194 \pm 0.043)$ GeV.
The extracted values
of \vub\, for each measurement along with their average are given in
Table~\ref{tab:bulnu} and illustrated in Figure~\ref{fig:DGE}.
The total error is $^{+4.8}_{-4.8}\%$, whose breakdown is:
statistics ($^{+2.0}_{-1.9}\%$),
detector ($^{+1.7}_{-1.7}\%$),
$B\to X_c \ell^+ \nul$ model ($^{+1.3}_{-1.3}\%$),
$B\to X_u \ell^+ \nul$ model ($^{+2.0}_{-1.9}\%$),
strong coupling $\alpha_s$ ($^{+0.5}_{-0.6}\%$),
$m_b$ ($^{+3.1}_{-2.8}\%$),
weak annihilation ($^{+0.0}_{-1.9}\%$),
DGE theory: matching scales ($^{+0.6}_{-0.3}\%$).
The largest contribution to the total error is due to the effect of the uncertainty 
on $m_b$. 
The uncertainty due to 
weak annihilation has been assumed to be asymmetric, i.e. it only tends to decrease \vub.

\subsubsection{GGOU}
\begin{figure}
\begin{center}
\includegraphics[width=0.48\textwidth]{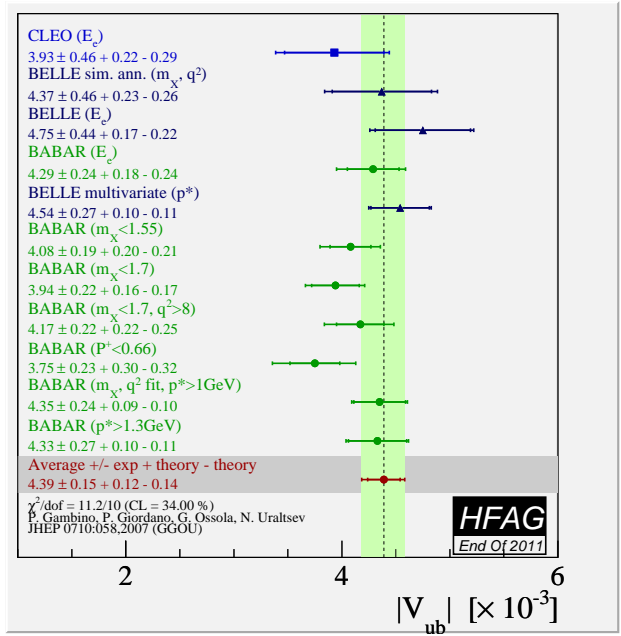}
\end{center}
\caption{Measurements of $\vub$ from inclusive semileptonic decays 
and their average based on the GGOU prescription.
``$E_e$'', ``$M_X$'', ``$(M_X,q^2)$'' `$P^+$'', ``$p^*$ and ``($E_e,s^{max}_h$)''  indicate the
analysis type and applied cut.}
\label{fig:GGOU}
\end{figure}

Gambino, Giordano, Ossola and Uraltsev (GGOU)~\cite{Gambino:2007rp} 
compute the triple differential decay rates of $B \to X_u \ell^+ \nul$, 
including all perturbative and non--perturbative effects through $O(\alphas^2 \beta_0)$ 
and $O(1/m_b^3)$. 
The Fermi motion is parameterized in terms of a single light--cone function 
for each structure function and for any value of $q^2$, accounting for all subleading effects. 
The calculations are performed in the kinetic scheme, a framework characterized by a Wilsonian 
treatment with a hard cutoff $\mu \sim 1 $ GeV.
GGOU have not included calculations for the ``($E_e,s^{max}_h$)'' analysis. 
The heavy quark parameters determined  
from the global fit in the kinetic scheme, described in \ref{globalfitsKinetic}, are used as inputs: 
$m_b(kin)=(4.560 \pm 0.023)$ GeV, 
$\mu_\pi^2(kin)=(0.453 \pm 0.036)$ GeV$^2$. 
The extracted values
of \vub\, for each measurement along with their average are given in
Table~\ref{tab:bulnu} and illustrated in Figure~\ref{fig:GGOU}.
The total error is $^{+4.4}_{-4.7}\%$ whose breakdown is:
statistics ($^{+2.0}_{-2.0}\%$),
detector ($^{+1.7}_{-1.7}\%$),
$B\to X_c \ell^+ \nul$ model ($^{+1.3}_{-1.3}\%$),
$B\to X_u \ell^+ \nul$ model ($^{+1.9}_{-1.9}\%$),
$\alpha_s$, $m_b$ and other non--perturbative parameters ($^{+1.9}_{-1.9}\%$), 
higher order perturbative and non--perturbative corrections ($^{+1.4}_{-1.4}\%$), 
modelling of the $q^2$ tail and choice of the scale $q^{2*}$ ($^{+1.3}_{-1.3}\%$), 
weak annihilations matrix element ($^{+0.0}_{-1.9}\%$), 
functional form of the distribution functions ($^{+0.2}_{-0.2}\%$), 
The leading uncertainties
on  \vub\ are both from theory, and are due to perturbative and non--perturbative
parameters and the modelling of the $q^2$ tail and choice of the scale $q^{2*}$. 
The uncertainty due to 
weak annihilation has been assumed to be asymmetric, i.e. it only tends to decrease \vub.

\subsubsection{ADFR}
\begin{figure}
\begin{center}
\includegraphics[width=0.48\textwidth]{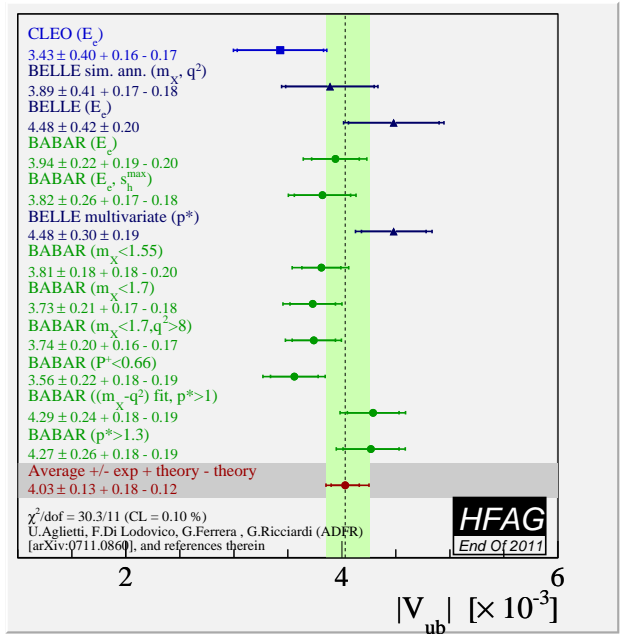}
\end{center}
\caption{Measurements of $\vub$ from inclusive semileptonic decays 
and their average based on the ADFR prescription.
``$E_e$'', ``$M_X$'', ``$(M_X,q^2)$'' `$P^+$'', ``$p^*$ and ``($E_e,s^{max}_h$)'' indicate the 
analysis type and applied cut.}
\label{fig:AC}
\end{figure}

Aglietti, Di Lodovico, Ferrera and Ricciardi (ADFR)~\cite{Aglietti:2007ik}
use an approach to extract \vub, which makes use of the ratio
of the  $B \to X_c \ell^+ \nul$ and $B \to X_u \ell^+ \nul$ widths. 
The normalized triple differential decay rate for 
$B \to X_u \ell^+ \nul$~\cite{Aglietti:2006yb,Aglietti:2005mb, Aglietti:2005bm, Aglietti:2005eq}
is calculated with a model based on (i) soft--gluon resummation 
to next--to--next--leading order and (ii) an effective QCD coupling without
Landau pole. This coupling is constructed by means of an extrapolation to low
energy of the high--energy behaviour of the standard coupling. More technically,
an analyticity principle is used.
The lower cut on the electron energy for the endpoint analyses is 2.3~GeV~\cite{Aglietti:2006yb}.
The ADFR calculation uses the $\overline{MS}$ renormalization scheme; the heavy quark parameters determined  
from the global fit in the kinetic scheme, described in \ref{globalfitsKinetic}, were therefore 
translated into the $\overline{MS}$ scheme by using a calculation by Gardi, giving 
$m_b({\overline{MS}})=(4.194 \pm 0.043)$ GeV.
The extracted values
of \vub\, for each measurement along with their average are given in
Table~\ref{tab:bulnu} and illustrated in Figure~\ref{fig:AC}.
The total error is $^{+5.4}_{-5.5}\%$ whose breakdown is:
statistics ($^{+1.9}_{-1.9}\%$),
detector ($^{+1.8}_{-1.8}\%$),
$B\to X_c \ell^+ \nul$ model ($^{+1.3}_{-1.3}\%$),
$B\to X_u \ell^+ \nul$ model ($^{+1.2}_{-1.4}\%$),
$\alpha_s$ ($^{+0.8}_{-1.2}\%$), 
$|V_{cb}|$ ($^{+1.7}_{-1.7}\%$), 
$m_b$ ($^{+0.7}_{-0.7}\%$), 
$m_c$ ($^{+1.3}_{-1.3}\%$), 
semileptonic branching fraction ($^{+0.7}_{-0.7}\%$), 
theory model ($^{+3.6}_{-3.6}\%$).
The leading
uncertainties, both from theory, are due to the $m_c$ mass and the theory model.

\subsubsection{BLL}
\begin{figure}
\begin{center}
\includegraphics[width=0.48\textwidth]{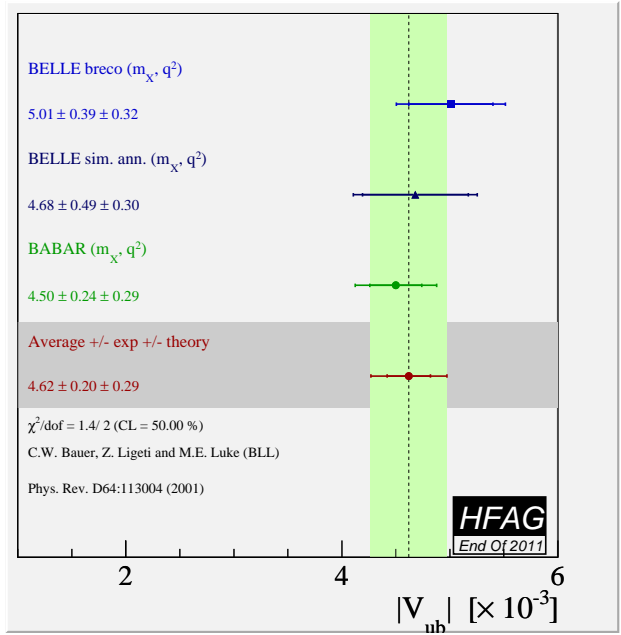}
\end{center}
\caption{Measurements of $\vub$ from inclusive semileptonic decays 
and their average in the BLL prescription.
``$(M_X, q^2)$'' indicates the analysis type.}
\label{fig:BLL}
\end{figure}

Bauer, Ligeti, and Luke (BLL)~\cite{ref:BLL} give a
HQET-based prescription that advocates combined cuts on the dilepton invariant mass, $q^2$,
and hadronic mass, $m_X$, to minimise the overall uncertainty on \vub.
In their reckoning a cut on $m_X$ only, although most efficient at
preserving phase space ($\sim$80\%), makes the calculation of the partial
rate untenable due to uncalculable corrections
to the $b$-quark distribution function or shape function. These corrections are
suppressed if events in the low $q^2$ region are removed. The cut combination used
in measurements is $M_x<1.7$ GeV/$c^2$ and $q^2 > 8$ GeV$^2$/$c^2$.  
The extracted values
of \vub\, for each measurement along with their average are given in
Table~\ref{tab:bulnu} and illustrated in Figure~\ref{fig:BLL}.
The total error is $^{+7.7}_{-7.7}\%$ whose breakdown is:
statistics ($^{+3.3}_{-3.3}\%$),
detector ($^{+3.0}_{-3.0}\%$),
$B\to X_c \ell^+ \nul$ model ($^{+1.6}_{-1.6}\%$),
$B\to X_u \ell^+ \nul$ model ($^{+1.1}_{-1.1}\%$),
spectral fraction ($m_b$) ($^{+3.0}_{-3.0}\%$),
perturbative : strong coupling $\alpha_s$ ($^{+3.0}_{-3.0}\%$),
residual shape function ($^{+2.5}_{-2.5}\%$),
third order terms in the OPE ($^{+4.0}_{-4.0}\%$),
The leading
uncertainties, both from theory, are due to residual shape function
effects and third order terms in the OPE expansion. The leading
experimental uncertainty is due to statistics. 

\subsubsection{Summary}
A summary of the averages presented in several different
frameworks and results by V.B.~Golubev, V.G.~Luth and Yu.I.~Skovpen~\cite{Golubev:2007cs},
based on prescriptions by LLR~\cite{Leibovich:1999xf} and LNP~\cite{Lange:2005qn} 
to reduce the leading shape function uncertainties are presented in 
Table~\ref{tab:vubcomparison}.
A value judgement based on a direct comparison should be
avoided at the moment, experimental and theoretical uncertainties play out
differently between the schemes and the theoretical assumptions for the
theory calculations are different.

\begin{table}[!htb]
\caption{\label{tab:vubcomparison}
Summary of inclusive determinations of $\vub$.
The errors quoted on \vub\ correspond to experimental and theoretical uncertainties, except for the last two 
measurements where the errors are due to the \babar\ endpoint analysis, the \babar $b\to s\gamma$ analysis~\cite{Aubert:2006qi}, 
the theoretical errors and $V_{ts}$ for the last averages. 
}
\begin{center}
\begin{small}
\begin{tabular}{|lc|}
\hline
Framework
&  $\Vub [10^{-3}]$\\
\hline\hline
BLNP
& $4.40 \pm 0.15 ^{+0.19}_{-0.21}$ \\ 
DGE
& $4.45 \pm 0.15 ^{+0.15}_{-0.16}$ \\
GGOU
& $4.39 \pm 0.15 ^{+0.12}_{-0.20}$ \\
ADFR
& $4.03 \pm 0.13 ^{+0.18}_{-0.12}$ \\
BLL ($m_X/q^2$ only)
& $4.62 \pm 0.20 \pm 0.29$ \\ 
LLR (\babar)~\cite{Aubert:2006qi}
& $4.43 \pm 0.45 \pm 0.29$ \\
LLR (\babar)~\cite{Golubev:2007cs}
& $4.28 \pm 0.29 \pm 0.29 \pm 0.26 \pm0.28$ \\
LNP (\babar)~\cite{Golubev:2007cs}
& $4.40 \pm 0.30 \pm 0.41 \pm 0.23$ \\
\hline
\end{tabular}\\
\end{small}
\end{center}
\end{table}

%
%

\clearpage
\mysection{$B$ decays to charmed hadrons
}
\label{sec:b2c}


\definecolor{hfviolet}{rgb}{0.5,0,0.5}
\definecolor{hflightcyan}{rgb}{0.1,1.0,1.0}

\def\hflmode{Mode}
\def\hflpdg#1{PDG #1}
\def\hflbabar{\mbox{BABAR}}
\def\hflbelle{\mbox{Belle}}
\def\hflcdf{\mbox{CDF}}
\def\hflcleo{\mbox{CLEO}}
\def\hfld0{\mbox{D0}}
\def\hfllhcb{\mbox{LHCb}}
\def\hflavg{Average}

\def\hfpubhotcolor{red}
\def\hfpubcolor{magenta}
\def\hfpuboldcolor{black}

\def\hfprehotcolor{blue}
\def\hfprecolor{cyan}
\def\hfpreoldcolor{hflightcyan}

\def\hfdefcolor{black}
\def\hferrcolor{yellow}
\def\hfsuperceededcolor{hfviolet}
\def\hfwaitingcolor{green}
\def\hfinactivecolor{hfviolet}
\def\hfnoquocolor{hfviolet}
\def\hfdeftext#1{\textcolor{\hfdefcolor}{#1}}
\def\hflabel#1{\textcolor{\hfdefcolor}{$#1$}}
\def\hfavg#1{\textcolor{\hfdefcolor}{$#1$}}
\def\hfnewavg#1{\textcolor{\hfdefcolor}{\boldmath$#1$}}
\def\hfdefault#1{\textcolor{\hfdefcolor}{$#1$}}
\def\hfpdg#1{\textcolor{\hfdefcolor}{$#1$}}
\def\hfwaiting#1{\textcolor{\hfwaitingcolor}{$#1$}}
\def\hfpubhot#1{\textcolor{\hfpubhotcolor}{$#1$}}
\def\hfprehot#1{\textcolor{\hfprehotcolor}{$#1$}}
\def\hfwaitingtext#1{\textcolor{\hfwaitingcolor}{#1}}
\def\hfpubhottext#1{\textcolor{\hfpubhotcolor}{#1}}
\def\hfprehottext#1{\textcolor{\hfprehotcolor}{#1}}
\def\hfpub#1{\textcolor{\hfpubcolor}{$#1$}}
\def\hfpre#1{\textcolor{\hfprecolor}{$#1$}}
\def\hfpubold#1{\textcolor{\hfpuboldcolor}{$#1$}}
\def\hfpreold#1{\textcolor{\hfpreoldcolor}{$#1$}}
\def\hfpubtext#1{\textcolor{\hfpubcolor}{#1}}
\def\hfpretext#1{\textcolor{\hfprecolor}{#1}}
\def\hfpuboldtext#1{\textcolor{\hfpuboldcolor}{#1}}
\def\hfpreoldtext#1{\textcolor{\hfpreoldcolor}{#1}}
\def\hferror#1{\textcolor{\hferrcolor}{$#1$}}
\def\hfsuperceeded#1{\textcolor{\hfsuperceededcolor}{$#1$}}
\def\hfinactive#1{\textcolor{\hfinactivecolor}{$#1$}}
\def\hfnoquo#1{\textcolor{\hfnoquocolor}{$#1$}}
\def\hferrortext#1{\textcolor{\hferrcolor}{#1}}
\def\hfsuperceededtext#1{\textcolor{\hfsuperceededcolor}{#1}}
\def\hfinactivetext#1{\textcolor{\hfinactivecolor}{#1}}
\def\hfnoquotext#1{\textcolor{\hfnoquocolor}{#1}}
\def\hfbb#1{#1M $B\bar{B}$ pairs}

\def\hffootnotemark#1{\tiny{$^{#1}$}}
\def\hffootnotetext#1{\tiny{#1}}
\def\hffootspacing{-5pt}
\def\hffootitemsep{-0.7}

\def\hfnewp{new particles }
\def\hfdstr{strange D mesons }
\def\hfbary{baryons }
\def\hfjpsi{$J/\psi(1S)$ }
\def\hfochm{charmonium other than $J/\psi(1S)$ }
\def\hfmuld{multiple $D$, $D^{*}$ or $D^{**}$ mesons }
\def\hfsgdx{a single $D^{*}$ or $D^{**}$ meson }
\def\hfsgld{a single D meson }
\def\hfothe{charmed particles } 

\def\hfsitebase{http://hfag.phys.ntu.edu.tw/b2charm/}
\def\hfurl#1{\href{\hfsitebase#1}{\hfsitebase#1}} 
\def\hfhref#1#2{\href{\hfsitebase#1}{#2}} 
\def\hftabletype{sidewaystable}
\def\hftableposn{!htbp}
\def\hfaftercapspace{\vspace*{2mm}}

\def\hfcaption#1#2#3#4#5{\textcolor{\hfdefcolor}{#1 of #2 modes producing #3 #4, #5.}}
\def\hfnewcaption#1#2#3#4#5#6#7{\caption{ #1 of #2 modes producing #3 #4, #5. The latest version is available at: \hfurl{#6} \label{#7}  }\hfaftercapspace}
\newcommand\hftabletlcell{\rule{0pt}{2.6ex}}  
\newcommand\hftableblcell{\rule[-1.2ex]{0pt}{0pt}}

\def\hfmetadata#1{}      

\def\hfcBR{{\cal{B}}}

%
%

This section reports the updated contribution to the HFAG report from the ``$B \to$ charm" group\footnote{The HFAG/BtoCharm group was formed in the spring of 2005; it performs its work using an XML database backed web application.}. 
The mandate of the group is to compile measurements and perform averages of all
available quantities related to $B$ decays to charmed particles, excluding $CP$
related quantities. To date the group has analyzed a total of 651 measurements
reported in 233 papers, principally branching fractions. The group aims to
organize and present the copious information on $b$-hadron decays to charmed
particles obtained from a combined sample of about two billion $B$-meson from
the BABAR, Belle Collaborations and data collected in hadronic colliders by the
CDF, D0 and LHCb Experiments. 

These huge samples of $b$-hadrons allow to measure decays to states with open or hidden charm content with unprecedented precision. 
Branching fractions for rare $b$-hadron decays or decay chains of a few $10^{-7}$ are 
being measured with statistical uncertainties typically below $30\%$, and new 
decay chains can be accessed with branching fractions down to $10^{-8}$. 
Results for more common decay chains, with branching fractions around $10^{-4}$, are becoming precision measurements, 
with uncertainties typically at the $3\%$ level. 


The measurements are classified according to the decaying particle: $B^{+}$, $B^{0}$, $B^{0}_{s}$, $B^{+}_{c}$, $\Lambda_b$ and Others ; the decay products 
and the type of quantity: branching fraction, product of branching fractions, ratio of branching fractions or other quantities. 
For the decay product classification the below precedence order is used to ensure that each measurement appears in only one category. 
\begin{itemize}\addtolength{\itemsep}{-0.4\baselineskip}
\item new particles
\item strange $D$ mesons
\item baryons
\item $J/\psi$
\item charmonium other than $J/\psi$
\item multiple $D$, $D^{*}$ or $D^{**}$ mesons
\item a single $D^{*}$ or $D^{**}$ meson
\item a single $D$ meson
\item other particles
\end{itemize}
  
Within each table the measurements are color coded according to the
publication status and age. Table~\ref{tab:hfc99999} provides a key to the
color scheme and categories used. When viewing the tables with most pdf
viewers every number, label and average provides hyperlinks to the corresponding 
reference and individual quantity web pages on the HFAG/BtoCharm group website
\hfhref{}{http://hfag.phys.ntu.edu.tw}.
The links provided in the captions of the table lead to the corresponding compilation
pages.  Both the individual and compilation webpages provide a graphical view
of the results, in a variety of formats.

Tables \ref{tab:hfc01101} to \ref{tab:hfc04300} provide either limits at 90\%
confidence level or measurements with statistical and systematic uncertainties 
and in some cases a third error corresponding to correlated systematics. 
For details on the meanings of the uncertainties and access to the references 
click on the numbers to visit the corresponding web pages.  Where there are
multiple determinations of the same quantity by one experiment the table
footnotes act to distinguish the methods or datasets used; such cases are
visually highlighted in the table by presenting the measurements on the lines
beneath the quantity label.
Where both limits and measured values of a quantity are available the limits 
are presented in the tables but are not used in the determination of the
average. Where only limits are available the most stringent is presented in
the Average column of the tables.
Where available the PDG 2010 result is also presented.

\clearpage 
\input{b2charm/b2charm_end_of_2011_v012_tables.tex}


\clearpage 
\mysection{$B$ decays to charmless final states}


\label{sec:rare}

The aim of this section is to provide the branching fractions, polarization 
fractions, and the partial rate asymmetries ($A_{CP}$) of charmless 
$B$ decays. The asymmetry is defined as 
$A_{CP} = \frac{N_{\Bbar} -N_B}{N_{\Bbar} +N_B}$, where $N_{\Bbar}$ 
and $N_B$ are respectively number of $\Bzb/\Bm$ and $\Bz/\Bp$ decaying
into a specific final state. 
Four different $B$ decay categories are considered: 
charmless mesonic, baryonic, radiative and leptonic. We also include
measurements of $B_s$ decays.
Measurements supported with  written documents are accepted in  
the averages; written documents include journal papers, 
conference contributed papers, preprints or conference proceedings.  
Results from  $A_{CP}$ measurements  obtained from time dependent analyses 
are listed and described in Sec.~\ref{sec:cp_uta}.

So far all branching fractions from \babar\ and Belle assume equal production 
of charged and neutral $B$ pairs.  The best measurements to date show that this
is still a reasonable approximation (see Sec.~\ref{sec:life_mix}).
For branching fractions, we provide either averages or the most stringent
90\% confidence level upper limits.  If one or more experiments have
measurements with $>$4$\sigma$ for a decay channel, all available central values
for that channel are used in the averaging.  We also give central values
and errors for cases where the significance of the average value is at
least $3 \sigma$, even if no single measurement is above $4 \sigma$. 
Since a few decay modes are sensitive to the contribution of
new physics and the current experimental upper limits are not far from the 
Standard Model expectation, we provide the combined upper limits or
averages in these cases.
Their upper limits can be estimated assuming that the errors are 
Gaussian.  For $A_{CP}$ we provide averages in all cases.  

Our averaging is performed by maximizing the likelihood,
   $\displaystyle {\mathcal L} = \prod_i {\mathcal P}_i(x),$  
where ${\mathcal P_i}$ is the probability density function (PDF) of the
$i$th  measurement, and $x$ is the branching fraction or $A_{CP}$.
The PDF is modeled by an asymmetric Gaussian function with the measured
central value as its mean and the quadratic sum of the statistical
and systematic errors as the standard deviations. The experimental
uncertainties are considered to be uncorrelated with each other when the 
averaging is performed. No error scaling is applied when the fit $\chi^2$ is 
greater than 1 since we believe that tends to overestimate the errors
except in cases of extreme disagreement (we have no such cases).
One exception to consider the correlated systematic errors is the inclusive
$B\to X_s\gamma$ mode, which is sensitive to physics beyond the Standard Model.
In this update, we have included new measurements from both Belle and \babar\
to perform the average. The detail is  
described  in Sec. ~\ref{sec:btosg}. 

At present, we have measurements of more than 400 decay modes, reported in
about 300 papers. Because the number of references is so large, we do
not include them with the tables shown here but the full set of
references is available quickly from active gifs at the 
``2011'' link on 
the rare web page: {\tt http://www.slac.stanford.edu/xorg/hfag/rare/index.html}.
The largest improvement since the last report has been inclusion of a
variety of new measurements from the LHC, especially LHCb.  The
measurements of $B_s$ decays are particularly noteworthy.

\mysubsection{Mesonic charmless decays}

\begin{table}
\begin{center}
\caption{Branching fractions (BF) of charmless mesonic  
$B^+$ decays with kaons (in units of $\times 10^6)$). Upper limits are
at 90\% CL. Values in {\red red} ({\blue blue}) are new {\red published}
({\blue preliminary}) results since PDG2010  [as of March 12, 2012].
}
\scriptsize
\begin{tabular}{|l@{}ccccccc|} 
\sgline
RPP\#   & Mode & PDG2010 Avg. & BABAR & Belle & CLEO & CDF & New avg. \\
\sglinespb
220                                               & 
$K^0 \pi^+$                                       & 
$23.1\pm 1.0$                                     & 
{$\err{23.9}{1.1}{1.0}$}                          & 
{\blue $\aerr{23.97}{0.53}{0.52}{0.69}$}          & 
$\aerrsy{18.8}{3.7}{3.3}{2.1}{1.8}$               & 
\nodata                                           & 
$23.80 \pm 0.74$                                  \\

221                                               & 
$K^+ \pi^0$                                       & 
$12.9\pm 0.6$                                     & 
{$\err{13.6}{0.6}{0.7}$}                          & 
{\blue $\err{12.62}{0.31}{0.56}$}                 & 
$\aerrsy{12.9}{2.4}{2.2}{1.2}{1.1}$               & 
\nodata                                           & 
$\cerr{12.94}{0.52}{0.51}$                        \\

222                                               & 
$\etapr K^+$                                      & 
$70.6\pm 2.5$                                     & 
{$\err{71.5}{1.3}{3.2}$}                          & 
{$\err{69.2}{2.2}{3.7}$}                          & 
$\aerr{80}{10}{9}{7}$                             & 
\nodata                                           & 
$71.1 \pm 2.6$                                    \\
\end{tabular}
\end{center}

\vspace{-0.4cm} 
\hspace*{-0.8cm} 
\hspace{0.4cm}
$\dag$Product BF - daughter BF taken to be 100\%
\end{table}

\begin{table}
\begin{center}
\caption{Branching Fractions (BF) of charmless mesonic $B^+$ decays 
 with kaons - part 2 (in units of $10^{-6}$). Upper limits are at 90\% CL.
Values in {\red red} ({\blue blue}) are new {\red published}
({\blue preliminary}) results since PDG2010  [as of March 12, 2012].
}
\scriptsize
\resizebox{0.99\textwidth}{!}{
\begin{tabular}{|l@{}ccccccc|} 
\sgline
RPP\#   & Mode & PDG2010 Avg. & BABAR & Belle & CLEO & CDF & New avg. \\
\sglinespb
269                                               & 
$a_1^+ K^0$                                       & 
$35\pm 7$                                         & 
{$\err{34.9}{5.0}{4.4}$}                          & 
\nodata                                           & 
\nodata                                           & 
\nodata                                           & 
$34.9 \pm 6.7$                                    \\

270                                               & 
$b_1^+ K^0$ \dag                                  & 
$9.6 \pm 1.9$                                     & 
{ $\err{9.6}{1.7}{0.9}$}                          & 
\nodata                                           & 
\nodata                                           & 
\nodata                                           & 
$9.6 \pm 1.9$                                     \\

271                                               & 
$K^{*0} \rho^+$                                   & 
$9.2 \pm 1.5$                                     & 
{$\err{9.6}{1.7}{1.5}$}                           & 
{$\err{8.9}{1.7}{1.2}$}                           & 
\nodata                                           & 
\nodata                                           & 
$9.2 \pm 1.5$                                     \\
\end{tabular}
}
\end{center}

\vspace{-0.4cm} 
\hspace*{-0.8cm} 
\hspace{0.4cm}
$\dag$Product BF - daughter BF taken to be 100\%; 
~\S $M_{\phi\phi}<2.85$ GeV/$c^2$
\end{table}

\begin{table}
\begin{center}
\caption{Branching Fractions (BF) of charmless mesonic $B^+$ decays 
 without kaons (in units of $10^{-6}$). Upper limits are at 90\% CL.
Values in {\red red} ({\blue blue}) are new {\red published}
({\blue preliminary}) results since PDG2010  [as of March 12, 2012].
}
\scriptsize
\begin{tabular}{|l@{}ccccccc|} 
\sgline
RPP\#   & Mode & PDG2010 Avg. & BABAR & Belle & CLEO & CDF & New avg. \\
\sglinespb
330                                               & 
$\pi^+ \pi^0$                                     & 
$5.7 \pm 0.5$                                     & 
{$\err{5.02}{0.46}{0.29}$}                        & 
{\blue $\err{5.86}{0.26}{0.38}$}                  & 
$\aerrsy{4.6}{1.8}{1.6}{0.6}{0.7}$                & 
\nodata                                           & 
$\cerr{5.48}{0.35}{0.34}$                         \\

331                                               & 
$\pi^+ \pi^+ \pi^-$                               & 
$15.2 \pm 1.5$                                    & 
{$\err{15.2}{0.6}{1.3}$}                          & 
\nodata                                           & 
\nodata                                           & 
\nodata                                           & 
$15.2 \pm 1.4$                                    \\

332                                               & 
$\rho^0 \pi^+$                                    & 
$8.3\pm 1.2$                                      & 
{$\berr{8.1}{0.7}{1.3}{1.6}$}                     & 
$\aerr{8.0}{2.3}{2.0}{0.7}$                       & 
$\aerr{10.4}{3.3}{3.4}{2.1}$                      & 
\nodata                                           & 
$\cerr{8.3}{1.2}{1.3}$                            \\
\end{tabular}
\end{center}

\vspace{-0.4cm} 
\hspace*{-0.8cm} 
\hspace{0.4cm}
$\dag$Product BF - daughter BF taken to be 100\%; 
\end{table}
\clearpage

\begin{table}
\begin{center}
\caption{Branching fractions of charmless mesonic $B^0$ decays with kaons 
(in units of $10^{-6}$).
Upper limits are at 90\% CL.
Values in {\red red} ({\blue blue}) are new {\red published}
({\blue preliminary}) results since PDG2010  [as of March 12, 2012].
}
\scriptsize
\begin{tabular}{|l@{}ccccccc|} 
\sgline
RPP\#   & Mode & PDG2010 Avg. & BABAR & Belle & CLEO & CDF & New avg. \\
\hline
210                                               & 
$K^+ \pi^-$                                       & 
$19.4\pm 0.6$                                     & 
{$\err{19.1}{0.6}{0.6}$}                          & 
{\blue $\err{20.0}{0.34}{0.63}$}                  & 
$\aerrsy{18.0}{2.3}{2.1}{1.2}{0.9}$               & 
\nodata                                           & 
$\cerr{19.55}{0.54}{0.53}$                        \\

211                                               & 
$K^0 \pi^0$                                       & 
$9.5\pm 0.8$                                      & 
\blue{$\err{10.1}{0.6}{0.4}$}                     & 
{\blue $\err{9.66}{0.46}{0.49}$}                  & 
$\aerrsy{12.8}{4.0}{3.3}{1.7}{1.4}$               & 
\nodata                                           & 
$\cerr{9.92}{0.49}{0.48}$                         \\

212                                               & 
$\etapr K^0$                                      & 
$66\pm 4$                                         & 
{$\err{68.5}{2.2}{3.1}$}                          & 
{$\aerr{58.9}{3.6}{3.5}{4.3}$}                    & 
$\aerr{89}{18}{16}{9}$                            & 
\nodata                                           & 
$66.1 \pm 3.1$                                    \\
\end{tabular}
\end{center}
\vspace{-0.4cm}
$\dag$Product BF - daughter BF taken to be 100\%, 
 $\ddag$Relative BF converted to absolute BF
 $^1 0.755<M(K\pi)<1.250$ GeV/$c^2$.
 $^2$Excludes $M(K_SK_S)$ regions [3.400,3.429] and [3.540,3.585] and $M(K_SK_L)<1.049$ GeV/$c^2$
 $^3$Includes $K\pi$ S-wave contribution and uncorrected for $K^*(1430)$ branching fraction
\end{table}
\clearpage

\begin{table}
\begin{center}
\caption{
Branching fractions of charmless mesonic $B^0$ decays with kaons - Part 2
(in units of $10^{-6}$).
Upper limits are at 90\% CL.
Values in {\red red} ({\blue blue}) are new {\red published}
({\blue preliminary}) results since PDG2010  [as of March 12, 2012].
}
\tiny
\resizebox{0.99\textwidth}{!}{
\begin{tabular}{|l@{}ccccccc|} 
\sgline
RPP\#   & Mode & PDG2010 Avg. & BABAR & Belle & CLEO & CDF & New avg. \\
\sglinespb
252                                               & 
$K^{*0} \pi^0$                                    & 
$\err{3.6}{0.7}{0.4}$                             & 
\red{$\err{3.3}{0.5}{0.4}$}                       & 
{$\aerr{0.4}{1.9}{1.7}{0.1}$}                     & 
$\aerrsy{0.0}{1.3}{0.0}{0.5}{0.0}$                & 
\nodata                                           & 
$2.5 \pm 0.6$                                     \\

253                                               & 
$K_2^*(1430)^+ \pi^-$                             & 
$<6.3$                                            & 
{$<16.2$}                                         & 
{$<6.3$}                                          & 
\nodata                                           & 
\nodata                                           & 
{$<6.3$}                                          \\

254                                               & 
$K^*(1680)^+ \pi^-$                               & 
$<10$                                             & 
{$<25$}                                           & 
{$<10.1$}                                         & 
\nodata                                           & 
\nodata                                           & 
{$<10.1$}                                         \\
\end{tabular}
}
\end{center}
\vspace{-0.4cm}
$\dag$Product BF - daughter BF taken to be 100\%, 
~\S $M_{\phi\phi}<2.85$ GeV/$c^2$
 $\ddag 0.55<M(\pi\pi)<1.42$ GeV/$c^2$ and $0.75<M(K\pi)<1.20$ GeV/$c^2$;
 $^1 0.55<M(\pi\pi)<1.42$ GeV/$c^2$;
 $^2 0.75<M(K\pi)<1.20$ GeV/$c^2$
\end{table}
\clearpage

\begin{table}
\begin{center}
\caption{
Branching fractions of charmless mesonic $B^0$ decays without kaons
(in units of $10^{-6}$).
Upper limits are at 90\% CL.
Values in {\red red} ({\blue blue}) are new {\red published}
({\blue preliminary}) results since PDG2010  [as of March 12, 2012].
}
\tiny
\begin{tabular}{|l@{}ccccccc|} 
\sgline
RPP\#   & Mode & PDG2010 Avg. & BABAR & Belle & CLEO & CDF & New avg. \\
\hline
315                                               & 
$\pi^+ \pi^-$                                     & 
$5.13 \pm 0.24$                                   & 
{$\err{5.5}{0.4}{0.3}$}                           & 
{\blue $\err{5.04}{0.21}{0.19}$}                  & 
$\aerrsy{4.5}{1.4}{1.2}{0.5}{0.4}$                & 
\red{$\err{5.02}{0.33}{0.35}$}\ddag               & 
$5.11 \pm 0.22$                                   \\

316                                               & 
$\pi^0 \pi^0$                                     & 
$1.62\pm 0.31$                                    & 
\blue{$\err{1.83}{0.21}{0.13}$}                   & 
{$\aerrsy{2.3}{0.4}{0.5}{0.2}{0.3}$}              & 
$<4.4$                                            & 
\nodata                                           & 
$\cerr{1.91}{0.22}{0.23}$                         \\

317                                               & 
$\eta \pi^0$                                      & 
$<1.5$                                            & 
{$<1.5$}                                          & 
{$<2.5$}                                          & 
$<2.9$                                            & 
\nodata                                           & 
{$<1.5$}                                          \\
\end{tabular}
\end{center}
\vspace{-0.4cm}
$\dag$Product BF - daughter BF taken to be 100\%, 
 $\ddag$Relative BF converted to absolute BF
\end{table}
\clearpage

\begin{table}
\begin{center}
\caption{
 Relative branching fractions of $B^0\to K^+K^-, K^+\pi^-, \pi^+\pi^-$.
Values in {\red red} ({\blue blue}) are new {\red published}
({\blue preliminary}) result since PDG2010  [as of March 12, 2012].
}
\footnotesize
\resizebox{0.99\textwidth}{!}{
\begin{tabular}{|lccccc|} \hline
RPP\# &Mode & PDG2010 Avg. &  CDF & D\O\ & New avg.  \\ \sglinespb
267                                               & 
$\mathcal{B}(B^0\rightarrow K^+K^-)/\mathcal{B}(B^0\rightarrow K^+\pi^-)$& 
\nodata                                           & 
\red{$0.020\pm0.008\pm0.006$}                     & 
\nodata                                           & 
$0.020 \pm 0.010$                                 \\

315                                               & 
$\mathcal{B}(B^0\rightarrow\pi^+\pi^-)/\mathcal{B}(B^0\rightarrow K^+\pi^-)$& 
\nodata                                           & 
\red{$0.259\pm0.017\pm0.016$}                     & 
\nodata                                           & 
$0.259 \pm 0.023$                                 \\

\hline
\end{tabular}
}
\end{center}
\end{table}
\clearpage


\mysubsection{Radiative and leptonic decays}

\begin{table}[!hb]
\begin{center}
\caption{Branching fractions of semileptonic and radiative $B^+$ decays
(in units of $10^{-6}$). Upper limits are at 90\% CL.
Values in {\red red} ({\blue blue}) are new {\red published}
({\blue preliminary}) results since PDG2010  [as of March 12, 2012].
}
\tiny
\begin{tabular}{|l@{}ccccccc|} \hline
RPP\# &Mode & PDG2010 Avg. & \babar  & Belle & CLEO & CDF
 & New Avg.  \\
 \sglinespb
314                                               & 
$K^{*+} \gamma$                                   & 
$42.1\pm 1.8$                                     & 
{$42.2\pm 1.4 \pm 1.6$}                           & 
{$42.5\pm 3.1\pm 2.4$}                            & 
$\aerr{37.6}{8.9}{8.3}{2.8} $                     & 
\nodata                                           & 
$42.1 \pm 1.8$                                    \\

315                                               & 
$K_1^+(1270) \gamma$                              & 
$43\pm13$                                         & 
\nodata                                           & 
{$\err{43}{9}{9}$}                                & 
\nodata                                           & 
\nodata                                           & 
$43 \pm 12$                                       \\

316                                               & 
$K^+ \eta \gamma$                                 & 
$7.9\pm 0.9$                                      & 
{$\err{7.7}{1.0}{0.4}$}                           & 
{$\aerr{8.4}{1.5}{1.2}{0.9}$}                     & 
\nodata                                           & 
\nodata                                           & 
$7.9 \pm 0.9$                                     \\
\hline
\end{tabular}
\vspace{-0.4cm}
\end{center}
~\dag $M_{K\pi\pi} < 1.8$ GeV/$c^2$;~\ddag~$1.0 < M_{K\pi\pi} < 2.0$ GeV/$c^2$;
~\S~$M_{K\pi\pi} < 2.4$ GeV/$c^2$
\end{table}
\clearpage

\begin{table}
\begin{center}
\caption{Branching fractions of semileptonic and radiative $B^0$ decays
(in units of $10^{-6}$).  Upper limits are at 90\% CL.
Values in {\red red} ({\blue blue}) are new {\red published}
({\blue preliminary}) results since PDG2010  [as of March 12, 2012].
}

\tiny
\begin{tabular}{|l@{}ccccccc|} \hline
RPP\# &Mode & PDG2010 Avg. & \babar  & Belle & CLEO & CDF
 & New Avg.  \\ \sglinespb
297                                               & 
$K^0 \eta \gamma$                                 & 
$7.6 \pm 1.8$                                     & 
{$\aerr{7.1}{2.1}{2.0}{0.4}$}                     & 
{$\aerrsy{8.7}{3.1}{2.7}{1.9}{1.6}$}              & 
\nodata                                           & 
\nodata                                           & 
$\cerr{7.6}{1.8}{1.7}$                            \\

298                                               & 
$K^0 \etapr \gamma$                               & 
{$<6.6$}                                          & 
{$<6.6$}                                          & 
\red{$<6.4$}                                      & 
\nodata                                           & 
\nodata                                           & 
{$<6.4$}                                          \\

299                                               & 
$K^0 \phi \gamma$                                 & 
$<2.7$                                            & 
{$<2.7$}                                          & 
\red{$2.74\pm 0.60\pm 0.32$}                      & 
\nodata                                           & 
\nodata                                           & 
$2.74 \pm 0.68$                                   \\
\hline
\end{tabular}
\vspace{-0.4cm}
\end{center}
~\dag $M_{K\pi\pi} < 1.8$ GeV/$c^2$;~\ddag~$1.0 < M_{K\pi\pi} < 2.0$ GeV/$c^2$;
~\S~$1.25$ GeV/$c^2 < M_{K\pi} < 1.6$ GeV/$c^2$
\end{table}
\clearpage

\begin{table}
\begin{center}
\caption{Branching fractions of  semileptonic and radiative $B$ decays
(in units of $10^{-6}$).  Upper limits are at 90\% CL.
Values in {\red red} ({\blue blue}) are new {\red published}
({\blue preliminary}) results since PDG2010  [as of March 12, 2012].
}
\scriptsize
\begin{tabular}{|lcccccc|} \hline
RPP\# &Mode & PDG2010 Avg. & \babar  & Belle & CLEO
 & New Avg.  \\ \sglinespb
~66                                               & 
$K \eta \gamma$                                   & 
$\cerr{8.5}{1.8}{1.6}$                            & 
\nodata                                           & 
{$\aerr{8.5}{1.3}{1.2}{0.9}$}                     & 
\nodata                                           & 
$\cerr{8.5}{1.6}{1.5}$                            \\

~68                                               & 
$K_2^*(1430) \gamma $                             & 
$\cerr{1.7}{0.6}{0.5}$                            & 
\nodata                                           & 
\nodata                                           & 
$\err{1.7}{0.6}{0.1}$                             & 
$1.7 \pm 0.6$                                     \\

~70                                               & 
$K_3^*(1780) \gamma $                             & 
$<37$                                             & 
\nodata                                           & 
{$<2.8$}                                          & 
\nodata                                           & 
{$<2.8$}                                          \\
\hline
\end{tabular}
\end{center}
~\dag $E_\gamma > 2.0$ GeV; ~\ddag $M(\ell^+\ell^-)>0.2$ GeV/$c^2$
\end{table}
\clearpage

\begin{table}
\begin{center}
\caption{ Isospin symmetry for various $B$ decays.  Values in {\red red} 
({\blue blue}) are new {\red published}
({\blue preliminary}) results since PDG2010  [as of March 12, 2012].
}

\vspace{0.3cm}
\footnotesize
\begin{tabular}{|lccccc|}
\sgline
RPP\# & Parameter & PDG2010 Avg. & \babar & Belle & New Avg. \\
\sglinespb
$~65$                                             & 
$\Delta_{0^-}(K^* \gamma)$                        & 
$0.066 \pm 0.030$                                 & 
{$\err{0.066}{0.021}{0.022}$}                     & 
{$\err{0.012}{0.044}{0.026}$}                     & 
$0.052 \pm 0.026$                                 \\

$~77$                                             & 
$\Delta_{0^-}(X_s\gamma)$                         & 
$-0.01 \pm 0.06$                                  & 
$-0.01 \pm 0.06$                                  & 
\nodata                                           & 
$-0.01 \pm 0.06$                                  \\

$~82$                                             & 
$\Delta_{\rho \gamma}$                            & 
$-0.46 \pm 0.17$                                  & 
{$\aerr{-0.43}{0.25}{0.22}{0.10}$}                & 
{$\aerrsy{-0.48}{0.21}{0.19}{0.08}{0.09}$}        & 
$\cerr{-0.46}{0.17}{0.16}$                        \\
\end{tabular}
\end{center}
\vspace{-0.3cm}
~$\dag~m_{\ell\ell} < m_{J/\psi}$
\end{table}

\begin{table}
\begin{center}
\caption{ Partial branching fractions for various $B$ decays.  Values in 
{\red red} ({\blue blue}) are new {\red published}
({\blue preliminary}) results since PDG2010  [as of March 12, 2012].
}

\vspace{0.3cm}
\tiny
\resizebox{0.99\textwidth}{!}{
\begin{tabular}{|llccccccc|}
\sgline
RPP\# & Mode & $q^2~[(\mathrm{GeV}/c^2)^2]$~\dag &
PDG2010 Avg. & \babar & Belle & CDF~\ddag & LHCb~\ddag & New Avg. \\
\sglinespb
121                                               & 
$K\ell^+\ell^-$                                   & 
$< 2.0$                                           & 
\nodata                                           & 
\nodata                                           & 
{$\aerr{0.81}{0.18}{0.16}{0.05}$}                 & 
{\red $\err{0.33}{0.10}{0.02}$}                   & 
\nodata                                           & 
$0.46 \pm 0.09$                                   \\

\nodata                                           & 
$K\ell^+\ell^-$                                   & 
$[2.0,4.3]$                                       & 
\nodata                                           & 
\nodata                                           & 
{$\aerr{0.46}{0.14}{0.12}{0.03}$}                 & 
{\red $\err{0.77}{0.14}{0.05}$}                   & 
\nodata                                           & 
$0.61 \pm 0.10$                                   \\
\end{tabular}
}
\end{center}
\vspace{-0.3cm}
\dag~see the original paper for the exact $q^2$ selection.~~
\ddag~muon mode only ($\ell = \mu$).
\end{table}

\begin{table}
\begin{center}
\caption{Forward-backward asymmetry for various $B$ decays.  Values in 
{\red red} ({\blue blue}) are new {\red published}
({\blue preliminary}) results since PDG2010  [as of March 12, 2012].
}
\label{tab:Kstarmumu-Afb}

\vspace{0.3cm}
\tiny
\begin{tabular}{|llcccccc|}
\sgline
RPP\# & Mode & $q^2~[(\mathrm{GeV}/c^2)^2]$~\dag &
PDG2010 Avg. & Belle & CDF~\ddag & LHCb~\ddag & New Avg. \\
\sglinespb
121                                               & 
$K\ell^+\ell^-$                                   & 
$< 2.0$                                           & 
{$\aerr{0.06}{0.32}{0.35}{0.02}$}                 & 
{$\aerr{0.06}{0.32}{0.35}{0.02}$}                 & 
{\blue $\aerr{0.13}{0.42}{0.43}{0.07}$}           & 
\nodata                                           & 
$\cerr{0.08}{0.20}{0.22}$                         \\

\nodata                                           & 
$K\ell^+\ell^-$                                   & 
$[2.0,4.3]$                                       & 
{$\aerr{-0.43}{0.38}{0.40}{0.09}$}                & 
{$\aerr{-0.43}{0.38}{0.40}{0.09}$}                & 
{\blue $\aerr{0.32}{0.15}{0.16}{0.05}$}           & 
\nodata                                           & 
$0.12 \pm 0.14$                                   \\

\nodata                                           & 
$K\ell^+\ell^-$                                   & 
$[4.3,8.68]$                                      & 
{$\aerr{-0.20}{0.12}{0.14}{0.03}$}                & 
{$\aerr{-0.20}{0.12}{0.14}{0.03}$}                & 
{\blue $\aerr{0.01}{0.13}{0.10}{0.01}$}           & 
\nodata                                           & 
$\cerr{-0.11}{0.06}{0.07}$                        \\
\end{tabular}
\end{center}
\vspace{-0.3cm}
\dag~see the original paper for the exact $q^2$ selection.~~
\ddag~muon mode only ($\ell = \mu$).
\end{table}

\begin{table}
\begin{center}
\caption{Fraction of the longitudinal polarization ($F_L$) for various $B$ 
decays.  Values in {\red red} ({\blue blue}) are new {\red published}
({\blue preliminary}) results since PDG2010  [as of March 12, 2012].
}

\vspace{0.3cm}
\tiny
\begin{tabular}{|llcccccc|}
\sgline
RPP\# & Mode & $q^2~[(\mathrm{GeV}/c^2)^2]$~\dag &
PDG2010 Avg. & Belle & CDF~\ddag & LHCb~\ddag & New Avg. \\
\sglinespb
122                                               & 
$K^*\ell^+\ell^-$                                 & 
$< 2.0$                                           & 
{$\aerr{0.29}{0.21}{0.18}{0.02}$}                 & 
{$\aerr{0.29}{0.21}{0.18}{0.02}$}                 & 
{\blue $\aerr{0.30}{0.16}{0.16}{0.02}$}           & 
{\blue $\aerr{0.03}{0.15}{0.03}{0.06}$}           & 
$\cerr{0.22}{0.09}{0.08}$                         \\

\nodata                                           & 
$K^*\ell^+\ell^-$                                 & 
$[2.0,4.3]$                                       & 
{$0.71 \pm 0.24 \pm 0.05$}                        & 
{$0.71 \pm 0.24 \pm 0.05$}                        & 
{\blue $\aerr{0.37}{0.25}{0.24}{0.10}$}           & 
{\blue $\aerr{0.84}{0.15}{0.13}{0.06}$}           & 
$0.73 \pm 0.10$                                   \\

\nodata                                           & 
$K^*\ell^+\ell^-$                                 & 
$[4.3,8.68]$                                      & 
{$\aerr{0.64}{0.23}{0.24}{0.07}$}                 & 
{$\aerr{0.64}{0.23}{0.24}{0.07}$}                 & 
{\blue $\aerr{0.68}{0.15}{0.17}{0.09}$}           & 
{\blue $\err{0.60}{0.07}{0.01}$}                  & 
$0.61 \pm 0.06$                                   \\

\nodata                                           & 
$K^*\ell^+\ell^-$                                 & 
$[10.09,12.86]$                                   & 
\nodata                                           & 
{$\aerr{0.17}{0.17}{0.15}{0.03}$}                 & 
{\blue $\aerr{0.47}{0.14}{0.14}{0.03}$}           & 
{\blue $\aerr{0.44}{0.12}{0.11}{0.02}$}           & 
$0.39 \pm 0.08$                                   \\

\nodata                                           & 
$K^*\ell^+\ell^-$                                 & 
$[14.18,16.00]$                                   & 
\nodata                                           & 
{$\aerr{-0.15}{0.27}{0.23}{0.07}$}                & 
{\blue $\aerr{0.29}{0.14}{0.13}{0.05}$}           & 
{\blue $\aerr{0.33}{0.11}{0.08}{0.04}$}           & 
$\cerr{0.28}{0.08}{0.07}$                         \\

\nodata                                           & 
$K^*\ell^+\ell^-$                                 & 
$>16.00$                                          & 
\nodata                                           & 
{$\aerr{0.12}{0.15}{0.13}{0.02}$}                 & 
{\blue $\aerr{0.20}{0.19}{0.17}{0.05}$}           & 
{\blue $\aerr{0.28}{0.10}{0.09}{0.04}$}           & 
$\cerr{0.22}{0.08}{0.07}$                         \\


\sglinespt
\end{tabular}
\end{center}
\vspace{-0.3cm}
\dag~see the original paper for the exact $q^2$ selection.~~
\ddag~muon mode only ($\ell = \mu$).
\end{table}

\begin{table}
\begin{center}
\caption{
Branching fractions of inclusive $B$ decays
(in units of $10^{-6}$). Values in {\red red} ({\blue blue}) 
are new {\red published}
({\blue preliminary}) results since PDG2010  [as of March 12, 2012].
}
\small
\vspace{0.3cm}
\begin{tabular}{|lcccccc|} \hline
RPP\# &Mode & PDG2010 Avg. & \babar  & Belle & CLEO & New Avg.  \\
\sglinespb
$~-$                                              & 
$K^+ X$                                           & 
New                                               & 
{\red $<187\dag$}                                 & 
\nodata                                           & 
\nodata                                           & 
{ $<187\dag$}                                     \\

$~-$                                              & 
$K^0 X$                                           & 
New                                               & 
{\red\aerr{195}{51}{45}{50}$\dag$}                & 
\nodata                                           & 
\nodata                                           & 
$\cerr{195}{71}{67}$                              \\

$~-$                                              & 
$\pi^+ X$                                         & 
New                                               & 
{\red\aerr{372}{50}{47}{59}$\dag$}                & 
\nodata                                           & 
\nodata                                           & 
$\cerr{372}{77}{75}$                              \\

$~80$                                             & 
$s \eta$                                          & 
$<440$                                            & 
\nodata                                           & 
{\red\berr{261}{30}{44}{74}} \S                   & 
$<440$                                            & 
$\cerr{261}{53}{79}$                              \\

$~81$                                             & 
$s \eta'$                                         & 
$420 \pm 90$                                      & 
\err{390}{80}{90}$\ddag$                          & 
\nodata                                           & 
\err{460}{110}{60}$\ddag$                         & 
$423 \pm 86$                                      \\

\hline
\end{tabular}
\end{center}
\vspace{-0.3cm}
~~~~~~~~~~~~~~~~~~\dag~$p^* > 2.34$ GeV;
~\S~$0.4 < M_{X_s} < 2.6$ GeV;
~\ddag~$2.0 < p^* < 2.7$ GeV
\end{table}

\begin{table}[!htb]
\begin{center}

\caption{
 Branching fractions of leptonic  $B$ decays
(in units of $10^{-6}$). Upper limits are at 90\% CL.
Values in {\red red} ({\blue blue}) are new {\red published}
({\blue preliminary}) results since PDG2010  [as of March 12, 2012].
}
\scriptsize
\resizebox{0.99\textwidth}{!}{
\begin{tabular}{|lcccccccccc|} \hline
RPP\# &Mode & PDG2010 Avg. & \babar  & Belle & CLEO &CDF & \dzero & LHCb & CMS
 & New Avg.  \\ \sglinespb
~24                                               & 
$e^+ \nu$                                         & 
$<1.9$                                            & 
{$<1.9$}                                          & 
{$<1.0$}                                          & 
$<15$                                             & 
\nodata                                           & 
\nodata                                           & 
\nodata                                           & 
\nodata                                           & 
{$<1.0$}                                          \\

~25                                               & 
$\mu^+ \nu$                                       & 
$<1.0$                                            & 
{$<1.0$}                                          & 
{$<1.7$}                                          & 
$<21$                                             & 
\nodata                                           & 
\nodata                                           & 
\nodata                                           & 
\nodata                                           & 
{$<1.0$}                                          \\

~26                                               & 
$\tau^+ \nu$                                      & 
$180\pm50$                                        & 
\blue{$176\pm49$}                                 & 
{\red $\aerrsy{162}{31}{30}{25}{26}$}\dag         & 
$<840$                                            & 
\nodata                                           & 
\nodata                                           & 
\nodata                                           & 
\nodata                                           & 
$167 \pm 30$                                      \\

~27                                               & 
$\ell^+ \nu_{\ell} \gamma$                        & 
$<15.6$                                           & 
{$<15.6$}                                         & 
\nodata                                           & 
\nodata                                           & 
\nodata                                           & 
\nodata                                           & 
\nodata                                           & 
\nodata                                           & 
{$<15.6$}                                         \\

~28                                               & 
$e^+ \nu_e \gamma$                                & 
$<17$                                             & 
{$<17$}                                           & 
\nodata                                           & 
$<200$                                            & 
\nodata                                           & 
\nodata                                           & 
\nodata                                           & 
\nodata                                           & 
{$<17$}                                           \\

~29                                               & 
$\mu^+ \nu_{\mu} \gamma$                          & 
$<24$                                             & 
{$<26$}                                           & 
\nodata                                           & 
$<52$                                             & 
\nodata                                           & 
\nodata                                           & 
\nodata                                           & 
\nodata                                           & 
{$<26$}                                           \\

412                                               & 
$\gamma \gamma$                                   & 
$<0.62$                                           & 
\red$<0.32$                                       & 
{$<0.62$}                                         & 
\nodata                                           & 
\nodata                                           & 
\nodata                                           & 
\nodata                                           & 
\nodata                                           & 
$<0.32$                                           \\

413                                               & 
$e^+ e^-$                                         & 
$<0.083$                                          & 
{$<0.113$}                                        & 
$<0.19$                                           & 
$<0.83$                                           & 
$<0.083$                                          & 
\nodata                                           & 
\nodata                                           & 
\nodata                                           & 
$<0.083$                                          \\

414                                               & 
$e^+ e^- \gamma$                                  & 
$<0.12$                                           & 
{$<0.12$}                                         & 
\nodata                                           & 
\nodata                                           & 
\nodata                                           & 
\nodata                                           & 
\nodata                                           & 
\nodata                                           & 
{$<0.12$}                                         \\

415                                               & 
$\mu^+ \mu^-$                                     & 
$<0.015$                                          & 
{$<0.052$}                                        & 
$<0.16$                                           & 
$<0.61$                                           & 
{\blue$<0.0050$}                                  & 
\nodata                                           & 
{\red$<0.0026$}                                   & 
\red$<0.0037$                                     & 
{$<0.0026$}                                       \\

416                                               & 
$\mu^+ \mu^- \gamma$                              & 
$<0.16$                                           & 
{$<0.16$}                                         & 
\nodata                                           & 
\nodata                                           & 
\nodata                                           & 
\nodata                                           & 
\nodata                                           & 
\nodata                                           & 
{$<0.16$}                                         \\

417                                               & 
$\tau^+ \tau^-$                                   & 
$<4100$                                           & 
{$<4100$}                                         & 
\nodata                                           & 
\nodata                                           & 
\nodata                                           & 
\nodata                                           & 
\nodata                                           & 
\nodata                                           & 
{$<4100$}                                         \\

432                                               & 
$e^\pm \mu^\mp$                                   & 
$<0.064$                                          & 
{$<0.092$}                                        & 
$<0.17$                                           & 
$<1.5$                                            & 
$<0.064$                                          & 
\nodata                                           & 
\nodata                                           & 
\nodata                                           & 
$<0.064$                                          \\

438                                               & 
$e^\pm \tau^\mp$                                  & 
$<28$                                             & 
{$<28$}                                           & 
\nodata                                           & 
{$<110$ }                                         & 
\nodata                                           & 
\nodata                                           & 
\nodata                                           & 
\nodata                                           & 
{$<28$}                                           \\

439                                               & 
$\mu^\pm \tau^\mp$                                & 
$<22$                                             & 
{$<22$}                                           & 
\nodata                                           & 
{$<38$}                                           & 
\nodata                                           & 
\nodata                                           & 
\nodata                                           & 
\nodata                                           & 
{$<22$}                                           \\

440                                               & 
$\nu \bar\nu$                                     & 
$<220$                                            & 
{$<220$}                                          & 
{\blue $<130$}                                    & 
\nodata                                           & 
\nodata                                           & 
\nodata                                           & 
\nodata                                           & 
\nodata                                           & 
{ $<130$}                                         \\

441                                               & 
$\nu \bar\nu \gamma$                              & 
$<47$                                             & 
{$<47$}                                           & 
\nodata                                           & 
\nodata                                           & 
\nodata                                           & 
\nodata                                           & 
\nodata                                           & 
\nodata                                           & 
{$<47$}                                           \\

\hline
\end{tabular}
}
~\dag This result has been averaged with the earlier PRL 97, 251802 (2006).
\end{center}
\end{table}

\mysubsection{$B\to X_s\gamma$}
\label{sec:btosg}
\newcommand{\BFcnv}{\mathcal{B}^{\mathrm{cnv}}}
\newcommand{\Egamma}{E_{\gamma}}
\newcommand{\Emin}{E_{\mathrm{min}}}
\newcommand{\bsgammaaverage}{355 \pm 24 \pm 9}

The decay $b \to s\gamma$ proceeds through a process of 
flavor changing neutral current. Since the charged Higgs or SUSY particles may
contribute in the penguin loop, the branching fraction is sensitive to physics
beyond the Standard Model. Experimentally, the branching fraction is measured
using either a semi-inclusive or an inclusive approach. A minimum 
photon energy requirement is applied in the analysis and the branching fraction
is corrected based on the theoretical model for the photon energy spectrum 
(shape function). Where there are multiple experimental results from an 
experiment, we use only the ones that are independent for \babar\ and Belle
to avoid dealing with correlated errors. Furthermore, the 
model uncertainties from the shape function should be highly 
correlated but no proper action was made in our older averages. 
To perform the average with better precision and good accuracy, it is 
important to use as many experimental 
results as possible and to handle the shape function issue in a proper
way. In this note, we report the updated average of $b\to s\gamma$ branching
fraction by implementing a common  shape function.  

Several shape function schemes are commonly used.
Usually one is chosen to obtain the extrapolation factor, 
defined as the ratio of the $b\to s\gamma$ branching fractions 
with minimum photon energies above and at 1.6 GeV,
and the difference between various schemes are treated as the 
model uncertainty. O. Buchm\"uller and H. Fl\"acher have calculated the
extrapolation factors \cite{Buchmuller:2005zv}.
Table \ref{tab:factor} lists the extrapolation factors with various photon
energy cuts for three different schemes and the average. The appropriate
approach to average the experimental results is to first convert them 
according to the average extrapolation factors and then perform the average,
assuming that the errors of the extrapolation factors are 100\% correlated. 

\begin{table}[!htb]
\caption{Extrapolation factor in various scheme with various minimum
   photon energy requirement (in GeV).}
\resizebox{0.99\textwidth}{!}{
\begin{tabular}{lccccc} \hline\hline
Scheme & $E_\gamma < 1.7$ & $E_\gamma < 1.8$ & $E_\gamma < 1.9$ & 
$E_\gamma < 2.0$ & $E_\gamma < 2.242$ \\ \hline
Kinetic & $0.986\pm 0.001$ & $0.968\pm 0.002$ & $0.939\pm 0.005$ & $0.903\pm0.009$ & $0.656\pm 0.031$ \\
Neubert SF & $0.982\pm 0.002$ & $0.962\pm 0.004$ & $0.930\pm 0.008$ & $0.888\pm 0.014$ & $0.665\pm 0.035$ \\
Kagan-Neubert & $0.988\pm 0.002$ & $0.970\pm 0.005$ & $0.940\pm 0.009$ & $0.892\pm 0.014$ & $0.643\pm 0.033$\\ \hline
Average & $0.985\pm 0.004$ & $0.967\pm 0.006$ & $0.936\pm 0.010$ & $0.894\pm 0.016$ &  $0.655\pm 0.037$ \\ \hline
\end{tabular}
}
\label{tab:factor}
\end{table}        
                  
After surveying all available experimental results, the six shown in 
Table \ref{tab:measurement} are selected for the average.  They
have provided in their papers either the $b\to s\gamma$ branching fraction at 
a certain photon energy cut or the extrapolation factor used.
Therefore we are able to convert them to the values at $E_{\rm min}= 1.6$ 
GeV using the information in Table \ref{tab:factor}.  
In the inclusive and full hadronic tag analysis, a possible $B\to X_d\gamma$
contamination has been considered according to the expectation
$(4.5\pm 0.3)$\%.  Compared to the other systematic uncertainties, the error 
that arises from the $B\to X_d\gamma$ fraction is too small to be considered.
We perform the average assuming that the systematic errors of the shape 
function and the $d\gamma$ fraction are correlated, and the other systematic
errors and the statistical errors are Gaussian and uncorrelated.     
The obtained average is 
${\cal B}(B\to X_s\gamma) = (\bsgammaaverage)\times 10^{-6}$ with
a $\chi^2$/DOF$= 0.85/5$, where the
errors are combined statistical and systematic, and systematic due to the shape 
function. The second error is estimated to
be the difference of the average after simultaneously varying the central
value of each experimental result by $\pm 1\sigma$. Although  a small fraction
of events was used in multiple analyses in the same experiment, 
we neglect their statistical correlations. Some
other correlated systematic errors, such as photon detection and the background
suppression, are not considered in our new average. 

\begin{table}[!htb]
\caption{Reported branching fraction, minimum photon energy, branching fraction
at minimum photon energy  and converted branching fraction $\BFcnv$ for the 
decay $b\to s\gamma$. All the branching fractions are in units of $10^{-6}$.
The errors are, in order, statistical, systematic and theoretical (if exists)
for $\cal{B}$, and statistical, systematic and shape-function systematic
for $\BFcnv$.
Theoretical errors in $\cal{B}(\Egamma>\Emin)$ are merged
into the systematic error of $\BFcnv$ during conversion.
The CLEO measurement on the branching fraction at $\Emin$ includes
$B\to X_d \gamma$ events.\newline
} 
\resizebox{0.99\textwidth}{!}{
\begin{tabular}{lccccc} \hline\hline
Mode & Reported $\cal{B}$ & $E_{\rm min}$ & $\cal{B}$ at $E_{\rm min}$ & Modified ${\cal{B}}~(E_{\rm min}=1.6)$ \\ \hline
   CLEO Inc. \cite{Chen:2001fja} & $321 \pm 43\pm 27^{+18}_{-10}$ & 2.0
           & $306\pm 41\pm 26$ & $327 \pm 44\pm 28 \pm 6$ \\
   Belle Semi.\cite{Abe:2001hk} & $336\pm 53 \pm 42^{+50}_{-54}$ & 2.24
           & $-$ & $369\pm 58 \pm 46^{+56}_{-60}$\\
   \babar\ Semi.\cite{Aubert:2005cua} & $335\pm 19^{+56+4}_{-41-9}$ & 1.9
           & $327\pm 18^{+55+4}_{-40-9}$ & $349\pm 20^{+59+4}_{-46-3}$ \\
   \babar\ Inc. \cite{Aubert:2006gg} & $-$ & 1.9
           & $367\pm 29\pm 34\pm 29$ & $390\pm 31\pm 47 \pm 4$ \\
   \babar\ Full \cite{Aubert:2007my} & $391\pm 91 \pm 64$ & 1.9
           & $366\pm 85\pm 60$ & $389 \pm 91\pm 64\pm 4$ \\
   Belle Inc.\cite{Limosani:2009qg} & $-$ & 1.7
           & $345\pm 15\pm 40$ & $347\pm 15 \pm 40\pm 1$ \\ \hline
   Average & & & & $\bsgammaaverage$ \\ \hline \hline

\end{tabular}
}
\label{tab:measurement}
\end{table}

\mysubsection{Baryonic decays}

\begin{table}
\begin{center}
\caption{
 Branching fractions of  baryonic $B^+$ decays (in units of $10^{-6}$).
Upper limits are at 90\% CL.
values in {\red red} ({\blue blue}) are new {\red published}
({\blue preliminary}) results since PDG2010  [as of March 12, 2012].
}
\end{center} 
\footnotesize
\begin{center}
\vspace{-0.3cm}
\resizebox{0.99\textwidth}{!}{
\begin{tabular}{|lcccccc|} 
\sgline
RPP\#   & Mode & PDG2010 Avg. & BABAR & Belle & CLEO & New Avg. \\
\sgline
368                                               & 
$p \overline p \pi^+$                             & 
$1.62\pm0.20$                                     & 
{$\err{1.69}{0.29}{0.26}~\dag$}                   & 
{$\aerr{1.57}{0.17}{0.15}{0.12}~\S$}              & 
$<160$                                            & 
$\cerr{1.60}{0.18}{0.17}$                         \\

371                                               & 
$p \overline p K^+$                               & 
$5.9\pm0.5$                                       & 
{$\err{6.7}{0.5}{0.4}~\dag$}                      & 
{$\aerr{5.00}{0.24}{0.22}{0.32}~\S$}              & 
\nodata                                           & 
$5.48 \pm 0.34$                                   \\

372                                               & 
$\Theta^{++} \overline p$ $^1$                    & 
$<0.091$                                          & 
{$<0.09$}                                         & 
{$<0.091$}                                        & 
\nodata                                           & 
{$<0.09$}                                         \\

373                                               & 
$f_J(2221) K^+$ $^2$                              & 
$<0.41$                                           & 
\nodata                                           & 
{$<0.41$}                                         & 
\nodata                                           & 
{$<0.41$}                                         \\

374                                               & 
$p \overline\Lambda(1520)$                        & 
$< 1.5$                                           & 
{$< 1.5$}                                         & 
\nodata                                           & 
\nodata                                           & 
{$< 1.5$}                                         \\

376                                               & 
$p \overline p K^{*+}$                            & 
$\cerr{3.6}{0.8}{0.7}$                            & 
{$\err{5.3}{1.5}{1.3}~\dag$}                      & 
{$\aerr{3.38}{0.73}{0.60}{0.39}~\ddag$}           & 
\nodata                                           & 
$\cerr{3.64}{0.79}{0.70}$                         \\

377                                               & 
$f_J(2221) K^{*+}$ $^2$                           & 
$<0.77$                                           & 
{$<0.77$}                                         & 
\nodata                                           & 
\nodata                                           & 
{$<0.77$}                                         \\

378                                               & 
$p \overline\Lambda$                              & 
$<0.32$                                           & 
\nodata                                           & 
{$< 0.32$}                                        & 
$< 1.5$                                           & 
{$< 0.32$}                                        \\

380                                               & 
$p \overline\Lambda \pi^0$                        & 
$\cerr{3.00}{0.7}{0.6}$                           & 
\nodata                                           & 
{$\aerr{3.00}{0.61}{0.53}{0.33}$}                 & 
\nodata                                           & 
$\cerr{3.00}{0.69}{0.62}$                         \\

381                                               & 
$p \overline\Sigma(1385)^0$                       & 
$<0.47$                                           & 
\nodata                                           & 
$<0.47$                                           & 
\nodata                                           & 
$<0.47$                                           \\

382                                               & 
$\Delta^+\overline \Lambda$                       & 
$<0.82$                                           & 
\nodata                                           & 
$<0.82$                                           & 
\nodata                                           & 
$<0.82$                                           \\

384                                               & 
$p \overline{\Lambda} \pi^+\pi^-$ (NR)            & 
$5.9\pm1.1$                                       & 
\nodata                                           & 
$\aerr{5.92}{0.88}{0.84}{0.69}$                   & 
\nodata                                           & 
$\cerr{5.92}{1.12}{1.09}$                         \\

385                                               & 
$p \overline{\Lambda} \rho^0$                     & 
$4.8\pm0.9$                                       & 
\nodata                                           & 
$\aerr{4.78}{0.67}{0.64}{0.60}$                   & 
\nodata                                           & 
$\cerr{4.78}{0.90}{0.88}$                         \\

386                                               & 
$p \overline{\Lambda} f_2(1270)$                  & 
$2.0\pm0.8$                                       & 
\nodata                                           & 
$\aerr{2.03}{0.77}{0.72}{0.27}$                   & 
\nodata                                           & 
$\cerr{2.03}{0.82}{0.77}$                         \\

387                                               & 
$\Lambda \overline{\Lambda} \pi^+$                & 
$<0.94$                                           & 
\nodata                                           & 
$<0.94~\S$                                        & 
\nodata                                           & 
$<0.94~\S$                                        \\

388                                               & 
$\Lambda \overline{\Lambda} K^+$                  & 
$3.4\pm0.6$                                       & 
\nodata                                           & 
$\aerr{3.38}{0.41}{0.36}{0.41}~\ddag$             & 
\nodata                                           & 
$\cerr{3.38}{0.58}{0.55}$                         \\

389                                               & 
$\Lambda \overline{\Lambda} K^{*+}$               & 
$\cerr{2.2}{1.2}{0.9}$                            & 
\nodata                                           & 
$\aerr{2.19}{1.13}{0.88}{0.33}~\S$                & 
\nodata                                           & 
$\cerr{2.19}{1.18}{0.94}$                         \\

390                                               & 
$\overline{\Delta}^0 p$                           & 
$<1.38$                                           & 
\nodata                                           & 
{$<1.38$} $\S$                                    & 
$<380$                                            & 
{$<1.38$} $\S$                                    \\

391                                               & 
$\Delta^{++} \overline p$                         & 
$<0.14$                                           & 
\nodata                                           & 
{$<0.14$} $\S$                                    & 
$<150$                                            & 
{$<0.14$} $\S$                                    \\


\hline
\end{tabular}
}
\end{center}
\S Di-baryon mass is less than 2.85 GeV/$c^2$; 
$\dag$ Charmonium decays to $p\bar p$ have been statistically subtracted;\\
$\ddag$ The charmonium mass region has been vetoed;
$^1~\Theta(1540)^{++}\to K^+p$ (pentaquark candidate); \\
$^2$ Product BF --- daughter BF taken to be 100\% \\
\end{table}

\begin{table}
\begin{center}
\caption{
 Branching fractions of  baryonic $B^+$ decays (in units of $10^{-6}$).
Upper limits are at 90\% CL.
values in {\red red} ({\blue blue}) are new {\red published}
({\blue preliminary}) results since PDG2010  [as of March 12, 2012].
}
\end{center} 
\footnotesize
\begin{center}
\vspace{-0.3cm}
\begin{tabular}{|lcccccc|} 
\sgline
RPP\#   & Mode & PDG2010 Avg. & BABAR & Belle & CLEO & New Avg. \\
\sgline
366                                               & 
$p \overline{p}$                                  & 
$< 0.11$                                          & 
{$<0.27$}                                         & 
{$<0.11$}                                         & 
$<1.4$                                            & 
{$<0.11$}                                         \\

368                                               & 
$p \overline{p} K^0$                              & 
$2.66\pm0.32$                                     & 
{$\err{3.0}{0.5}{0.3}~\dag$}                      & 
$\aerr{2.51}{0.35}{0.29}{0.21}~\ddag$             & 
\nodata                                           & 
$\cerr{2.66}{0.34}{0.32}$                         \\

369                                               & 
$\Theta^+ \overline{p}$~$^1$                      & 
$<0.05$                                           & 
{$<0.05$}                                         & 
{$<0.23$}                                         & 
\nodata                                           & 
{$<0.05$}                                         \\

370                                               & 
$f_J(2221) K^0$ $^2$                              & 
$<0.45$                                           & 
{$<0.45$}                                         & 
\nodata                                           & 
\nodata                                           & 
{$<0.45$}                                         \\

371                                               & 
$p \overline{p} K^{*0}$                           & 
$\cerr{1.24}{0.28}{0.25}$                         & 
{$\err{1.47}{0.45}{0.40}~\dag$}                   & 
$\aerr{1.18}{0.29}{0.25}{0.11}~\ddag$             & 
\nodata                                           & 
$\cerr{1.24}{0.28}{0.25}$                         \\

372                                               & 
$f_J(2221) K^{*0}$ $^2$                           & 
$<0.15$                                           & 
{$<0.15$}                                         & 
\nodata                                           & 
\nodata                                           & 
{$<0.15$}                                         \\

373                                               & 
$p \overline\Lambda \pi^-$                        & 
$3.14\pm0.29$                                     & 
$\err{3.07}{0.31}{0.23}$                          & 
{$\aerr{3.23}{0.33}{0.29}{0.29}$}                 & 
$<13$                                             & 
$\cerr{3.14}{0.29}{0.28}$                         \\

374                                               & 
$p \overline\Sigma(1385)^-$                       & 
$<0.26$                                           & 
\nodata                                           & 
$<0.26$                                           & 
\nodata                                           & 
$<0.26$                                           \\

375                                               & 
$\Delta^0 \overline\Lambda$                       & 
$<0.93$                                           & 
\nodata                                           & 
$<0.93$                                           & 
\nodata                                           & 
$<0.93$                                           \\

376                                               & 
$p \overline\Lambda K^-$                          & 
$<0.82$                                           & 
\nodata                                           & 
$< 0.82$                                          & 
\nodata                                           & 
$< 0.82$                                          \\

377                                               & 
$p \overline\Sigma^0 \pi^-$                       & 
$<3.8$                                            & 
\nodata                                           & 
$< 3.8$                                           & 
\nodata                                           & 
$< 3.8$                                           \\

340                                               & 
$\overline\Lambda \Lambda$                        & 
$<0.32$                                           & 
\nodata                                           & 
{$<0.32$}                                         & 
$<1.2$                                            & 
{$<0.32$}                                         \\

379                                               & 
$\overline\Lambda \Lambda K^0$                    & 
$\cerr{4.8}{1.0}{0.9}$                            & 
\nodata                                           & 
$\aerr{4.76}{0.84}{0.68}{0.61}~\ddag$             & 
\nodata                                           & 
$\cerr{4.76}{1.04}{0.91}$                         \\

380                                               & 
$\Lambda \overline{\Lambda} K^{*0}$               & 
$\cerr{2.5}{0.9}{0.8}$                            & 
\nodata                                           & 
$\aerr{2.46}{0.87}{0.72}{0.34}~\ddag$             & 
\nodata                                           & 
$\cerr{2.46}{0.93}{0.80}$                         \\

            
\hline
\end{tabular}
\end{center}
$\dag$ Charmonium decays to $p\bar p$ have been statistically subtracted;
$\ddag$ The charmonium mass region has been vetoed;
$^1~\Theta(1540)^+\to p K^0$ (pentaquark candidate);
$^2$ Product BF --- daughter BF taken to be 100\%.
\end{table}
 
\clearpage

\mysubsection{$B_s$ decays}

\begin{table}[h]
\begin{center}
\caption{
 $B_s$  branching fractions (in units of $10^{-6}$). 
Upper limits are at 90\% CL.
Values in {\red red} ({\blue blue}) are new {\red published}
({\blue preliminary}) results since PDG2010  [as of March 12, 2012].
}
\vskip 0.25cm
\scriptsize
\resizebox{0.99\textwidth}{!}{
\begin{tabular}{|lcccccccc|} \hline
RPP\# &Mode & PDG2010 Avg. &  Belle & CDF & D0 & LHCb & CMS & New Avg.  \\ \sglinespb
$15$                                              & 
$\pi^+\pi^-$                                      & 
$<1.2 $                                           & 
\red$<12$                                         & 
\blue$\err{0.57}{0.15}{0.10}$$\dag$               & 
\nodata                                           & 
\blue{$\aerr{0.98}{0.23}{0.19}{0.11}$}            & 
\nodata                                           & 
$0.73 \pm 0.14$                                   \\

$21$                                              & 
$\phi\phi$                                        & 
$14 \pm 8$                                        & 
\nodata                                           & 
\red $23.2\pm1.8\pm8.2$ $\dag$                    & 
\nodata                                           & 
\nodata                                           & 
\nodata                                           & 
$23.2 \pm 8.4$                                    \\

$22$                                              & 
$\pi^+K^-$                                        & 
$4.9\pm1.0$                                       & 
\red$<26$                                         & 
$5.0\pm0.7\pm0.8$$\dag$                           & 
\nodata                                           & 
\nodata                                           & 
\nodata                                           & 
$5.0 \pm 1.1$                                     \\
\hline
\end{tabular}
}
\end{center}
\vspace{-0.3cm}
$\dag$Relative BF converted to absolute BF
\end{table}

\begin{table}[h]
\begin{center}

\caption{
 $B_s$ rare relative branching fractions.
Values in {\red red} ({\blue blue}) are new {\red published}
({\blue preliminary}) results since PDG2010  [as of March 12, 2012].
}
\vskip 0.25cm

\scriptsize

\resizebox{0.99\textwidth}{!}{
\begin{tabular}{|lccccc|} \hline
RPP\# & Mode & PDG2010 Avg. &  CDF & D0 & New Avg.  \\ \sglinespb
$13$                                              & 
$\it{f}_s\mathcal{B}(B^0_s\rightarrow\pi^+\pi^-)/\it{f}_d\mathcal{B}(B^0\rightarrow K^+\pi-)$& 
\nodata                                           & 
\blue$0.008\pm0.002\pm0.001$                      & 
\nodata                                           & 
$0.008 \pm 0.002$                                 \\

$19$                                              & 
$\mathcal{B}(B^0_s\rightarrow\phi\phi)/\mathcal{B}(B^0_s\rightarrow J/\psi\phi)$& 
\nodata                                           & 
\red$(1.78\pm0.14\pm0.20)\times 10^{-2} $         & 
\nodata                                           & 
$1.78 \pm 0.24$                                   \\

$20$                                              & 
$\it{f}_s\mathcal{B}(B^0_s\rightarrow K^+\pi^-)/\it{f}_d\mathcal{B}(B^0_d\rightarrow K^+\pi^-)$& 
\nodata                                           & 
$0.071\pm0.010\pm0.007 $                          & 
\nodata                                           & 
$0.071 \pm 0.012$                                 \\

$21$                                              & 
$\it{f}_s\mathcal{B}(B^0_s\rightarrow K^+K^-)/\it{f}_d\mathcal{B}(B^0_d\rightarrow K^+\pi^-)$& 
\nodata                                           & 
\red$0.347\pm0.020\pm0.021 $                      & 
\nodata                                           & 
$0.347 \pm 0.029$                                 \\
\hline
\end{tabular}
}
\end{center}
\end{table}

\mysubsection{Charge asymmetries}

\begin{sidewaystable}[!htbp]
\parbox{8.2in}{\caption{
 $CP$ asymmetries for charmless hadronic charged $B$ decays (part I).
Values in {\red red} ({\blue blue}) are 
new {\red published}
({\blue preliminary}) results since PDG2010  [as of March 12, 2012].
}}

\scriptsize
\begin{tabular}{|lccccccc|} 
\hline
RPP\#   & Mode & PDG2010 Avg. & BABAR & Belle & CLEO & CDF & New Avg. \\
\sglinespb
220                                               & 
$K^0 \pi^+$                                       & 
$0.009\pm 0.029$                                  & 
{$\err{-0.029}{0.039}{0.010}$}                    & 
{\blue $\err{-0.014}{0.012}{0.006}$}              & 
$\err{0.18}{0.24}{0.02}$                          & 
\nodata                                           & 
$-0.015 \pm 0.012$                                \\

221                                               & 
$K^+ \pi^0$                                       & 
$0.051\pm 0.025$                                  & 
{$\err{0.030}{0.039}{0.010}$}                     & 
{$\err{0.043}{0.024}{0.002}$}                     & 
$\err{-0.29}{0.23}{0.02}$                         & 
\nodata                                           & 
$0.037 \pm 0.021$                                 \\
\end{tabular}
\end{sidewaystable}

\begin{sidewaystable}[!htbp]
\begin{center}
\parbox{7.5in}{\caption{$CP$ asymmetries for 
charmless hadronic charged  $B$ decays (part II).
Values in {\red red} ({\blue blue}) are new {\red published}
({\blue preliminary}) results since PDG2010  [as of March 12, 2012].
}}

\scriptsize
\begin{tabular}{|lccccccc|} 
\sgline
RPP\#   & Mode & PDG2010 Avg. & BABAR & Belle & CLEO & CDF & New Avg. \\
\sglinespb
329                                               & 
$\rho^+ \gamma$                                   & 
$-0.11 \pm 0.33$                                  & 
\nodata                                           & 
{$\err{-0.11}{0.32}{0.09}$}                       & 
\nodata                                           & 
\nodata                                           & 
$-0.11 \pm 0.33$                                  \\

330                                               & 
$\pi^+ \pi^0$                                     & 
$0.06\pm 0.05$                                    & 
{$\err{0.03}{0.08}{0.01}$}                        & 
{\blue $\err{0.025}{0.043}{0.007}$}               & 
\nodata                                           & 
\nodata                                           & 
$0.026 \pm 0.039$                                 \\
\end{tabular}
\end{center}
\end{sidewaystable}

\begin{sidewaystable}[!htbp]
\begin{center}
\parbox{7.5in}{\caption{$CP$ asymmetries for 
charmless hadronic neutral $B$ decays.
Values in {\red red} ({\blue blue}) are new {\red published}
({\blue preliminary}) results since PDG2010  [as of March 12, 2012].
}}

\scriptsize
\begin{tabular}{|lccccccc|} 
\sgline
RPP\#   & Mode & PDG2010 Avg. & BABAR & Belle & CDF & LHCb & New Avg. \\
\sglinespb
210                                               & 
$K^+ \pi^-$                                       & 
$-0.098\pm 0.013$                                 & 
{\blue $\berr{-0.107}{0.016}{0.006}{0.004}$}      & 
{$\err{-0.069}{0.014}{0.007}$}                    & 
{\red $\err{-0.086}{0.023}{0.009}$}               & 
{\blue $\err{-0.088}{0.011}{0.008}$}              & 
$-0.087 \pm 0.008$                                \\

213                                               & 
$\etapr K^{*0}$                                   & 
$\err{0.08}{0.25}{0.02}$                          & 
\red{$\err{0.02}{0.23}{0.02}$}                    & 
\nodata                                           & 
\nodata                                           & 
\nodata                                           & 
$0.02 \pm 0.23$                                   \\
\end{tabular}
\end{center}

\vspace{-0.3cm}
\dag~Measurements of time-dependent $CP$ asymmetries are listed in the section
of the Unitarity Triangle.

\end{sidewaystable}

\begin{sidewaystable}[!htbp]
\begin{center}

\parbox{7.5in}{\caption{
Charmless hadronic $CP$ asymmetries for $B^\pm/B^0$ admixtures.
Values in {\red red} ({\blue blue}) are new {\red published}
({\blue preliminary}) results since PDG2010  [as of March 12, 2012].
}}

\vspace*{ 0.4cm}
\scriptsize
\hspace*{-1.4cm}
\begin{tabular}{|lcccccc|}
\hline
RPP\#    & Mode & PDG2010 Avg. & BABAR & Belle & CLEO & New Avg. \\
\sglinespb
~65                                               & 
$K^* \gamma$                                      & 
$-0.003\pm 0.017$                                 & 
{$-0.003\pm 0.017\pm 0.007$}                      & 
{$-0.015\pm 0.044\pm 0.012$}                      & 
$0.08\pm 0.13\pm 0.03$                            & 
$-0.003 \pm 0.017$                                \\

~7                                                & 
$s \gamma$                                        & 
$0.014\pm0.028$                                   & 
$\err{-0.011}{0.030}{0.014}$                      & 
$\err{0.002}{0.050}{0.030}$                       & 
$\err{-0.079}{0.108}{0.022}$                      & 
$-0.012 \pm 0.028$                                \\

$~-$                                              & 
$(s+d) \gamma$                                    & 
$\err{-0.110}{0.115}{0.017}$                      & 
$\err{-0.11}{0.12}{0.02}$                         & 
\nodata                                           & 
\nodata                                           & 
$-0.11 \pm 0.12$                                  \\
\hline
\end{tabular}
\end{center}
\vspace{-0.3cm}
~~~~~~~~~~~~~~~~~~$\dag~p^* > 2.34$ GeV;
~$\S~0.4 < M_{X_s} < 2.6$ GeV;
\vspace{1cm}
\begin{center}
\parbox{7.5in}{\caption{
 $CP$ asymmetries for charmless hadronic $B_s$ decays.
Values in {\red red} ({\blue blue}) are new {\red published}
({\blue preliminary}) results since PDG2010  [as of March 12, 2012].
}}

\vspace*{ 0.4cm}
\begin{tabular}{|lcccccc|}
\hline
RPP\#    & Mode & PDG2010 Avg. & Belle & CDF & LHCb & New Avg. \\
\sglinespb
~22                                               & 
$K^+ \pi^-$                                       & 
New                                               & 
\nodata                                           & 
\red$\err{0.39}{0.15}{0.08}$                      & 
\blue$\err{0.27}{0.08}{0.02}$                     & 
$0.29 \pm 0.07$                                   \\

\hline
\end{tabular}
\end{center}
\end{sidewaystable}

\clearpage

\mysubsection{Polarization measurements}

\begin{table}[!htbp]
\caption{
 Longitudinal polarization fraction $f_L$ for $B^+$ decays.
Values in {\red red} ({\blue blue}) are new {\red published}
({\blue preliminary}) results since PDG2010 [as of March 12, 2012].
\vspace{0.3cm}
}
\small
\begin{center}
\resizebox{0.99\textwidth}{!}{
\begin{tabular}{|lccccc|} 
\sgline
RPP\#   & Mode & PDG2010 Avg. & BABAR & Belle & New Avg. \\
\sglinespb
237                                               & 
$\omega K^{*+}$                                   & 
$\err{0.41}{0.18}{0.05}$                          & 
$\err{0.41}{0.18}{0.05}$                          & 
\nodata                                           & 
$0.41 \pm 0.19$                                   \\

240                                               & 
$\omega K_2^*(1430)^+$                            & 
$\err{0.56}{0.10}{0.04}$                          & 
$\err{0.56}{0.10}{0.04}$                          & 
\nodata                                           & 
$0.56 \pm 0.11$                                   \\

267                                               & 
$K^{*+} \rho^0$                                   & 
\nodata                                           & 
\red{$\err{0.78}{0.12}{0.03}$}                    & 
\nodata                                           & 
$0.78 \pm 0.12$                                   \\

271                                               & 
$K^{*0} \rho^+$                                   & 
$0.48\pm0.08$                                     & 
{$\err{0.52}{0.10}{0.04}$}                        & 
{$\berr{0.43}{0.11}{0.05}{0.02}$}                 & 
$0.48 \pm 0.08$                                   \\
\end{tabular}
}
\end{center}
\end{table}

\begin{table}[!htbp]
\caption{
Full angular analysis of $B^+ \to \phi K^{*+}$.
Values in {\red red} ({\blue blue}) are new {\red published}
({\blue preliminary}) results since PDG2010 [as of March 12, 2012].
\vspace{0.3cm}
}

\begin{center}
\begin{tabular}{|lccccc|} 
\sgline
\RPP & Parameter & PDG2010 Avg. & BABAR & Belle & New Avg. \\
\sglinespb
\nodata                                           & 
$f_\perp = \Lambda_{\perp\perp}$                  & 
$0.20\pm0.05$                                     & 
{$\err{0.21}{0.05}{0.02}$}                        & 
{$\err{0.19}{0.08}{0.02}$}                        & 
$0.20 \pm 0.05$                                   \\

\nodata                                           & 
$\phi_\parallel$                                  & 
$2.34\pm0.18$                                     & 
{$\err{2.47}{0.20}{0.07}$}                        & 
{$\err{2.10}{0.28}{0.04}$}                        & 
$2.34 \pm 0.17$                                   \\

\nodata                                           & 
$\phi_\perp$                                      & 
$2.58\pm0.17$                                     & 
{$\err{2.69}{0.20}{0.03}$}                        & 
{$\err{2.31}{0.30}{0.07}$}                        & 
$2.58 \pm 0.17$                                   \\

\nodata                                           & 
$\delta_0$                                        & 
{$\err{3.07}{0.18}{0.06}$}                        & 
{$\err{3.07}{0.18}{0.06}$}                        & 
\nodata                                           & 
$3.07 \pm 0.19$                                   \\
\end{tabular}
\end{center}
\vspace*{-0.3cm}
\hspace*{1.0cm}BR, $f_L$ and $A_{CP}$ are tabulated separately.
\end{table}
\clearpage

\begin{table}[!htbp]
\caption{
Longitudinal polarization fraction $f_L$ for $B^0$ decays.
Values in {\red red} ({\blue blue}) are new {\red published}
({\blue preliminary}) results since PDG2010  [as of March 12, 2012].
\vspace{0.3cm}
}

\begin{center}
\footnotesize
\begin{tabular}{|lccccc|} 
\sgline
RPP\#   & Mode & PDG2010 Avg. & BABAR & Belle & New Avg. \\
\hline
227                                               & 
$\omega K^{*0}$                                   & 
$0.69\pm0.13$                                     & 
$\err{0.72}{0.14}{0.02}$                          & 
$\berr{0.56}{0.29}{0.18}{0.08}$                   & 
$0.70 \pm 0.13$                                   \\

230                                               & 
$\omega K_2^*(1430)^0$                            & 
$\err{0.45}{0.12}{0.02}$                          & 
$\err{0.45}{0.12}{0.02}$                          & 
\nodata                                           & 
$0.45 \pm 0.12$                                   \\

260                                               & 
$K^{*0} \rho^0$                                   & 
$\err{0.57}{0.09}{0.08}$                          & 
\blue{$\err{0.40}{0.08}{0.11}$}                   & 
\nodata                                           & 
$0.40 \pm 0.14$                                   \\
\end{tabular}
\end{center}
\end{table}

\begin{table}[!htbp]
\caption{
Full angular analysis of $B^0 \to \phi K^{*0}$.
Values in {\red red} ({\blue blue}) are new {\red published}
({\blue preliminary}) results since PDG2010  [as of March 12, 2012].
\vspace{0.3cm}
}

\begin{center}
\small
\begin{tabular}{|lccccc|} 
\sgline
\RPP & Parameter & PDG2010 Avg. & BABAR & Belle & New Avg. \\
\sglinespb
\nodata                                           & 
$f_\perp = \Lambda_{\perp\perp}$                  & 
$0.24 \pm0.05$                                    & 
{$\err{0.212}{0.032}{0.013}$}                     & 
{$\aerr{0.31}{0.06}{0.05}{0.02}$}                 & 
$0.241 \pm 0.029$                                 \\

\nodata                                           & 
$\phi_\parallel$                                  & 
$2.40\pm 0.13$                                    & 
{$\err{2.40}{0.13}{0.08}$}                        & 
{$\aerr{2.40}{0.28}{0.24}{0.07}$}                 & 
$\cerr{2.40}{0.14}{0.13}$                         \\

\nodata                                           & 
$\phi_\perp$                                      & 
$2.39\pm0.13$                                     & 
{$\err{2.35}{0.13}{0.09}$}                        & 
{$\err{2.51}{0.25}{0.06}$}                        & 
$2.39 \pm 0.13$                                   \\

\nodata                                           & 
$\delta_0$                                        & 
$\err{2.82}{0.15}{0.09}$                          & 
{$\err{2.82}{0.15}{0.09}$}                        & 
\nodata                                           & 
$2.82 \pm 0.17$                                   \\
\end{tabular}
\end{center}
\vspace{-0.3cm}
\hspace{1.3cm}
BR, $f_L$ and $A_{CP}$ are tabulated separately.
\end{table}

\begin{table}[!htbp]
\caption{
Full angular analysis of $B^0 \to \phi K_2^*(1430)^0$.
Values in {\red red} ({\blue blue}) are new {\red published}
({\blue preliminary}) results since PDG2010  [as of March 12, 2012].
\vspace{0.3cm}
}

\begin{center}
\begin{tabular}{|lccccc|} 
\sgline
\RPP & Parameter & PDG2010 Avg. & BABAR & Belle & New Avg. \\
\sglinespb
\nodata                                           & 
$f_\perp = \Lambda_{\perp\perp}$                  & 
$\aerr{0.002}{0.018}{0.002}{0.031}$               & 
{$\aerr{0.002}{0.018}{0.002}{0.031}$}             & 
\nodata                                           & 
$\cerr{0.002}{0.036}{0.031}$                      \\

\nodata                                           & 
$\phi_\parallel$                                  & 
$\err{3.96}{0.38}{0.06}$                          & 
{$\err{3.96}{0.38}{0.06}$}                        & 
\nodata                                           & 
$3.96 \pm 0.39$                                   \\

\nodata                                           & 
$\delta_0$                                        & 
$\err{3.41}{0.13}{0.13}$                          & 
{$\err{3.41}{0.13}{0.13}$}                        & 
\nodata                                           & 
$3.41 \pm 0.18$                                   \\
\end{tabular}
\end{center}
\vspace{-0.3cm}
\hspace{2.3cm}
BR, $f_L$ and $A_{CP}$ are tabulated separately.
\end{table}

\begin{table}[!htbp]
\caption{
 Longitudinal polarization fraction $f_L$ for $B_s$ decays.
Values in {\red red} ({\blue blue}) are new {\red published}
({\blue preliminary}) results since PDG2010 [as of March 12, 2012].
\vspace{0.3cm}
}

\begin{center}
\small
\begin{tabular}{|lccccc|} 
\sgline
RPP\#   & Mode & PDG2010 Avg. & CDF & LHCB & New Avg. \\
\hline
$21$                                              & 
$\phi\phi$                                        & 
New                                               & 
\red{$\err{0.348}{0.041}{0.021}$}                 & 
\nodata                                           & 
$0.348 \pm 0.046$                                 \\

$25$                                              & 
$K^{*0}\overline{K}^{*0}$                         & 
New                                               & 
\nodata                                           & 
\red{$\err{0.38}{0.11}{0.04}$}                    & 
$0.38 \pm 0.12$                                   \\

\sglinespt
\end{tabular}
\end{center}
\end{table}

\clearpage
\section{$D$ decays}
\label{sec:charm_physics}

\def\kbar{\overline{K}{}^{\,0}}
\def\dbar{\overline{D}{}^{\,0}}
\def\bbar{\overline{B}{}^{\,0}}
\def\cp{$CP$}
\def\cpv{$CPV$}
\def\ra{\!\rightarrow\!}
\def\ddbar{$D^0$-$\dbar$}
\def\ycp{$y^{}_{\rm CP}$}

\def\dklnu{$D^0\ra K^+\ell^-\nu$}
\def\dkpi{$D^0\ra K^+\pi^-$}
\def\dkk{$D^0\ra K^+K^-$}
\def\dpipi{$D^0\ra\pi^+\pi^-$}
\def\dkkpp{$D^0\ra K^+K^-/\pi^+\pi^-$}
\def\dkspp{$D^0\ra K^0_S\,\pi^+\pi^-$}
\def\dkskk{$D^0\ra K^0_S\,K^+ K^-$}

\def\dsphipi{$D^+_s\ra\phi\,\pi^+$}
\def\dsmunu{$D^+_s\ra\mu^+\nu$}
\def\dstaunu{$D^+_s\ra\tau^+\nu$}

\def\gevm{~GeV/$c^2$}
\def\gevp{~GeV/$c$}
\def\geve{~GeV}
\def\mevm{~MeV/$c^2$}
\def\meve{~MeV}


\def\simge{\mathrel{%
   \rlap{\raise 0.511ex \hbox{$>$}}{\lower 0.511ex \hbox{$\sim$}}}}
\def\simle{\mathrel{
   \rlap{\raise 0.511ex \hbox{$<$}}{\lower 0.511ex \hbox{$\sim$}}}}

\newcommand{\Dnan}{\ensuremath{D_0^\ast(2400)^0}}
\newcommand{\Dtan}{\ensuremath{D_2^\ast(2460)^0}}
\newcommand{\Don}{\ensuremath{D_1(2420)^{0}}}
\newcommand{\Dopn}{\ensuremath{D_1(2430)^{0}}}
\newcommand{\Dnap}{\ensuremath{D_0^\ast(2400)^\pm}}
\newcommand{\Dtap}{\ensuremath{D_2^\ast(2460)^\pm}}
\newcommand{\Dop}{\ensuremath{D_1(2420)^{\pm}}}
\newcommand{\Dopp}{\ensuremath{D_1(2430)^{\pm}}}

\newcommand{\Dsa}{\ensuremath{D_s^{\ast\pm}}}
\newcommand{\Dsna}{\ensuremath{D_{s0}^\ast(2317)^{\pm}}}
\newcommand{\Dsop}{\ensuremath{D_{s1}(2460)^{\pm}}}
\newcommand{\Dso}{\ensuremath{D_{s1}(2536)^{\pm}}}
\newcommand{\Dst}{\ensuremath{D_{s2}(2573)^{\pm}}}
\newcommand{\Dsts}{\ensuremath{D_{sJ}(2700)^{\pm}}}
\newcommand{\Dste}{\ensuremath{D_{sJ}(2860)^{\pm}}}
\newcommand{\Dstsi}{\ensuremath{D_{sJ}(2632)^{\pm}}}

\newcommand{\citep}{\cite}

\newcommand{\kst}{K^*(892)^0}
\newcommand{\akst}{\overline{K}^*(892)^0}
\newcommand{\kstp}{K^*(1410)^0}
\newcommand{\akstp}{\overline{K}^*(1410)^0}
\newcommand{\kstd}{K^*_2(1430)^0}
\newcommand{\akstd}{\overline{K}^*_2(1430)^0}

\newcommand{\ksts}{K^*_0(1430)^0}
\newcommand{\aksts}{\overline{K}^*_0(1430)^0}

\subsection{\emph{$D^0$-$\dbar$} mixing and \emph{\cp}\ violation}
\label{sec:charm:mixcpv}

\subsubsection{Introduction}

In 2007 Belle~\cite{Staric:2007dt} and \babar~\cite{Aubert:2007wf} 
obtained the first evidence for $D^0$-$\dbar$ mixing, which 
had been searched for for more than two decades. 
These results were later confirmed by CDF~\cite{Aaltonen:2007uc}.
There are now numerous measurements of $D^0$-$\dbar$ mixing 
with various levels of sensitivity. All the results are
input into a global fit to determine
world averages of mixing parameters, \cp-violation (\cpv) 
parameters, and strong phases.

Our notation is as follows.
The mass eigenstates are denoted 
$D^{}_1 = p|D^0\rangle-q|\dbar\rangle$ and
$D^{}_2 = p|D^0\rangle+q|\dbar\rangle$, 
where we use the convention 
$CP|D^0\rangle=-|\dbar\rangle$ and 
$CP|\dbar\rangle=-|D^0\rangle$. Thus in the absence of 
\cp\ violation, $D^{}_1$ is \cp-even and $D^{}_2$ is \cp-odd.
The weak phase $\phi\equiv {\rm Arg}(q/p)$.
The mixing parameters are defined as 
$x\equiv(m^{}_1-m^{}_2)/\Gamma$ and 
$y\equiv (\Gamma^{}_1-\Gamma^{}_2)/(2\Gamma)$, where 
$m^{}_1,\,m^{}_2$ and $\Gamma^{}_1,\,\Gamma^{}_2$ are
the masses and decay widths for the mass eigenstates,
and $\Gamma\equiv (\Gamma^{}_1+\Gamma^{}_2)/2$.

The global fit determines central values and errors for
ten underlying parameters. These consist of mixing parameters
$x$ and $y$; a parameter describing the ratio of decay rates
$R^{}_D\equiv\left|{\cal A}(D^0\ra K^+\pi^-)/
              {\cal A}(\dbar\ra K^+\pi^-)\right|^2$;
\cpv\ parameters $|q/p|$, $\phi$, and
$A^{}_D\equiv (R^+_D-R^-_D)/(R^+_D+R^-_D)$, where the $+\,(-)$
superscript corresponds to $D^0\,(\dbar)$ decays; 
{\it direct\/} \cpv\ parameters $A^{}_{KK}$ and 
$A^{}_{\pi\pi}$ (discussed below); 
the strong phase difference
$\delta$ between $\dbar\ra K^-\pi^+$ and 
$D^0\ra K^-\pi^+$ amplitudes; and 
the strong phase difference $\delta^{}_{K\pi\pi}$ between 
$\dbar\ra K^-\rho^+$ and $D^0\ra K^-\rho^+$ amplitudes. 

The fit uses 38 observables taken from 
measurements of \dklnu, \dkk\ and \dpipi, \dkpi, 
$D^0\ra K^+\pi^-\pi^0$, 
\dkspp, and \dkskk\ decays,\footnote{Charge-conjugate modes
are implicitly included.} and from double-tagged branching 
fractions measured at the $\psi(3770)$ resonance. Correlations 
among observables are accounted for by using covariance matrices 
provided by the experimental collaborations. Errors are assumed
to be Gaussian, and systematic errors among different experiments 
are assumed uncorrelated unless specific correlations have been 
identified.
We have checked this method with a second method that adds
together three-dimensional log-likelihood functions 
for $x$, $y$, and $\delta$ obtained from several analyses;
this combination accounts for non-Gaussian errors.
When both methods are applied to the same set of 
measurements, equivalent results are obtained.

Mixing in heavy flavor systems such as those of $B^0$ and $B^0_s$ 
is governed by a short-distance box diagram. In the $D^0$ system,
however, this diagram is doubly-Cabibbo-suppressed relative to 
amplitudes dominating the decay width, and it is also GIM-suppressed.
Thus the short-distance mixing rate is tiny, and $D^0$-$\dbar$ 
mixing is expected to be dominated by long-distance processes. 
These are difficult to calculate reliably, and theoretical estimates 
for $x$ and $y$ range over two-three orders of 
magnitude~\cite{Bigi:2000wn,Petrov:2003un,Petrov:2004rf,Falk:2004wg}.

With the exception of $\psi(3770)\ra DD$ measurements, all methods 
identify the flavor of the $D^0$ or $\dbar$ when produced by 
reconstructing the decay $D^{*+}\ra D^0\pi^+$ or $D^{*-}\ra\dbar\pi^-$. 
The charge of the pion, which has low momentum and is usually 
referred to as the ``soft'' pion~$\pi^{}_s$,
identifies the $D$ flavor. For signal 
decays, $M^{}_{D^*}-M^{}_{D^0}-M^{}_{\pi^+}\equiv Q\approx 6$\meve, 
which is close to the threshold; thus analyses typically
require that the reconstructed $Q$ be small to suppress backgrounds. 
For time-dependent measurements, the $D^0$ decay time is 
calculated as $(d/p)\times M^{}_{D^0}$, where $d$ is
the distance between the $D^*$ and $D^0$ decay vertices and 
$p$ is the $D^0$ momentum. The $D^*$ vertex position is 
taken to be at the primary vertex for $\bar{p}p$ collider 
experiments~\cite{Aaltonen:2007uc}, and at the intersection 
of the $D^0$ momentum vector with the beamspot profile for 
$e^+e^-$ experiments.

\subsubsection{Input observables}

The global fit determines central values and errors for
the underlying parameters using a $\chi^2$ statistic.
The fitted parameters are 
$x$, $y$, $R^{}_D$, $A^{}_D$, $|q/p|$, $\phi$, $\delta$, $\delta^{}_{K\pi\pi}$,
$A^{}_{KK}$ and $A^{}_{\pi\pi}$.
The parameter $\delta^{}_{K\pi\pi}$ is the strong phase 
difference between the amplitudes ${\cal A}(\dbar\ra K^+\rho^-)$ 
and ${\cal A}(D^0\ra K^+\rho^-)$. In the $D\ra K^+\pi^-\pi^0$ 
Dalitz plot analysis that provides sensitivity to $x$ and $y$, 
the $\dbar\ra K^+\pi^-\pi^0$ isobar phases are determined 
relative to that for ${\cal A}(\dbar\ra K^+\rho^-)$, and 
the $D^0\ra K^+\pi^-\pi^0$ isobar phases are determined 
relative to that for ${\cal A}(D^0\ra K^+\rho^-)$. 
As the $\dbar$ and $D^0$ Dalitz plots are fit independently, 
the phase difference $\delta^{}_{K\pi\pi}$ between the
two ``normalizing'' amplitudes cannot be determined
from these fits.

\begin{figure}
\begin{center}
\includegraphics[width=4.2in]{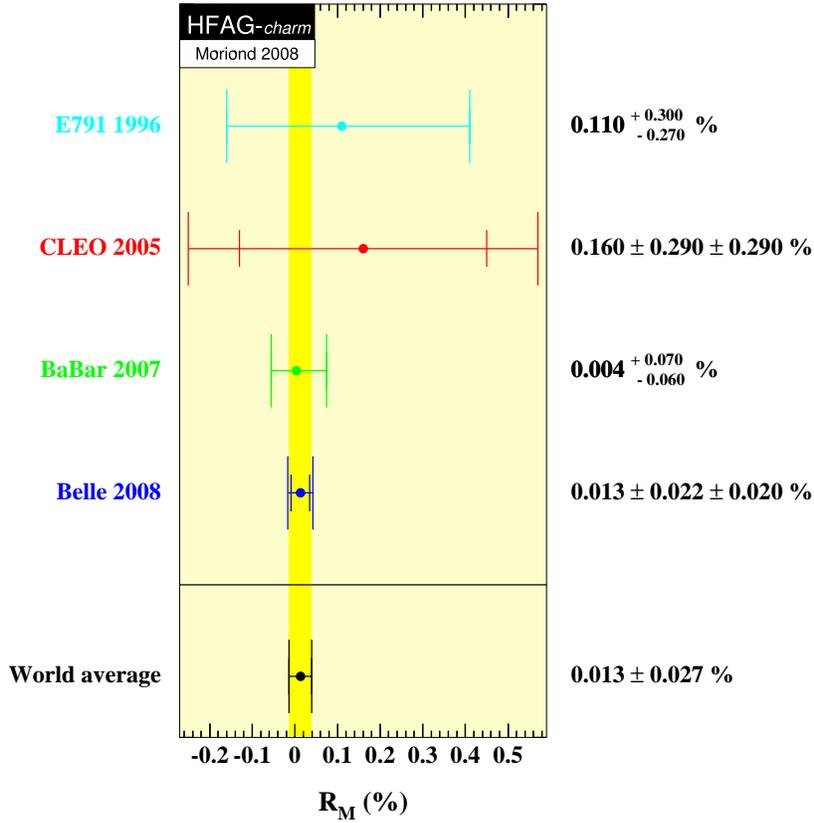}
\end{center}
\vskip-0.20in
\caption{\label{fig:rm_semi}
World average value of $R^{}_M$ from Ref.~\cite{HFAG_charm:webpage},
as calculated from $D^0\ra K^+\ell^-\nu$ 
measurements~\cite{Aitala:1996vz,Cawlfield:2005ze,Aubert:2007aa,Bitenc:2008bk}. }
\end{figure}

\begin{figure}
\begin{center}
\includegraphics[width=4.2in]{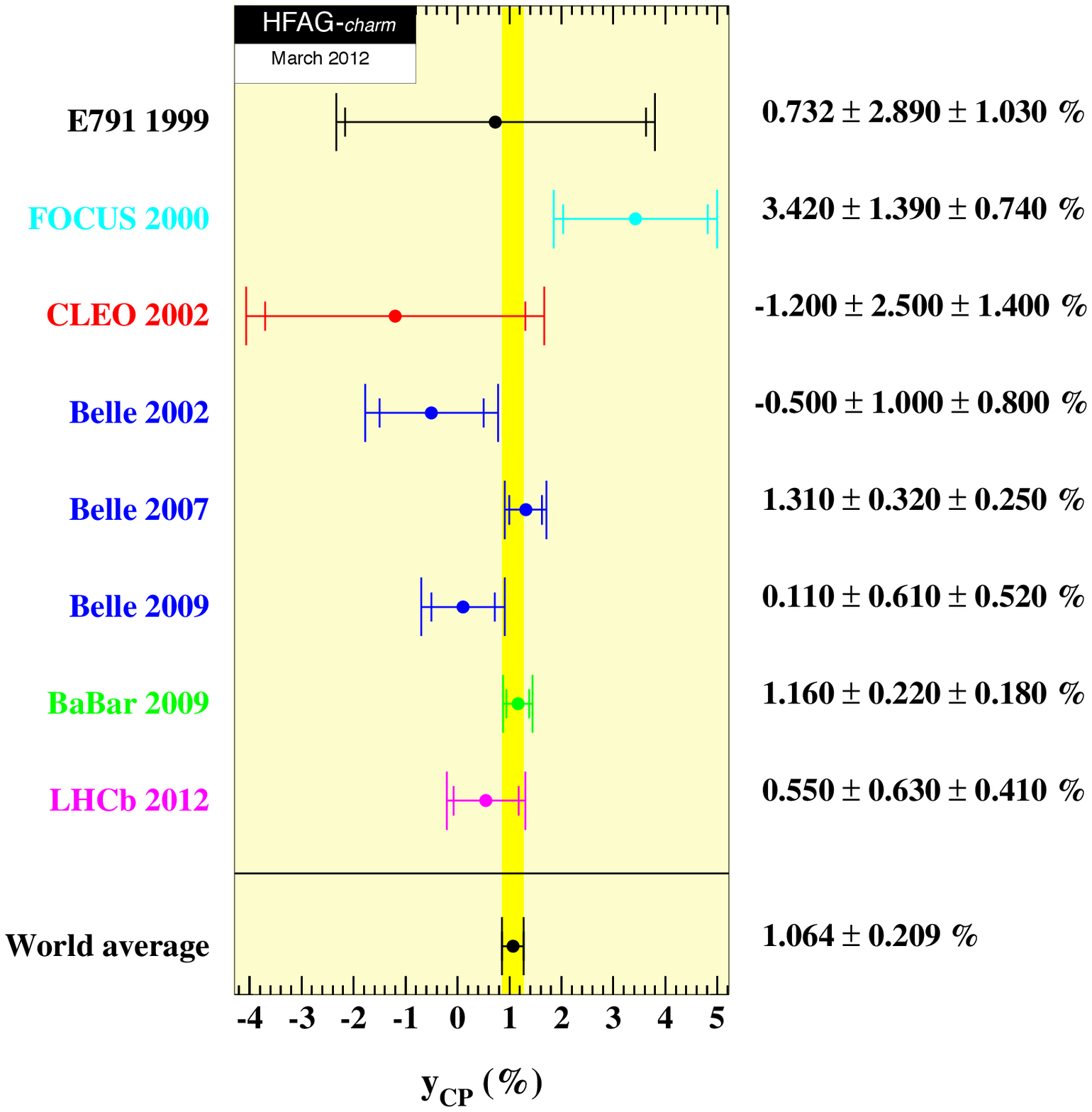}
\end{center}
\vskip-0.20in
\caption{\label{fig:ycp}
World average value of $y^{}_{CP}$ from Ref.~\cite{HFAG_charm:webpage}, 
as calculated from \dkkpp\ 
measurements~\cite{Staric:2007dt,Aitala:1999dt,Link:2000cu,
Csorna:2001ww,Aubert:2007en,Zupanc:2009sy,Aaij:2011ad}.  }
\end{figure}

\begin{figure}
\begin{center}
\includegraphics[width=4.2in]{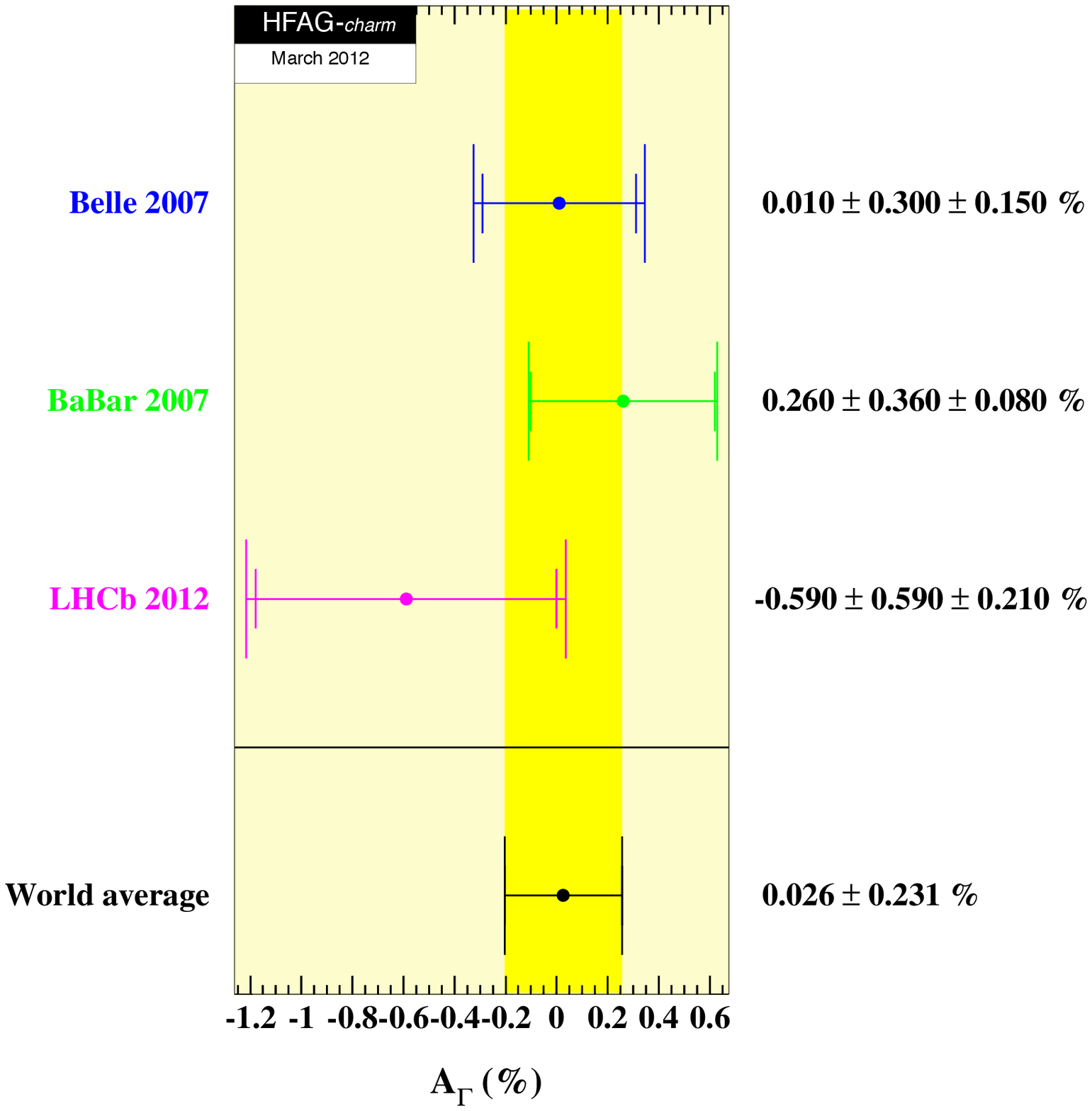}
\end{center}
\vskip-0.20in
\caption{\label{fig:Agamma}
World average value of $A^{}_\Gamma$ from Ref.~\cite{HFAG_charm:webpage}, 
as calculated from \dkkpp\ 
measurements~\cite{Staric:2007dt,Aubert:2007en,Aaij:2011ad}.  }
\end{figure}

All input measurements are listed in 
Tables~\ref{tab:observables1}-\ref{tab:observables3}. 
The observable $R^{}_M=(x^2+y^2)/2$ is calculated from \dklnu\ 
decays~\cite{Aitala:1996vz,Cawlfield:2005ze,Aubert:2007aa,Bitenc:2008bk}
and is the world average (WA) value calculated by HFAG~\cite{HFAG_charm:webpage}.
The inputs used for these averages are plotted in Fig.~\ref{fig:rm_semi}. 
The observables $y^{}_{CP}$ and $A^{}_\Gamma$ are also HFAG WA 
values~\cite{HFAG_charm:webpage}; the inputs used for these 
averages are plotted in Figs.~\ref{fig:ycp} and \ref{fig:Agamma}.
The \dkpi\ observables used are from Belle~\cite{Zhang:2006dp}, 
\babar~\cite{Aubert:2007wf}, and CDF~\cite{Aaltonen:2007uc};
earlier measurements have much less precision and are not used.
The observables from \dkspp\ decays for no-\cpv\ are from 
Belle~\cite{Abe:2007rd} and \babar~\cite{delAmoSanchez:2010xz}, 
but for the \cpv-allowed case only Belle measurements~\cite{Abe:2007rd} 
are available. The $D^0\ra K^+\pi^-\pi^0$ results are from 
\babar~\cite{Aubert:2008zh}, and the $\psi(3770)\ra\overline{D}D$ 
results are from CLEOc~\cite{Sun:2010zz}.

The relationships between the observables and the fitted
parameters are listed in Table~\ref{tab:relationships}. 
For each set of correlated observables we construct a
difference vector $\vec{V}$; e.g., for 
$D^0\ra K^0_S\,\pi^+\pi^-$ decays,
$\vec{V}=(\Delta x,\Delta y,\Delta |q/p|,\Delta \phi)$
where $\Delta$ represents the difference between the 
measured value and the fitted value. The 
contribution of a set of observables to the $\chi^2$ 
is calculated as $\vec{V}\cdot (M^{-1})\cdot\vec{V}^T$, 
where $M^{-1}$ is the inverse of the covariance matrix 
for the measurement. Covariance matrices are constructed 
from the correlation coefficients among the measured observables.
These coefficients (where applicable) are also listed in 
Tables~\ref{tab:observables1}-\ref{tab:observables3}. 

\begin{table}
\renewcommand{\arraystretch}{1.3}
\renewcommand{\arraycolsep}{0.02in}
\renewcommand{\tabcolsep}{0.05in}
\caption{\label{tab:observables1}
All observables used in the global fit except those from \dkpi\ 
and those used for measuring direct \cpv, from
Refs.~\cite{Staric:2007dt,
Aitala:1996vz,
Cawlfield:2005ze,
Aubert:2007aa,
Bitenc:2008bk,
Abe:2007rd,
delAmoSanchez:2010xz,
Aubert:2008zh,
Sun:2010zz,
Aitala:1999dt,
Link:2000cu,
Csorna:2001ww,
Aubert:2007en}.}
\vspace*{6pt}
\footnotesize
\resizebox{0.99\textwidth}{!}{
\begin{tabular}{l|ccc}
\hline
{\bf Mode} & \textbf{Observable} & {\bf Values} & {\bf Correlation coefficients} \\
\hline
\begin{tabular}{l}  
$D^0\ra K^+K^-/\pi^+\pi^-$, \\
\hskip0.30in $\phi\,K^0_S$~\cite{HFAG_charm:webpage} 
\end{tabular}
&
\begin{tabular}{c}
 $y^{}_{CP}$  \\
 $A^{}_{\Gamma}$
\end{tabular} & 
$\begin{array}{c}
(1.064\pm 0.209)\% \\
(0.026\pm 0.231)\% 
\end{array}$   & \\ 
\hline
\begin{tabular}{l}  
$D^0\ra K^0_S\,\pi^+\pi^-$~\cite{HFAG_charm:webpage} \\
\ (Belle+CLEO WA: \\
\ \ no \cpv\ or \\
\ \ no direct \cpv)
\end{tabular}
&
\begin{tabular}{c}
$x$ \\
$y$ \\
$|q/p|$ \\
$\phi$ 
\end{tabular} & 
\begin{tabular}{c}
 $(0.811\pm 0.334)\%$ \\
 $(0.309\pm 0.281)\%$ \\
 $0.95\pm 0.22^{+0.10}_{-0.09}$ \\
 $(-0.035\pm 0.19\pm 0.09)$ rad
\end{tabular} &  \\ 
 & & \\
\begin{tabular}{l}  
$D^0\ra K^0_S\,\pi^+\pi^-$~\cite{Abe:2007rd} \\
\ (Belle: \\
\ \ \cpv-allowed)
\end{tabular}
&
\begin{tabular}{c}
$x$ \\
$y$ \\
$|q/p|$ \\
$\phi$  
\end{tabular} & 
\begin{tabular}{c}
 $(0.81\pm 0.30^{+0.13}_{-0.17})\%$ \\
 $(0.37\pm 0.25^{+0.10}_{-0.15})\%$ \\
 $0.86\pm 0.30^{+0.10}_{-0.09}$ \\
 $(-0.244\pm 0.31\pm 0.09)$ rad
\end{tabular} &
\begin{tabular}{l} \hskip-0.15in
$\left\{ \begin{array}{cccc}
 1 &  -0.007 & -0.255\alpha & 0.216  \\
 -0.007 &  1 & -0.019\alpha & -0.280 \\
 -0.255\alpha &  -0.019\alpha & 1 & -0.128\alpha  \\
  0.216 &  -0.280 & -0.128\alpha & 1 
\end{array} \right\}$ \\
\hskip0.10in ($\alpha=(|q/p|+1)^2/2$ is a \\ 
\hskip0.30in transformation factor)
\end{tabular} \\
 & & \\
\begin{tabular}{l}  
$D^0\ra K^0_S\,\pi^+\pi^-$~\cite{delAmoSanchez:2010xz} \\
\hskip0.30in $K^0_S\,K^+ K^-$ \\
\ (\babar: no \cpv) 
\end{tabular}
&
\begin{tabular}{c}
$x$ \\
$y$ 
\end{tabular} & 
\begin{tabular}{c}
 $(0.16\pm 0.23\pm 0.12\pm 0.08)\%$ \\
 $(0.57\pm 0.20\pm 0.13\pm 0.07)\%$ 
\end{tabular} &  $0.0615$ \\ 
\hline
\begin{tabular}{l}  
$D^0\ra K^+\ell^-\nu$~\cite{HFAG_charm:webpage}
\end{tabular} 
  & $R^{}_M$  & $(0.0130\pm 0.0269)\%$  &  \\ 
\hline
\begin{tabular}{l}  
$D^0\ra K^+\pi^-\pi^0$ 
\end{tabular} 
&
\begin{tabular}{c}
$x''$ \\ 
$y''$ 
\end{tabular} &
\begin{tabular}{c}
$(2.61\,^{+0.57}_{-0.68}\,\pm 0.39)\%$ \\ 
$(-0.06\,^{+0.55}_{-0.64}\,\pm 0.34)\%$ 
\end{tabular} & $-0.75$ \\
\hline
\begin{tabular}{c}  
$\psi(3770)\ra\overline{D}D$ \\
(CLEOc)
\end{tabular}
&
\begin{tabular}{c}
$x^2$ \\
$y$ \\
$R^{}_D$ \\
$2\sqrt{R^{}_D}\cos\delta$ \\
$2\sqrt{R^{}_D}\sin\delta$ 
\end{tabular} & 
\begin{tabular}{c}
$(0.1549 \pm 0.2223)\%$ \\
$(2.997 \pm 2.293)\%$ \\
$(0.4118 \pm 0.0948)\%$ \\
$(12.64 \pm 2.86)\%$ \\
$(-0.5242 \pm 6.426)\%$ 
\end{tabular} &
$\left\{ \begin{array}{ccccc}
1 & -0.6217 & -0.00224 &  0.3698 &  0.01567 \\
  &  1 	    &  0.00414 & -0.5756 & -0.0243 \\
  &         &  1       &  0.0035 &  0.00978 \\
  &         &          &  1 	 &  0.0471 \\
  &         &          &         &  1    
\end{array} \right\}$ \\
\hline
\end{tabular}
}
\end{table}

\begin{table}
\renewcommand{\arraystretch}{1.3}
\renewcommand{\arraycolsep}{0.02in}
\caption{\label{tab:observables2}
\dkpi\ observables used for the global fit, from
Refs.~\cite{Aubert:2007wf,Aaltonen:2007uc,Zhang:2006dp}.}
\vspace*{6pt}
\footnotesize
\begin{center}
\begin{tabular}{l|ccc}
\hline
{\bf Mode} & \textbf{Observable} & {\bf Values} & {\bf Correlation coefficients} \\
\hline
\begin{tabular}{c}  
$D^0\ra K^+\pi^-$ \\
(\babar)
\end{tabular}
&
\begin{tabular}{c}
$R^{}_D$ \\
$x'^{2+}$ \\
$y'^+$ 
\end{tabular} & 
\begin{tabular}{c}
 $(0.303\pm 0.0189)\%$ \\
 $(-0.024\pm 0.052)\%$ \\
 $(0.98\pm 0.78)\%$ 
\end{tabular} &
$\left\{ \begin{array}{ccc}
 1 &  0.77 &  -0.87 \\
0.77 & 1 & -0.94 \\
-0.87 & -0.94 & 1 
\end{array} \right\}$ \\ \\
\begin{tabular}{c}  
$\dbar\ra K^-\pi^+$ \\
(\babar)
\end{tabular}
&
\begin{tabular}{c}
$A^{}_D$ \\
$x'^{2-}$ \\
$y'^-$ 
\end{tabular} & 
\begin{tabular}{c}
 $(-2.1\pm 5.4)\%$ \\
 $(-0.020\pm 0.050)\%$ \\
 $(0.96\pm 0.75)\%$ 
\end{tabular} & same as above \\
\hline
\begin{tabular}{c}  
$D^0\ra K^+\pi^-$ \\
(Belle)
\end{tabular}
&
\begin{tabular}{c}
$R^{}_D$ \\
$x'^{2+}$ \\
$y'^+$ 
\end{tabular} & 
\begin{tabular}{c}
 $(0.364\pm 0.018)\%$ \\
 $(0.032\pm 0.037)\%$ \\
 $(-0.12\pm 0.58)\%$ 
\end{tabular} &
$\left\{ \begin{array}{ccc}
 1 &  0.655 &  -0.834 \\
0.655 & 1 & -0.909 \\
-0.834 & -0.909 & 1 
\end{array} \right\}$ \\ \\
\begin{tabular}{c}  
$\dbar\ra K^-\pi^+$ \\
(Belle)
\end{tabular}
&
\begin{tabular}{c}
$A^{}_D$ \\
$x'^{2-}$ \\
$y'^-$ 
\end{tabular} & 
\begin{tabular}{c}
 $(2.3\pm 4.7)\%$ \\
 $(0.006\pm 0.034)\%$ \\
 $(0.20\pm 0.54)\%$ 
\end{tabular} & same as above \\
\hline
\begin{tabular}{c}  
$D^0\ra K^+\pi^-$ \\
\ \ \ \ \ + c.c. \\
(CDF)
\end{tabular}
&
\begin{tabular}{c}
$R^{}_D$ \\
$x'^{2}$ \\
$y'$ 
\end{tabular} & 
\begin{tabular}{c}
 $(0.304\pm 0.055)\%$ \\
 $(-0.012\pm 0.035)\%$ \\
 $(0.85\pm 0.76)\%$ 
\end{tabular} & 
$\left\{ \begin{array}{ccc}
 1 &  0.923 &  -0.971 \\
0.923 & 1 & -0.984 \\
-0.971 & -0.984 & 1 
\end{array} \right\}$ \\ 
\hline
\end{tabular}
\end{center}
\end{table}

\begin{table}
\renewcommand{\arraystretch}{1.3}
\renewcommand{\arraycolsep}{0.02in}
\caption{\label{tab:observables3}
Measurements of direct \cpv, from
Refs.~\cite{Aubert:2007if,Staric:2008rx,Aaij:2011in,Aaltonen:2011se,
cdf_public_note_10784}.
The parameter $A^{}_{CP}(f)$ is defined as
$[\Gamma(D^0\ra f)-\Gamma(\dbar\ra f)]/
[\Gamma(D^0\ra f)+\Gamma(\dbar\ra f)]$.}
\vspace*{6pt}
\footnotesize
\begin{center}
\begin{tabular}{l|ccc}
\hline
{\bf Mode} & \textbf{Observable} & {\bf Values} & 
                  {\boldmath $\Delta\langle t\rangle/\tau$} \\
\hline
\begin{tabular}{c}
$D^0\ra K^+K^-/\pi^+\pi^-$ \\
(\babar)
\end{tabular} & 
\begin{tabular}{c}
$A^{}_{CP}(K^+K^-)$ \\
$A^{}_{CP}(\pi^+\pi^-)$ 
\end{tabular} & 
\begin{tabular}{c}
$(0.00 \pm 0.34 \pm 0.13)\%$ \\
$(-0.24 \pm 0.52 \pm 0.22)\%$ 
\end{tabular} &
0 \\
\hline
\begin{tabular}{c}
$D^0\ra K^+K^-/\pi^+\pi^-$ \\
(Belle)
\end{tabular} & 
\begin{tabular}{c}
$A^{}_{CP}(K^+K^-)$ \\
$A^{}_{CP}(\pi^+\pi^-)$ 
\end{tabular} & 
\begin{tabular}{c}
$(-0.43 \pm 0.30 \pm 0.11)\%$ \\
$(0.43 \pm 0.52 \pm 0.12)\%$ 
\end{tabular} &
0 \\
\hline
\begin{tabular}{c}
$D^0\ra K^+K^-/\pi^+\pi^-$ \\
(LHCb 37~pb$^{-1}$)
\end{tabular} & 
$A^{}_{CP}(K^+K^-)-A^{}_{CP}(\pi^+\pi^-)$  &
$(-0.82 \pm 0.21 \pm 0.11)\%$ &  
$0.0983 \pm 0.00291$ \\
\hline
\begin{tabular}{c}
$D^0\ra K^+K^-/\pi^+\pi^-$ \\
(CDF 9.7~fb$^{-1}$ prelim.)
\end{tabular} & 
$A^{}_{CP}(K^+K^-)-A^{}_{CP}(\pi^+\pi^-)$  &
$(-0.62 \pm 0.21 \pm 0.10)\%$ & 
$0.26 \pm 0.01$ \\ 
\hline
(CDF 5.9~fb$^{-1}$ not used)  & 
\begin{tabular}{c}
$A^{}_{CP}(K^+K^-)$ \\
$A^{}_{CP}(\pi^+\pi^-)$ 
\end{tabular} & 
\begin{tabular}{c}
$(-0.24 \pm 0.22 \pm 0.09)\%$ \\
$(0.22 \pm 0.24 \pm 0.11)\%$ 
\end{tabular} &
\begin{tabular}{c}
$2.65 \pm 0.03$ \\
$2.40 \pm 0.03$ 
\end{tabular}
\\
\hline
\end{tabular}
\end{center}
\end{table}

\begin{table}
\renewcommand{\arraycolsep}{0.02in}
\renewcommand{\arraystretch}{1.3}
\begin{center}
\caption{\label{tab:relationships}
Left: decay modes used to determine fitted parameters 
$x,\,y,\,\delta,\,\delta^{}_{K\pi\pi},\,R^{}_D,\,A^{}_D,\,|q/p|$, and $\phi$.
Middle: the observables measured for each decay mode. Right: the 
relationships between the observables measured and the fitted parameters.}
\vspace*{6pt}
\footnotesize
\resizebox{0.99\textwidth}{!}{
\begin{tabular}{l|c|l}
\hline
\textbf{Decay Mode} & \textbf{Observables} & \textbf{Relationship} \\
\hline
$D^0\ra K^+K^-/\pi^+\pi^-$  & 
\begin{tabular}{c}
 $y^{}_{CP}$  \\
 $A^{}_{\Gamma}$
\end{tabular} & 
$\begin{array}{c}
2y^{}_{CP} = 
\left(\left|q/p\right|+\left|p/q\right|\right)y\cos\phi - \\
\hskip0.50in \left(\left|q/p\right|-\left|p/q\right|\right)x\sin\phi \\
2A^{}_\Gamma = 
\left(\left|q/p\right|-\left|p/q\right|\right)y\cos\phi - \\
\hskip0.50in \left(\left|q/p\right|+\left|p/q\right|\right)x\sin\phi
\end{array}$   \\
\hline
$D^0\ra K^0_S\,\pi^+\pi^-$ & 
$\begin{array}{c}
x \\ 
y \\ 
|q/p| \\ 
\phi
\end{array}$ &   \\ 
\hline
$D^0\ra K^+\ell^-\nu$ & $R^{}_M$  & $R^{}_M = (x^2 + y^2)/2$ \\
\hline
\begin{tabular}{l}
$D^0\ra K^+\pi^-\pi^0$ \\
(Dalitz plot analysis)
\end{tabular} & 
$\begin{array}{c}
x'' \\ 
y''
\end{array}$ &
$\begin{array}{l}
x'' = x\cos\delta^{}_{K\pi\pi} + y\sin\delta^{}_{K\pi\pi} \\ 
y'' = y\cos\delta^{}_{K\pi\pi} - x\sin\delta^{}_{K\pi\pi}
\end{array}$ \\
\hline
\begin{tabular}{l}
``Double-tagged'' \\
branching fractions \\
measured in \\
$\psi(3770)\ra DD$ decays
\end{tabular} & 
$\begin{array}{c}
R^{}_M \\
y \\
R^{}_D \\
\sqrt{R^{}_D}\cos\delta
\end{array}$ &   $R^{}_M = (x^2 + y^2)/2$ \\
\hline
$D^0\ra K^+\pi^-$ &
$\begin{array}{c}
x'^2,\ y' \\
x'^{2+},\ x'^{2-} \\
y'^+,\ y'^-
\end{array}$ & 
$\begin{array}{l}
x' = x\cos\delta + y\sin\delta \\ 
y' = y\cos\delta - x\sin\delta \\
A^{}_M\equiv (|q/p|^4-1)/(|q/p|^4+1) \\
x'^\pm = [(1\pm A^{}_M)/(1\mp A^{}_M)]^{1/4} \times \\
\hskip0.50in (x'\cos\phi\pm y'\sin\phi) \\
y'^\pm = [(1\pm A^{}_M)/(1\mp A^{}_M)]^{1/4} \times \\
\hskip0.50in (y'\cos\phi\mp x'\sin\phi) \\
\end{array}$ \\
\hline
\begin{tabular}{c}
$D^0\ra K^+\pi^-/K^-\pi^+$ \\
(time-integrated)
\end{tabular} & 
\begin{tabular}{c}
$\frac{\displaystyle \Gamma(D^0\ra K^+\pi^-)+\Gamma(\dbar\ra K^-\pi^+)}
{\displaystyle \Gamma(D^0\ra K^-\pi^+)+\Gamma(\dbar\ra K^+\pi^-)}$  \\ \\
$\frac{\displaystyle \Gamma(D^0\ra K^+\pi^-)-\Gamma(\dbar\ra K^-\pi^+)}
{\displaystyle \Gamma(D^0\ra K^+\pi^-)+\Gamma(\dbar\ra K^-\pi^+)}$ 
\end{tabular} & 
\begin{tabular}{c}
$R^{}_D$ \\ \\ \\
$A^{}_D$ 
\end{tabular} \\
\hline
\begin{tabular}{c}
$D^0\ra K^+K^-/\pi^+\pi^-$ \\
(time-integrated)
\end{tabular} & 
\begin{tabular}{c}
$\frac{\displaystyle \Gamma(D^0\ra K^+K^-)-\Gamma(\dbar\ra K^+K^-)}
{\displaystyle \Gamma(D^0\ra K^+K^-)+\Gamma(\dbar\ra K^+K^-)}$    \\ \\
$\frac{\displaystyle \Gamma(D^0\ra\pi^+\pi^-)-\Gamma(\dbar\ra\pi^+\pi^-)}
{\displaystyle \Gamma(D^0\ra\pi^+\pi^-)+\Gamma(\dbar\ra\pi^+\pi^-)}$ 
\end{tabular} & 
\begin{tabular}{c}
$A^{}_K  + \frac{\displaystyle \langle t\rangle}
{\displaystyle \tau^{}_D}\,{\cal A}_{CP}^{\rm indirect}$ 
\ \ (${\cal A}_{CP}^{\rm indirect}\approx -A^{}_\Gamma$)
\\ \\ \\
$A^{}_\pi + \frac{\displaystyle \langle t\rangle}
{\displaystyle \tau^{}_D}\,{\cal A}_{CP}^{\rm indirect}$ 
\ \ (${\cal A}_{CP}^{\rm indirect}\approx -A^{}_\Gamma$)
\end{tabular} \\
\hline
\end{tabular}
}
\end{center}
\end{table}

\subsubsection{Fit results}

The global fit uses MINUIT with the MIGRAD minimizer, 
and all errors are obtained from MINOS~\cite{MINUIT:webpage}. 
Four separate fits are performed: 
{\it (a)}\ assuming \cp\ conservation, i.e., fixing
$A^{}_D\!=\!0$, $A_K\!=\!0$, $A^{}_\pi\!=\!0$, $\phi\!=\!0$, 
and $|q/p|\!=\!1$;
{\it (b)}\ assuming no direct \cpv\ and fitting for
parameters $x$, $y$, and $\phi$; 
{\it (c)}\ assuming no direct \cpv\ and fitting for
parameters $x^{}_{12}= 2|M^{}_{12}|/\Gamma$, 
$y^{}_{12}= \Gamma^{}_{12}/\Gamma$, and 
$\phi^{}_{12}= {\rm Arg}(M^{}_{12}/\Gamma^{}_{12})$,
where $M^{}_{12}$ and $\Gamma^{}_{12}$ are the off-diagonal
elements of the $D^0$-$\dbar$ mass and decay matrices, respectively; and
{\it (d)}\ allowing full \cpv, i.e., floating all parameters. 

For the no-direct-\cpv\ fits, we set direct-\cpv\ parameters 
$A^{}_D\!=\!0$, $A_K\!=\!0$, and $A^{}_\pi\!=\!0$. In addition, for the 
first fit {\it (b)\/} we impose the relation~\cite{Ciuchini:2007cw,Kagan:2009gb}
$\tan\phi = (1-|q/p|^2)/(1+|q/p|^2)\times (x/y)$; this reduces 
four independent parameters to 
three.\footnote{One can also use Eq.~(15) of Ref.~\cite{Grossman:2009mn}
to reduce four parameters to three.} 
We impose this relationship in two ways:
first we float parameters $x$, $y$, and $\phi$ and from them derive $|q/p|$;
then we repeat the fit floating $x$, $y$, and $|q/p|$ and from them derive 
$\phi$. The central values returned by the two fits are identical, but the 
first fit yields MINOS errors for $\phi$, while the second fit 
yields MINOS errors for $|q/p|$. For no-direct-\cpv\ fit 
{\it (c)}, we fit for the underlying parameters $x^{}_{12}$, $y^{}_{12}$, 
and $\phi^{}_{12}$, from which parameters $x$, $y$, $|q/p|$, and $\phi$ 
are derived. 

All fit results are listed in 
Table~\ref{tab:results}. For the \cpv-allowed fit,
individual contributions to the $\chi^2$ are listed 
in Table~\ref{tab:results_chi2}. The total $\chi^2$ 
is 35.6 for $37-10=27$ degrees of freedom; this 
corresponds to a confidence level of~0.124, which 
is satisfactory.

\begin{figure}
\begin{center}
\includegraphics[width=4.2in]{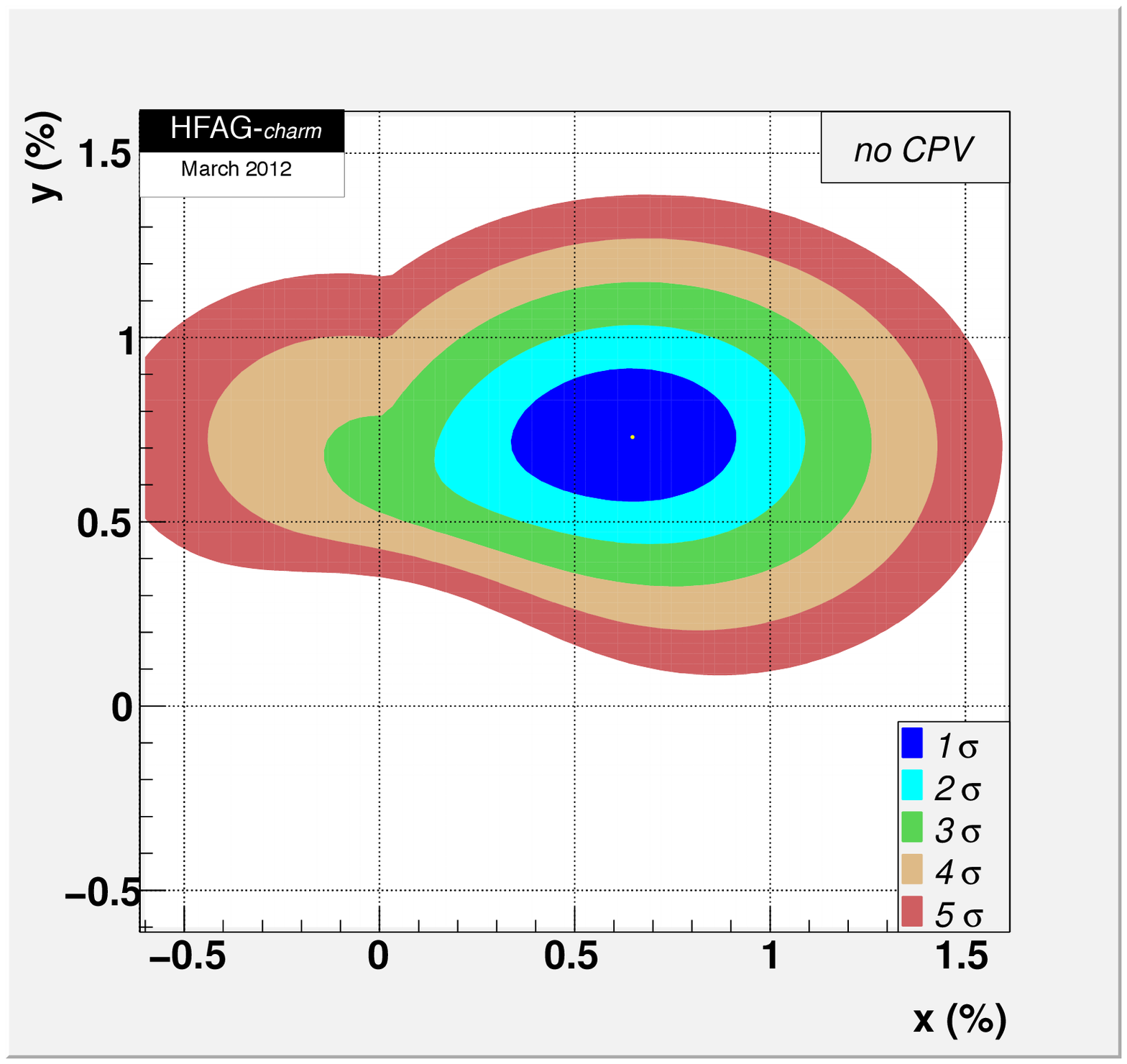}
\end{center}
\vskip-0.20in
\caption{\label{fig:contours_ncpv}
Two-dimensional contours for mixing parameters $(x,y)$, for no \cpv. }
\end{figure}

\begin{figure}
\begin{center}
\vbox{
\includegraphics[width=84mm]{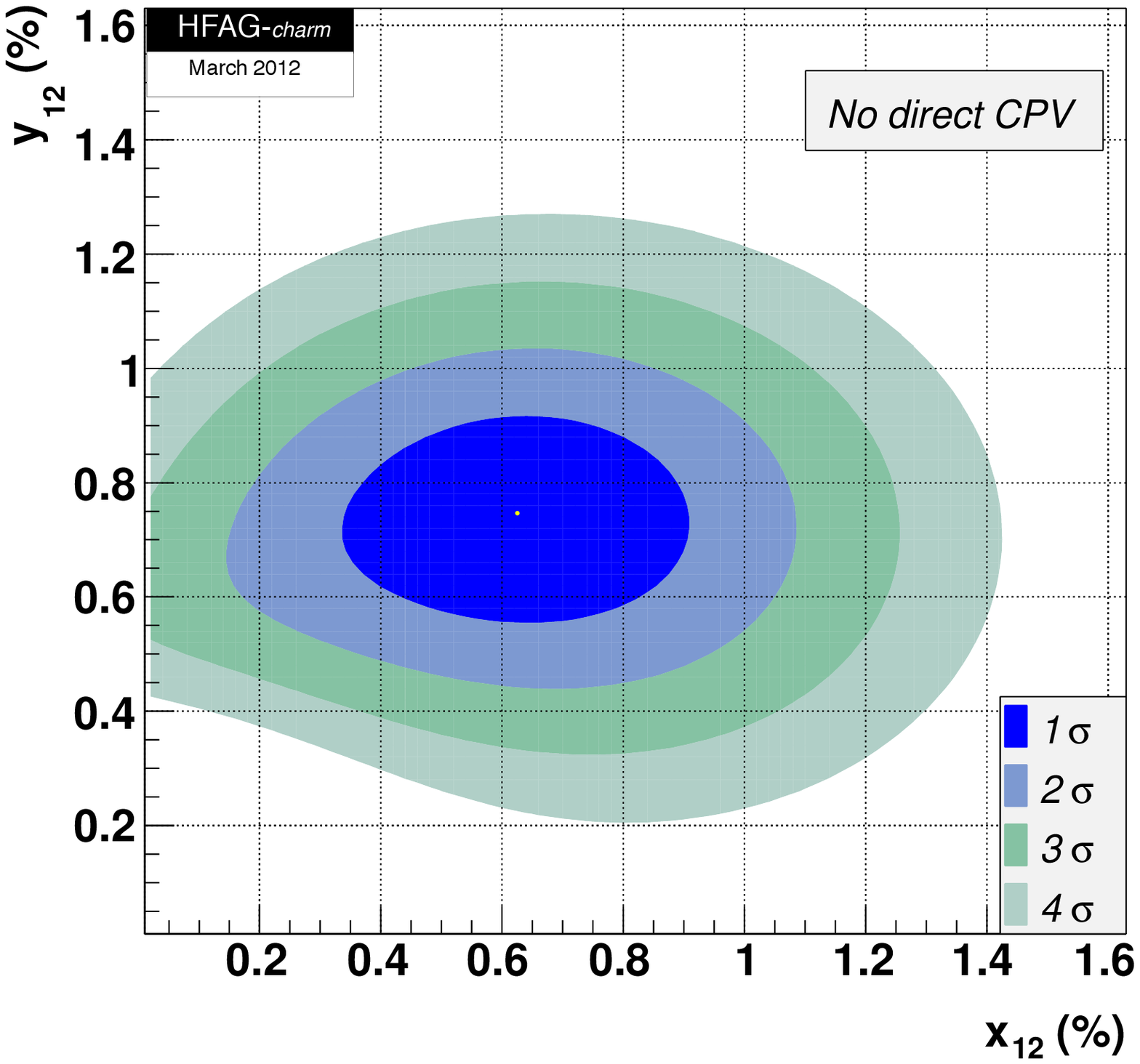}
\includegraphics[width=84mm]{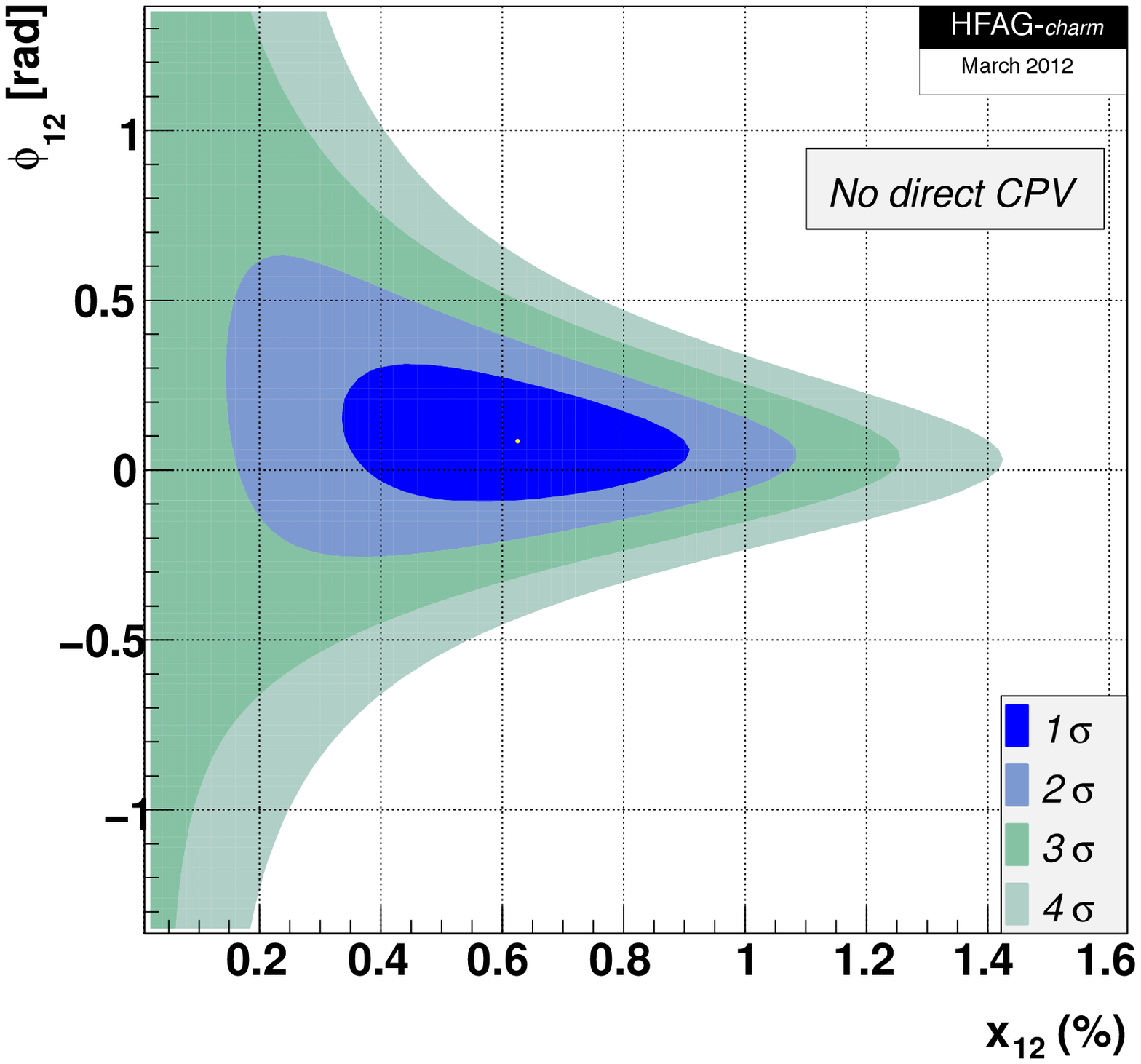}
\vskip0.30in
\includegraphics[width=84mm]{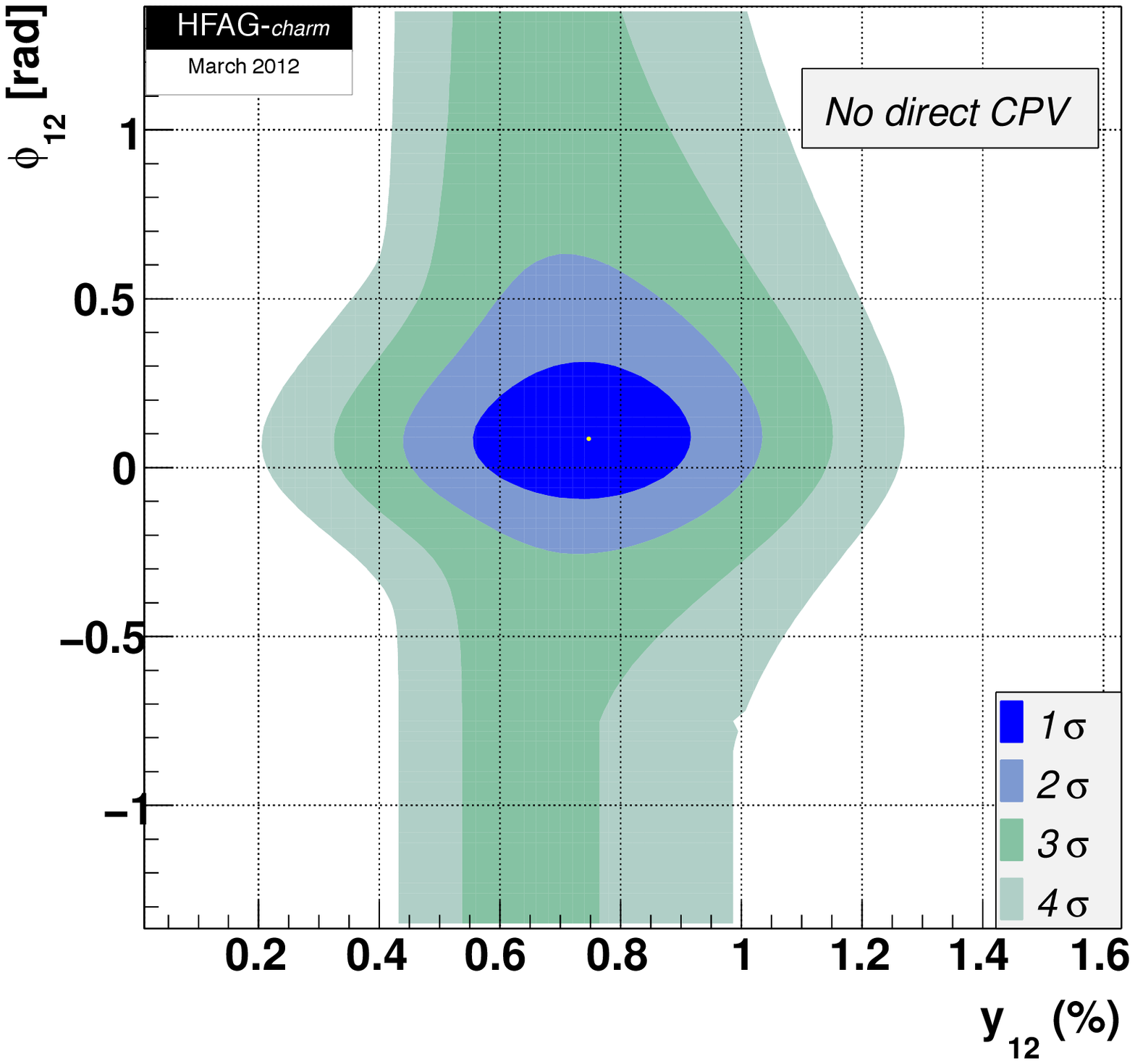}
}
\end{center}
\vskip-0.10in
\caption{\label{fig:contours_ndcpv}
Two-dimensional contours for theoretical parameters 
$(x^{}_{12},y^{}_{12})$ (top left), 
$(x^{}_{12},\phi^{}_{12})$ (top right), and 
$(y^{}_{12},\phi^{}_{12})$ (bottom), 
for no direct \cpv.}
\end{figure}

\begin{figure}
\begin{center}
\vbox{
\includegraphics[width=4.2in]{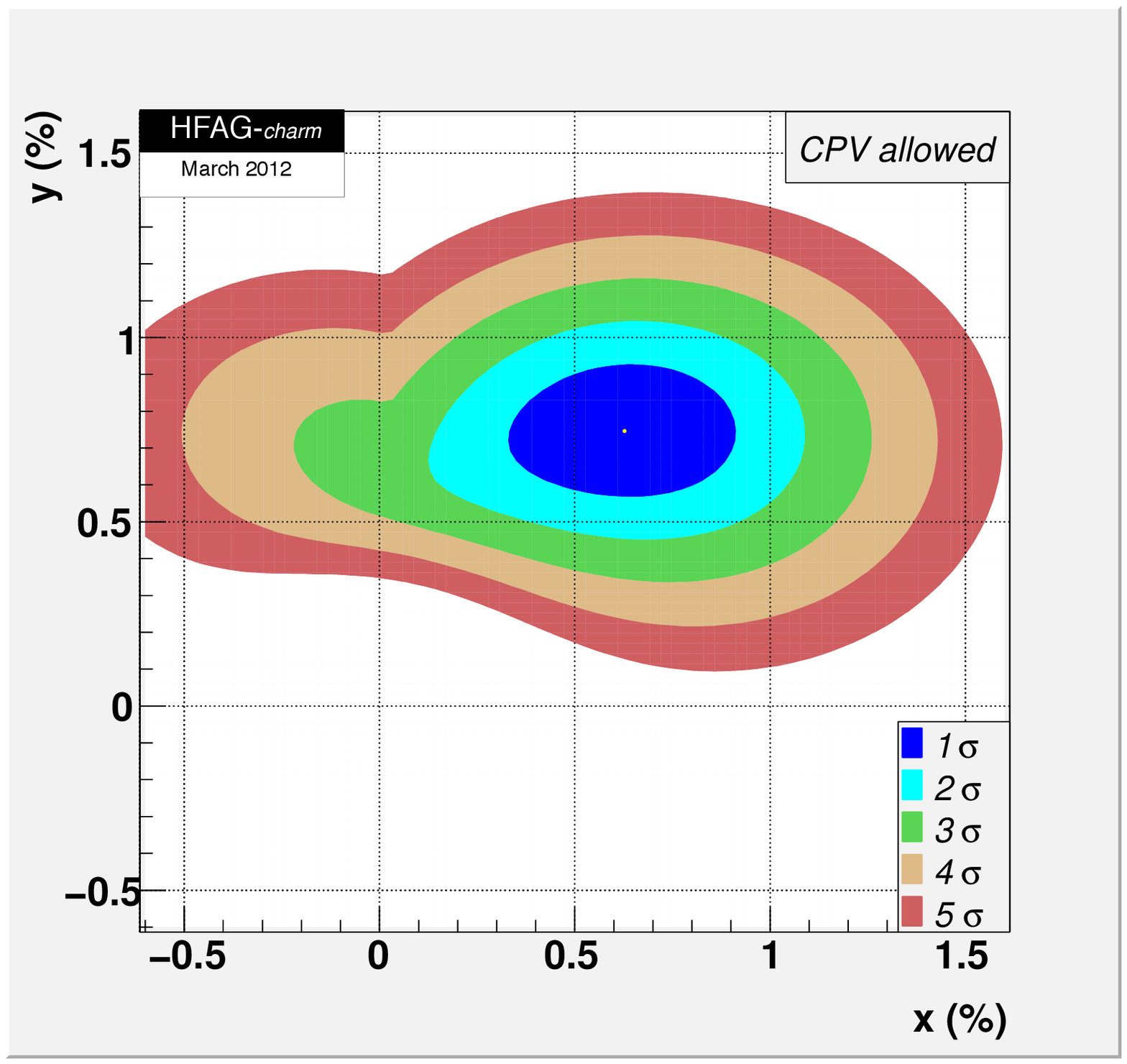}
\vskip0.10in
\includegraphics[width=4.2in]{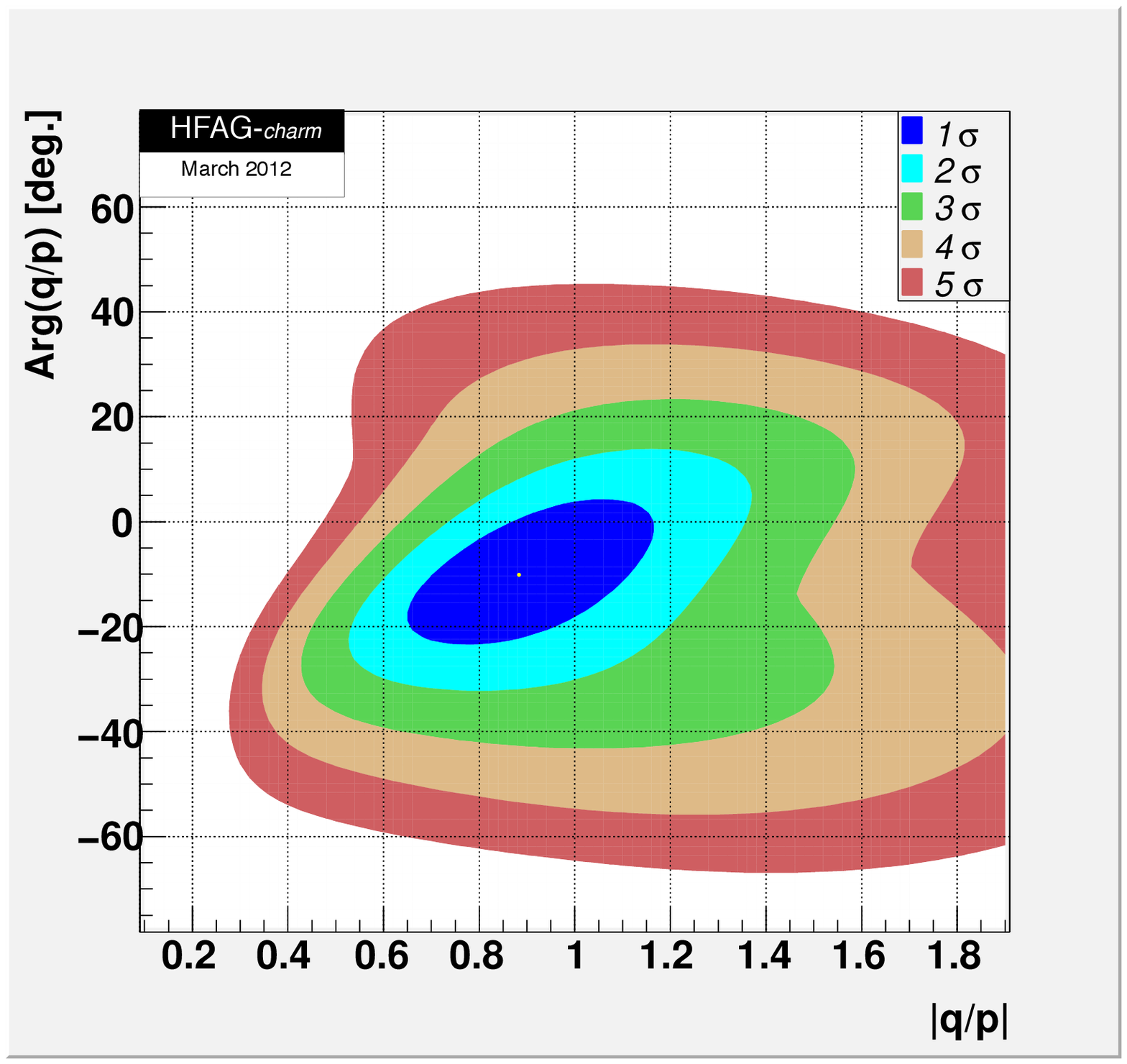}
}
\end{center}
\vskip-0.10in
\caption{\label{fig:contours_cpv}
Two-dimensional contours for parameters $(x,y)$ (top) 
and $(|q/p|,\phi)$ (bottom), allowing for \cpv.}
\end{figure}

Confidence contours in the two dimensions $(x,y)$ or 
in $(|q/p|,\phi)$ are obtained by letting, for any point in the
two-dimensional plane, all other fitted parameters take their 
preferred values. The resulting $1\sigma$-$5\sigma$ contours 
are shown 
in Fig.~\ref{fig:contours_ncpv} for the \cp-conserving case, 
in Fig.~\ref{fig:contours_ndcpv} for the no-direct-\cpv\ case, 
and in Fig.~\ref{fig:contours_cpv} for the \cpv-allowed 
case. The contours are determined from the increase of the
$\chi^2$ above the minimum value.
One observes that the $(x,y)$ contours for the no-\cpv\ fit 
are very similar to those for the \cpv-allowed fit. 
In the latter fit, the
$\chi^2$ at the no-mixing point $(x,y)\!=\!(0,0)$ is 110 units above 
the minimum value; for two degrees of freedom this has a confidence 
level corresponding to $10.2\sigma$. Thus, no mixing is excluded 
at this high level. In the $(|q/p|,\phi)$ plot, the point $(1,0)$ 
is within the $1\sigma$ contour; thus the data is consistent 
with \cp\ conservation.

\begin{figure}
\begin{center}
\hbox{\hskip0.50in
\includegraphics[width=72mm]{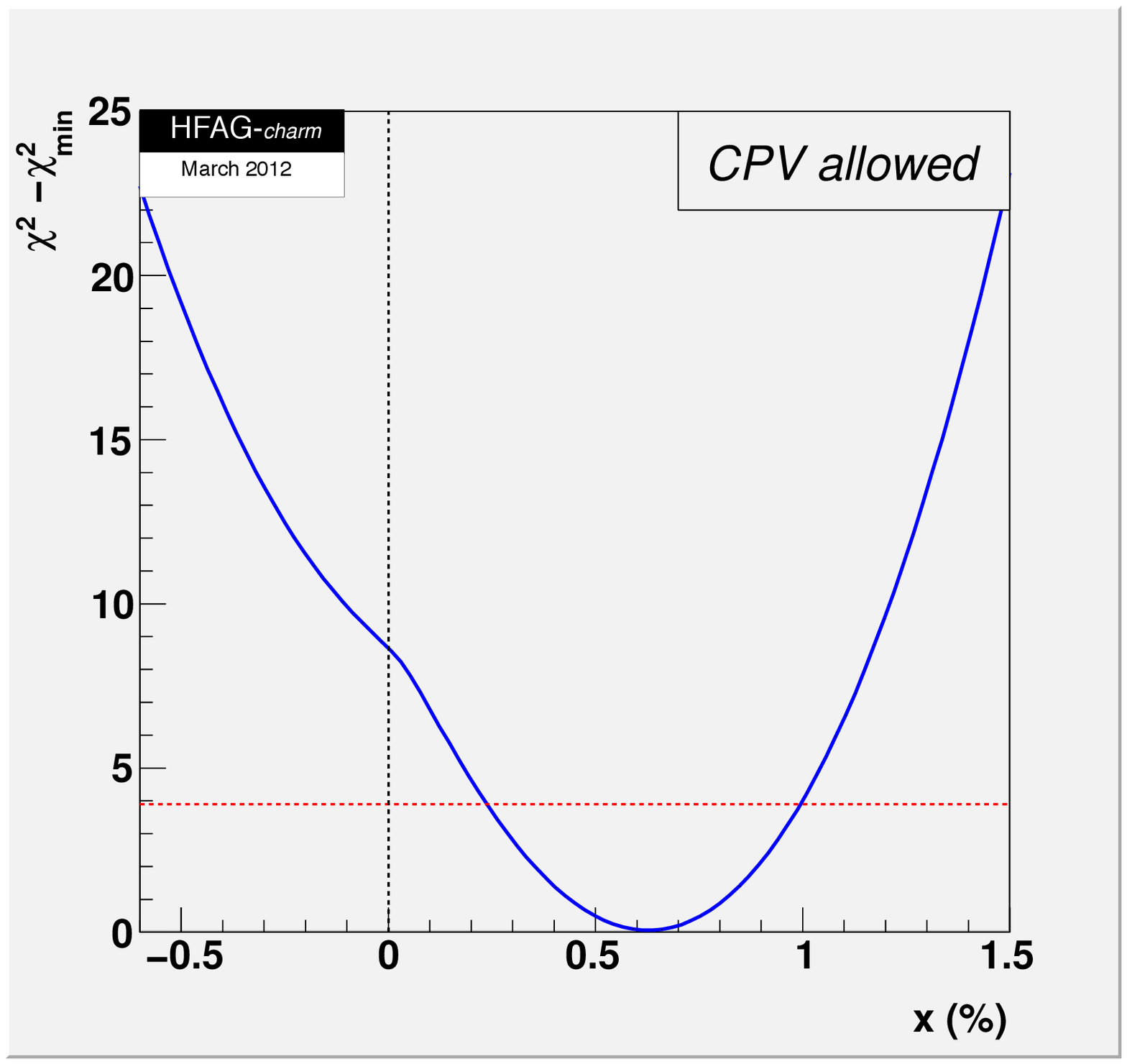}
\hskip0.20in
\includegraphics[width=72mm]{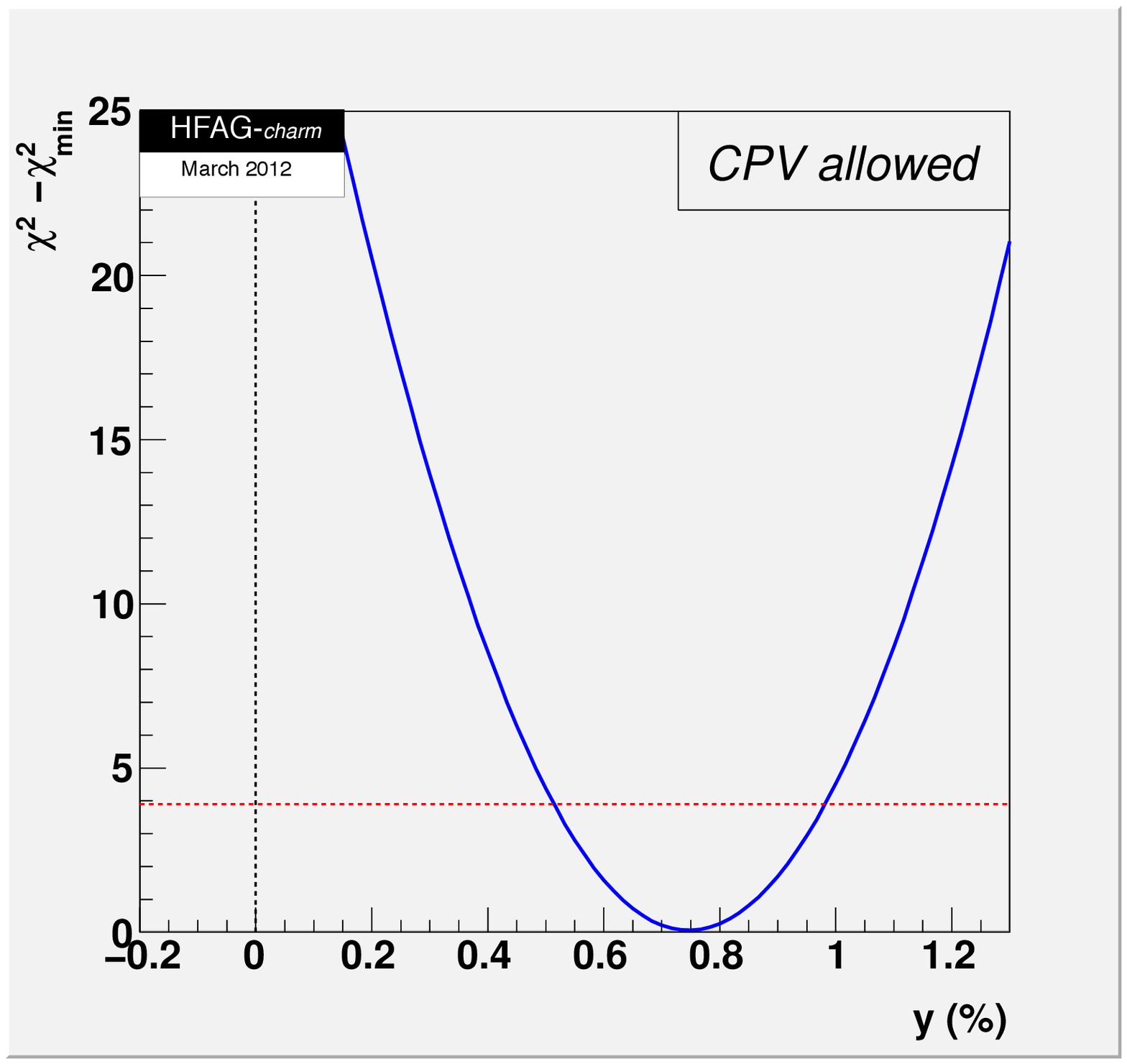}}
\hbox{\hskip0.50in
\includegraphics[width=72mm]{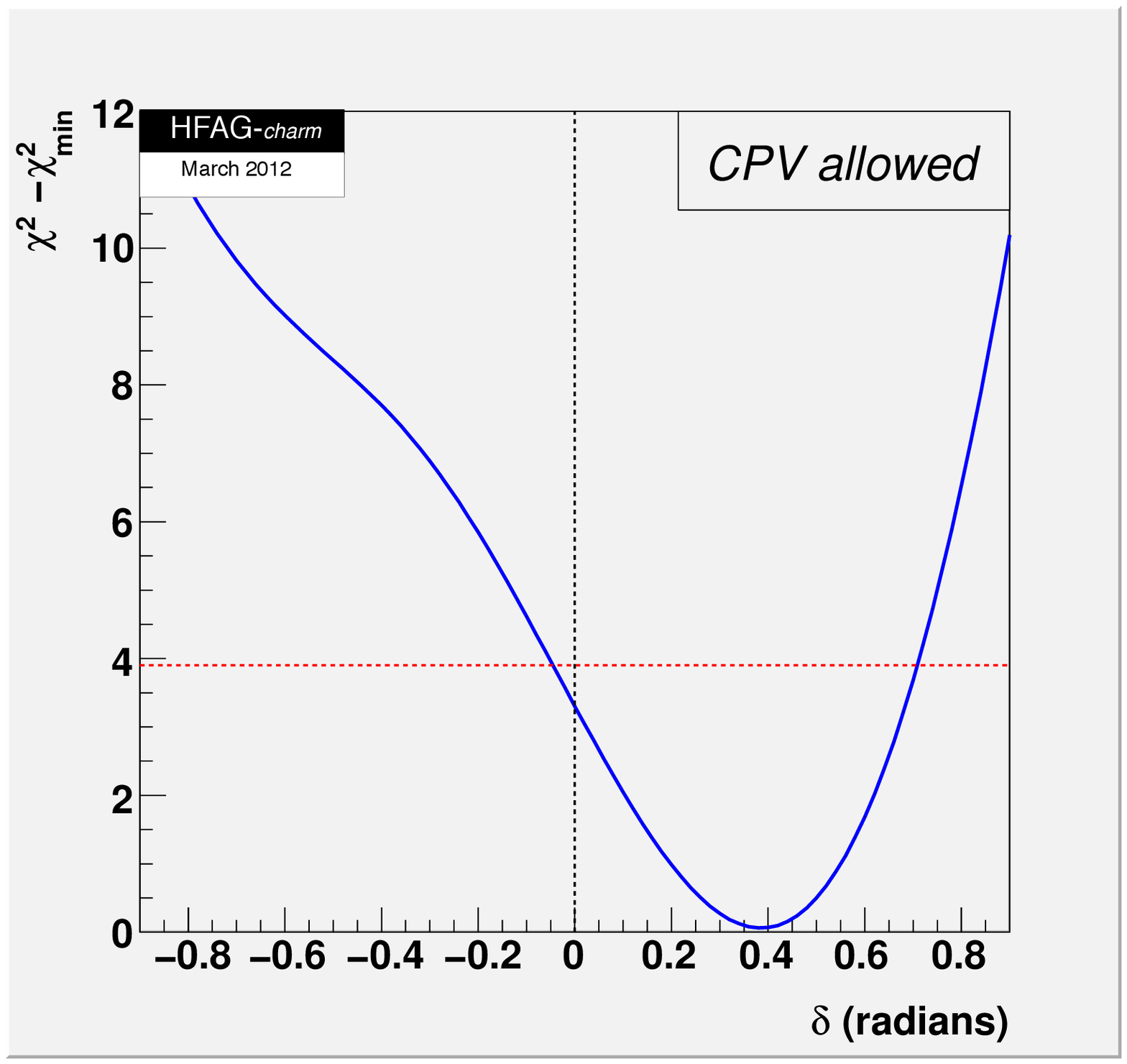}
\hskip0.20in
\includegraphics[width=72mm]{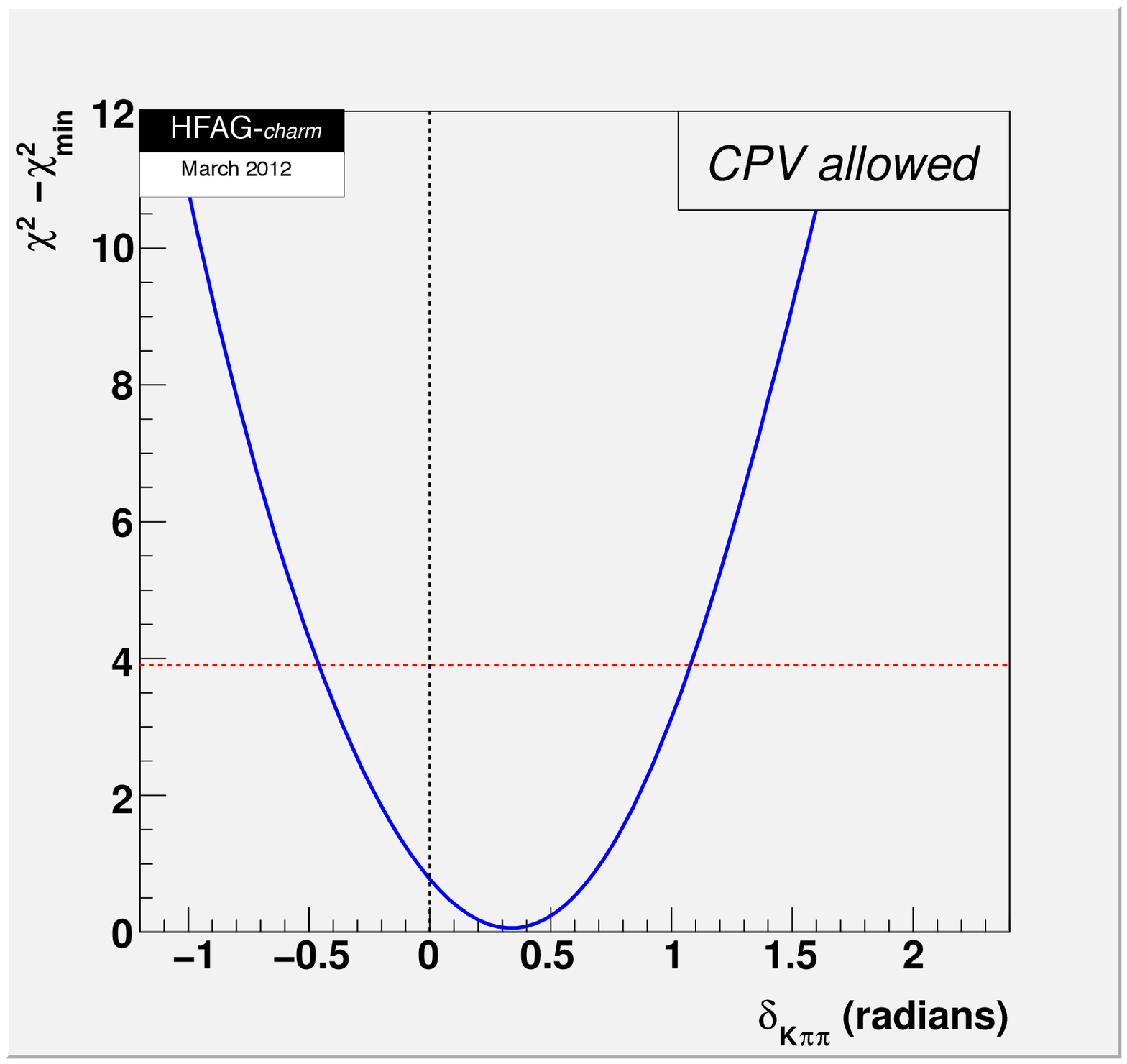}}
\hbox{\hskip0.50in
\includegraphics[width=72mm]{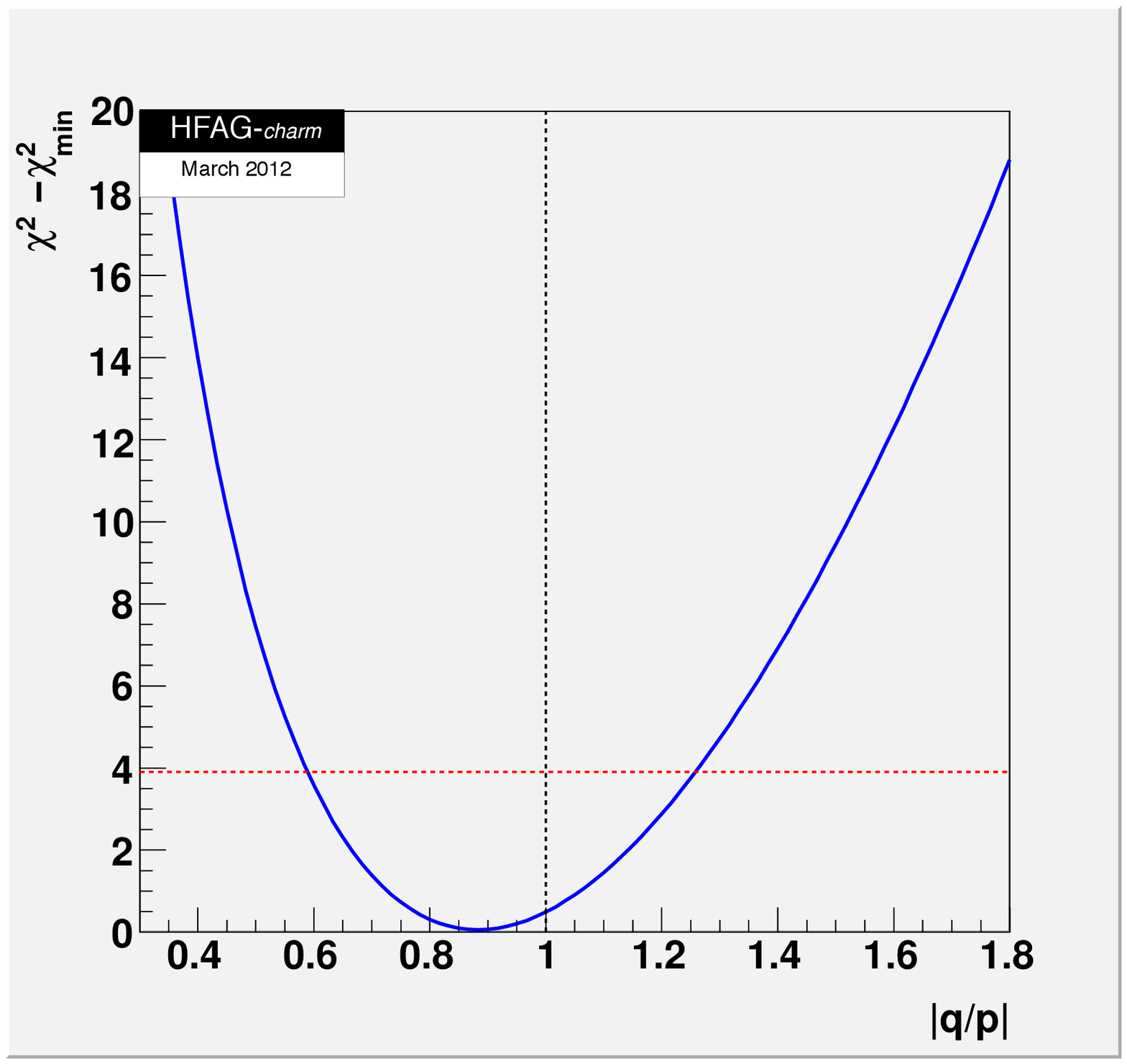}
\hskip0.20in
\includegraphics[width=72mm]{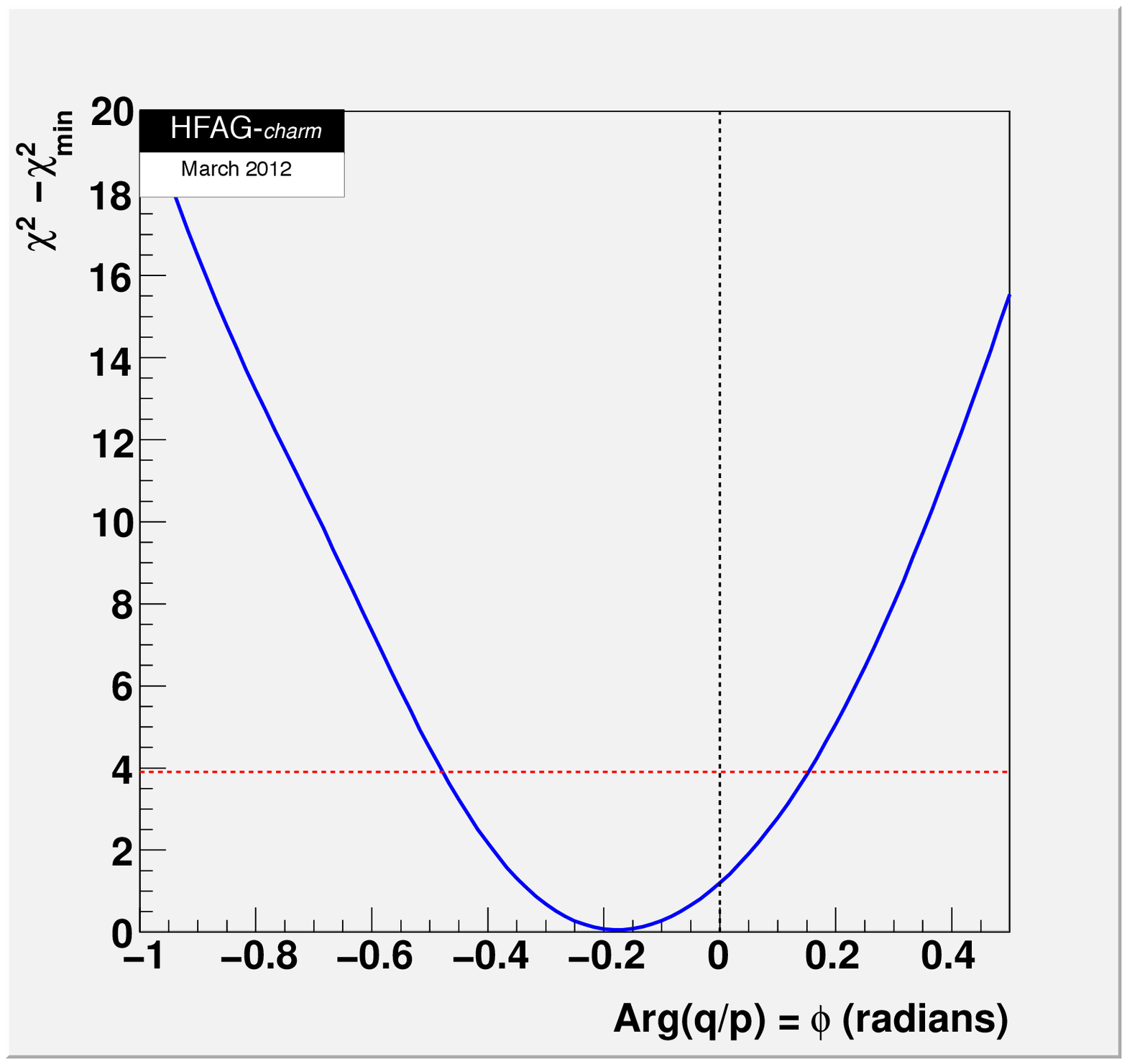}}
\end{center}
\vskip-0.30in
\caption{\label{fig:1dlikelihood}
The function $\Delta\chi^2=\chi^2-\chi^2_{\rm min}$ 
for fitted parameters
$x,\,y,\,\delta,\,\delta^{}_{K\pi\pi},\,|q/p|$, and $\phi$.
The points where $\Delta\chi^2=3.84$ (denoted by dashed 
horizontal lines) determine 95\% C.L. intervals. }
\end{figure}

One-dimensional confidence curves for individual parameters 
are obtained by letting, for any value of the parameter, all other 
fitted parameters take their preferred values. The resulting
functions $\Delta\chi^2=\chi^2-\chi^2_{\rm min}$ ($\chi^2_{\rm min}$
is the minimum value) are shown in Fig.~\ref{fig:1dlikelihood}.
The points where $\Delta\chi^2=3.84$ determine 95\% C.L. intervals 
for the parameters; these intervals are listed in Table~\ref{tab:results}.

\begin{table}
\renewcommand{\arraystretch}{1.4}
\begin{center}
\caption{\label{tab:results}
Results of the global fit for different assumptions concerning~\cpv.}
\vspace*{6pt}
\footnotesize
\begin{tabular}{c|cccc}
\hline
\textbf{Parameter} & \textbf{\boldmath No \cpv} & \textbf{\boldmath No direct \cpv} 
& \textbf{\boldmath \cpv-allowed} & \textbf{\boldmath \cpv-allowed 95\% C.L.}  \\
\hline
$\begin{array}{c}
x\ (\%) \\ 
y\ (\%) \\ 
\delta\ (^\circ) \\ 
R^{}_D\ (\%) \\ 
A^{}_D\ (\%) \\ 
|q/p| \\ 
\phi\ (^\circ) \\
\delta^{}_{K\pi\pi}\ (^\circ)  \\
A^{}_{\pi} \\
A^{}_K \\
x^{}_{12}\ (\%) \\ 
y^{}_{12}\ (\%) \\ 
\phi^{}_{12} (^\circ)
\end{array}$ & 
$\begin{array}{c}
0.65\,^{+0.18}_{-0.19} \\
0.73\,\pm 0.12 \\
21.0\,^{+9.8}_{-11.0} \\
0.3307\,\pm 0.0080 \\
- \\
- \\
- \\
17.8\,^{+21.7}_{-22.8} \\
- \\
- \\
- \\
- \\
- 
\end{array}$ &
$\begin{array}{c}
0.62\,\pm 0.19 \\
0.75\,\pm 0.12 \\
22.2\,^{+9.9}_{-11.2} \\
0.3305\,\pm 0.0080 \\
- \\
1.04\,^{+0.07}_{-0.06} \\ 
-2.02\,^{+2.67}_{-2.74} \\ 
19.4\,^{+21.8}_{-22.9} \\
- \\
- \\
0.62\,\pm 0.19 \\
0.75\,\pm 0.12 \\
4.9\,^{+7.7}_{-6.5} 
\end{array}$ &
$\begin{array}{c}
0.63\,^{+0.19}_{-0.20}  \\
0.75\,\pm 0.12 \\
22.1\,^{+9.7}_{-11.1} \\
0.3311\,\pm 0.0081 \\
-1.7\,\pm 2.4 \\
0.88\,^{+0.18}_{-0.16} \\
-10.1\,^{+9.5}_{-8.9} \\ 
19.3\,^{+21.8}_{-22.9} \\
0.36\,\pm 0.25 \\
-0.31\,\pm 0.24 \\
- \\
- \\
- 
\end{array}$ &
$\begin{array}{c}
\left[0.24 ,\, 0.99\right] \\
\left[0.51 ,\, 0.98\right] \\
\left[-2.6 ,\, 40.6\right] \\
\left[0.315 ,\, 0.347\right] \\
\left[-6.4 ,\, 3.0\right] \\
\left[0.59 ,\, 1.26\right] \\
\left[-27.4 ,\, 8.7\right] \\
\left[-26.3 ,\, 61.8\right] \\ 
\left[-0.13 ,\, 0.86\right] \\
\left[-0.78 ,\, 0.15\right] \\
\left[0.25 ,\, 0.99\right] \\
\left[0.51 ,\, 0.98\right] \\
\left[-8.4 ,\, 24.6\right] \\
\end{array}$ \\
\hline
\end{tabular}
\end{center}
\end{table}

\begin{table}
\renewcommand{\arraystretch}{1.4}
\begin{center}
\caption{\label{tab:results_chi2}
Individual contributions to the $\chi^2$ for the \cpv-allowed fit.}
\vspace*{6pt}
\footnotesize
\begin{tabular}{l|rr}
\hline
\textbf{Observable} & \textbf{\boldmath $\chi^2$} & \textbf{\boldmath $\sum\chi^2$} \\
\hline
$y^{}_{CP}$                      & 2.61 & 2.61 \\
$A^{}_\Gamma$                    & 0.00 & 2.61 \\
\hline
$x^{}_{K^0\pi^+\pi^-}$ Belle       & 0.28 & 2.88 \\
$y^{}_{K^0\pi^+\pi^-}$ Belle       & 1.65 & 4.54 \\
$|q/p|^{}_{K^0\pi^+\pi^-}$ Belle   & 0.01 & 4.54 \\
$\phi^{}_{K^0\pi^+\pi^-}$  Belle   & 0.51 & 5.05 \\
\hline
$x^{}_{K^0 h^+ h^-}$ \babar         & 2.97 & 8.02 \\
$y^{}_{K^0 h^+ h^-}$ \babar         & 0.37 & 8.38 \\
\hline
$R^{}_M(K^+\ell^-\nu)$           & 0.09 & 8.48 \\
\hline
$x^{}_{K^+\pi^-\pi^0}$ \babar       & 5.71 & 14.19 \\
$y^{}_{K^+\pi^-\pi^0}$ \babar       & 2.22 & 16.40 \\
\hline
CLEOc                           &      &       \\
($x/y/R^{}_D/\sqrt{R^{}_D}\cos\delta/\sqrt{R^{}_D}\sin\delta$) 
                                & 7.28 & 23.68 \\
\hline
$R^+_D/x'{}^{2+}/y'{}^+$ \babar   & 2.34 & 26.02    \\
$R^-_D/x'{}^{2-}/y'{}^-$ \babar   & 1.30 & 27.31    \\
$R^+_D/x'{}^{2+}/y'{}^+$ Belle   & 4.12 & 31.44    \\
$R^-_D/x'{}^{2-}/y'{}^-$ Belle   & 1.35 & 32.79    \\
$R^{}_D/x'{}^{2}/y'$ CDF         & 0.39 & 33.17    \\
\hline
$A^{}_{KK}/A^{}_{\pi\pi}$  \babar  & 1.89 & 35.06  \\
$A^{}_{KK}/A^{}_{\pi\pi}$  Belle  & 0.12 & 35.18  \\
$A^{}_{KK}/A^{}_{\pi\pi}$  CDF    & 0.06 & 35.25  \\
$A^{}_{KK}-A^{}_{\pi\pi}$  LHCb   & 0.37 & 35.62  \\
\hline
\end{tabular}
\end{center}
\end{table}


\subsubsection{Conclusions}

From the fit results listed in Table~\ref{tab:results}
and shown in Figs.~\ref{fig:contours_cpv} and \ref{fig:1dlikelihood},
we conclude the following:
\begin{itemize}
\item the experimental data consistently indicate that 
$D^0$ mesons undergo mixing. The no-mixing point $x=y=0$
is excluded at $10.2\sigma$. The parameter $x$ differs from
zero by $2.7\sigma$, and $y$ differs from zero by
$6.0\sigma$. This mixing is presumably dominated 
by long-distance processes, which are difficult to calculate.
Unless it turns out that $|x|\gg |y|$~\cite{Bigi:2000wn},
which is not indicated, it will probably be difficult to 
identify new physics from $(x,y)$ alone.
\item Since \ycp\ is positive, the \cp-even state is shorter-lived
as in the $K^0$-$\kbar$ system. However, since $x$ also appears
to be positive, the \cp-even state is heavier, 
unlike in the $K^0$-$\kbar$ system.
\item The LHCb and CDF experiments have obtained first evidence
for {\it direct\/} \cpv\ in $D^0$ decays. Higher statistics 
measurements should be able to clarify this effect. There is 
no evidence for \cpv\ arising from $D^0$-$\dbar$ mixing 
($|q/p|\neq 1$) or from a phase difference between the 
mixing amplitude and a direct decay amplitude ($\phi\neq 0$). 
\end{itemize}

\clearpage
\subsection{Semileptonic decays}

\subsubsection{Introduction}

Semileptonic decays of $D$ mesons involve the interaction of a leptonic
current with a hadronic current. The latter is nonperturbative
and cannot be calculated from first principles; thus it is usually
parameterized in terms of form factors. The transition matrix element 
is written
\begin{eqnarray}
  {\cal M} & = & -i\,\frac{G_F}{\sqrt{2}}\,V^{}_{cq}\,L^\mu H_\mu\,,
  \label{Melem}
\end{eqnarray}
where $G_F$ is the Fermi constant and $V^{}_{cq}$ is a CKM matrix element.
The leptonic current $L_\mu$ is evaluated directly from the lepton spinors 
and has a simple structure; this allows one to extract information about 
the form factors (in $H^{}_\mu$) from data on semileptonic decays~\cite{Becher:2005bg}.  
Conversely, because there are no final-state interactions between the
leptonic and hadronic systems, semileptonic decays for which the form 
factors can be calculated allow one to 
determine~$V^{}_{cq}$~\cite{Kobayashi:1973fv}.

\subsubsection{$D\ra P \overline \ell \nu_{\ell}$ decays}

When the final state hadron is a pseudoscalar, the hadronic 
current is given by
\begin{eqnarray}
H_\mu & = & \left< P(p) | \bar{q}\gamma^\mu c | D(p') \right> \ =\  
f_+(q^2)\left[ (p' + p)^\mu -\frac{m_D^2-m_P^2}{q^2}q^\mu\right] + 
 f_0(q^2)\frac{m_D^2-m_P^2}{q^2}q^\mu\,,
\label{eq:hadronic}
\end{eqnarray}
where $m_D$ and $p'$ are the mass and four momentum of the 
parent $D$ meson, $m_P$ and $p$ are those of the daughter meson, 
$f_+(q^2)$ and $f_0(q^2)$ are form factors, and $q = p' - p$.  
Kinematics require that $f_+(0) = f_0(0)$.
The contraction $q^\mu L_\mu$ results in terms proportional 
to $m^{}_\ell$\cite{Gilman:1989uy}, and thus for $\ell=e $
the last two terms in Eq.~(\ref{eq:hadronic}) are negligible. 
Considering that, only the $f_+(q^2)$ form factor 
is relevant, the differential partial width is
\begin{eqnarray}
\frac{d\Gamma(D \to P \bar \ell \nu_\ell)}{dq^2\, d\cos\theta_\ell} & = & 
   \frac{G_F^2|V_{cq}|^2}{32\pi^3} p^{*\,3}|f_{+}(q^2)|^2\sin\theta^2_\ell\,,
\label{eq:dGamma}
\end{eqnarray}
where ${p^*}$ is the magnitude of the momentum of the final state hadron
in the $D$ rest frame.


\subsubsection{Form factor parameterizations} 

The form factor is traditionally parametrized with an explicit pole 
and a sum of effective poles:
\begin{eqnarray}
f_+(q^2) & = & \frac{f_+(0)}{1-\alpha}
\left(\frac{1}{1- q^2/m^2_{\rm pole}}\right)\ +\ 
\sum_{k=1}^{N}\frac{\rho_k}{1- q^2/(\gamma_k\,m^2_{\rm pole})}\,,
\label{eqn:expansion}
\end{eqnarray}
where $\rho_k$ and $\gamma_k$ are expansion parameters. The parameter
$m_{{\rm pole}}$ is the mass of the lowest-lying $c\bar{q}$ resonance
with the appropriate quantum numbers; this is expected to provide the
largest contribution to the form factor for the $c\ra q$ transition.  
For example, for $D\to\pi$ transitions the dominant resonance is
expected to be $D^*$, and thus $m^{}_{\rm pole}=m^{}_{D^*}$.

\subsubsubsection{Simple pole}

Equation~(\ref{eqn:expansion}) can be simplified by neglecting the 
sum over effective poles, leaving only the explicit vector meson pole. 
This approximation is referred to as ``nearest pole dominance'' or 
``vector-meson dominance.''  The resulting parameterization is
\begin{eqnarray}
  f_+(q^2) & = & \frac{f_+(0)}{(1-q^2/m^2_{\rm pole})}\,. 
\label{SimplePole}
\end{eqnarray}
However, values of $m_{{\rm pole}}$ that give a good fit to the data 
do not agree with the expected vector meson masses~\cite{Hill:2006ub}. 
To address this problem, the ``modified pole'' or Becirevic-Kaidalov~(BK) 
parameterization~\cite{Becirevic:1999kt} was introduced.
$m_{\rm pole} /\sqrt{\alpha_{\rm BK}}$
is interpreted as the mass of an effective pole, higher than $m_{\rm pole}$, thus it is expected that $\alpha_{\rm BK}<1$.


The parameterization takes the form
\begin{eqnarray}
f_+(q^2) & = & \frac{f_+(0)}{(1-q^2/m^2_{\rm pole})}
\frac{1}{\left(1-\alpha^{}_{\rm BK}\frac{q^2}{m^2_{\rm pole}}\right)}\,.
\end{eqnarray}
 

This parameterization has been used by several experiments to 
determine form factor parameters.
Measured values of $m^{}_{\rm pole}$ and $\alpha^{}_{\rm BK}$ are
listed Tables~\ref{kPseudoPole} and~\ref{piPseudoPole} for
$D\to K\ell\nu_{\ell}$ and $D\to\pi\ell\nu_{\ell}$ decays, respectively.


\subsubsubsection{$z$ expansion}

Several groups have advocated an alternative series 
expansion around some value $q^2=t_0$ to parameterize 
$f^{}_+$~\cite{Boyd:1994tt,Boyd:1997qw,Arnesen:2005ez,Becher:2005bg}.
This expansion is given in terms of a complex parameter $z$,
which is the analytic continuation of $q^2$ into the
complex plane:
\begin{eqnarray}
z(q^2,t_0) & = & \frac{\sqrt{t_+ - q^2} - \sqrt{t_+ - t_0}}{\sqrt{t_+ - q^2}
	  + \sqrt{t_+ - t_0}}\,, 
\end{eqnarray}
where $t_\pm \equiv (m_D \pm m_P)^2$ and $t_0$ is the (arbitrary) $q^2$ 
value corresponding to $z=0$. The physical region corresponds to $|z|<1$.

The form factor is expressed as
\begin{eqnarray}
f_+(q^2) & = & \frac{1}{P(q^2)\,\phi(q^2,t_0)}\sum_{k=0}^\infty
a_k(t_0)[z(q^2,t_0)]^k\,,
\label{z_expansion}
\end{eqnarray}
where the $P(q^2)$ factor accommodates sub-threshold resonances via
\begin{eqnarray}
P(q^2) & \equiv & 
\begin{cases} 
1 & (D\to \pi) \\
z(q^2,M^2_{D^*_s}) & (D\to K)\,. 
\end{cases}
\end{eqnarray}
The ``outer'' function $\phi(t,t_0)$ can be any analytic function,
but a preferred choice (see, {\it e.g.}
Refs.~\cite{Boyd:1994tt,Boyd:1997qw,Bourrely:1980gp}) obtained
from the Operator Product Expansion (OPE) is
\begin{eqnarray}
\phi(q^2,t_0) & =  & \alpha 
\left(\sqrt{t_+ - q^2}+\sqrt{t_+ - t_0}\right) \times  \nonumber \\
 & & \hskip0.20in \frac{t_+ - q^2}{(t_+ - t_0)^{1/4}}\  
\frac{(\sqrt{t_+ - q^2}\ +\ \sqrt{t_+ - t_-})^{3/2}}
     {(\sqrt{t_+ - q^2}+\sqrt{t_+})^5}\,,
\label{eqn:outer}
\end{eqnarray}
with $\alpha = \sqrt{\pi m_c^2/3}$.
The OPE analysis provides a constraint upon the 
expansion coefficients, $\sum_{k=0}^{N}a_k^2 \leq 1$.
These coefficients receive $1/M_D$ corrections, and thus
the constraint is only approximate. However, the
expansion is expected to converge rapidly since 
$|z|<0.051\ (0.17)$ for $D\ra K$ ($D\ra\pi$) over 
the entire physical $q^2$ range, and Eq.~(\ref{z_expansion}) 
remains a useful parameterization.

\subsubsection{Experimental techniques and results}
 Different techniques by several experiments have been used to measure D meson semileptonic decays with a pseudoscalar 
particle in the final state. The most recent results are provided by the  
Belle \cite{Widhalm:2006wz}, \babar \cite{Aubert:2007wg} and CLEO-c \cite{Besson:2009uv},\cite{Dobbs:2007aa} collaborations. 
The BES III experiment has also reported preliminary results at the CHARM 2012 conference with 923 $\pb^{-1}$ \cite{BESIII}.  
The Belle experiment fully reconstruct the D events from the continuum under the $\Upsilon(4S)$ resonance, 
achieving a very good $q^2$ resolution ($\Delta q^2 = 15 MeV^2$) and low background level, 
but having a low efficiency. Using 282 $\fb^{-1}$, about 1300 and 115 signal semileptonic decays are 
isolated for each lepton flavour (e and $\mu$). 
The \babar experiment uses a partial reconstruction technique where the semileptonic decays are tagged through the 
$ D^{\ast +}\to D^0\pi^+$ decay. The D direction and neutrino energy is obtained using information of the rest of the event. 
With 75 $\fb^{-1}$ 74000 signal events in the $D^0 \to {K}^- e^+ \nu$ mode are obtained. The measurement of the Cabibbo suppressed 
mode has not been published yet. This technique provides larger statistics but higher background level 
and poorer $q^2$ resolution ($\Delta q^2$ ranges from 66 to 219 $MeV^2$). In this case the measurement of the branching fraction is obtained by normalizing to the $D^0 \to K^- \pi^+$ decay channel and can benefit from future improvements 
in the determination of this reference channel. 
The CLEO-c experiment uses two different methods to measure charm semileptonic decays. Tagged analyses \cite{Besson:2009uv} 
rely on the full reconstruction of $\Psi(3770)\to D {\overline D}$ events. One of the D mesons is reconstructed in a hadronic decay 
mode, the other in the semileptonic channel. The only missing particle is the neutrino so the $q^2$ resolution is very good and
the background level very low.   
With the entire CLEO-c data sample, 818 $\pb^{-1}$, 14123 and 1374 signal events are reconstructed for 
the $D^0 \to K^{-} e^+\nu$ and $D^0\to \pi^{-} e^+\nu$ channels, respectively, and 8467 and 838 for the 
$D^+\to {\overline K}^{0} e^+\nu$ and $D^+\to \pi^{0} e^+\nu$ decays. 
Another technique without tagging the D meson in a hadronic mode (``untagged'' in the following) has been also 
used by CLEO-c \cite{Dobbs:2007aa}. This method rests upon the association of the missing energy and momentum in 
an event with the neutrino four momentum, with the penalty of larger background as compared to the tagged method. 

  Previous measurements were also performed by CLEO~III and FOCUS experiments. 
Events registered at the $\Upsilon (4S)$ energy corresponding to an integrated luminosity of 7 $\fb^{-1}$ were 
analyzed by CLEO~III \cite{Huang:2004fra}. 
In the FOCUS fixed target photo-production experiment $D^0$ semileptonic events were obtained from the decay of 
a $D^{\ast +}$, and the kaon or pion was reconstructed in the muon channel. 

 Results of the hadronic form factor parameters by the different groups are given in Tables
\ref{kPseudoPole} and \ref{piPseudoPole} for $m_{pole}$ and $\alpha_{BK}$. 
\begin{table}[htbp]
\caption{Results for $m_{\rm pole}$ and $\alpha_{\rm BK}$ from various
  experiments for $D^0\to K^-\ell^+\nu$ and $D^+\to K_S\ell^+\nu$
  decays. The last entry is a lattice QCD prediction (errors have been increased as compared to the publication 
to take into account remaining systematic uncertainties in Lattice calculations, as advised by the authors).
\label{kPseudoPole}}
\begin{center}
\begin{tabular}{cccc}
\hline
\vspace*{-10pt} & \\
 $D\to K\ell\nu_\ell$ Expt. & Ref.  & $m_{\rm pole}$ (GeV$/c^2$) 
& $\alpha^{}_{\rm BK}$       \\
\vspace*{-10pt} & \\
\hline
 \omit        & \omit                         & \omit                                  & \omit                  \\
 CLEO III     & \cite{Huang:2004fra}          & $1.89\pm0.05^{+0.04}_{-0.03}$          & $0.36\pm0.10^{+0.03}_{-0.07}$ \\
 FOCUS        & \cite{Link:2004dh}            & $1.93\pm0.05\pm0.03$                   & $0.28\pm0.08\pm0.07$     \\
 Belle        & \cite{Widhalm:2006wz}         & $1.82\pm0.04\pm0.03$                   & $0.52\pm0.08\pm0.06$     \\
 \babar        & \cite{Aubert:2007wg}          & $1.889\pm0.012\pm0.015$                & $0.366\pm0.023\pm0.029$  \\

 CLEO-c (tagged)   &\cite{Besson:2009uv}      & $1.93\pm0.02\pm0.01$                   & $0.30\pm0.03\pm0.01$     \\
 CLEO-c (untagged, $D^0$) &\cite{Dobbs:2007aa}       & $1.97 \pm0.03 \pm 0.01 $ & $0.21 \pm 0.05 \pm 0.03 $  \\
 CLEO-c (untagged, $D^+$) &\cite{Dobbs:2007aa}       & $1.96 \pm0.04 \pm 0.02 $ & $0.22 \pm 0.08 \pm 0.03$  \\
 BESIII (prel)     &\cite{BESIII}                     & $1.943 \pm 0.025 \pm 0.003$ & $ 0.265 \pm 0.045 \pm 0.006$   \\

\hline
 Fermilab lattice/MILC/HPQCD & \cite{Aubin:2004ej}            & --                                &  $0.50\pm0.04\pm0.07$         \\
\vspace*{-10pt} & \\
\hline
\end{tabular}
\end{center}
\end{table}

\begin{table}[htbp]
\caption{Results for $m_{\rm pole}$ and
  $\alpha_{\rm BK}$ from various experiments for 
  $D^0\to \pi^-\ell^+\nu$ and $D^+\to \pi^0\ell^+\nu$ decays.  The last 
  entry is a lattice QCD prediction (errors have been increased as compared to the publication 
to take into account remaining systematic uncertainties in Lattice calculations, as advised by the authors).
\label{piPseudoPole}}
\begin{center}
\begin{tabular}{cccc}
\hline
\vspace*{-10pt} & \\
 $D\to \pi\ell\nu_\ell$ Expt. & Ref.               & $m_{\rm pole}$ (GeV$/c^2$) & $\alpha_{\rm BK}$ \\
\vspace*{-10pt} & \\
\hline
 \omit        & \omit                         & \omit                                  & \omit                  \\
 CLEO III     & \cite{Huang:2004fra}          & $1.86^{+0.10+0.07}_{-0.06-0.03}$       & $0.37^{+0.20}_{-0.31}\pm0.15$         \\
 FOCUS        & \cite{Link:2004dh}            & $1.91^{+0.30}_{-0.15}\pm0.07$          & --                                    \\
 Belle        & \cite{Widhalm:2006wz}         & $1.97\pm0.08\pm0.04$                   & $0.10\pm0.21\pm0.10$                  \\
 CLEO-c (tagged)   &\cite{Besson:2009uv}      & $1.91\pm0.02\pm0.01$                   & $0.21\pm0.07\pm0.02$     \\
 CLEO-c (untagged, $D^0$) &\cite{Dobbs:2007aa}       & $1.87 \pm0.03 \pm 0.01 $ & $0.21 \pm 0.05 \pm 0.03 $  \\
 CLEO-c (untagged, $D^+$) &\cite{Dobbs:2007aa}       & $1.97 \pm0.07 \pm 0.02 $ & $0.22 \pm 0.08 \pm 0.03$  \\
 BESIII (prel)     &\cite{BESIII}                    & $1.876 \pm 0.023 \pm 0.004$ & $ 0.315 \pm 0.071 \pm 0.012$   \\

\hline
 Fermilab lattice/MILC/HPQCD & \cite{Aubin:2004ej}            & --                              & $0.44\pm 0.04\pm 0.07$         \\
\vspace*{-10pt} & \\
\hline
\end{tabular}
\end{center}
\end{table}

The $z$-expansion formalism has been used by 
\babar~\cite{Aubert:2007wg} and CLEOc~\cite{Besson:2009uv},~\cite{Dobbs:2007aa}. 
BES III has also shown preliminary results\cite{BESIII} . 
Their fits used the first three terms of the expansion,
and the results for the ratios $r_1\equiv a_1/a_0$ and $r_2\equiv a_2/a_0$ are 
listed in Tables~\ref{KPseudoZ} and~\ref{piPseudoZ}. The CLEO~III\cite{Huang:2004fra} and FOCUS\cite{Link:2004dh} results
listed are obtained by refitting their data using the full
covariance matrix. The \babar correlation coefficient listed is 
obtained by refitting their published branching fraction using 
their published covariance matrix.  

These measurements correspond to using the standard 
outer function $\phi(q^2,t_0)$ of Eq.~(\ref{eqn:outer}) and 
$t_0=t_+\left(1-\sqrt{1-t_-/t_+}\right)$. This choice of $t^{}_0$
constrains $|z|$ to vary between $\pm z_{max.}$

\begin{table}[htbp]
\caption{Results for $r_1$ and $r_2$ from various experiments, for 
$D\to K\ell\nu_{\ell}$. The correlation coefficient listed is 
for the total uncertainties (statistical $\oplus$ systematic) on $r^{}_1$ and~$r^{}_2$.}
\label{KPseudoZ}
\begin{center}
\begin{tabular}{cccccc}
\hline
\vspace*{-10pt} & \\
Expt. $D\to K\ell\nu_{\ell}$     & mode &  Ref.                         & $r_1$               & $r_2$               & $\rho$        \\
\hline
 \omit    & \omit         & \omit                & \omit               & \omit               & \omit         \\
 CLEO III & \omit  & \cite{Huang:2004fra} & $0.2^{+3.6}_{-3.0}$ & $-89^{+104}_{-120}$ & -0.99         \\
 FOCUS    & \omit                & \cite{Link:2004dh}   & $-2.54\pm0.75$  & $7\pm 13$       & -0.97 \\
 \babar    & \omit        & \cite{Aubert:2007wg} & $-2.5\pm0.2\pm0.2$  & $2.5\pm6.0\pm5.0$     & -0.64         \\
 CLEO-c (tagged)     & $D^0\to K^-$    & \cite{Besson:2009uv}          & $-2.65\pm0.34\pm0.08$  & $13\pm9\pm1$       & -0.82 \\
 CLEO-c (tagged)     & $D^+\to \overline K^0$   & \cite{Besson:2009uv} & $-1.66\pm0.44\pm0.10$  & $-14\pm11\pm1$       & -0.82 \\
 CLEO-c (untagged)   & $D^0\to K^-$           &\cite{Dobbs:2007aa}     & $-2.4\pm0.4\pm0.1$  & $21\pm11\pm2$     & -0.81    \\
 CLEO-c (untagged)   & $D^+\to \overline K^0$ & \cite{Dobbs:2007aa}    & $-2.8\pm6\pm2$      & $32\pm18\pm4$       & -0.84         \\
 BES III  & \omit         & \cite{BESIII}        & $-2.18\pm0.36\pm0.05$ & $5\pm 9\pm 1$  &            \\
\hline
\hline
 Combined  & \omit         &  \omit               & $-2.39\pm0.17$       & $6.2\pm3.8$         & -0.82        \\ 
\hline
\end{tabular}
\end{center}
\end{table}

\begin{table}[htbp]
\caption{Results for $r_1$ and $r_2$ from various experiments, for 
$D\to \pi \ell\nu_{\ell}$. The correlation coefficient listed is 
for the total uncertainties (statistical $\oplus$ systematic) on 
$r^{}_1$ and~$r^{}_2$.}
\label{piPseudoZ}
\begin{center}
\begin{tabular}{cccccc}
\hline
\vspace*{-10pt} & \\
Expt. $D\to \pi\ell\nu_{\ell}$     & mode &  Ref.                         & $r_1$               & $r_2$               & $\rho$        \\
\hline
 \omit    & \omit         & \omit                & \omit               & \omit               & \omit         \\
\hline
\hline
 CLEO-c (tagged)     & $D^0\to\pi^+$ & \cite{Besson:2009uv}      &  $-2.80\pm0.49\pm0.04$ & $6\pm 3\pm$ 0 & -0.94 \\            
 CLEO-c (tagged)     & $D^+\to\pi^0$ & \cite{Besson:2009uv}      &  $-1.37\pm0.88\pm0.24$ & $-4\pm 5\pm$ 1 & -0.96 \\            
 CLEO-c  (untagged)  & $D^0\to\pi^+$ & \cite{Dobbs:2007aa}  & $-2.1\pm0.7\pm0.3$      & $-1.2\pm4.8\pm1.7$  & -0.96         \\
 CLEO-c   (untagged) & $D^+\to\pi^0$ & \cite{Dobbs:2007aa}  & $-0.2\pm1.5\pm0.4$    & $-9.8\pm9.1\pm2.1$  & -0.97         \\
 BES III  & \omit    & \cite{BESIII} & $-2.73\pm0.48\pm0.08$ & $4\pm 3\pm$ 1&            \\
 \hline 
 \hline
 Combined  & \omit         &  \omit               & $-2.69\pm 0.32$       & $ 4.18\pm 2.16$         & -0.95        \\ 
\vspace*{-10pt} & \\
\hline
\end{tabular}
\end{center}
\end{table}

Tables~\ref{KPseudoZ} and~\ref{piPseudoZ} also list average values for $r_1$ and $r_2$  
obtained from a 3D fit, taking the full correlations between $|V_{cq}|f_+(0)$, $r_1$ and $r_2$ into account, 
to CLEO~III, FOCUS, \babar, CLEO-c, and BES III data. Only the $D^0$ channels are entering in the fit. 
The fit is constrained by the branching fractions measured at Belle \cite{Widhalm:2006wz}. 



In the values quoted in Tables~\ref{kPseudoPole} and~\ref{KPseudoZ} 
the effect of radiative events has been taken into account slightly modifying the values from \babar by 
correcting the numbers given in Tab. III of Ref.~\cite{Aubert:2007wg} 
by the shifts quoted in the last column of Tab. IV given in Ref.~\cite{Aubert:2007wg}. 


The $\chi^2/d.o.f$ of the combined fits are $16/22$ and $6.2/10$ for $D^0\to K^-\ell^+\nu_{\ell}$ 
and $D^0\to \pi^-\ell^+\nu_{\ell}$, respectively. The correlation matrices are given in Tables \ref{tab:corrpi} and \ref{tab:corrK}.

\begin{table} 
\begin{center}
\caption{Correlation matrix for the combined fit for the $D^0\to \pi^-\ell^+\nu_\ell$ channel}
\label{tab:corrpi}
\begin{tabular}{c  c c c } \\
\hline
 \omit & $|V_{cd}|f_{+}^{\pi}(0)$ & $r_1$ &  $r_2$ \\
\hline
 $|V_{cd}|f_{+}^{\pi}(0)$ & 1.000 & -0.446 & 0.672 \\
 $r_1$                    & -0.446 & 1.000 & -0.946 \\
 $r_2$                    & 0.672 & -0.946 & 1.000 \\
\hline
\end{tabular}
\end{center}
\end{table}

\begin{table} 
\begin{center}
\caption{Correlation matrix for the combined fit for the $D^0\to K^-\ell^+\nu_\ell$ channel}
\label{tab:corrK}
\begin{tabular}{c  c c c }
\hline
 \omit & $|V_{cs}|f_{+}^{K}(0)$ & $r_1$ &  $r_2$ \\
\hline 
$|V_{cs}|f_{+}^{K}(0)$ & 1.000 & -0.088 & 0.433 \\
                 $r_1$ & -0.088 & 1.000 &-0.824 \\
                 $r_2$ & 0.433 & -0.824 & 1.000 \\
\hline
\end{tabular}
\end{center}
\end{table}


\begin{figure}[p]
\begin{center}
\includegraphics[width=0.47\textwidth]{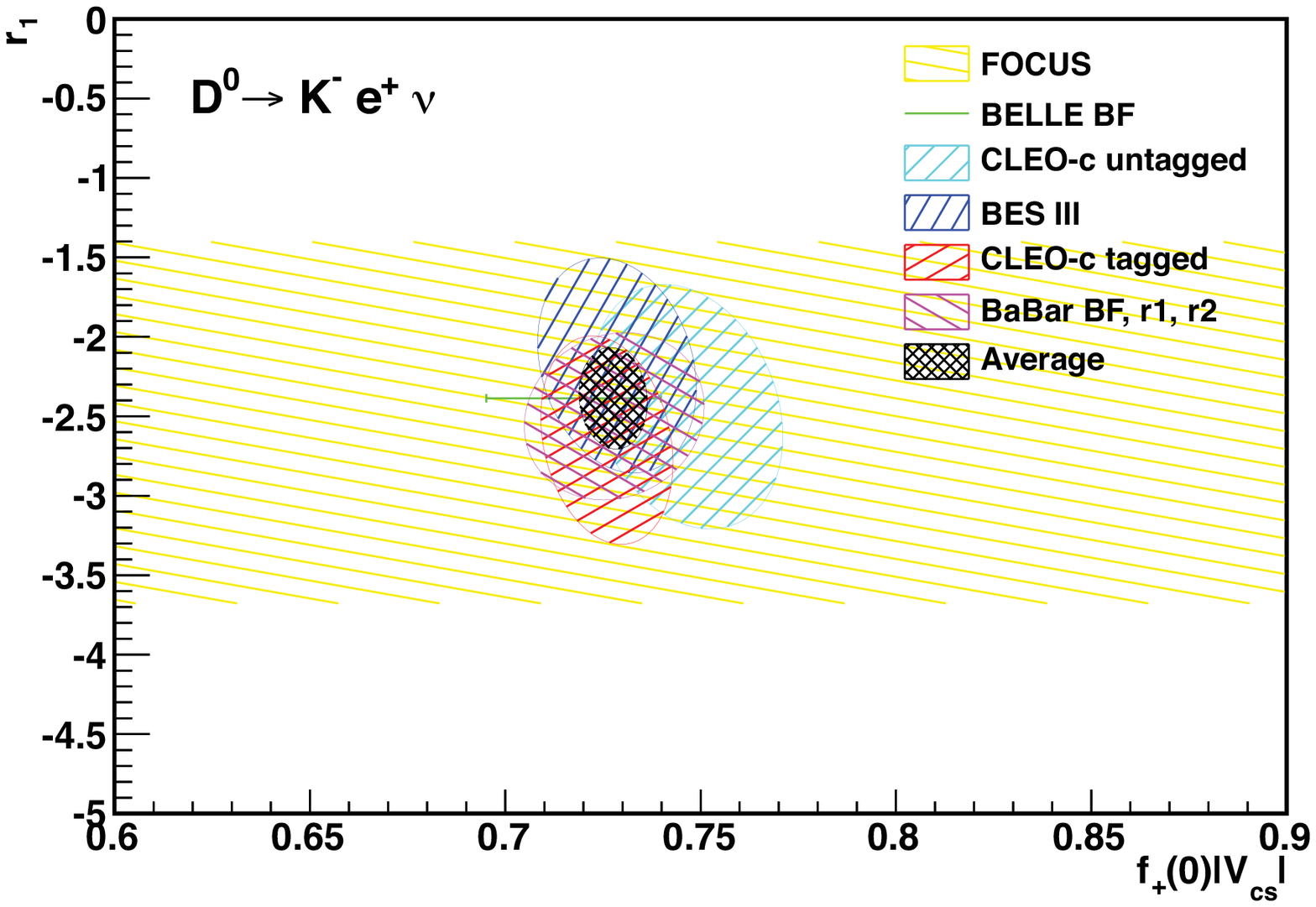}\hfill
\includegraphics[width=0.47\textwidth]{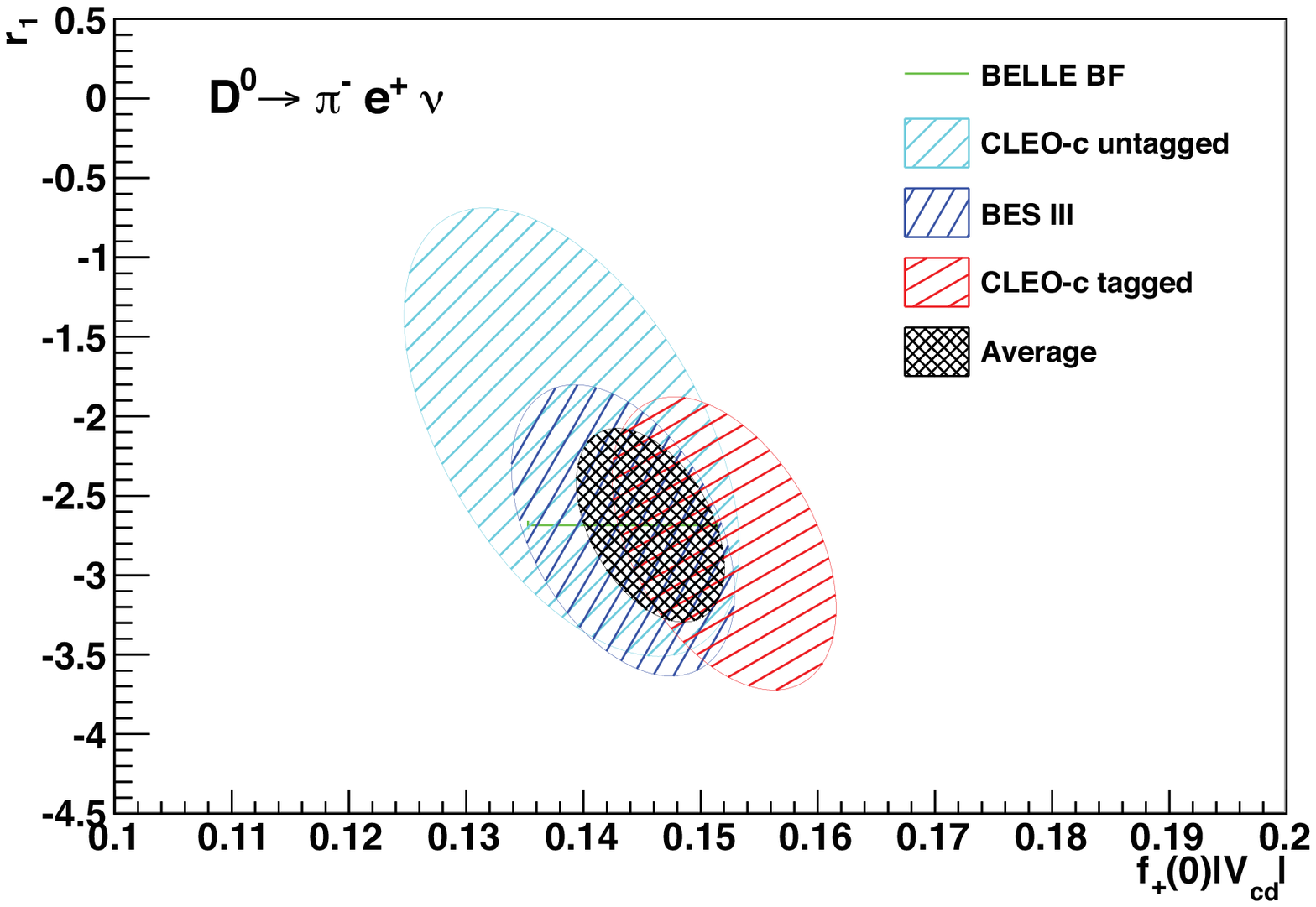}
\includegraphics[width=0.47\textwidth]{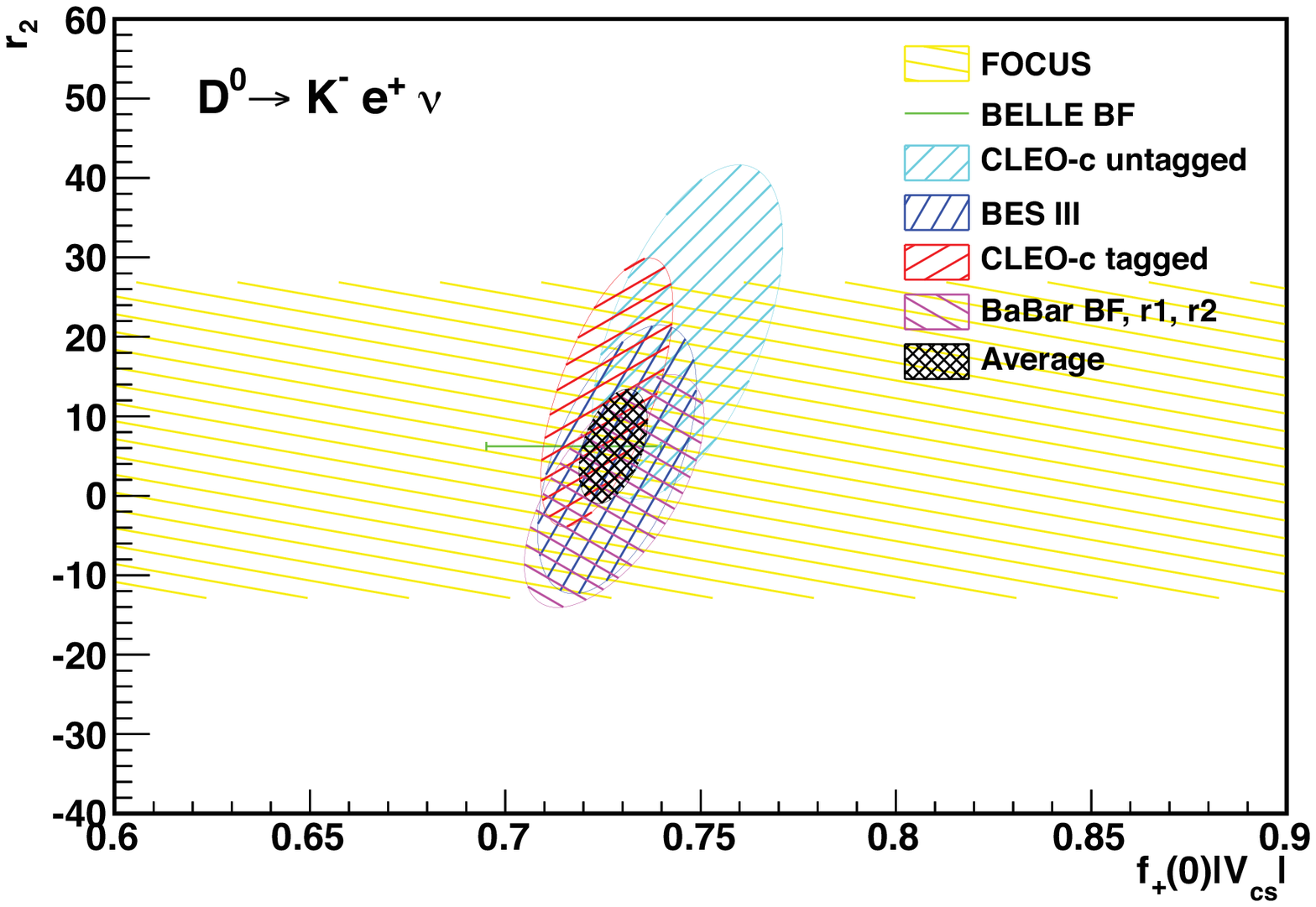}\hfill
\includegraphics[width=0.47\textwidth]{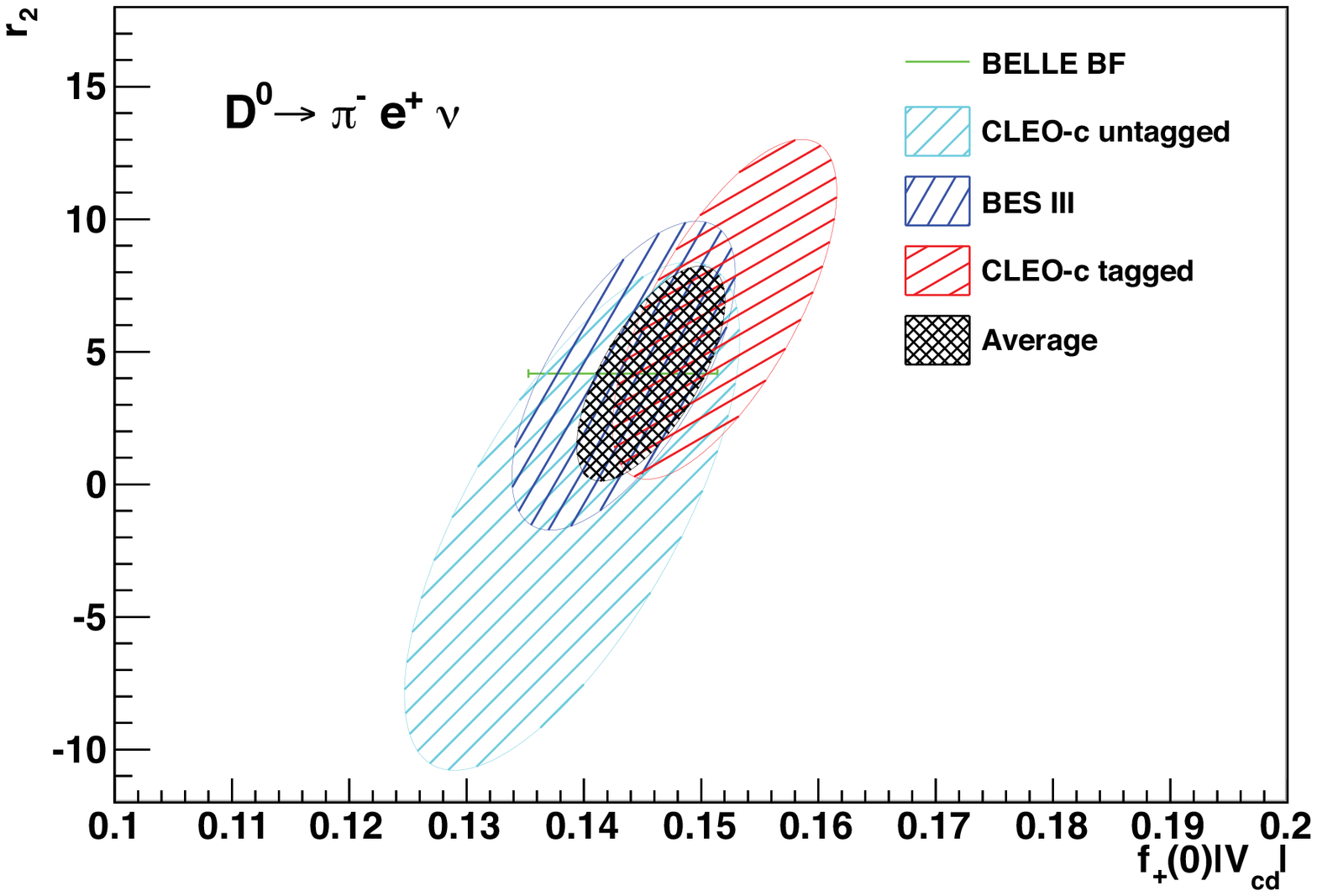}
\includegraphics[width=0.47\textwidth]{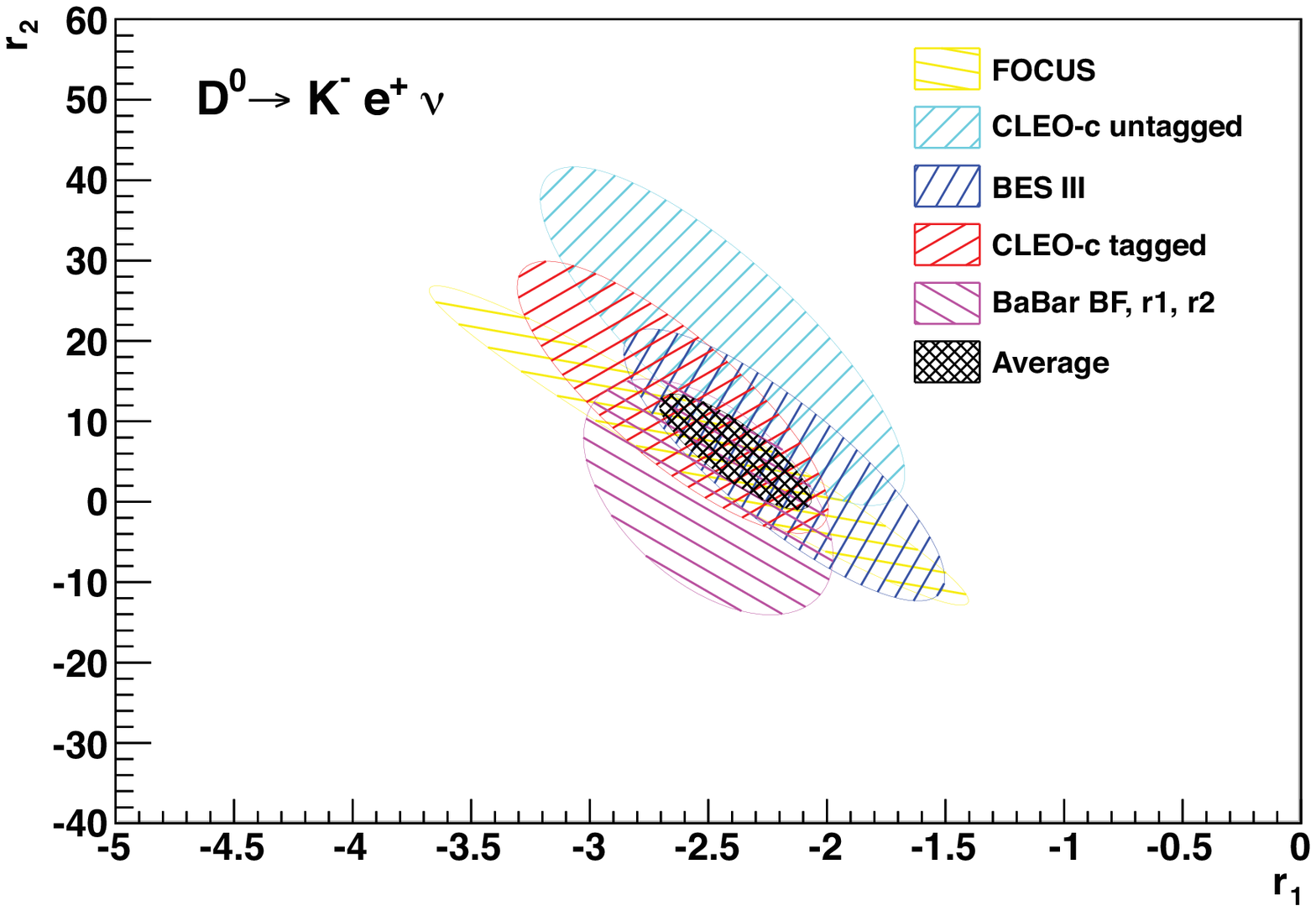}\hfill
\includegraphics[width=0.47\textwidth]{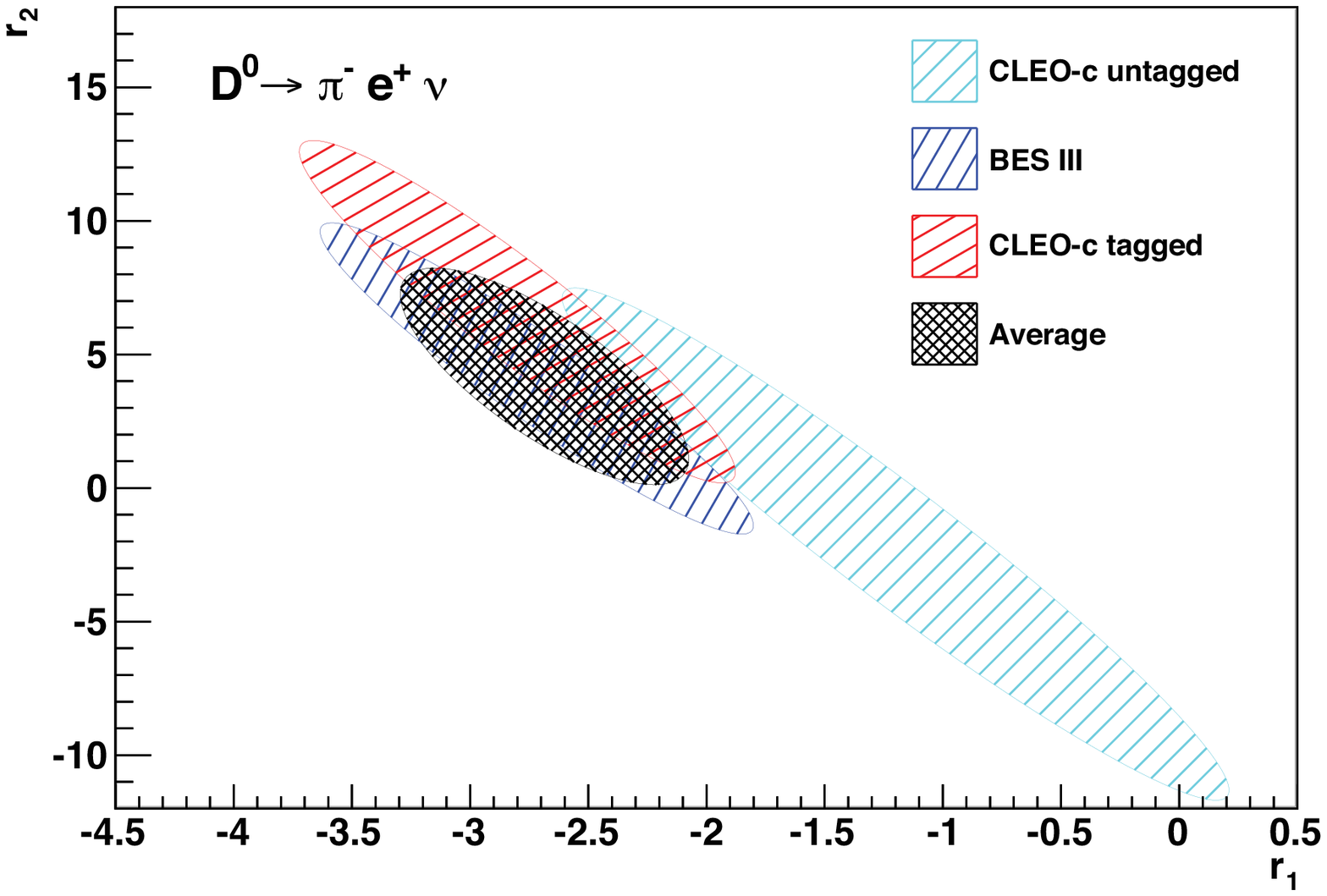}
\caption{The $D^0\to K^-e^+\nu$ (left) and $D^0\to \pi^-e^+\nu$ (right) 68\% C.L. error ellipses from the
average fit of the 3-parameter $z$-expansion results. 
\label{fig:fitellipse}}
\end{center}
\end{figure}




\begin{table}[!htb]
\begin{center}
 \caption[]{{Results for the form factor normalization
$f_+^K(0)|V_{cs}|$ and $f_+^{\pi}(0)|V_{cd}|$. Results from the
different collaborations have been corrected, if needed, using
values from PDG 2010. Prior to 2006, and apart for BESII, experiments
measure the ratio $(f_+^{\pi}(0)|V_{cd}| / f_+^K(0)|V_{cs}|)^2$. 
Corresponding values given in this Table for $f_+^{\pi}(0)|V_{cd}|$
are obtained by assuming that $f_+^K(0)|V_{cs}|=0.714\pm 0.009$.
Results of the combined fit include measurements from 2006 and later
for experiments measuring $f_+(0)|V_{cq}|$, $r_1$ and $r_2$.
Results of CLEO (2008) (untagged) only refer to the $D^0$ channel.
Results from LQCD are given in the last line \cite{Na:2010uf} \cite{Na:2011mc}.}
Results quoted for LQCD are obtained by multiplying the values
computed for $f_+^K(0)$ and $f_+^{\pi}(0)$ from lattice by
$|V_{cs}|=0.9729$ and $|V_{cd}|=0.2253$ respectively. These
values of $|V_{cs(d)}|$ correspond to present estimates assuming
the unitarity of the CKM matrix.
Values entering in the combination explained before are marked with a $\ast$.

  \label{tab:fomfactors}}
\begin{tabular}{c c c c}
\vspace*{0pt} & \\
\hline
Experiment & Ref. & $f_+^K(0)|V_{cs}|$ & $f_+^{\pi}(0)|V_{cd}|$ \\
\hline\hline 
E691 (1989)&\cite{Anjos:1988ue} & $0.69 \pm 0.05\pm 0.05 $ &  \\
CLEO (1991) &\cite{Crawford:1991zd} & $$ &  \\
CLEOII (1993) &\cite{Bean:1993zv}  & $0.76 \pm 0.01 \pm 0.04$ &  \\
CLEOII (1995) &\cite{Butler:1995ca}  & &$0.163 \pm 0.031 \pm 0.011$  \\
E687 (1995) &\cite{Frabetti:1995xq} & $0.69\pm 0.03\pm 0.03 $ &  \\
E687 (1996) &\cite{Frabetti:1996jj} &  &$0.160 \pm 0.018 \pm 0.004$  \\
BESII (2004) &\cite{Ablikim:2004ej} &$0.78 \pm 0.04 \pm 0.03$ & $0.164 \pm 0.032 \pm0.014$   \\
CLEOIII (2005) $\ast$ &\cite{Huang:2004fra} & &$0.139^{+0.011~+0.009}_{-0.013~-0.006}$ \\
FOCUS (2005)  & \cite{Link:2004dh} & &$0.137 \pm 0.008 \pm 0.008$ \\
\hline
Belle (2006) $\ast$ & \cite{Widhalm:2006wz}& $0.692 \pm 0.007 \pm 0.022$ & $0.140\pm 0.004\pm0.007$ \\ 
\babar (2007) $\ast$ & \cite{Aubert:2007wg}& $0.720 \pm 0.007\pm 0.007$ &  \\
CLEO-c (2008)(untagged) $\ast$ &\cite{Dobbs:2007aa} & $0.747 \pm 0.009\pm 0.009$ & $0.139 \pm 0.007\pm 0.003$ \\
CLEO-c (2009) (tagged) $\ast$ &\cite{Besson:2009uv} & $0.719 \pm 0.006\pm 0.005$ & $0.150 \pm 0.004\pm 0.001$ \\
BESIII (2012)(prel.) $\ast$ &\cite{BESIII} & $0.729 \pm 0.008\pm 0.007$ & $0.144 \pm 0.005\pm 0.002$ \\
\hline\hline
 Combined fit                 & \omit      & $ 0.728 \pm 0.005 $               &       $ 0.146 \pm 0.003$   \\
\hline
\hline
HPQCD & \cite{Na:2011mc}\cite{Na:2010uf}             & $0.727\pm0.018$         & $0.150 \pm 0.007$      \\
\vspace*{-10pt} & \\
\hline
\end{tabular}
\end{center}
\end{table}

\subsubsection{$D\ra V\overline \ell \nu_\ell$ decays}

\begin{figure}[htbp]
  \begin{center}
\includegraphics[width=3.50in]{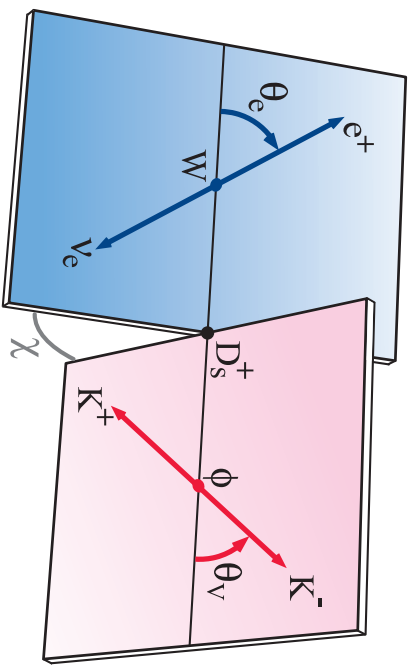}
  \end{center}
  \caption{
    Decay angles $\theta_V$, $\theta_\ell$ 
    and $\chi$. Note that the angle $\chi$ between the decay
    planes is defined in the $D$-meson reference frame, whereas
    the angles $\theta^{}_V$ and $\theta^{}_\ell$ are defined
    in the $V$ meson and $W$ reference frames, respectively.}
  \label{DecayAngles}
\end{figure}

When the final state hadron is a vector meson, the decay can proceed through
both vector and axial vector currents, and four form factors are needed.
The hadronic current is $H^{}_\mu = V^{}_\mu + A^{}_\mu$, 
where~\cite{Gilman:1989uy} 
\begin{eqnarray}
V_\mu & = & \left< V(p,\varepsilon) | \bar{q}\gamma^\mu c | D(p') \right> \ =\  
\frac{2V(q^2)}{m_D+m_V} 
\varepsilon_{\mu\nu\rho\sigma}\varepsilon^{*\nu}p^{\prime\rho}p^\sigma \\
 & & \nonumber\\
A_\mu & = & \left< V(p,\varepsilon) | -\bar{q}\gamma^\mu\gamma^5 c | D(p') \right> 
 \ =\  -i\,(m_D+m_V)A_1(q^2)\varepsilon^*_\mu \nonumber \\
 & & \hskip2.10in 
  +\ i \frac{A_2(q^2)}{m_D+m_V}(\varepsilon^*\cdot q)(p' + p)_\mu \nonumber \\
 & & \hskip2.30in 
+\ i\,\frac{2m_V}{q^2}\left(A_3(q^2)-A_0(q^2)\right)[\varepsilon^*\cdot (p' + p)] q_\mu\,.
\end{eqnarray}
In this expression, $m_V$ is the daughter meson mass and
\begin{eqnarray}A_3(q^2) & = & \frac{m_D + m_V}{2m_V}A_1(q^2)\ -\ \frac{m_D - m_V}{2m_V}A_2(q^2)\,.
\end{eqnarray}
Kinematics require that $A_3(0) = A_0(0)$.
The differential partial width is
\begin{eqnarray}
\frac{d\Gamma(D \to V \overline \ell \nu_\ell)}{dq^2\, d\cos\theta_\ell} & = & 
  \frac{G_F^2\,|V_{cq}|^2}{128\pi^3m_D^2}\,p^*\,q^2 \times \nonumber \\
 & &  
\left[\frac{(1-\cos\theta_\ell)^2}{2}|H_-|^2\ +\  
\frac{(1+\cos\theta_\ell)^2}{2}|H_+|^2\ +\ \sin^2\theta_\ell|H_0|^2\right]\,,
\end{eqnarray}
where $H^{}_\pm$ and $H^{}_0$ are helicity amplitudes given by
\begin{eqnarray}
H_\pm & = & \frac{1}{m_D + m_V}\left[(m_D+m_V)^2A_1(q^2)\ \mp\ 
      2m^{}_D\,p^* V(q^2)\right] \\
 & & \nonumber \\
H_0 & = & \frac{1}{|q|}\frac{m_D^2}{2m_V(m_D + m_V)}\ \times\ \nonumber \\
 & & \hskip0.01in \left[
    \left(1- \frac{m_V^2 - q^2}{m_D^2}\right)(m_D + m_V)^2 A_1(q^2) 
    \ -\ 4{p^*}^2 A_2(q^2) \right]\,.
\label{HelDef}
\end{eqnarray}
$p^*$ is the three-momentum of the $K \pi$ system, measured in the $D$ rest frame.
The left-handed nature of the quark current manifests itself as
$|H_-|>|H_+|$. The differential decay rate for $D\ra V\ell\nu$ 
followed by the vector meson decaying into two pseudoscalars is

\begin{eqnarray}
\frac{d\Gamma(D\ra V\ell\nu, V\ra P_1P_2)}{dq^2 d\cos\theta_V d\cos\theta_\ell d\chi} 
 &  = & \frac{3G_F^2}{2048\pi^4}
       |V_{cq}|^2 \frac{p^*(q^2)q^2}{m_D^2} {\cal B}(V\to P_1P_2)\ \times \nonumber \\ 
 & & \hskip0.10in \big\{ (1 + \cos\theta_\ell)^2 \sin^2\theta_V |H_+(q^2)|^2 \nonumber \\
 & & \hskip0.20in +\ (1 - \cos\theta_\ell)^2 \sin^2\theta_V |H_-(q^2)|^2 \nonumber \\
 & & \hskip0.30in +\ 4\sin^2\theta_\ell\cos^2\theta_V|H_0(q^2)|^2 \nonumber \\
 & & \hskip0.40in +\ 4\sin\theta_\ell (1 + \cos\theta_\ell) 
             \sin\theta_V \cos\theta_V \cos\chi H_+(q^2) H_0(q^2) \nonumber \\
 & & \hskip0.50in -\ 4\sin\theta_\ell (1 - \cos\theta_\ell) 
          \sin\theta_V \cos\theta_V \cos\chi H_-(q^2) H_0(q^2) \nonumber \\
 & & \hskip0.60in -\ 2\sin^2\theta_\ell \sin^2\theta_V 
                \cos 2\chi H_+(q^2) H_-(q^2) \big\}\,,
\label{eq:dGammaVector}
\end{eqnarray}
where the angles $\theta^{}_\ell$, $\theta^{}_V$, and $\chi$ are defined
in Fig.~\ref{DecayAngles}. 

\begin{figure}[htbp]
  \begin{center}
    \includegraphics[width=6.5in,angle=0]{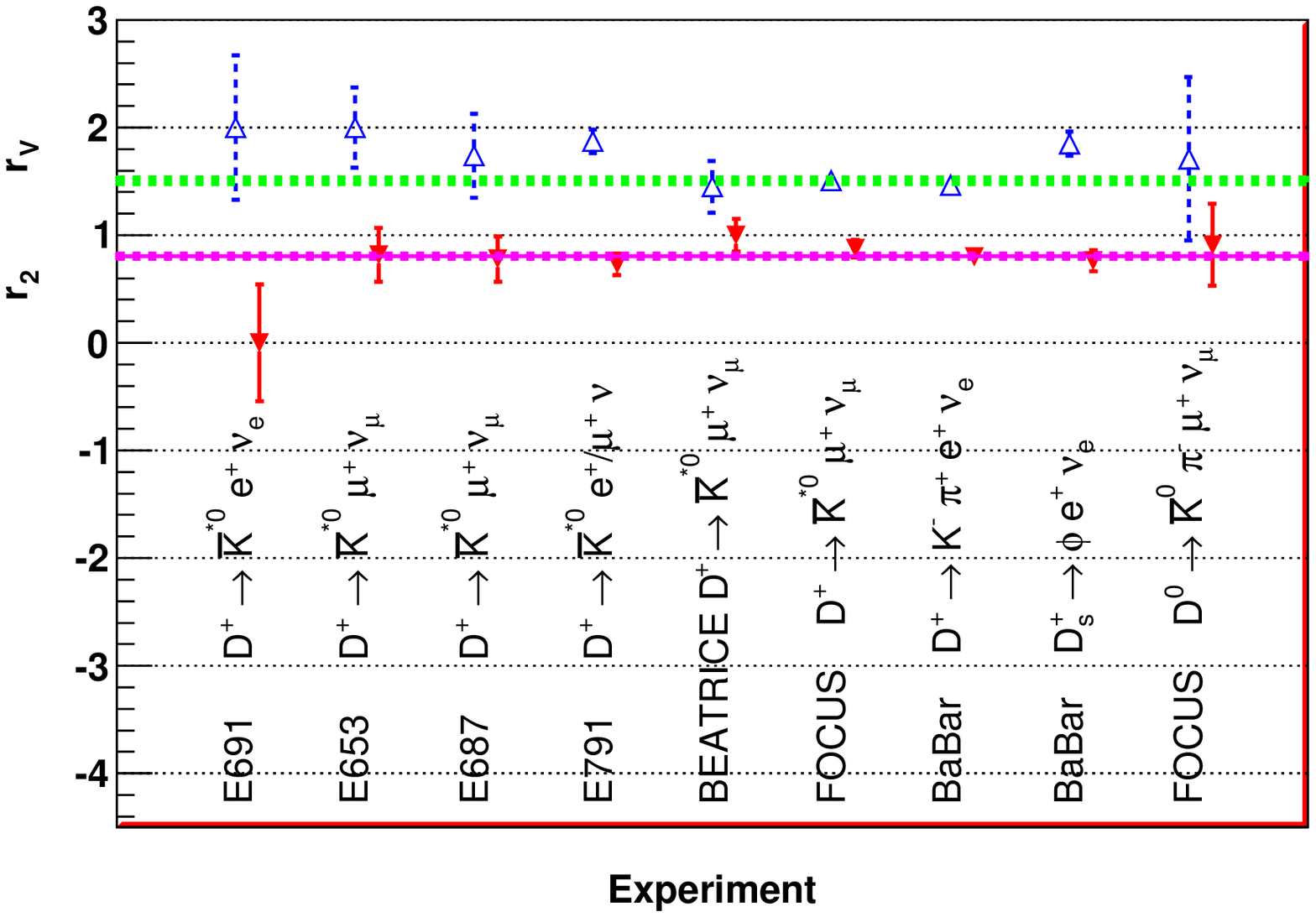}
  \end{center}
\vskip0.10in
  \caption{A comparison of $r_2$ and $r_V$ values 
    from various experiments. The first seven measurements are for $D^+
    \to K^-\pi^+ l^+\nu_l$ decays. Also shown as a line with
    1-$\sigma$ limits is the average of these. The last two points are
    $D_s^+$ decays and Cabibbo-suppressed $D$ decays. 
  \label{fig:r2rv}}
\end{figure}

Ratios between the values of the hadronic form factors expressed at $q^2=0$ are usually introduced:
\begin{eqnarray}
r_V \equiv V(0) / A_1(0), & &  r_2 \equiv A_2(0) / A_1(0) \label{rVr2_eq}\,.
\end{eqnarray}
Table \ref{Table1} lists measurements of $r_V$ and $r_2$ from several
experiments. Most of measurements assume that the $q^2$ dependence of hadronic form factors 
is given by the simple pole anzats.
The measurements are plotted in
Fig.~\ref{fig:r2rv} which shows that they are all consistent.

\begin{table}[htbp]
\caption{Results for $r_V$ and $r_2$ from various experiments.
\label{Table1}}
\begin{center}
\begin{tabular}{cccc}
\hline
\vspace*{-10pt} & \\
Experiment & Ref. & $r_V$ & $r_2$ \\
\vspace*{-10pt} & \\
\hline
\vspace*{-10pt} & \\
$D^+\to \overline{K}^{*0}l^+\nu$ & \omit & \omit & \omit         \\
E691         & \cite{Anjos:1990pn}     & 2.0$\pm$  0.6$\pm$  0.3  & 0.0$\pm$  0.5$\pm$  0.2    \\
E653         & \cite{Kodama:1992tn}     & 2.00$\pm$ 0.33$\pm$ 0.16 & 0.82$\pm$ 0.22$\pm$ 0.11   \\
E687         & \cite{Frabetti:1993jq}     & 1.74$\pm$ 0.27$\pm$ 0.28 & 0.78$\pm$ 0.18$\pm$ 0.11   \\
E791 (e)     & \cite{Aitala:1997cm}    & 1.90$\pm$ 0.11$\pm$ 0.09 & 0.71$\pm$ 0.08$\pm$ 0.09   \\
E791 ($\mu$) & \cite{Aitala:1998ey}    & 1.84$\pm$0.11$\pm$0.09   & 0.75$\pm$0.08$\pm$0.09     \\
Beatrice     & \cite{Adamovich:1998ia} & 1.45$\pm$ 0.23$\pm$ 0.07 & 1.00$\pm$ 0.15$\pm$ 0.03   \\
FOCUS        & \cite{Link:2002wg}   & 1.504$\pm$0.057$\pm$0.039& 0.875$\pm$0.049$\pm$0.064  \\
\hline
$D^0\to \overline{K}^0\pi^-\mu^+\nu$ & \omit & \omit & \omit         \\
FOCUS        & \cite{Link:2004uk}    & 1.706$\pm$0.677$\pm$0.342& 0.912$\pm$0.370$\pm$0.104 \\
\babar        & \cite{delAmoSanchez:2010fd} & $1.493 \pm 0.014 \pm 0.021$ & $0.775 \pm 0.011 \pm 0.011$ \\
\hline
$D_s^+ \to \phi\,e^+ \nu$ &\omit  &\omit     & \omit                  \\
\babar        & \cite{Aubert:2008rs}    & 1.636$\pm$0.067$\pm$0.038& 0.705$\pm$0.056$\pm$0.029 \\
\hline
$D^0, D^+\to \rho\,e \nu$ & \omit  & \omit    & \omit                 \\
CLEO         & \cite{Mahlke:2007uf}    & 1.40$\pm$0.25$\pm$0.03   & 0.57$\pm$0.18$\pm$0.06    \\
\hline
\end{tabular}
\end{center}
\end{table}

\subsubsection{$S$-wave component}

In 2002 FOCUS reported~\cite{Link:2002ev} an asymmetry in
the observed $\cos(\theta_V)$ distribution. This is interpreted as
evidence for an $S$-wave component in the decay amplitude as follows. 
Since $H_0$ typically dominates over $H_{\pm}$, the distribution given 
by Eq.~(\ref{eq:dGammaVector}) is, after integration over $\chi$,
roughly proportional to $\cos^2\theta_V$. 
Inclusion of a constant $S$-wave amplitude of the form $A\,e^{i\delta}$ 
leads to an interference term proportional to 
$|A H_0 \sin\theta_\ell \cos\theta_V|$; this term causes an asymmetry 
in $\cos(\theta_V)$.
When FOCUS fit their data including this $S$-wave amplitude, 
they obtained $A = 0.330 \pm 0.022 \pm 0.015$ GeV$^{-1}$ and 
$\delta = 0.68 \pm 0.07 \pm 0.05$~\cite{Link:2002wg}. 

More recently, both \babar~\cite{Aubert:2008rs} and 
CLEO-c~\cite{Ecklund:2009fia} have also found evidence 
for an $f^{}_0$ component in semileptonic $D^{}_s$ decays.

\subsubsection{Model-independent form factor measurement}

\begin{figure}[htb]
  \begin{center}
    \vskip0.20in
    \includegraphics[width=4.75in,angle=0.]{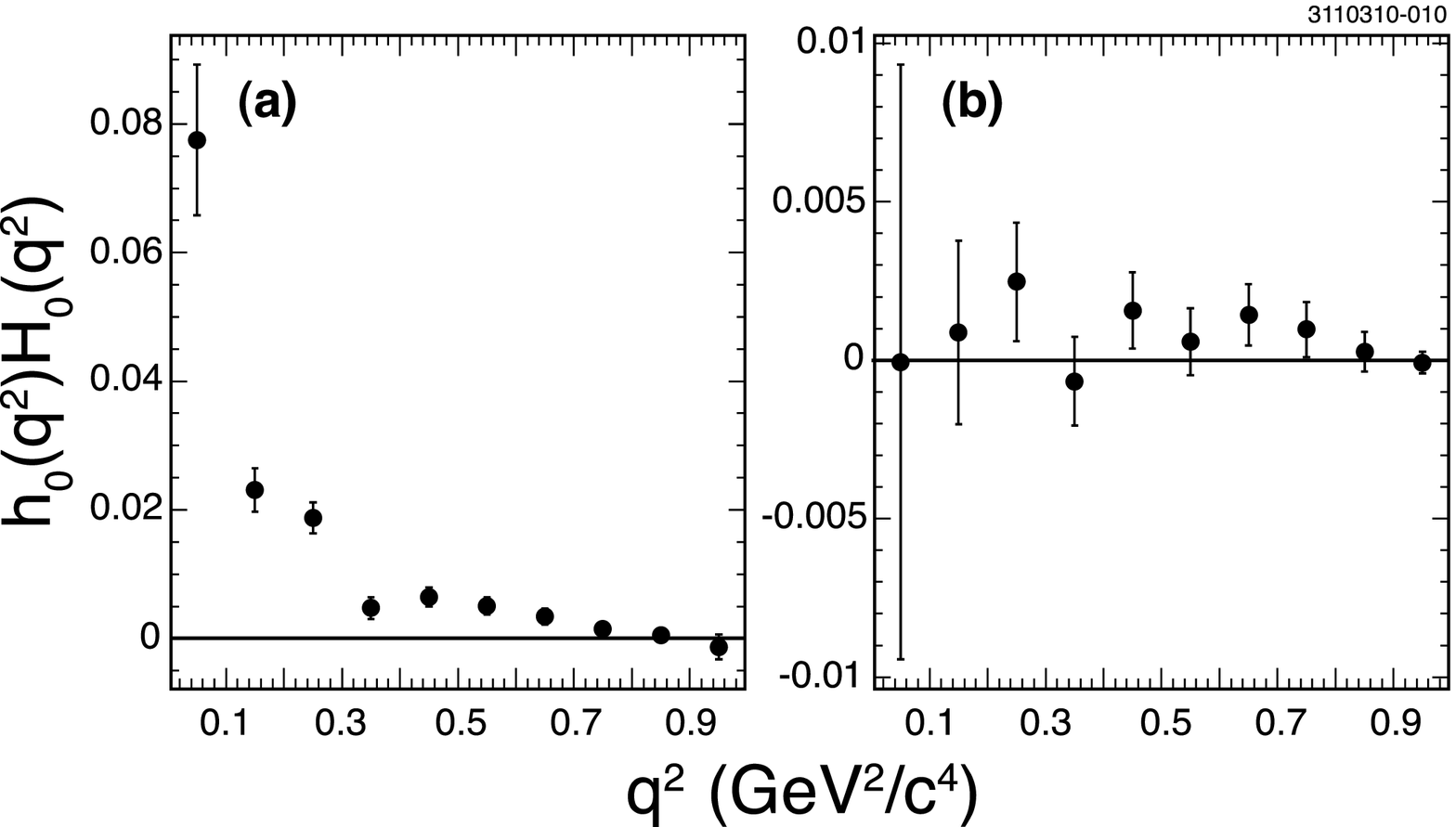}
  \end{center}
  \begin{center}
    \vskip0.20in
    \includegraphics[width=4.75in,angle=0.]{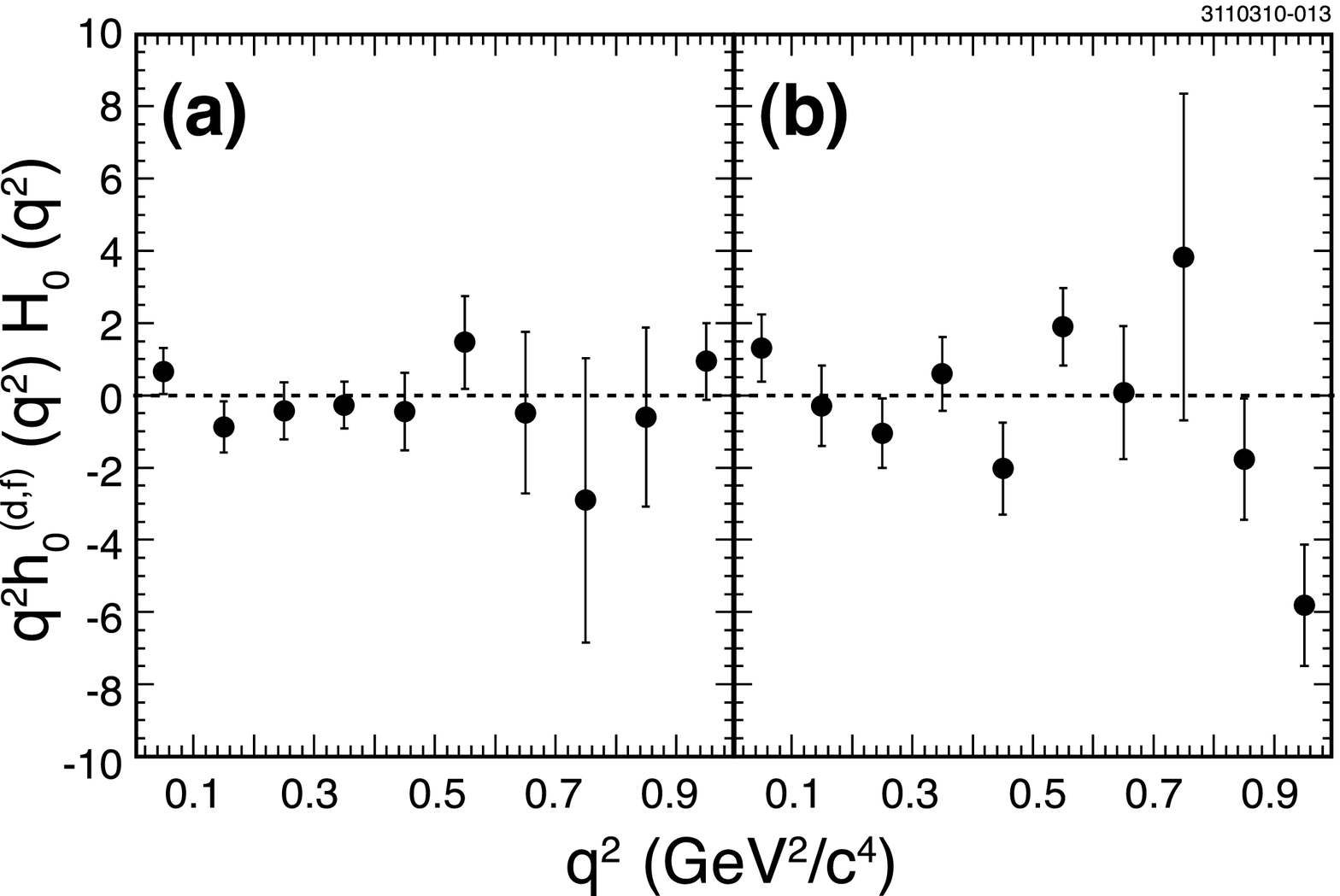}
  \end{center}
\vskip-0.20in
  \caption{Model-independent form factors $h_0(q^2)$ measured by 
    CLEO-c~\cite{Briere:2010zc}.
  \label{fig:cleoc_h0}}
\end{figure}

\begin{figure}[htb]
  \begin{center}
    \vskip0.20in
    \includegraphics[width=4.75in,angle=0.]{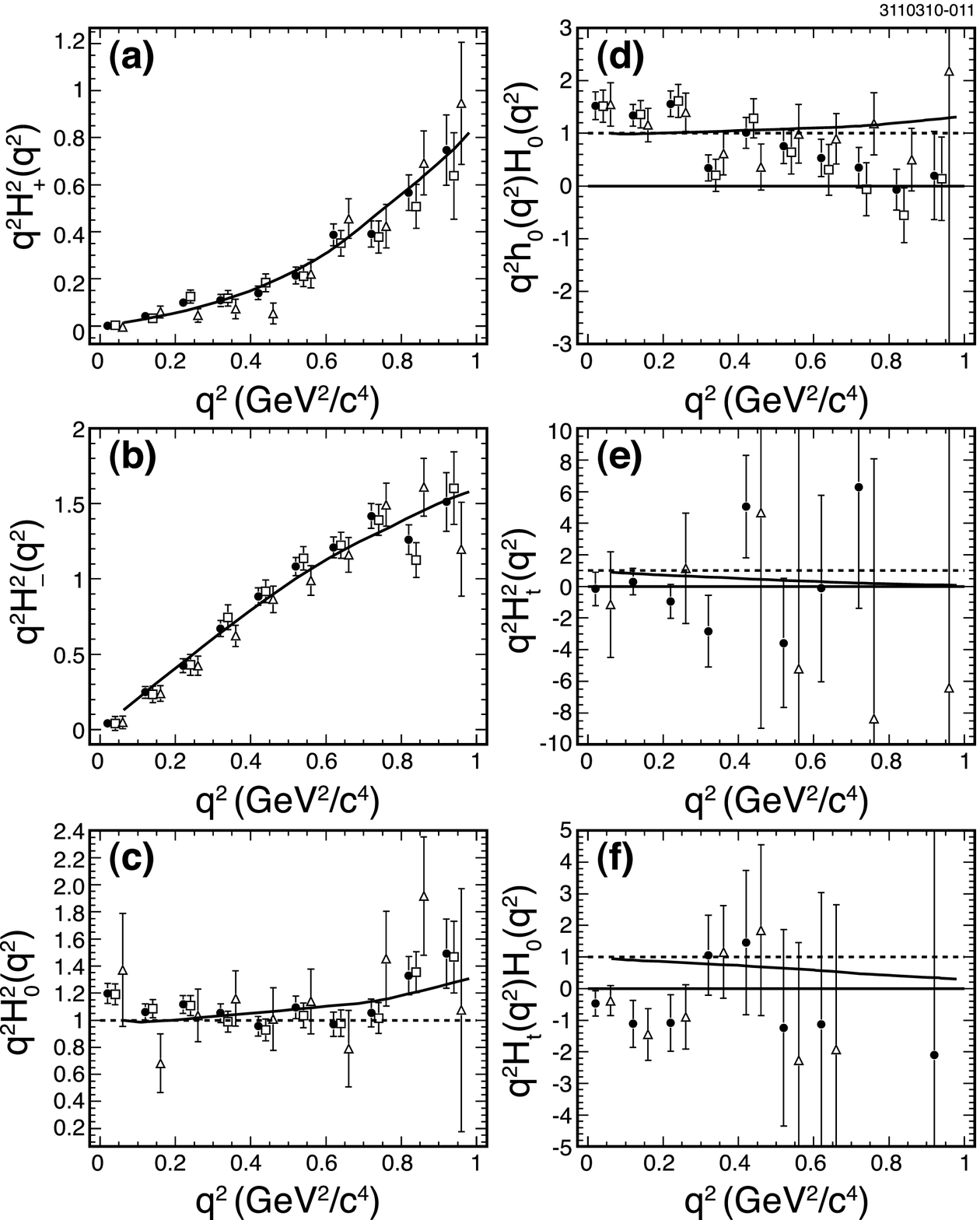}
  \end{center}
\vskip-0.20in
  \caption{Model-independent form factors $H(q^2)$ measured by 
    CLEO-c~\cite{Briere:2010zc}.
  \label{fig:cleoc_H}}
\end{figure}

Subsequently the CLEO-c collaboration extracted the form factors
$H_+(q^2)$, $H_-(q^2)$, and $H_0(q^2)$ in a model-independent fashion
directly as functions of $q^2$~\cite{Briere:2010zc} and also determined the
$S$-wave form factor $h_0(q^2)$ via the interference term, despite the
fact that the $K\pi$ mass distribution appears dominated by the vector
$K^*(892)$ state. Their results are shown in Figs.~\ref{fig:cleoc_H} and
\ref{fig:cleoc_h0}.  Plots in Fig.~\ref{fig:cleoc_H} clearly show that
$H_0(q^2)$ dominates over essentially the full range of $q^2$, but
especially at low $q^2$. They also show that the transverse form factor
$H_t(q^2)$ (which can be related to $A_3(q^2)$ is small (compared to
Lattice Gauge Theory calculations) and suggest that the form factor
ratio $r_3 \equiv A_3(0) / A_1(0)$ is large and negative.

The product $H_0(q^2)\times h_0(q^2)$ is shown in
Fig.~\ref{fig:cleoc_h0} and clearly indicates the existence of
$h_0(q^2)$, although it seems to fall faster with $q^2$ than $H_0(q^2)$.
The other plots in that figure show that $D$- and $F$-wave versions of
the $S$-wave $h_0(q^2)$ are not significant.

\subsubsection{Detailed measurements of the $D^+ \rightarrow K^- \pi^+ e^+ \nu_e$ 
decay channel}

\babar \cite{delAmoSanchez:2010fd} has selected a large sample of $244\times 10^3$ signal events with a ratio $S/B\sim 2.3$ from an analyzed integrated
luminosity of $347~\fb^{-1}$. With four particles emitted in the final state, 
the differential decay rate depends on five variables.
In addition to the four variables defined in previous sections there is 
$m^2$, the mass squared of the $K\pi$ system.
Apart for this last variable, the reconstruction algorithm does not provide 
a high resolution on the other measured quantities 
and a multi-dimensional unfolding procedure
is not used to correct for efficiency and resolution effects. Meanwhile these
limitations still allow an essentially model independent measurement of
the differential decay rate. This is because, apart for the $q^2$
and mass dependence of the form factors, angular distributions are fixed by
kinematics. In addition, present accurate measurements of 
$D \rightarrow P \overline{\ell}\nu_{\ell}$ decays have shown that the 
$q^2$ dependence of the form factors can be well described by several models
as long as the corresponding model parameter(s) are fitted on data.
This is even more true in $D \rightarrow V \overline{\ell}\nu_{\ell}$ decays
because the $q^2$ range is reduced. To analyze the
$D^+ \rightarrow K^- \pi^+ e^+ \nu_e$ decay channel it is assumed
that all form factors have a $q^2$ variation given by
the simple pole model and the effective pole mass value,
 $m_A=(2.63 \pm 0.10 \pm 0.13)~GeV/c^2$,
is fitted
for the axial vector form factors. This value is compatible
with expectations when comparing with the mass
of $J^P=1^+$ charm mesons. Data are not sensitive to the effective mass
of the vector form factor for which $m_V=(2.1 \pm 0.1)~GeV/c^2$ is used,
nor to the effective pole mass of the scalar component for which $m_A$ is used.
 For the mass dependence of the form factors,
a Breit-Wigner with a mass dependent width and a Blatt-Weisskopf damping factor
is used. For the S-wave amplitude, considering
what was measured in $D^+ \rightarrow K^- \pi^+\pi^+$ decays,
a polynomial variation below the $\overline{K}^*_0(1430)$
and a Breit-Wigner distribution, above are assumed. For the polynomial part, 
a linear term is sufficient to fit data.

\begin{figure}[!htb]
	\centering
\includegraphics[width=0.7\textwidth]{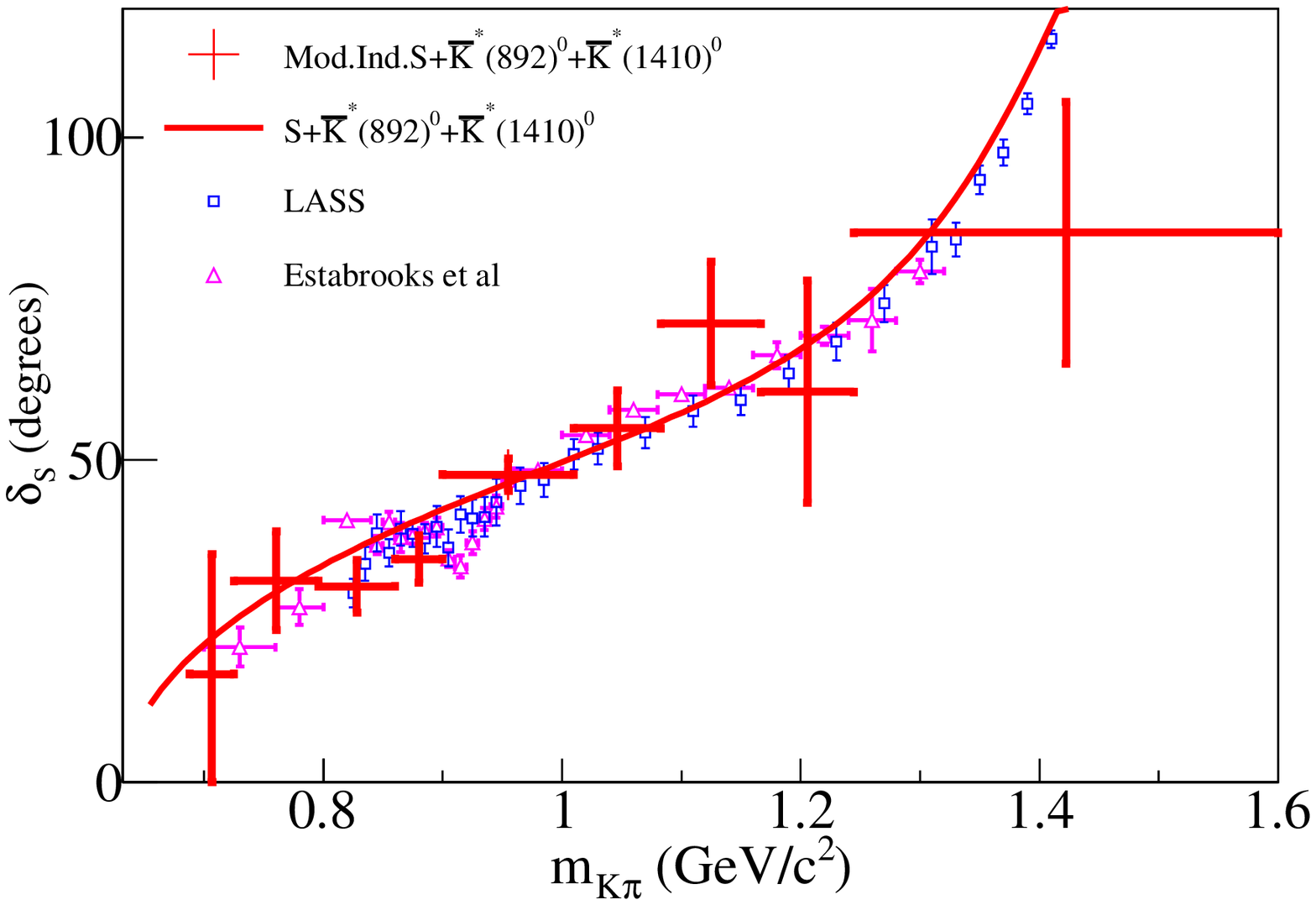}
\caption{Points (full circles) give the \babar $S$-wave phase variation 
assuming a signal containing $S$-wave, $\akst$ and $\akstp$ components. 
Error bars include systematic uncertainties.
The full line corresponds to a parameterized $S$-wave phase variation 
fitted on \babar data.
The phase variation measured in $K\pi$ scattering
by Ref. \cite{Estabrooks:1977xe} (triangles) and LASS \cite{Aston:1987ir} (squares), 
after correcting 
for $\delta^{3/2}$, are given.}
\label{fig:swave_phase}
\end{figure}

It is verified that the variation of the S-wave phase is compatible 
with expectations from elastic $K\pi$ scattering, according
to the Watson theorem. At variance with elastic scattering, 
a negative relative sign between the S- and P-waves is measured; 
this is compatible with 
the previous theorem. In Fig. \ref{fig:swave_phase}, the measured
S-wave phase is compared with the phase of the elastic, $I=1/2$,
$K\pi$ elastic phase for different values of the $K\pi$ mass.

Contributions from other resonances decaying into $K^-\pi^+$ are considered.
A small signal from the $\overline{K}^*(1410)$ is observed, compatible
with expectations from $\tau$ decays and this component is included in the
nominal fit. In total, 11 parameters are fitted in addition to the total
number of signal events. They give a detailed description of the differential
decay rate versus the 5 variables and corresponding matrices for 
statistical and systematic uncertainties are provided allowing to 
evaluate the compatibility of data with future theoretical expectations.

\begin{figure}[htbp!]
  \begin{center}
\mbox{\epsfig{file=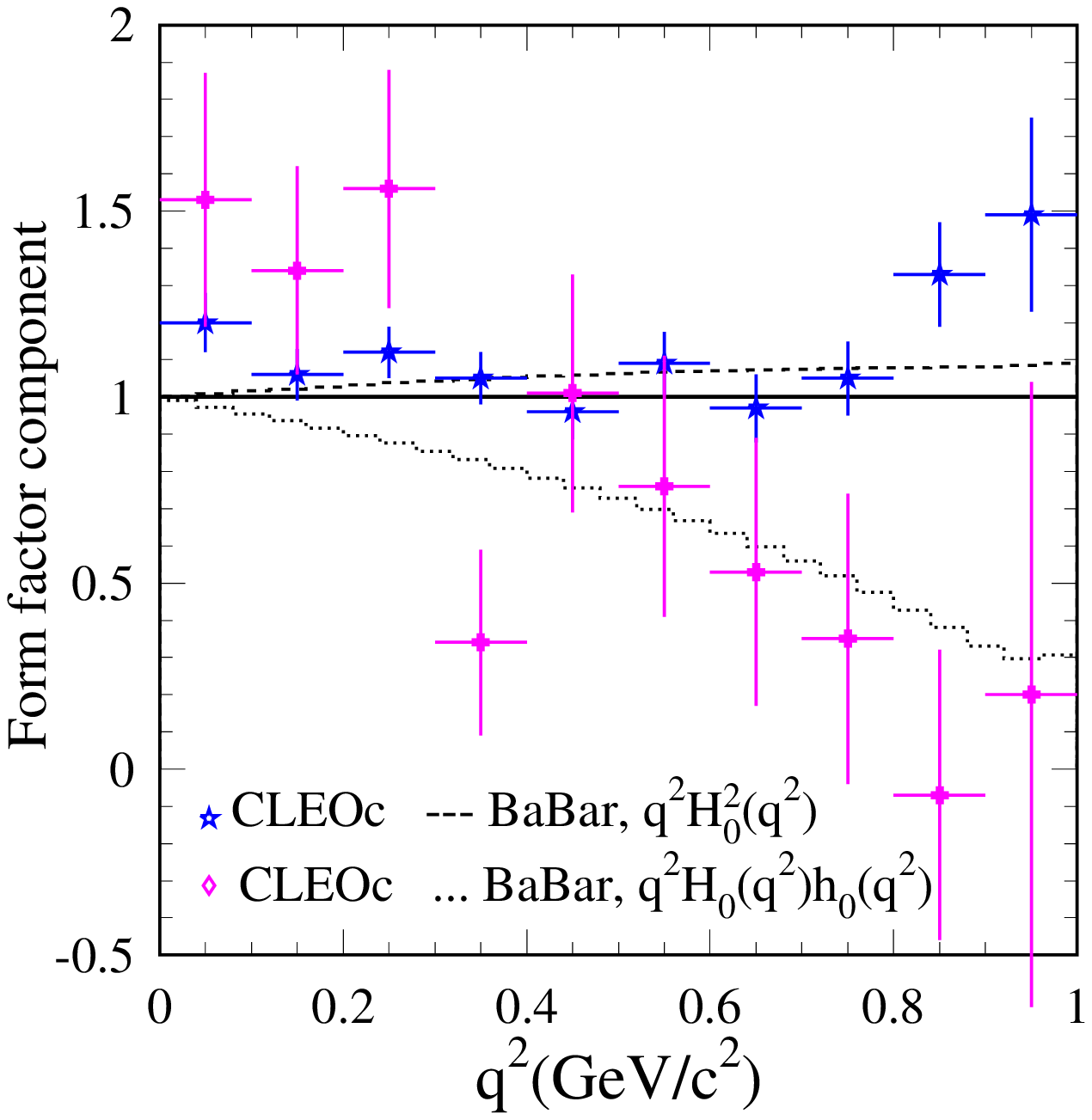,width=.60\textwidth}
}
  \end{center}
  \caption[]{{Comparison between CLEO-c measurements and \babar results
for the quantities $q^2H_0^2(q^2)$ and
 $q^2H_0(q^2)h_0(q^2)$.}
   \label{fig:h0FF}}
\end{figure}

In Fig. \ref{fig:h0FF} are compared measured values from CLEO-c
of the products $q^2H_0^2(q^2)$ and $q^2h_0(q^2)H_0(q^2)$ with 
corresponding results from \babar illustrating the difference in behaviour
of the scalar $h_0$ component and the helicity zero $H_0$ P-wave form factor.
For this comparison, plotted values from \babar for the two distributions
are equal to 1 at $q^2=0$. The different behaviour of $h_0(q^2)$
and $H_0(q^2)$ can be explained by they different dependence in the 
$p^*$ variable.
Results of this analysis for the rates and few characteristics 
for S, P and D-waves are given in Table \ref{tab:comparison}.

\begin{table}[!htb]
\begin{center}
 \caption[]{{Detailed determination of the properties of the 
$D^+ \rightarrow K^-\pi^+ e^+ \nu_e$ decay channel from \babar. 
Values for ${\cal B}(D^+\rightarrow \akstp / \akstd e^+\nu_e)$ are corrected for their respectivebranching fractions into $K^-\pi^+$.}\hspace{1cm}
  \label{tab:comparison}}
\begin{tabular}{c c}
\vspace*{-10pt} & \\
\hline
\textbf{Measurement} & \textbf{\babar result} \\
\hline\hline
{ $m_{\kst}(~MeV/c^2)$} & {$895.4\pm{0.2}\pm0.2$}\\
{ $\Gamma^0_{\kst}(~MeV/c^2)$} & {$46.5\pm{0.3}\pm0.2$} \\
{$r_{BW}(~GeV/c)^{-1}$ }& {$2.1\pm{0.5}\pm 0.5$} \\
\hline
{$r_{V}$} & {$1.463\pm{0.017}\pm 0.031$} \\
{$r_{2}$} & {$0.801\pm{0.020}\pm 0.020$}  \\
{ $m_{A} (~GeV/c^2)$} & {$2.63\pm 0.10 \pm 0.13$}\\
\hline
${\cal B}(D^+ \rightarrow K^-\pi^+ e^+ \nu_e)(\%)$ &$4.04 \pm 0.03 \pm 0.04 \pm 0.09$ \\
${\cal B}(D^{+}\rightarrow K^- \pi^+ e^{+} \nu_{e})_{\overline{K}^{*0}}(\%)$ & $3.80\pm0.04\pm0.05 \pm0.09 $  \\ 
${\cal B}(D^+ \rightarrow K^-\pi^+ e^+ \nu_e)_{S-wave}(\%)$ &$0.234 \pm0.007  \pm0.007  \pm0.005  $ \\
${\cal B}(D^{+}\rightarrow \akstp e^{+} \nu_{e})(\%)$ & $0.30\pm0.12\pm0.18\pm0.06$ ($<0.6$ at 90$\%$ C.L.) \\ 
${\cal B}(D^{+}\rightarrow \akstd e^{+} \nu_{e})(\%)$ & $0.023\pm0.011\pm0.011\pm0.001$ ($<0.05$ at 90$\%$ C.L.)  \\ 
\hline\hline
\end{tabular}
\end{center}
\end{table}


\clearpage
\subsection{\emph{CP} asymmetries}

\emph{CP} violation occurs if the decay rate for a particle differs 
from that of its \emph{CP}-conjugate\cite{Bigi:2000yz}. 
In general there are two classes of \emph{CP} violation, termed
{\it indirect\/} and {\it direct\/}\cite{Nir:1999mg}. Indirect \emph{CP} 
violation refers to $\Delta C\!=\!2$ processes and 
arises in $D^0$ decays due to $D^0$-$\dbar$ mixing. 
It can occur as an asymmetry in the mixing itself, or it can 
result from interference between a decay 
amplitude arising via mixing and a non-mixed amplitude. 
Direct \emph{CP} violation refers to
$\Delta C\!=\!1$ processes and occurs in both charged and neutral 
$D$ decays. It results from interference between two different decay
amplitudes (e.g., a penguin and tree amplitude) that have
different weak (CKM) and strong phases\footnote{The weak 
phase difference will have opposite signs for $D\ra f$ and 
$\overline{D}\ra\bar{f}$ decays, while the strong phase difference 
will have the same sign. As a result, squaring the total amplitudes 
to obtain the decay rates gives interference terms having 
opposite sign, i.e., non-identical decay rates.}.
A difference in strong phases typically arises due to 
final-state interactions (FSI)\cite{Buccella:1994nf}. A difference
in weak phases arises from different CKM vertex couplings, as 
is often the case for spectator and penguin diagrams.

\vspace{0.8cm}
The \emph{CP} asymmetry is defined as the difference between 
$D$ and $\overline{D}$ partial widths divided by their sum:
\begin{eqnarray}  
A_{CP} & = & \frac{\Gamma(D)-\Gamma(\overline{D})}
{\Gamma(D)+\Gamma(\overline{D})}\,.
\end{eqnarray}
However, to take into account differences in production rates between 
$D$ and $\overline{D}$ (which would affect the number of respective 
decays observed), experiments usually normalize to a Cabibbo-favored 
mode. In this case there is the additional benefit that most corrections 
due to inefficiencies cancel out, reducing systematic uncertainties. An 
implicit assumption is that there is no measurable \emph{CP} 
violation in the Cabibbo-favored normalizing mode. 
The \emph{CP} asymmetry is calculated as
\begin{eqnarray}
A_{CP} & = & \frac{\eta(D)-\eta(\overline{D})}{\eta(D)+\eta(\overline{D})}\,,
\end{eqnarray}
where (considering, for example, $D^0 \to K^-K^+$)
\begin{eqnarray}
 \eta(D) & = & \frac{N(D^0 \rightarrow K^-K^+)}{N(D^0 \rightarrow K^-\pi^+)}\,, \\
 \eta(\overline{D}) & = & \frac{N(\dbar\rightarrow K^-K^+)}
{N(\dbar\rightarrow K^+\pi^-)}\,.
\end{eqnarray}
In the case of $D^+$ and $D^+_s$ decays, $A^{}_{CP}$ measures 
direct \emph{CP} violation; in the case of $D^0$ decays, $A^{}_{CP}$ 
measures direct and indirect \emph{CP} violation combined.
Values of $A^{}_{CP}$ for $D^+$, $D^0$ and $D_s^+$ decays are listed in
Tables~\ref{tab:cp_charged}, \ref{tab:cp_neutral} and \ref{tab:cp_ds} respectively.

\begin{table}
\renewcommand{\arraystretch}{1.4}
\caption{\cp\ asymmetries 
$A^{}_{CP}= [\Gamma(D^+)-\Gamma(D^-)]/[\Gamma(D^+)+\Gamma(D^-)]$
for $D^\pm$ decays.
\label{tab:cp_charged}}
\footnotesize
\begin{center}
\begin{tabular}{|l|c|c|c|} 
\hline
{\bf Mode} & {\bf Year} & {\bf Collaboration} & {\boldmath $A^{}_{CP}$} \\
\hline
{\boldmath $D^+ \to \mu^+ \nu$} &
  2008 & CLEOc~\cite{Eisenstein:2008sq} &  $ +0.08  \pm 0.08 $ \\
\hline
{\boldmath $D^+ \to \pi^+ \pi^0$} &
  2010 & CLEOc~\cite{Mendez:2009aa} &  $ +0.029  \pm 0.029 \pm 0.003 $ \\
\hline
{\boldmath $D^+ \to \pi^+ \eta$} &
  2010 & CLEOc~\cite{Mendez:2009aa} &  $ -0.020  \pm 0.023 \pm 0.003 $ \\
\hline
{\boldmath $D^+ \to \pi^+ \eta^\prime$} &
  2010 & CLEOc~\cite{Mendez:2009aa} &  $ -0.040  \pm 0.034 \pm 0.003 $ \\
\hline
{\boldmath $D^+ \to K^+ \pi^0$} &
  2010 & CLEOc~\cite{Mendez:2009aa} &  $ -0.035  \pm 0.107 \pm 0.009 $ \\
\hline
{\boldmath $D^+ \to K^0_s\pi^+$}   &
   2011 & \babar~\cite{Amo:2011ab} &  $ -0.0044 \pm 0.0013 \pm 0.0010 $ \\
&  2010 & Belle~\cite{Ko:2010ng} &  $ -0.0071 \pm 0.0019 \pm 0.0020 $ \\
&  2010 & CLEOc~\cite{Mendez:2009aa} &  $ -0.013  \pm 0.007  \pm 0.003  $ \\
&  2002 & FOCUS~\cite{Link:2001zj}   &  $ -0.016  \pm 0.015  \pm 0.009  $ \\
&       & COMBOS average           &  $ -0.0054 \pm 0.0014 $ \\
\hline
{\boldmath $D^+ \to K^0_sK^+$} &
  2010 & Belle~\cite{Ko:2010ng} &  $ -0.0016 \pm 0.0058 \pm 0.0025 $ \\
& 2010 & CLEOc~\cite{Mendez:2009aa} &  $ -0.002  \pm 0.015  \pm 0.009  $ \\  
& 2002 & FOCUS~\cite{Link:2001zj}   &  $ +0.071  \pm 0.061  \pm 0.012 $ \\
&      & COMBOS average           &  $ -0.0010 \pm 0.0059 $ \\
\hline
{\boldmath $D^+ \to \pi^+\pi^-\pi^+$} &
  1997 & E791~\cite{Aitala:1996sh}    &  $ -0.017  \pm 0.042  $ (stat.) \\
\hline
{\boldmath $D^+ \to K^-\pi^+\pi^+$} &
  2010 & CLEOc~\cite{Mendez:2009aa} &  $ -0.001  \pm  0.004 \pm 0.009  $ \\
\hline
{\boldmath $D^+ \to K^0_s\pi^+\pi^0$} &
  2007 & CLEO-c~\cite{Dobbs:2007zt} &  $ +0.003  \pm 0.009  \pm 0.003  $ \\
\hline
{\boldmath $D^+ \to K^+K^-\pi^+$} &
   2008 & CLEO-c~\cite{Rubin:2008zi} &  $ -0.0003 \pm 0.0084 \pm 0.0029 $ \\
&  2005 & \babar~\cite{Aubert:2005gj}     &  $ +0.014  \pm 0.010  \pm 0.008  $ \\
&  2000 & FOCUS~\cite{Link:2000aw}     &  $ +0.006  \pm 0.011  \pm 0.005  $ \\
&  1997 & E791~\cite{Aitala:1996sh}       &  $ -0.014  \pm 0.029  $ (stat.)    \\
&  1994 & E687~\cite{Frabetti:1994kv}       &  $ -0.031  \pm 0.068  $ (stat.)    \\
&       & COMBOS average             &  $ +0.0039 \pm 0.0061 $            \\
\hline
{\boldmath $D^+ \to K^-\pi^+\pi^+\pi^0$} &
  2007 & CLEOc~\cite{Dobbs:2007zt}  &  $ +0.010  \pm 0.009  \pm 0.009  $ \\
\hline
{\boldmath $D^+ \to K^0_s\pi^+\pi^+\pi^-$} &
  2007 & CLEOc~\cite{Dobbs:2007zt}  &  $ +0.001  \pm 0.011  \pm 0.006  $ \\
\hline
{\boldmath $D^+ \to K^0_sK^+\pi^+\pi^-$} &
  2005 & FOCUS~\cite{Link:2005th}  &  $ -0.042  \pm 0.064  \pm 0.022  $ \\
\hline 
\end{tabular}
\end{center} 
\end{table}

\begin{table}
\renewcommand{\arraystretch}{1.3}
\caption{\cp\ asymmetries 
$A^{}_{CP}=[\Gamma(D^0)-\Gamma(\dbar)]/[\Gamma(D^0)+\Gamma(\dbar)]$
for $D^0,\dbar$ decays.
\label{tab:cp_neutral}}
\footnotesize
\begin{center}
\begin{tabular}{|l|c|c|c|} 
\hline
{\bf Mode} & {\bf Year} & {\bf Collaboration} & {\boldmath $A^{}_{CP}$} \\
\hline
{\boldmath $D^0 \to \pi^+\pi^-$} &
  2012 & CDF~\cite{Aaltonen:2012ab}  & $ +0.0022  \pm 0.0024  \pm 0.0011  $ \\
& 2008 & Belle~\cite{Staric:2008rx}& $ +0.0043 \pm 0.0052 \pm 0.0012 $ \\
& 2008 & \babar~\cite{Aubert:2007if}  & $ -0.0024 \pm 0.0052 \pm 0.0022 $ \\
& 2002 & CLEO~\cite{Csorna:2001ww}    & $ +0.019  \pm 0.032  \pm 0.008  $ \\
& 2000 & FOCUS~\cite{Link:2000aw}  & $ +0.048  \pm 0.039  \pm 0.025  $ \\
& 1998 & E791~\cite{Aitala:1997ff}    & $ -0.049  \pm 0.078  \pm 0.030  $ \\
&      & COMBOS average          & $ +0.0020 \pm 0.0022 $ \\
\hline
{\boldmath $D^0 \to \pi^0\pi^0$} &
  2001 & CLEO~\cite{Bonvicini:2000qm}  & $ +0.001  \pm 0.048 $ (stat. and syst. combined) \\
\hline
{\boldmath $D^0 \to K_s^0\pi^0$} &
  2011 & Belle~\cite{Ko:2011ab}        & $ -0.0028 \pm 0.0019 \pm 0.0010 $ \\
& 2001 & CLEO~\cite{Bonvicini:2000qm}  & $ +0.001  \pm  0.013 $ (stat. and syst. combined) \\
&      & COMBOS average                & $ -0.0027 \pm 0.0021 $ \\
\hline
{\boldmath $D^0 \to K_s^0\eta$} &
  2011 & Belle~\cite{Ko:2011ab}        & $ +0.0054 \pm 0.0051 \pm 0.0016 $ \\
\hline
{\boldmath $D^0 \to K_s^0\eta^\prime$} &
  2011 & Belle~\cite{Ko:2011ab}        & $ +0.0098 \pm 0.0067 \pm 0.0014 $ \\  
\hline
{\boldmath $D^0 \to K^0_sK^0_s$} &
 2001 & CLEO~\cite{Bonvicini:2000qm}   & $ -0.23  \pm 0.19  $ (stat. and syst. combined) \\
\hline
{\boldmath $D^0 \to K^+K^-$} &
  2012 & CDF~\cite{Aaltonen:2012ab}& $ -0.0024  \pm 0.0022  \pm 0.0009  $ \\
& 2008 & Belle~\cite{Staric:2008rx}& $ -0.0043 \pm 0.0030 \pm 0.0011 $ \\
& 2008 & \babar~\cite{Aubert:2007if}  & $ +0.0000 \pm 0.0034 \pm 0.0013 $ \\
& 2002 & CLEO~\cite{Csorna:2001ww}    & $ +0.000  \pm 0.022  \pm 0.008  $ \\
& 2000 & FOCUS~\cite{Link:2000aw}  & $ -0.001  \pm 0.022  \pm 0.015  $ \\
& 1998 & E791~\cite{Aitala:1997ff}    & $ -0.010  \pm 0.049  \pm 0.012  $ \\
& 1995 & CLEO~\cite{Bartelt:1995vr}    & $ +0.080  \pm 0.061             $ (stat.) \\
& 1994 & E687~\cite{Frabetti:1994kv}    & $ +0.024  \pm 0.084             $ (stat.) \\
&      & COMBOS average          & $ -0.0023 \pm 0.0017            $ \\
\hline
{\boldmath $D^0 \to \pi^+\pi^-\pi^0$} &
   2008 & \babar~\cite{Aubert:2008yd} & $ -0.0031 \pm  0.0041 \pm  0.0017$ \\
&  2008 & Belle~\cite{Arinstein:2008zh}  & $ +0.0043 \pm  0.0130 $ \\
&  2005 & CLEO~\cite{CroninHennessy:2005sy}  & $ +0.001^{+0.09}_{-0.07} \pm  0.05 $ \\
&       & COMBOS average         & $ -0.0023 \pm 0.0042 $ \\
\hline
{\boldmath $D^0 \to K^-\pi^+\pi^0$} &
  2007 & CLEOc~\cite{Dobbs:2007zt} & $  +0.002  \pm 0.004  \pm 0.008 $ \\
& 2001 & CLEO~\cite{Kopp:2000gv}  & $ -0.031   \pm 0.086 $ (stat.) \\
&       & COMBOS average         & $ +0.0016 \pm 0.0089 $ \\
\hline   
{\boldmath $D^0 \to K^+\pi^-\pi^0$} &
  2005 & Belle~\cite{Tian:2005ik} & $ -0.006  \pm 0.053  $ (stat.) \\
& 2001 & CLEO~\cite{Brandenburg:2001ze}  & $ +0.09^{+0.25}_{-0.22}  $ (stat.) \\
&       & COMBOS average         & $ -0.0014 \pm 0.0517 $ \\
\hline
{\boldmath $D^0 \to K^0_s\pi^+\pi^-$} &
 2004 & CLEO~\cite{Asner:2003uz}    & $ -0.009  \pm 0.021^{+0.016}_{-0.057} $ \\
\hline
{\boldmath $D^0 \to K^+ K^-\pi^0$} &
   2008 & \babar~\cite{Aubert:2008yd} & $ 0.0100 \pm  0.0167 \pm  0.0025$ \\ 
\hline
{\boldmath $D^0 \to K^+\pi^-\pi^+\pi^-$} &
  2005 & Belle~\cite{Tian:2005ik} & $ -0.018  \pm 0.044  $ (stat.) \\
\hline
{\boldmath $D^0 \to K^+K^-\pi^+\pi^-$} &
  2005 & FOCUS~\cite{Link:2005th} & $ -0.082  \pm 0.056  \pm .047  $ \\
\hline                   
\end{tabular}
 \end{center} 
\end{table}

\begin{table}
\renewcommand{\arraystretch}{1.4}
\caption{\cp\ asymmetries 
$A^{}_{CP}= [\Gamma(D_s^+)-\Gamma(D_s^-)]/[\Gamma(D_s^+)+\Gamma(D_s^-)]$
for $D_s^\pm$ decays.
\label{tab:cp_ds}}
\footnotesize
\begin{center}
\begin{tabular}{|l|c|c|c|} 
\hline
{\bf Mode} & {\bf Year} & {\bf Collaboration} & {\boldmath $A^{}_{CP}$} \\
\hline
{\boldmath $D_s^+ \to \mu^+ \nu$} &
  2009 & CLEOc~\cite{Alexander:2009ux} &  $ +0.048  \pm 0.061 $ \\
\hline
{\boldmath $D_s^+ \to \pi^+ \eta$} &
  2010 & CLEOc~\cite{Mendez:2009aa} &  $ -0.046  \pm 0.029 \pm 0.003 $ \\
\hline
{\boldmath $D_s^+ \to \pi^+ \eta^\prime$} &
  2010 & CLEOc~\cite{Mendez:2009aa} &  $ -0.061  \pm 0.030 \pm 0.003 $ \\
\hline
{\boldmath $D_s^+ \to K^0_s\pi^+$}   &
   2010 & Belle~\cite{Ko:2010ng}     &  $ +0.0545 \pm 0.0250 \pm 0.0033 $ \\
&  2010 & CLEOc~\cite{Mendez:2009aa} &  $ +0.163  \pm 0.073  \pm 0.003  $ \\
&       & COMBOS average             &  $ +0.066 \pm 0.024 $            \\
\hline
{\boldmath $D_s^+ \to K^+ \pi^0$}   &
  2010 & CLEOc~\cite{Mendez:2009aa} &  $ +0.266 \pm 0.228 \pm 0.009 $ \\
\hline
{\boldmath $D_s^+ \to K^+ \eta$}  &
  2010 & CLEOc~\cite{Mendez:2009aa} &  $ +0.093 \pm 0.152 \pm 0.009 $ \\
\hline
{\boldmath $D_s^+ \to K^+ \eta^\prime$}  &
  2010 & CLEOc~\cite{Mendez:2009aa}      &  $ +0.060 \pm 0.189 \pm 0.009 $ \\
\hline
{\boldmath $D_s^+ \to K^+K^0_s$}   &
   2010 & Belle~\cite{Ko:2010ng}     &  $ +0.0012 \pm 0.0036 \pm 0.0022 $ \\
&  2010 & CLEOc~\cite{Mendez:2009aa} &  $ +0.047  \pm 0.018  \pm 0.009  $ \\
&       & COMBOS average           &  $ +0.0031 \pm 0.0041 $            \\
\hline
{\boldmath $D_s^+ \to \pi^+ \pi^+ \pi^-$} &
  2008 & CLEOc~\cite{Alexander:2008cqa} &  $ +0.020  \pm 0.046 \pm 0.007 $ \\
\hline
{\boldmath $D_s^+ \to K^+ \pi^+ \pi^-$} &
  2008 & CLEOc~\cite{Alexander:2008cqa} &  $ +0.112  \pm 0.070 \pm 0.009 $ \\
\hline
{\boldmath $D_s^+ \to K^+ K^- \pi^+$} &
  2008 & CLEOc~\cite{Alexander:2008cqa} &  $ +0.003  \pm 0.011 \pm 0.008 $ \\  
\hline
{\boldmath $D_s^+ \to K^0_s K^- \pi^+\pi^+$} &
  2008 & CLEOc~\cite{Alexander:2008cqa} &  $ -0.007  \pm 0.036 \pm 0.011 $ \\  
\hline
{\boldmath $D_s^+ \to K^+ K^- \pi^+\pi^0$} &
  2008 & CLEOc~\cite{Alexander:2008cqa} &  $ -0.059  \pm 0.042 \pm 0.012 $ \\  
\hline 
\end{tabular}
\end{center} 
\end{table}

\clearpage
\subsection{\emph{$T$}-violating asymmetries}
                                               
$T$-violating asymmetries are measured using triple-product
correlations and assuming the validity of the $CPT$ theorem.
Triple-product correlations of the form 
$\vec{a}\cdot(\vec{b}\times\vec{c})$, 
where $a$, $b$, and $c$ are spins or momenta, are odd 
under time reversal~(\emph{T}).
For example, for $D^0 \to K^+K^-\pi^+\pi^-$ decays, 
$C_T \equiv \vec{p}^{}_{K^+}\cdot(\vec{p}_{\pi^+}\times \vec{p}_{\pi^-})$  
changes sign (i.e., is odd) under a \emph{T} transformation.
The corresponding quantity for $\dbar$ is
$\overline{C}_T \equiv 
      \vec{p}^{}_{K^-}\cdot(\vec{p}_{\pi^-}\times \vec{p}_{\pi^+})$.
Defining  
\begin{eqnarray}
 A_{T} & = &
    \frac{\Gamma(C_T>0)-\Gamma(C_T<0)}{\Gamma(C_T>0)+\Gamma(C_T<0)}
\end{eqnarray}
for $D^0$ decay and
\begin{eqnarray}
\overline{A}_{T} & = & 
   \frac{\Gamma(-\overline{C}_T>0)-\Gamma(-\overline{C}_T<0)}
                        {\Gamma(-\overline{C}_T>0)+\Gamma(-\overline{C}_T<0)}
\end{eqnarray} 
for $\dbar$ decay, in the absence of strong phases
either $A^{}_T\neq 0$ or $\overline{A}^{}_T\neq 0$ indicates
$T$ violation. In these expressions the $\Gamma$'s are partial widths. 
The asymmetry
\begin{eqnarray}
A^{}_{T\,{\rm viol}} & \equiv & \frac{A_{T}-\overline{A}_{T}}{2}
\end{eqnarray}
tests for $T$ violation even with nonzero strong phases (see 
Refs.~\cite{Golowich:1988ig,Bigi:2001sg,Bensalem:2002ys,Bensalem:2000hq,Bensalem:2002pz}).
Values of $A_{T\,{\rm viol}}$ for some $D^+$, $D^+_s$, and
$D^0$ decay modes are listed in Table~\ref{tab:t_viol}.

\begin{table}[h]
\renewcommand{\arraystretch}{1.4}
\caption{$T$-violating asymmetries 
$A^{}_{T\,{\rm viol}} = (A_{T}-\overline{A}_{T})/2$.
\label{tab:t_viol}}
\footnotesize
\begin{center}
\begin{tabular}{|l|c|c|c|} 
\hline
{\bf Mode} & {\bf Year} & {\bf Collaboration} & {\boldmath $A^{}_{T\,{\rm viol}}$} \\
\hline
{\boldmath $D^0 \to K^+K^-\pi^+\pi^-$} &
   2010 & \babar~\cite{Sanchez:2010xj}  &  $ +0.0010 \pm 0.0051 \pm 0.0044 $ \\
&  2005 & FOCUS~\cite{Link:2005th}     &  $ +0.010  \pm 0.057  \pm 0.037  $ \\
&       & COMBOS average               &  $ +0.0011 \pm 0.0067            $ \\  
\hline
{\boldmath $D^+ \to K^0_sK^+\pi^+\pi^-$} &
  2010 & \babar~\cite{Lees:2011ab}  &  $ -0.0120 \pm 0.0100 \pm 0.0046 $ \\
& 2005 & FOCUS~\cite{Link:2005th}  &  $ +0.023  \pm 0.062  \pm 0.022  $ \\
&      & COMBOS average            &  $ -0.0110 \pm 0.0109            $ \\
\hline
{\boldmath $D^+_s \to K^0_sK^+\pi^+\pi^-$} &
  2010 & \babar~\cite{Lees:2011ab}  &  $ -0.0136 \pm 0.0077 \pm 0.0034 $ \\
& 2005 & FOCUS~\cite{Link:2005th}  &  $ -0.036  \pm 0.067  \pm 0.023  $ \\
&      & COMBOS average            &  $ -0.0139 \pm 0.0084            $ \\
\hline                    
\end{tabular}
\end{center} 
\end{table}

\vskip0.30in
\begin{center}  ---------------  \end{center}
\vskip0.30in

\clearpage
\subsection{World average for the \emph{$D^+_s$} decay constant~\fds}

The Heavy Flavor Averaging Group has used 
measurements of the branching fractions 
${\cal B}(D^+_s\ra\mu^+\nu)$~\cite{Alexander:2009ux,delAmoSanchez:2010jg,Zupanc:2012}
and 
${\cal B}(D^+_s\ra\tau^+\nu)$~\cite{Alexander:2009ux,
Onyisi:2009th,Naik:2009tk,delAmoSanchez:2010jg,Lees:2010qj,
Zupanc:2012} from Belle, \babar, and CLEO to calculate a 
world average (WA) value for the $D^+_s$ decay 
constant~\fds. We do not use older results from 
the ALEPH~\cite{Heister:2002fp}, 
BEATRICE~\cite{Alexandrov:2000ns}, 
OPAL~\cite{Abbiendi:2001nb}, and 
L3~\cite{Acciarri:1996bv} experiments as
the errors are large and these measurements 
have some unknown systematic errors. 

\begin{figure}
\begin{center}
\includegraphics[width=5.0in]{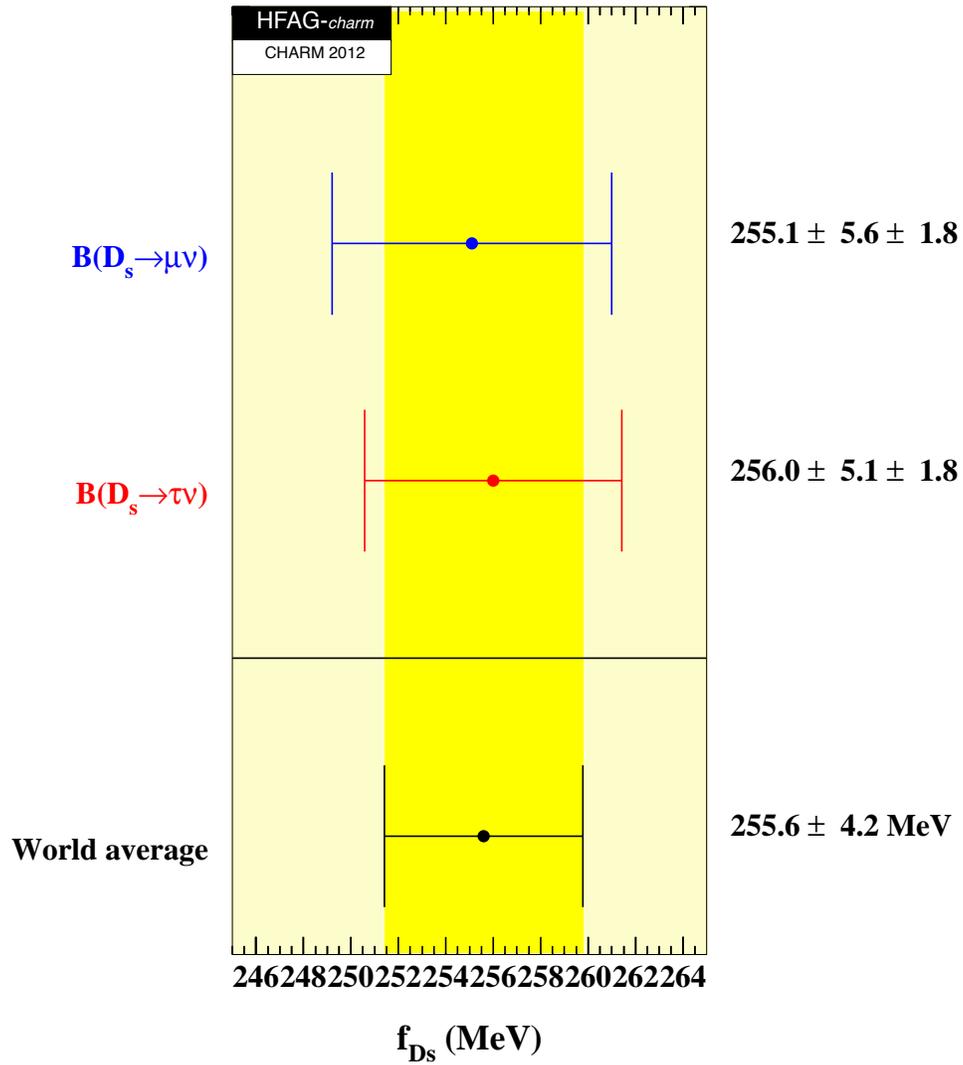}
\end{center}
\vskip-0.20in
\caption{\label{fig:charm_fds}
WA value for \fds. For each measurement, the first error listed 
is the total uncorrelated error, and the second error is the total
correlated error (mostly from $\tau^{}_{D^{}_s}$).}
\end{figure}

The value for \fds\ is calculated using the formula
\begin{eqnarray}
f^{}_{D^{}_s} & = & \frac{1}
{G^{}_F |V^{}_{cs}| m^{}_{\ell}
\biggl( 1-\frac{\displaystyle m^2_{\ell}}{\displaystyle m^2_{D^{}_s}}\biggr)}
\sqrt{\frac{8\pi\,{\cal B}(D^+_s\ra\ell^+\nu)}{m^{}_{D^{}_s} \tau^{}_{D^{}_s}}}\,,
\label{eqn:fds_inverted}
\end{eqnarray}
where $\ell^+=\mu^+$ or~$\tau^+$.
For ${\cal B}(D^+_s\ra\ell^+\nu)$ we use the WA values 
obtained below. The error on \fds\ is calculated as follows:
values for variables on the right-hand-side of Eq.~(\ref{eqn:fds_inverted}) 
are sampled from Gaussian distributions having mean values equal to the 
central values and standard deviations equal to their respective 
errors. The resulting values of \fds\ are plotted, and the 
r.m.s. of the distribution is taken as the $\pm 1\sigma$ errors
on \fds. 
The procedure is done separately for 
${\cal B}(D^+_s\ra\mu^+\nu)$ and ${\cal B}(D^+_s\ra\tau^+\nu)$;
the resulting two values for \fds\ are averaged together using 
COMBOS~\cite{Combos:1999}, which accounts for correlations such 
as the values of $|V^{}_{cs}|$, $m^{}_{D^{}_s}$, and $\tau^{}_{D^{}_s}$ 
used\footnote{These values (taken from the PDG~\cite{PDG_2012}) are
$|V^{}_{cs}|=  0.97345^{+0.00015}_{-0.00016}$; 
$m^{}_\tau = (1.77682\pm 0.00016)$~GeV/$c^2$;
$m^{}_{D^{}_s}= (1.96847\pm 0.00033)$~GeV/$c^2$; and 
$\tau^{}_{D^{}_s}= (500\pm 7)\times 10^{-15}$~s.}
in Eq.~(\ref{eqn:fds_inverted}). 
The result is plotted in Fig.~\ref{fig:charm_fds}. The WA value is
\begin{eqnarray}
f^{}_{D^{}_s} & = & 255.6\,\pm 4.2 {\rm\ \ MeV},
\end{eqnarray}
where the statistical and systematic errors are combined.

\begin{figure}
\begin{center}
\includegraphics[width=5.0in]{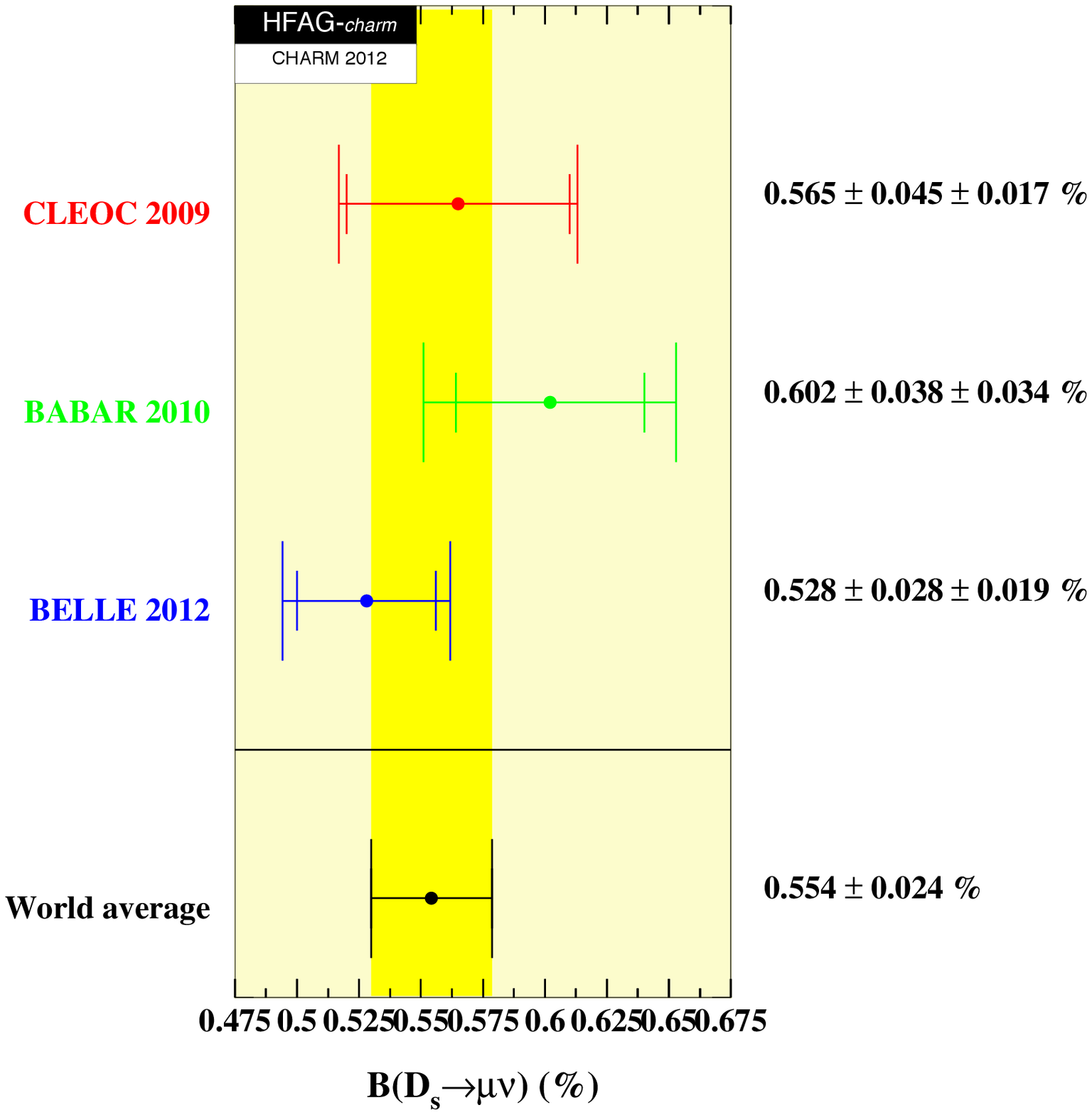}
\end{center}
\vskip-0.20in
\caption{\label{fig:charm_fds_mu}
WA value for ${\cal B}(D^+_s\ra\mu^+\nu)$, as
calculated from Refs.~\cite{Alexander:2009ux,delAmoSanchez:2010jg,Zupanc:2012}.
When two errors are listed, the first one is statistical and 
the second is systematic.}
\end{figure}

The WA value for ${\cal B}(D^+_s\ra\mu^+\nu)$ is calculated from
CLEOc~\cite{Alexander:2009ux}, Belle~\cite{Zupanc:2012}, and 
\babar~\cite{delAmoSanchez:2010jg} measurements of absolute branching
fractions. These measurements are {\it not\/} normalized to 
$D^+_s\ra\phi\pi^+$ decays as was done for earlier measurements,
and thus they do not have uncertainties due to the non-resonant 
$D^+_s\ra K^+K^-\pi^+$ contribution. All input values and the 
result are plotted in Fig.~\ref{fig:charm_fds_mu}. The WA 
value is ${\cal B}(D^+_s\ra\mu^+\nu)= (0.554\pm 0.024)\%$.

\begin{figure}
\begin{center}
\includegraphics[width=5.0in]{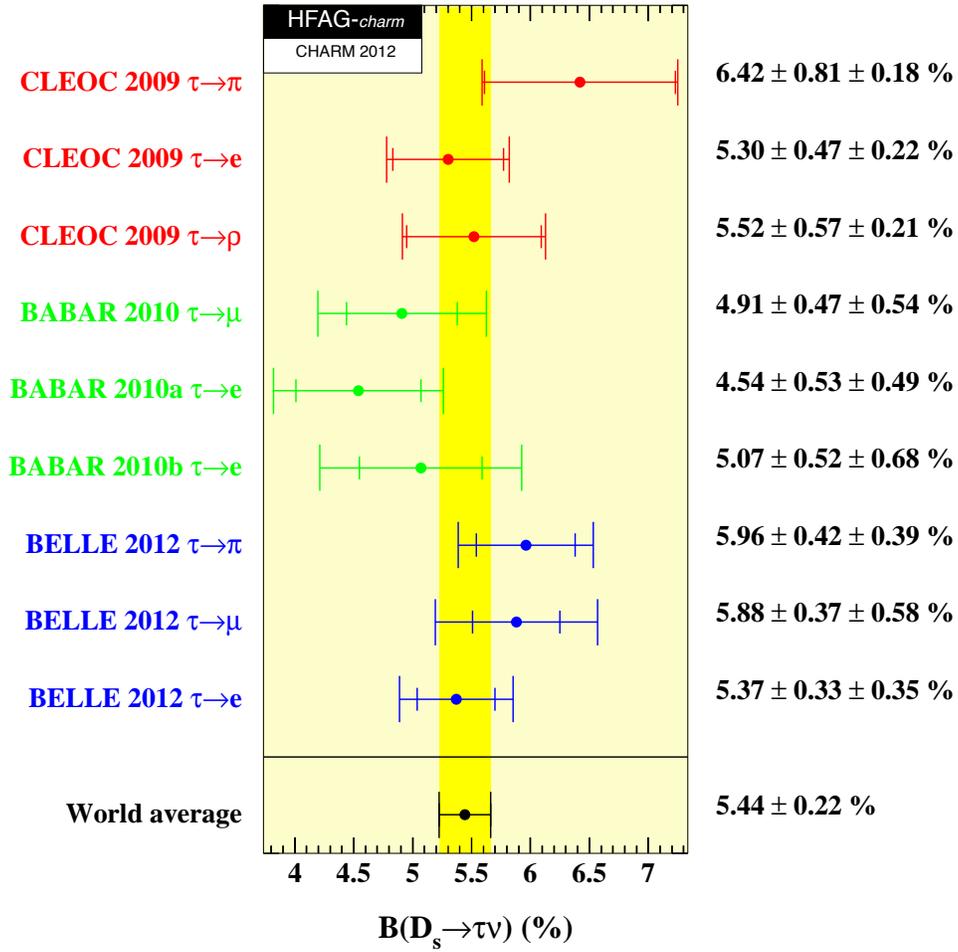}
\end{center}
\vskip-0.20in
\caption{\label{fig:charm_fds_tau}
WA value for ${\cal B}(D^+_s\ra\tau^+\nu)$, as calculated from 
Refs.~\cite{Alexander:2009ux,Onyisi:2009th,Naik:2009tk,delAmoSanchez:2010jg,
Lees:2010qj,Zupanc:2012}. When two errors are listed, the first one 
is statistical and the second is systematic.}
\end{figure}

The WA value for ${\cal B}(D^+_s\ra\tau^+\nu)$ is also calculated 
from CLEOc, Belle, and \babar\ measurements. 
CLEOc made separate measurements for 
$\tau^+\ra e^+\nu\bar{\nu}$~\cite{Naik:2009tk},
$\tau^+\ra\pi^+\nu$~\cite{Alexander:2009ux}, and
$\tau^+\ra\rho^+\nu$~\cite{Onyisi:2009th};
\babar\ made separate measurements for 
$\tau^+\ra \mu^+\nu\bar{\nu}$~\cite{delAmoSanchez:2010jg} and 
$\tau^+\ra e^+\nu\bar{\nu}$~\cite{delAmoSanchez:2010jg,Lees:2010qj}; and
Belle made separate measurements for 
$\tau^+\ra \mu^+\nu\bar{\nu}$, $\tau^+\ra e^+\nu\bar{\nu}$,
and $\tau^+\ra\pi^+\nu$~\cite{Zupanc:2012}.
All input values and the result are plotted in 
Fig.~\ref{fig:charm_fds_tau}. The WA value is 
${\cal B}(D^+_s\ra e^+\nu)= (5.44\pm 0.22)\%$.

\clearpage

\subsection{Two-body hadronic $D^0$ decays and final state radiation}

\begin{figure}[bt]
\begin{center}
\includegraphics[width=0.5\textwidth,angle=-90.]{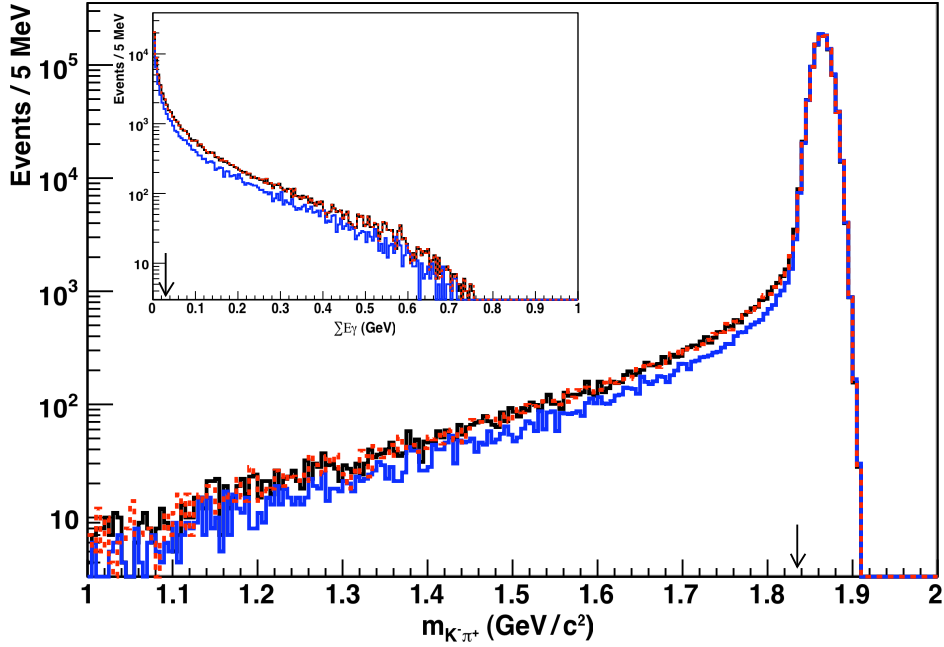}
\caption{The $K\pi$ invariant mass distribution for 
$D^0\to K^-\pi^+ (n\gamma)$ decays. The 3 curves correspond 
to three different configurations of PHOTOS for modeling FSR: 
version 2.02 without interference (blue), version 2.02 with 
interference (red dashed) and version 2.15 with interference (black).  
The true invariant mass has been smeared with a typical experimental 
resolution of 10 MeV${}/c^2$.  Inset: The corresponding spectrum of 
total energy radiated per event.  The arrow indicates the $E_\gamma$ 
value that begins to shift kinematic quantities outside of the range 
typically accepted in a measurement.}
\label{fig:FSR_mass_shift}
\end{center}
\end{figure}

Branching fractions measurements for $D^0\to K^-\pi^+$, $D^0\to \pi^+\pi^-$ 
and $D^0\to K^+ K^-$ have reached sufficient precision to allow averages 
with ${\cal O}(1\%)$ relative uncertainties. At these precisions, Final 
State Radiation (FSR) must be treated correctly and consistently across 
the input measurements for the accuracy of the averages to match the 
precision.  The sensitivity of measurements to FSR arises because of 
a tail in the distribution of radiated energy that extends to the 
kinematic limit.  The tail beyond $E_\gamma \approx 30$ MeV causes 
typical selection variables like the hadronic invariant mass to 
shift outside the selection range dictated by experimental 
resolution (see Fig.~\ref{fig:FSR_mass_shift}).  While the 
differential rate for the tail is small, the integrated rate 
amounts to several percent of the total $h^+ h^-(n\gamma)$ 
rate because of the tail's extent.  The tail therefore 
translates directly into a several percent loss in 
experimental efficiency.

All measurements that include an FSR correction have a correction 
based on use of 
PHOTOS~\cite{Barberio:1990ms,Barberio:1993qi,Golonka:2005pn,Golonka:2006tw} 
within the experiment's Monte Carlo simulation.  PHOTOS itself, however, 
has evolved, over the period spanning the set of measurements.  In 
particular, incorporation of interference between radiation off of 
the two separate mesons has proceeded in stages: it was first available 
for particle--antiparticle pairs in version 2.00 (1993), and extended 
to any two body, all charged, final states in version 2.02 (1999).  
The effects of interference are clearly visible 
(Figure~\ref{fig:FSR_mass_shift}), and cause a 
roughly 30\% increase in the integrated rate into 
the high energy photon tail.  To evaluate the FSR 
correction incorporated into a given measurement, 
we must therefore note whether any correction was 
made, the version of PHOTOS used in correction, 
and whether the interference terms in PHOTOS were 
turned on.  

\subsubsection{Branching fraction corrections}

Before averaging the measured branching fractions, the published 
results are updated, as necessary, to the FSR prediction of 
PHOTOS~2.15 with interference included.  The correction will 
always shift a branching fraction to a higher value: with no 
FSR correction or with no interference term in the correction, 
the experimental efficiency determination will be biased high, 
and therefore the branching fraction will be biased low.

Most of the branching fraction analyses used the kinematic quantity sensitive to FSR in the candidate selection criteria.  For the analyses at the $\psi(3770)$, the variable was $\Delta E$, the difference between the candidate $D^0$ energy and the beam energy ({\em e.g.}, $E_K + E_\pi - E_{\rm beam}$ for $D^0\to K^-\pi^+$).  In the remainder of the analyses, the relevant quantity was the reconstructed hadronic two-body mass $m_{h^+h^-}$.  To correct we need only to evaluate the fraction of decays that FSR moves outside of the range accepted for the analysis.  

The corrections were evaluated using an event generator (EvtGen \cite{Ryd:2005zz}) that incorporates PHOTOS to simulate the portions of the decay process most relevant to the correction.  We compared corrections determined both with and without smearing to account for experimental resolution.  The differences were negligible, typically of order of a 1\% of the correction itself.  The immunity of the correction to resolution effects comes about because most of the long FSR-induced tail in, for example, the $m_{h^+h^-}$ distribution resides well away from the selection boundaries.  The smearing from resolution, on the other hand, mainly affects the distribution of events right at the boundary.  

For measurements incorporating an FSR correction that did not include interference, we update by assessing the FSR-induced efficiency loss for both the PHOTOS version and configuration used in the analysis and our nominal version 2.15 with interference.  For measurements that published their sensitivity to FSR, our generator-level predictions for the original efficiency loss agreed to within a few percent (of the correction).  This agreement lends additional credence to the procedure.

Once the event loss from FSR in the most sensitive kinematic quantity is accounted for, the event loss from other quantities is very small.  Analyses using $D^*$ tags, for example, showed little sensitivity to FSR in the reconstructed $D^*-D^0$ mass difference: for example, in $m_{K^-\pi^+\pi^+}-m_{K^-\pi^+}$. Because the effect of FSR tends to cancel in the difference of the reconstructed masses, this difference showed a much smaller sensitivity than the two body mass even before a two body mass requirement. In the $\psi(3770)$ analyses, the beam-constrained mass distributions ($\sqrt{E_{\rm beam}^2 - |\vec{p}_K + \vec{p}_\pi|^2}$)  showed little further sensitivity.

\begin{figure}
\begin{center}
\includegraphics[width=0.23\textwidth,angle=-90.]{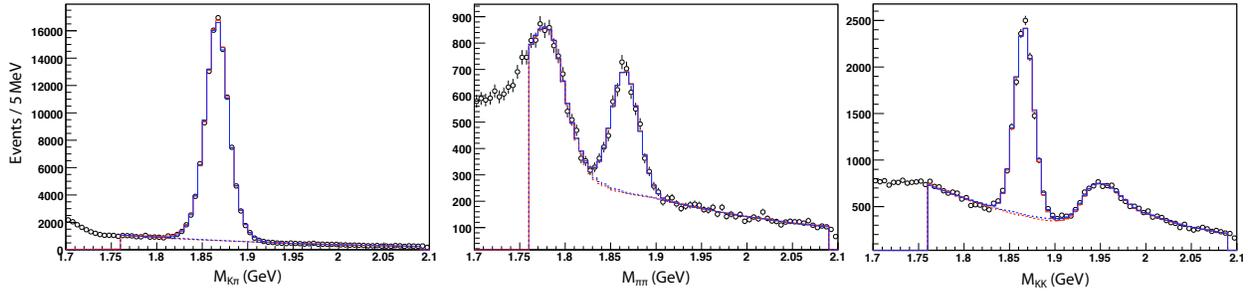}
\caption{FOCUS data (dots), original fits (blue) and 
toy MC parameterization (red) for $D^0\to K^-\pi^+$ (left) , 
$D^0\to \pi^+\pi^-$ (center) and $D^0\to \pi^+\pi^-$ (right).}
\label{fig:FocusFits}
\end{center}
\end{figure}

The FOCUS \cite{Link:2002hi} analysis of the branching ratios ${\cal B}(D^0\to \pi^+\pi^-)/{\cal B}(D^0\to K^-\pi^+)$ and ${\cal B}(D^0\to K^+ K^-)/{\cal B}(D^0\to K^-\pi^+)$ obtained yields using fits to the two body mass distributions.  FSR will both distort the low end of the signal mass peak, and will contribute a signal component to the low side tail used to estimate the background.  The fitting procedure is not sensitive to signal events out in the FSR tail, which would be counted as part of the background.

A more complex toy Monte Carlo procedure was required to analyze the effect of FSR on the fitted yields, which were published with no FSR corrections applied.  A detailed description of the procedure and results is available on the HFAG web page, and a brief summary is provided here.  Determining the correction involved an iterative procedure in which samples of similar size to the FOCUS sample were generated and then fit using the FOCUS signal and background parameterizations.  The MC parameterizations were tuned based on differences between the fits to the toy MC data and the FOCUS fits, and the procedure was repeated. These steps were iterated until the fit parameters matched the original FOCUS parameters.  

\begin{table}
  \centering 
  \caption{The experimental measurements relating to ${\cal B}(D^0\to K^-\pi^+)$, ${\cal B}(D^0\to \pi^+\pi^-)$ and ${\cal B}(D^0\to K^+ K^-)$ after correcting to the common version and configuration of PHOTOS.  The uncertainties are statistical and total systematic, with the FSR-related systematic estimated in this procedure shown in parentheses.  Also listed are the percent shifts in the results from the correction, if any, applied here, as well as the original PHOTOS and interference configuration for each publication.}
  \label{tab:FSR_corrections}
\begin{tabular}{lccc}
\hline \hline
Experiment & result (rescaled) & correction [\%] & PHOTOS \\ \hline
\multicolumn{4}{l}{$D^{0} \to K^{-} \pi^{+}$} \\
      CLEO-c 07  (CC07) \cite{Dobbs:2007zt}         & $3.891 \pm 0.035 \pm 0.065(27)\%$ & --   & 2.15/Yes \\      
      \babar 07   (BB07) \cite{Aubert:2007wn}   & $4.035 \pm 0.037 \pm 0.074(24)\%$ & 0.69 & 2.02/No \\
      CLEO II 98 (CL98) \cite{Artuso:1997mc}   & $3.920 \pm 0.154 \pm 0.168(32)\%$ & 2.80 & none \\
      ALEPH 97   (AL97) \cite{Barate:1997mm}   & $3.930 \pm 0.091 \pm 0.125(32)\%$ & 0.79 & 2.0/No \\
      ARGUS 94   (AR94) \cite{Albrecht:1994nb} & $3.490 \pm 0.123 \pm 0.288(24)\%$ & 2.33 & none \\
      CLEO II 93 (CL93) \cite{Akerib:1993pm}   & $3.960 \pm 0.080 \pm 0.171(15)\%$ & 0.38 & 2.0/No \\
      ALEPH 91   (AL91) \cite{Decamp:1991jw}   & $3.730 \pm 0.351 \pm 0.455(34)\%$ & 3.12 & none \\
\multicolumn{4}{l}{$D^{0} \to \pi^{+}\pi^{-} / D^{0} \to K^{-} \pi^{+}$} \\
      CLEO-c 10  (CC10) \cite{Mendez:2009aa}   & $0.0370  \pm 0.0006  \pm 0.0009(02)$  & --   & 2.15/Yes \\
      CDF 05     (CD05) \cite{Acosta:2004ts}   & $0.03594 \pm 0.00054 \pm 0.00043(15)$ & --   & 2.15/Yes \\
      FOCUS 02   (FO02) \cite{Link:2002hi}     & $0.0364  \pm 0.0012  \pm 0.0006(02)$  & 3.10 & none \\
\multicolumn{4}{l}{$D^{0} \to K^{+}K^{-} / D^{0} \to K^{-} \pi^{+}$} \\
      CLEO-c 10   \cite{Mendez:2009aa}         & $0.1041 \pm 0.0011 \pm 0.0012(03)$ & --    & 2.15/Yes \\ 
      CDF 05      \cite{Acosta:2004ts}         & $0.0992 \pm 0.0011 \pm 0.0012(01)$ & --    & 2.15/Yes \\
      FOCUS 02    \cite{Link:2002hi}           & $0.0982 \pm 0.0014 \pm 0.0014(01)$ & -1.12 & none \\ \hline
\end{tabular}
\end{table}

The toy MC samples for the first iteration were based on the generator-level distribution of $m_{K^-\pi^+}$, $m_{\pi^+\pi^-}$ and $m_{K^+K^-}$, including the effects of FSR, smeared according to the original FOCUS resolution function, and on backgrounds thrown using the parameterization from the final FOCUS fits.  For each iteration, 400 to 1600 individual data-sized samples were thrown and fit. The means of the parameters from these fits determined the corrections to the generator parameters for the following iteration.  The ratio between the number of signal events generated and the final signal yield provides the required FSR correction in the final iteration.  Only a few iterations were required in each mode.  Figure ~\ref{fig:FocusFits} shows the FOCUS data, the published FOCUS fits, and the final toy MC parameterizations.  The toy MC provides an excellent description of the data.

The corrections obtained to the individual FOCUS yields were $1.0298\pm 0.0001$ for $K^-\pi^+$, $1.062 \pm 0.001$ for $\pi^+\pi^-$, and $1.0183 \pm 0.0003$ for $K^+K^-$.  These corrections tend to cancel in the branching ratios, leading to corrections of 1.031 to  ${\cal B}(D^0\to \pi^+\pi^-)/{\cal B}(D^0\to K^-\pi^+)$, and 0.9888 for ${\cal B}(D^0\to K^+ K^-)/{\cal B}(D^0\to K^-\pi^+)$.

Table~\ref{tab:FSR_corrections} summarizes the corrected branching fractions.  The published FSR-related modeling uncertainties have been replaced by with a new, common, estimate based on the assumption that the dominant uncertainty in the FSR corrections come from the fact that the mesons are treated like structureless particles. No contributions from structure-dependent terms in the decay process (eg. radiation off individual quarks) are included in PHOTOS. Internal studies done by various experiments have indicated that in $K\pi$ decay, the PHOTOS corrections agree with data at the 20-30\% level. We therefore attribute a 25 uncertainty to the FSR prediction from potential structure-dependent contributions. For the other two modes, the only difference in structure is the final state valence quark content. While radiative corrections typically come in with a $1/M$ dependence, one would expect the additional contribution from the structure terms to come in on time scales shorter than the hadronization time scale. In this case, you might expect LambdaQCD to be the relevant scale, rather than the quark masses, and therefore that the amplitude is the same for the three modes. In treating the correlations among the measurements this is what we assume. We also assume that the PHOTOS amplitudes and any missing structure amplitudes are relatively real with constructive interference.  The uncertainties largely cancel in the branching fraction ratios. For the final average branching fractions, the FSR uncertainty on $K\pi$ dominates. Note that because of the relative sizes of FSR in the different modes, the $\pi\pi/K\pi$ branching ratio uncertainty from FSR is positively correlated with that for $K\pi$ branching, while the $KK/K\pi$ branching ratio FSR uncertainty is negatively correlated.

The ${\cal B}(D^0\to K^-\pi^+)$ measurement of reference~\cite{Coan:1997ye}, the  
${\cal B}(D^0\to \pi^+\pi^-)/{\cal B}(D^0\to K^-\pi^+)$ measurements of references~\cite{Aitala:1997ff} 
and~\cite{Csorna:2001ww} and the ${\cal B}(D^0\to K^+ K^-)/{\cal B}(D^0\to K^-\pi^+)$ measurement
of reference~\cite{Csorna:2001ww} are excluded from the branching fraction averages presented here.
The measurements appear not to have incorporated any FSR corrections, and insufficient information
is available to determine the 2-3\% corrections that would be required.

\begin{sidewaystable}[p]
  \centering 
  \caption{The correlation matrix corresponding to the covariance matrix from the sum of statistical,
  systematic and FSR covariances.}\label{tab:correlations}
  \small
\begin{tabular}{lr@{.}lr@{.}lr@{.}lr@{.}lr@{.}lr@{.}lr@{.}lr@{.}lr@{.}lr@{.}lr@{.}lr@{.}lr@{.}l}
\hline\hline
           & \multicolumn{2}{c}{CC07}
                   & \multicolumn{2}{c}{BB07}
                           & \multicolumn{2}{c}{CL98}
                                   & \multicolumn{2}{c}{AL97}
                                           & \multicolumn{2}{c}{AR94} 
                                                   & \multicolumn{2}{c}{CL93} 
                                                           & \multicolumn{2}{c}{AL91} 
                                                                   & \multicolumn{2}{c}{FO02} 
                                                                           & \multicolumn{2}{c}{CD05} 
                                                                                   & \multicolumn{2}{c}{CC10} 
                                                                                           & \multicolumn{2}{c}{FO02}
                                                                                                    & \multicolumn{2}{c}{CD05} 
                                                                                                            & \multicolumn{2}{c}{CC10} \\ \hline
CC07 & 1&000 & 0&106 & 0&044 & 0&064 & 0&023 & 0&025 & 0&018 & 0&053 & 0&078 & 0&052 &-0&015 &-0&025 &-0&065 \\
BB07 & 0&106 & 1&000 & 0&035 & 0&051 & 0&019 & 0&020 & 0&014 & 0&042 & 0&062 & 0&041 &-0&012 &-0&019 &-0&051 \\
CL98 & 0&044 & 0&035 & 1&000 & 0&021 & 0&008 & 0&298 & 0&006 & 0&017 & 0&026 & 0&017 &-0&005 &-0&008 &-0&021 \\
AL97 & 0&064 & 0&051 & 0&021 & 1&000 & 0&011 & 0&012 & 0&116 & 0&025 & 0&038 & 0&025 &-0&007 &-0&012 &-0&031 \\
AR94 & 0&023 & 0&019 & 0&008 & 0&011 & 1&000 & 0&004 & 0&003 & 0&009 & 0&014 & 0&009 &-0&003 &-0&004 &-0&011 \\
CL93 & 0&025 & 0&020 & 0&298 & 0&012 & 0&004 & 1&000 & 0&003 & 0&010 & 0&015 & 0&010 &-0&003 &-0&005 &-0&012 \\
AL91 & 0&018 & 0&014 & 0&006 & 0&116 & 0&003 & 0&003 & 1&000 & 0&007 & 0&010 & 0&007 &-0&002 &-0&003 &-0&009 \\
FO02 & 0&053 & 0&042 & 0&017 & 0&025 & 0&009 & 0&010 & 0&007 & 1&000 & 0&031 & 0&021 &-0&006 &-0&010 &-0&026 \\
CD05 & 0&078 & 0&062 & 0&026 & 0&038 & 0&014 & 0&015 & 0&010 & 0&031 & 1&000 & 0&031 &-0&009 &-0&014 &-0&038 \\
CC10 & 0&052 & 0&041 & 0&017 & 0&025 & 0&009 & 0&010 & 0&007 & 0&021 & 0&031 & 1&000 &-0&006 &-0&010 &-0&025 \\
FO02 &-0&015 &-0&012 &-0&005 &-0&007 &-0&003 &-0&003 &-0&002 &-0&006 &-0&009 &-0&006 & 1&000 & 0&003 & 0&007 \\
CD05 &-0&025 &-0&019 &-0&008 &-0&012 &-0&004 &-0&005 &-0&003 &-0&010 &-0&014 &-0&010 & 0&003 & 1&000 & 0&012 \\
CC10 &-0&065 &-0&051 &-0&021 &-0&031 &-0&011 &-0&012 &-0&009 &-0&026 &-0&038 &-0&025 & 0&007 & 0&012 & 1&000 \\

\hline
\end{tabular}
\end{sidewaystable}

\subsubsection{Average branching fractions}

\begin{figure}
\begin{center}
\includegraphics[width=0.55\textwidth,angle=90.]{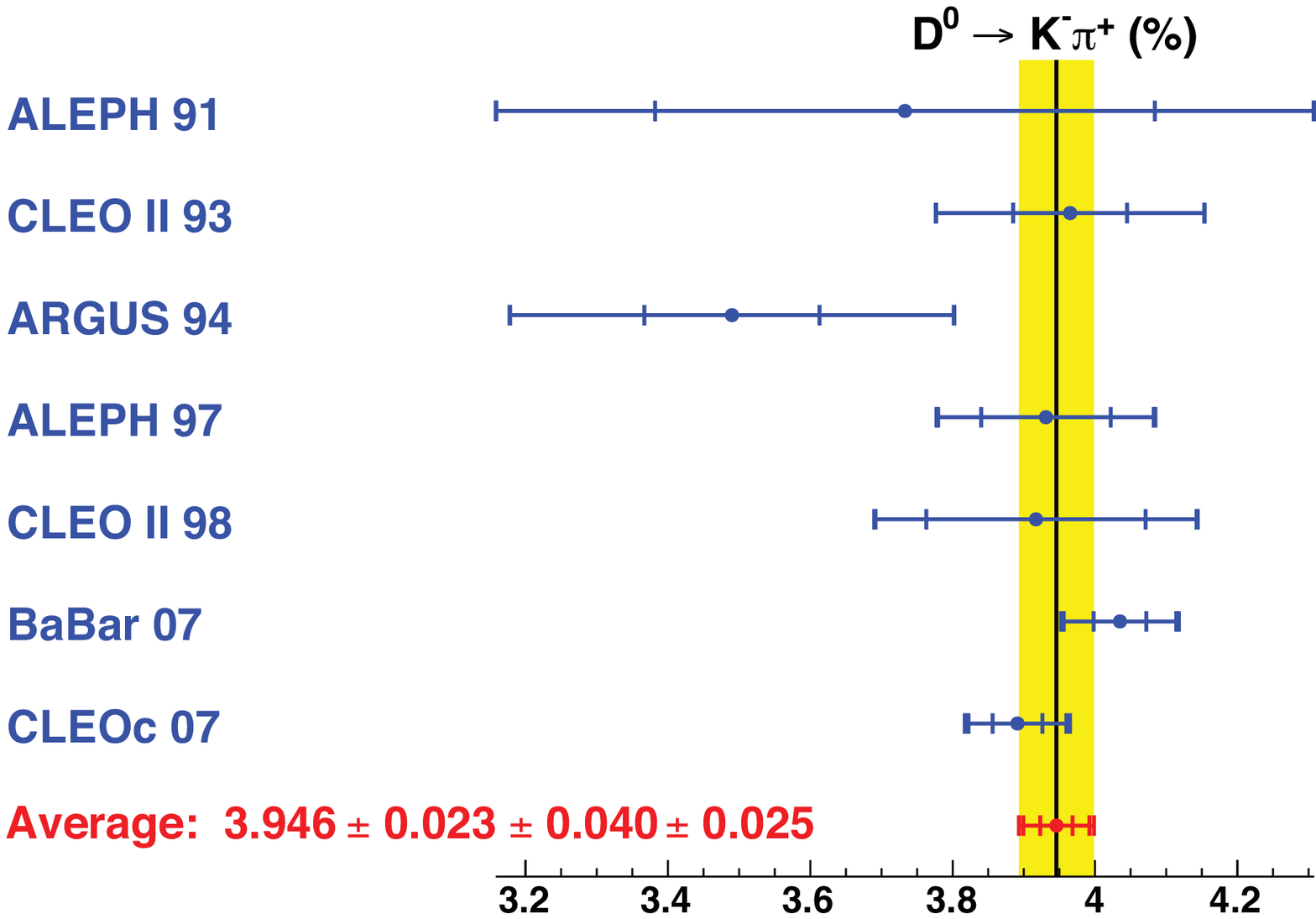}
\caption{Comparison of measurements of 
${\cal B}(D^0\to K^-\pi^+)$ (blue) with the average 
branching fraction obtained here (red, and yellow band).}
\label{D0bfs}
\end{center}
\end{figure}

The average branching fractions for $D^0\to K^-\pi^+$, $D^0\to \pi^+\pi^-$ and $D^0\to K^+ K^-$ are obtained from
a single $\chi^2$ minimization procedure,  in which the three branching fractions are floating parameters.  The
central values derive from a fit in which the covariance matrix is the sum of the covariance matrices for the
statistical, systematic (excluding FSR) and FSR uncertainties.  The statistical uncertainties are obtained from
a fit using only the statistical covariance matrix.  The systematic uncertainties are obtained from the
quadrature uncertainties from a fit with statistical-only and statistical+systematic covariance matrices, and
the FSR uncertainties on the averages from the quadrature differences in the uncertainties obtained from the
nominal fit and a fit excluding the FSR uncertainties.

In forming the covariance matrix for the FSR uncertainties, the FSR
uncertainties are treated as fully correlated (or anti-correlated) as described above.  For the
systematic covariance matrix, ALEPH's systematic uncertainties in the $\theta_{D^*}$ parameter are treated
as fully correlated between the ALEPH 97
and ALEPH 91 measurements.  Similarly, the tracking efficiency uncertainties in the CLEO II 98 and the
CLEO II 93 measurements are treated as fully
correlated.  
Table~\ref{tab:correlations} presents the correlation matrix for the nominal fit (stat.+syst.+FR).

The averaging procedure results in a final $\chi^2$ of 11.6  for 13-3 degrees of freedom.  The branching
fractions obtained are
\begin{eqnarray*}
  {\cal B}(D^0\to K^-\pi^+)   & = & 3.946 \pm 0.023 \pm 0.040 \pm 0.025 \\
  {\cal B}(D^0\to \pi^+\pi^-) & = & 0.143 \pm 0.002 \pm 0.002 \pm 0.002 \\
  {\cal B}(D^0\to K^+ K^-)    & = & 0.398 \pm 0.004 \pm 0.005 \pm 0.002. \\
\end{eqnarray*}
The uncertainties, estimated as described above, are statistical, systematic (excluding FSR), and
FSR modeling.  The correlation coefficients from the fit using the total uncertainties are
\begin{center}
\begin{tabular}{llll}
               & $K^-\pi^+$ & $\pi^+\pi^-$ & $K^+ K^-$ \\
$K^-\pi^+$     &  1.00 & 0.72 & 0.78  \\
$\pi^+\pi^-$   &  0.72 & 1.00 & 0.55  \\
$K^+ K^-$      &  0.78 & 0.55 & 1.00  \\
\end{tabular}
\end{center}

\begin{table}[b]
  \centering 
  \caption{Evolution of the $D^0\to K^-\pi^+$ branching fraction from a fit with no FSR corrections or correlations (similar  to the average in the PDG 2011 update~\cite{PDG_2008}) to the nominal fit presented here.}\label{tab:fit_evolution}
\begin{tabular}{cccll}
\hline\hline
Modes &  description                       & ${\cal B}(D^0\to K^-\pi^+)$ (\%)           & $\chi^2$ / (d.o.f.) \\
fit        &                               &                                       & \\ \hline
$K^-\pi^+$ & PDG summer 2011 equivalent    & $3.913 \pm 0.022 \pm 0.043 $          & 6.0 / (8-1)\\
$K^-\pi^+$ & drop Ref.~\cite{Coan:1997ye}  & $3.921 \pm 0.023 \pm 0.044$           & 4.8 / (7-1)\\
$K^-\pi^+$ & add FSR corrections           & $3.940 \pm 0.023 \pm 0.041 \pm 0.015$ & 4.0 / (7-1)\\
$K^-\pi^+$ & add FSR correlations          & $3.940 \pm 0.023 \pm 0.041 \pm 0.025$ & 4.2 / (7-1)\\
all        & --                            & $3.946 \pm 0.023 \pm 0.040 \pm 0.025$ &11.6 /(13-3) \\
\hline
\end{tabular}
\end{table}

\begin{figure}
\begin{center}
\includegraphics[width=0.3\textwidth,angle=90.]{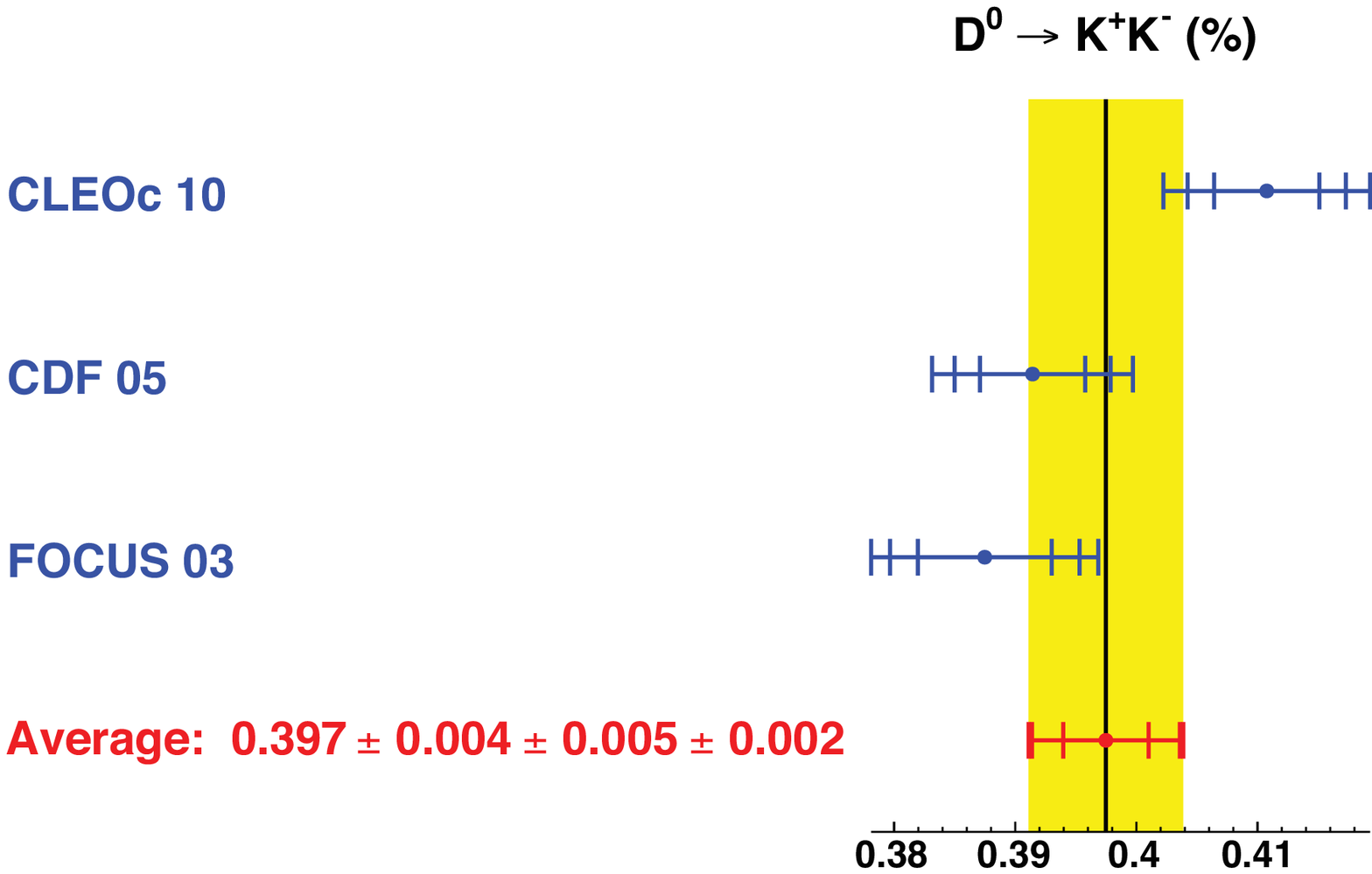}\hfill
\includegraphics[width=0.3\textwidth,angle=90.]{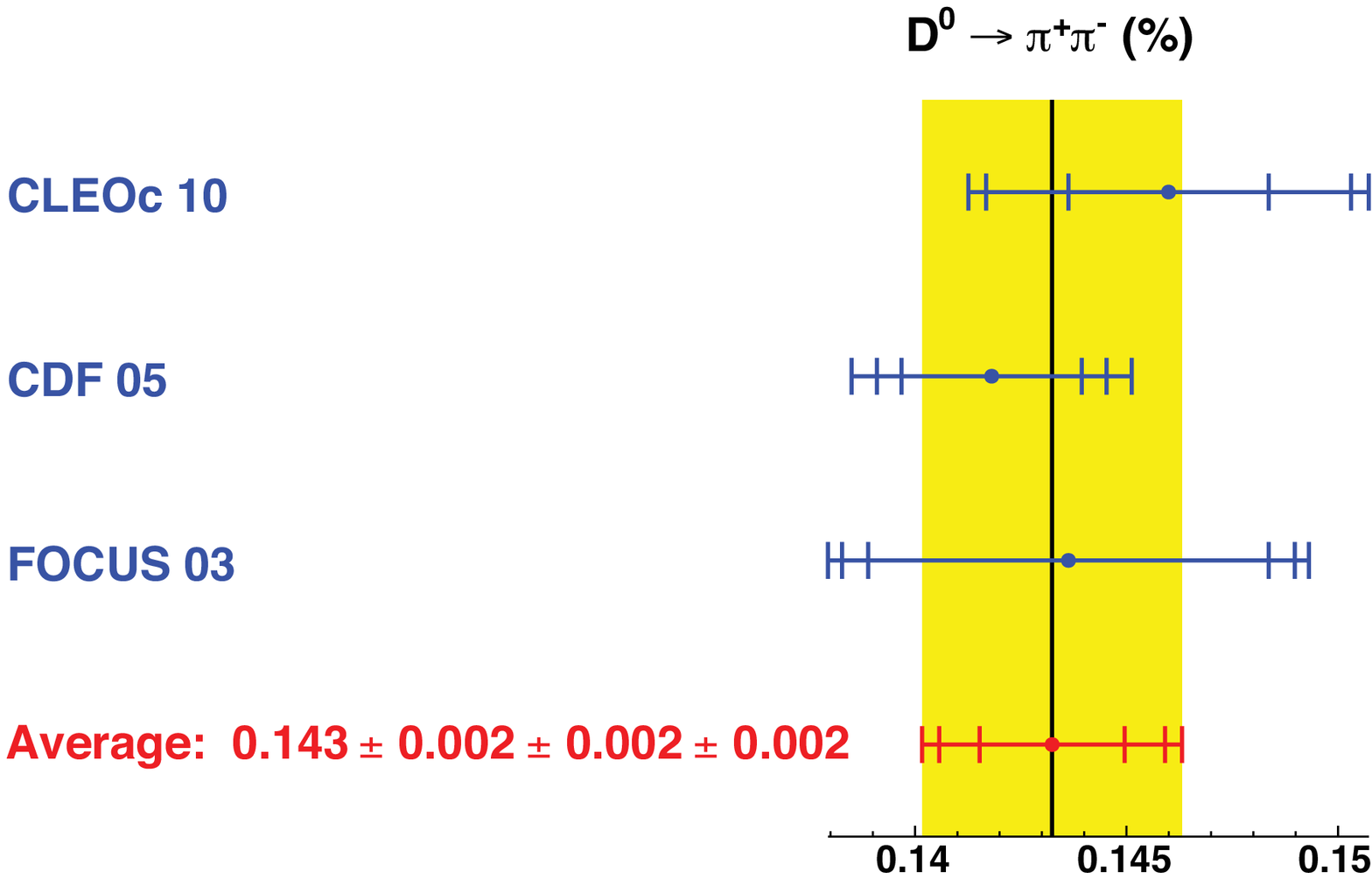}
\caption{The ${\cal B}(D^0\to K^+K^-)$ (left) and ${\cal B}(D^0\to \pi^+\pi^-)$ (right) 
values obtained by scaling the measured branching ratios with the ${\cal B}(D^0\to K^-\pi^+)$ branching fraction
average obtained here.  For the measurements (blue points), the error bars correspond to the statistical, systematic
and $K\pi$ normalization uncertainties.  The average obtained here (red point, yellow band) lists the statistical,
systematics excluding FSR, and the FSR systematic.
\label{fig:kkpipi}}
\end{center}
\end{figure}

As the $\chi^2$ would suggest and Fig.~\ref{D0bfs} shows, the average value for ${\cal B}(D^0\to K^-\pi^+)$ and
the input branching fractions agree very well.  With the estimated uncertainty in the FSR modeling used here,
the FSR uncertainty dominates the statistical uncertainty in the average, suggesting that experimental
work in the near future should focus on verification of FSR with 
$E_\gamma \simge 100$ MeV.  The ${\cal B}(D^0\to K^+K^-)$ and ${\cal B}(D^0\to \pi^+\pi^-)$ measurements inferred
from the branching ration measurements also agree well (Fig.~\ref{fig:kkpipi}).

The ${\cal B}(D^0\to K^-\pi^+)$ average obtained here is approximately one statistical standard deviation higher than the
2011 PDG update average \cite{PDG_2010}.  Table~\ref{tab:fit_evolution} shows the evolution from a
fit similar to the PDG's (no FSR corrections or correlations, reference~\cite{Coan:1997ye}
included) to the average presented here.  
There are two main contributions to the difference. The branching fraction in reference~\cite{Coan:1997ye} is
low, and its exclusion shifts the result upwards.  The FSR corrections  also
shift the result upwards, as expected, and  contribute the dominant shift of $+0.019\%$.  


\clearpage
\subsection{Direct \cp\ violation}
\label{sec:charm:cpvdir}

In decays of $D^0$ mesons, \cp\ asymmetry measurements have contributions from 
both direct and indirect \cp\ violation as discussed in Sec.~\ref{sec:charm:mixcpv}.
The contribution from indirect \cp\ violation depends on the decay-time distribution 
of the data sample~\cite{Kagan:2009gb}. This section describes a combination of 
measurements that allows the extraction of the individual contributions of the 
two types of \cp\ violation.
At the same time, the level of agreement for a no-\cp-violation hypothesis is 
tested. The observables are: 
\begin{equation}
A_{\Gamma} \equiv \frac{\tau(D^0 \ra h^+ h^-) - \tau(\dbar \ra h^+ h^- )}
{\tau(D^0 \ra h^+ h^-) + \tau(\dbar \ra h^+ h^- )},
\end{equation}
where $h^+ h^-$ can be $K^+ K^-$ or $\pi^+\pi^-$, and 
\begin{equation}
\Delta A_{\rm CP}   \equiv A_{\rm CP}(K^+K^-) - A_{\rm CP}(\pi^+\pi^-),
\end{equation}
where $A_{\rm CP}$ are time-integrated \cp\ asymmetries. The underlying 
theoretical parameters are: 
\begin{eqnarray}
a_{\rm CP}^{\rm dir} & \equiv & \frac{|A_{D^0\ra f} |^2 - |A_{\dbar\ra f} |^2} 
{|A_{D^0\ra f} |^2 + |A_{\dbar\ra f} |^2} ,\nonumber\\ 
a_{\rm CP}^{\rm ind}  & \equiv & \frac{1}{2} 
\left[ \left(\left|\frac{q}{p}\right| + \left|\frac{p}{q}\right|\right) x \sin \phi - 
\left(\left|\frac{q}{p}\right| - \left|\frac{p}{q}\right|\right) y \cos \phi \right] ,
\end{eqnarray}
where $A_{D\ra f}$ is the amplitude for $D\ra f$~\cite{Grossman:2006jg}. 
We use the following relations 
between the observables and the underlying parameters~\cite{Gersabeck:2011xj}: 
\begin{eqnarray}
A_{\Gamma} & = & - a_{\rm CP}^{\rm ind} - a_{\rm CP}^{\rm dir} y_{\rm CP},\nonumber\\ 
\Delta A_{\rm CP} & = &  \Delta a_{\rm CP}^{\rm dir} \left(1 + y_{\rm CP} 
\frac{\langle t\rangle}{\tau} \right)   +   
   a_{\rm CP}^{\rm ind} \frac{\Delta\langle t\rangle}{\tau}   +   
  a_{\rm CP}^{\rm dir} y_{\rm CP} \frac{\Delta\langle t\rangle}{\tau},\\ 
& ≈ & \Delta a_{\rm CP}^{\rm dir} \left(1 + y_{\rm CP} 
\frac{\langle t\rangle}{\tau} \right)   +   a_{\rm CP}^{\rm ind} 
\frac{\Delta\langle t\rangle}{\tau}.
\end{eqnarray}
The first relation constrains mostly indirect \cp\ violation, and the 
direct \cp\ violation contribution can differ for different final states. 
In the second relation, $\langle t\rangle/\tau$ denotes the mean decay 
time in units of the $D^0$ lifetime; $\Delta X$ denotes the difference 
in quantity $X$ between $K^+K^-$ and $\pi^+\pi^-$ final states; and $X$ 
denotes the average for quantity $X$. 
We neglect the last term in this relation as all three factors are 
$\mathcal{O}(10^{-2})$ or smaller, and thus this term is negligible 
with respect to the other two terms. 
Note that $\Delta\langle t\rangle/\tau \ll\langle t\rangle/\tau$, and 
it is expected that $|a_{\rm CP}^{\rm dir}| < |\Delta a_{\rm CP}^{\rm dir}|$ 
because $a_{\rm CP}^{\rm dir}(K^+K^-)$ and $a_{\rm CP}^{\rm dir}(\pi^+\pi^-)$ 
are expected to have opposite signs. 

A $\chi^2$ fit is performed in the plane $\Delta a_{\rm CP}^{\rm dir}$ 
vs. $a_{\rm CP}^{\rm ind}$. 
For the \babar result the difference of the quoted values for 
$A_{\rm CP}(K^+K^-)$ and $A_{\rm CP}(\pi^+\pi^-)$ is calculated, 
adding all uncertainties in quadrature. 
This may overestimate the systematic uncertainty for the difference 
as it neglects correlated errors; however, the result is conservative 
and the effect is small as all measurements are statistically limited. 
For all measurements, statistical and systematic uncertainties are added 
in quadrature when calculating the $\chi^2$. 
We use the current world average value $y_{\rm CP} = (1.064 \pm 0.209)\%$ 
(see Sec.~\ref{sec:charm:mixcpv}) and the measurements listed in 
Table~\ref{tab:charm:dir_indir_comb}. 

\begin{table}
\centering 
\caption{Inputs to the fit for direct and indirect \cp\ violation. 
The first uncertainty listed is statistical, and the second is systematic.}
\label{tab:charm:dir_indir_comb}
\vspace{3pt}
\begin{tabular}{ll|ccccc}
\hline \hline
Year & 	Experiment	& Results	
& $\Delta \langle t\rangle/\tau$ & $\langle t\rangle/\tau$ & Reference\\
\hline
2007	& Belle	& $A_\Gamma = (0.01 \pm 0.30 \pm 0.15 )\%$ &	-&	-&	 
\cite{Staric:2007dt}\\
2008	& \babar	& $A_\Gamma = (0.26 \pm 0.36 \pm 0.08 )\%$ &	-&	-&	 
\cite{Aubert:2007en}\\
2011	& LHCb	& $A_\Gamma = (−0.59 \pm 0.59 \pm 0.21 )\%$ &	-&	-&	 
\cite{Aaij:2011ad}\\
2008	& \babar	& $A_{\rm CP}(KK) = (0.00 \pm 0.34 \pm 0.13 )\%$&&&\\ 
& & $A_{\rm CP}(\pi\pi) = (−0.24 \pm 0.52 \pm 0.22 )\%$ &	$0.00$ &	
$1.00$ &	 \cite{Aubert:2007if}\\
2008	& Belle	& $\Delta A_{\rm CP} = (−0.86 \pm 0.60 \pm 0.07 )\%$ &	
$0.00$ &	$1.00$ &	 \cite{Staric:2008rx}\\
2011	& LHCb	& $\Delta A_{\rm CP} = (−0.82 \pm 0.21 \pm 0.11 )\%$ &	
$0.10$ &	$2.08$ &	 \cite{Aaij:2011in}\\
2012	& CDF Prelim.	 & $\Delta A_{\rm CP} = (−0.62 \pm 0.21 \pm 0.10 )\%$ &	
$0.25$ &	$2.58$ &	 \cite{cdf_public_note_10784}\\
\hline
\end{tabular}
\end{table}

The combination plot shows the measurements listed in 
Table~\ref{tab:charm:dir_indir_comb} for
$\Delta A_{\rm CP}$ and $A_\Gamma$, where the bands represent $\pm1\sigma$ 
intervals.  The point of no \cp\ violation (0,0) is shown as a filled circle, 
and two-dimensional $68\%$ CL, $95\%$ CL, and $99.7\%$ CL regions are plotted 
as ellipses. The best fit value is indicated by a cross showing the
one-dimensional errors.

\begin{figure}
\begin{center}
\includegraphics[width=0.90\textwidth]{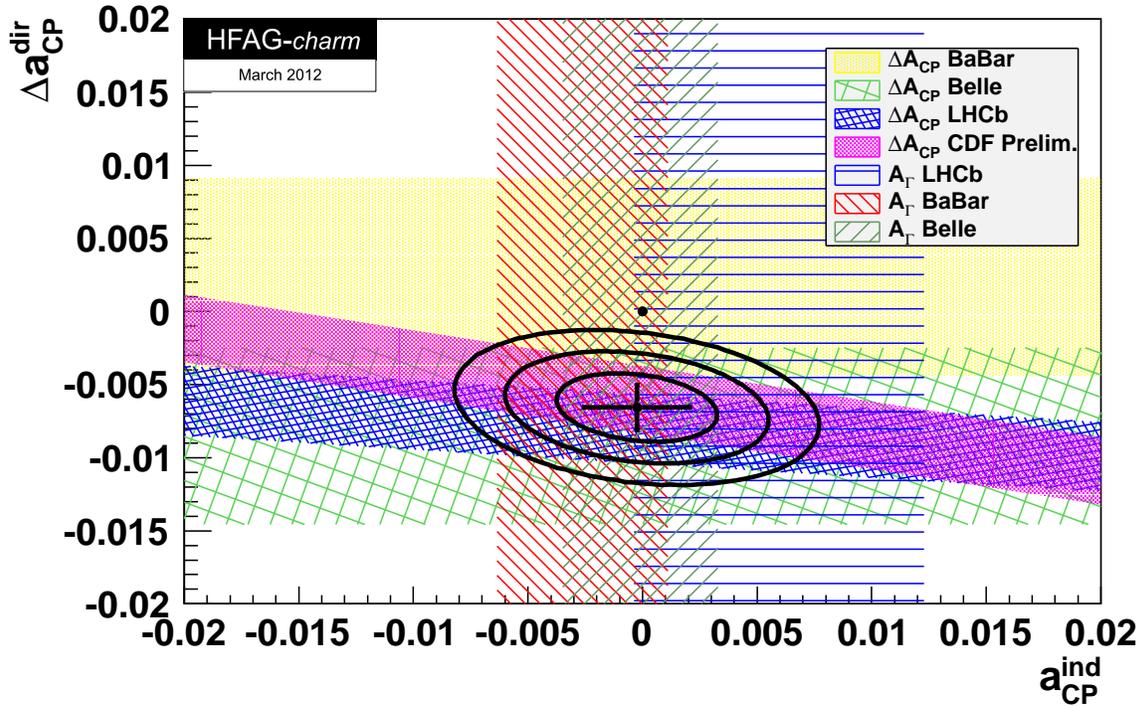}
\caption{Plot of all data and the fit result. Individual 
measurements are plotted as bands showing their $\pm1\sigma$ range. 
The no-\cpv\ point (0,0) is shown as a filled circle, and the best 
fit value is indicated by a cross showing the one-dimensional errors. 
Two-dimensional $68\%$ CL, $95\%$ CL, and $99.7\%$ CL regions are 
plotted as ellipses. }
\label{fig:charm:dir_indir_comb}
\end{center}
\end{figure}

From the fit, the change in $\chi^2$ from the minimum value for the no-\cpv\ 
point (0,0) is $19.4$, which corresponds to a CL of $6.1\times 10^{-5}$ for 
two degrees of freedom. Thus the data are consistent with the no-\cp-violation 
hypothesis at only $0.006\%$ CL. The central values and $\pm1\sigma$ errors for 
the individual parameters are
\begin{eqnarray}
a_{\rm CP}^{\rm ind} & = & (-0.025 \pm 0.231 )\% \nonumber\\
\Delta a_{\rm CP}^{\rm dir} & = & (-0.656 \pm 0.154 )\%.
\end{eqnarray}
These results indicate that the origin of this \cp\ violation lies in the 
difference between direct \cp\ violation in the two final states, rather 
than in a common indirect \cp\ violation.

\clearpage
\subsection{Charm baryons}

Here we summarizes present status of excited charm baryons, 
decaying strongly or electromagnetically: their masses (or 
mass difference between excited baryon and the corresponding 
ground state), natural widths, decay modes and presumably 
assigned quantum numbers. 
Table~\ref{sumtable1} summarizes the excited $\Lambda_c^+$'s.  
First two states, $\Lambda_c(2595)^+$ and $\Lambda_c(2625)^+$ 
are well established. 
Based on measured masses they are believed to be orbitally 
excited $\Lambda_c^+$'s with total momentum of light quarks 
L=1. Therefore, it's quantum numbers are assigned to be 
$J^P=(\frac{1}{2})^-$ and $J^P=(\frac{3}{2})^-$. Recently, 
their masses 
were precisely measured by CDF~\cite{Aaltonen:2011sf}: 
$M(\Lambda_c(2595)^+)=2592.25\pm 0.24\pm 0.14$~MeV/c$^2$, 
$M(\Lambda_c(2625)^+)=2628.11\pm 0.13\pm 0.14$~MeV/c$^2$. 
Next two states, $\Lambda_c(2765)^+$ and $\Lambda_c(2880)^+$, 
were discovered by CLEO~\cite{Artuso:2000xy} in $\Lambda_c^+\pi^+\pi^-$ 
final state. 
They found that $\Lambda_c(2880)^+$ decays also through the 
$\Sigma_c(2445)^{++/0}\pi^{-/+}$ mode. 
Later, \babar~\cite{Aubert:2006sp} 
observed that this state has also $D^0 p$ decay mode. It is the 
first example where excited charm baryon decays into charm meson 
and light baryon. (Usually, excited charm baryons decay onto charm 
baryon and light mesons.) In that analysis \babar 
observed for the first time else one state, $\Lambda_c(2940)^+$, 
decaying into $D^0 p$. By looking at $D^+ p$ final state, they found 
no signals which results in the conclusion that the $\Lambda_c(2880)^+$ 
and $\Lambda_c(2940)^+$ are really 
$\Lambda_c^+$ excited states, not $\Sigma_c$. 
Belle reported the result of an angular analysis that favors the $5/2$ 
for the $\Lambda_c(2880)^+$ spin hypothesis. 
Moreover, the measured ratio of branching fractions 
${\cal B}(\Lambda_c(2880)^+\rightarrow \Sigma_c(2520)\pi^{\pm})/{\cal B}(\Lambda_c(2880)^+\rightarrow \Sigma_c(2455)\pi^{\pm})=(0.225\pm 0.062\pm 0.025)$ combined 
with theoretical predictions based on HQS~\cite{Isgur:1991wq,Cheng:2006dk} 
favoring even parity.     

The open questions in the present excited $\Lambda_c^+$ family are the 
experimental determination of quantum numbers for almost all states and 
the nature of $\Lambda_c(2765)^+$ state: whether it is excited $\Sigma_c^+$ 
or $\Lambda_c^+$.

\begin{table}[t]
\caption{Summary of excited $\Lambda_c^+$ baryons family.} 
\resizebox{0.99\textwidth}{!}{
\begin{tabular}{c|c|c|c|c}
Charmed Baryon   & Mode  & Mass  or $\Delta M$, & Natural Width,  & $J^P$  \\
Excited State &  &  MeV/c$^2$ & MeV/c$^2$  \\
\hline
$\Lambda_c(2595)^+$ & $\Lambda_c^+\pi^+\pi^-$, $\Sigma_c\pi$ &  $2595.4\pm 0.6$ & $3.6^{+2.0}_{-1.3}$  & $1/2^-$  \\
\hline
$\Lambda_c(2625)^+$ & $\Lambda_c^+\pi^+\pi^-$, $\Sigma_c\pi$ & $2628.1\pm 0.6$ & $<1.9$ & $3/2^-$  \\
\hline
$\Lambda_c(2765)^+$ & $\Lambda_c^+\pi^+\pi^-$, $\Sigma_c\pi$ & $2766.6\pm 2.4$ & $50$ & ??  \\
\hline
$\Lambda_c(2880)^+$ & $\Lambda_c^+\pi^+\pi^-$, $\Sigma_c\pi$,  &$2881.53\pm 0.35$ & $5.8\pm 1.1$ & $5/2^+$ \\
 &  $\Sigma_c(2520)\pi$, $D^0p$     & & & (experimental evidence) \\
\hline
$\Lambda_c(2940)^+$ & $D^0p$, $\Sigma_c\pi$ & $2939.3^{+1.4}_{-1.5}$ & $17^{+8}_{-6}$  & ??  \\
\end{tabular}
}
\label{sumtable1} 
\end{table}

Table~\ref{sumtable2} summarizes the excited $\Sigma_c^{++,+,0}$ baryons.
Triplet of $\Sigma_c(2520)^{++,+,0}$ baryons is well established. 
Recently CDF~\cite{Aaltonen:2011sf} precisely measured the masses 
and widths of the double charged and neutral members of this triplet 
to be $M(\Sigma_c(2520)^{++})=(2517.19\pm 0.46\pm 0.14)$~MeV/c$^2$, 
$\Gamma(\Sigma_c(2520)^{++})=(15.03\pm 2.52)$~MeV/c$^2$ and 
$M(\Sigma_c(2520)^{0})=(2519.34\pm 0.58\pm 0.14)$~MeV/c$^2$, 
$\Gamma(\Sigma_c(2520)^{0})=(12.51\pm 2.28)$~MeV/c$^2$, respectively. 
The short list of excited $\Sigma_c$ baryons completes the triplet 
of $\Sigma_c(2800)$ states observed by Belle~\cite{Mizuk:2004yu}. Based 
on measured mass and theoretical predictions~\cite{Copley:1979wj,Pirjol:1997nh} 
one can tentatively identify these states as members of the predicted 
$\Sigma_{c2}$ $3/2^-$ triplet. From the study of resonant substructure 
in $B^-\rightarrow \Lambda_c^+\bar{p}\pi^-$ decays, \babar found 
significant signal in $\Lambda_c^+\pi^-$ with the mean value 
higher by about $3~\sigma$ from obtained by Belle (see 
Table~\ref{sumtable2}). The widths from measurements, 
Belle and \babar, are consistent.

\begin{table}[!htb]
\caption{Summary of excited $\Sigma_c^{++,+,0}$ baryons family.} 
 \begin{tabular}{c|c|c|c|c}
Charmed Baryon   & Mode  & Mass  or $\Delta M$, & Natural Width,  & $J^P$  \\
Excited State &  &  MeV/c$^2$ & MeV/c$^2$  \\
\hline
$\Sigma_c(2520)^{++}$ &$\Lambda_c^+\pi^+$  & $231.9\pm 0.6$ & $14.9\pm 1.9$ & $3/2^+$   \\
$\Sigma_c(2520)^{+}$ &$\Lambda_c^+\pi^+$  & $231.0\pm 2.3$ & $<17$~@~90$\%$~CL & $3/2^+$ \\
$\Sigma_c(2520)^{0}$ &$\Lambda_c^+\pi^+$  & $231.6\pm 0.5$ & $16.1\pm 2.1$ & $3/2^+$    \\
\hline
$\Sigma_c(2800)^{++}$ & $\Lambda_c^+\pi^{+}$ & $514.5^{+3.4+2.8}_{-3.1-4.9}$ & $75^{+18+12}_{-13-11}$ & tentatively identified      \\
$\Sigma_c(2800)^{+}$ & $\Lambda_c^+\pi^{0}$&$505.4^{+5.8+12.4}_{-4.6-2.0}$ &$62^{+37+52}_{-23-38}$ & as members of the predicted  \\
$\Sigma_c(2800)^{0}$ & $\Lambda_c^+\pi^{-}$&$515.4^{+3.2+2.1}_{-3.1-6.0}$ & $61^{+18+22}_{-13-13}$ &$\Sigma_{c2}$ $3/2^-$ isospin triplet  \\
 & $\Lambda_c^+\pi^{-}$ & $560\pm 8\pm 10$ & $86^{+33}_{-22}$  \\

\end{tabular}
\label{sumtable2} 
\end{table}

Table~\ref{sumtable3} summarizes the excited $\Xi_c^{+,0}$ and $\Omega_c^0$ 
baryons. Recently, the list of excited $\Xi_c$ baryons have enriched by 
several states with masses above 2900~MeV/c$^2$ and decaying into 
$\Lambda_c^+ K^-$  and $\Lambda_c^+ K^{-/0}\pi^{+/-}$. 
Some of these states are seen by both Belle~\cite{Chistov:2006zj} 
and \babar~\cite{Aubert:2007dt} and are believed to be well-established, 
these are $\Xi_c(2980)^+$ and $\Xi_c(3080)^{+,0}$. 
All others need to be confirmed or studied in more depth. 
These are $\Xi_c(2930)^0$ seen in $\Lambda_c^+ K^-$ final state, 
$\Xi_c(3055)^+$ found in $\Sigma_c(2455)^{++}\pi^-$ final state, 
and $\Xi_c(3123)^+$ claimed by \babar~\cite{Aubert:2007dt} in 
$\Sigma_c(2520)^{++}\pi^-$ final state.   

The excited $\Omega_c^0$ double charm baryon are seen by both 
\babar~\cite{Aubert:2006je} and Belle~\cite{Solovieva:2008fw}, the 
$\delta M=M(\Omega_c^{*0})-M(\Omega_c^0)$ are in good agreement 
agreement in both experiments and consistent with most theoretical 
predictions~\cite{Rosner:1995yu,Glozman:1995xy,Jenkins:1996de,
Burakovsky:1997vm}.

\begin{table}[b]
\caption{Summary of excited $\Xi_c^{+,0}$ and $\Omega_c^0$ baryons families.} 
\resizebox{0.99\textwidth}{!}{
\begin{tabular}{c|c|c|c|c}
Charmed Baryon   & Mode  & Mass  or $\Delta M$, & Natural Width,  & $J^P$  \\
Excited State &  &  MeV/c$^2$ & MeV/c$^2$  \\
\hline
$\Xi_c'^+$ & $\Xi_c^+\gamma$ & $2575.6\pm 3.1$  &  & $1/2^+$    \\
$\Xi_c'^0$ & $\Xi_c^0\gamma$ & $2577.9\pm 2.9$  &  & $1/2^+$   \\
\hline
$\Xi_c(2645)^+$ & $\Xi_c^0\pi^+$ & $2645.9^{+0.6}_{-0.5}$  & $<3.1$ & $3/2^+$   \\
$\Xi_c(2645)^0$ & $\Xi_c^+\pi^-$ & $2645.9\pm 0.5$  & $<5.5$ & $3/2^+$   \\
\hline
$\Xi_c(2790)^+$ &$\Xi_c'^0\pi^+$ & $2789.1\pm 3.2$ & $<15$ & $1/2^-$   \\
$\Xi_c(2790)^0$ &$\Xi_c'^+\pi^-$ & $2791.8\pm 3.3$ & $<12$ & $1/2^-$   \\
\hline
$\Xi_c(2815)^+$ &$\Xi_c^+\pi^+\pi^-$, $\Xi_c(2645)^0\pi^+$ & $2816.6\pm 0.9$ & $<3.5$ &  $3/2^-$  \\
$\Xi_c(2815)^0$ & $\Xi_c^0\pi^+\pi^-$, $\Xi_c(2645)^+\pi^-$& $2819.6\pm 1.2$  & $<6.5$  & $3/2^-$   \\
\hline
$\Xi_c(2930)^0$ & $\Lambda_c^+ K^-$ & $2931.6\pm 6$ & $36\pm 13$ & ??     \\
\hline
$\Xi_c(2980)^+$ & $\Lambda_c^+K^-\pi^+$, $\Sigma_c^{++}K^-$, $\Xi_c(2645)^0\pi^+$
 &  $2971.4\pm 3.3$  & $26\pm 7$ & ??     \\
$\Xi_c(2980)^0$ & $\Xi_c(2645)^+\pi^-$
&  $2968.0\pm 2.6$ &$20\pm 7$ & ??       \\
\hline
$\Xi_c(3055)^+$ & $\Sigma_c^{++}K^-$ & 	$3054.2\pm 1.3$  &     	$17\pm 13$ & ??   \\
\hline
$\Xi_c(3080)^+$ & $\Lambda_c^+K^-\pi^+$, $\Sigma_c^{++}K^-$, $\Sigma_c(2520)^{++}K^-$ & $3077.0\pm 0.4$ & $5.8\pm 1.0$ & ??   \\
$\Xi_c(3080)^0$ &$\Lambda_c^+ K^0_S\pi^-$, $\Sigma_c^0K^0_S$, $\Sigma_c(2520)^{0}K^0_S$ & $3079.9\pm 1.4$ & $5.6\pm 2.2$ & ??   \\
\hline
$\Xi_c(3123)^+$ &$\Sigma_c(2520)^{++}K^-$ & $3122.9\pm 1.3$ & $4\pm 4$ & ??   \\
\hline
\\
\hline
$\Omega_c(2770)^0$ & $\Omega_c^0\gamma$& $2765.9\pm 2.0$  & $70.7^{+0.8}_{-0.9}$ & $3/2^+$  \\
\end{tabular}
}
\label{sumtable3} 
\end{table}

\begin{figure}[!htb]
\includegraphics[width=1.0\textwidth]{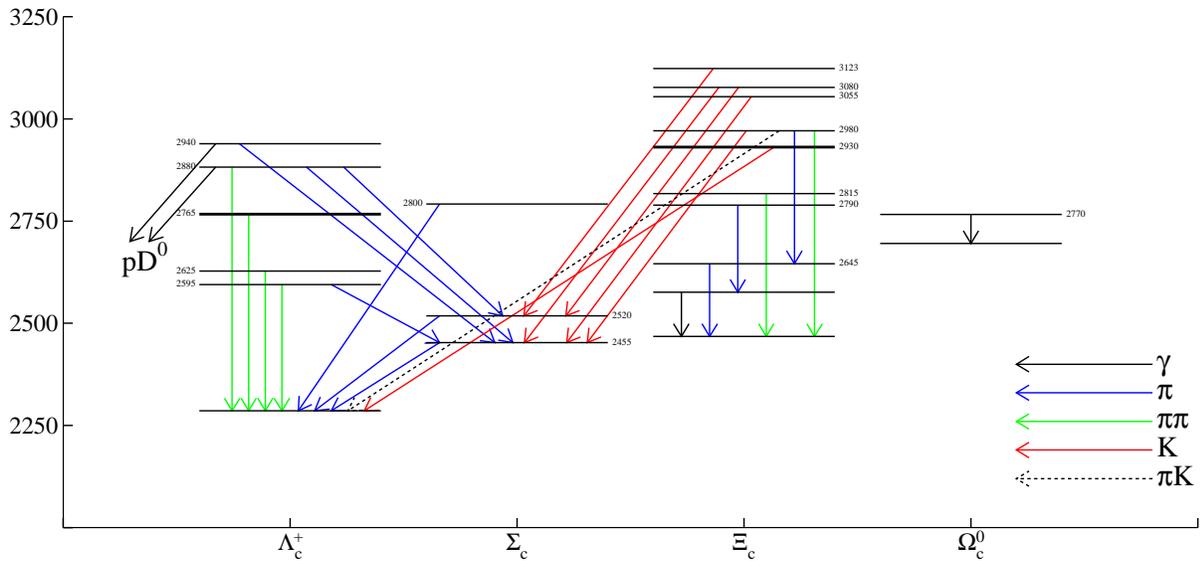}
 \caption{Level diagram for excited charm baryons.}  
\label{leveldiagram}
\end{figure}

Figure~\ref{leveldiagram} shows the levels of excited charm baryons with 
the corresponding transitions between them or to the charm baryon ground 
states. Interesting feature recently discovered by \babar and Belle is that 
now we know that transitions between families are possible (between 
$\Xi_c$ and $\Lambda_c^+$ families of excited baryons). Also, highly 
excited $\Lambda_c^+$ baryons can decay into charm meson and proton.

\clearpage
\subsection{Rare and forbidden decays}
\label{sec:charm:rare}

\begin{figure}
\begin{center}
\includegraphics[width=6.0in]{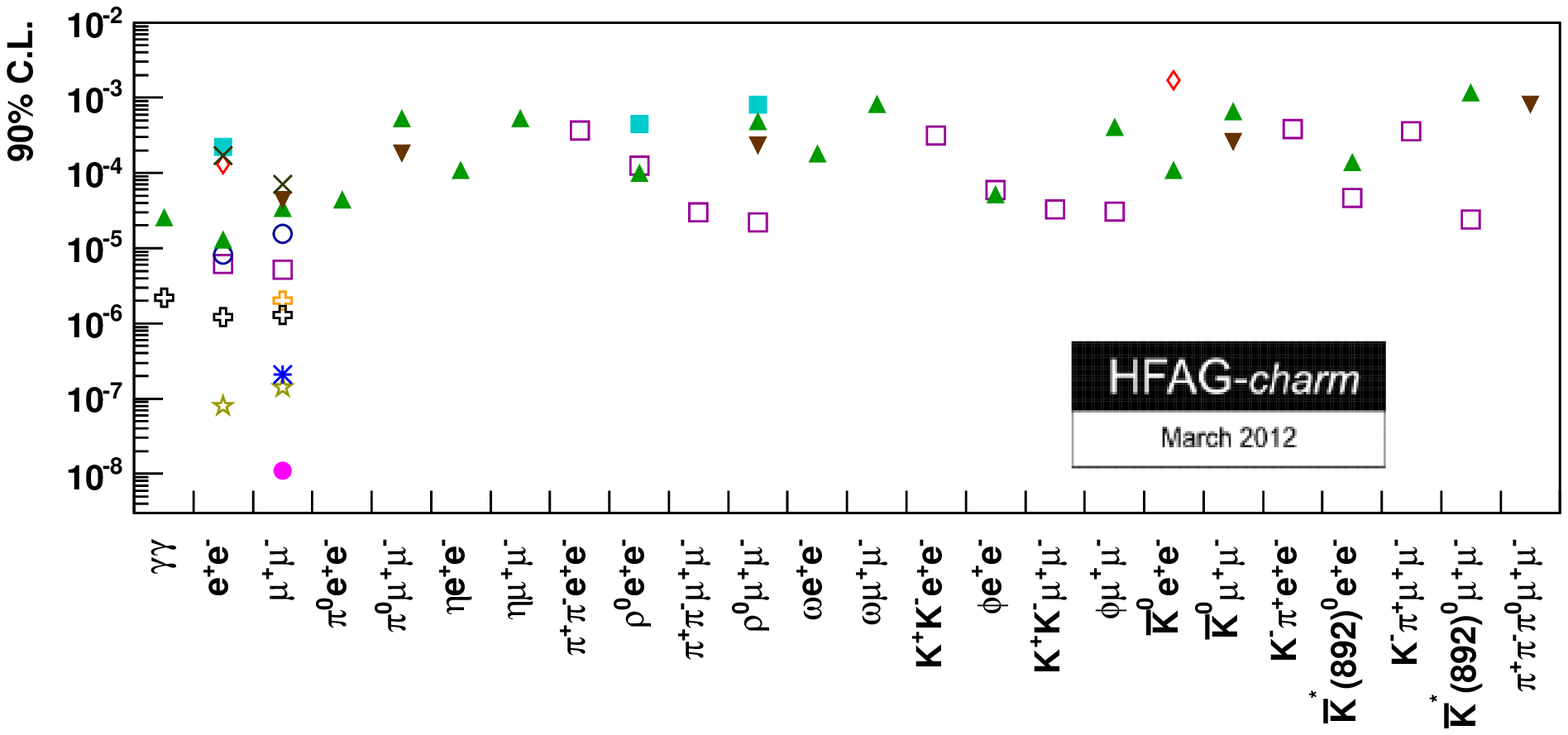}
\vskip-0.10in
\includegraphics[width=6.0in]{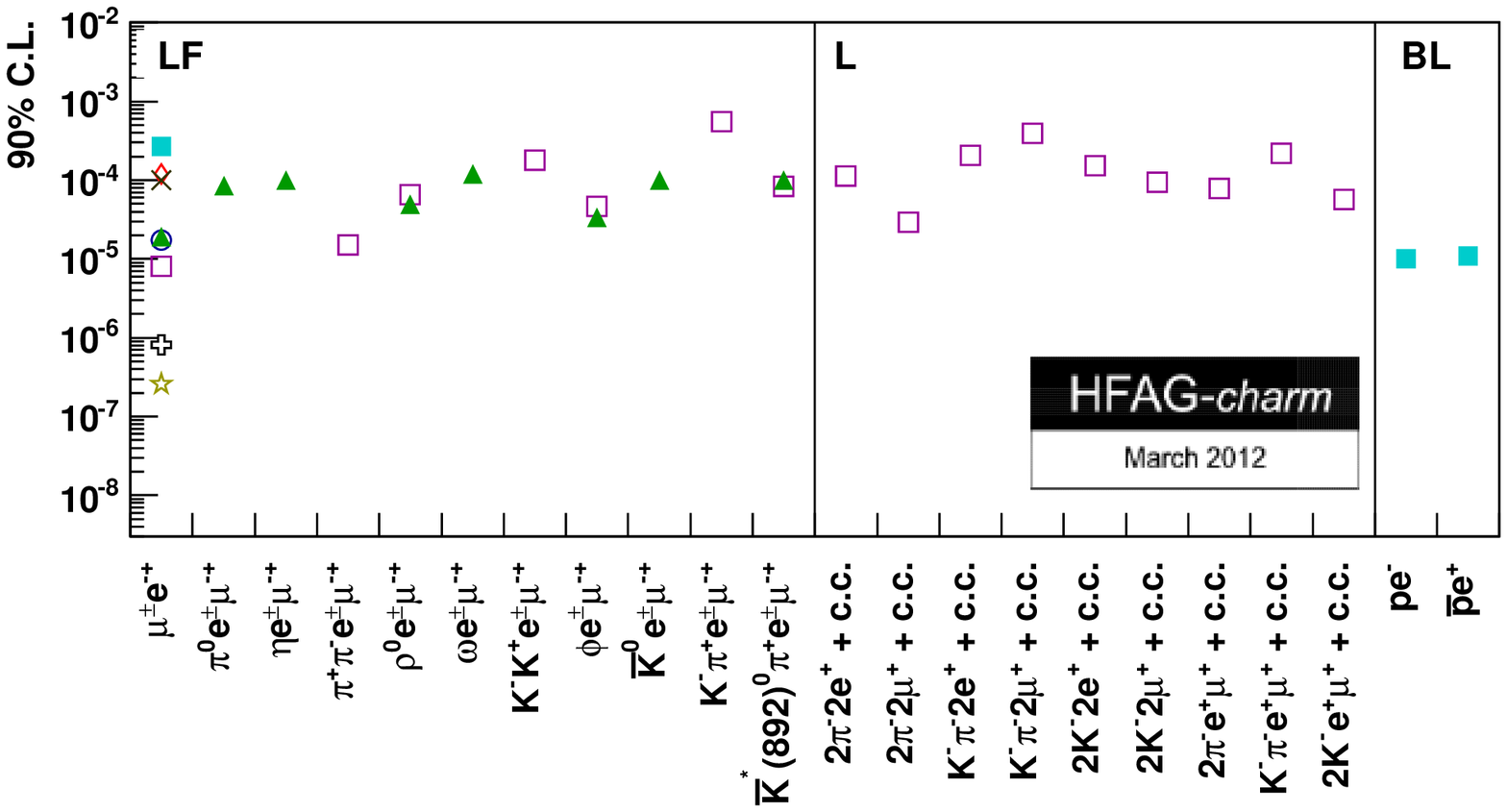}
\caption{Upper limits at $90\%$ CL for $D^0$ decays. The top plot
shows flavor-changing neutral current decays, and the bottom plot
shows lepton-flavor-changing (LF), lepton-number-changing (L), and 
both baryon- and lepton-number-changing (BL) decays.
The legend is given in Fig.~\ref{fig:charm:lambdac}.}
\label{fig:charm:rare_d0}
\end{center}
\end{figure}

\begin{figure}
\begin{center}
\includegraphics[width=5.0in]{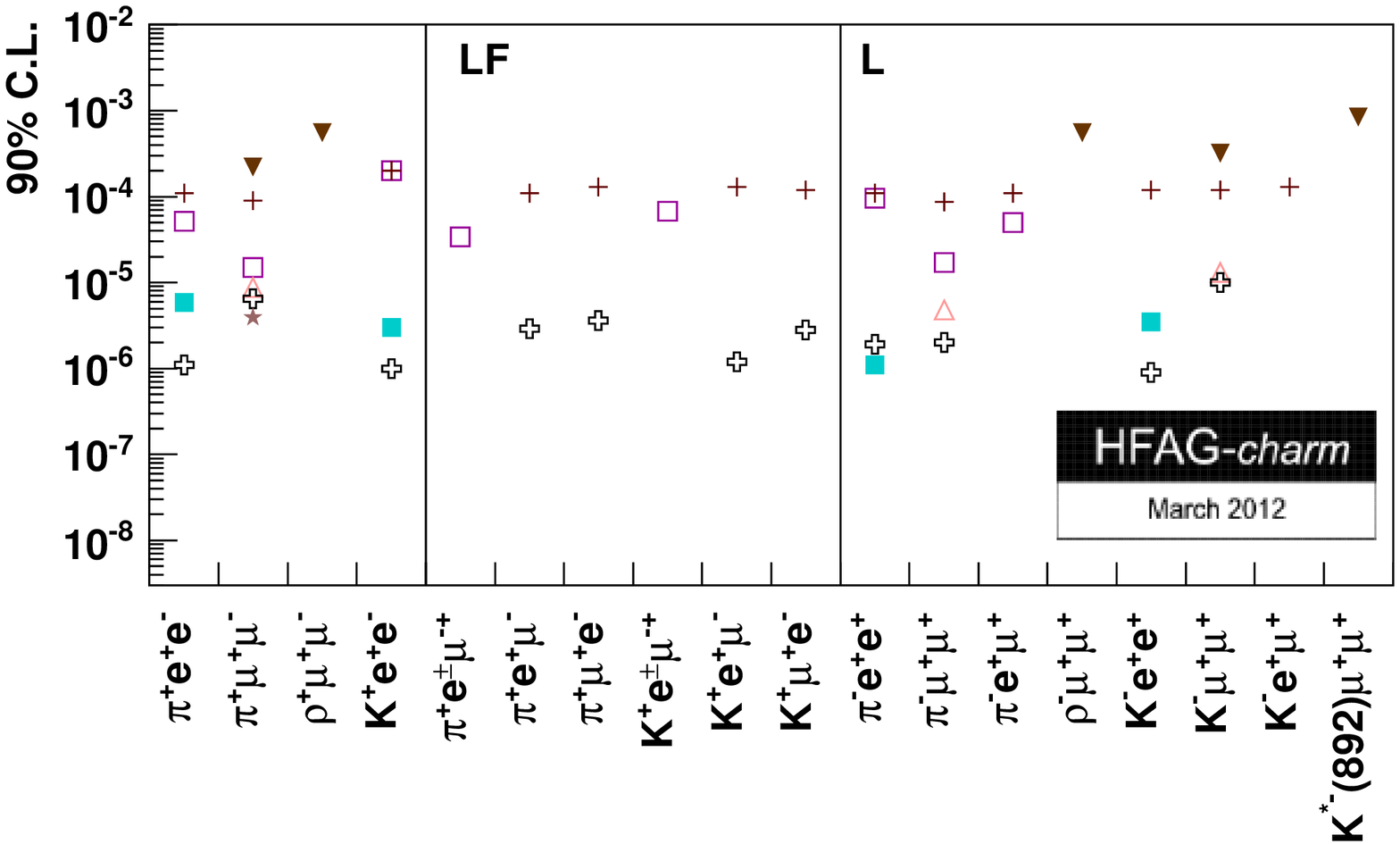}
\vskip0.10in
\includegraphics[width=5.0in]{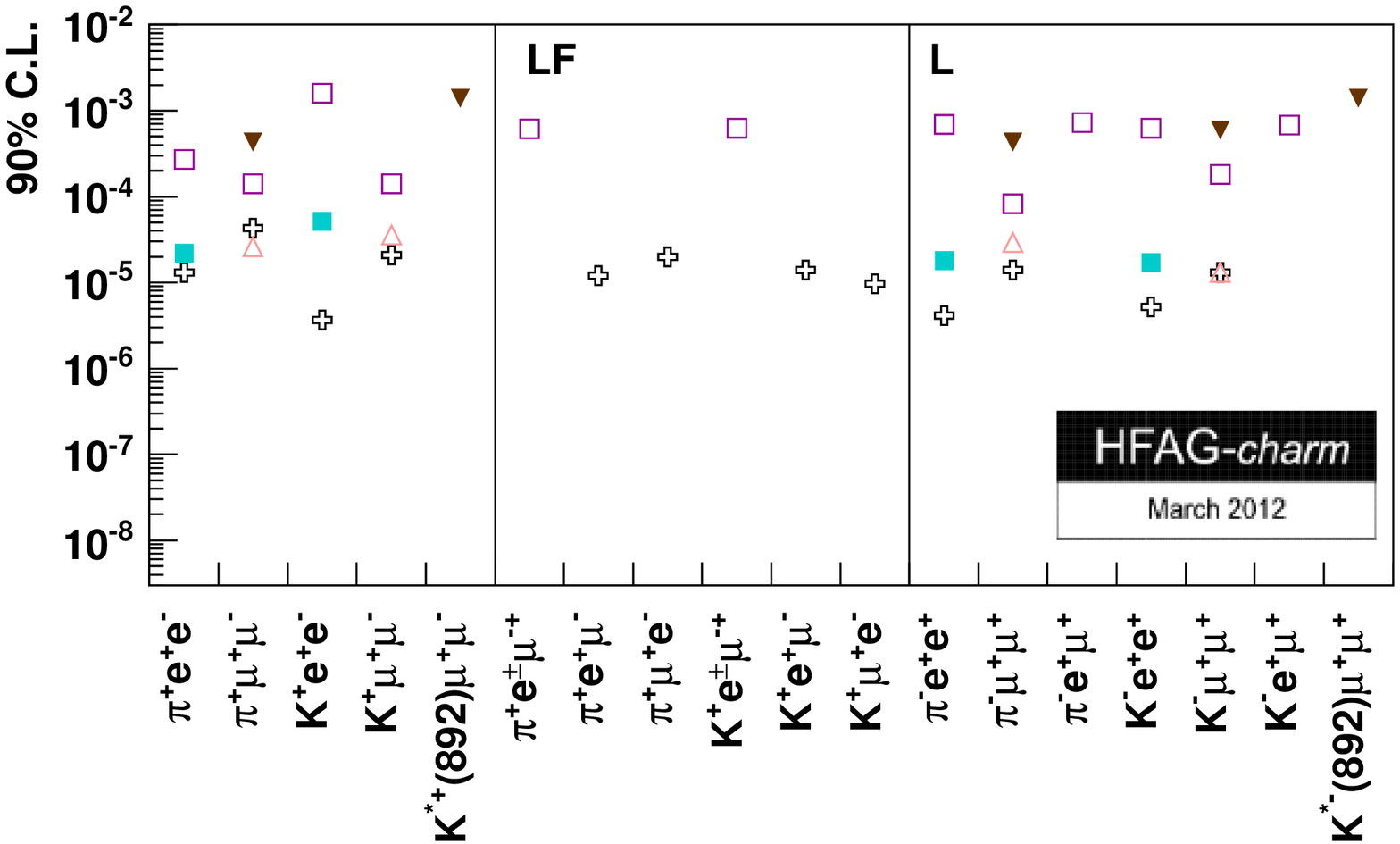}
\caption{Upper limits at $90\%$ CL for $D^+$ (top) and $D_s^+$ (bottom) 
decays. Each plot shows flavor-changing neutral current decays, 
lepton-flavor-changing decays (LF), and lepton-number-changing (L) decays. 
The legend is given in Fig.~\ref{fig:charm:lambdac}.}
\label{fig:charm:rare_charged}
\end{center}
\end{figure}

\begin{figure}
\begin{center}
\hbox{
\includegraphics[width=3.0in]{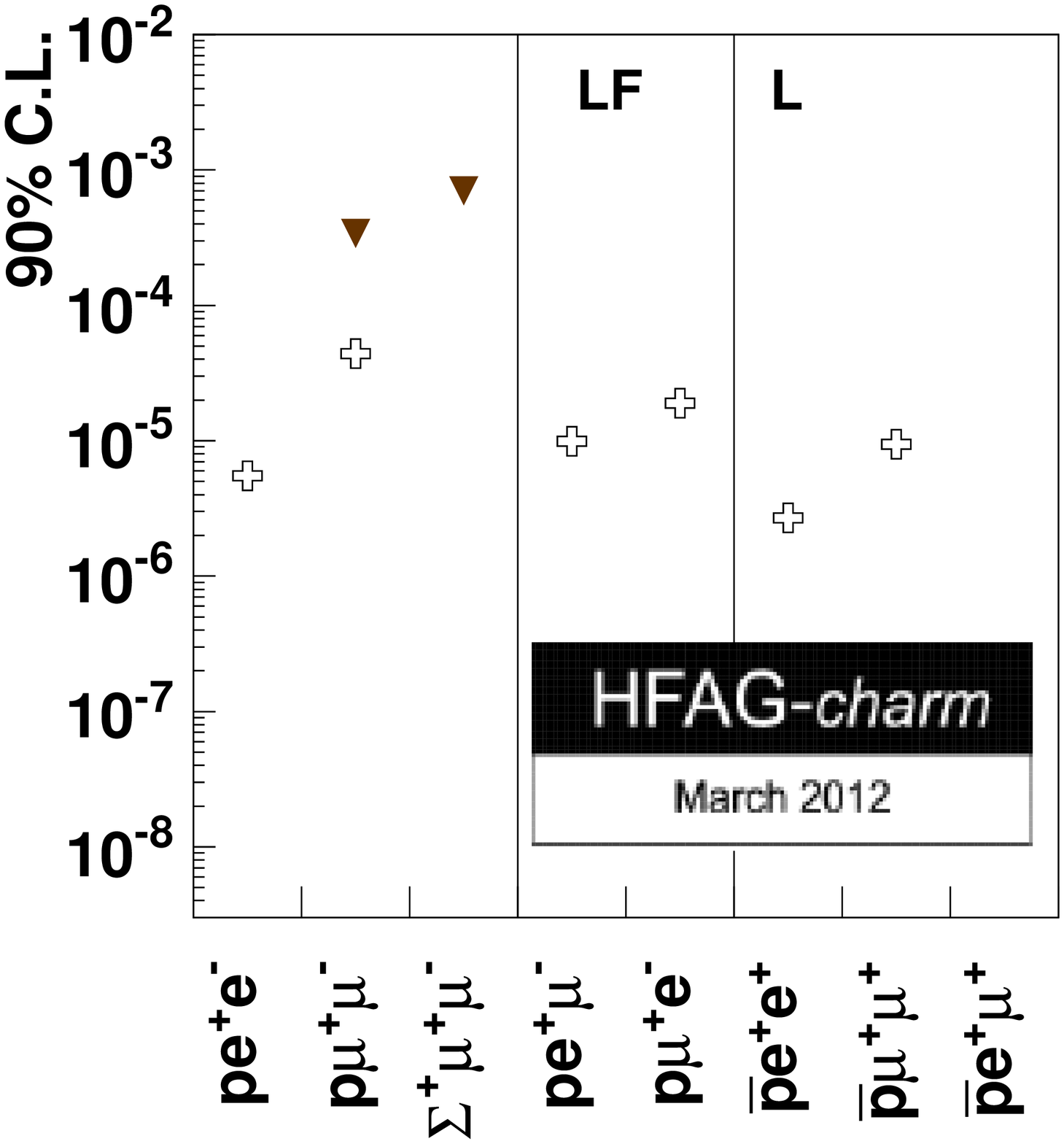}
\hskip-1.80in
\vbox{
\includegraphics[width=2.5in]{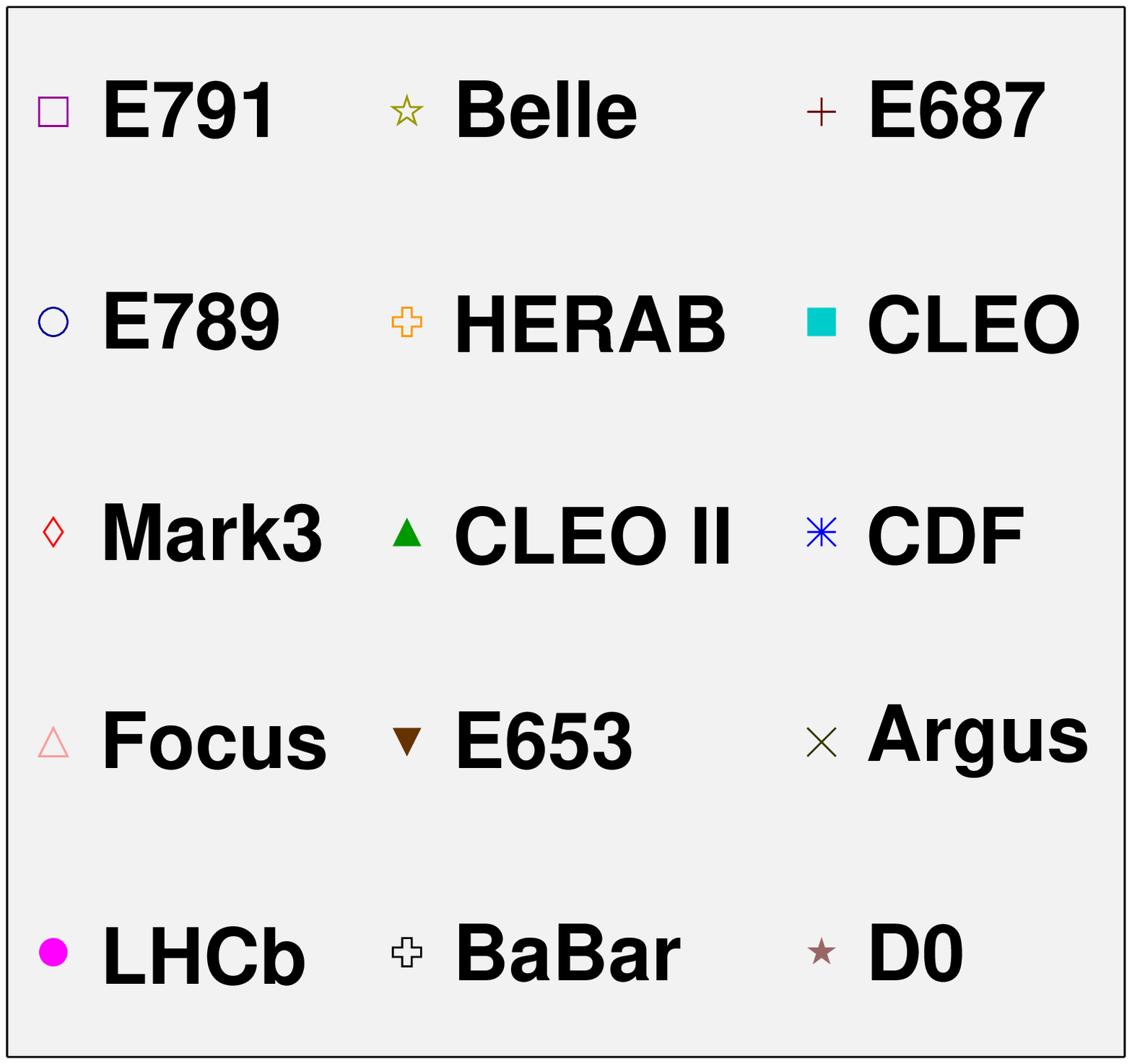}
\vskip1.2in
}}
\vskip-0.20in
\caption{Upper limits at $90\%$ CL for $\Lambda_c^+$ decays. Shown are 
flavor-changing neutral current decays, lepton-flavor-changing (LF) 
decays, and lepton-number-changing (L) decays. }
\label{fig:charm:lambdac}
\end{center}
\end{figure}

This section provides a summary of rare and forbidden charm decays
in tabular form. The decay modes can be categorized as 
flavor-changing neutral currents, lepton-flavor-violating, 
lepton-number-violating, and both baryon- and lepton-number-violating decays.
Figures~\ref{fig:charm:rare_d0}-\ref{fig:charm:lambdac} plot the 
upper limits for $D^0$, $D^+$, $D_s^+$, and $\Lambda_c^+$ decays. 
Tables~\ref{tab:charm:rare_d0}-\ref{tab:charm:rare_lambdac} give the 
corresponding numerical results. Some theoretical predictions are given in 
Refs.~\cite{Burdman:2001tf,Fajfer:2002bu,Fajfer:2007dy,Golowich:2009ii,Paul:2010pq,Borisov:2011aa}.

\begin{longtable}{l|ccc}
\caption{Upper limits at $90\%$ CL for $D^0$ decays.
\label{tab:charm:rare_d0}
}\\

\hline\hline
Decay & Limit $\times10^6$ & Experiment & Reference\\
\endfirsthead
\multicolumn{4}{c}{\tablename\ \thetable{} -- continued from previous page} \\ \hline
Decay & Limit $\times10^6$ & Experiment & Reference\\
\endhead
\hline
$\gamma{}\gamma{}$ & 26.0 & CLEO II & \cite{Coan:2002te}\\
& 2.2 & \babar Preliminary & \cite{Lees:2011qz}\\
\hline
$e^+e^-$ & 220.0 & CLEO & \cite{Haas:1988bh}\\
& 170.0 & Argus & \cite{Albrecht:1988ge}\\
& 130.0 & Mark3 & \cite{Adler:1987cp}\\
& 13.0 & CLEO II & \cite{Freyberger:1996it}\\
& 8.19 & E789 & \cite{Pripstein:1999tq}\\
& 6.2 & E791 & \cite{Aitala:1999db}\\
& 1.2 & \babar & \cite{Aubert:2004bs}\\
& 0.079 & Belle & \cite{Petric:2010yt}\\
\hline
$\mu{}^+\mu{}^-$ & 70.0 & Argus & \cite{Albrecht:1988ge}\\
& 44.0 & E653 & \cite{Kodama:1995ia}\\
& 34.0 & CLEO II & \cite{Freyberger:1996it}\\
& 15.6 & E789 & \cite{Pripstein:1999tq}\\
& 5.2 & E791 & \cite{Aitala:1999db}\\
& 2.0 & HERAb & \cite{Abt:2004hn}\\
& 1.3 & \babar & \cite{Aubert:2004bs}\\
& 0.21 & CDF & \cite{Aaltonen:2010hz}\\
& 0.14 & Belle & \cite{Petric:2010yt}\\
& 0.011 & LHCb Preliminary & \cite{LHCb-CONF-2012-005}\\
\hline
$\pi{}^0e^+e^-$ & 45.0 & CLEO II & \cite{Freyberger:1996it}\\
\hline
$\pi{}^0\mu{}^+\mu{}^-$ & 540.0 & CLEO II & \cite{Freyberger:1996it}\\
& 180.0 & E653 & \cite{Kodama:1995ia}\\
\hline
$\eta{}e^+e^-$ & 110.0 & CLEO II & \cite{Freyberger:1996it}\\
\hline
$\eta{}\mu{}^+\mu{}^-$ & 530.0 & CLEO II & \cite{Freyberger:1996it}\\
\hline
$\pi{}^+\pi{}^-e^+e^-$ & 370.0 & E791 & \cite{Aitala:2000kk}\\
\hline
$\rho{}e^+e^-$ & 450.0 & CLEO & \cite{Haas:1988bh}\\
& 124.0 & E791 & \cite{Aitala:2000kk}\\
& 100.0 & CLEO II & \cite{Freyberger:1996it}\\
\pagebreak
\hline
$\pi{}^+\pi{}^-\mu{}^+\mu{}^-$ & 30.0 & E791 & \cite{Aitala:2000kk}\\
\hline
$\rho{}\mu{}^+\mu{}^-$ & 810.0 & CLEO & \cite{Haas:1988bh}\\
& 490.0 & CLEO II & \cite{Freyberger:1996it}\\
& 230.0 & E653 & \cite{Kodama:1995ia}\\
& 22.0 & E791 & \cite{Aitala:2000kk}\\
\hline
$\omega{}e^+e^-$ & 180.0 & CLEO II & \cite{Freyberger:1996it}\\
\hline
$\omega{}\mu{}^+\mu{}^-$ & 830.0 & CLEO II & \cite{Freyberger:1996it}\\
\hline
$K^+K^-e^+e^-$ & 315.0 & E791 & \cite{Aitala:2000kk}\\
\hline
$\phi{}e^+e^-$ & 59.0 & E791 & \cite{Aitala:2000kk}\\
& 52.0 & CLEO II & \cite{Freyberger:1996it}\\
\hline
$K^+K^-\mu{}^+\mu{}^-$ & 33.0 & E791 & \cite{Aitala:2000kk}\\
\hline
$\phi{}\mu{}^+\mu{}^-$ & 410.0 & CLEO II & \cite{Freyberger:1996it}\\
& 31.0 & E791 & \cite{Aitala:2000kk}\\
\hline
$\overline{K}^0e^+e^-$ & 1700.0 & Mark3 & \cite{Adler:1988es}\\
& 110.0 & CLEO II & \cite{Freyberger:1996it}\\
\hline
$\overline{K}^0\mu{}^+\mu{}^-$ & 670.0 & CLEO II & \cite{Freyberger:1996it}\\
& 260.0 & E653 & \cite{Kodama:1995ia}\\
\hline
$K^-\pi{}^+e^+e^-$ & 385.0 & E791 & \cite{Aitala:2000kk}\\
\hline
$\overline{K}^{*0}(892)e^+e^-$ & 140.0 & CLEO II & \cite{Freyberger:1996it}\\
& 47.0 & E791 & \cite{Aitala:2000kk}\\
\hline
$K^-\pi{}^+\mu{}^+\mu{}^-$ & 360.0 & E791 & \cite{Aitala:2000kk}\\
\hline
$\overline{K}^{*0}(892)\mu{}^+\mu{}^-$ & 1180.0 & CLEO II & \cite{Freyberger:1996it}\\
& 24.0 & E791 & \cite{Aitala:2000kk}\\
\hline
$\pi{}^+\pi{}^-\pi{}^0\mu{}^+\mu{}^-$ & 810.0 & E653 & \cite{Kodama:1995ia}\\
\pagebreak
\hline
$\mu{}^{\pm}e^{\mp}$ & 270.0 & CLEO & \cite{Haas:1988bh}\\
& 120.0 & Mark3 & \cite{Becker:1987mu}\\
& 100.0 & Argus & \cite{Albrecht:1988ge}\\
& 19.0 & CLEO II & \cite{Freyberger:1996it}\\
& 17.2 & E789 & \cite{Pripstein:1999tq}\\
& 8.1 & E791 & \cite{Aitala:1999db}\\
& 0.81 & \babar & \cite{Aubert:2004bs}\\
& 0.26 & Belle & \cite{Petric:2010yt}\\
\hline
$\pi{}^0e^{\pm}\mu{}^{\mp}$ & 86.0 & CLEO II & \cite{Freyberger:1996it}\\
\hline
$\eta{}e^{\pm}\mu{}^{\mp}$ & 100.0 & CLEO II & \cite{Freyberger:1996it}\\
\hline
$\pi{}^+\pi{}^-e^{\pm}\mu{}^{\mp}$ & 15.0 & E791 & \cite{Aitala:2000kk}\\
\hline
$\rho{}e^{\pm}\mu{}^{\mp}$ & 66.0 & E791 & \cite{Aitala:2000kk}\\
& 49.0 & CLEO II & \cite{Freyberger:1996it}\\
\hline
$\omega{}e^{\pm}\mu{}^{\mp}$ & 120.0 & CLEO II & \cite{Freyberger:1996it}\\
\hline
$K^+K^-e^{\pm}\mu{}^{\mp}$ & 180.0 & E791 & \cite{Aitala:2000kk}\\
\hline
$\phi{}e^{\pm}\mu{}^{\mp}$ & 47.0 & E791 & \cite{Aitala:2000kk}\\
& 34.0 & CLEO II & \cite{Freyberger:1996it}\\
\hline
$\overline{K}^0e^{\pm}\mu{}^{\mp}$ & 100.0 & CLEO II & \cite{Freyberger:1996it}\\
\hline
$K^-\pi{}^+e^{\pm}\mu{}^{\mp}$ & 550.0 & E791 & \cite{Aitala:2000kk}\\
\hline
$K^{*0}(892)e^{\pm}\mu{}^{\mp}$ & 100.0 & CLEO II & \cite{Freyberger:1996it}\\
& 83.0 & E791 & \cite{Aitala:2000kk}\\
\hline
$\pi{}^{\mp}\pi{}^{\mp}e^{\pm}e^{\pm}$ & 112.0 & E791 & \cite{Aitala:2000kk}\\
\hline
$\pi{}^{\mp}\pi{}^{\mp}\mu{}^{\pm}\mu{}^{\pm}$ & 29.0 & E791 & \cite{Aitala:2000kk}\\
\hline
$K^{\mp}\pi{}^{\mp}e^{\pm}e^{\pm}$ & 206.0 & E791 & \cite{Aitala:2000kk}\\
\hline
$K^{\mp}\pi{}^{\mp}\mu{}^{\pm}\mu{}^{\pm}$ & 390.0 & E791 & \cite{Aitala:2000kk}\\
\hline
$K^{\mp}K^{\mp}e^{\pm}e^{\pm}$ & 152.0 & E791 & \cite{Aitala:2000kk}\\
\hline
$K^{\mp}K^{\mp}\mu{}^{\pm}\mu{}^{\pm}$ & 94.0 & E791 & \cite{Aitala:2000kk}\\
\hline
$\pi{}^{\mp}\pi{}^{\mp}e^{\pm}\mu{}^{\pm}$ & 79.0 & E791 & \cite{Aitala:2000kk}\\
\hline
$K^{\mp}\pi{}^{\mp}e^{\pm}\mu{}^{\pm}$ & 218.0 & E791 & \cite{Aitala:2000kk}\\
\hline
$K^{\mp}K^{\mp}e^{\pm}\mu{}^{\pm}$ & 57.0 & E791 & \cite{Aitala:2000kk}\\
\hline
$pe^-$ & 10.0 & CLEO & \cite{Rubin:2009aa}\\
\hline
$\overline{p}e^+$ & 11.0 & CLEO & \cite{Rubin:2009aa}\\
\end{longtable}

\pagebreak

\begin{longtable}{l|ccc}
\caption{Upper limits at $90\%$ CL for $D^+$ decays.\label{tab:charm:rare_dplus}}\\
\hline\hline
Decay & Limit $\times10^6$ & Experiment & Reference\\
\endfirsthead
\multicolumn{4}{c}{\tablename\ \thetable{} -- continued from previous page} \\ \hline
Decay & Limit $\times10^6$ & Experiment & Reference\\
\endhead

\hline
$\pi{}^+e^+e^-$ & 110.0 & E687 & \cite{Frabetti:1997wp}\\
& 52.0 & E791 & \cite{Aitala:1999db}\\
& 5.9 & CLEO & \cite{Rubin:2010cq}\\
& 1.1 & \babar & \cite{Lees:2011hb}\\
\hline
$\pi{}^+\mu{}^+\mu{}^-$ & 220.0 & E653 & \cite{Kodama:1995ia}\\
& 89.0 & E687 & \cite{Frabetti:1997wp}\\
& 15.0 & E791 & \cite{Aitala:1999db}\\
& 8.8 & FOCUS & \cite{Link:2003qp}\\
& 6.5 & \babar & \cite{Lees:2011hb}\\
& 3.9 & D0 & \cite{Abazov:2007aj}\\
\hline
$\rho{}^+\mu{}^+\mu{}^-$ & 560.0 & E653 & \cite{Kodama:1995ia}\\
\hline
$K^+e^+e^-$ & 200.0 & E687 & \cite{Frabetti:1997wp}\\
& 3.0 & CLEO & \cite{Rubin:2010cq}\\
& 1.0 & \babar & \cite{Lees:2011hb}\\
\hline
$\pi{}^+e^{\pm}\mu{}^{\mp}$ & 34.0 & E791 & \cite{Aitala:1999db}\\
\hline
$\pi{}^+e^+\mu{}^-$ & 110.0 & E687 & \cite{Frabetti:1997wp}\\
& 2.9 & \babar & \cite{Lees:2011hb}\\
\hline
$\pi{}^+\mu{}^+e^-$ & 130.0 & E687 & \cite{Frabetti:1997wp}\\
& 3.6 & \babar & \cite{Lees:2011hb}\\
\hline
$K^+e^{\pm}\mu{}^{\mp}$ & 68.0 & E791 & \cite{Aitala:1999db}\\
\hline
$K^+e^+\mu{}^-$ & 130.0 & E687 & \cite{Frabetti:1997wp}\\
& 1.2 & \babar & \cite{Lees:2011hb}\\
\hline
$K^+\mu{}^+e^-$ & 120.0 & E687 & \cite{Frabetti:1997wp}\\
& 2.8 & \babar & \cite{Lees:2011hb}\\
\hline
$\pi{}^-e^+e^+$ & 110.0 & E687 & \cite{Frabetti:1997wp}\\
& 96.0 & E791 & \cite{Aitala:1999db}\\
& 1.9 & \babar & \cite{Lees:2011hb}\\
& 1.1 & CLEO & \cite{Rubin:2010cq}\\
\hline
$\pi{}^-\mu{}^+\mu{}^+$ & 87.0 & E687 & \cite{Frabetti:1997wp}\\
& 17.0 & E791 & \cite{Aitala:1999db}\\
& 4.8 & FOCUS & \cite{Link:2003qp}\\
& 2.0 & \babar & \cite{Lees:2011hb}\\
\hline
$\pi{}^-e^+\mu{}^+$ & 110.0 & E687 & \cite{Frabetti:1997wp}\\
& 50.0 & E791 & \cite{Aitala:1999db}\\
\hline
$\rho{}^-\mu{}^+\mu{}^+$ & 560.0 & E653 & \cite{Kodama:1995ia}\\
\hline
$K^-e^+e^+$ & 120.0 & E687 & \cite{Frabetti:1997wp}\\
& 3.5 & CLEO & \cite{Rubin:2010cq}\\
& 0.9 & \babar & \cite{Lees:2011hb}\\
\hline
$K^-\mu{}^+\mu{}^+$ & 320.0 & E653 & \cite{Kodama:1995ia}\\
& 120.0 & E687 & \cite{Frabetti:1997wp}\\
& 13.0 & FOCUS & \cite{Link:2003qp}\\
& 10.0 & \babar & \cite{Lees:2011hb}\\
\hline
$K^-e^+\mu{}^+$ & 130.0 & E687 & \cite{Frabetti:1997wp}\\
\hline
$K^{*-}(892)\mu{}^+\mu{}^+$ & 850.0 & E653 & \cite{Kodama:1995ia}\\

\end{longtable}

\begin{longtable}{l|ccc}
\caption{Upper limits at $90\%$ CL for $D_s^+$ decays.\label{tab:charm:rare_dsplus}}\\
\hline\hline
Decay & Limit $\times10^6$ & Experiment & Reference\\
\endfirsthead
\multicolumn{4}{c}{\tablename\ \thetable{} -- continued from previous page} \\ \hline
Decay & Limit $\times10^6$ & Experiment & Reference\\
\endhead

\hline
$\pi{}^+e^+e^-$ & 270.0 & E791 & \cite{Aitala:1999db}\\
& 22.0 & CLEO & \cite{Rubin:2010cq}\\
& 13.0 & \babar & \cite{Lees:2011hb}\\
\hline
$\pi{}^+\mu{}^+\mu{}^-$ & 430.0 & E653 & \cite{Kodama:1995ia}\\
& 140.0 & E791 & \cite{Aitala:1999db}\\
& 43.0 & \babar & \cite{Lees:2011hb}\\
& 26.0 & FOCUS & \cite{Link:2003qp}\\
\hline
$K^+e^+e^-$ & 1600.0 & E791 & \cite{Aitala:1999db}\\
& 52.0 & CLEO & \cite{Rubin:2010cq}\\
& 3.7 & \babar & \cite{Lees:2011hb}\\
\hline
$K^+\mu{}^+\mu{}^-$ & 140.0 & E791 & \cite{Aitala:1999db}\\
& 36.0 & FOCUS & \cite{Link:2003qp}\\
& 21.0 & \babar & \cite{Lees:2011hb}\\
\hline
$K^{*+}(892)\mu{}^+\mu{}^-$ & 1400.0 & E653 & \cite{Kodama:1995ia}\\
\hline
$\pi{}^+e^{\pm}\mu{}^{\mp}$ & 610.0 & E791 & \cite{Aitala:1999db}\\
\hline
$\pi{}^+e^+\mu{}^-$ & 12.0 & \babar & \cite{Lees:2011hb}\\
\hline
$\pi{}^+\mu{}^+e^-$ & 20.0 & \babar & \cite{Lees:2011hb}\\
\hline
$K^+e^{\pm}\mu{}^{\mp}$ & 630.0 & E791 & \cite{Aitala:1999db}\\
\hline
$K^+e^+\mu{}^-$ & 14.0 & \babar & \cite{Lees:2011hb}\\
\hline
$K^+\mu{}^+e^-$ & 9.7 & \babar & \cite{Lees:2011hb}\\
\hline
$\pi{}^-e^+e^+$ & 690.0 & E791 & \cite{Aitala:1999db}\\
& 18.0 & CLEO & \cite{Rubin:2010cq}\\
& 4.1 & \babar & \cite{Lees:2011hb}\\
\hline
$\pi{}^-\mu{}^+\mu{}^+$ & 430.0 & E653 & \cite{Kodama:1995ia}\\
& 82.0 & E791 & \cite{Aitala:1999db}\\
& 29.0 & FOCUS & \cite{Link:2003qp}\\
& 14.0 & \babar & \cite{Lees:2011hb}\\
\hline
$\pi{}^-e^+\mu{}^+$ & 730.0 & E791 & \cite{Aitala:1999db}\\
\hline
$K^-e^+e^+$ & 630.0 & E791 & \cite{Aitala:1999db}\\
& 17.0 & CLEO & \cite{Rubin:2010cq}\\
& 5.2 & \babar & \cite{Lees:2011hb}\\
\hline
$K^-\mu{}^+\mu{}^+$ & 590.0 & E653 & \cite{Kodama:1995ia}\\
& 180.0 & E791 & \cite{Aitala:1999db}\\
& 13.0 & FOCUS & \cite{Link:2003qp}\\
\hline
$K^-e^+\mu{}^+$ & 680.0 & E791 & \cite{Aitala:1999db}\\
\hline
$K^{*-}(892)\mu{}^+\mu{}^+$ & 1400.0 & E653 & \cite{Kodama:1995ia}\\

\end{longtable}

\begin{longtable}{l|ccc}
\caption{Upper limits at $90\%$ CL for $\Lambda_c^+$ decays.\label{tab:charm:rare_lambdac}}\\
\hline\hline
Decay & Limit $\times10^6$ & Experiment & Reference\\
\endfirsthead
\multicolumn{4}{c}{\tablename\ \thetable{} -- continued from previous page} \\ \hline
Decay & Limit $\times10^6$ & Experiment & Reference\\
\endhead

\hline
$pe^+e^-$ & 5.5 & \babar & \cite{Lees:2011hb}\\
\hline
$p\mu{}^+\mu{}^-$ & 340.0 & E653 & \cite{Kodama:1995ia}\\
& 44.0 & \babar & \cite{Lees:2011hb}\\
\hline
$\sigma{}^+\mu{}^+\mu{}^-$ & 700.0 & E653 & \cite{Kodama:1995ia}\\
\hline
$pe^+\mu{}^-$ & 9.9 & \babar & \cite{Lees:2011hb}\\
\hline
$p\mu{}^+e^-$ & 19.0 & \babar & \cite{Lees:2011hb}\\
\hline
$\overline{p}e^+e^+$ & 2.7 & \babar & \cite{Lees:2011hb}\\
\hline
$\overline{p}\mu{}^+\mu{}^+$ & 9.4 & \babar & \cite{Lees:2011hb}\\

\end{longtable}

\clearpage
\input{tau/tau_all.tex}
 

\def\babar{\mbox{\slshape B\kern-0.1em{\smaller A}\kern-0.1em
    B\kern-0.1em{\smaller A\kern-0.2em R}}\xspace}

\clearpage

\section{Summary}
\label{sec:summary}

This article provides updated world averages for 
$b$-hadron properties using results available before the end of 2011. 
In some sections, results that appeared before the end of April 2012 are also included.

\begin{table}
\caption{Selected world averages 
from Chapters~\ref{sec:life_mix} and~\ref{sec:cp_uta}.}
\label{tab_summary1}
\renewcommand{\arraystretch}{1.15}
\begin{center}
\begin{tabular}{|l|c|}
\hline
 {\bf\boldmath \b-hadron lifetimes} &   \\
 ~~$\tau(\Bd)$  & \hfagTAUBD \\
 ~~$\tau(\Bu)$  & \hfagTAUBU \\
 ~~$\bar{\tau}(\Bs) = 1/\Gs$  & \hfagTAUBSMEANC \\
 ~~$\tau(\Bc)$  & \hfagTAUBC \\
 ~~$\tau(\Lb)$  & \hfagTAULB \\
\hline
 {\bf\boldmath \b-hadron fractions} &   \\
 ~~$f^{+-}/f^{00}$ in \Ups decays  & \hfagFF \\ 
 ~~\fBs in \Upsfive decays & \hfagFSFIVE \\
 ~~\fBs, \fbb in $Z$ decays & \hfagZFBS, \hfagZFBB \\
 ~~\fBs, \fbb at Tevatron & \hfagTFBS, \hfagTFBB \\
\hline
 {\bf\boldmath \Bd\ and \Bs\ mixing / \cpv\ parameters} &   \\
 ~~\dmd &  \hfagDMDWU \\
 ~~$|q/p|_{\particle{d}}$ & \hfagQPDB  \\
 ~~\dms  &  \hfagDMS \\
 ~~$\DGs = \Gamma_{\rm L} - \Gamma_{\rm H}$ & \hfagDGSCON \\
 ~~$|q/p|_{\particle{s}}$ & \hfagQPS   \\
 ~~\phiccbars  & \hfagPHISCOMB \\
\hline
{\bf Measurements related to Unitarity Triangle angles} & \\
 ~~ $\stwob \equiv \sin\! 2\phi_1$ & $0.679 \pm 0.020$ \\
 ~~ $\beta \equiv \phi_1$          & $\left( 21.4 \pm 0.8 \right)^\circ$ \\
 ~~ $-\etacp S_{\phi \KS}$       & $0.74\,^{+0.11}_{-0.13}$ \\
 ~~ $-\etacp S_{\etapr \Kz}$       & $0.59 \pm 0.07$ \\
 ~~ $-\etacp S_{\KS \KS \KS}$       & $0.72 \pm 0.19$ \\
 ~~ $-\etacp S_{\Kp \Km \KS}$       & $0.68\,^{+0.09}_{-0.10}$ \\
 ~~ $-\etacp S_{\jpsi \piz}$       & $0.93 \pm 0.15$ \\
 ~~ $S_{K^* \gamma}$       & $-0.16 \pm 0.22$ \\
 ~~ $S_{\pi^+\pi^-}$               & $-0.65 \pm 0.07$ \\  
 ~~ $C_{\pi^+\pi^-}$               & $-0.36 \pm 0.06$ \\  
 ~~ $S_{\rho^+\rho^-}$       & $-0.05 \pm 0.17$ \\
 ~~ $a(D^{*\pm}\pi^{\mp})$       & $-0.039 \pm 0.010$ \\
 ~~ $A^{}_{CP}(B\ra D^{}_{CP+}K)$       & $0.19 \pm 0.03$ \\
 ~~ $A_{\rm ADS}(B\ra D^{}_{K\pi}K)$       & $-0.54 \pm 0.12$ \\
 ~~ $R_{\rm ADS}(B\ra D^{}_{K\pi}K)$       & $0.0153 \pm 0.0017$ \\
\hline
\end{tabular}
\end{center}
\end{table}
\begin{table}
\caption{Selected world averages at the end of 2011
from Chapters~\ref{sec:slbdecays}--\ref{sec:rare}.}
\label{tab_summary2}
\renewcommand{\arraystretch}{1.15}
\begin{center}
\begin{tabular}{|l|c|}
\hline
{\bf\boldmath Semileptonic \B decay parameters} & \\
 ~~${\cal B}(\Bzb\to D^{*+}\ell^-\nub)$ & $(4.95\pm 0.11)\%$\\
 ~~${\cal B}(\B^-\to D^{*0}\ell^-\nub)$ & $(5.70\pm 0.19)\%$\\
 ~~${\cal F}(1)\vcb$ & $(35.90\pm 0.45)\times 10^{-3}$\\
 ~~$\vcb$ from $\bar B\to D^*\ell^-\bar\nu_\ell$ & $(39.54\pm
 0.50_{\rm exp}\pm 0.74_{\rm th})\times 10^{-3}$\\
\hline
 ~~${\cal B}(\Bzb\to D^+\ell^-\nub)$ & $(2.18\pm 0.12)\%$\\
 ~~${\cal B}(\B^-\to D^0\ell^-\nub)$ & $(2.26\pm 0.11)\%$\\
 ~~${\cal G}(1)\vcb$ & $(42.64 \pm 1.53)\times 10^{-3}$\\
 ~~$\vcb$ from $\bar B\to D\ell^-\bar\nu_\ell$ & $(39.70\pm 1.42_{\rm
 exp}\pm 0.89_{\rm th})\times 10^{-3}$\\
\hline
 ~~${\cal B}(\bar B\to X_c\ell^-\bar\nu_\ell)$ & $(10.51\pm 0.13)\%$\\
 ~~${\cal B}(\bar B\to X\ell^-\bar\nu_\ell)$ & $(10.72\pm 0.13)\%$\\
 ~~$\vcb$ from $\bar B\to X\ell^-\bar\nu_\ell$ & $(41.88\pm
 0.73)\times 10^{-3}$\\
\hline
 ~~${\cal B}(\Bb\to\pi\ell^-\nub)$ & $(1.42\pm 0.05)\times 10^{-4}$\\
 ~~$\vub$ from $\Bb\to\pi\ell^-\nub$ & $(3.23\pm 0.30)\times
 10^{-3}$\\
 ~~$\vub$ from $\Bb\to X_u\ell^-\nub$ & $(4.40\pm 0.15_{\rm exp}\pm
 0.20_{\rm th})\times 10^{-3}$\\
\hline
{\bf\boldmath Rare \B decays} &   \\
~~ ${\cal B}(B \to X_s \gamma)$ & $(3.55 \pm 0.24 \pm 0.09) \times 10^{-4}$ \\
~~ ${\cal B}(\Bp \to \tau^+ \nu)$ & $(1.67 \pm 0.30) \times 10^{-4}$ \\
~~ $A_{\rm FB}(\Bz \to K^{*0}\mu^+\mu^-)$ in bins of $q^2 = m^2(\mu^+\mu^-)$ & see Table~\ref{tab:Kstarmumu-Afb} \\
~~ ${\cal B}(\Bs \to \mu^+\mu^-)$ & $<1.2 \times 10^{-8}$ (90\,\% C.L.) \\
~~$A_{\CP}(\particle{\Bd\to K^+\pi^-})$ & $(-0.087 \pm 0.008)$\\
~~$A_{\CP}(\particle{B^+\to K^+\pi^0})$ & $(0.037 \pm 0.021)$ \\
~~$A_{\CP}(\particle{\Bs\to K^-\pi^+})$ & $(0.29 \pm 0.07)$ \\
\hline
\end{tabular}
\end{center}
\end{table}
\begin{table}
\caption{Selected world averages at the end of 2011
from Chapters~\ref{sec:charm_physics} and~\ref{sec:tau}.}
\label{tab_summary3}
\renewcommand{\arraystretch}{1.15}
\begin{center}
\begin{tabular}{|l|c|}
\hline
 {\bf\boldmath $D^0$ mixing and \cpv\ parameters} &   \\
 ~~$x$ &  $(0.63\,^{+0.19}_{-0.20})\%$  \\
 ~~$y$ &  $(0.75\,\pm 0.12)\%$  \\
 ~~$A^{}_D$ &  $(-1.7\,\pm 2.4)\%$  \\
 ~~$|q/p|$ & $0.88\,^{+0.18}_{-0.16}$  \\
 ~~$\phi$ &  $(-10.1\,^{+9.5}_{-8.9})^\circ$  \\
\hline
 ~~$x^{}_{12}$ (no direct \cpv) &  $(0.62\,\pm 0.19)\%$  \\
 ~~$y^{}_{12}$ (no direct \cpv) &  $(0.75\,\pm 0.12)\%$  \\
 ~~$\phi^{}_{12}$ (no direct \cpv) &  $(4.9\,^{+7.7}_{-6.5})^\circ$  \\
\hline
~~$a^{\rm ind}_{CP}$ & $(-0.02 \pm 0.23)\%$ \\
~~$\Delta a^{\rm dir}_{CP}$ & $(-0.66 \pm 0.15)\%$ \\
\hline
{\bf\boldmath $\tau$ parameters, Lepton Universality, and $|V_{us}|$} &   \\
 ~~ $g^{}_\mu/g^{}_e$ & \quantgmubygeUtau \\
 ~~ $g^{}_{\tau}/g^{}_{\mu}$ & \quantgtaubygmuUtau \\
 ~~ $g^{}_{\tau}/g^{}_{e}$ & \quantgtaubygeUtau \\
 ~~ ${\cal B}_e^{\text{uni}}$ & $(\quantBeUuniv)\%$ \\
 ~~ $R_{\text{had}}$ & \quantRUtau \\
 ~~ $|V_{us}|$ from ${\cal{B}}(\tau^-\to K^-\nu^{}_\tau)$ &  \quantVusUtauKnu \\
 ~~ $|V_{us}|$ from ${\cal{B}}(\tau^- \to K^-\nu^{}_\tau)/ {\cal{B}}(\tau^- \to \pi^-\nu^{}_\tau)$ & \quantVusUtauKpi \\ 
 ~~ $|V_{us}|$ from inclusive sum of strange branching fractions & \quantVus \\
 ~~ $|V_{us}|$ tau average & \quantVusUtau \\
\hline
\end{tabular}
\end{center}
\end{table}

Concerning $b$-hadron lifetime and mixing averages,
the most significant changes in the past two years
are due to new results from the CDF, \dzero and LHCb experiments, 
mainly in the \Bs sector. While the Tevatron 
experiments have updated some of their analyses with the 
full Run II data sample, LHCb has just entered the game 
and is taking the lead already 
with results based on the 2010--2011 data samples collected 
at the LHC.
While the updated \dzero like-sign dimuon asymmetry 
still deviates from the Standard Model prediction
(with a significance increased to $3.9\,\sigma$), 
there is still no evidence of \CP violation in either 
\Bd or \Bs mixing, with precisions on the semileptonic asymmetries 
reaching below the 1\% level. 
However, the most impressive progress was achieved in the 
analysis of $\Bs \to\jpsi\phi$ decays, 
where new or significantly improved results became recently 
available from CDF, \dzero and LHCb. 
The non-zero decay width difference in the $\Bs-\Bsbar$ system 
is now firmly established, with a relative difference of
$(14\pm2)\%$.
Its sign has also been determined by LHCb: 
the heavy state of the  $\Bs-\Bsbar$ system lives longer 
than the light state, as expected in the Standard Model.
In contrast, and despite the 
recent efforts from Belle, the relative decay width difference in 
the $\Bd-\Bdbar$ system,
which has momentarily reached a slightly better absolute precision,
is still consistent with zero. 
One the other hand, a quantum step
has been achieved in the measurement of 
mixing-induced \CP violation in \Bs decays proceeding through 
the $b\to c\bar{c}s$ transition: the corresponding weak phase 
has been pinned down to a precision below 0.1 radian 
and is so far compatible with the Standard Model expectation.

The measurement of $\sin 2\beta \equiv \sin 2\phi_1$ from $b \to
c\bar{c}s$ transitions such as $\Bz \to \jpsi\KS$ has reached $<3\,\%$
precision: $\sin 2\beta \equiv \sin 2\phi_1 = 0.679 \pm 0.020$.
Measurements of the same parameter using different quark-level processes
provide a consistency test of the Standard Model and allow insight into
possible new physics.  Recent improvements include the use of
time-dependent Dalitz plot analyses of $\Bz \to \KS\Kp\Km$ and $\Bz \to
\KS\pip\pim$ to obtain \CP\ violation parameters for $\phi\KS$,
$f_0(980)\KS$ and $\rho\KS$.  All results among hadronic $b \to s$
penguin dominated decays are currently consistent with the Standard
Model expectations.  Among measurements related to the Unitarity
Triangle angle $\alpha \equiv \phi_2$, results from the
$\rho\rho$ system allow constraints at the level of $\approx
6^\circ$.  Knowledge of the third angle $\gamma \equiv \phi_3$ also
continues to improve.  Notwithstanding the well-known statistical issues
in extracting the value of the angle itself, the world average values of
the parameters in $B \to DK$ decays now show significant direct 
\CP\ violation effects.

Regarding semileptonic $B$~meson decays, the $B$ factories Belle
and \babar\ continue to dominate the field and a number of results
have appeared since the last update. Semileptonic decays remain a
focus of interest for theorists: New lattice QCD and light-cone sum
rule results help to understand exclusive transitions. Inclusive
semileptonic decays are understood at full ${\cal
O}(\alpha^2_s)$. Still, the experimental situation is not satisfactory:
While inclusive and exclusive determinations of $\vcb$ agree at the
level of $2\sigma$, inclusive and exclusive measurements of $\vub$
differ by three standard deviations. Clearly more effort on the
experimental and theory side is required in the future.

The most important new measurements of rare decays are coming from the
LHC.  CMS and LHCb both have restrictive limits for the decays
$B\to\mumu$ and $B_s\to\mumu$.  The sensitivity is approaching the SM
expectations with no significant signals seen yet.  LHCb has already
produced many other results on a wide variety of decays as indicated in 
the tables in Sec.~\ref{sec:rare}.  Belle and \babar\ continue to
produce new results though their rates are dwindling.  It will still
be some years before we see new results from upgraded $B$ factories.

Many $b$ to charm results from LHCb are included in our report for the  
first time this year, combining with
results from \babar, Belle and CDF to yield a total of 632  
measurements reported in 216 papers.
The huge combined sample of $b$ hadrons allows measurements of decays  
to states with open or hidden charm
content with unprecedented precision.

In the charm sector, $D^0$-$\dbar$ mixing is now well-established.
Measurements of 38 separate observables from five experiments are input 
into a global fit for 10 underlying parameters, and the no-mixing 
hypothesis is excluded at a confidence level corresponding to 
$10.2\sigma$. The mixing parameters
$x$ and $y$ (see Table~\ref{tab_summary3}) differ from zero by 
$2.7\sigma$ and $6.0\sigma$, respectively. The central values are 
consistent with mixing arising from long-distance processes, as
predicted by theory; thus it will probably be difficult to identify 
new physics from mixing alone. The WA value for the observable \ycp\ 
is positive, which indicates that the \cp-even state is shorter-lived 
as in the $K^0$-$\kbar$ system. However, $x$ also appears to be 
positive, which implies that the \cp-even state is heavier, 
unlike in the $K^0$-$\kbar$ system. 
%
In the $D^0$-$\dbar$ system, 
there is no evidence for \cpv\ arising from mixing ($|q/p|\neq 1$) or 
from a phase difference between the mixing amplitude and 
a direct decay amplitude ($\phi\neq 0$). However, both the LHCb 
and CDF experiments have obtained evidence for direct
\cpv\ in $D^0\ra K^+K^-$ and $D^0\ra \pi^+\pi^-$ decays. These
experiments measure nonzero values for the {\it difference\/} 
in direct \cpv\ between $K^+K^-$ and $\pi^+\pi^-$ modes, which 
requires that direct \cpv\ exists in at least
one of them. Inputting these measurements into a global 
fit and also including measurements from Belle and \babar\ 
gives $\Delta a^{\rm dir}_{CP}\neq 0$ with a significance 
greater than $4\sigma$.

Concerning tau decays, in this report we include three new 
tau branching fraction measurements from the $B$-factories, 
and we provide more information on the tau branching fraction 
fit. The \Vus calculation uses now a more complete set of tau
branching fractions to strange final states, and thanks primarily to
improvements in QCD lattice predictions, two tau determinations of \Vus
have reduced errors. For the first time, we compute an average of all \Vus
determinations with tau data.

\section{Acknowledgments}

We are grateful for the strong support of the \belle, 
\babar, CLEO, CDF, \dzero\ and LHCb collaborations, without whom 
this compilation of results and world averages would not have 
been possible. The success of these experiments in turn would 
not have been possible without the excellent operations of the 
KEKB, PEP-II, CESR, Tevatron and LHC accelerators, and fruitful 
collaborations between the accelerator groups and the experiments.

Our averages and this compilation have benefitted greatly from 
contributions to the Heavy Flavor Averaging Group from numerous
individuals. We especially thank David Kirkby, Yoshihide Sakai, 
Simon Eidelman, Soeren Prell and Gianluca Cavoto for their
past leadership of HFAG. 
We are grateful to Paolo Gambino for assistance with averages that appear in Chapter~\ref{sec:slbdecays},
to David Asner, David Cassel and Milind Purohit 
for significant contributions to Chapter~\ref{sec:charm_physics}, 
and to Michel Davier for providing valuable input to Chapter~\ref{sec:tau}.

\clearpage


\bibliographystyle{HFAGutphys}
\raggedright
\bibliography{EndOfYear11,life_mix,cp_uta,slbdecays/slb_ref,rare/rare_refs,charm/charm_refs,tau/pdg11-tau-refs,tau/hfag-tau-refs}

\begin{mcitethebibliography}{100}
\bibitem{Aaltonen:2011qsa_cont}
 we consider that these results supersede those from the old CDF note
  7386~\cite{CDFnote7386:2005}, although the lifetime analysis of one of the
  modes ($B_s \to D_s\pi\pi\pi$) has not been updated\relax
\mciteBstWouldAddEndPuncttrue
\mciteSetBstMidEndSepPunct{\mcitedefaultmidpunct}
{\mcitedefaultendpunct}{\mcitedefaultseppunct}\relax
\EndOfBibitem
\end{mcitethebibliography}

\end{document}